\documentclass[12pt,a4paper,openany]{book}
\usepackage[utf8x]{inputenc}
\usepackage{ucs}
\usepackage{amsmath}
\usepackage{xspace} 
\usepackage{cite}
\usepackage{amsfonts}
\usepackage{amssymb}
\usepackage[T1]{fontenc}
\usepackage{graphicx}
\usepackage{slashed}
\usepackage[nohints]{minitoc}
\usepackage[titletoc]{appendix}
\usepackage{pdftexcmds}
\usepackage[colorlinks=true,linkcolor=blue,citecolor=blue,urlcolor=blue,hypertexnames=false]{hyperref}
\usepackage{csquotes}
\usepackage{float} 
\usepackage[rightcaption]{sidecap}
\usepackage{xstring}
\usepackage{color}
\usepackage{ifthen}
\usepackage{etoolbox}
\usepackage{enumerate}

\sidecaptionvpos{figure}{c}

\providecommand{\abs}[1]{\left|#1\right|}
\providecommand{\ep}[1]{{\rm e}^{#1}}

\providecommand{\Texp}[0]{\mathcal{T}\!\!\exp}
\providecommand{\bk}[0]{\mathbf{k}}
\providecommand{\bx}[0]{\mathbf{x}}
\providecommand{\bz}[0]{\mathbf{z}}
\providecommand{\bp}[0]{\mathbf{p}}
\providecommand{\bv}[0]{\mathbf{v}}
\newcommand{\bP}[0]{\mathbf{P}}

\makeatletter
\newcounter{subsubsubsection}[subsubsection]
\renewcommand\section{\@startsection {section}{1}{\z@}%
                                   {-3.5ex \@plus -1ex \@minus -.2ex}%
                                   {2.3ex \@plus.2ex}%
                                   {\normalfont\Large\bfseries}}
\renewcommand\subsection{\@startsection{subsection}{2}{\z@}%
                                     {-3.25ex\@plus -1ex \@minus -.2ex}%
                                     {1.5ex \@plus .2ex}%
                                     {\normalfont\large\bfseries}}
\renewcommand\subsubsection{\@startsection{subsubsection}{3}{\z@}%
                                     {-3.25ex\@plus -1ex \@minus -.2ex}%
                                     {1.5ex \@plus .2ex}%
                                     {\normalfont\normalsize\em\underline}}
\newcommand{\subsubsubsection}{\@startsection{subsubsubsection}{4}{\z@}%
                          {-3.25ex\@plus -1ex \@minus -.2ex}%
                          {1.5ex \@plus .2ex}%
                          {\normalfont\normalsize\em}}

\renewcommand{\thesection}{\@Roman\c@section }
\renewcommand{\thesubsection}{\@Alph\c@subsection }
\renewcommand{\thesubsubsection}{\@arabic\c@subsubsection }
\renewcommand{\thesubsubsubsection}{\@alph\c@subsubsubsection }

\newcommand*{\l@subsubsubsection}{\@dottedtocline{3}{10em}{5em}}
\newcommand*{\subsubsubsectionmark}[1]{}
\def\toclevel@subsubsubsection{4}

\newcommand\ifnumber[1]{%
        \begingroup
        \edef\temp{#1}%
        \expandafter\ifstrempty\expandafter{\temp}
                {\endgroup\@secondoftwo}
                {\expandafter\ifnumber@i\temp\@nnil}%
}
\def\ifnumber@i#1#2\@nnil{%
        \if-#1%
                \ifstrempty{#2}
                        {\def\temp{X}}
                        {\def\temp{#2}}%
        \else
                \def\temp{#1#2}%
        \fi
        \afterassignment\ifnumhelper
        \count@0\temp\relax\@nnil
        \endgroup
}

\def\numrelax{\relax}%
\def\ifnumhelper#1\@nnil{%
        \def\temp{#1}%
        \ifx\temp\numrelax
                \aftergroup\@firstoftwo
        \else
                \aftergroup\@secondoftwo
        \fi
}

\def\p@part{Part.~}
\def\thearchapter{\@arabic\c@chapter}
\def\chaporapp{\ifnum\pdf@strcmp{\thechapter}{\thearchapter}=\z@ Chap.\nobreakspace  {}\else App.\nobreakspace  {}\fi}
\def\p@chapter{\chaporapp}
\def\p@section{\chaporapp\thechapter,\ Sec.\nobreakspace  {}}
\def\p@subsection{\chaporapp\thechapter,\ Sec.\nobreakspace  {}\thesection.\,}
\def\p@subsubsection{\chaporapp\thechapter,\ Sec.\nobreakspace  {}\thesection.\,\thesubsection.\,}
\renewcommand{\@seccntformat}[1]{\csname the#1\endcsname.\quad}

\def\@firstoffivecutted#1#2#3#4#5{\StrDel[1]{#1}{\chaporapp\thechapter,\ }}
\RequirePackage{nameref}
\newcommand{\reftempxav}[1]{%
     \StrLeft{\chaporapp\thechapter,\ }{14}[\firstchar]%
     \@safe@activestrue%
     \StrLeft{\expandafter\real@setref\csname r@#1\endcsname\@firstoffive{#1}}{14}[\secondchar]%
     \@safe@activesfalse%
     \IfStrEq%
          {\firstchar}%
          {\secondchar}%
          {\@safe@activestrue
               \expandafter\@setref\csname r@#1\endcsname\@firstoffivecutted{#1}%
               \@safe@activesfalse}%
          {\@safe@activestrue%
               \expandafter\@setref\csname r@#1\endcsname\@firstoffive{#1}%
               \@safe@activesfalse}%
}

\newcommand{\partimage}[1]{\gdef\@partimage{#1}}
\newcommand{\partcitation}[1]{\gdef\@partcitation{#1}}
\renewcommand\part{%
  \if@openright
    \cleardoublepage
  \else
    \clearpage
  \fi
  \thispagestyle{plain}%
  \if@twocolumn
    \onecolumn
    \@tempswatrue
  \else
    \@tempswafalse
  \fi
  \null\vfill
  \secdef\@part\@spart}
\def\@part[#1]#2{%
    \ifnum \c@secnumdepth >-2\relax
      \refstepcounter{part}%
      \addcontentsline{toc}{part}{\thepart\hspace{1em}#1}%
    \else
      \addcontentsline{toc}{part}{#1}%
    \fi
    \markboth{}{}%
    {\flushright 
\begin{minipage}{0.37\linewidth}
              \small\@partcitation
\end{minipage} 
\vfill
    \centering
     \interlinepenalty \@M
     \normalfont
     \ifnum \c@secnumdepth >-2\relax
       \huge\bfseries \partname\nobreakspace\thepart
       \par
       \vskip 20\p@
     \fi
     \Huge \bfseries #2 %
          \vskip 80\p@
          \includegraphics[height=250\p@]{\@partimage} \par \vfill}%
    \@endpart}

\newcommand\fauxchap{\global\@topnum\z@
                    \@afterindentfalse
                    \secdef\@chapter\@schapter}

\makeatother

\AtBeginDocument{\DeclareRobustCommand\ref{\reftempxav}}

\newcommand{\hiddensubsubsection}[1]{
    \stepcounter{subsubsection}
    \subsubsection*{\thesubsubsection.\hspace{1em}{#1}}
}

\usepackage[Lenny]{fncychap}
\begin{document}

\dominitoc

\frontmatter 
\thispagestyle{empty}
\begin{figure}[H] 
\begin{minipage}{1\linewidth}
\begin{center}
\includegraphics[height=2.5cm]{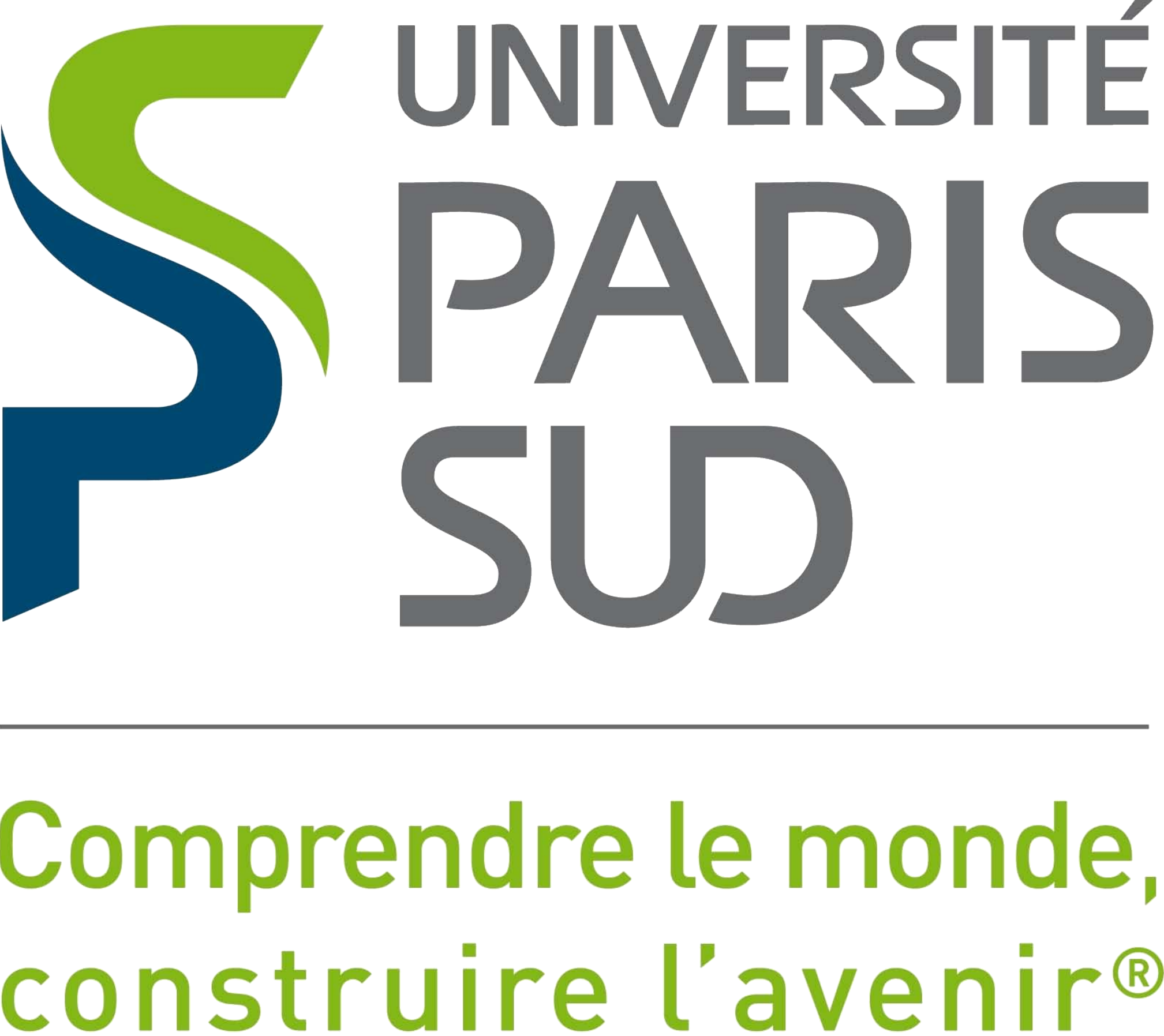} \qquad\quad
\includegraphics[height=2.5cm]{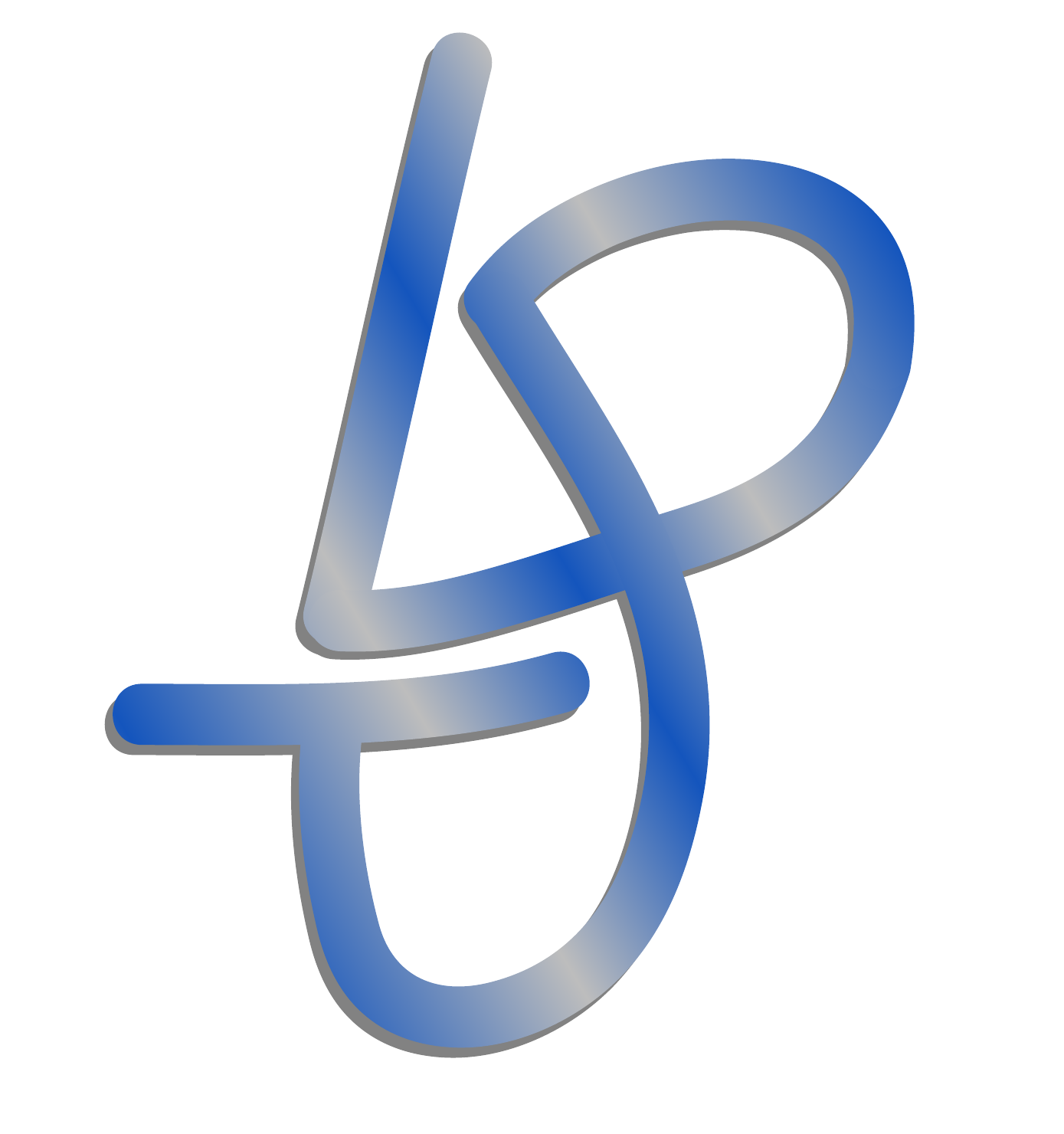} \qquad\quad
\includegraphics[height=2.5cm]{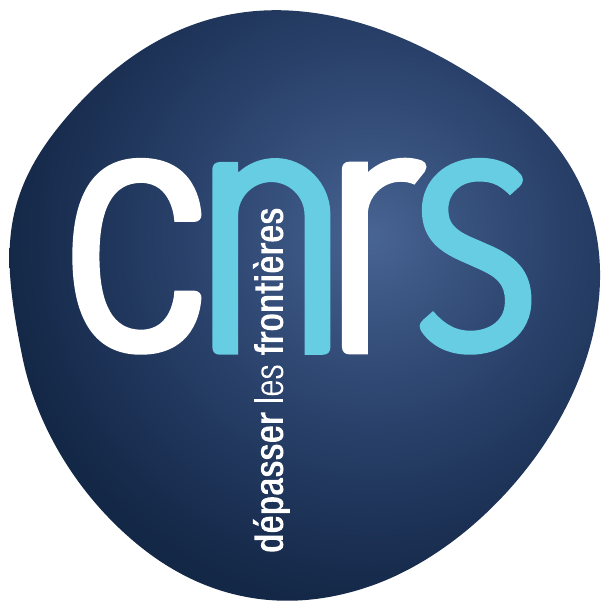} \qquad\quad
\includegraphics[height=2.5cm]{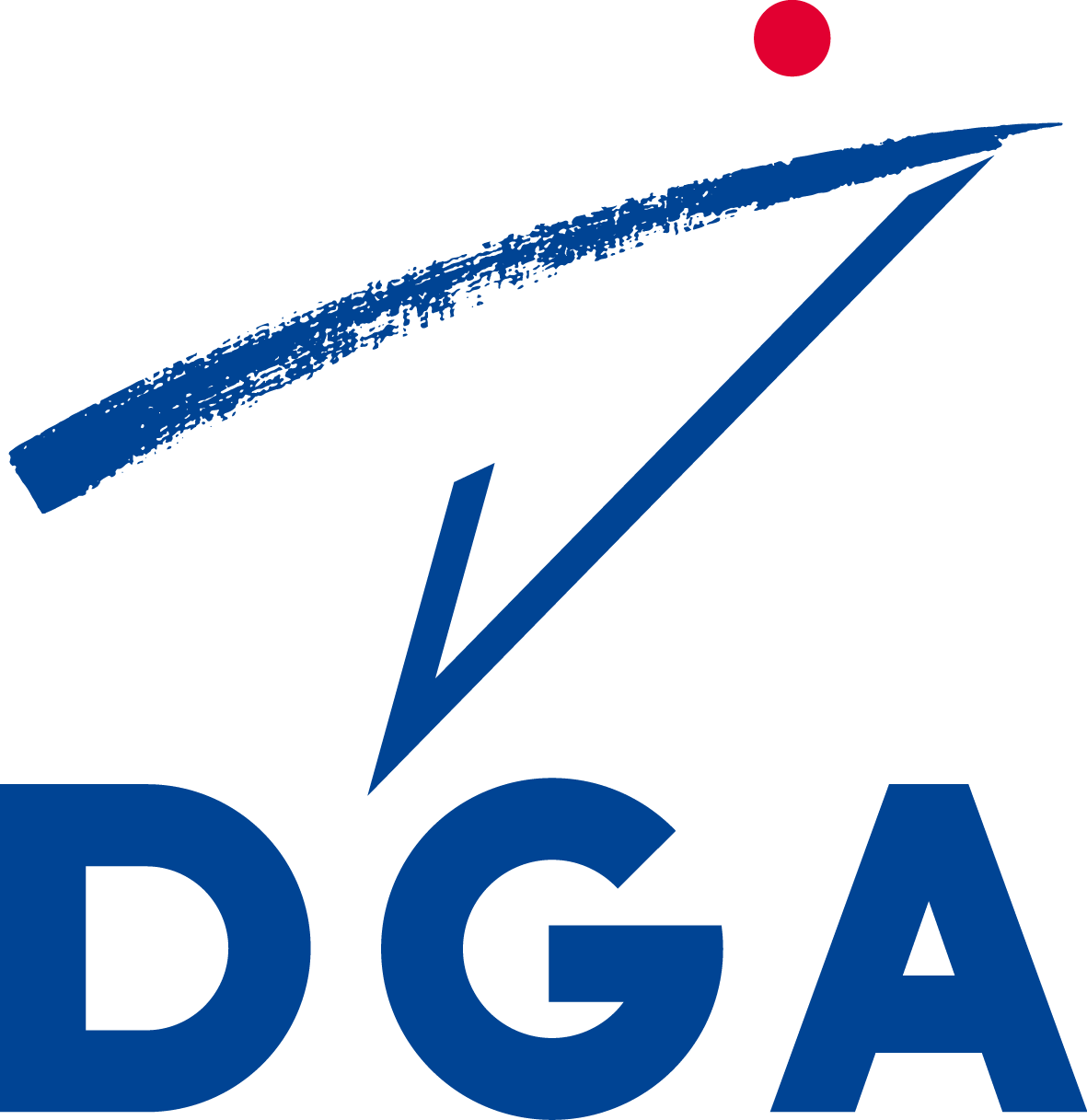} 
\end{center} \end{minipage} \end{figure}
\bigskip
\begin{center} \textbf{PhD thesis} \end{center} \medskip
\begin{center} \textbf{UNIVERSITY PARIS-SUD} \end{center} \medskip
\begin{center} \'Ecole doctorale : ED 107 (Physique de la R\'egion Parisienne) \end{center} \medskip
\begin{center} Laboratoire de Physique th\'eorique (LPT) \end{center} \medskip
\begin{center} Discipline : Physique \end{center} \medskip
\bigskip\bigskip\bigskip\bigskip
\begin{center} \textbf{\textit{\LARGE{Dispersive and dissipative effects in quantum field theory in curved space-time to model condensed matter systems}}} \end{center}
\bigskip\bigskip\bigskip\bigskip\bigskip\bigskip
\begin{center} by \end{center} \medskip
\begin{center} \textbf{Xavier Busch} \end{center}
\bigskip\bigskip\bigskip\bigskip
\begin{center} Defended on September $26^{th}, 2014$ before the jury : \end{center}
$$\begin{array}{l l}
\text{Theodore A. Jacobson} & \text{Jury member} \\
\text{Ulf Leonhardt} & \text{Referee} \\
\text{Stefano Liberati} & \text{Referee} \\
\text{Renaud Parentani} & \text{PhD advisor} \\
\text{Daniele Sanvitto} & \text{Jury member} \\
\text{Chris Westbrook} & \text{Jury member} \\
\end{array}$$

\newpage\thispagestyle{plain}
\null
\vfill
\begin{flushright}
\begin{minipage}{0.57\linewidth}
Thèse préparée dans le cadre de l'Ecole Doctorale 107 au Laboratoire de Physique Théorique d'Orsay (UMR 8627) Bât. 210, Université Paris-Sud 11, 91405 Orsay Cedex
\end{minipage}
\end{flushright}

\newpage\thispagestyle{plain}\chapter*{Notations}
\phantomsection
\addstarredchapter{Notations}

In this thesis we shall use the following conventions and notations:

\hiddensubsubsection{General relativity}
\begin{itemize}
\item
The signature of space-time is $(-,+,+,+)$.

\item
The metric is noted $\mathtt{g_{\mu \nu}}$ and the covariant derivative is $\nabla_\mu$. A point in space-time is $\mathtt{x}$. The volume element is $d \mathtt{x}\sqrt{\mathtt{-g}}$.

\item
Implicit summation is made on repeated indexes. It runs on space and time for Greek variables ($\mu, \nu \cdots$), on space only for Latin variables ($i,j,\cdots$)

\end{itemize}

\hiddensubsubsection{Quantum mechanics}

\begin{itemize}
\item
Expectation value of an operator $\hat O$ in a mixed state $\hat \rho$ is noted $\left <\hat O\right >_{\hat\rho} \doteq {\rm Tr }\left ( \hat \rho \hat O  \right )$, or when no confusion can be made, only $\left <\hat O\right >$. For pure states $\left | \Psi \right >$, the state will be explicitly put in the formula as $\left < \Psi \right | \hat O\left | \Psi \right >$.

\item
Hamiltonian is $\hat H$ in Schrödinger representation and $\tilde H$ in Heisenberg representation.

\end{itemize}

\hiddensubsubsection{Miscellaneous}

\begin{itemize}
\item
Bold notations designate vectors (in space only). In one dimension, it designates algebraic quantity.

\item
Commutators and anticommutators are defined by $[a,b] = ab-ba$,  $\{ a,b\} = \frac{ab + ba}{2}$.

\item
Fourier transformation is normalized as $f(x) = \int \frac{dk}{\sqrt{2 \pi}} \ep{ i k x} f(k)$, and when integrating over functions depending only on $x-x'$, we have the coherent normalization $ \int\frac{ dx dx'}{2\pi} \ep{i k x - i k' x'} f(x-x') = \int d(x-x') \ep{i k (x- x')}f(x-x') \delta(k-k')$

\end{itemize}

\tableofcontents

\mainmatter

\chapter*{Introduction}
\phantomsection
\addstarredchapter{Introduction}
\markboth{\MakeUppercase{introduction}}{\MakeUppercase{introduction}}
\label{chap:introen}

During the \textsc{\romannumeral 20}\textsuperscript{th}~century, two fundamental theories were proposed. They describe the world in a dual and apparently incompatible way. These are quantum mechanics and general relativity. Their merging could provide us a unified theory of the universe. Unfortunately, Planck scale --i.e., the scale built with the fundamental constants of quantum mechanics and general relativity-- physics is today experimentally out of reach. However, quantum particles propagating in a classical gravitational field is a first step towards this theory. Such a semi-classical theory is similar to what was done in the beginning of quantum mechanics, when quantum electrons were propagating in a classical electromagnetic field. Such a theory was necessarily incomplete but gave birth to many interesting features and in particular to the laser effect.

During the early years of studying quantum particles in the presence of classical gravitation, one of the most counter intuitive phenomenon that was predicted was
the Hawking radiation~\cite{Hawking:1974sw}. It comes from the creation of pairs of quantum particles out of vacuum in the vicinity of an event horizon, which delimits a black hole. Instead of annihilating, as it would be the case in the absence of a black hole, it happens that one of the two particles has a positive energy and is emitted towards infinity, while the other one has a negative energy (total energy is conserved) and falls into the black hole. The black hole thus effectively emits a thermal flux of particles (if these are massive, the ones with insufficient energy fall back into the black hole, while the others reach infinity).

It turns out that this phenomenon is not specific to black holes. Indeed, as was shown by W. Unruh in 1976~\cite{Unruh:1976db}, if a particle detector is placed in vacuum and accelerated, it will see its environment as if it were thermally populated. Moreover, if a second particle detector is placed at rest, it will interpret each particle detection of the first detector as a particle emission. This phenomenon can be understood quite as Hawking radiation: the vacuum contains virtual pairs of particles. The one of negative rest-energy is absorbed by the detector (in his frame, this particle have positive energy) and the one of positive rest-energy is emitted towards infinity.

A third similar example is pair particle production~\cite{Parker:1969au} in a non stationary space-time (e.g., in expansion). The modulations of the geometry brings energy that allows virtual pairs of particles present in the vacuum to become real. Contrary to the previous cases, in this case, both particles have a positive energy.

From an experimental point of view, the three effects are far out of range. For example, a black hole emitting a lot of particles would emit a thermal flux with a temperature of the order of $\mu K$. It is thus indistinguishable from the cosmological background at a temperature of $2.7K$. For the Unruh effect, given a humanly bearable acceleration, the expected temperature should be of the order of $10^{-20} K$. For a proton at the Large Hadron Collider, taking into account the acceleration in the RF cavities (with field strength of $5 MV/m$), it should be of the order of a few $\mu K $.

The possibility of experimentally verifying these theories was raised by W. Unruh in 1981~\cite{Unruh:1980cg}. He noticed that sound waves in (non-rotational) fluid are driven by the same equations of motion as a scalar field in curved space-time. If the fluid possesses a sonic horizon, (place separating subsonic from supersonic regions), then this horizon should behave as a relativistic event horizon.

As all analogies, this one is limited. The first limit concerns the dispersion relation: in relativity, any particle has its energy and momentum linked by $E^2 = m^2 c^4 + c^2 P^2$. Even though this expression remains (approximately) true for low momenta waves (the previous expression is just a Taylor expansion), at large momenta, its breaks down, as the wavelength becomes so low that the discrete character of the fluid is manifest. The second limit comes from quantum mechanics itself: a sound wave in a fluid like water is a purely classical object.

This second point has been partially solved by the profusion of analogue models that were proposed (see Ref.~\cite{Barcelo:2005fc} for an explicit list), and which properties were closer to quantum mechanics than sound waves were. As an example, we shall study Bose-Einstein condensates and polaritons, see \ref{part:analoguegravity}.

Concerning the modification of the dispersion relation, no analogue medium has a purely relativistic dispersion relation, mainly because of the fact that it is composed of atoms. It was thus natural to introduce a non conventional dispersion relation in gravitational context, and thus to break the local Lorenz invariance. This was done in a covariant manner by T. Jacobson~\cite{Corley:1996ar} introducing a vector field, the vacuum expectation value of which is timelike. It thus introduces a preferred time which corresponds in analogue gravity to laboratory time. From a relativistic point of view, the next step is to consider such a field as dynamical, as is the metric~\cite{Jacobson:2000xp}.

From the point of view of condensed matter, the next step was to introduce dissipation on top of dispersion. To do so, a simplified model was introduced by R. Parentani~\cite{Parentani:2007uq}. From a logical point of view, any interacting field with a non relativistic dispersion relation should get dissipated. However, most of the studies are made in free (non interacting) theories (the only interacting theories in the presence of gravitation are not yet well known~\cite{Serreau:2011fu}). It was thus necessary to introduce dissipation in a phenomenological way, so as not to introduce interactions.

On the other hand, introduction of dispersion is motivated by relativity itself. Indeed, when taking the Hawking reasoning, and when considering the past of an emitted quantum, this quantum stayed in the vicinity of the black hole since its creation. And during all this time it went through an exponential redshift. This redshift is so huge that one second before its emission, the wavelength of the quantum is a million orders of magnitude smaller than its final wavelength (for a black hole of the size of the sun). This length is thus way smaller than Planck length and general relativity can not be considered as it stands and must be quantized. However, as announced in the first paragraph, there is no theory of quantum gravitation. This is the transplanckian problem. It turns out that it is solved by dispersion: the quantum leaves the vicinity of the black hole and thus stops being blueshifted (when going to the past) when its wavelength is such that the dispersion relation is no longer relativistic.This point shall be discussed in more details in \ref{part:desitter}.

In this thesis, we shall study the robustness of the Hawking process when introducing dispersion and dissipation in relativity. In \ref{part:basics}, we shall present the fundamental concepts for the following parts. To do so, we first consider the classical parametric amplifier and quantize it until we arrive to pair particle production in cosmology. Then, in \ref{part:desitter}, we shall look at the de Sitter space-time, introducing dispersion and dissipation. We shall then draw generic conclusions for Hawking radiation and (analogue) black holes. Finally, in \ref{part:analoguegravity}, we shall turn towards analogue models and study the influence of dissipation on dynamical Casimir effect and the consequence of initial conditions on (analogue) Hawking radiation.

\partimage{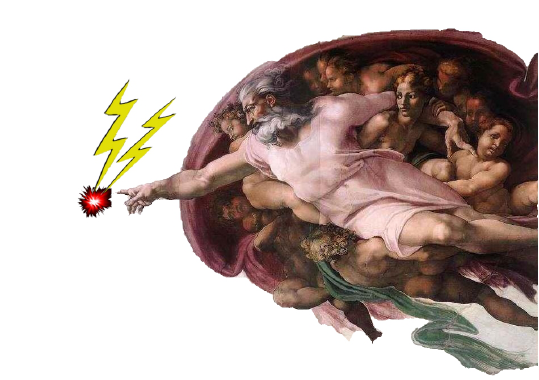}
\partcitation{\textit{Your chemistry high school teacher lied to you when they told you that there was such a thing as a vacuum, that you could take space and move every particle out of it.} \\ Adam Riess}
\part{About particle production}
\label{part:basics}
\chapter*{Introduction}
\phantomsection
\addstarredchapter{Introduction}
\markboth{INTRODUCTION}{INTRODUCTION}

When studying quantum effects in a curved background, or their analogue experimentally accessible counterparts, one notices that most of the effects are already present in much simpler contexts. Indeed, Hawking radiation and pair particle production are spontaneous emissions due to the amplification of quantum noise present in the vacuum. A huge simplification can be made when the initial state is not vacuum. The effect becomes amplification of initially present (eventually classical) noise. If these stimulated processes dominate, a classical description of the theory is enough. As a consequence, some classical systems contain the precursor of Hawking radiation of pair particle production. This is the essence of the experiment of Ref.~\cite{Weinfurtner:2010nu}. In such experiment, the classical equivalent of Hawking radiation is claimed to be observed as the amplification by an accelerating (water) flow of an incoming wave. The outgoing waves are produced in pairs as would be the Hawking flux. From the point of view of pair particle production, classical analogue are even simpler since a swing in a garden is enough, see \ref{sec:classicparamamp}.

In the first part of this thesis we use this simplification to develop, for the non specialized readership, the formalism and the tools that shall be used in the rest of this thesis.

\ref{chap:paramamptopartprod} progressively introduces notion of quantum field theory in curved space-time and gives an overview of the introduction of dispersion and dissipation. This is really the bedrock of the thesis. The goal of the chapter is not to give an overview of field theory nor relativity. Rather, it focuses on concepts that will be of great use in the following. The readership interested by a deeper introduction to these subjects may read Refs.~\cite{landau1975classical,Misner1973} for relativity or~\cite{weinberg1996quantum,peskin1995introduction} for an introduction to quantum field theory in flat space-time. For quantum field theory in curved space-time, a rather complete and clear overview is given by~\cite{wald1994quantum,Brout:1995rd}.

\ref{chap:separability} introduces the notion of separability of a quantum state and gives tools allowing to characterize separable states. Here again, the goal is not to present each notion in details. Some of the concepts are introduced rapidly and the interested reader can consult~\cite{leonhardt1997measuring,mandel1995optical}.

\chapter{From the parametric amplifier to pair particle production}
\chaptermark{Parametric amplifier \& particle production}
\label{chap:paramamptopartprod}

\section*{Introduction}

When considering a non stationary system, the temporal modifications of its parameters break energy conservation. As an example, a child sitted on a swing, when he moves his legs, makes the period of oscillations vary. He hence brings energy to the swing that projects him higher. The simplest non stationary system is the parametric amplifier. It is an harmonic oscillator with a time-varying frequency.

We shall here consider first the classical version of the parametric amplifier, then quantize it. After introducing to relativity, in \ref{sec:fields}, we shall give a quick view of the pair particle production occurring in quantum field theory. In each case, we shall see that the formalism is similar but that physical interpretations differ.

\minitoc
\vfill

\section{Classical parametric amplifier}
\label{sec:classicparamamp}

\subsection{General statements}
\subsubsection{Hamiltonian mechanics}
\label{sec:Hamiltondynamics}

Hamiltonian mechanics is a way to derive equations of motion from first principles in conservative systems. The process consists of two steps. The first one fixes the kinematics and the second one the dynamics of the system.

To fix the kinematics of the system, one first needs canonical quantities that describe completely the system. If the system has $n$ degrees of freedom, there are $2 n$ quantities, namely $q_i$ and $p_i$ corresponding to the positions and momenta of all degrees of freedom. The space generated by these variables is called phase space when it is endowed with a bilinear antisymmetric form called Poisson brackets. When the Poisson brackets take the form $\{A,B\} = \sum_i \partial_{q_i} A \partial_{p_i} B- \partial_{p_i} A \partial_{q_i} B $ where $A$ and $B$ are any functions of the phase space, $q_i$ and $p_i$ are called canonical. One can show that any other choice of $q'_i$ and $p'_i$ is canonical if and only if $\{q'_i,q'_j\}=\{p'_i,p'_j\}=0 $ and $\{q'_i,p'_j\}=\delta_{ij}$.

The dynamics of the system is determined by the evolution in time of the different $q_i$ and $p_i$. It is represented by a parametric curve $( q_i(t),p_i(t) )$ in the phase space, for which the curvilinear coordinate is time. Experiments have shown that equations of motion are of order two. When one knows all the $q_i$ and $p_i$ of a system at instant $t_0$, its evolution is thus fully determined. Therefore, there exists $2n$ functions $f_i,g_i$ which depend on phase space, such that any curve passing through $( q_i,p_i )$ at time $t$ should pass at $( q_i + f_i dt,p_i+ g_i dt )$ at time $t+dt$. In general, these functions evolve in time. When it is not the case, the system is called \textit{stationary}. A system is called \textit{conservative} when the canonicity of the variables is conserved through time, i.e., if $(q_i(t),p_i(t))$ is canonical, then $(q_i(t+dt),p_i(t+dt))$ also is. In such a case, one can show that there exists a function $H$ called the Hamiltonian of the system such that $\dot{q_i} = \{q_i,H\}$, $\dot{p_i} = \{p_i,H\}$. These are the equations of motion of the system. This implies that for any observable $O$ depending on canonical variables and time explicitly,
\begin{equation}
\label{eq:eomclassgeneral}
\begin{split}
\frac{dO}{dt} = \partial_t O + \{O,H\}.
\end{split}
\end{equation}
In general, the Hamiltonian $H(q_i,p_i,t)$ depends explicitly on time. However, when the system is stationary, it depends on time only through the phase space. In such a case, because of Eq.~\eqref{eq:eomclassgeneral} applied to $O=H$, the restriction of the Hamiltonian to the solution of the equation of motion $H(q_i(t),p_i(t))$ is a constant. This physically corresponds to the energy of the system. By extension, the energy of a non stationary system is the value of the Hamiltonian $H(q_i(t),p_i(t);t)$.

One can solve the equation $\dot{q_i} = \{q_i,H\}$ to extract $p_i(q_i,\dot{q_i})$. One then defines the Lagrangian by making the Legendre transformation of the Hamiltonian: $L(q_i,\dot{q_i}) \doteq \sum_i p_i \dot{q_i} - H $. One can show that the equations of motion are then equivalent to 
\begin{equation}
\label{eq:lagrangianeom}
\begin{split}
\frac{d}{dt} \frac{\partial L}{\partial \dot q_i} = \frac{\partial L}{\partial q_i}.
\end{split}
\end{equation}
The form of Eq.~\eqref{eq:lagrangianeom} is called Lagrange equations of motion. It is also equivalent to extremizing the action $S = \int L(q_i,\dot{q_i}) dt$. The advantage of the action is that it is unchanged when one parametrizes time differently.

\subsubsection{The system}

Consider first a mass $m$ at the extremity of a massless spring. Suppose also that the rate of the spring depends on time, so that the system is a parametric amplifier. Such a system has only one degree of freedom and is completely described by the position of the mass $q$ and its momentum $p$. The Hamiltonian hence decomposes into the kinetic part $p^2/2m$ and the potential one $k(t) q^2/2$ (the origin of the axis is such that $q=0$ is the equilibrium). It hence reads, with $\omega = \sqrt{k/m}$:
\begin{equation}
\label{eq:hamiltonien}
\begin{split}
H(t) = \frac{p^2}{2m } + \frac{m \omega^2(t)}{2} q^2.
\end{split}
\end{equation}
the equations of motion deduced from such a Hamiltonian are then
\begin{equation}
\label{eq:classeom}
\begin{split}
\dot{p} = - m \omega^2(t) q, \quad m \dot{q} = p.
\end{split}
\end{equation}
This system being linear and of order two, the space of solutions is of dimension two. It is then engendered by two pairs $(q_1,p_1)$ et $(q_2,p_2)$. Given two of those, the quantity
\begin{equation}
\label{eq:Wronskianconserved}
\begin{split}
W \doteq q_1 p_2 - q_2 p_1
\end{split}
\end{equation}
is called Wronskian of the system and does not depend on time\footnote{This is not a constant of motion since it mixes two different solutions}.

\subsection{Particular cases}
\subsubsection{Adiabatic case}

We shall consider here the adiabatic case, i.e., the case in which $\omega$ varies slowly : $\dot{\omega}\ll \omega^2$. Then, the real and imaginary parts of the complex solution
\begin{equation}
\label{eq:adiabaticsol}
\begin{split}
q(t) = N \frac{\ep{ -i \int \omega}}{\sqrt{m \omega}}\left [ 1+ \mathcal{O}\left (\frac{\dot{\omega}}{\omega^2} \right )\right ], \quad p(t) = -i N \sqrt{m \omega}\ep{ -i \int \omega}\left [ 1+ \mathcal{O}\left (\frac{\dot{\omega}}{\omega^2} \right )\right ]
\end{split}
\end{equation}
is a base of solutions. In Eq.~\eqref{eq:adiabaticsol}, N is a normalization constant. The energy associated to the two solutions is identical and varies with time. It is equal to $H(t) = N^2 \omega(t) /2$. The ratio
\begin{equation}
\begin{split}
J \doteq H(t) / \omega(t) = N^2/2 \left [1+ \mathcal{O}\left (\frac{\dot{\omega}}{\omega^2} \right ) \right ].
\end{split}
\end{equation}
is constant in the adiabatic limit $\dot{\omega}\ll \omega^2$. It is called an adiabatic invariant.

\subsubsection{Asymptotic case}

We consider here the important case where $\omega$ is constant for $t \to \pm \infty$. The limits are called respectively $\omega_{\rm in}$ and $\omega_{\rm f}$. By choosing the origin of times, we can write the solution for initial times as
\begin{equation}
\label{eq:qptneg}
\begin{split}
(q,p)\underset{t\to -\infty}\sim \left (\frac{A}{\sqrt{ m \omega_{\rm in}}} \sin[\omega_{\rm in} t], A \sqrt{ m \omega_{\rm in}} \cos[\omega_{\rm in} t]\right ),
\end{split}
\end{equation}
where $A$ is a normalization constant. Then, at final times, we will have
\begin{equation}
\label{eq:qptpos}
\begin{split}
(q,p)\underset{t\to +\infty}\sim \left (\frac{B}{\sqrt{m \omega_{\rm f}}} \sin[\omega_{\rm f} t+ \varphi], B \sqrt{m \omega_{\rm f}} \cos[\omega_{\rm f} t+ \varphi]\right ).
\end{split}
\end{equation}
The energy is then initially $E_{\rm in} = {A^2 \omega_{\rm in}}/{2} $ and in the end $E_{\rm f} = {B^2 \omega_{\rm f}}/{2} $. Hence, the adiabatic invariant is asymptotically
\begin{equation}
\begin{split}
J_{\rm in} = \frac{A^2}{2} , \quad J_{\rm f} = \frac{B^2}{2} 
\end{split}
\end{equation}
The ratio $R \doteq J_{\rm f}/J_{\rm in}$ only depends on the detail of the evolution of $\omega(t)$. 
The modification of the energy of the system is hence due to both the change in frequency (adiabatic part) and to the change in $R$ (non adiabatic part).

\subsubsection{Example}

In this paragraph, we consider the case where $\omega$ is a constant function for $t\neq t_0$, but has a sudden change at time $t=t_0$. Then, for $t<t_0$, Eq.~\eqref{eq:qptneg} and for $t>t_0$, Eq.~\eqref{eq:qptpos} apply. At $t=t_0$, equations of motion implies that both $q$ and $p$ are continuous. One deduces
\begin{equation}
\begin{split}
R = \frac{\omega_{\rm in}^2 \cos^2\omega_{\rm in} t_0 + \omega_{\rm f}^2 \sin^2\omega_{\rm in} t_0}{\omega_{\rm f}\omega_{\rm in}}.
\end{split}
\end{equation}
This means that the mean amplification (when $t_0$ is random) is $(\omega_{\rm in}^2 + \omega_{\rm f}^2 )/2\omega_{\rm f}\omega_{\rm in} > 1$. There is in general amplification for the adiabatic invariant. The time-reversibility is here lost by taking the average on $t_0$ even though the system is time reversible. Indeed, for each solution $(q(t),p(t))$, $(q(-t),p(-t))$ is solution of the reverse time evolution of the system $\omega(-t)$. However, for both evolutions, the adiabatic invariant increases when time evolves. It can hence be conceived as some sort of entropy.

One can note that if one makes a second sudden change after a time $\Delta t$ taking back the value of $\omega$ to its initial value, the mean value (with respect to $t_0$) of the amplification of the adiabatic invariant is for any value of $\Delta t$:
\begin{equation}
\begin{split}
\left < R \right >_{t_0} = \frac{\left (\omega_{\rm f}^2+\omega_{\rm in}^2\right )^2}{4 \omega_{\rm f}^2 \omega_{\rm in}^2} >1 .
\end{split}
\end{equation}
For a collection of non correlated parametric amplifiers, such a \textit{cyclic} operation leads to energy amplification.

\subsubsection{Phase space diagram}

\begin{figure}[ht]
\begin{minipage}{0.45\linewidth}
\includegraphics[width= 1 \linewidth]{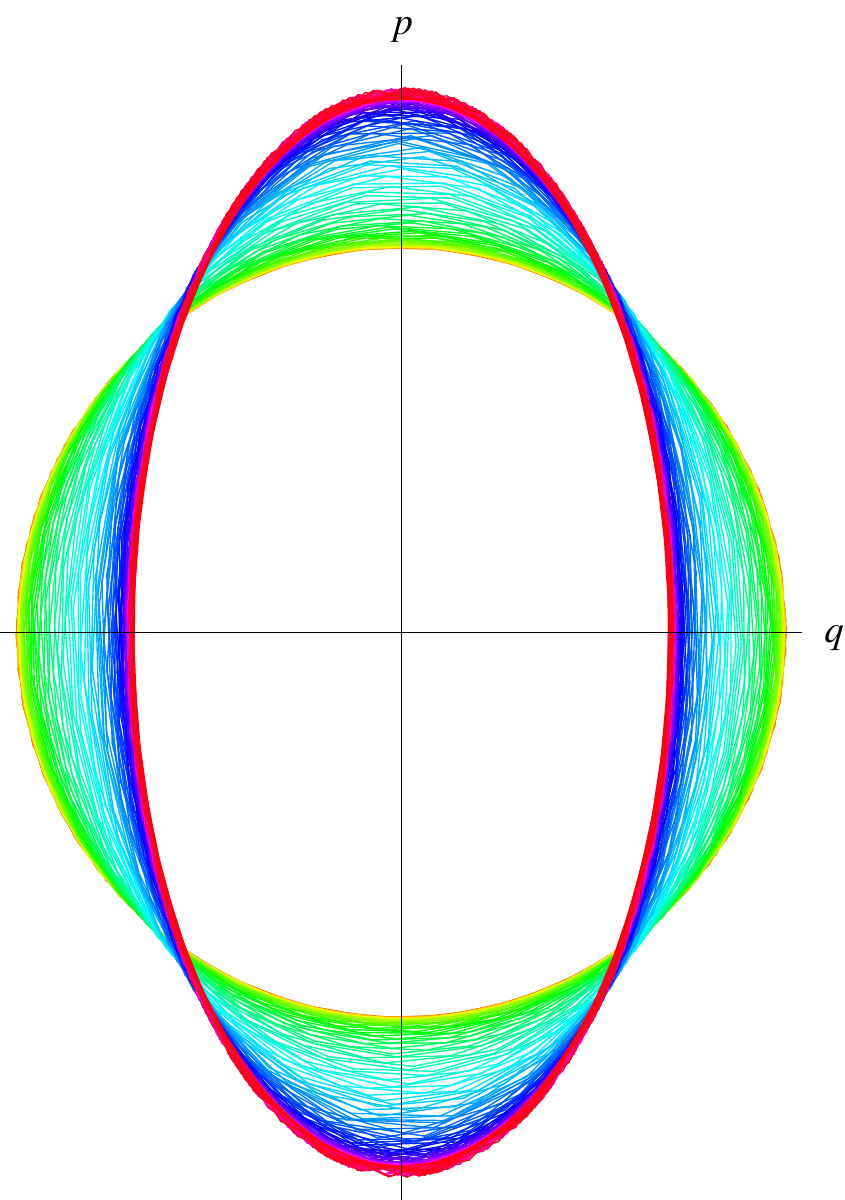}
\caption{Evolution of the phase space in the case of an adiabatic evolution of $\omega$. State goes slowly from the yellow ellipse to the red one, conserving the total area.}
\label{fig:adiabaticphasespace}
\end{minipage}
\hspace{0.05\linewidth}
\begin{minipage}{0.45\linewidth}
\includegraphics[width= 1 \linewidth]{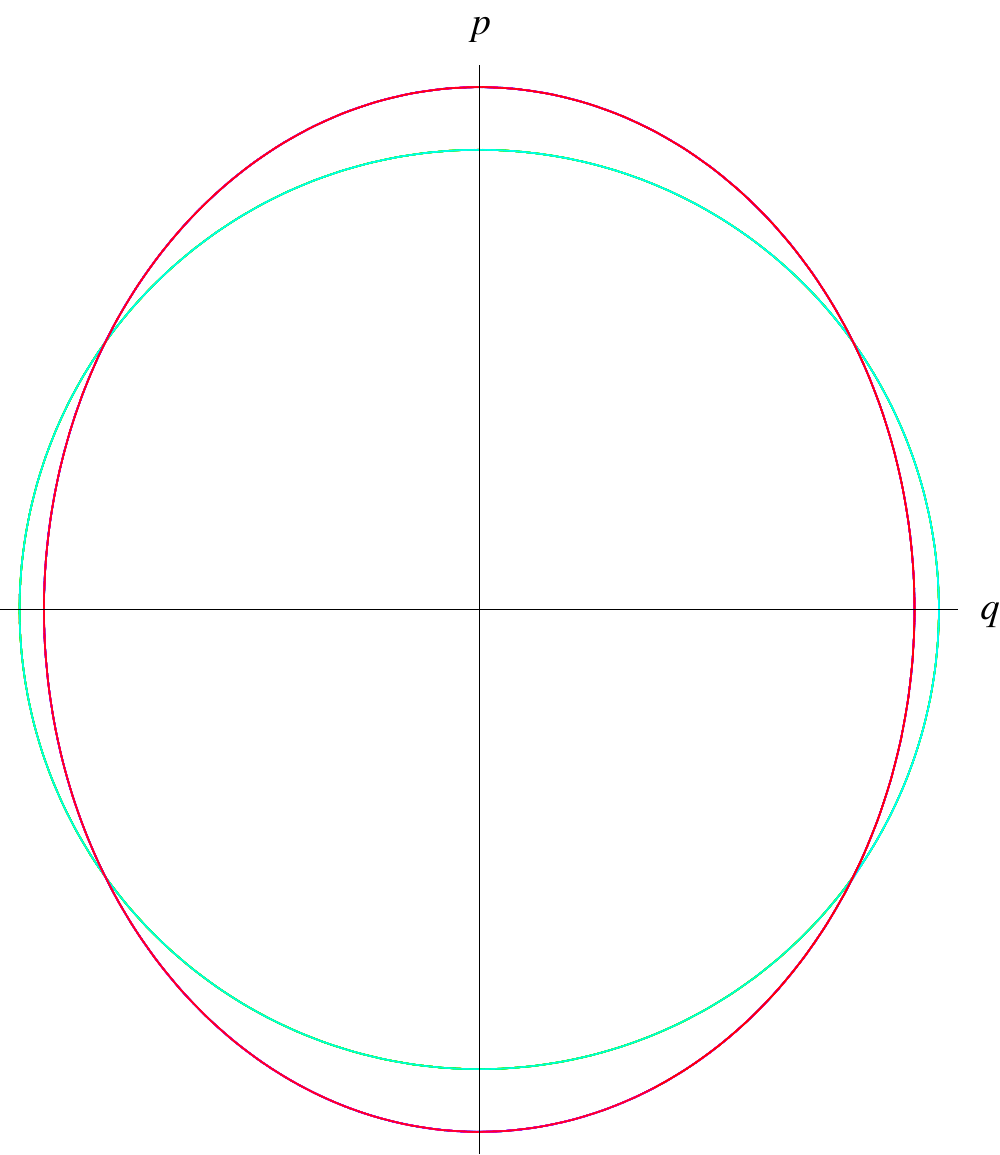}
\caption{ Evolution of the phase space in the case of a sudden change of $\omega$. State changes suddenly from one ellipse to the other.} 
\label{fig:gradinophasespace}
\end{minipage}
\end{figure}

Because the system is completely described by two quantities $q,p$, we can plot the trajectory in the plane $q,p$. As said before, the evolution of the system is a curve labeled by time. In the case of the harmonic oscillator, the curve is simply an ellipse and the time necessary to make one turn is the period of the system $2 \pi / \omega$. When considering the adiabatic limit, see Fig.~\ref{fig:adiabaticphasespace}, the ellipse is slowly deformed and its area remains constant. This is because the area of the ellipse is proportional to the adiabatic invariant. In the case $\omega$ is asymptotically constant, the phase space will be composed of two ellipses corresponding to initial and final states and some path linking one to the other. When the change is sudden, the length of this path is null and the two ellipses touch each other, see Fig.~\ref{fig:gradinophasespace}.

\subsubsection{Canonic base and Bogoliubov transformation.}

We shall here make a completely formal study in order to prepare quantification. Rather that decomposing $q$ on trigonometric functions, we shall consider complex $q$ and define canonical base through the positivity of the frequency in the exponential. Hence, for harmonic oscillator ($\omega$ is constant),
\begin{equation}
\label{eq:defbasecanon}
\begin{split}
(q,p)_{\rm can} \doteq \left (\frac{\ep{-i \omega t}}{\sqrt{2 m \omega}} , -i \sqrt{ m \omega/2} \ep{-i \omega t}\right ).
\end{split}
\end{equation}
The other element of the base will be the complex conjugate of the first. This base is such that the associated Wronskian is equal to $-i$.

When considering the parametric amplifier, and in the case where it is asymptotically stationary, we can define two privileged basis according to whether they are canonical for asymptotically low or large times:
\begin{equation}
\begin{split}
(q,p)_{\rm in} \underset{t\to -\infty}{\sim} (q,p)_{\rm can},\quad (q,p)_{\rm f} \underset{t\to +\infty}{\sim} (q,p)_{\rm can}
\end{split}
\end{equation}
These two basis are then linked by a Bogoliubov transformation~\cite{Bogolubov}:
\begin{equation}
\label{eq:modebogotf}
\begin{split}
(q,p)_{\rm in} = \alpha (q,p)_{\rm f} + \beta \left [(q,p)_{\rm f}\right ]^*.
\end{split}
\end{equation}
Wronskian conservation then implies $\abs{\alpha}^2 - \abs{\beta}^2 =1$. Moreover, Wronskian can be viewed as a bilinear form
\begin{equation}
\begin{split}
\left < (q_1,p_1), (q_2,p_2) \right >_{\rm KG} = -i (q_1 p_2^* - q_2^* p_1).
\end{split}
\end{equation}
This form has all the properties of a scalar product, except for positivity. As an example, the product of two solutions is time independent and canonical modes form an orthonormal base. Hence, we find back the Bogoliubov coefficients through:
\begin{equation}
\begin{split}
\alpha = \left < (q,p)_{\rm in}, (q,p)_{\rm f} \right >_{\rm KG} ,\quad \beta = \left < (q,p)_{\rm in}, \left [(q,p)_{\rm f}\right ]^* \right >_{\rm KG}.
\end{split}
\end{equation}

The notion of Bogoliubov transformation is thus entirely classical. Moreover, it is possible to show that during time evolution, the amplification of the adiabatic invariant is bounded by
\begin{equation}
\begin{split}
\left (\abs{\alpha}- \abs{\beta} \right )^2 \leq R \leq \left (\abs{\alpha}+ \abs{\beta} \right )^2.
\end{split}
\end{equation}

\section{Parametric amplifier in quantum mechanics}
\label{sec:Quantumparamampli}

We shall consider now the quantization of the parametric amplifier. The canonical quantization procedure of a system consist in converting all observable and canonical variables to operators acting on some state. Moreover, the canonicity of variables now translates in commutators ($[\hat q,\hat p]= i \hbar $) instead of Poisson Brackets. More details will be given in \ref{sec:QMalgapproach}.

\subsection{Schrödinger representation}

\subsubsection{General considerations }

As for its classical version, the quantum parametric amplifier has only one degree of freedom. Its Hamiltonian is a Hermitian operator quadratic in the operators position $\hat q$ and momentum $\hat p$. Its expression is identical to Eq.~\eqref{eq:hamiltonien}. The Schrödinger equation reads
\begin{equation}
\label{eq:Schodinger}
\begin{split}
i \hbar \partial_t \left | \Psi(t)\right >= \hat H(t) \left | \Psi(t)\right >= \left (\frac{\hat p^2}{2m } + \frac{m \omega^2(t)}{2} \hat q^2 \right ) \left | \Psi(t)\right >.
\end{split}
\end{equation}
Since the Hamiltonian explicitly depends on time, there exist no stationary state. It is then necessary to use a different approach from the usual one using decomposition into stationary states. To do so, let us consider the matrix element of canonical operators
\begin{equation}
\begin{split}
q_{\rm el}(t) = \left < \Psi_1(t) \right | \hat q \left | \Psi_2(t) \right >, \quad p_{\rm el}(t) = \left < \Psi_1(t) \right | \hat p \left | \Psi_2(t) \right >,
\end{split}
\end{equation}
where $ \left | \Psi_{1,2}(t) \right >$ are two solutions of Eq.~\eqref{eq:Schodinger}. Because of the Schrödinger equation, these quantities are solution of the classical equations of motion [see Eq.~\eqref{eq:classeom}], i.e.,
\begin{equation}
\label{eq:quantumeomqp}
\begin{split}
m \dot{q_{\rm el}} = p_{\rm el} ,\quad \dot{p_{\rm el}}= - m \omega^2(t) q_{\rm el}.
\end{split}
\end{equation}
All the conclusions of \ref{sec:classicparamamp} hence apply. In particular, the adiabatic solutions and the amplification are the same.

\subsubsection{Harmonic oscillator}

When the frequency $\omega$ is constant in time, we are facing the usual harmonic oscillator. We can then decompose the state on a basis of stationary states
\begin{equation}
\label{eq:generalquantumstate}
\begin{split}
\left | \Psi(t)\right > = \sum_n c_n \ep{-i E_n t/\hbar}\left | \Psi_n\right >
\end{split}
\end{equation}
where $\left | \Psi_n\right >$ is the solution of $\hat H \left | \Psi_n\right > = E_n \left | \Psi_n\right >$, and $E_n = \hbar \omega (n+1/2)$ with $n \in \mathbb{N}$. 
We note here that energy is not discrete, since it can reach any value. Only the ratio $E / \omega$, i.e., the adiabatic invariant is discrete. Hence, in the adiabatic limit of the system the value of $n$ will be constant. In classical mechanics, this result was non trivial. Here, it comes from the fact that $n$ is an integer that varies slowly.

To make link with the previous paragraph, we consider the matrix elements of $(\hat q, \hat p)$ between eigen states of the Hamiltonian:
\begin{equation}
\begin{split}
q_{k,l}(t) = \ep{i (E_k-E_l)t/\hbar } \left < \Psi_k \right | \hat q \left | \Psi_l \right >,\quad p_{k,l}(t) = \ep{i (E_k-E_l)t/\hbar } \left < \Psi_k \right | \hat p \left | \Psi_l \right >.
\end{split}
\end{equation}
because they are solution of Eq.~\eqref{eq:quantumeomqp}, if $\abs{E_k - E_l} \neq \hbar \omega $, then $q_{k,l}(t)=p_{k,l}(t)=0$. Using the Hermitian character of $\hat q$ ($q_{k,l} = q_{l,k}^*$) and the canonical commutators $\sum_l q_{k,l} p_{l,m}-p_{k,l} q_{l,m} = -i \hbar \delta_{k,m}$, we show that up to a phase defining origin of times, $q_{k,l}(t) = \sqrt{\frac{\hbar k}{2 m \omega}} \delta_{k-l-1} \ep{i \omega t} + \sqrt{\frac{\hbar l}{2 m \omega}} \delta_{k-l+1} \ep{-i \omega t}$.
We now define the creation and annihilation operators through their action of stationary states:
\begin{equation}
\label{eq:defcreatannihilop}
\begin{split}
\hat a^\dagger \left | \Psi_n \right > = \sqrt{\hbar (n+1)} \left | \Psi_{n+1} \right > ,\quad \hat a \left | \Psi_n \right > = \sqrt{\hbar n} \left | \Psi_{n-1} \right >.
\end{split}
\end{equation}
$q_{k,l}$ then simply reads in term of those operators and the functions defined in Eq.~\eqref{eq:defbasecanon}:
\begin{equation}
\label{eq:qkloftosharm}
\begin{split}
q_{k,l}(t) = (\hat a^\dagger)_{k,l} q^*_{\rm can} + \hat a_{k,l} q_{\rm can},\quad p_{k,l}(t) = (\hat a^\dagger)_{k,l} p^*_{\rm can} + \hat a_{k,l} p_{\rm can}.
\end{split}
\end{equation}
When choosing non stationary states $ \left | \Psi_{1,2}(t) \right >$ of the form of Eq.~\eqref{eq:generalquantumstate}, $q_{\rm el}$ and $p_{\rm el}$ then decompose on the basis of $q_{k,l}$ and $p_{k,l}$. Each term is then either proportional to $q_{\rm can}$ and corresponds to lowering the energy of the system (it only contains operators $\hat a$), or it is proportional to $q^*_{\rm can}$ and corresponds to increasing the energy of the system. This justifies the definition we used in Eq.~\eqref{eq:defbasecanon}.

Moreover, the operators $\hat a$ can be expressed in term of canonical operators, as $\hat a = \sqrt{\frac{m \omega}{2}} \hat q + i \frac{\hat p }{\sqrt{2 m \omega}} $. The Hamiltonian thus reads in term of these operators
\begin{equation}
\begin{split}
\hat H = \hbar \omega \frac{\hat a^\dagger \hat a +\hat a \hat a^\dagger}{2}.
\end{split}
\end{equation}

\subsubsection{Representation position}

For completeness, we give here for the interested reader the expressions of operators and states in the representation position.

In this representation, the state is described by a space dependent wave function $\left | \Psi(t)\right > \equiv \psi(q,t)$. The operator $\hat q$ corresponds to multiplication by $q$ and the operator $\hat p$ corresponds to derivation with respect to $q$:
\begin{equation}
\begin{split}
\hat q\left | \Psi(t)\right > \equiv q \psi(q,t), \quad \hat p\left | \Psi(t)\right > \equiv - i \hbar \partial_q. \psi(q,t)
\end{split}
\end{equation}
The scalar product is the integration on all real values of $q$. Eigen states $\left | \Psi_n\right >$ are then proportional to Hermite polynomials $H_n$. They are equal to
\begin{equation}
\begin{split}
\psi_n(q,t) = \frac{1}{\sqrt{2^n n!}} (\frac{m \omega}{\pi \hbar})^{1/4} H_n\left ( \sqrt{\frac{m\omega}{\hbar}} q \right ) \ep{- \frac{m\omega}{2\hbar} q^2} \ep{- i \omega(n+1/2) t}.
\end{split}
\end{equation} 
Moreover, the operator $\hat a$ is:
\begin{equation}
\begin{split}
\hat a &\equiv \sqrt{\frac{m \omega}{2}} q + \frac{\hbar \partial_q }{\sqrt{2 m \omega}}= \frac{-\hbar }{\sqrt{2 m \omega}} \ep{ -\frac{m\omega}{2\hbar} q^2} \partial_q \ep{ \frac{m\omega}{2\hbar} q^2}, \\
\hat a^\dagger &\equiv \sqrt{\frac{m \omega}{2}} q - \frac{\hbar \partial_q }{\sqrt{2 m \omega}} = \frac{\hbar }{\sqrt{2 m \omega}} \ep{ \frac{m\omega}{2\hbar} q^2} \partial_q \ep{ -\frac{m\omega}{2\hbar} q^2}.
\end{split}
\end{equation}

\subsection{Heisenberg representation}

In some cases, it is useful to modify the way we see operators and states. To do so, we notice that neither states nor operators have any physical meaning. Only the combination $\left < \Psi(t) \right | \hat O \left | \Psi(t) \right >$, where $\hat O$ is some Hermitian operator, has a physical meaning and can be measured. It is then possible to transfer the time dependence of the state to the operators without modifying any physical prediction of the theory. This is the idea of the Heisenberg representation

\subsubsection{Evolution operator}

We should first notice that the Schrödinger Eq.~\eqref{eq:Schodinger} is a first order linear equation. Hence, it is analytically solvable and the solution reads formally
\begin{equation}
\begin{split}
\left | \Psi(t) \right > \doteq \hat U(t;t_0)\left | \Psi(t_0) \right >.
\end{split}
\end{equation}
Here, $\hat U$ is the evolution operator. An explicit calculation gives $\hat U(t;t_0) = \Texp\left ( \frac{i}{\hbar} \int_{t_0}^t \hat H \right )$ where $\Texp$ is the time ordered exponential:
\begin{equation}
\begin{split}
\Texp\left ( \frac{i}{\hbar}\int_{t_0}^t \hat H \right ) = \sum_n \left (\frac{i}{\hbar}\right )^n \int_{t>t_1> \cdots > t_n>t_0} dt_1\cdots dt_n \hat H(t_1)\cdots \hat H(t_n) .
\end{split}
\end{equation}
The operator $\hat U$ is then unitary because the Hamiltonian is hermitian. The expression of a physical quantity can hence be decomposed as:
\begin{equation}
\begin{split}
\left < \Psi(t) \right | \hat O \left | \Psi(t) \right > = \left < \Psi(t_0) \right | \hat U^\dagger(t;t_0) \hat O \hat U(t;t_0)\left | \Psi(t_0) \right > = \left < \Psi(t_0) \right | \hat O(t;t_0) \left | \Psi(t_0) \right >
\end{split}
\end{equation}
where $\hat O(t;t_0) = \hat U^\dagger(t;t_0) \hat O\hat U(t;t_0)$. This operator is a solution of the differential equation
\begin{equation}
\label{eq:eomheisenberggen}
\begin{split}
 i \hbar \frac{d \hat O(t)}{dt} = [ \hat O(t),\tilde H],
\end{split}
\end{equation}
where $\tilde H \doteq \hat U^\dagger(t;t_0) \hat H(t)\hat U(t;t_0)$ is a modified version of the Hamiltonian. We formally pass from $\hat H$ to $\tilde H$ by replacing every operator by its version in Heisenberg representation. Moreover, the commutators of $\hat q(t)$ and $\hat p(t')$ are no longer canonical. Only the equal time commutators are canonical $[\hat q(t),\hat p(t)] = i \hbar$.

The main advantage of Heisenberg representation is its similarity with classical theory. Compare for instance Eq.~\eqref{eq:eomheisenberggen} with Eq.~\eqref{eq:eomclassgeneral}. Moreover, operators obey the same equations as the matrix elements [see Eq.~\eqref{eq:quantumeomqp}].

\subsubsection{Back to the parametric amplifier}

For the parametric amplifier, the Hamiltonian in Heisenberg representation $\tilde H $ reads [see Eq.~\eqref{eq:hamiltonien}]
\begin{equation}
\begin{split}
\tilde H = \frac{\hat p(t)^2}{2m} + \frac{m \omega^2}{2} \hat q(t)^2.
\end{split}
\end{equation}
The equations of motion in Heisenberg representation [see Eq.~\eqref{eq:eomheisenberggen}] then read for the parametric amplifier
\begin{equation}
\label{eq:eomoscharmQ}
\begin{split}
\dot{ \hat q}(t) = \frac{1}{i \hbar} [ \hat q(t),\tilde H(t)] = \frac{\hat p(t)}{m},\quad \dot{\hat p}(t)= \frac{1}{i \hbar} [ \hat p(t),\tilde H(t)] = - m \omega^2 \hat q(t).
\end{split}
\end{equation}
which is the operator version of Eq.~\eqref{eq:quantumeomqp}.

In the case of the harmonic oscillator, the solutions are simply the exponentials introduced in Eq.~\eqref{eq:defbasecanon}. Using Eq.~\eqref{eq:qkloftosharm} to fix the normalization, we get
\begin{equation}
\begin{split}
\hat q(t) = \hat a q_{\rm can}(t) + \hat a^\dagger q_{\rm can}^*(t), \quad \hat p(t) = \hat a p_{\rm can}(t) + \hat a^\dagger p_{\rm can}^*(t).
\end{split}
\end{equation}
We emphasize that operators $\hat a$ and $\hat a^\dagger$ do not depend on time and were defined in Eq.~\eqref{eq:defcreatannihilop}.

For the parametric amplifier, it is impossible to define meaningful creation and annihilation operators. However, if $\omega$ is asymptotically constant in time, it is possible to define two basis of operators: one for small times, and one for large times. We have then two possible decompositions:
\begin{equation}
\label{eq:qdecompcanon}
\begin{split}
(\hat q(t),\hat p(t)) = \hat a_{\rm in} (q,p)_{\rm in} + \hat a^\dagger_{\rm in} \left [(q,p)_{\rm in}\right ]^* = \hat a_{\rm f} (q,p)_{\rm f} + \hat a^\dagger_{\rm f} \left [(q,p)_{\rm f}\right ]^*
\end{split}
\end{equation}
The Bogoliubov transformation of canonical modes Eq.~\eqref{eq:modebogotf} thus transcripts to the creation and annihilation operators:
\begin{equation}
\label{eq:bogosuroperateur}
\begin{split}
\hat a_{\rm f} = \alpha \hat a_{\rm in} + \beta^* \hat a_{\rm in}^\dagger.
\end{split}
\end{equation} 
This transformation acquires here all its physical sense. It is similar to the transformation occurring during reflexion/transmission. Indeed, in such a case, we have $\hat a_{\rm in} = r \hat a_{\rm reflected} + t \hat a_{\rm transmitted}$. The incoming ray then cuts into a reflected part and a transmitted part. Since all operators are destruction operators, unitarity imposes $\abs{r}^2 + \abs{t}^2 = 1$. Here, on the other hand because Eq.~\eqref{eq:bogosuroperateur} mixes creation and annihilation operators, unitarity imposes $\abs{\alpha}^2 - \abs{\beta}^2 =1$. As a consequence, the outgoing flux contains more quanta of energy than the incoming flux.

\subsubsection{Ground state}

We now have every tool to express the initial ground state $\left | \Psi_0\right >_{\rm in}$ after the evolution of the system. The ground state is characterized by
\begin{equation}
\label{eq:defetatfond}
\begin{split}
\hat a_{\rm in} \left | \Psi_0\right >_{\rm in} = 0.
\end{split}
\end{equation}
It then decomposes on the basis of final eigen-states (i.e., $\left | \Psi_n\right >_{\rm f} = (\hat a_{\rm f}^\dagger)^n \left | \Psi_0\right >_{\rm f})$ as $\left | \Psi_0\right >_{\rm in} = \sum_n c_n (\hat a_{\rm f}^\dagger)^n \left | \Psi_0\right >_{\rm f} = \hat S \left | \Psi_0\right >_{\rm f}$ where $\hat S$ is some unitary operator. Using Bogoliubov transformation to express $\hat a_{\rm in}$ in term of final operators, Eq.~\eqref{eq:defetatfond} gives $\left (\alpha^* \hat a_{\rm f} - \beta^* \hat a_{\rm f}^\dagger \right ) \hat S \left | \Psi_0\right >_{\rm f} = 0$. Using $\hat a_{\rm f} \left | \Psi_0\right >_{\rm f}=0$, it simplifies into
\begin{equation}
\label{eq:initialstateasfinal}
\begin{split}
\left ([\hat a_{\rm f},\hat S] - \frac{\beta^*}{\alpha^*} \hat a_{\rm f}^\dagger \hat S \right )\left | \Psi_0\right >_{\rm f} = 0.
\end{split}
\end{equation}
We can consider $\hat S$ as a function of $\hat a_{\rm f}^\dagger$. We then have $[\hat a_{\rm f},\hat S] =\sum_n n c_n (\hat a_{\rm f}^\dagger)^{n-1} =S'(\hat a_{\rm f}^\dagger)$ is the derivative of $\hat S$. Hence $\hat S'(\hat a_{\rm f}^\dagger)- \frac{\beta^*}{\alpha^*} \hat a_{\rm f}^\dagger \hat S(\hat a_{\rm f}^\dagger)$ is function of $\hat a_{\rm f}^\dagger$ and is identically $0$ because of Eq.~\eqref{eq:initialstateasfinal}. Hence, we have $\hat S(\hat a_{\rm f}^\dagger) \propto \ep{\frac{\beta^*}{2\alpha^*} \hat a_{\rm f}^\dagger \hat a_{\rm f}^\dagger}$. The proportionality coefficient is then fixed by the normalization of the state. To summarize, we have
\begin{equation}
\label{eq:videinsurbaseout}
\begin{split}
\left | \Psi_0\right >_{\rm in} = \frac{1}{\sqrt{\abs{\alpha}}} \ep{\frac{\beta^*}{2\alpha^*}\hat a_{\rm f}^\dagger \hat a_{\rm f}^\dagger}\left | \Psi_0\right >_{\rm f}.
\end{split}
\end{equation}
The ground state then gets excited by the time evolution. Its mean energy level is after evolution 
\begin{equation}
\begin{split}
n_{\rm f} \doteq \left < \Psi_0\right|_{\rm in} \hat a_{\rm f}^\dagger \hat a_{\rm f}\left | \Psi_0\right >_{\rm in} = \abs{\beta}^2.
\end{split}
\end{equation}
For an initially excited state, we have
\begin{equation}
\label{eq:nfQMofninbeta}
\begin{split}
n_{\rm f}= \left < \Psi_n\right|_{\rm in} \hat a_{\rm f}^\dagger \hat a_{\rm f}\left | \Psi_n\right >_{\rm in} = n_{\rm in} + 2 \abs{\beta}^2 n_{\rm in} + \abs{\beta}^2.
\end{split}
\end{equation}
The third term is the spontaneous excitation from vacuum. The first term is the initial value of the adiabatic invariant. The middle term then corresponds to stimulated emission: Each initial quantum of energy has a probability $\abs{\beta}^2$ to create two quanta of energy. These ones are necessarily produced by pairs. Indeed, we can show that if $n-m$ is odd, $\left < \Psi_m\right|_{\rm f} \left | \Psi_n\right >_{\rm in}=0$. The vacuum behaves as if it can lend $1/2$ quantum of energy and get it back after the time evolution. This behavior is commonly interpreted as creation of virtual quantum.

\begin{SCfigure}
\includegraphics[width=0.5\linewidth]{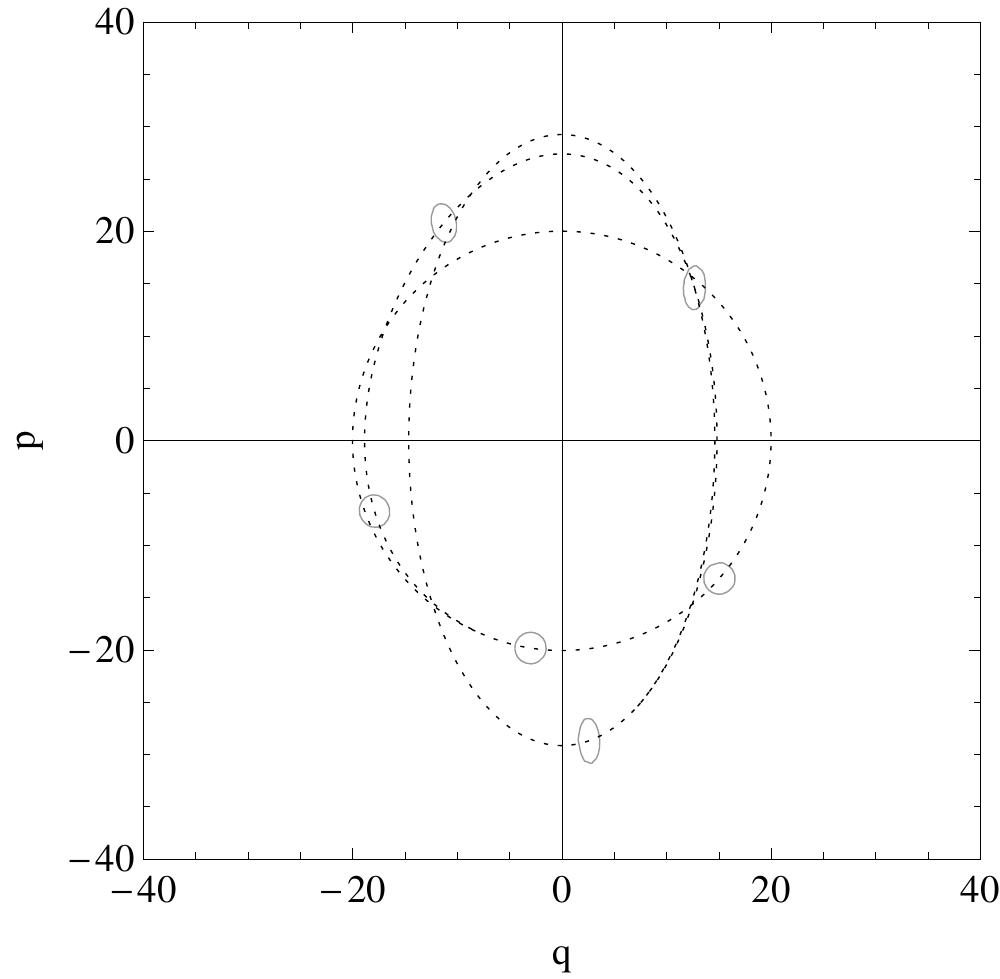}
\caption{Representation of the Wigner function for different times. The initial ellipse corresponding to $W(q,p) = 1/10 W_{\max}$ is moved and deformed as time evolves. The classical trajectory (dotted) corresponds to $W(q,p) = W_{\max}$. 
}
\label{fig:Gaussianphasespace}
\end{SCfigure}

\subsubsection{Wigner function}

To describe some quantum state, there exists an equivalent to the phase space diagrams that is representation of the Wigner function. Before defining the Wigner function, we need first to define the quadrature states $\left | q\right >$ that are the eigenstate of the operator $\hat q$ and associated to the eigenvalue $q$ : $\hat q \left | q\right > = q \left | q\right >$\footnote{These states are represented by $\delta$ functions in representation position. Moreover, $\psi(q,t) = \left <q | \Psi(t) \right >$.}. Then, we define the Wigner function of a state $\left | \Psi(t)\right >$ in Schrödinger representation through
\begin{equation}
\label{eq:defWigner}
W(t;q,p) \doteq \frac{1}{2\pi}\int dx  \left < q-x/2 | \Psi(t)\right > \left < \Psi(t)| q+x/2 \right > \ep{i p x}
\end{equation}
Physically, this function corresponds to quasi-probability distribution in phase space. 

In Fig.~\ref{fig:Gaussianphasespace} we represented an initially Gaussian coherent state and its time evolution. At each time, the Wigner function is a Gaussian function. We represented the contour line corresponding to $W(q,p) = W_{\max}/10$ at different times. The dotted line correspond to classical motion. We see that the contour line is an ellipse that gets deformed as time passes. Because it gets squeezed, the final state is called squeezed coherent state. 

\subsubsection{Dissipation}

In this subsection we briefly introduce dissipation in a manner similar to~\cite{Unruh:1989dd}. To do so by conserving the linearity of equations of motion, we couple the parametric amplifier to some continuum of harmonic oscillators. When doing so, the total Hamiltonian reads in Heisenberg representation
\begin{equation}
\label{eq:dissipHforQM}
\begin{split}
\tilde H = \frac{\hat p(t)^2}{2m} + \frac{m \omega^2}{2} \hat q(t)^2 + \int d\zeta \frac{\hat p_\zeta(t)^2}{2} + \frac{\omega_\zeta ^2}{2}\hat q_\zeta(t)^2 + g \frac{\omega_\zeta}{\sqrt{\pi d\omega_\zeta/d\zeta}} \hat q_\zeta \hat q
\end{split}
\end{equation}
Equations of motion for the harmonic oscillator read
\begin{equation}
\begin{split}
\ddot{\hat{q}}_\zeta + \omega_\zeta^2 \hat q_\zeta = -g \omega_\zeta \hat q.
\end{split}
\end{equation}
Assuming that $q_\zeta^0$ is a solution of the homogeneous corresponding equation ($g=0$) that matches the solution initially, the general solution reads
\begin{equation}
\begin{split}
\hat q_\zeta = \hat q_\zeta^0 - \int dt' \theta(t-t') \frac{\sin \omega_\zeta (t-t') }{\omega_\zeta} g \frac{\omega_\zeta}{\sqrt{\pi d\omega_\zeta/d\zeta}} \hat q(t')
\end{split}
\end{equation}
When we inject this in equations of motion for the parametric amplifier, we get
\begin{equation}
\begin{split}
m \dot{\hat q} = \hat p, \quad \dot{\hat p} = -m \omega^2 q - \int d\zeta g \frac{\omega_\zeta}{\sqrt{\pi d\omega_\zeta/d\zeta}} \hat q_\zeta^0 - g^2 \dot{\hat q}.
\end{split}
\end{equation}
The $g^2 \dot{\hat q}$ term is the term that induces dissipation. It is local because the coupling is chosen proportional to $\omega_\zeta/\sqrt{d\omega_\zeta/d\zeta}$. In classical mechanics, this corresponds to friction force: when $ q_\zeta^0=0$ is the trivial solution, the system is then equivalent to some non conservative system. In quantum mechanics, one can no longer assume $\hat q_\zeta^0=0$, since the commutator $[\hat q_\zeta^0,p_\zeta^0]= - i \hbar$. The term proportional to $\hat q_\zeta^0$ is the term that is necessary so that unitarity of the whole system is preserved.

\subsubsection{ Coupled parametric amplifiers }

In order to prepare \ref{sec:fields}, we consider here a system of coupled parametric amplifiers. For simplicity, we put them on the same line so that we have a one dimensional system. The Hamiltonian of such system reads
\begin{equation}
\begin{split}
H = \sum_i \frac{p_i^2}{2m } + \frac{m \omega^2(t)}{2} q_i^2 + m \sum_{i,j} k_{i,j}(t) (q_i - q_j)^2.
\end{split}
\end{equation}
In the most general case, the Bogoliubov transformation between initial and final state reads, in term of matrices
\begin{equation}
\begin{split}
 a_{i,\rm f} = \sum_j \alpha_{i,j}a_{j,\rm in} + \beta_{i,j} a_{j,\rm in}^\dagger,
\end{split}
\end{equation}
and unitarity is expressed though matrix equation $\alpha \alpha^\dagger - \beta \beta^\dagger = 1$.

When each of these oscillators is only coupled to its closest neighbor, in the continuum limit, and with a parametric amplifier's density $ \rho(z) dz $, Hamiltonian becomes
\begin{equation}
\begin{split}
H = \int dz \rho(z) \left [ \frac{p(z)^2}{2m } + \frac{m \omega^2(t)}{2} q(z)^2 + m c^2(z) (\partial_z q)^2\right ].
\end{split}
\end{equation}
It reads, when making the canonical transformation $\phi = \sqrt{m} q$, $\pi = p/ \sqrt{m}$:
\begin{equation}
\label{eq:Hampparamcouple}
\begin{split}
H = \int dz \rho(z) \left [ \frac{\pi(z)^2}{2} + \frac{\omega^2(t)}{2} \phi(z)^2 + c^2 (\partial_z \phi)^2\right ].
\end{split}
\end{equation}

\section{Relativity}

We make here a break in our study of parametric amplifiers to study the basics of general relativity (GR). This is not GR in the strict sense since we do not study the influence of matter on geometry. Rather, we wish to consider curved geometry and matter living on it. 

\subsection{Special relativity}

\subsubsection{Notion of space-time}

Special relativity is based on two assumptions. The first one is the relativity principle. It states that the results of an experiment do not depend on the \textit{inertial} reference frame it is made in. A reference frame is called inertial when every free particle has a constant speed. A reference frame moving at constant speed with respect to some inertial referential is hence itself inertial. The second assumption is the fact that there exist a maximum speed in the universe. We call $c$ this maximum speed. Because of the first assumption, it has the same value in all inertial referential.

As a result, if in an inertial frame, two points $\mathsf x = (t,\bx)$ and $ \mathsf x'= (t',\bx')$ are such that $(\bx-\bx')^2 - c^2 (t-t')^2=0$, it will remain so in any inertial frame (since we go from one point to the other using maximal speed). We name distance between two infinitely close points $ds$ defined by
\begin{equation}
\label{eq:Minkowskimetric}
\begin{split}
ds^2 = d\bx^2 - c^2 dt^2.
\end{split}
\end{equation}
One can show from homogeneity of space that the distance is the same in every inertial reference frame. We then have three types of distance : \textit{space-like} if $ds^2>0$, \textit{time-like} if $ds^2<0$ and \textit{light-like} if $ds^2 =0$. The assumption that there is a maximal velocity translates in the fact that the trajectory of any particle is necessarily time-like or light-like. Let $\bx(t)$ be the (time-like) trajectory of a particle in some inertial reference frame, and $\bv \doteq d \bx/dt$ its velocity in this frame. Then, along the trajectory, $ds^2 = -(c^2 - v^2) dt^2$. 

\subsubsection{Action, energy, momentum}

For a relativistic particle, the action does not depend on the reference frame. This is a generalization of the independence of the action through reparametrization of time, see \ref{sec:Hamiltondynamics}, and a consequence of the relativity principle. Because $ds$ is the only invariant, it reads $S = -m c \int \sqrt{-ds^2} $ where $m$ is a constant and the integral runs along the trajectory of the particle. The action hence has the form in some inertial frame $S = -m c \int \sqrt{c^2 -v^2} dt$. Hence, momentum and Hamiltonian read in that frame
\begin{equation}
\begin{split}
\bp = \frac{\delta S}{\delta \bv } = m \frac{\bv}{\sqrt{1 -v^2/c^2 }},\quad H = \bp \bv + m c \sqrt{c^2 -v^2} = \frac{m c^2}{\sqrt{1 -v^2/c^2}}, 
\end{split}
\end{equation}
and the equations of motion are $m \dot \bp =0$. In the limit of low velocity, $E \sim m c^2 + p^2 /2m$. We here recognize that $m$ is the mass of the particle. Moreover, we see that
\begin{equation}
\begin{split}
E^2 - p^2 c^2 = m^2 c^4.
\end{split}
\end{equation}
This is the relation between energy and momentum in special relativity. It remains true in every inertial frame. For massless particles, the relation becomes $E^2 = p^2 c^2$, which implies that $v=c$ and $ds=0$. They thus follow light-like trajectories.

\subsection{General relativity}

\subsubsection{Manifold}

In general relativity, the relativistic principle is extended to the equivalence principle. It states that the effects of gravitation are \textit{locally} not different than effect of acceleration. In an accelerating frame, the distance between two points is no longer given by Eq.~\eqref{eq:Minkowskimetric}, but is still a quadratic function of 1-forms $d\bx$ and $dt$. The right concept for general relativity is hence the concept of manifold. A manifold is some space that looks locally like Minkowski space [described by the metric of Eq.~\eqref{eq:Minkowskimetric}]. The distance is generally given with the help of the metric $\mathsf g_{\mu \nu}$ though 
\begin{equation}
\begin{split}
ds^2 = \mathsf g_{\mu \nu} d\mathsf x^\mu d\mathsf x^\nu,
\end{split}
\end{equation}
where summation over all repeated indexes $\mu, \nu$ is implicit and where $x^\mu = (t,\bx)$. 

A coordinate transformation should not change $ds^2$. But since $d\mathsf x'^\mu = \Lambda_\rho^\mu d\mathsf x^\rho$ (with $\Lambda_\rho^\mu = \partial_{\mathsf x^\rho} \mathsf x'^\mu$), we also need to have $\mathsf g'_{\mu \nu}(\mathsf x') \Lambda_\rho^\mu \Lambda_\sigma^\nu = \mathsf g_{\rho \sigma}(\mathsf x)$, or equivalently, $ \mathsf g'_{\mu \nu}(\mathsf x')  =  \mathsf g_{\rho \sigma} (\mathsf x) (\Lambda^{-1})^\rho_\mu (\Lambda^{-1})^\sigma_\nu$. Objects that transform like $d\mathsf x^\mu$ are called vectors. Objects that transform like a product of $d\mathsf x$ and $\mathsf g$ (eventually with contractions) are called tensors. The rank of the tensor is the number of subscript indexes (covariant) and of superscript indexes (contravariant). The product or contraction of tensors is also a tensor.

\subsubsection{Covariance}

In order to build a theory that do not depend on coordinate system that we chose, we need it to be \textit{covariant}, so that its equations of motions are scalars (tensors of rank $0$). Since they generally contain derivatives, we need the notion of covariant derivative, i.e., such that the derivative of a tensor is still a tensor. We will note $\nabla_\mu$ the covariant derivative. It shall have all properties of derivatives (Leibniz rule and linearity). Ordinary derivatives of a scalar is a tensor. Hence, on scalars, we chose $\nabla_\mu =\partial_\mu $.\footnote{
Because $\partial_\mu$ is a tensor when acting on scalars, a particular representation of a vector is $V = V^\mu \partial_\mu$. A particular basis of vectors will then be identified to the different derivatives along each coordinates. Similarly, the covariant equivalent of vectors (or 1-form) are represented as $U = U_\mu dx^\mu$. Hence, the differentials are a basis of $1$-forms.}
The covariant derivative of any tensor can then be defined from the one of a vector using Leibniz rule ($\forall e, e^\nu \nabla_\mu T_\nu = \partial_\mu (T_\nu e^\nu) - T_\nu \nabla_\mu e^\nu $ for example). Moreover, given a basis constant $e_i^\mu$ (such that $\partial_\nu e_i^\mu = 0$, the covariant derivative of any vector ($V^\mu = \sum_i a_i e_i^\mu$) reads $\nabla_\nu V^\mu = \partial_\nu V^\mu + \sum_i a_i \nabla_\nu e_i^\mu$. Defining the Christoffel symbols as $\forall i, \Gamma_{\nu \rho}^\mu e_i^\rho = \nabla_\nu e_i^\mu $, we get $\nabla_\nu V^\mu = \partial_\nu V^\mu + \Gamma_{\nu \rho}^\mu V^\rho$.

Two more conditions will be imposed to our covariant derivative, so that it is unique: It shall be metric (i.e., $\nabla_\mu \mathsf g_{\nu \rho}=0$) and without torsion (i.e., for all scalar $\varphi$, $\nabla_\mu \nabla_\nu \varphi = \nabla_\nu \nabla_\mu \varphi$). These two conditions impose that 
\begin{equation}
\begin{split}
\Gamma_{\nu \rho}^\mu = \frac{1}{2} \mathsf g^{\mu \sigma} \left (\partial_{\nu} \mathsf g_{\rho \sigma} + \partial_{\rho} \mathsf g_{\nu \sigma} -\partial_{\sigma} \mathsf g_{\nu \rho}\right ). 
\end{split}
\end{equation}

An other notion that is needed is the notion of covariant volume. Indeed the naive volume $d\mathsf x^1\cdots d\mathsf x^n$ (where $n$ is the dimentionality of space-time) is not conserved when making a coordinate transformation. On the contrary, the invariant volume element is defined by $dV = \sqrt{-\mathsf g} d\mathsf x^1\cdots d\mathsf x^n$ where $\mathsf g = \mathsf g_{\mu_{1}\nu_{1}} \cdots \mathsf g_{\mu_{n}\nu_{n}} \epsilon^{\mu_1\cdots \mu_n} \epsilon^{\nu_1\cdots \nu_n} $ is the determinant of the metric and $\epsilon$ is the fully antisymmetric symbol with $\epsilon_{1\cdots n } = 1$. The interested reader can check that for all scalar function $f$, $\int dV f$ does not depend on the chosen coordinate system.

\subsubsection{Test particle}

We consider here the trajectory of a free particle inside a curved geometry. As in special relativity, its action reads $S = - m c \int \sqrt{-ds^2}$. The trajectory is parametrized by $\mathsf x^\mu(\lambda)$ where $\lambda$ is a parameter of the trajectory and the $4$-velocity of the particle is a vector defined by $u^\mu = d\mathsf x^\mu / d\lambda$. Using these notations, the action reads $S = - m c \int d\lambda \sqrt{- \mathsf g_{\mu \nu} u^\mu u^\nu }$. The Lagrange equation of motion hence reads 
\begin{equation}
\begin{split}
\frac{d}{d\lambda} \frac{\mathsf g_{\mu \nu} u^\mu}{\sqrt{-ds^2}} = 0,
\end{split}
\end{equation}
or equivalently, noticing that $\frac{d}{d\lambda} = u^\rho \partial_\rho$, 
\begin{equation}
\label{eq:eomtestpartnaive}
\begin{split}
u^\rho \partial_\rho u^\mu + u^\sigma u^\rho \mathsf g^{\mu \nu}\partial_\rho \mathsf g_{\sigma \nu} = u^\mu \frac{d}{d\lambda} \log(\sqrt{-ds^2}).
\end{split}
\end{equation}
We choose $\lambda$ to be the proper time $\tau$ of the particle (defined by $ds^2= -d\tau^2$) so that $\mathsf g_{\mu \nu} u^\mu u^\nu = 1 $. The rhs of Eq.~\eqref{eq:eomtestpartnaive} is then $0$. Moreover, writing $\partial_\rho \mathsf g_{\sigma \nu}$ in term of Christoffel symbols and recognizing covariant derivative, equation of motion takes the form of a geodesic equation:
\begin{equation}
\begin{split}
u^\rho \nabla_\rho u^\mu =0.
\end{split}
\end{equation}
Free particles in curved space-time hence follow geodesic. When choosing a coordinate system, the term $du/d\tau$ corresponds to the acceleration of the particle and the term $ \Gamma^\mu _{\nu \rho} u^\nu u^\rho$ corresponds to the gravitational force divided by the inertial mass. We see here that both inertial and gravitational mass are necessarily equal.

\section{Field theory}
\label{sec:fields}

\subsection{Local Lorentz invariant theory}

\subsubsection{Free field in flat space-time}

Before considering field theory, let us consider a system of $N$ identical free particles (relativistic or not). The Hamiltonian of such a system is $H_N = \sum E_n$, where $E_n$ is a function of $p_n$, momentum of the $n^{\text th}$ particle. It is its kinetic energy ($E_n = p_n^2/2m$ if the particle is not relativistic, and $E_n = mc^2 \sqrt{1 + p_n^2 / m^2c^2}$ if it is relativistic). A convenient basis to describe (a representation of) the quantum states is the eigenvectors of momentum operator $\hat p_n \left | \Psi_n^{p_n}\right > = p_n \left | \Psi_n^{p_n}\right >$. A basis for states of the whole system is then a tensor product of such $1$-particle state. 

Because all particle are identical, a state should be identified to itself when inverting two of the particles. They are thus linked by a phase. The only representations from permutation group to phase $U(1)$ are the trivial one and the signature. There will then be two kind of particles. The ones for which the representation is signature are called fermions. Their state is anti-symmetric when inverting to particles. When the representation is the trivial one, they are called bosons. Their state is symmetric when inverting two particles. From now on, we shall only consider bosons. The Hilbert space describing $N$ bosons is $\mathcal{H}_N$, engendered by symmetrical tensor product of $\left | \Psi_n^{p_n}\right >$

When the number of particles is not fixed, the total Hilbert space describing the state is thus $\oplus_N \mathcal{H}_N$, and the Hamiltonian is the sum of $H_N$, each of which act on $\mathcal{H}_N$ only. However, this description is not convenient and not practical. For this reason, we introduce creation operator $a^\dagger_{p_n}$ that \enquote{creates a particle of momentum $p_n$} by $a^\dagger_{p_n} \left | \Psi\right > =  \sqrt{\hbar (n_{p_n}+1)} \left | \Psi_n^{p_n}\right > \otimes \left | \Psi\right > $, where the symmetrization is implicit and $n_{p_n}$ is the number of particles of momentum $p_n$. Compare for example with Eq.~\eqref{eq:defcreatannihilop}. Using such definition, one can show that the Hamiltonian reads (in all $\mathcal{H}_N$ and thus in the total Hilbert space) $\hat H = \sum_{p_n} E(p_n) \hat a^\dagger_{p_n} \hat a_{p_n}$. To avoid ponderous occurences of $\hbar$, we introduce $k = p / \hbar$ and $a^\dagger_k = a^\dagger_p / \sqrt{\hbar}$. Using such operators, the quantum Hamiltonian for a system of free identical bosons, the number of which is not fixed reads, in the limit of infinite volume (so that momentum is continuous parameter)
\begin{equation}
\begin{split}
\hat H = \int d\bk E(k) \hat a^\dagger_{\bk} \hat a_{\bk}.
\end{split}
\end{equation}

In order to recover canonical formalism, inspired from Eq.~\eqref{eq:qkloftosharm}, we introduce canonical operators $\hat \phi_\bk \doteq \frac{\hat a^\dagger_{-\bk} + \hat a_{\bk}}{\sqrt{2 E(k)}}, \hat \pi_\bk \doteq i \frac{\hat a^\dagger_{-\bk} - \hat a_{\bk}}{\sqrt{2 / E(k)}}$. Their canonical commutator is $[\hat \phi_\bk,\hat \pi_{\bk'}] = i  \delta(\bk - \bk')$. Using such canonical fields, the Hamiltonian reads, up to the addition of a constant,
\begin{equation}
\begin{split}
\hat H = \int d\bk \frac12 \hat \pi_\bk^\dagger \hat \pi_\bk + \frac{E(k)^2 }{2 } \hat \phi_\bk^\dagger \hat \phi_\bk .
\end{split}
\end{equation}
To switch to representation position, one introduces the fields $\hat \phi(z) = \int \frac{d\bk}{\sqrt{2\pi}^d} \ep{i \bk \bz} \hat \phi_\bk$, and identically for $\hat \pi$. Using these fields, the Hamiltonian for a free particle reads $\hat H =\frac{1}{2} \int d\bz \hat \pi^2 + \phi E^2(i\partial_\bz) \hat \phi$. If the particle is relativistic, the Hamiltonian thus reads
\begin{equation}
\begin{split}
\hat H = \int d\bz \frac{\hat \pi(\bz)^2}{2} + \frac{m^2 c^4}{2 } \hat \phi(\bz)^2 + \frac{\hbar^2 c^2}{2 } (\partial_\bz \hat \phi)^2.
\end{split}
\end{equation}
Because when considering relativistic settings, one can in principle create particle out of energy, the total number of particle in relativistic quantum theory is not conserved and the field theory becomes very relevant. Because of similarities with Eq.~\eqref{eq:Hampparamcouple}, such a system can be considered as a bunch of harmonic oscillators coupled to their closest neighbors.

\subsubsection{In a general space-time}

In general relativity, equations of motion must be written in a covariant way (independently of any choice of coordinate system). We already saw that the Hamiltonian is not covariant but that the action is. Because fields are functions of space-time (the action is only defined in Heisenberg representation), we write the action as $S = \int d\mathsf x \sqrt{-\mathsf g} \mathcal{L}$ where $\mathcal{L}$ is the Lagrangian density and is a scalar. As in the Legendre transformation, we have $\mathcal{L} = \pi \partial_t \phi - \mathcal{H}$, where $\mathcal{H}$ is the Hamiltonian density. The Lagrangian density then reads for a free field in flat space-time
\begin{equation}
\begin{split}
\mathcal{L} = \frac{(\partial_t \hat \phi)^2}{2} - \frac{c^2}{2} (\partial_\bz \hat \phi)^2 - \frac{m^2 c^4}{2 \hbar^2} \hat \phi^2,
\end{split}
\end{equation}
and in a covariant way as
\begin{equation}
\begin{split}
\mathcal{L} = \frac{-c^2}{2} \mathsf g^{\mu\nu}\partial_\mu \hat \phi \partial_\nu \hat \phi - \frac{m^2 c^4}{2\hbar^2} \hat \phi^2,
\end{split}
\end{equation}
where $\mathsf g^{\mu \nu}$ is the inverse of the metric $\mathsf g_{\mu \nu } \mathsf g^{\nu \rho} = \delta_\mu^\rho$, and $\mathsf g_{\mu \nu } d\mathsf x^\mu d\mathsf x^\nu = -c^2 dt^2 + d\bz^2$ [see Eq.~\eqref{eq:Minkowskimetric}].

Such an expression allows to immediately generalize to curved space-time. In the case of minimal coupling\footnote{The field could be coupled to the curvature, in which case, such term disappears in flat space. We assume here it is not the case.}, the action for a free scalar field in a curved metric reads
\begin{equation}
\begin{split}
S = \frac{-1}{2 \hbar}\int c dt d\bz \sqrt{-\mathsf g} \left [ \mathsf g^{\mu\nu}\partial_\mu \hat \phi \partial_\nu \hat \phi +\frac{ m^2 c^2}{\hbar^2} \hat \phi^2 \right ].
\end{split}
\end{equation}
From such an action, the equations of motion read
\begin{equation}
\begin{split}
\frac{1}{\sqrt{-\mathsf g}} \partial_\nu\sqrt{-\mathsf g} \mathsf g^{\mu\nu}\partial_\mu \hat \phi +m^2 c^2 \hat \phi =0.
\end{split}
\end{equation}
$(1/\sqrt{-\mathsf g}) \partial_\nu\sqrt{-\mathsf g} \mathsf g^{\mu\nu}\partial_\mu = \nabla^\mu \nabla_\mu$ (where $\nabla$ is the covariant derivative) is the D'Alembert operator associated to the curved space-time. Under this form, the equations of motion are trivially covariant.

Moreover, after we choose a coordinate system, the metric decomposes into $ds^2 = - N^2 c^2 dt^2 + g_{ij}(dX^i + N^i dt) (dX^j + N^j dt)$, where we introduced the lapse $N$ and the shift $N^i$, and where indexes $i,j$ run on space variables. The canonical momentum associated to $\phi$ is then
\begin{equation}
\begin{split}
\hat \pi = \frac{\delta \mathcal{L}}{\delta \partial_t \hat \phi} = \frac{1}{N^2}\partial_t \hat \phi - \frac{N^i}{N^2}\partial_{i} \hat \phi
\end{split}
\end{equation}
and the Hamiltonian reads
\begin{equation}
\label{eq:hamiltoniengeneralRG}
\begin{split}
H = \frac{1}{2} \int d\bz \frac{\sqrt{-\mathsf g}}{c} \left [ (N \hat \pi + N^i /N \partial_{i} \hat \phi )^2 + c^2 g^{i j} \partial_{i} \hat \phi \partial_{j} \hat \phi + \frac{m^2 c^4}{\hbar^2} \hat \phi^2 \right ]
\end{split}
\end{equation}
where $g^{i j}$ is the space component of $\mathsf g^{\mu \nu}$ (it is not the inverse of $g_{ij}$). 

For the Hamiltonian defined in Eq.~\eqref{eq:hamiltoniengeneralRG} to be a conserved energy, space-time needs to be stationary (i.e., there needs to exist a time-like Killing vector\footnote{
Killing vectors are generators of symmetries of space-time. See \ref{sec:Killingalgebra} for a more accurate definition.
}) and that the Killing vector reads in the chosen coordinate system $K_t = \partial_t$. When this Killing vector is time-like in all space-time, we can chose the space coordinates such as $g^{ij}$ be positive definite. This implies that the Hamiltonian is bounded from below and the system is stable. When it is no longer the case, the hyper-surface defined by $K_t^\mu K_t^\nu \mathsf g_{\mu \nu}=0$ (i.e., such that $K_t$ is no longer time-like) is called a Killing horizon. This geometry is the one for stationary black hole. In this case, the Hamiltonian is no longer positive operator and an \textit{energetic} instability can occur, i.e., there exist some negative energy states. This \textit{energetic} instability is responsible for the emission of a particle flux by the black hole. This is Hawking radiation.

\subsubsection{In cosmology}
\label{sec:encosmo}

An other case where we can define a preferred time is when space is homogeneous and isotropic, as in cosmology. In such a case, homogeneity implies the existence of $3$ Killing vectors generating the translations in space. Time is then generated by the vector orthogonal to these three. In cosmology, space is filled by a fluid (the particles of which being clusters). The preferred referential is then the one in which the fluid is at rest. Hence, the (comobile) time is the time for an observer at rest with respect to the fluid. Assuming one dimensional space, the space-time metric reads
\begin{equation}
\label{eq:FLRWspacetime}
\begin{split}
ds^2 = -c^2 dt^2 + a(t)^2 dz^2 ,
\end{split}
\end{equation}
where $a$ is the scale factor and is a function of time. Eq.~\eqref{eq:FLRWspacetime} defines a Friedman-Lemaître-Robertson-Walker (FLRW) space-time. Because there exists a preferred time, we can express the Hamiltonian and canonical momentum associated to it. It does not have exactly the form of a parametric amplifier because of the prefactor $\sqrt{-\mathsf g} = a c$. It is however possible to consider the renormalized field $\tilde \phi = \sqrt{a} \hat \phi$. The Lagrangian for such field reads
\begin{equation}
\begin{split}
L = \frac{1}{2} \int dz (\partial_t \tilde \phi)^2 - \frac{c^2}{a^2}(\partial_z \tilde \phi)^2 - \frac{\tilde m^2 c^4}{\hbar^2} \tilde \phi^2 - \partial_t\left ( H \tilde \phi^2/2 \right ),
\end{split}
\end{equation}
where $\tilde m^2 c^4 = m^2 c^4 - (H/2)^2 - \partial_t (H/2)$ and $H = \dot a /a$ is the Hubble parameter. The last term is a total derivative and shall play no role in the equations of motion. The canonical momentum associated to $\tilde \phi$ is then
\begin{equation}
\label{eq:HamiltoniantildephiFLRW}
\begin{split}
\tilde \pi = \frac{\delta L}{\delta \partial_t \tilde \phi} = \partial_t \tilde \phi
\end{split}
\end{equation}
and the Hamiltonian $H = \frac{1}{2}\int dz \left [ \tilde \pi^2 + \frac{c^2}{a^2}(\partial_z \tilde \phi)^2 + \tilde m^2 c^4 \tilde \phi^2 \right ]$ reads in Fourier components:
\begin{equation}
\label{eq:Hamiltonienrescaled}
\begin{split}
H = \frac{1}{2}\int d\bk \left [ \abs{ \tilde \pi_\bk}^2 +\omega_k^2 \abs{\tilde \phi_\bk}^2 \right ].
\end{split}
\end{equation}
Here, $\omega_k^2 = \tilde m^2 c^4/\hbar^2 + c^2 k^2/a^2$ depends on time. We hence have a collection of (complex) parametric amplifier labeled by $\bk$. Hence, by a similar reasoning than the one leading to Eq.~\eqref{eq:qdecompcanon}, we get
\begin{equation}
\begin{split}
\tilde \phi_\bk = \hat a_\bk^{\rm in} \phi_k^{\rm in} + (\hat a_{-\bk}^{\rm in})^\dagger \left [\phi_k^{\rm in}\right ]^*=\hat a_\bk^{\rm f} \phi_k^{\rm f} + (\hat a_{-\bk}^{\rm f})^\dagger \left [\phi_k^{\rm f}\right ]^*,
\end{split}
\end{equation}
where $\phi_k^{\rm in/f} \underset{t\to \pm \infty}\sim \ep{- i \omega_k t}/ \sqrt{2\omega_k}$ are the canonical modes for initial and final times. The Bogoliubov~\eqref{eq:bogosuroperateur} transformation then reads
\begin{equation}
\begin{split}
\hat a_\bk^{\rm f} = \alpha_k \hat a_\bk^{\rm in} + \beta_k^* (\hat a_{-\bk}^{\rm in})^\dagger.
\end{split}
\end{equation} 
Moreover, the initial vacuum is expressed in term of final states [see Eq.~\eqref{eq:videinsurbaseout}]:
\begin{equation}	
\begin{split}
\left | 0\right >_{\rm in} = \prod_{k>0} \frac{1}{\sqrt{\abs{\alpha_k}}} \ep{\frac{\beta_k^*}{\alpha_k^*} a_\bk^\dagger a_{-\bk}^\dagger}\left | 0\right >_{\rm f}.
\end{split}
\end{equation}
There have been here production of pairs of particles preserving homogeneity and isotropy of space. The density of particles of momentum between $\bk$ and $\bk+d\bk$ is $n_k d\bk$ where
\begin{equation}
\begin{split}
n_\bk \doteq \left < 0\right |_{\rm in} (\hat a_\bk^{\rm f})^\dagger \hat a_\bk^{\rm f} \left | 0\right >_{\rm in} = \abs{\beta_k}^2.
\end{split}
\end{equation}
These particles have been created by the evolution of space-time. The second quantity that characterizes the state and that shall be of great use in \ref{chap:separability} is the term defined by
\begin{equation}
\begin{split}
c_k \doteq \left < 0\right |_{\rm in} \hat a_{-\bk}^{\rm f} \hat a_\bk^{\rm f} \left | 0\right >_{\rm in} = \alpha_{k} \beta_k^*.
\end{split}
\end{equation}
It quantifies the correlation between particles of opposite momenta.

\subsection{Local Lorentz symmetry breaking}

\subsubsection{Introduction of dispersion}
\label{sec:dispfield}

In this section we wish to break local Lorentz invariance by introducing a non relativistic dispersion relation, e.g., $E^2 = m^2 c^4 + c^2 \hbar^2 k^2 + k^4 / \Lambda^2$. The motivations have been given in the introduction and come both from analogue gravity side and from the transplanckian problem. However, any non relativistic dispersion relation breaks Lorentz invariance, implying the existence of a privileged frame. In order to keep the full covariance of the theory, we introduce a time-like unit vector field $u$. To keep second order in time and linear equations of motion, the action for the scalar field is necessarily of the form, see e.g.,~\cite{Jacobson:2000gw},
\begin{equation}
\begin{split}
S = \frac{1}2 \int c d\mathsf x \sqrt{-\mathsf g} \left [ (u^\mu \partial_\mu \phi)^2 - \phi f( \nabla_\mu \perp^{\mu\nu} \nabla_\nu, [\nabla_\mu, u^\mu] , (u^\mu \nabla_\mu u^\nu) \nabla_\nu) \phi \right ]
\end{split}
\end{equation}
where $ \perp^{\mu\nu} \doteq g^{\mu\nu}- u^\mu u^\nu$ is the space part of the metric and $\nabla_\mu$ is the covariant derivative. In the relativistic case, $f(x,y,z) = m^2 - c^2 x$. The first argument of $f$ is a term introducing dispersion in $k^2$. The second in independent of $k$. It introduces a non minimal coupling to the field with its environment. The third one is proportional to the acceleration of $u$. Because $u$ is a unit-norm field, its acceleration is orthogonal to it. It hence defines a preferred direction in space, and odd terms in the dispersion relation.

In the full theory, the field $u$ has to be considered as dynamic as well as the metric. The theory that fixes the mean value of the field $u$ is unknown to us. But because this theory mixes the vector field and the metric, we shall suppose that the symmetries of this two fields are similar. In particular, the Killing fields commute with $u$. In a FLRW space-time, $u$ is then expressed in the frame of Eq.~\eqref{eq:FLRWspacetime} as $u = c \partial_t$ and hence $[D_\mu, u^\mu]=c H$, $u^\mu D_\mu u^\nu=0$. Neglecting non minimal coupling, the Hamiltonian reads in term of renormalized fields [see discussion before Eq.~\eqref{eq:Hamiltonienrescaled}]
\begin{equation}
\begin{split}
H = \frac{1}{2}\int dz \left [ \tilde \pi^2 + \tilde\phi \tilde f\left (\frac{1}{a^2}\partial_z^2\right ) \tilde\phi \right ],
\end{split}
\end{equation}
where $\tilde f$ is linked to $f$ the same way $\tilde{m}$ was linked to $m$. The only modification with respect to \ref{sec:encosmo} is hence at the level of definition of $\omega_k^2 = \tilde f\left (\frac{-k^2}{a^2} \right ) $. In \ref{sec:pfour}, we shall compute $\beta_k$ in a simple example of dispersive quantum field theory.

\subsubsection{kinematics}
\label{sec:kinematicsofdissip}

We here consider the kinematics for an interaction between particles having a superluminal dispersion relation ($\omega^2 = k^2 + k^4 / \Lambda^2$) in a flat space-time of any dimension. We show that such a particle is instable and can decompose into two other particles, i.e., the full theory necessarily contains dissipation.

We first consider the reaction $1 \to 2 + 3$ where each number $i$ represent a particle of energy $\omega_i$ and momentum $\bk_i$. Total energy and impulsion are conserved quantities:
\begin{equation}
\begin{split}
\omega_1 = \omega_2 + \omega_3, \quad \bk_1 = \bk_2 + \bk_3.
\end{split}
\end{equation}
Using dispersion relation and considering the square of these equalities, we get
\begin{equation}
\begin{split}
[ u_2 u_3+ \frac{ 2 k_2 k_3(u_2 u_3)^2 + 2 (k_2^2 + k_3^2) u_2 u_3 +  k_2 k_3 }{\Lambda^2} ]^2 = ( 1+ \frac{k_2 ^2}{\Lambda^2})(1 + \frac{k_3^2}{\Lambda^2}),
\end{split}
\end{equation}
where $\bk_i = k_i u_i$. In the limit $k_2,k_3\ll \Lambda$, it is solved by
\begin{equation}
\begin{split}
u_2 u_3 =  1- \frac{3(k_2 + k_3)^2}{2\Lambda^2} .
\end{split}
\end{equation}
In the relativistic case, $\Lambda \to \infty$, we necessarily have $u_2 u_3 =  1$. The two final particles propagate in the same direction with the same speed. The final state is hence identical to the initial one: reaction $1 \to 2+3$ is impossible. In the presence of dispersion, however, $u_2 u_3 \neq 1$. The final particles hence propagate in different directions. Reaction $1 \to 2+3$ is hence kinetically possible. When considering only one momentum $k$, the other ones act as an environment and particles are dissipated.

\subsubsection{Free field with dissipation}

When considering free field, no interaction occur and reaction $1 \to 2+3$ is impossible. If we wish to get dissipation without introducing interaction (theory for an interacting field in curved space-time is not yet known), it is necessary to introduce by hand environmental degrees of freedom and to couple our field to it as we did in Eq.~\eqref{eq:dissipHforQM}. If the goal is to get dissipation, it is also necessary to have, for each momentum $k$, a continuous set of environmental degrees of freedom. The environment then lives in a space-time having one more dimension. The most general quadratic action then decomposes $ S_T = S_\phi + S_\Psi + S_{\rm int}$ where $S_{\phi}$ is the free dispersive field action of \ref{sec:dispfield}, $S_\Psi$ is the action for the free environment $\Psi$, and $S_{\rm int}$ is responsible for the interaction. More details will be given in \ref{chap:dissipdS} and in \ref{part:analoguegravity}.

\section*{Conclusion}
\addcontentsline{toc}{section}{Conclusions}

In this chapter, we introduced the notion of parametric amplifier both in classical and quantum mechanics. We observed that many notions are common to both theories. In particular, the notion of Bogoliubov transformation and the increase of the adiabatic invariant are already present in classical mechanics. The latter translates in quantum mechanics by an excitation of the system. The main difference between classical and quantum theories is at the level of the ground state. Indeed, in classical mechanics, the ground state remains the ground state as time evolves. On the contrary, in quantum mechanics, the ground state of a parametric amplifier gets excited. In fact, the final level is increased by a multiple of two and the transition probability is given by the Bogoliubov transformation. 

We then introduced the notion of quantum field theory and observed that for free fields in homogeneous curved space-time, one can consider the field as a collection of non coupled parametric amplifiers. The quantum excitation of vacuum then translates for the fields as spontaneous creation of pairs of particles. On the other hand, initial particles also induce creation of pairs as in a classical (wave amplification) process. The study of these pairs of particles in many systems is the main subject of this thesis. In the next chapter, we shall see a tool that allows us to determine whether spontaneous or induced pair creation contributed the most.

\chapter{Notion of separability}
\label{chap:separability}

\section*{Introduction}

When studying quantum field theory in analogue gravity, a key challenge is to identify whether the produced particles are due to spontaneous or induced emission. 
Indeed, the created particles come both from amplification of initially present (thermal) noise and from the quantum process of vacuum amplification. To answer this question, the right notions are entanglement and separability.

In this chapter, we first review some basics of quantum mechanics and entanglement together with the definition of the separability criterion. As a second step, we recall some useful tools to determine whether a state is separable or not. We also show that for Gaussian states, the separability criterion often used is also a non separability criterion, i.e., that \textit{only} separable states verify that condition. 

An even more practical tool for Bose Einstein condensates using directly observable quantities will be derived in \ref{sec:BECseparability}. No dynamics is considered in this chapter.

\minitoc
\vfill

\section{Quantum mechanics}

\subsection{Algebraic approach}
\label{sec:QMalgapproach}

We recall first the basics of quantum mechanics. Given a phase space, the different observables are functions of the canonical quantities, e.g., $\hat q, \hat q^2, \hat p, \cdots$. Observables can be summed, multiplied by (complex) scalars or multiplied to each other. In mathematical term, they form an algebra $\mathcal{A}$. A state $\omega$ is then an application which associates a number to an observable. This number physically corresponds to the mean value of the observable (when making many experiments with the same state $\omega$). The state is supposed to be a linear function of observables: $\omega( \lambda \hat O_1 + \mu \hat O_2)= \lambda \omega(\hat O_1 )+ \mu \omega(\hat O_2) $. We also impose $\omega(\hat 1) = 1$, which translates the fact that the mean value of a number is this number. In the following, we shall drop the notation $\hat 1$. 

In classical mechanics, we also supposed that $ \omega( \hat O_1 \hat O_2) = \omega( \hat O_1) \omega( \hat O_2)$. This implies that mean values can be identified with the observables [$q \equiv \omega(\hat q)$]. In particular, this means that the spread in measurement of the canonical quantities is null. This fails to be true in the real world because of Heisenberg inequalities, and this is the basics of quantum mechanics. As a consequence, one can no longer suppose that $\omega$ commutes with multiplication. As an example, given some observable $\hat O$, the quantity
\begin{equation}
\Delta O^2 \doteq \omega\left (\left [\hat O- \omega(\hat O)\right ]^2   \right )  = \omega(\hat O^2) - \omega(\hat O)^2
\end{equation}
gives the variance of the measurement. This implies some probabilistic interpretation for the observables.

Before considering equations of motion, two additional steps need to be taken. First, a $*$ structure is added to the observables by defining an antilinear operator [$(\hat A+ \lambda \hat B)^\dagger = \hat A^\dagger+ \lambda^* \hat B^\dagger$] such that its square is identity [$(\hat A^\dagger)^\dagger =\hat  A$] and that it acts as reflection [$(\hat A\hat B)^\dagger = \hat B^\dagger \hat A^\dagger$]. This means that $\dagger$ is an adjoint operation. To fix uniquely this quantity, one supposes that canonical coordinates are self adjoint ($\hat q^\dagger  = \hat q$, $\hat p^\dagger  = \hat p$). Second, the state is supposed to be positive, i.e., $\omega(\hat O^\dagger ) = \omega(\hat O)^*$ and $\omega(\hat O^\dagger \hat O)\geq 0$. This is to recover that the mean values of position and momenta are real values, that their variances are positive. 

Given all these properties, one verifies that if $\omega_1$ and $\omega_2$ are two states, and if $\alpha\in ]0,1[$, then $\alpha \omega_1 + (1- \alpha)\omega_2$ is also a state. States for which such a decomposition (other than the trivial one $\omega_1 = \omega_2$) exists are called mixed states. Others are called pure states.

One can show that when phase space is finite dimensional, there exists a unique Hilbert space $\mathcal{H}$ (up to unitary transformation) and an irreducible representation $\pi$\footnote{
$\pi(\hat  O_1 \hat O_2)=\pi(\hat  O_1)\pi( \hat O_2)$, $ \pi$ is linear, $\pi(\hat O^\dagger) = \pi(\hat O)^\dagger$. For mathematical construction of this Hilbert space, see the GNS construction~\cite{wald1994quantum}. The uniqueness follows from the Stone-von Neumann theorem.} 
of observables [such that $\pi(\hat O)$ is an operator acting on the Hilbert space] such that for any pure state $\omega$, there exists a vector in the Hilbert space $\left |\Psi \right >$ such that for all observable $\hat O$, $\omega(\hat O) = \left < \Psi \right | \pi(\hat O) \left | \Psi \right >$:
\begin{equation}
\begin{split}
\forall \mathcal{A},\ \exists \mathcal{H},\ \exists \pi : \mathcal{A} \to \mathcal{L} (\mathcal{H}),\ \forall \omega,\ \exists \left |\Psi \right >\in \mathcal{H}, \quad 
\omega(\hat O) = \left < \Psi \right | \pi(\hat O) \left | \Psi \right >.
\end{split}
\end{equation}
For mixed states, such a vector does not exist, but there exists an operator $\hat \rho \in \mathcal{L} (\mathcal{H})$ such that $\omega(\hat O) = {\rm Tr} \left [ \hat \rho \pi(\hat O) \right ]$. Because of the assumptions made on the state, this operator is Hermitian, positive, of trace unity. It is called the density matrix of the state. When phase space is of infinite dimension, there may exist many different Hilbert spaces, corresponding to different ground states of the theory. When restricting the class of states (i.e., specifying the ground state), the Hilbert space becomes uniquely defined. In the following, we shall identify observables with their representations, and states with the corresponding vector in Hilbert space. 

In the kinematic description of a system, the canonicity of variables now translates in commutators ($[\hat q,\hat p]= i \hbar $) instead of Poisson Brackets. Using Cauchy Schwartz inequalities, one shows that $\sqrt{\Delta q^2 \Delta p^2} \geq \sqrt{\abs{[\hat q,\hat p]}^2/4  + \left <\{\hat q-\left <q\right >,\hat p-\left <p\right >\}\right > ^2 } \geq \hbar$, which is the Heisenberg uncertainty relation.

When treating the dynamics of the system, one needs the time evolution of $\left <\Psi \right | \hat O \left | \Psi\right >$. To fix it one requires that this coincides with classical equations of motion for the canonical coordinates. One also requires the superposition principle, i.e., equation of motion is linear in $\left | \Psi\right >$. In Schrödinger representation, the observable is supposed to be time independent while the state vary. One can then show that the equation of motion is the Schrödinger equation, Eq.~\eqref{eq:Schodinger} as we saw in Eq.~\eqref{eq:quantumeomqp}. The possibility of writing equations of motion in term of pure state follows from unitarity which imposes that a pure state remains so after time evolution.

To summarize, quantum mechanics follows from three assumptions: unitarity of the evolution, the superposition principle, and the Heisenberg uncertainty principle (at the level of canonical commutators). The fact that time evolution of mean values of canonical coordinate coincide with the classical mechanics predictions fixes the equivalence between classical and quantum system.

\subsection{Mixed states}

\subsubsection{Natural light}

To show the interest of mixed states, we consider in this section the polarization of natural light. If one measures polarization of light in any direction, one will observe that half of the photons pass through the polarizer and half are blocked. One can show that such a system can not be described in term of a pure state $\left | \Psi \right >$. Indeed, vector space of this state is dimension $2$ and an orthonormal basis is $\left | \rightarrow \right >$,$\left | \uparrow \right >$, i.e., states polarized horizontally and vertically. Hence, any state $\left | \Psi \right >$ is written
\begin{equation}
\begin{split}
\left | \Psi \right > = \ep{i \alpha}\cos\theta \left | \rightarrow \right > + \ep{-i \alpha}\sin\theta\left | \uparrow \right >.
\end{split}
\end{equation}
Because state is not polarized when measured with respect to horizontal axes, one deduces that $\theta = \pi /4$. If we rotate the axis and make the same remark, it comes that $\alpha = \pi /4$. But this state has a circular polarization, which is not the case for natural light. QED.

\subsubsection{Density matrix}

To describe general system (including natural light), we then need to represent mixed states. To do so, we use the density matrix of the state. It can be written as statistical superposition of pure states
\begin{equation}
\begin{split}
\hat\rho = \sum_n p_n \left | \Psi_n \right > \left < \Psi_n \right |
\end{split}
\end{equation}
where $p_n \geq 0$ corresponds to some (statistical) probability. $p_n$ can also be considered as the eigenvalues of the density matrix. The mean value of an observable $\hat O$ is then the statistical superposition of the different mean values
\begin{equation}
\begin{split}
 \left < \hat O \right > = {\rm Tr}\left (\hat \rho \hat O\right ) = \sum_n p_n \left < \Psi_n \right | \hat O \left | \Psi_n \right > .
\end{split}
\end{equation}
Pure states can be recognized as they are the only states for which the density matrix satisfies ${\rm Tr}\left (\hat \rho ^2 \right )=1$, or equivalently, they are the only states for which the entropy is $0$, where the entropy of a mixed state is defined by
\begin{equation}
\begin{split}
S= -\sum_n p_n \log p_n = -{\rm Tr}\left ( \hat \rho \log \hat \rho\right ).
\end{split}
\end{equation}
The states with the maximum of entropy are states for which all probabilities equal. They are proportional to identity.

For natural light, the state $\hat \rho_{NP}$ is easily written as 
\begin{equation}
\label{eq:nonpolarizeddensitymatrix}
\begin{split}
\hat \rho_{NP} = \frac{1}{2}\left [ \left | \rightarrow \right >\left < \rightarrow \right | + \left | \uparrow \right > \left < \uparrow \right | \right ].
\end{split}
\end{equation}
Because the matrix is proportional to identity, it reads identically in every orthonormal basis. Hence, for all projector $\hat O = \left | \Psi \right > \left < \Psi \right |$, we have ${\rm Tr}\left (\hat \rho_{NP} \hat O\right ) = 1/2$. QED.

\subsubsection{Reduced state}

When considering a subsystem of some large system, the subsystem is generally characterized by a mixed state. We obtain the corresponding density matrix by tracing all other degree of freedom. The new state is called reduced state. To be explicit, consider an orthonormal basis of states that is factorized $\left | \Psi_A \right >\otimes \left | \Psi_B \right >$, the first part describing the subsystem and the second describing the rest of the system. Then, for any observable $\hat O_A$ on the subsystem $A$, we have
\begin{equation}
\begin{split}
{\rm Tr}\left ( \hat \rho O_A \right ) = \sum_{A,B} \left < \Psi_A \right |\otimes \left < \Psi_B \right | \hat\rho O_A \left | \Psi_A \right >\otimes \left | \Psi_B \right > = \sum_A \left < \Psi_A \right | \hat\rho_A O_A \left | \Psi_A \right > = {\rm Tr}_A\left ( \hat \rho_A O_A \right )
\end{split}
\end{equation}
where $ \hat\rho_A = \sum_B \left < \Psi_B \right | \hat\rho \left | \Psi_B \right >$ is the reduced state. Even in the case where $\hat \rho$ was pure, there is no reason why $\hat \rho_A$ should be also pure. As an example, consider the two photon polarization pure state $\left | \Psi \right > = (\left | \rightarrow \right > \otimes \left | \rightarrow \right > + \left | \uparrow \right > \otimes \left | \uparrow \right >)\sqrt{2}$. After tracing out the second photon, the density matrix describing the state of the first photon is $\hat \rho_{NP}$ of Eq.~\eqref{eq:nonpolarizeddensitymatrix}. We already showed that this state is not pure.

\subsection{Separability}

\subsubsection{Definition}

The notion of separability is linked to the notion of entanglement. We consider here a system composed of two subsystems $A$ and $B$. 

Two kinds of pure state can exist. Either it can be factorized $\left | \Psi \right > =\left | \Psi_A \right >\otimes \left | \Psi_B \right >$, or it cannot. 
When it is factorized, the mean value of any factorized observable $\hat O_A \otimes \hat O_B$ is the product of the two mean values. This case corresponds to having disconnected experiments that were prepared independently. 
When the pure state can not be factorized, it can violate some Bell's inequality, and hence cannot be described by statistical ensemble.

In mixed states, R.F. Werner introduced the notion of separable state~\cite{Werner:1989}. He considered the following apparatus: imagine some random generator picking number $n$ with probability $p_n$. Given the value of $n$, an experimenter prepares the system in the factorized state $\rho_n^{(A)} \otimes \rho_n^{(B)}$. Because the random generator can be considered as classical, the state is \textit{classically entangled}. It reads explicitly
\begin{equation}
\begin{split}
\hat \rho = \sum_n p_n \rho_n^{(A)} \otimes \rho_n^{(B)}.
\end{split}
\end{equation}
States that cannot be written in such a way are called \textit{non separable}, or \textit{quantum mechanically entangled}. 

A state is not separable in itself. The notions depends on the two subsystems $A,B$ we consider. For example, consider some homogeneous state. When considered in Heisenberg representation, the subsystem of given wave number $k$ can be initially separable [when considered the two Hilbert spaces generated by $(\hat a_{\pm \bk}^{in})^\dagger$] and non separable after time evolution [when considered the two Hilbert spaces generated by $(\hat a_{\pm \bk}^{out})^\dagger$]. On the other hand, because we are in Heisenberg representation, the state is the same.

\subsubsection{Interest of the concept}

The concept of separability is a key concept in analogue gravity when one tries to identify some intrinsically quantum phenomenon such as pair particle production from vacuum. Indeed, in analogue gravity, one has three different phenomena that resemble much. The first one is the amplification of some (macroscopic) incident wave. It is not different from the parametric amplification of the classical harmonic oscillator of \ref{sec:classicparamamp}. The second is the amplification of inherent thermal noise. It leads to induced particle production. The third one is the amplification of quantum vacuum leading to spontaneous particle production. When we have only the second channel, the final state is always separable, even though there is coherence. When only the third channel is present, the final state is (maximally) non separable. The fact that the final state is non separable (when the initial state has no coherence) is then the sign that spontaneous particle production from vacuum significantly occurred. 

\section{Peres Horodecki criterion}

We consider here a two mode state (it can be the reduced state of some larger state) engendered by creation operators $a^\dagger$ and $b^\dagger$.

\subsection{Partial transpose}

One of the most general criterion to determine whether our state is separable or not is the Peres-Horodecki criterion. This criterion needs the notion of partial transposed state. It is the density matrix one obtains when transposing only the part corresponding to the subsystem $a$. Formally, this means
\begin{equation}
\begin{split}
\left < i_a \right |\otimes \left < \mu_b \right | \hat\rho \left | j_a \right >\otimes \left | \nu_b \right > = \left < j_a \right |\otimes \left < \mu_b \right | \hat\rho^{T_a}\left | i_a \right >\otimes \left | \nu_b \right > ,
\end{split}
\end{equation}
where $i_a,j_a$ (resp $\mu_b,\nu_b$) are any vectors depending only on $\hat a^\dagger$ (resp $\hat b^\dagger$). If the state is separable, since the (total) transpose of a physical state is a physical state, it is obvious that the partial transpose is a physical state\footnote{A physical state $\hat \rho$ is a hermitian operator of trace unity and such that for all pure state $\left |\Psi\right >$, $\left <\Psi\right | \hat \rho \left |\Psi\right > \geq 0$, or equivalently with all eigenvalues positive. }.

The equivalence statement is not true in general but only when Hilbert space is dimension $2 \times 2$ or $2 \times 3$. In a more general case, one has to generalize the partial transpose, see Ref.~\cite{Horodecki19961}. The generalization is the following: a state $\hat \rho$ is separable if and only if for all positive linear maps $\Lambda$ (from first subspace to the other), $(\Lambda \otimes I) \hat \rho$ is positive. When $\hat \rho$ is separable, because $\Lambda(\rho_n^{(A)}) $ and $ \rho_n^{(B)}$ are positive, the statement is trivially true. The equivalence necessitates more tricky demonstration. We shall not present it here, since we won't use it in the rest of the thesis.

An other case where the equivalence is true is in continuous variables (bipartite) Gaussian system. Because this is a useful case for this thesis, we treat it in the next subsection. It is largely inspired from Ref.~\cite{Simon:2000zz} and includes more details.

\subsection{Gaussian case}

\subsubsection{Wigner and \texorpdfstring{$P$}{P} function}

We consider in this subsection a Gaussian bipartite state. It is then characterized by a Gaussian Wigner function, see Eq.~\eqref{eq:defWigner}. For a bipartite state, there are $4$ canonical variables, namely $Q \doteq (q^a,p^a,q^b,p^b)$ which form a vector. The covariance matrix is then
\begin{equation}
\begin{split}
V \doteq {\rm Tr} \hat \rho Q Q.
\end{split}
\end{equation}
One can show that in term of this, the Wigner function for a Gaussian state is determined only by the covariance matrix and the first momenta $Q_0 \doteq {\rm Tr} \hat \rho Q$.
\begin{equation}
\begin{split}
W(Q) = \frac{\ep{- (Q-Q_0) V^{-1} (Q-Q_0) /2}}{\sqrt{(2\pi)^4 \det V}}.
\end{split}
\end{equation}

In some cases, we can define the $P$ representation of the state as
\begin{equation}
\begin{split}
W(Q) = \int d\tilde Q P(\tilde Q) \frac{\ep{- (Q- \tilde Q)^2}}{\pi^2}.
\end{split}
\end{equation}
Because the Wigner function of a coherent state $\hat a \left | \alpha \right > = (q_0 + i p_0 )\left | \alpha \right > $ has the Gaussian form $W_{\alpha}(q,p)=\ep{- (q- \tilde q)^2- (p- \tilde p)^2} / \pi$, this decomposition implies that
\begin{equation}
\begin{split}
\hat \rho = \int d\tilde Q P(\tilde Q) \left | \alpha_a \right > \left < \alpha_a \right | \left | \alpha_b \right > \left < \alpha_b \right |.
\end{split}
\end{equation}
In the case one can define a $P$ representation that is positive, the state is separable.

We shall here inspire from this $P$ representation to try to decompose our state in term of coherent state in (Gaussian) superposition of Gaussian factorized states, i.e., $W_{(\rho_a \otimes \rho_b)(\bar Q,D)}(Q)=\ep{- (Q-\bar Q) D^{-1} (Q-\bar Q) /2 } / 4 \pi^2 \sqrt{\det D}$ where $D$ is $2 \times 2$ block diagonal, each block being of determinant larger than $1/4$. The state
\begin{equation}
\begin{split}
\hat \rho(Q_0,Z) = \int d \bar Q \frac{\ep{- (\bar Q- Q_0) Z^{-1} (\bar Q-Q_0) /2 } }{4 \pi \sqrt{\det Z}} (\rho_a \otimes \rho_b)(\bar Q,D),
\end{split}
\end{equation}
where $Z$ is a positive symmetric matrix, is the separable state with $ {\rm Tr} \hat \rho Q = Q_0 $ and covariance matrix $V = Z + D$. From this, we deduce that any Gaussian state for which the covariance matrix reads $Z+D$ with $Z$ positive and $D$ block diagonal, each block being of determinant larger than $1/4$ is separable. By writing $Z = \sqrt{D} (D^{-1/2} V D^{-1/2} -1) \sqrt{D}$, we get that if the covariance matrix is such that $D^{-1/2} V D^{-1/2} -1$ is positive, then the state is separable. 

Physically, $D^{-1/2} V D^{-1/2} -1 \geq 0$ means that there exist some canonical transformation on both sectors separately such that the subfluctuant mode (i.e., the smaller eigenvalue of the full covariance matrix) is larger than $1/2$.

\subsubsection{Gauge choice}

We notice that making any Bogoliubov transformation on $a$ and $b$ separately does not modify the separability of the system, since it is equivalent to multiplying the state by a block diagonal operator on the right and on the left\footnote{This operation is often called LOCC: Local Operations and Classical Communication}. Moreover, it does not spoil either the fact that $\hat \rho^{T_a}$ is positive operator. We can then chose a gauge by imposing
\begin{equation}
\begin{split}
{\rm Tr} \hat \rho \hat a \hat a =0, \quad {\rm Tr} \hat \rho \hat b \hat b = 0, \quad {\rm Tr} \hat \rho \hat a^\dagger \hat b \in \mathbb R, \quad {\rm Tr} \hat \rho \hat a \hat b \in \mathbb R , 
\end{split}
\end{equation}
This implies in particular that ${\rm Tr} \hat \rho q_i p_j =0$. With such a choice, the covariance matrix reads
\begin{equation}
\begin{split}
V = \left (\begin{array}{llll}
n & 0 & c & 0\\
0 & n & 0 & c'\\
c & 0 & m & 0\\
0 & c'& 0 & m
\end{array} \right )
\end{split}
\end{equation}
In this gauge, we look for a matrix $D$ diagonal, with each of the two submatrix of determinant $1/4$. It explicitly reads $ 2 D = {\rm diag}\left ( x^2 y^2 ,1/x^2 y^2 , y^2 /x^2,x^2/y^2 \right )$ where $x,y$ are free parameters. This implies that 
\begin{equation}
\begin{split}
 D^{-1/2} V D^{-1/2} = 2 \left (\begin{array}{llll}
nx^2 y^2 & 0 & c y^2& 0\\
0 & n/ x^2 y^2 & 0 & c'/y^2\\
cy^2 & 0 & m y^2 /x^2 & 0\\
0 & c'/y^2 & 0 & m x^2/y^2
\end{array} \right ).
\end{split}
\end{equation}
The eigenvalues of this matrix are
\begin{equation}
\begin{split}
y^2 &\left ( n x^2 + m x^{-2} \pm \sqrt{(n x^2 - m x^{-2} )^2 + 4 c^2} \right ), \\
y^{-2} &\left ( n x^{-2} + m x^{2} \pm \sqrt{(n x^{-2} - m x^{2} )^2 + 4 c'^2} \right ).
\end{split}
\end{equation}
The two with minus sign can be made equal to each other by choice of $y$. Moreover, their product is
\begin{equation}
\begin{split}
P_{\rm low} \doteq &\left ( n x^2 + m x^{-2} - \sqrt{(n x^2 - m x^{-2} )^2 + 4 c^2} \right )\\
&\times \left ( n x^{-2} + m x^{2} - \sqrt{(n x^{-2} - m x^{2} )^2 + 4 c'^2} \right ).
\end{split}
\end{equation}
This, seen as a function of $x$ is extremal for $x^4 = \frac{\abs{c} n+m \abs{c'}}{\abs{c} m+n \abs{c'}}$, and at this value, it is
\begin{equation}
\begin{split}
P_{\rm low}^{\rm max} = 2m^2+2n^2 + 4 \abs{c c'} - 2 \sqrt{4 (\abs{c} n+m \abs{c'})(\abs{c} m+n \abs{c'})+\left(m^2-n^2\right)^2}.
\end{split}
\end{equation}
When this quantity is larger than $1$, $ D^{-1/2} V D^{-1/2} -1 $ is positive and the state is separable. Moreover, 
\begin{equation}
\begin{split}
P_{\rm low}^{\rm max} >1 \Leftrightarrow \left(\frac{1}{4}- \abs{c c'}\right)^2+m^2 n^2 > \frac{m^2 + n^2}{4} +m n \left(c^2+c'^2\right).
\end{split}
\end{equation}
The right hand side of this equivalence is the Peres Horodecki criterion defined in Ref.~\cite{Simon:2000zz}. Under gauge invariant form, it reads in one of the two equivalent forms
\begin{subequations}
\label{eq:pereshoro}
\begin{align}
\label{eq:pereshorosimple}
\det V - \frac{\det A + \det B}{4} - \frac{\abs{\det C}}{2} + \frac{1}{16} &\geq 0,\\
\left (\frac{1}{4}- \abs{\det C}\right )^2 + \det A \det B - \frac{\det A + \det B}{4} - {\rm tr} \left (A J C J B J C^T J \right )&\geq 0,
\end{align}
\end{subequations}
where the different $2\times 2$ matrices are defined by 
\begin{equation}
\begin{split}
V = \left (\begin{array}{ll}
A & C\\
C^T & B
\end{array}\right ), 
\end{split}
\end{equation}
and $J$ is the symplectic matrix.

The conclusion of this subsection is that if Eq.~\eqref{eq:pereshoro} is fulfilled, then there exists some operator $D$ such that $ D^{-1/2} V D^{-1/2} -1 \geq 0$, and hence the state is separable.

\subsubsection{Positivity of the state}

From the positivity of the state, we get the positivity of
\begin{equation}
\begin{split}
f(\alpha,\beta,\gamma,\delta) &\doteq {\rm Tr}\hat \rho \left (\alpha q_a - i \beta p_a + \gamma q_b - i \delta p_b \right )\left (\alpha q_a + i \beta p_a + \gamma q_b + i \delta p_b \right ).
\end{split}
\end{equation}
This is a quadratic function of $\alpha$ that is always positive. The associated discriminant is hence negative. The discriminant is still a quadratic function of $\beta$ and of constant sign. Its own discriminant is hence negative. Repeating once again the operation with $\gamma$, we obtain that\footnote{To obtain this equation in a general gauge, it is necessary to treat both real and imaginary parts of the $4$ coefficients. In the gauge of the previous section, this result falls even when we assume the $4$ coefficients to be real.}
\begin{equation}
\label{eq:positivestate}
\begin{split}
\det V - \frac{\det A + \det B}{4} - \frac{\det C}{2} + \frac{1}{16} &\geq 0.
\end{split}
\end{equation}
The computation is long but not difficult. It won't be presented here.

The covariance matrix of $\hat \rho^{T_a}$ is obtained from $V$ by multiplying by $-1$ the second line and the second column. Hence, $\det V , \det A $ and $\det B$ are unchanged while $\det C$ is multiplied by $-1$. In the case $\hat \rho^{T_a}$ corresponds to a physical state, we then have
\begin{equation}
\label{eq:positivepartialtranspose}
\begin{split}
\det V - \frac{\det A + \det B}{4} + \frac{\det C}{2} + \frac{1}{16} \geq 0.
\end{split}
\end{equation}
Eqs.~\eqref{eq:positivepartialtranspose} and~\eqref{eq:positivestate} imply Eq.~\eqref{eq:pereshoro}. 

To conclude this section, we have that a Gaussian state such that $\hat \rho^{T_a}$ is physical is separable, then implying the equivalence. Moreover, a practical criterion is given by Eq.~\eqref{eq:pereshoro}. 

\subsection{State with symmetries}
\label{sec:homosepcriterion}

In many physical systems, some symmetries occur and we have $\det V = (\sqrt{\det A \det B} - \abs{ \det C})^2$. This happens for example, in homogeneous or stationary systems. In this case, Eq.~\eqref{eq:pereshorosimple} reduces to
\begin{equation}
\begin{split}
\left [ \sqrt{\det V}+\frac{1}{4} - \frac{\sqrt{\det A} + \sqrt{ \det B}}{2} \right ] \left [ \sqrt{\det V } +\frac{1}{4} + \frac{\sqrt{\det A}+ \sqrt{ \det B}}{2} \right ] \geq 0.
\end{split}
\end{equation}
The second term is always positive, and hence, separability is equivalent to 
\begin{equation}
\label{eq:Campocritcovariant}
\begin{split}
\left (\sqrt{\det A} -1/2 \right ) \left ( \sqrt{ \det B}-1/2\right ) \geq \abs{ \det C}.
\end{split}
\end{equation}
In what follows, to fix ideas, we assume homogeneity of the state. It translates identically to stationary systems. The creation and annihilation operators are then labeled by momentum $\bk$. If both sectors are labeled by momenta of different absolute value, then no entanglement can occur. Hence, two interesting cases can occur: Either the state is an entanglement of two co-propagating modes $a^\dagger_\bk$, $b^\dagger_\bk$ or two counter-propagating modes $a^\dagger_\bk$, $b^\dagger_{-\bk}$. Because of homogeneity, in both cases, covariance matrices $A,B,C$ are diagonal. Moreover, $A$ and $B$ are proportional to identity and we have $A = n_a +1/2$, $B = n_b +1/2$ where $n_a = \left < a^\dagger_\bk a_\bk \right >$ and similarly for $n_b$.

In the case where modes are co-propagating, the matrix $C$ is such that $\det C = \abs{d_\bk}^2$, where $d_\bk = \left < a^\dagger_\bk b_\bk\right >$. Because $\det C \geq 0$, positivity of the state [see Eq.~\eqref{eq:positivestate}] implies separability [see Eq.~\eqref{eq:pereshoro}]. Eq.~\eqref{eq:Campocritcovariant} reduces to $\abs{d_\bk}^2 \leq n_a n_b$ is always fulfilled.

In the case modes are counter-propagating, the matrix $C$ is such that $\det C = -\abs{c_\bk}^2$, where $c_\bk = \left < a_\bk b_{-\bk}\right >$. The positivity of the state then implies 
\begin{equation}
\label{eq:Campotoujoursvrai}
\begin{split}
\abs{c_\bk}^2 \leq n_a (n_b +1).
\end{split}
\end{equation}
For separable states, Eq.~\eqref{eq:Campocritcovariant} reduces to
\begin{equation}
\label{eq:Camposeparable}
\begin{split}
\abs{c_\bk}^2 \leq n_a n_b .
\end{split}
\end{equation}
Hence, separable states are such that $\abs{c_k}^2 \leq n_a n_b $ and \textit{Gaussian} homogeneous states such that $\abs{c_k}^2 \leq n_a n_b $ are separable. This criterion was first used in Ref.~\cite{Campo:2005sy} in the case $n_a = n_b$.

To get a measure of the non separability of a state, three different parameters can be introduced and are used in different systems. The first one was introduced in Ref.~\cite{Campo:2008ju} and is $\delta_\bk$ defined by
\begin{equation}
\label{eq:defdeltacampo}
\begin{split}
\abs{c_\bk}^2 = n_a (n_b +1 - \delta_\bk).
\end{split}
\end{equation}
This parameter belongs to $[0,n_b+1]$. $\delta_\bk = 0$ corresponds to maximally entangled states whereas $\delta_\bk = n_b+1$ corresponds to states with no coherence (i.e., thermal)~\cite{Campo:2005sy}. The limit of non separability is given by $\delta_\bk = 1$. 

The second parameter comes from a more pragmatic point of view and is defined by 
\begin{equation}
\label{eq:defDeltasquares}
\begin{split}
\Delta_\bk \doteq n_a n_b - \abs{c_\bk}^2 .
\end{split}
\end{equation}
The third one replaces the definition of Eq.~\eqref{eq:defDeltasquares} in the case where $n_a = n_b = n_k$. In such a case, because the equation factorizes and because $ n + \abs{c}\geq 0$, we define
\begin{equation}
\label{eq:defDeltalinear}
 \begin{split}
\Delta_\bk \doteq n_k - \abs{c_\bk} .
 \end{split}
 \end{equation} 
We use the same notation for the second and the third definition since they do not apply in the same case; no misunderstanding can thus occur. One can show that the isotropic $\Delta_\bk$ obeys $-1/2 < n_k-\sqrt{n_k(n_k+1)} \leq\Delta_\bk \leq n_k$, where the minimal and maximum values characterize, respectively, the pure (squeezed) state and the incoherent thermal state. In addition, whenever $\Delta < 0$, the two-mode state is nonseparable. Notice that $\Delta$ is linked to the logarithmic negativity $E_\mathcal{N}$ introduced in Ref.~\cite{Vidal:2002zz} and used in Refs.~\cite{Horstmann:2009yh,Horstmann:2010xd}. Indeed, for Gaussian two-mode states with $n_\bk = n_{-\bk}$, one finds $E_\mathcal{N} = \max[-\log_2(1+ 2\Delta),0]$ which is positive only if $\Delta <0$.

\section*{Conclusion}
\addcontentsline{toc}{section}{Conclusions}

We saw here an unambiguous way to determine the existence of spontaneous processes, even when $\beta$ and the initial temperature are not known. This allows experiments to test the quantum nature of the vacuum. We also saw a quantifiable criterion that implies non separability and that is equivalent to it in many cases.
\partimage{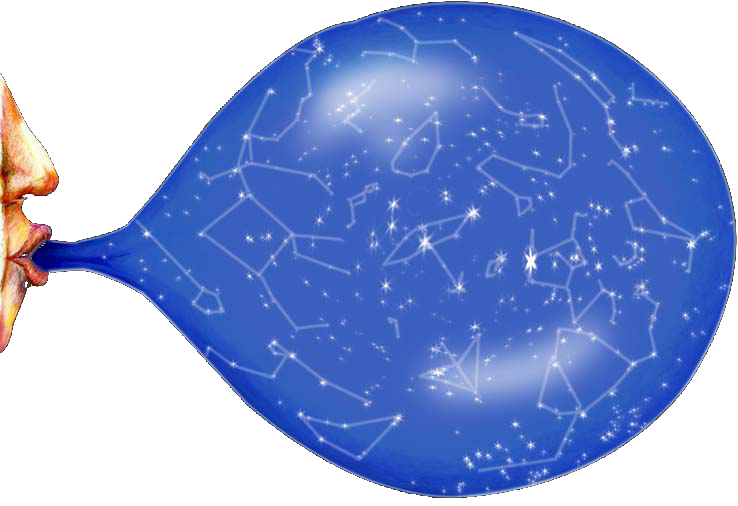}
\partcitation{\textit{We have simply arrived too late in the history of the universe to see this primordial simplicity easily...} \\ Steven Weinberg}
\part{Quantum effects in de Sitter space}
\label{part:desitter}
\chapter*{Introduction}
\phantomsection
\addstarredchapter{Introduction}
\markboth{INTRODUCTION}{INTRODUCTION}

The two main predictions of quantum field theory in curved space, namely black hole \linebreak radiation~\cite{Hawking:1974sw,birrell1984quantum,Brout:1995rd} and primordial spectra in inflation~\cite{Starobinsky:1982ee,Mukhanov:1982nu,Mukhanov:1990me} share many properties. In particular, both spectra stem from vacuum fluctuations with extremely short wave lengths~\cite{Jacobson:1999zk}. They are therefore in principle sensitive to the ultrahigh frequency behavior of the theory. To check this sensitivity, following~\cite{Unruh:1994je}, nonlinear dispersion relations, which break the local Lorentz invariance, have been used in the context of black holes~\cite{Brout:1995wp,Corley:1996ar,Balbinot:2006ua,Unruh:2004zk,Coutant:2011in} and in cosmology~\cite{Martin:2000xs,Niemeyer:2000eh,Niemeyer:2001qe,Macher:2008yq}. In homogeneous cosmology, time dependent modes with a fixed comoving wave vector have been used, whereas for black holes, the analysis was based on stationary modes. In spite of this, the two cases are unexpectedly similar, as we shall show.

Moreover, for free fields, relativistic or dispersive, this pair creation (also called the dynamical Casimir effect in condensed matter physics, see e.g., Refs.~\cite{Carusotto:2009re,PhysRevLett.109.220401}) is associated with the building up of nonlocal correlations that lead to quantum mechanically entangled states~\cite{Campo:2003pa,Campo:2005sv}, see \ref{chap:separability}. For dissipative fields, i.e., fields coupled to an environment, there is a competition between the squeezing of the state, which increases the strength of the correlations, and the coupling to the external bath, which reduces it~\cite{Campo:2005sy,Prokopec:2006fc,Campo:2008ju,Campo:2008ij}. 

In the present part, we analyze dispersive and dissipative fields in de Sitter space for two reasons. First, since de Sitter endowed with a cosmological preferred frame is both homogeneous and stationary, high frequency dispersion can be studied along both approaches. This will allow us to relate them in a very precise way. We shall see that their compatibility relies on a two-dimensional symmetry group which is a subgroup of the de Sitter isometry group~\cite{Eling:2006xg}. Because the generators of the two symmetries do not commute, in each approach only one symmetry is manifest, while the other is somehow hidden. In fact, this extra symmetry has been exploited in the black hole near horizon approximation of Refs.~\cite{Brout:1995wp,Corley:1996ar,Balbinot:2006ua,Unruh:2004zk,Coutant:2011in}, but without noticing (in general) that it relies on properties that are exact in de Sitter space.

Second, the main consequence of high frequency dispersion, that is the loss of the thermality of the spectrum, has raised deep questions concerning the relationships between Lorentz symmetry and black hole thermodynamics~\cite{Jacobson:1991gr,Jacobson:1993hn,Jacobson:1996zs}. It has been claimed that this loss should lead to violations of the second law~\cite{Dubovsky:2006vk,Eling:2007qd,Jacobson:2008yc}. These issues are particularly relevant when working with extended theories of gravity, such as Einstein-aether~\cite{Eling:2004dk} or Horava gravity~\cite{Horava:2009uw}, see Ref.~\cite{Blas:2011ni}. 

The part is organized as follows: in \ref{chap:geom}, we introduce de Sitter space and its symmetries. We also introduce the affine group and show that theories invariant under this group are dispersive and dissipative theories. In \ref{chap:dispdS}, we consider only dispersion. We demonstrate that the Bunch-Davies vacuum is no longer thermal for any (superluminal) dispersion relation. We also show that it is the only stationary stable state. The departures from thermality and the $S$-matrix are then exactly calculated for a quartic superluminal dispersion relation. In \ref{chap:dissipdS}, we present the action which engenders dissipative effects while conserving covariance. Exploiting the homogeneity and stationarity of de Sitter, we compute the spectral properties and the correlations of pairs with opposite momenta and the deviations with respect to the Gibbons-Hawking temperature. We then apply our model to black holes in \ref{chap:BHdesitter} and show that the main result apply in black hole geometries. We work in units where $\hbar =c=1$. 

\chapter{Geometry}
\label{chap:geom}

\section*{Introduction}

De Sitter space-time was first discovered as a solution of the equations of motion for gravitation in vacuum, in the presence of a (positive) cosmological constant. It is thus a maximally symmetric solution with a non vanishing (and positive) curvature. When completing the space time so that the particle reach infinity in infinite proper time, it has been shown that the de Sitter space time is embedded in a flat space-time

We here define the de Sitter space from this embedding and identify its invariances. We then break minimally the symmetry so that local Lorentz invariance is broken. In the last section, we identify the simplest representation for fields and observables on de Sitter that are invariant under the remaining symmetry. 

\minitoc
\vfill

\section{de Sitter space}
\label{sec:desitterspace}

\subsection{Definition}

Consider Minkowski (flat) space-time in $2+1$ dimensions in the Cartesian frame so that the metric reads $ds^2 = -dx_0^2 + dx_1 ^2 +d x_2^2$. The $1+1$ dimensional de Sitter space-time is the hyperboloid of one sheet described by the equation $-x_0^2 + x_1^2 + x_2^2 = H^{-2}$. A coordinate set describing the whole space is given by ($\zeta,\theta$) defined by
\begin{equation}
\begin{split}
H x_0 = \sinh \zeta , \quad H x_1 = \cosh\zeta \cos\theta,\quad Hx_2 = \cosh\zeta \sin\theta.
\end{split}
\end{equation}
Using these coordinates, de Sitter space time's metric is
\begin{equation}
\begin{split}
H^2 ds^2 = - d\zeta^2 + \cosh^2\zeta d\theta^2 ,
\end{split}
\end{equation}
and the whole space-time is covered by $(\zeta,\theta)\in \mathbb{R}\times [-\pi,\pi[$, with $(\zeta,-\pi)$ identified to ($\zeta,\pi$). It is then geodesically complete, in the sense that any geodesic reaches infinity with an infinite proper time. This property is inherited from the Minkowski space time. 

Given this definition, the de Sitter space-time has constant curvature and is maximally symmetric, i.e., the Riemann curvature tensor reads $R_{\mu \nu \rho \sigma} \propto \mathsf g_{\mu \rho } \mathsf g_{\nu \sigma} - \mathsf g_{\nu \rho } \mathsf g_{\mu \sigma} $.

\subsection{Poincaré patch}

\begin{figure}[ht]
\begin{center}
\includegraphics[width = 0.8 \linewidth]{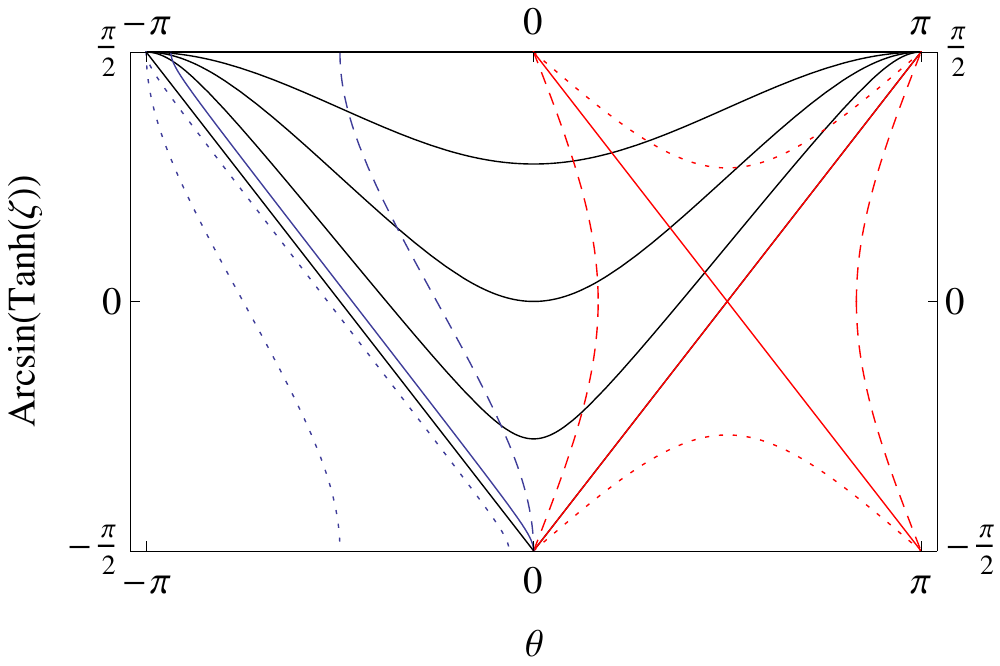}
\end{center}
\caption{Penrose diagram of de Sitter space. The light like curves correspond to diagonals with a $45$ degrees angle with respect to vertical. In black are the surfaces $t=cte$. They span a triangle which constitutes the Poincaré patch. In red are the surfaces $X=cte$. The plain line is $H X= 1$, the dashed line is $HX <1$ and the dotted line is $HX >1$. In blue are the surfaces $z=cte$. The dotted curves correspond to $z>0$ outside the Poincaré patch (the $z>0$ inside Poincaré patch have not been represented). Solid line correspond to large (negative) value of $z$ and dashed to lower value of $z$.}
\label{fig:poincaredesitter}
\end{figure}

It can be useful to introduce two other sets of coordinates $(\tau,X)$ and $(t,z)$ defined by
\begin{equation}
\begin{split}
\ep{H \tau} =\ep{H t} &= \sinh \zeta + \cosh\zeta \cos\theta,\quad H X = \cosh\zeta \sin \theta, \quad z= X\ep{-H \tau}.
\end{split}
\end{equation}
The inverse transformation is
\begin{equation}
\begin{split}
\sinh \zeta &= \frac{H^2X^2-1 + \ep{2H \tau}}{2\ep{H \tau}}, \\
 \tan\frac{\theta}{2} &= \frac{\sqrt{(1- H^2 X^2 + \ep{2 H t})^2 + 4 H^2 X^2 \ep{2H \tau}} - (1- H^2 X^2 + \ep{2 H \tau})}{2 H X \ep{H \tau}}.
\end{split}
\end{equation}
This inverse transformation is only defined as long as $\abs{\theta} \leq \arccos\left ( - \tanh\zeta \right )$. This subpart of de Sitter space-time is called Poincaré patch. The complementary part is obtained by symmetry $PT$, i.e., replacing $(\ep{H t },X)$ by $(-\ep{H t },-X)$ in the definition, or equivalently $(\zeta,\theta)$ by $(-\zeta, \theta+ \pi)$. The Poincaré patch and the surfaces of constant $t,X,z$ have been represented in Fig.~\ref{fig:poincaredesitter}. The axes have been chosen so that diagonals are light-like curves. Indeed, the light-like curves are defined by $\zeta = L(\theta)$ with $L' = \pm \cosh(L)$. This differential equation is solved by $ \arcsin \tanh \zeta = \pm (\theta -\theta_0) $.

With such definitions, the metric of the space time restricted to Poincaré patch reads
\begin{equation}
\label{eq:dSmetricintwoways}
\begin{split}
ds^2 = -dt^2 + \ep{2 H t} dz^2 = - d\tau^2+ \left [dX - v(X) d\tau\right ]^2,
\end{split}
\end{equation}
with $v(X) = HX$. The Poincaré patch can hence either be viewed as a spatially homogeneous isotropic (i.e., FLRW) expanding space-time, or a stationary space-time with two black hole horizons (in higher dimension, the horizon becomes a sphere). However, it can not be viewed at the same time as homogeneous and stationary for algebraic reasons, see \ref{sec:Killingalgebra}.

\section{de Sitter group}
\label{sec:desittergroup}

\subsection{Killing Algebra}
\label{sec:Killingalgebra}

A Killing vector $K$ is the generator of a symmetry of space-time. It is such that the metric is the same at two point infinitely close with a distance proportional to $K$, i.e., by a coordinate transformation $\mathsf x'^\mu = \mathsf x^\mu + \epsilon K^\mu $, we have at first order in $\epsilon$ $\mathsf g'_{\mu\nu} (\mathsf x') = \mathsf g_{\mu \nu} (\mathsf x)$. Because the metric is a tensor, we have $\mathsf g'_{\mu\nu} (\mathsf x') = (\delta_\mu^\rho + \epsilon \partial_\mu K^\rho)(\delta_\nu^\sigma +\epsilon \partial_\nu K^\sigma) \mathsf g_{\rho \sigma} (\mathsf x + \epsilon K) \sim \mathsf g_{\mu\nu}(\mathsf x) + \epsilon \left ( \partial_\mu K^\rho \mathsf g_{\nu\rho} + \partial_\nu K^\rho \mathsf g_{\mu\rho} + K^\rho \partial_\rho \mathsf g_{\mu \nu} \right )$. The equation for a Killing vector is the cancellation of the last parenthesis, of equivalently, using covariant derivatives $\nabla_\mu$: 
\begin{equation}
\begin{split}
\nabla_\mu K_\nu + \nabla_\nu K_\mu = 0.
\end{split}
\end{equation}
If in a coordinate system, the metric is invariant under one of its coordinate, say $x^i$ then the vector $\partial_{x^i}$ is a Killing vector field.

The three different coordinate systems introduced in \ref{sec:desitterspace}. Each emphasizes one Killing vector. We express them as
\begin{equation}
\begin{split}
K_\theta = \partial_{\theta\vert \zeta}, \quad K_t = \partial_{\tau \vert X}, \quad K_z = \partial_{z\vert t}.
\end{split}
\end{equation}
It can be shown that these $3$ Killing vector fields engender the whole Killing algebra of the space-time. It corresponds to $SO(1,2)$ algebra and we have
\begin{equation}
\begin{split}
[K_\theta,K_t] = H K_\theta - K_z,\quad [K_z,K_\theta] = -K_t, \quad [K_t,K_z] = H K_z.
\end{split}
\end{equation}

\subsection{Invariant quantum field theory}
\label{sec:invariantQFT}

We here investigate which of the free scalar local quantum field theories (QFT) would be invariant under the full de Sitter group $SO(1,2)$. Because we restrict ourselves to free scalar and local QFT, the equations of motion are of the form $\hat O \phi = 0$ where $\hat O$ is some differential operator that commutes with the three Killing vectors. The most general local operator acting on scalar fields can be written in the homogeneous frame ($t,z$) as
\begin{equation}
\begin{split}
\hat O = \sum_{n,m=0}^\infty \alpha_{n,m}(t,z) \partial_z^n \partial_t^m. 
\end{split}
\end{equation}
Imposing that $\hat O$ commutes with $K_z$, implies that the $\alpha$ functions depend only on $t$. Imposing that it also commutes with $K_t$ implies $\alpha_{n,m}(t)=\alpha_{n,m} \ep{- n Ht}$, where $\alpha_{n,m}$ are constants. Hence, $\hat O$ is necessarily of the form
\begin{equation}
\label{eq:OcomutingwithKtKz}
\begin{split}
\hat O = \sum_{n,m=0}^\infty \alpha_{n,m} \ep{- n Ht} \partial_z^n \partial_t^m. 
\end{split}
\end{equation}
When we impose the invariance under the third generator, namely $K_\theta$, the only invariant operator are necessarily functions of the Casimir of the group, i.e., $\hat O = -\partial_t^2 - H \partial_t + ( \ep{-H t}\partial_z)^2 = - K_t^2 + K_z K_\theta + K_\theta K_z - K_z^2 $. This is simply the d'Alembertian of the space-time. Hence, no dispersion can occur.

\section{Affine group}
\label{sec:affinegroup}

\subsection{Algebra}

We here suppose that the de Sitter group is minimally broken. 
We hence need to pick two independent vectors in the three dimensional space of Killing vectors so that their commutator linearly depends on the two. These engender the affine group. Such a pair is necessarily of the form
\begin{equation}
\begin{split}
K_1 &= \sin(\theta - \theta_0) \partial_\zeta + \left [ 1+\tanh \zeta \cos(\theta - \theta_0) \right ]\partial_\theta, \\
K_2 &= \cos(\theta - \theta_0) \partial_\zeta + \tanh \zeta \sin(\theta - \theta_0)\partial_\theta.
\end{split}
\end{equation}
Choosing the origins of $\theta$ such that $\theta_0 = 0$, the affine group becomes engendered by $K_t, K_z$. The affine algebra is then
\begin{equation}
\label{eq:alg}
\begin{split}
[K_t,K_z] = H K_z.
\end{split}
\end{equation}
We notice that the affine group has no Casimir operator in the sense that the universal enveloping algebra --i.e., the polynomials in $K_t$, $K_z$-- of the affine group has no element, but the identity, that commute with the affine group. However, the local operator invariant under this affine group are more general and given by Eq.~\eqref{eq:OcomutingwithKtKz}. In fact, defining $u_{\mathrm{ff}}$ and $s_{\mathrm{ff}}$ as the two orthonormal vectors that commute with $K_t, K_z$ and such that $u_{\mathrm{ff}}$ is freely falling\footnote{
this is unique up to changing $s$ to its opposite $-s$.}, [i.e., $s_{\mathrm{ff}}^\mu \partial_\mu = \ep{-H t}\partial_z $ and $u_{\mathrm{ff}}^\mu \partial_\mu = -\partial_t $] Eq.~\eqref{eq:OcomutingwithKtKz} means that the most general affine group invariant operator is a function of $u_{\mathrm{ff}}^\mu \partial_\mu$ and $s_{\mathrm{ff}}^\mu \partial_\mu$. In addition we notice that $u_{\mathrm{ff}}$ and $s_{\mathrm{ff}}$ obey 
\begin{equation}
\begin{split}
\left[u_{\mathrm{ff}},s_{\mathrm{ff}}\right] = H s_{\mathrm{ff}} ,
\label{eq:usc}
\end{split}
\end{equation}
which is the affine algebra of Eq.~\eqref{eq:alg}. Hence this algebra is intrinsic to the cosmological frame on de Sitter space. Eq.~\eqref{eq:usc} follows from the fact that the commutator $[u_{\mathrm{ff}},s_{\mathrm{ff}}]$ must be a linear combination of $u_{\mathrm{ff}}$ and $s_{\mathrm{ff}}$ since they are the only fields that commute with $K_t$ and $K_z$. It is equal to $s_{\mathrm{ff}}$ because $u_{\mathrm{ff}}$ is freely falling. The preferred time associated to $u_{\mathrm{ff}}$ is the cosmological time. 

\subsection{Field theory}
\label{sec:affinefieldth}

Using the operator $\hat O$ of \ref{sec:invariantQFT}, our program is to treat $\hat O \hat \phi =0$ as defining the general field equation. To this end, we impose that $\hat O$ be second order in some (preferred) time derivative. We name $u$ the unit time-like linear combination of $u_{\mathrm{ff}}$ and $s_{\mathrm{ff}}$ such that $\hat O$ is quadratic in $u$. $s$ is the unit space-like linear combination of $u_{\mathrm{ff}}$ and $s_{\mathrm{ff}}$ that is orthogonal to $u$. Because $s$ is space-like and $u$ is time-like, $\hat P = -i s^\mu \partial_\mu$ is the preferred momentum operator, and $\hat \Omega = - i u^\mu \partial_\mu$ the preferred frequency. The general form of the operator $\hat O$ is
\begin{equation}
\begin{split}
\hat O = - \hat \Omega^2 + g(\hat P) \hat \Omega +h(\hat P).
\end{split}
\end{equation}
Next we impose the invariance under the discrete parity symmetry $s\rightarrow -s$. This implies that $g$ and $h$ are even functions of $\hat P$. In higher dimensions, this condition would follow from the requirement of isotropy (in fact, isotropy also imposes that $u$ is freely falling since its acceleration is orthogonal to $u$ and would define a preferred direction). The last important condition is that $\hat O$ be compatible with a unitary evolution~\cite{Parentani:2007uq}. The proper way to specify this condition is the following: the part of $\hat O$ that is even in $\hat \Omega$ describes dispersive effects and should be self-adjoint, whereas the odd part describes dissipative effects and should be anti-self-adjoint, where the adjoint is defined by
\begin{equation}
\begin{split}
 \int d^2\mathsf x \sqrt{-\mathsf g} \Phi^* (\hat O \Psi ) &= \int d^2\mathsf x \sqrt{-\mathsf g} (\hat O^\dagger \Phi)^* \Psi .
\end{split}
\end{equation}
To sort out the contributions which are due to the expansion, it is useful to introduce the self-adjoint operator $\hat \Omega_{\rm sa} = \frac12\left (\hat \Omega + \hat \Omega^\dagger\right )$. Then, the \enquote{unitary} operators are given by
\begin{equation}
\begin{split}
\label{eq:hatOf}
\hat O &= -\hat \Omega_{\rm sa}^2 - i \left ( \gamma_{\rm sa} \hat \Omega_{\rm sa} + \hat \Omega_{\rm sa} \gamma_{\rm sa} \right ) + F_{\rm sa}^2,
\end{split}
\end{equation}
where $\gamma_{sa} $ and $F_{sa}$ are both real functions of $\hat P^\dagger \hat P$.

The function $F_{sa}$ describes the dispersive effects compatible with the affine group. Instead, the function $\gamma_{sa} $, which multiplies the odd term in $\Omega$, describes the dissipative effects compatible with it. It precisely matches the set of $\gamma$ functions introduced in~\cite{Parentani:2007uq,Adamek:2008mp} to describe dissipative effects that are local in time, and that obey the generalized equivalence principle (GEP), which states that the action must be a sum of scalars under general coordinate transformations which reproduce those one had in Minkowski space-time endowed with a homogeneous static $u$ field. This agreement is nontrivial and follows from the fact that, on one side, the GEP implies that the field equation can only depend on the two {\it scalars} $\hat \Omega$ and $\hat P$ defined by the metric $g$ and the $u$ field, whereas, on the other side, $\hat \Omega$ and $\hat P$ are the two {\it invariants} under the generators $K_t$ and $K_z$ of the affine group. 

We note that the different frequencies and momenta are not independent quantities. The freely falling momentum $P_\mathrm{ff} = s_\mathrm{ff}^\mu \partial_\mu$ and frequency $\Omega_\mathrm{ff}= u_\mathrm{ff}^\mu \partial_\mu$ are linked to the conserved momentum $k = K_z^\mu \partial_\mu $ and frequency $\omega = K_t^\mu \partial_\mu$ by
\begin{equation}
\label{eq:PandOmff}
\begin{split}
P_\mathrm{ff} = k \ep{-H t}, \quad \Omega_\mathrm{ff} = \omega - H X P_\mathrm{ff}.
\end{split}
\end{equation}

As a final comment, we notice that the affine group is closely related to Fourier and Mellin analysis~\cite{morse1953methods}. When working on the representation of the affine group $\mathcal{L}^2(\mathbb{R})$, the eigenmodes of $-i K_z = -i \partial_z$ of eigenvalue $k$ are the plane waves $\ep{ikz}$, whereas those of $-iK_t=-i H (z\partial_z+1/2)$ of eigenvalue $\omega$, are $\phi_\omega^\pm = \theta(\pm z)(\pm z)^{i \omega/H-1/2}$. The latter live on either side of $z= 0$ and they correspond to Mellin modes. They are complete for $\omega \in \mathbb{R}$ (since invertible) on $\mathcal{L}^2(\mathbb{R}^+)$. Hence, to have completeness on functions of $\mathbb R$, one needs two families of Mellin modes, on either side of $z=0$, given by $\phi_\omega^\pm$. 

\begin{figure}[ht]
\begin{minipage}{0.47\linewidth}
\includegraphics[width=1\linewidth]{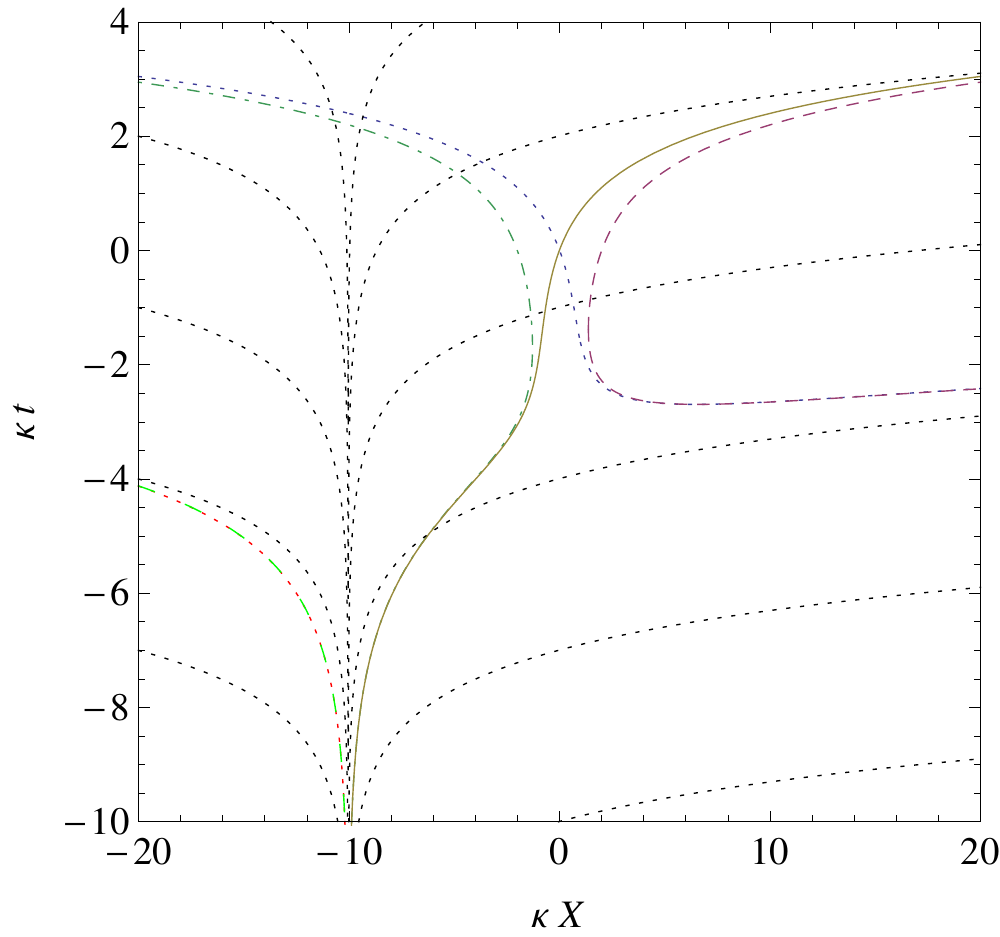}
\end{minipage}
\hspace{0.03\linewidth}
\begin{minipage}{0.47\linewidth}
\includegraphics[width=1\linewidth]{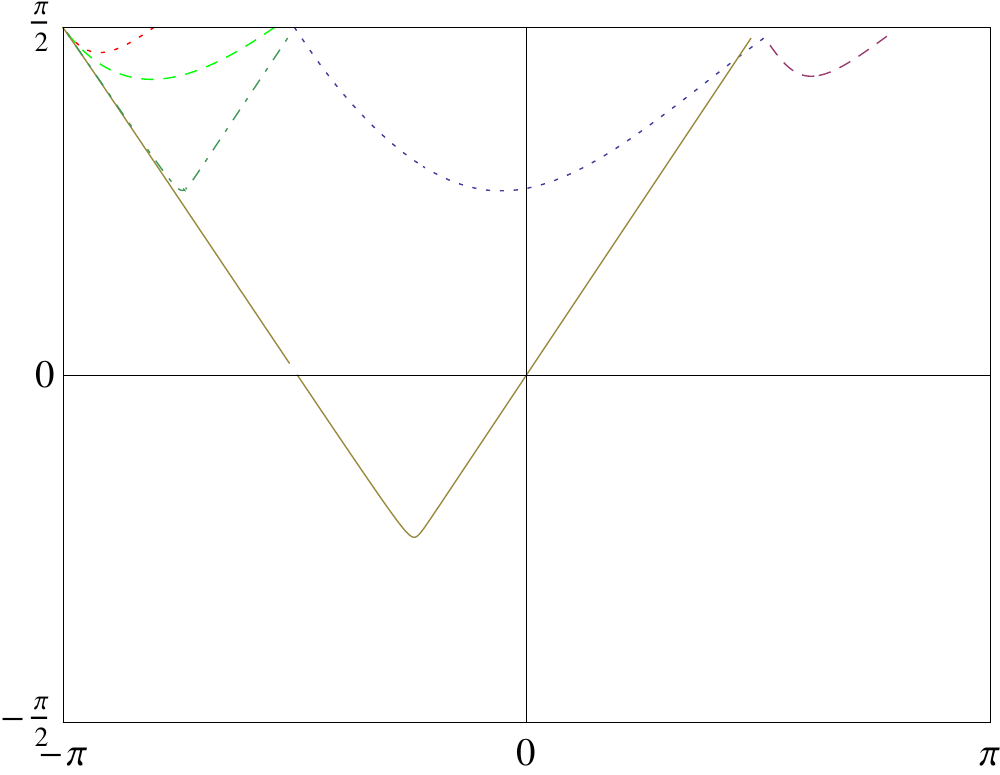}
\end{minipage}
\caption{We draw here the characteristics of a dispersive field with a superluminal dispersion relation $F(P^2) = P^2(1+P^2/2/\Lambda)^2$ with $\Lambda = 10 H$, $\omega=H$. The acceleration of the $u$ field is fixed by $\xi = -0.1$ in $t,X$ coordinates (left panel) and $\xi =-1.5$ in a Penrose diagram (right panel). In blue dotted line, we represent the positive norm $V$ mode and in purple dashed line its negative energy partner. When they reach the universal horizon, they are so redshifted that they behave relativistically. In their past, however, they seem to appear at some initial $t_{\rm pref}$ and were absent at $t\to - \infty$. The green dashed and red dotted curve (hardly distinguishable on the left plot) are trapped $V$ modes inside the universal horizon. They seem to exist only for a finite lifetime. The yellow solid and green dot-dashed curve are the $U$ modes. In the presence of acceleration, they emerge from the universal horizon. We also note that all mode is included in the Poincaré patch, justifying the left panel representation. In black dots are represented curves of constant $t_{\rm pref}$.}
\label{fig:characdsaccel}
\end{figure}

\subsection{Characteristics of the modes}

When $u$ is no longer freely falling, because the only vectors that commute with $K_t$ and $K_z$ are the linear combinations of $u_{\rm ff} $ and $s_{\rm ff}$, so is $u$. O	ne generalizes Eq.~\eqref{eq:PandOmff} by writing the orthonormal character of $u,s$ under the existence of some real parameter $\xi$ so that
\begin{equation}
 \begin{split}
 u = \cosh\xi u_{\rm ff} + \sinh\xi s_{\rm ff}, \quad s = \sinh\xi u_{\rm ff} + \cosh\xi s_{\rm ff}. 
 \end{split}
 \end{equation} 
The acceleration of $u$ is then constant and one has $u^\mu D_\mu s_\nu = \sinh^2\xi H u_{\rm ff} $. The same equalities read at the level of frequencies and moment $\Omega,P$ using Eq.~\eqref{eq:PandOmff}: 
\begin{equation}
\label{eq:PandOmnotff}
\begin{split}
\Omega &= \cosh\xi \omega +(\sinh\xi - \cosh\xi H X ) k \ep{-H t} ,\\
 P &=\sinh\xi \omega + ( \cosh\xi - \sinh\xi H X ) k \ep{-H t}.
\end{split}
\end{equation}
Writing the dispersion relation (without dissipation) $\Omega^2 = F^2(P^2)$, one gets an implicit version of the characteristics equation. 

Moreover, the preferred time, defined by the assumptions that equations of motion are second order in time (i.e., $s^\mu$ contains only space derivatives, or equivalently $u_\mu dx^\mu \propto dt_{\rm pref} $) is $t_{\rm pref} = t - \log\abs{\cosh\xi -H X \sinh \xi}/H $. This time is ill defined on the universal\footnote{
In the presence of dispersion, even though the group velocity of the modes can be infinite, if the vector field $u$ is not freely falling, there exists a region in space from where modes can not escape. The limit of this region is called universal horizon~\cite{Berglund:2012bu}. It is situated at $x= -10$ on Fig.~\ref{fig:characdsaccel}. This concept was introduced by Sibiryakov in a talk, and further described in \cite{Barausse:2011pu,Blas:2011ni}.} 
horizon $\cosh\xi -H X \sinh \xi =0$. With such definition, $u_\mu dx^\mu = -(\cosh\xi -H X \sinh \xi ) dt_{\rm pref} $. Because the prefactor changes its sign inside the universal horizon, the flow of particles goes backwards (in preferred time) within this region.

In Fig.~\ref{fig:characdsaccel}, we represent the different characteristics of the mode. The dimensionality of modes seems at first glance to be $6$. There are two $U$ modes represented in black and dashed red (the red one has negative energy and the black one has positive energy), two conventional $V$ modes in blue and magenta and two modes trapped inside the universal horizon. The fact that the two (low energy) $out$ modes can traverse the universal horizon is not a surprise. On the first hand, since they are extremely redshifted, they behave relativistically and do not feel the universal horizon. On the second hand, when they reach the universal horizon, they reach $t_{\rm perf} \to \infty$. Inside the horizon, they propagate under decaying $t_{\rm perf}$.

An analytical analysis shows that at large values of $X$, the modes follow curves of constant $H X \ep{-H t}\sim \ep{- H t_{\rm pref}} $. The value of this constant is $\pm \omega/k$ for the outgoing relativistic modes and $ \pm \omega/k \pm \Lambda/k \sqrt{2 \ep{-\abs{\xi}}/ \sinh^3\abs{\xi}} $ for the other ones. Because of this similar behavior, the $8$ roots of these modes take their origin on the same limit in a Penrose diagram, i.e., $\zeta\to \infty$ with $\theta$ finite (and different from $\pi$). The four remaining curves take their origin at finite value of $X$ but infinite time. This corresponds to $\zeta \to \infty$, $\theta= {\rm sgn} (\xi) \pi $.

The fact that incoming and outgoing modes have the same asymptotic behavior is an indication that the theory may be ill defined. In particular, the identification of a Cauchy surface where initial conditions have to be chosen is highly non trivial.

When the $u$ field is not accelerated, the Cauchy surface is much easier to identify. Indeed, in such a case, there is no universal horizon and all the modes originate from $t_{\rm pref} = t \to -\infty$. In the following, for reasons of simplicity, we shall consider the case where $u$ is freely falling and drop the subscript $\mathrm{ff}$.

\section{P-representation}
\label{sec:Prep}

\subsection{One point functions}

Given the two symmetries engendered by $K_t, K_z$, the scalar fields can be decomposed either as 
\begin{equation}
\Phi = \int_{-\infty}^{\infty} \frac{d{\bf{k}}}{H \sqrt{2\pi}} \ep{i {\bf{k}}z} \phi_{\bf{k}}(t), 
\label{eq:kdec}
\end{equation}
or as
\begin{equation}
\Phi = \int_{-\infty} ^{\infty} \frac{ d\omega}{H \sqrt{2\pi }} \ep{- i \omega t} \phi_\omega(X). 
\label{eq:omdec}
\end{equation}
In the $k$-representation, Eq.~\eqref{eq:hatOf} gives the second order equation\footnote{
In a theory with dissipation, ($\gamma \neq 0$), unitarity imposes that the rhs is different from $0$. In this section, we neglect the rhs and we only consider the differential operator of the lhs. For the construction of a unitary dissipative theory, we refer to \ref{chap:dissipdS}.}
\begin{equation}
\bigg [ \frac{1}{a(t)} \partial_t a(t) \partial_t +\left ( \gamma_{\rm sa}(\frac{k^2}{a(t)^2}) \partial_t + \partial_t \gamma_{\rm sa}(\frac{k^2}{a(t)^2}) + H \gamma_{\rm sa}(\frac{k^2}{a(t)^2}) \right ) + F_{\rm sa}^2(\frac{k^2}{a(t)^2}) + \frac{H^2}{4} \bigg ] \phi_{\bf{k}}(t) =0 .
\label{eq:kmodeq}
\end{equation}
As in all cosmological spaces, see e.g., Ref.~\cite{Macher:2008yq}, the general (homogeneous) solution thus lives in a two-dimensional space and takes the form
\begin{equation}
\phi_{\bf{k}}(t) = A_{\bf{k}} \phi_k(t) + ( B_{\bf{-k}} \phi_k(t))^*. 
\end{equation}
In de Sitter, because of the affine group invariance, the $k$ and $t$ dependences in Eq.~\eqref{eq:kdec} can be combined in a single variable $P= H k \ep{-H t}$. Indeed Eq.~\eqref{eq:kmodeq} can be rewritten as 
\begin{equation}
\label{eq:Pmodeeq}
\left[H^2 P^2\partial_P^2 - H P^2 \left ( \frac{\gamma_{\rm sa}(P^2)}{P}\partial_P + \partial_P \frac{\gamma_{\rm sa}(P^2)}{P} \right ) + F_{\rm sa}^2(P^2) + H^2/4 \right] \chi(P) =0 .
\end{equation}
This possibility is due to the presence of the \enquote{spectator} Killing field $K_t$. Whereas $K_z$ guarantees that $\bf k$-modes separate, $K_t$ tells us that Eq.~\eqref{eq:kmodeq} is invariant under 
\begin{equation}
\label{eq:repar}
t\rightarrow t+T,\quad k \rightarrow k \ep{H T}, \quad P\rightarrow P.
\end{equation}
This implies that $\phi_k(t)$ only depends on $t$ only through $P = k\ep{-H t}$. Eq.~\eqref{eq:Pmodeeq} can be further simplified by extracting a decaying factor from $\chi$: if we pose $\chi = \ep{ \int dP \gamma_{\rm sa}(P^2) /P H} \bar \chi$, Eq.~\eqref{eq:Pmodeeq} translates to $\bar \chi$ as
\begin{equation}
\label{eq:generaleomchibar}
\begin{split}
\left [ H^2 P^2\partial_P^2 + F_{\rm sa}^2(P^2) + H^2/4 - \gamma_{\rm sa}^2(P^2)\right ] \bar\chi =0.
\end{split}
\end{equation}

On the other hand, in the $\omega$-representation, the spatial modes obey the higher order equation
\begin{equation}
\Big[ -(\omega+i \partial_{X} v)(\omega + i v \partial_{X}) - i \{\gamma_{\rm sa}(-\partial_X^2), (\omega + i v \partial_{X} + i H/2)\} + F_{\rm sa}^2(-\partial_X^2 ) +H^2 /4 \Big] \phi_\omega(X) = 0. 
\label{eq:ommodeeq}
\end{equation}
Unlike what is found in the $k$-representation, at fixed $\omega$, the dimensionality of the space of solutions now depends on the dispersion relation: it is $2n$ when the highest power of $P$ in $F_{\rm sa}^2(P^2)$ is $2n$ or $2n+1$ when the highest power of $P$ in $\gamma_{\rm sa}(P^2)$ is $2n$. In spite of this it is possible, and very instructive, to relate the solutions of Eq.~\eqref{eq:ommodeeq} to those of Eq.~\eqref{eq:kmodeq}. To this end, it is useful to consider the Fourier transform,
\begin{equation}
\tilde \phi_\omega({\bf{P}})=\int_{-\infty}^\infty \frac{d{{X}}}{\sqrt{2\pi}} \ep{-i {\bf{P}}{{X}}} \phi_\omega ({{X}}),
\end{equation}
where ${\bf P}$ designates the wave vector, and $P > 0$ its norm. In the $(\omega,P)$-representation, Eq.~\eqref{eq:ommodeeq} becomes
\begin{equation} 
\label{eq:ommodeeqinP}
\Big[-(\omega-i H {{P}} \partial_{{P}})(\omega - i H \partial_{{P}} {{P}}) - i \{\gamma_{\rm sa}(P^2), (\omega - i H \partial_P P + i H/2)\} + F_{\rm sa}^2(P^2) +H^2 /4 \Big] \tilde \phi_\omega = 0 .
\end{equation}
As in the $k$-representation, this is a second order equation (in $\partial_P$). Moreover one verifies, that $\tilde \phi_\omega$ exactly factorizes as~\cite{Brout:1995wp}
\begin{equation}
\tilde \phi_\omega({\bf{P}}) = P^{-i \frac{\omega}{H}-1}\times \tilde \chi({\bf{P}}) , 
\label{eq:facto}
\end{equation}
where $ \tilde \chi$ is independent of $\omega$. In addition, one also verifies that $ \tilde \chi$ obeys Eq.~\eqref{eq:Pmodeeq}. These unusual properties are due to the other Killing field $K_z$. In fact because of Eq.~\eqref{eq:repar}, in $(t,P)$ representation, irrespectively of Lorentz violating terms $F_{\rm sa}$ and $\gamma_{\rm sa}$, the modes trivially depend on $t$ through a delta-function: $\delta(P-H k\ep{H t})$. When working in the $(\omega,P)$ representation, this implies both the factorization of Eq.~\eqref{eq:facto}, and the fact that $\tilde \chi$ obeys Eq.~\eqref{eq:Pmodeeq}. The extra factor of $1/P$ in Eq.~\eqref{eq:facto} is due the Jacobian $dt/dP = -1/HP$. 

Since Eq.~\eqref{eq:ommodeeqinP} is second order and singular at $P=0$, the dimensionality of the space of solutions of $\tilde \phi_\omega({\bf P})$ is $4$, because $\bf P$ has both signs. The physical meaning of this four-dimensional space, and its relation with the two-dimensional one found in the $k$-representation, are given in \ref{sec:scalarpdt}. It is mathematically linked to the difference between Mellin and Fourier transform.

\subsection{Two point functions}

When working with states $\hat \rho_{\rm tot}$ that are invariant under the affine group (this is analogous to the restriction to the so-called $\alpha-$vacua which are invariant under the full de Sitter group~\cite{Schomblond:1976xc,Mottola:1984ar}), the two point functions are invariant under both $K_t$ and $K_z$. However, because the commutator $ [ K_z , K_t] = - H K_z$ does not vanish, one cannot simultaneously diagonalize $K_z$ and $K_t$. This leads to two different ways to express the two-point functions, either at fixed wave number $\bk = -i \partial_{z\vert t}$, or at fixed frequency $\omega = i\partial_{t\vert X} $. Explicitly, one has 
\begin{subequations}
\begin{align}
\label{eq:fourierk}
 G_{\rm any}^k(t,t')&\doteq \int d\Delta z \ep{-i \bk \Delta z } G_{\rm any}(\Delta z ,t,t'),\\
\label{eq:fouriert}
G_{\rm any}^{\omega}(X,X') &\doteq \int d\Delta t \ep{i \omega \Delta t } G_{\rm any}(\Delta t ,X,X'), 
\end{align}
\end{subequations}
where $k = \abs{\bk} $, and where the \enquote{any} subscript indicates that these Fourier transforms apply to any two-point function which is invariant under the affine group. (In Eq.~\eqref{eq:fourierk}, $G^k$ only depends on $k$ because we impose isotropy.) 

In direct space, the invariance under the affine group translates in the fact that the two-point correlation functions only depend on geometrical invariants evaluated between the two points. In de Sitter space, there are two geometrical invariants under the affine group. Using the coordinates $t,X$, they read 
\begin{subequations}
\begin{align}
\Delta_1 &= \ep{H(t-t')} ,\\
\Delta_2 &= X\ep{-H(t-t')/2} -X'\ep{H(t-t')/2}. 
\end{align}
\end{subequations}
They are linked to the de Sitter (full group) invariant distance by 
\begin{equation}
\label{eq:distdesitter}
\Delta^2 = \Delta_2^2 - (\Delta_1-\frac{1}{\Delta_1})^2 . 
\end{equation}
The distances $\Delta_1, \Delta_2$ can be also defined in a coordinate invariant manner. The interested reader will find the expressions at the end of this subsection. 

Hence, any 2-point function $G_\mathrm{any}(\mathsf x,\mathsf x')$ can be written as $ \tilde G_\mathrm{any} \left (\Delta_1(\mathsf x,\mathsf x'),\Delta_2(\mathsf x,\mathsf x') \right )$. 
The fact that only two variables are needed is not a surprise, given the homogeneity and stationarity of the setting. However, because $\Delta_1, \Delta_2$ mix the properties of the two different points, they are not convenient to use. We Fourier transform the function with respect to $\Delta_2$ and rename variables, so that it reads
\begin{equation}
\begin{split}
\label{eq:Prepresentation}
G_\mathrm{any}(\bP,\bP') &=\theta(\bP \bP') \frac{\sqrt{P P'}}{H} \int d\Delta_2\ep{-i \sqrt{P P'} \rm{sgn}(\bP) \Delta_2} \tilde G_\mathrm{any} \left (\Delta_1 = \frac{P'}{P},\Delta_2 \right ).
\end{split}
\end{equation}
$G_\mathrm{any}$ only depends on $\bP$ and $\bP'$. This is the essence of the $P$ representation and its physical meaning is explicated below.

First, the momentum conjugated to $\Delta_2$, defined by $\bar \bP \doteq \partial_{\Delta_2 \vert \Delta_1} $, is given by the geometrical mean
\begin{equation}
\bar \bP = \mathrm{sgn}(\bP) \sqrt{P P'} = \bP \sqrt{\Delta_1}.
\end{equation}
On the other hand, since in homogeneous coordinates, $\Delta_2 = (z-z')\ep{H(t+t')/2} $, it is linked to the conserved momentum $\bk \doteq \partial_{(z-z')\vert (t,t')} $ by 
\begin{equation}
\begin{split}
\bP = \bk \ep{-H t}, \quad \bP' = \bk \ep{-H t'}.
\end{split}
\end{equation}
Hence, $\bP = s_\mathrm{ff}^\mu \partial_\mu$ is the same operator as in the previous section, seen as its eigenvalue. When considering the space-Fourier transform in the homogeneous frame of the two-point function, momentum $\bk$ can then be absorbed by a redefinition of the times, because
\begin{equation}
\label{eq:Prep2ptfn}
\begin{split}
G_\mathrm{any}(\bP,\bP') = \frac{k }{H} G_\mathrm{any}^k (t,t') .
\end{split}
\end{equation}
 $t$ is then viewed as a function of $\bP$.

Second, when we use stationarity to define the Fourier transform in time $G_{\mathrm{any}}^{\omega}(X,X') $, its Fourier transform in $X,X'$ defined by
\begin{equation}
G_{\mathrm{any}}^{\omega}(\bP,\bP') = \int \frac{dX dX'}{2\pi } \ep{-i \bP X + i \bP' X'} G_{\mathrm{any}}^{\omega}(X,X')
\end{equation}
has automatically the following structure
\begin{equation}
\begin{split}
\label{eq:prepomega}
G_{\mathrm{any}}^{\omega}(\bP,\bP') &= \frac{(P/P')^{-i \omega/H}}{P P' }G_\mathrm{any} (\bP,\bP'), 
\end{split}
\end{equation}
where $G_\mathrm{any} (\bP,\bP')$ is given by Eq.~\eqref{eq:Prepresentation}. This follows from $\bP = \partial_{X \vert t,t',X'}$

Together with Eq.~\eqref{eq:Prep2ptfn}, Eqs.~\eqref{eq:Prepresentation} and~\eqref{eq:prepomega} are the key equations of this section: Whenever a 2-point function $G_\mathrm{any} (\mathsf x,\mathsf x')$ is invariant under the affine group, its Fourier transforms $G_\mathrm{any}^k (t,t')$ and $G_\mathrm{any}^\omega (\bP,\bP')$ are related to $G_\mathrm{any} (\bP,\bP')$ of Eq.~\eqref{eq:Prepresentation} by Eq.~\eqref{eq:Prep2ptfn} and Eq.~\eqref{eq:prepomega} respectively. 

To conclude this section, we express $\Delta_1$ and $\Delta_2$ in covariant terms. The log of $\Delta_1$ is given by the line integral of $u^{\mathrm{ff}}$ from $\mathsf x$ to $\mathsf x'$, that is
\begin{equation}
\ln \Delta_1= - H \int_{\mathsf x}^{\mathsf x'} u^{\mathrm{ff}}_{\mu} dx^\mu .
\end{equation}
This is a well defined expression. Indeed, since $u^{\mathrm{ff}}$ is geodesic, $u^{\mathrm{ff}}_{\mu} dx^\mu$ is an exact 1-form and the above integral does not depend on the path. Since, Eq.~\eqref{eq:distdesitter} gives $\Delta_2$ as a combination of $\Delta_1$ and $\Delta$ which are both invariantly defined, so is $\Delta_2$.~\footnote{If one wishes, $\Delta_2$ can also be seen as the integral of $s^{\mathrm{ff}}_{\mu} dx^\mu $, the 1-form associated to the vector orthogonal to $u^{\mathrm{ff}}$. Since this form is not exact, one has to specify the contour from $\mathsf x$ to $\mathsf x'$. Using the $t,z$ coordinates, one should go at fixed $z$ from $t$ to $(t+t')/2$, then vary $z$ at fixed time until $((t+t')/2,z')$, and vary $t$ at fixed $z$ until $(t',z')$. Any different contour would give some combination of $\Delta_1$ and $\Delta_2$.}

To summarize, we showed that firstly, $P$ is invariantly defined; secondly, $\Delta_1$ is easily expressed in $P,P'$ space; thirdly, so is the variable conjugated to $\Delta_2$; and fourthly, $P$ can be attributed to the field itself, so that one can easily take the even (anti-commutator) and the odd part of the 2-point functions. These are the reasons that make the $P$-representation a very convenient representation of affine invariant functions.

\section*{Conclusions}
\addcontentsline{toc}{section}{Conclusions}

In this chapter, we considered de Sitter space both as some geometrical object, and as a manifold on which is defined some field theory. In \ref{sec:desitterspace}, we defined the geometrical object as an hypersurface in Minkowski (flat) space time. We identified Killing vectors forming an algebra in \ref{sec:desittergroup} and showed that the only invariant field theory in de-Sitter is the relativistic one. In \ref{sec:affinegroup}, we broke minimally the de Sitter group so as to obtain the affine group. We showed that the affine group is defined uniquely up to a rotation and that the only field theories that are invariant are dispersive and dissipative ones. In \ref{sec:Prep}, we defined the $P$ representation for the fields that are invariant under the affine group.

\chapter{Dispersive fields in de Sitter space}
\label{chap:dispdS}

\section*{Introduction}

The quantum field theories in curved space time are mathematically well defined in hyperbolic space-times~\cite{wald1994quantum} --i.e., in space times with well defined initial Cauchy surface-- in the future of this Cauchy surface. For a dispersive field in de Sitter with a non accelerated preferred frame, the surface $t= - \infty$ is a Cauchy surface and its future is the Poincarré patch.

We here consider free dispersive theory defined on the Poincaré patch of de Sitter. We identify the general consequences of local Lorentz violation and then compute observables in an explicit solvable model of quartic superluminal dispersion. This chapter is mainly based on~\cite{Busch:2012ne}.

\minitoc
\vfill

\section{Fields in de Sitter space} 
\label{sec:setting}

\subsection{Ultraviolet dispersion} 
\label{sec:dSspace}

We chose to neglect all dissipative effect in this chapter, and to make dispersion only in the ultraviolet. From Eq.~\eqref{eq:hatOf}, with $\gamma_{\rm sa} =0$, the dispersion relation reads
\begin{equation}
\begin{split}
\label{eq:disprel}
\Omega^2 = F^2(P^2) = m^2 + P^2 + f(P^2) .
\end{split}
\end{equation}
We suppose that $f$ vanishes faster than $P^2$ for $P \to 0$, so as to recover a relativistic relation for $ P \ll \Lambda$, where $\Lambda$ gives the ultraviolet dispersive scale. As such, Eq.~\eqref{eq:disprel} can be viewed as the Hamilton-Jacobi equation for the corresponding dispersive particle~\cite{Brout:1995wp,Balbinot:2006ua}. Using $\mathsf g_{\mu\nu} + u_\mu u_\nu = s_\mu s_\nu $, this equation reads 
\begin{equation}
\mathsf g^{\mu \nu} \partial_\mu S \partial_\nu S +m^2 + f\left( (s^\mu \partial_\mu S )^2 \right) = 0 ,
\label{eq:dispHJeq}
\end{equation}
where $S(t,z)$ is the action of the particle. On the other hand, Eq.~\eqref{eq:disprel} can also be viewed as the dispersion relation governing some field. However there is some ambiguity because of the ordering of the differential operators, and nonminimal couplings. In this chapter, we work with~\cite{Jacobson:2000gw,Lemoine:2001ar} 
\begin{equation}
\left[ - \mathsf g^{\mu \nu} \nabla_\mu \nabla_\nu +m^2 + f( - s^\mu \nabla_\mu s^\nu \nabla_\nu ) \right] \Phi = 0 , 
\label{eq:dispKGeq}
\end{equation}
where $\nabla_\mu$ is the covariant derivative. Such an equation may come from the action
\begin{equation}
\label{eq:Actiondesitterdisp}
\begin{split}
S = \frac{-1}{2}\int d\mathtt{x} \sqrt{-\mathtt{g}} \left [ \mathsf g^{\mu \nu} \partial_\mu \Phi \partial_\nu \Phi + m^2 \Phi^2 + \Phi f( - s^\mu \nabla_\mu s^\nu \nabla_\nu )\Phi \right ].
\end{split}
\end{equation}
For other approaches based on condensed matter models, see Refs.~\cite{Schutzhold:2002rf,Macher:2009nz,Unruh:2012ve} and \ref{part:analoguegravity}. One can check that with the decomposition of Eqs.~\eqref{eq:kdec} and~\eqref{eq:omdec} this equation of motion is of the kind of Eqs.~\eqref{eq:kmodeq} and~\eqref{eq:ommodeeq}, with $F_{\rm sa}^2 + H^2/4 = F^2= m^2 + P^2 +f $. We see that $F_{sa}=0$ corresponds to a mass $m = H/2$. This is the algebraic origin of the problem that arises for $m < H/2$, see \ref{sec:pfour}.

\subsection{Scalar product and BD vacuum} 
\label{sec:scalarpdt}

To identify a basis of solutions for Eqs.~\eqref{eq:kmodeq} and~\eqref{eq:ommodeeq}, we consider the conserved scalar product [see Eq.~\eqref{eq:Wronskianconserved}]. It is given by~\cite{Unruh:1994je}
\begin{equation}
(\Phi_1,\Phi_2) = i \int dl \left( \Pi_1^* \Phi_2 - \Phi_1^* \Pi_2\right) , 
\label{eq:scalpr}
\end{equation}
where $\Pi = - u^\mu \partial_\mu \Phi$ is the momentum conjugated to $\Phi$. The integral must be evaluated along $u_\mu dx^\mu = dt = 0$, and the line element is $dl = dX = a(t) dz$.

In the $k$-representation, for $ \Phi_{\bf{k}}= \ep{i{\bf{k}}z} \phi_{\bf{k}} $ and $\Phi_{\bf{k'}} = \ep{i{\bf{k'}}z} \phi_{\bf{k'}}$, one has 
\begin{equation}
( \Phi_{\bf{k}}, \Phi_{\bf{k'}}) = 2\pi \delta({\bf{k}}-{\bf{k'}}) \times \left[ a(t) \left( \phi^*_{\bf{k}} i\partial_t \phi_{\bf{k'}}-\phi_{\bf{k'}} i\partial_t \phi^*_{\bf{k}}\right) \right] .
\end{equation}
The standard normalization $( \Phi_{\bf{k}}, \Phi_{\bf{k'}}) =2 \pi H \delta({\bf{k}}-{\bf{k'}})$ imposes to work with modes $\phi_{\bf{k}}$ that have a unit positive current with respect to $a(t)/H i\partial_t$. When considering $\chi$ of Eq.~\eqref{eq:Pmodeeq}, it is convenient to reexpress this condition as
\begin{equation}
 H\chi i\partial_P \chi^* -H\chi^* i\partial_P \chi =1 . 
\label{eq:unitW}
\end{equation}
That is, the $\chi$ mode is imposed to be of unit positive Wronskian. However, because Eq.~\eqref{eq:Pmodeeq} is second order, $\chi$ is not completely fixed by Eq.~\eqref{eq:unitW}. To identify the $in$ mode which describes particles at early time, one has to impose that it behaves as the positive frequency WKB mode at early time~\cite{birrell1984quantum}. Using Eq.~\eqref{eq:repar} to reexpress this condition in terms of $P$, the $in$ mode $\chi_{\rm BD}$ must obey [compare with Eq.~\eqref{eq:adiabaticsol}]
\begin{equation}
\label{eq:chiWKB}
\chi_{\rm BD}\underset{P\to \infty}{\sim} \frac{\ep{i \int^P {F\left(P^{\prime 2}\right)} \frac{dP'}{P' H}}}{\sqrt{2 \frac{F(P^2)}{ P}}}.
\end{equation}
Then the modes $\phi_{\bf k}$ with unit positive norm can all be written as 
\begin{equation}
\label{eq:kp}
\phi_{\bf{k}}(t) = \sqrt{\frac{H}{k}}\left[A_{\bf{k}} \chi_{\rm BD}(P) + (B_{-\bf{k}} \chi_{\rm BD}(P))^*\right],
\end{equation}
where $A_{\bf{k}}$ and $B_{-\bf{k}}$ satisfy $\abs{A_{\bf{k}}}^2-\abs{B_{-\bf{k}}}^2=1 $, and where the extra factor of $\sqrt{H/k}$ ensures that $( \Phi_{\bf{k}}, \Phi_{\bf{k'}}) =2 \pi H \delta({\bf{k}}-{\bf{k'}})$ is found when $\chi_{\rm BD}$ obeys Eq.~\eqref{eq:unitW}. The state which is vacuum with respect to $\chi_{\rm BD}$ for all values of $\bf k$, i.e., $B_{\bf k} = 0$ for all $\bf k$, is the Bunch-Davies (BD) vacuum~\cite{Schomblond:1976xc,Bunch:1977sq}.

To handle the mode identification in the $\omega$-representation, it is appropriate to work with the Fourier mode of Eq.~\eqref{eq:facto} and to separate solutions with positive and negative values of $\bf P$. In the WKB approximation, positive norm solutions describe right moving ($U$) particles for ${\bf P} > 0$, left moving ($V$) particles for ${\bf P} < 0$, and vice versa for negative norm solutions. However, the exact solutions of Eq.~\eqref{eq:ommodeeqinP} mix $U$ and $V$ modes. The general solution should thus be decomposed as 
\begin{equation}
\label{eq:omegaP}
\begin{split}
\tilde \phi_\omega({\bf{P}}) = &\left(\frac{P}{H}\right )^{-i \frac{\omega}{H}-1} \left\{ \frac{\theta({\bf{P}})}{H } \left[A_\omega^U \chi_{\rm BD} + \left (B^{V}_{-\omega} \chi_{\rm BD} \right )^*\right] + \frac{\theta(-{\bf{P}})}{H } \left[ A^V_\omega \chi_{\rm BD} + \left ( B^{U}_{-\omega} \chi_{\rm BD} \right )^*\right] \right\},
\end{split}
\end{equation}
where the $4$ coefficients weigh the initial (BD) contributions with positive (negative) norm $A$ ($B$), and with $U$ or $V$ content. In fact, the scalar product of two such modes $\Phi_\omega ={\ep{- i\omega t}} \phi_\omega, \Phi_{\omega'} = {\ep{- i\omega' t}} \phi_{\omega'}$ is 
\begin{equation}
\begin{split}
(\Phi_{\omega} , \Phi_{\omega'} )&=2 \pi H \delta(\omega-\omega') \left( \abs{A^U_\omega}^2-\abs{B^U_{-\omega}}^2+\abs{A^V_\omega}^2- \abs{B^{V}_{-\omega}}^2\right) .
\end{split}
\end{equation}
This is exact and can be verified be expressing Eq.~\eqref{eq:scalpr} in the $P$-representation, see Ref.~\cite{Balbinot:2006ua}. Imposing the positive norm condition $(\Phi_{\omega} , \Phi_{\omega'} ) = 2 \pi H \delta(\omega-\omega')$ on the mode basis constraints the above parenthesis to be unity. Hence, in de Sitter, irrespectively of the dispersion relation of Eq.~\eqref{eq:disprel}, the complete set of positive norm stationary modes contains $2$ modes $\phi_\omega^U,\phi_\omega^V$ for $\omega \in (-\infty, \infty)$. One verifies that $2n-4$ solutions of Eq.~\eqref{eq:ommodeeq} are not asymptotically bounded in $X$, and cannot be normalized. These modes should not be used when decomposing the canonical field obeying Eq.~\eqref{eq:dispKGeq}~\cite{Macher:2009tw}. It is interesting to notice that the completeness of the stationary modes follows from the completeness of the Mellin transform~\cite{morse1953methods}, in a manner similar that the completeness of the homogeneous mode basis follows from that of the Fourier transform, see the end of \ref{sec:affinefieldth}. With this remark, we have verified that the set of the asymptotically bounded solutions of Eq.~\eqref{eq:ommodeeq} matches that of the solutions of Eq.~\eqref{eq:kmodeq}. 

To conclude this section, we point out that the $S$-matrix in the $k$-representation factorizes into 2-mode sectors containing particles with opposite wave vectors $\bf k$, because the space is homogeneous. Instead, in the $\omega$-representation, the $S$-matrix factorizes in different sectors with $\omega > 0$, each of them being a 4-mode sector which contains two $U$ modes $\phi_\omega^U, (\phi_{-\omega}^U)^*$, and two $V$ modes $\phi_\omega^V, (\phi_{-\omega}^V)^*$. The $4 \times 4$ character of the $S$-matrix in this representation results from the composition of the cosmological mixing of $U$ and $V$ modes with the stationary mixing of modes of opposite frequency, see \ref{sec:Smatrix} for details.

\subsection{The two Hamiltonians} 

In preparation for the analysis of the stability of the BD vacuum, we study the Hamiltonian of our dispersive field. We first point out that the fields $u$ and $K_t$ define two different Hamiltonian functions, that we call respectively, $H_u$ and $H_t$. Using the conjugated momentum $\Pi = - u^\mu \partial_\mu \phi $ and the Lagrangian density $\mathcal{L}$ of Eq.~\eqref{eq:Actiondesitterdisp}, they are respectively given by 
\begin{equation}
\label{eq:prefE1}
\begin{split}
H_u \doteq \int dl ( \Pi \partial_t|_z \phi - \mathcal{L}), \quad
H_t \doteq \int dl ( \Pi \partial_t |_X \phi - \mathcal{L}),
\end{split}
\end{equation}
where we recall that $\partial_t |_z= -u^\mu \partial_\mu $ and $\partial_t |_X = -K_t^\mu \partial_\mu $. $H_u$ thus engenders time translations at fixed $z$, while $H_t$ does it at fixed $X$. In de Sitter space, $H_u$ and $H_t$ differ because $u = K_t - v s \neq K_t$ since the flow $v= HX$ does not vanish. In Minkowski space endowed with a Cartesian $u$ field, which is obtained in the limiting case $H \to 0$, the two Hamiltonians coincide since $u \to K_t$ when $H \to 0$. This implies that in de Sitter $H_t $ and $H_u$ share the properties that the Hamiltonian possess in Minkowski space. 

On the one hand, the stationary $ H_t$ 
\begin{equation}
\begin{split}
 H_t= \frac12 \int_{-\infty}^\infty dX &\left\{ (\partial_t\phi)^2 + m^2\phi^2 + (1 - v^2) (\partial_X \phi)^2 + \phi f(- \partial_X^2) \phi \right\} ,
\end{split}
\end{equation}
is conserved for all dispersion relations. However, for both Lorentz invariant theories and dispersive ones with $f(P^2)\geq 0$, it is {\it not} positive definite precisely because $K_t$ is space like outside the horizons. (Recall that its norm is $- K_t^2 = 1 - v^2$.) Notice also that when working in the $P$ representation one easily verifies that the last term in $H_t$ is positive definite for $f> 0$. For dispersive theories with $f < 0$, such as phonons in Helium$^4$, the density of $H_t$ becomes negative where $v$ reaches the critical Landau velocity~\cite{Jacobson:1999zk,Pitaevskii1984}. In any case, in de Sitter, when using the stationary modes of Eq.~\eqref{eq:omegaP}, $H_t$ can be decomposed as $\int_{0}^\infty d\omega H_\omega $, where
\begin{equation}
\label{eq:omE}
\begin{split}
 H_\omega &= \frac{\omega}{H} \left(\abs{A_\omega^U}^2 + \abs{A_\omega^V}^2 - \abs{B_{-\omega}^U}^2 - \abs{B_{-\omega}^V}^2 \right) .
\end{split}
\end{equation}
$H_t$ is thus manifestly conserved and not positive definite.

On the other hand, the cosmological $H_u$ can be decomposed as $H_u =\int_{-\infty} ^\infty d{\bf k} H_{\bf k}$ where
\begin{equation}
\begin{split}
 H_{\bf k}(t) &= \frac{a(t)}{ 2H^2 } \left( \abs{\partial_t \phi_{\bf k}(t)}^2 + F^2(\frac{k^2}{ a(t)^2}) {\abs{\phi_{\bf k} (t)}}^2 \right) .
\end{split}
\end{equation}
$H_u$ is thus positive definite whenever $F^2 > 0$. (Notice that theories with $F^2 < 0$ are dynamically unstable even in Minkowski space.) However, $H_t$ is not conserved because $ d \ln a /dt = H \neq 0$. The nonconservation of $H_{\bf k}(t)$ engenders nonadiabatic transitions~\cite{Massar:1997en} which describe pair creation of quanta of opposite $\bf k$, see \ref{sec:betacosmo} for a particular example. When using Eq.~\eqref{eq:kp}, the time dependence of $H_k$ can be entirely expressed in the $P$ representation as
\begin{equation}
\begin{split}
 H_{\bf k}(t) &= \frac{ \abs{ A_{\bf k}}^2+ \abs{ B_{ -\bf k}}^2 }{2 H P } \left({\abs{HP\partial_P \chi_{\rm BD}}}^2 + F^2(P^2) {\abs{\chi_{\rm BD}}}^2 \right) \\
 &+ {\rm Re}\left\{ \frac{A_{\bf k} B_{ -\bf k}}{H P } \left( (HP\partial_P \chi_{\rm BD})^2 + F^2(P^2) \chi_{\rm BD}^2 \right) \right\}. 
\end{split}
\end{equation}
We also see that when imposing $\abs{A_{\bf{k}}}^2-\abs{B_{-\bf{k}}}^2=1 $, the minimization of $H_k$ (more precisely, its integral over one period) implies $B_{-\bf{k}} = 0$. This is the classical equivalent of saying that the BD vacuum is the lowest energy state with respect to the preferred frame field $u$.

To conclude, we clearly see the complementary roles played by the Hubble constant $H$. In the stationary representation, it is responsible for an {\it energetic instability}, i.e., for a conserved Hamiltonian $H_t$ unbounded from below. Instead in the homogeneous representation, $H$ is responsible for the time dependence of the positive definite $H_u$, which engenders pair creation, i.e., a {\it vacuum instability}. These two properties are valid for all dispersion relations, and therefore they also apply to Lorentz invariant theories. This is a reminder that field theories in de Sitter space, and in black hole metrics, are threatened by dynamical instabilities, i.e., complex frequency modes~\cite{Damour:1976kh,Coutant:2009cu}. In addition, as argued below, violations of thermodynamical laws are also related to an energetic instability.

\section{The consequences of Lorentz violations} 
\label{sec:thermal}

In de Sitter space, when considering Lorentz invariant fields, the Bunch-Davies vacuum possesses many remarkable properties. On the one hand, it is homogeneous and stationary, and on the other hand, it is an Hadamard state. In fact, it is the only stationary Hadamard state~\cite{Schomblond:1976xc,Joung:2006gj}. In addition, it can be shown that all other Hadamard states flow towards the BD vacuum. By this we mean that the $n$-point functions evaluated in these states flow towards the corresponding one evaluated in the BD vacuum. In this sense, the BD vacuum is the only stable regular state. Finally, when evaluated in the static patch $\vert HX\vert < 1$, the $n$-point functions are all thermal~\cite{Kay:1988mu}. They indeed obey the double KMS condition: they are periodic in imaginary time with period ${2\pi}/{H}$, and they are analytic in the strip $0<\Im (t) < {2\pi}/{H}$. 

When considering dispersive fields, we shall see that, for free fields at least, the BD vacuum still satisfies all these properties, save the very last. In fact, even though the periodicity in $\Im (t)$ is still exactly found, the analyticity in the strip is always lost when there is high frequency dispersion. This means that the BD vacuum is no longer a thermal state. 

\subsection{Stationarity and periodicity}

Since we work with free fields and since the BD vacuum is a Gaussian state, we only need to consider the $2$-point function. When using the settings of the former section, the Wightman function in the BD vacuum can be written as~\cite{birrell1984quantum}
\begin{equation}
\begin{split}
G_{\rm BD}({{z}},t,{{z}}_0,t_0) =\int_{-\infty} ^{\infty} d{\bf{k}} \frac{\ep{i{\bf{k}}(z-z_0)}}{2 \pi k} \chi_{\rm BD}\left(k H\ep{-Ht}\right) \chi^{*}_{\rm BD}\left(k H \ep{-Ht_0}\right) .
\end{split}
\end{equation}
When considered at fixed $t_0$ and $t$, this function is manifestly homogeneous. It is also stationary, when considered at fixed $X_0 = a(t_0) z_0 $ and $X = a(t) z $. Indeed in terms of $P = k/a(t)$, one gets 
\begin{equation}
\begin{split}
G_{\rm BD}({{X}},t,{{X}}_0,t_0) &= \int_0^\infty dP \frac{i \sin(PX - PX_0 \ep{H(t-t_0)})}{ \pi P}\chi_{\rm BD}(P) \chi_{\rm BD}^{*}(P \ep{H(t-t_0)}) ,
\label{eq:gfP}
\end{split}
\end{equation}
which is a function of $t - t_0$ only. Hence, for all dispersion relations imposed in the cosmological frame, the BD vacuum is both stationary and homogeneous, i.e., invariant under the affine group, as it is for relativistic fields. More surprisingly, $G_{\rm BD}$ is also periodic in the imaginary time lapse, with the usual period $2\pi/H$, exactly as for Lorentz invariant fields, and for thermal functions. 

We expect that homogeneity, stationarity and the above periodicity will be exactly preserved when considering the $n$-point functions of interacting fields evaluated in the BD vacuum, because these properties are protected by the affine group of Eq.~\eqref{eq:alg}. In other words, the $n$-points functions will always be invariant under this subgroup. 

\subsection{Thermality} 
\label{sec:thermbroke}

For Lorentz invariant theories, it has been shown by Gibbons and Hawking~\cite{Gibbons:1977mu} that freely falling observers immersed in the BD vacuum detect a thermal bath with a temperature $k_B T_H =\frac{H}{2\pi }$. It was also understood that, when restricted to the static patch $-1 < HX < 1$, the reduced density matrix of any quantum field theory (interacting or not) is a thermal state at that temperature. Interestingly, this result is always violated for dispersive fields. 

To demonstrate this, we consider particle detectors which follow orbits that are stationary with respect to the Killing field $K_t$. Because $K_t^2 = - 1 + H^2 X^2$, the only stationary orbits are at fixed $X$, and with $\vert HX\vert < 1$ when they are timelike. The detector transition rate of spontaneous excitation $R_-$ (de-excitation $R_+$) is proportional to~\cite{Unruh:1976db,Brout:1995rd} 
\begin{equation}
\begin{split}
R_{\pm} (\omega,{{X}}) \propto & \int_{-\infty}^{\infty} H dt \ep{\pm i \omega t} G_{\rm BD}({{X}},t; {{X}}, 0).
\label{eq:Rs}
\end{split}
\end{equation}
To relate the detector energy gap $\Delta E > 0$ to the Killing frequency $\omega$, one must take into account the $X$ dependent redshift factor ($\Delta E = \omega / \sqrt{1-H^2 X^2}$) coming from the detector's kinematics, which also enters in Tolman law $T_{\rm loc} (X)= T_{\rm gl} / \sqrt{1-H^2 X^2}$ relating the local temperature to the globally defined one~\cite{Misner1973,Tolman:1930zza}. In what follows, we shall work with the globally defined temperature and with $\omega$. To study the deviations from thermality, it is convenient to use the temperature function $T_{\rm gl}(\omega,X)$ defined by
\begin{equation}
\begin{split}
\label{eq:tempdef}
\frac{R_-(\omega,X)}{R_+(\omega,X)}&= \ep{ - {\omega}/{k_B T_{\rm gl}(\omega,X)}}.
\end{split}
\end{equation}
For relativistic fields, one has $k_B T_{\rm gl}(\omega,X)= k_B T_H = H/2\pi$, for all $\vert HX\vert < 1$ and for $0< \omega < \infty$, in accord with Tolman law and the Planck spectrum. To compute $T_{\rm gl}$ in the presence of dispersion, we shall use the fact that the rates are given by 
\begin{equation}
\begin{split}
R_\pm(\omega,{{X}}) = \abs{\phi^{{\rm BD}, U}_{\pm\omega}({{X}})}^2+\abs{\phi^{{\rm BD}, U}_{\pm\omega}(-{X})}^2, 
\end{split}
\label{eq:Rs2}
\end{equation}
where $\phi^{{\rm BD}, U}_{\omega}$ is the positive norm BD mode that is initially right moving, i.e., $A_\omega^U = 1, A_\omega^V = B_\omega^U = B_\omega^V = 0$ in Eq.~\eqref{eq:omegaP}. Similarly, $(\phi^{{\rm BD}, U}_{-\omega})^*$ is given by the negative norm, negative frequency, mode: $B_\omega^U= 1$. In the above equation we have used the symmetry $X \to - X$ to express the contribution of the left moving $V$-mode evaluated at $X$ as that of the right $U$-moving one at $-X$. Explicitly, $\phi^{{\rm BD}, U}_{\omega}$ is 
\begin{equation}
\label{eq:amp0}
\begin{split}
\phi^{{\rm BD}, U}_{\omega}({{X}})= \int_0^\infty \frac{dP}{\sqrt{2 \pi} H} \ep{i P X} \left (\frac{P}{H}\right )^{- i\omega/H -1} \chi_{\rm{BD}}(P) ,
\end{split}
\end{equation}
where $\chi_{\rm{BD}}$ obeys Eq.~\eqref{eq:Pmodeeq}. To prove that thermality is violated it is sufficient to work with $X= 0$ and to consider very high frequencies $\omega/\Lambda \gg 1$. In this limit the integral is dominated by high values of $P$, and therefore by the leading term of the dispersion relation, that we parametrize here by 
\begin{equation}
f_n(P^2)=\frac{P^{2n}}{\Lambda^{2n-2}}. 
\label{eq:fn}
\end{equation}
In the high $P$ regime, the WKB expression of Eq.~\eqref{eq:chiWKB} offers a reliable approximation of $ \chi_{\rm{BD}}$. Hence, up to irrelevant constants, one gets
\begin{equation}
\phi^{{\rm BD}, U}_{\omega}({{X}}= 0) \approx \int \frac{dP}{P} P^{- i\frac{\omega}{H} - \frac{n-1}{2}} \ep{i P^n} . 
\end{equation}
Using $Q= P^n$ as integration variable, one obtains a $\Gamma$ function, namely
\begin{equation}
\begin{split}
\phi^{{\rm BD}, U}_{\omega}({{X}}= 0) &\approx{\ep{{{ \omega \pi }/{2nH} }}} \times \Gamma \left( {-\frac {i\omega}{n H}} - \frac{n-1}{2n} \right) .
\end{split}
\end{equation}
Using this result in Eq.~\eqref{eq:Rs2}, Eq.~\eqref{eq:tempdef} gives
\begin{equation}
\begin{split}
k_B T_{\rm gl}(\omega \gg\Lambda, X = 0) = n \frac{H}{2\pi},
\end{split}
\end{equation}
i.e., $n$ times the standard temperature $T_H$. 

Instead, for $\omega/\Lambda \ll 1$ and $H/\Lambda \ll 1$, $T_{\rm gl}(\omega, X = 0)$ reduces approximatively to the standard temperature $T_H$~\cite{Coutant:2011in}. Hence the BD vacuum is no longer thermal. This result is nontrivial since $G_{\rm BD}$ of Eq.~\eqref{eq:gfP} is still periodic in imaginary time, with the standard period. From the above equations and from $P \propto \ep{-Ht}$, one understands that the power $n$ of Eq.~\eqref{eq:fn} reduces the domain of analyticity of $G_{\rm BD}$ in $\Im (t)$ by a factor of $n$. Indeed since $\chi_{\rm BD} \sim \ep{i P^n }$ for large $P$, the integral in Eq.~\eqref{eq:gfP} contributes as ${1}/({1-\ep{n H t}})$ which is analytic in the reduced strip $0< \Im (t)< 2\pi /n H$ only. The observation that $T_{\rm gl}(\omega, X = 0) = n T_H$ for $\omega/\Lambda \gg 1$ shall be verified, for $n=2$, in a exactly solvable model in \ref{sec:thermidev}.

\subsection{Regular states and stability}

In this section, we show that some of the ingredients of event horizon thermodynamics~\cite{Gibbons:1977mu,Jacobson:2003wv} are still present when adding high frequency dispersion. Namely, we show, firstly that the BD vacuum is the only stationary state which is regular and, second, that the other regular states flow towards the BD state. (As explained below, the notion of regular states should be understood as the generalization of Hadamard states in the presence of short distance dispersion.) Hence for free fields at least, the BD state is the only stable state. To prove these claims we shall use concepts that are common to Lorentz invariant and dispersive fields. 

Before proceeding, let us discuss our criterion of stability. We say that the BD vacuum is stable because, at large time, observables computed in nearby states converge towards those evaluated in the BD vacuum. Hence, for these observables, the perturbed states will be asymptotically indistinguishable from the BD vacuum. This flow is often referred to as a cosmic no hair theorem~\cite{Marolf:2010nz,Marolf:2011sh,Hollands:2010pr} as it closely follows the Price's no hair theorem~\cite{Misner1973}. We adopted this criterion because there is no stationary Killing field which is globally timelike in de Sitter. As a consequence, there is an {\it energetic} instability, see Eq.~\eqref{eq:omE}, which means that stability cannot be deduced from a spectrum bounded from below. It is worth mentioning that to study the thermalization in interacting quantum field theories, the flow towards stationary thermal states is established in Ref.~\cite{Giraud:2009tn} by studying some $n$-point functions. Even though the purity of the initial state is preserved by the Hamiltonian evolution, after a while, these functions become indistinguishable from thermal ones. In that case as well, the stability of the state is thus inferred from the flow of some observables, rather than from the evolution of the state itself. 

Since high frequency dispersion modifies the short distance behavior of the $2$-point function, we first need to define what we mean by \enquote{regular states} because the standard definition of Hadamard state is precisely based on this behavior~\cite{birrell1984quantum}. In homogeneous cosmological spaces, this difficulty can be overcome because one can rephrase the standard definition in terms of an adiabatic expansion of the solutions of Eq.~\eqref{eq:kmodeq} at fixed $\bf k$. Since these new terms are common to both Lorentz invariant and dispersive field, one can implement the subtraction procedure to dispersive fields. Let us recall the key elements, for more details, see Ref.~\cite{LopezNacir:2005db}. In de Sitter, because of Eq.~\eqref{eq:alg}, the adiabatic expansion can be done in terms of a single mode $\chi$, solution of Eq.~\eqref{eq:Pmodeeq}, and of unit Wronskian, see Eq.~\eqref{eq:unitW}. This expansion generalizes Eq.~\eqref{eq:chiWKB} and is best expressed as 
\begin{equation}
 \chi_{\rm adiab}= \sqrt{\frac{1}{2 W(P)}} \ep{i\frac{1}{H} \int^{P} W(P') dP'}, 
\end{equation}
where $W$ obeys the nonlinear equation
\begin{equation}
W^2 = \frac{F^2(P^2)}{P^2}-\frac{H^2}{2} \left( \frac{\partial_P^2 W}{W}-\frac{3}{2} \frac{(\partial_P W)^2}{W^2} \right),
\label{eq:WF}
\end{equation}
and where $F^2$ determines the dispersion relation in Eq.~\eqref{eq:disprel}.

When working with Lorentz invariant fields in $D$ dimensions, the first $1+ D/2$ terms in a iterative solution of Eq.~\eqref{eq:WF} should be taken into account when determining the $1+ D/2$ quantities that need to be subtracted. This guarantees that the renormalized stress tensor evaluated in the BD vacuum is finite in cosmological spaces, and thus in de Sitter. This is not a surprise since $\chi_{\rm BD}$ and $\chi_{\rm adiab}$ obey the same condition for $P\to \infty$. Hence their differences develop at finite $P$, and because of the expansion $H$. When working with dispersive fields, this finiteness is still found when $F^2$ is positive, sufficiently regular, and grows faster that $P^2$ for $P \to \infty$. Indeed the higher the power $n$ in the leading term of Eq.~\eqref{eq:WF}, the more suppressed are the next order terms in the adiabatic expansion~\cite{LopezNacir:2005db}. For instance, in two dimensions, for $F^2_n \sim P^{2n}/\Lambda^{2n -2}$, with $n\geq 2$, the second quantity which is usually subtracted in the stress tensor is already finite. It can thus be either subtracted or not.\footnote{
As a result, in Ref.~\cite{LopezNacir:2005db}, it is proposed to subtract only the first term. We claim instead that the first two terms should be subtracted, as done when dealing with Lorentz invariant fields. Indeed only this choice guarantees that the stress tensor would remain finite when taking the limit $\Lambda\to \infty$. In addition, in our proposal, the two manners to consider $\Lambda$ become compatible. $\Lambda$ can be either seen as a (Lorentz violating) regulator to be sent at $\infty$ when computing observables, or as a physical finite ultraviolet parameter, but which enters suppressed in observables.}
In either case, in the BD vacuum of de Sitter, the renormalized values of $\rho = u^\mu u^\nu T_{\mu \nu}$ and $\Pi= s^\mu s^\nu T_{\mu \nu}$ are constant in space and time, while the flow $J = u^\mu s^\nu T_{\mu \nu}$ vanishes. Explicitly, they are (we neglect the part coming from dynamics of the unit time-like vector field $u$), with $W_0 = F / P$~\cite{Lemoine:2001ar}
\begin{equation}
\begin{split}
 \rho   &=\int_0^\infty \frac{P  d P}{2\pi}  \left[ \frac{W_0^2}{W} -  W_0 - \frac{H^2 }{4W_0}  \left( \frac{\partial_P^2 W_0}{W_0}-\frac{3}{2} \frac{(\partial_P W_0)^2}{W_0^2} \right) \right]\\
 \Pi  &= \int_0^\infty \frac{P  d P}{2\pi}  \frac{d P W_0}{d P}  \left[   \frac{ W_0}{W}   - 1 - \frac{H^2 }{4W_0^2}  \left( \frac{\partial_P^2 W_0}{W_0}-\frac{3}{2} \frac{(\partial_P W_0)^2}{W_0^2} \right) \right],
\end{split}
\end{equation}
where the last two term come from the subtraction procedure.

We now consider the change of the stress tensor with respect to that of the BD vacuum when working with some (possibly mixed) state $\Psi$ described by the density matrix $\hat \rho_\Psi$. For free fields, this change is determined by the difference of the $2$-point functions $\delta G_\Psi= G_\Psi - G_{\rm BD}$. This difference can be expressed in terms of the positive norm BD modes $\phi_{\bf k}^{\rm BD}$ as 
\begin{equation}
\label{eq:green}
\begin{split}
\delta G_\Psi&= 2 {\rm Re}\int \frac{{d{\bf{k}} d{\bf{k'}}}}{2\pi} \left[ \ep{i ({\bf{k}} {{z}}-{\bf{k'}} {{z'}})} \phi_{\bf k}^{\rm BD}(t) \left\lbrace n_\Psi({{\bf{k}},{\bf{k'}}}) (\phi_{\bf k'}^{\rm BD}(t'))^* +c_\Psi({{\bf{k}},{\bf{k'}}}) \phi_{\bf k'}^{\rm BD}(t')\right\rbrace \right] , 
\end{split}
\end{equation}
where $n_\Psi({{\bf{k}},{\bf{k'}}})$ and $c_\Psi({{\bf{k}},{\bf{k'}}})$ are expectation values of normal ordered products of BD destruction and creation operators $a_{\bf{k}}, a^{\dagger}_{\bf{k'}}$: 
\begin{equation}
\begin{split}
n_\Psi({{\bf{k}},{\bf{k'}}} )=  \left< a^{\dagger}_{\bf{k'}} a_{\bf{k}} \right >_{\hat \rho_\Psi},
\quad
c_\Psi({{\bf{k}},{\bf{k'}}}) =  \left< a_{\bf{k}} a_{-{\bf{k'}}}\right >_{\hat \rho_\Psi}. 
\end{split}
\end{equation}
They respectively encode the power spectrum and the coherence of $\hat \rho$ at the Gaussian level~\cite{Campo:2005sy}.

To establish the stability of the BD vacuum, we first point out that the other stationary states are all singular. The reason comes from the fact that the stationarity of $G_\Psi$ implies that, irrespectively of $c_\Psi$, ${\bf{k}}\times n_\Psi({{\bf{k}},{\bf{k'}}} )$ only depends on the ratio ${{\bf{k}}/{\bf{k'}}}$. Therefore, the change of the expectation value of $H_u$ of Eq.~\eqref{eq:prefE1} with respect to the BD vacuum, necessarily diverges because the contribution of high ${\bf k}$ is not suppressed enough, as $ n_\Psi({{\bf{k}},{\bf{k}}} )\propto 1/k$. This is true for dispersion functions $F(P) \geq \epsilon >0 $ for $P\to \infty$, and therefore true for Lorentz invariant theories. This generalizes the fact~\cite{Joung:2006gj} that the $\alpha$-vacua, which are invariant under the full de Sitter group and therefore stationary, are all singular, save the BD vacuum. 

In our second step we consider states that describe, at some initial time, {\it local} perturbations containing a {\it finite} number of BD particles: $N_{\rm tot} = \int d{\bf k} n_\Psi({{\bf k},{\bf k}}) < \infty$. Moreover, to be able to handle all dispersion relations at once, we suppose that there exists a cut off wave number $k_{\max}$ above which the number of particles decreases exponentially, i.e.
\begin{equation}
\label{eq:nottoofar}
n_\Psi({{\bf k},{\bf k}}) \leq \ep{- b k }, \quad \forall k>k_{\max},
\end{equation}
with $b > 0$. Then Schwartz inequalities and the hermiticity of $\hat \rho_\Psi$ implies the following inequalities generalizing those of Ref.~\cite{Campo:2005sy} 
\begin{equation}
\abs{n_\Psi({{\bf k'},{\bf k}})}^2 \leq \abs{n_\Psi({{\bf k},{\bf k}})}
\abs{n_\Psi({{\bf k'},{\bf k'}})},
\end{equation} 
and 
\begin{equation}
\begin{split}
 &\abs{\int d{\bf k_1} f_{{\bf k_1}} c_\Psi({{\bf k_1},{\bf k}})}^2 
\leq {n_\Psi({{\bf k},{\bf k}})} \left [\int d{\bf k_1} d{\bf k_2} 
f_{{\bf k_1}} f_{{\bf k_2}}^* (n_\Psi({{\bf k_1},{\bf k_2}})+\delta({\bf k_1}-{\bf k_2}))\right ],
\end{split}
\end{equation}
for all test functions $f_{\bf k} \in \mathbb{C}$.

Using these inequalities one can study the behavior of Eq.~\eqref{eq:green} at large time. Since there is a momentum cutoff $k_{\max}$, at large time only low momenta $P$ matter. Hence for all dispersion relations of Eq.~\eqref{eq:disprel}, the dominant term is the mass term. One should then distinguish massive fields with $m > H/2$ [This means that $F_{\rm sa}$ of Eq.~\eqref{eq:hatOf} is positive in the $P\to 0$ limit ; see Eq.~\eqref{eq:mum} for the origin of this condition] from massless fields. At this point, one also needs to consider the pair creation amplitudes relating the initial BD mode $\phi^{\rm BD}_{\bf k}$ [obeying Eq.~\eqref{eq:chiWKB}] to the out mode $\phi^{out}_{\bf k}$ defined at low momentum. Using the techniques of Ref.~\cite{Massar:1997en} and the fact that Eq.~\eqref{eq:Pmodeeq} is second order for all $F^2$, we can verify that for both $m > H/2$ and $m = 0$, the $\alpha_k, \beta_k$ coefficients of Eq.~\eqref{eq:Bogok} are bounded for dispersion relations with $F^2 > 0$. Using this result, at large time and for massive fields, one finds that 
\begin{equation}
\delta_\Psi G < \ep{-H (t+t')/2},
\label{eq:expondec}
\end{equation}
 i.e., $\delta_\Psi G$ decreases exponentially in time. This implies that the changes of density, current and pressure with respect to the BD vacuum $\delta \rho_\Psi = u^\mu u^\nu \delta T_{\mu \nu}^\Psi$, $\delta J_\Psi = u^\mu s^\nu \delta T_{\mu \nu}^\Psi$ and $\delta \Pi_\Psi = s^\mu s^\nu \delta T_{\mu \nu}^\Psi$ also flow exponentially to $0$. On the other hand, when $m=0$, at large times the dispersive modes become conformally invariant, i.e., proportional to $\ep{i {\bf k} z - i k \eta}$ where $\eta \propto \ep{-Ht} $ is the conformal time. As a result, both scalar derivatives $u^\mu \partial_\mu \delta G_\Psi$ and $s^\mu \partial_\mu \delta G_\Psi$ flow to $0$ as in Eq.~\eqref{eq:expondec}. This implies that $\delta \rho_\Psi$, $\delta J_\Psi$ and $\delta \Pi_\Psi$ also flow exponentially fast to $0$.

In conclusion, we have shown that for all dispersion relations, the mean stress tensor $\left < T_{\mu\nu}\right >_{\hat \rho_\Psi} $ computed with an arbitrary localized state containing a finite number of BD quanta flows towards that computed in the BD vacuum. This follows from the cosmological expansion $a \sim \ep{Ht}$ which redshifts the momenta $P \sim k \ep{-Ht}$, and dilutes the particles. In our proof we have used the condition of Eq.~\eqref{eq:nottoofar} because, for all polynomial dispersion relations, it guarantees that the change of the stress tensor with respect to its value in the BD vacuum is finite. Less restrictive conditions, and therefore larger set of states, can certainly be used once having chosen some class of dispersion relations. One could also relax the condition that the perturbation is local. However a detailed study of these extensions goes beyond the scope of this thesis. 

\section{Quartic superluminal dispersion}
\label{sec:pfour}

It is of value to explicitly compute the modifications of the observables which are due to high frequency dispersion. In de Sitter there are \textit{a priori} two types of observables: first, the pair creation rates which are due to cosmological expansion, and second the thermal-like response of stationary particle detectors. In \ref{sec:Smatrix} we shall study a third type of observables, namely asymptotic pair creation rates in the $\omega$-representation, which combines the former two phenomena. This has no physical meaning in Lorentz invariant theory. However, because of Lorentz violation, nothings forbids observers to follow space-like trajectories. Moreover, in analogue gravity experiments, this is the measurable quantity since it is the flux that is ejected from the sample.

To get analytical expressions, we consider the quartic superluminal dispersion, i.e., $f = {P^4}/{\Lambda^2}$. In this case, the general solution of Eq.~\eqref{eq:Pmodeeq} is given by 
\begin{equation}
\begin{split}
\chi &= \frac{C}{\sqrt {p}} {{\rm \bf M}\left(\frac{-i\lambda}{4} ,\frac{i \mu}{2}, {\frac {i p^{2}}{ \lambda }}\right)} +\frac{D}{\sqrt {p}} {{\rm \bf W}\left(\frac{-i\lambda}{4} ,\frac{i \mu}{2}, {\frac {i p^{2}}{\lambda }}\right)},
\end{split}
\end{equation}
where $C$ and $D$ are the two integration constants, and where ${\rm \bf M}$ and ${\rm \bf W}$ are the two Whittaker functions, see Chap.~13 in Ref.~\cite{Abramowitz}. For simplicity, we introduced the adimensional quantities $ \mu= \sqrt {\frac{m^2}{H^2} -\frac14} = \frac{F_{\rm sa}(P=0)}{H}$, $\lambda=\Lambda/H$, $p=P/H$. Using Eq.~\eqref{eq:chiWKB} to characterize the initial large $p$ behavior, the unit Wronskian BD mode, when complex conjugated, is given by~\cite{Macher:2008yq,LopezNacir:2005db} 
\begin{equation}
\begin{split}
\label{eq:chist}
\chi_{\rm BD}^* = \sqrt{\frac{\lambda}{2 p}} \ep{\frac{-\pi \lambda}{8 }} {{\rm \bf W}\left(\frac{-i\lambda}{4} ,\frac{i \mu}{2}, {\frac {i p^{2}}{\lambda }}\right)}.
\end{split}
\end{equation}

\subsection{Cosmological pair creation rates} 
\label{sec:betacosmo}

To get the pair creation rates, we need to identify the combination of ${\rm \bf M}$ and ${\rm \bf W}$ that corresponds to the final mode $\chi_{out}$. As Eq.~\eqref{eq:chiWKB} does not offer a reliable approximation for $P \to 0$, the identification should be done using the cosmological time $t$. Using Eq.~\eqref{eq:kmodeq}, one finds that asymptotic positive norm solutions are proportional to $\ep{- i \mu H t}$ at large $t$. When $m < \frac{H}{2} $, $\mu$ is imaginary and the modes grow or decay at large time~\cite{Joung:2006gj}. Hence it is not possible to define asymptotic $out$ modes. When $m > \frac{H}{2} $, there is no difficulty: when reexpressing $\ep{- i \mu H t}$ in terms of $P \propto \ep{-Ht}$, one gets 
\begin{equation}
\begin{split}
\chi _{out}& \underset{p\to 0}{\sim}\frac{p^{\frac12+i{\mu}}}{\sqrt{2 {\mu} }}.
\label{eq:mum}
\end{split}
\end{equation}
Using this behavior, the positive unit Wronskian $out$ mode is found to be
\begin{equation}
\label{eq:chioupfour}
\begin{split}
\chi _{out}= {\left(-i\lambda\right)^{\frac{1+ i \mu }{2}} } \sqrt{\frac{1}{2 \mu p}} {{{\rm \bf M}\left(\frac{-i\lambda}{4} ,\frac{i \mu}{2}, {\frac {i p^{2}}{\lambda}}\right)}}.
\end{split}
\end{equation}
The $in-out$ Bogoliubov transformation is given by; see Eq.~\eqref{eq:modebogotf}
\begin{equation}
\chi_{out}= \alpha_{k} \chi_{\rm BD} +\beta_{k} \chi^*_{\rm BD} .
\label{eq:Bogok}
\end{equation} 
We put a subscript $k$ to the above ($k$-independent) coefficients to remind the reader that all these calculations are done in the $k$-representation. Using Sec 13.1 in Ref~\cite{Abramowitz}, one finds, 
see Appendix~B.2 in Ref.~\cite{Macher:2008yq}, 
\begin{equation}
\label{eq:betakmassive}
\begin{split}
\abs{\beta_k(\mu,\lambda)}^2 
&= \frac{1}{\ep{2\pi \mu}-1} \left( 1 + \ep{-\frac{\lambda \pi}{2}} \times \ep{\pi \mu}\right).
\end{split}
\end{equation}
For $\lambda \gg 1$, up to exponentially small correction, one recovers the relativistic result, i.e., the first term in the above equation. For $\lambda \ll 1$, there is an enhancement of the pair creation probability by a factor equal to $\ep{\pi \mu}$. 

Even though the asymptotic $out$ modes cannot be defined for $0 < m \leq H/2$, when $m=0$, it is again possible to define these modes since for $t\to \infty$, they behave as $\ep{- i k \eta }$ where $d \eta = dt/a(t)$ is the conformal time. It is then possible to identify the massless $out$ combination of ${\rm \bf M}$ and ${\rm \bf W}$, and to extract the Bogoliubov coefficients. In this case, the norm of $\beta_k$ is
\begin{equation}
\label{eq:betakmassless}
\begin{split}
\abs{\beta_k(\lambda)}^2 & = \frac{\pi}{\sqrt{\lambda} \abs{\Gamma\left(\frac{1}{4}+i\frac{\lambda}{4}\right)}^2 \ep{\pi \lambda /4}}+ \frac{\pi \sqrt{\lambda}}{4 \abs{\Gamma\left(\frac{3}{4}+i \frac{\lambda}{4}\right)}^2 \ep{\pi \lambda/4}}-\frac{1}{2} .
\end{split}
\end{equation}
For $\lambda \gg 1$, one gets $\abs {\beta_k}^2 \sim {1}/({64 \lambda^4})$, i.e., a power law decrease, unlike what we found above for the massive case. Eq.~\eqref{eq:betakmassless} corrects an error in Eq.~(131) of Ref.~\cite{Balbinot:2006ua} but without altering the conclusions of that section.

\subsection{Deviations from thermality}
\label{sec:thermidev}

Following \ref{sec:thermbroke}, our aim is to exactly compute $T_{\rm gl}(\omega,X)$ of Eq.~\eqref{eq:tempdef} using Eq.~\eqref{eq:Rs2}. To this end, we need to evaluate Eq.~\eqref{eq:amp0} for quartic dispersion. Using Eq.~\eqref{eq:chist}, we get
\begin{equation}
\label{eq:AMP1}
\begin{split}
(\phi^{{\rm BD}, U}_{ - \omega} (X) )^* = \sqrt{\frac{\lambda}{4 \pi }} \ep{\frac{-\pi \lambda}{8}} \int_{0}^{\infty}d{p} \ep{- i p H X } p^{ - i \frac{\omega}{H}-\frac32 } {{\rm \bf W}\left(\frac{-i\lambda}{4} ,\frac{i \mu }{2}, {\frac {i p^{2}}{\lambda }}\right)}.
\end{split}
\end{equation}
Surprisingly, it turns out that this integral can be exactly done, as shown in the next subsection. Since the final expression is a sum that converges as $2^{-n}$ for large $n$, one can accurately numerically compute the ratio of Eq.~\eqref{eq:tempdef} in terms of known hypergeometric functions. To study the consequences of quartic dispersion, we plot the temperature function $T_{\rm gl}(\omega,X)$ in various cases, see \ref{sec:numTgl}. 

\subsubsection{Evaluation of Eq.~\texorpdfstring{\eqref{eq:AMP1}}{}} 

In this technical section, we derive explicit expression for $(\phi^{{\rm BD}, U}_{- \omega})^*$ of Eq.~\eqref{eq:AMP1}. To do so, we shall compute a more general function $A(z)$ to be able to exploit some analytical property in $z$. It is defined by
\begin{equation}
A(z) \doteq \ep{i\pi(\alpha+1)/4} \int_{0}^{\infty} d{p} p^{\alpha} \ep{- i p x } {\rm \bf W}\left( \kappa, \nu, {{i z p^{2}} }\right) \frac{\ep{i(z-1)p^2}}{ z^{\frac12}}, 
\end{equation}
and which is related to the BD mode by
\begin{equation}
(\phi^{{\rm BD}, U}_{ - \omega}(X))^* = \ep{-i\pi(\alpha+1)/4} \sqrt{\frac{\lambda}{4 \pi }} \ep{\frac{-\pi \lambda}{8}} \lambda^{(\alpha+1)/2} A(z =1).
\end{equation}
To simplify notations, we introduced $\alpha=-3/2 - i {\omega}/{H}$, $\xi = {-i\lambda}/{4} $, $\nu ={i \mu }/{2}$, $x=H X\sqrt{\lambda}$ and rescaled $p\to p\sqrt{\lambda}$. Making a rotation in complex $p$ plane of angle $\pi/4$, one gets
\begin{equation}
\begin{split}
A(z)&=\int_{0}^{\infty}d{p} p^{\alpha} \ep{- p x\ep{i\pi/4} } {\rm \bf W}\left( \xi, \nu, {z p^{2}}\right) \frac{\ep{(z-1)p^2}}{ z^{\frac12}}.
\end{split}
\end{equation}
The Whittaker is then expressed as a sum ${\rm \bf W}\left( \xi, \nu, {z p^{2}}\right)= B(\nu)+B(-\nu)$ where~\cite{Abramowitz} 
\begin{equation}
\begin{split}
B(\nu)&=\frac{-\pi}{\sin 2 \pi \nu} \frac{\ep{-zp^2/2 } z^{1/2+\nu} p^{1+2\nu} }{\Gamma(1/2-\nu -\xi) \Gamma (1/2+\nu -\xi)} \sum_{n=0}^\infty\frac{\Gamma(1/2+\nu -\xi+n)}{n! \Gamma(1+2\nu +n)} (z p^2)^n . 
\end{split}
\end{equation}
Then the amplitude $A$ is expressed as
\begin{equation}
A(z) =  \frac{ -\pi}{ \sin 2 \pi \nu }  \frac{ z^\nu A_{+\nu}(z)- z^{-\nu} A_{-\nu}(z) }{\Gamma(1/2-\nu -\xi) \Gamma (1/2+\nu -\xi)} , 
\label{eq:Anu}
\end{equation}
where 
\begin{equation}
\begin{split}
A_{+\nu}(z)& \doteq \int_{0}^{\infty}d{p} p^{\beta} \ep{- p x\ep{i\pi/4} } \ep{-p^2/2 } \sum_{n=0}^\infty\frac{\Gamma(1/2+\nu -\xi+n)}{n! \Gamma(1+2\nu +n)} (z p^2)^n\\
&=\sum_{n=0}^\infty\frac{\Gamma(1/2+\nu -\xi+n)}{n! \Gamma(1+2\nu +n)} z^n \int_{0}^{\infty}d{p} p^{\beta+2n} \ep{- p x\ep{i\pi/4} } \ep{-p^2/2 },
\end{split}
\end{equation}
and where $\beta =\alpha +1+2 \nu$. The last equality is valid only inside the radius of convergence of the power series -- i.e., $\abs z < 1/2$. We notice that $z=1$ is not in the radius, this is why we introduced the extra variable $z$. Expanding the oscillating exponential in $x$ as a series, we get
\begin{equation}
\begin{split}
\int_{0}^{\infty}d{p} p^{\beta+2n} &\ep{- p x\ep{i\pi/4} } \ep{-p^2/2 } = \sum_{k=0}^\infty \frac{(-1)^k x^k \ep{i k \pi/4}}{k! } 2^{(\beta -1+k)/2+n} \Gamma\left (\frac{\beta+1+k+2n}{2}\right ) .
\end{split}
\end{equation}
Using this expression, the sum over $n$ gives 
\begin{equation}
\begin{split}
A_{+\nu}(z)=\sum_{k=0}^\infty &\frac{(-\sqrt{2i} x)^k \Gamma(1/2+\nu -\xi) \Gamma((k+1+\beta)/2)}{k! \Gamma(1+2 \nu) } 2^{(\beta-1)/2} \\
&\times _2F_1 \left (\frac12+\nu -\xi,\frac{k+1+\beta}{2};1+2\nu ; 2z \right )
\end{split}
\end{equation}
Using Eq. (15.3.8) of Ref.~\cite{Abramowitz} and Eq.~\eqref{eq:Anu}, one obtains 
\begin{equation}
\label{eq:Asumk}
\begin{split}
{A(z) } =\sum_{k=0}^\infty &\frac{\Gamma(1+(k+\alpha)/2+\nu)\Gamma(1+(k+\alpha)/2-\nu)}{ \Gamma(-\xi+{(k+3+\alpha)}/{2})} \frac{(-\sqrt{2i} x)^k }{k! } 2^{\alpha /2} \\
& \times _2F_1 \left (\frac12+\nu -\xi,\frac12-\nu -\xi; -\xi+\frac{k+3+\alpha}{2} ;1-\frac{1}{ 2z} \right )\\
\end{split}
\end{equation}

The above expressions are all valid for $\abs z <1/2$. However, since both $A$ and the sum are analytic on $\mathbb{C}$, the result is still valid at $z=1$.

It is then convenient to express the $ _2F_1$ as 
\begin{equation}
_2F_1 (a,b;c;u) = \sum_{n=0}^\infty \frac{\Gamma(a+n) \Gamma(b+n) \Gamma(c)}{\Gamma(a) \Gamma(b) \Gamma(c+n) n!} u^n ,
\end{equation}
 to split the sum between odd and even $k$, and to notice that each sum over $k$ gives an hypergeometric function
 
\begin{equation}
\label{eq:ampfinal}
\begin{split}
A(1) = \sum_{n=0}^\infty & 2^{-n} \frac{\Gamma(1/2+ \nu -\xi +n)\Gamma(1/2- \nu -\xi +n)}{n! \Gamma(1/2+ \nu -\xi )\Gamma(1/2-\nu -\xi) } 2^{\alpha /2} \\
&\times \left [ B\left (\alpha,\frac12\right ) -\ep{i\pi/4} x \sqrt{2} B\left (\alpha+\frac12,\frac32\right ) \right ] ,
\end{split}
\end{equation}
where
\begin{equation}
\begin{split}
 &B(\alpha,\epsilon) = \Gamma\left (1+\frac{\alpha}{2}+\nu\right )\Gamma\left (1+\frac{\alpha}{2}-\nu\right ) \frac{ _2F_2 \left (1+\frac{\alpha}{2}+\nu,1+\frac{\alpha}{2}-\nu;\epsilon;n-\xi+\frac{3+\alpha}{2};i\frac{x^2}{2}\right )}{\Gamma\left (n-\xi+{(3+\alpha)}/{2}\right )}.
 \end{split}
\end{equation}
We verified that the above expression is a solution of the $4^{th}$ order differential Eq.~\eqref{eq:ommodeeq}. From the symmetries of Eq.~\eqref{eq:ommodeeq}, four independent solutions are 
\begin{equation}
(\phi^{{\rm BD}, U}_{- \omega}(X))^*, \phi^{{\rm BD}, U}_{+ \omega}(X), (\phi^{{\rm BD}, U}_{- \omega}(-X))^*, \phi^{{\rm BD}, U}_{+ \omega}(-X) .
\end{equation}
The last two ones give the $V$ modes evaluated at $X$. These four functions are independent because they are orthogonal to each other when using the scalar product of Eq.~\eqref{eq:scalpr}. 

In addition, to validate this long calculation, we compared the final expression of Eq.~\eqref{eq:ampfinal} with the original integral of Eq.~\eqref{eq:AMP1} that we evaluated numerically with Mathematica\textsuperscript{\textregistered}. We found a perfect agreement. 

\subsubsection{Study of \texorpdfstring{$T_{\rm gl}(\omega,X)$}{Tgl(X)}}
\label{sec:numTgl}

\begin{figure}[htb] 
\begin{minipage}[t]{0.45 \linewidth}
\includegraphics[width=1\linewidth]{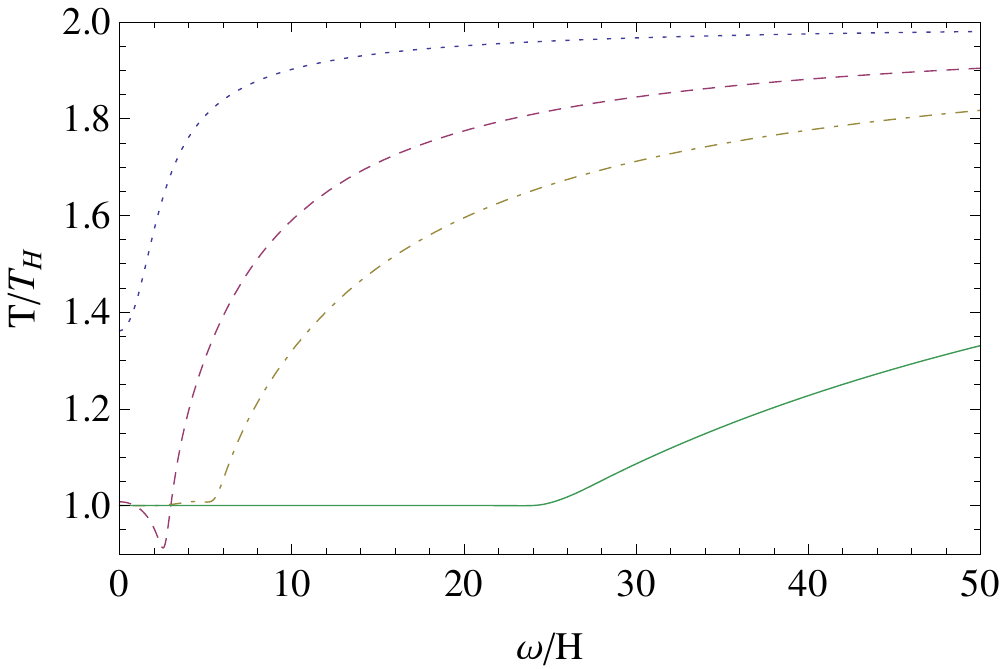} 
\end{minipage}
\hspace{0.05\linewidth}
\begin{minipage}[t]{0.45 \linewidth}
\includegraphics[width=1\linewidth]{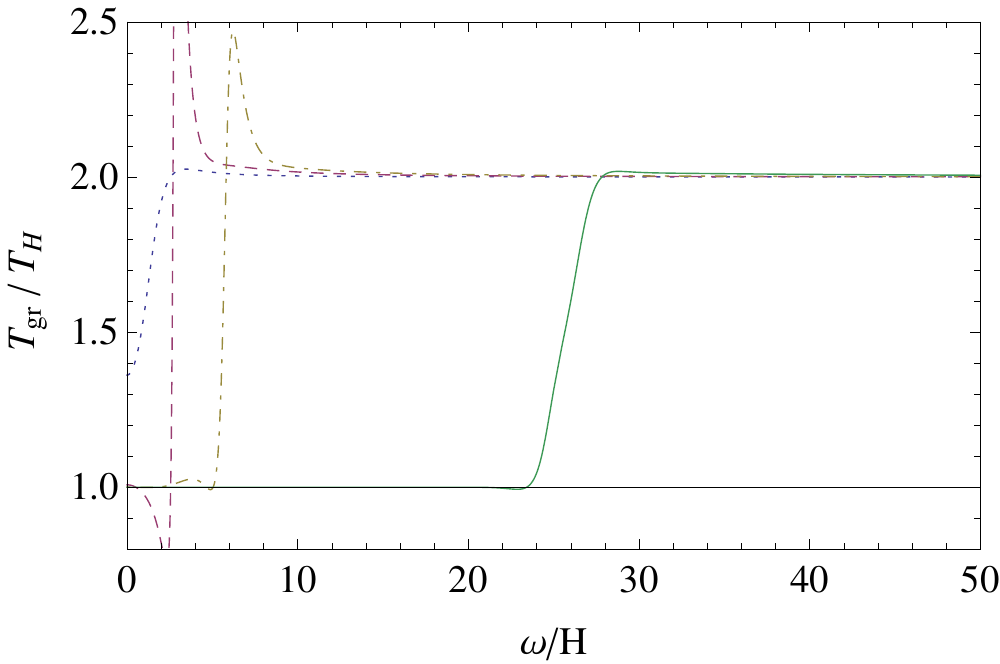}
\end{minipage}
\caption{The ratio of $T_{\rm gl}(\omega, X)$ of Eq.~\eqref{eq:tempdef} over the standard relativistic temperature ${T_H}$ as a function of $\frac{\omega}{H}$, for $X=0$ and $m=0$, and for four values of $\lambda$, namely $1$ (dotted blue), $5$ (dashed purple), $10$ (dot-dashed yellow), and $50$ (green solid line). One clearly sees that for large values $\lambda$, the spectrum is accurately Planckian and at the standard temperature, until $\omega$ reaches a certain critical value $\omega_{\rm crit}$, which is equal to $H \lambda/2$. For $\omega > \omega_{\rm crit}$, $T(\omega,X=0)$ increases sharply and reaches $2 T_H$. This figure is essentially unchanged when we use a massive field with $\mu<\lambda/2$. In the right panel, we represent the slope of $\ln(R_+/R_-)$ defining a \enquote{group temperature} $T_{\rm gr}$ as $1/T_{\rm gr} = d(\omega/T_{\rm gl})/d\omega $ (which is similar to the link between phase and group velocity), or equivalently $1/ T_{\rm gr} = d \ln(R_+/R_-) / d\omega$.  We observe that $T_{\rm gr}$ is subject to a sudden change when $\omega$ crosses $\omega_{\rm crit}$. } 
\label{fig:fig1}
\end{figure}

In Fig.~\ref{fig:fig1}, we plot ${T_{\rm gl}(\omega)}/{T_H}$ as a function of $\omega/H$, for various values of $\lambda$, and evaluated at $X = 0$, i.e., for an inertial detector. First, when $\omega/\Lambda$ and $1/\lambda$ are both much smaller than $1$, we see that this ratio is very close to 1, as expected from former analysis~\cite{Brout:1995wp,Corley:1996ar,Balbinot:2006ua,Unruh:2004zk,Coutant:2011in}. In this robust regime, the detector will perceive a Planck law at the standard temperature, up to negligible corrections. Second, in the high frequency limit, for $\omega/\Lambda \gg 1$, in agreement with the analysis of \ref{sec:thermbroke}, the ratio goes to $2$ irrespectively of the value of $\lambda$. This last point is not clear from the figure but can be verified analytically from the expressions of Eq.~\eqref{eq:Asumk} and the fact that $\abs{A_\omega / A_{-\omega} } \to 1$ when $\lambda \to 0^+$. Third, we see that there is a sharp transition from the robust relativistic regime to a new regime. An examination of Eq.~\eqref{eq:Asumk} confirms that the transition occurs at a critical frequency $\omega_{\rm crit} = \Lambda/2$. 

\begin{SCfigure}[2]
\includegraphics[width=0.5\linewidth]{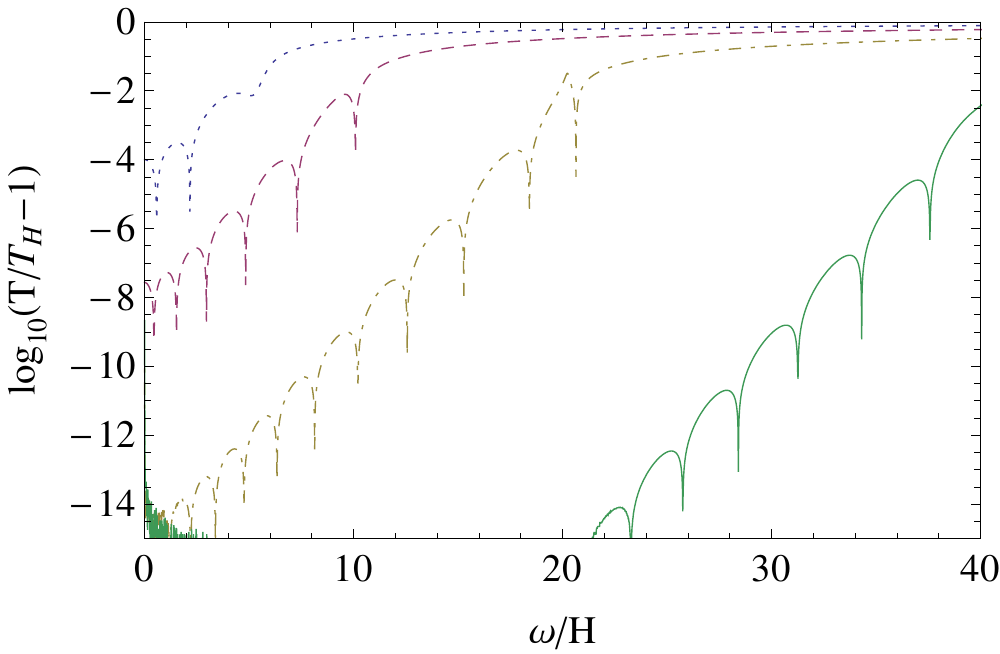}
\caption{The $\log_{10}$ of the temperature difference $ \vert T(\omega) /T_H - 1 \vert $ as a function of $\omega$, for $X=0$ and $m=0$, and for four values of $\lambda$, namely $10$ (dotted blue), $20$ (dashed purple), $40$ (dot-dashed yellow), and $80$ (green solid line). We see that $ \vert T(\omega) /T_H - 1 \vert $ increases exponentially in $\omega$ until $\omega$ reaches $\omega_{\rm crit}$. It can be shown analytically that $ \vert T(\omega) /T_H - 1 \vert $ follows Eq.~\eqref{eq:pexp}.}
\label{fig:figlog} 
\end{SCfigure}

In Fig.~\ref{fig:figlog}, we plot $\log_{10} \vert T_{\rm gl}(\omega) /T_H - 1 \vert $ to study the small deviations from the relativistic regime for $\omega < \omega_{\rm crit} $. We first notice that the sharp peaks are due to the fact that $T_{\rm gl}(\omega) /T_H - 1$ crosses $0$ while decreasing for $\omega \to 0$. A careful examination of the envelope reveals that
\begin{equation}
\vert T_{\rm gl}(\omega) /T_H - 1 \vert \sim \ep{-\pi \lambda/4+\pi \omega/2 H}. 
\label{eq:pexp}
\end{equation}
Hence, at fixed $\omega$, the deviations decrease exponentially with $\lambda$, whereas, at fixed $\lambda$, they grow exponentially till $\omega$ reaches $\omega_{\rm crit} $. 

\begin{figure}[htb] 
\centering
\includegraphics[width=0.47\linewidth]{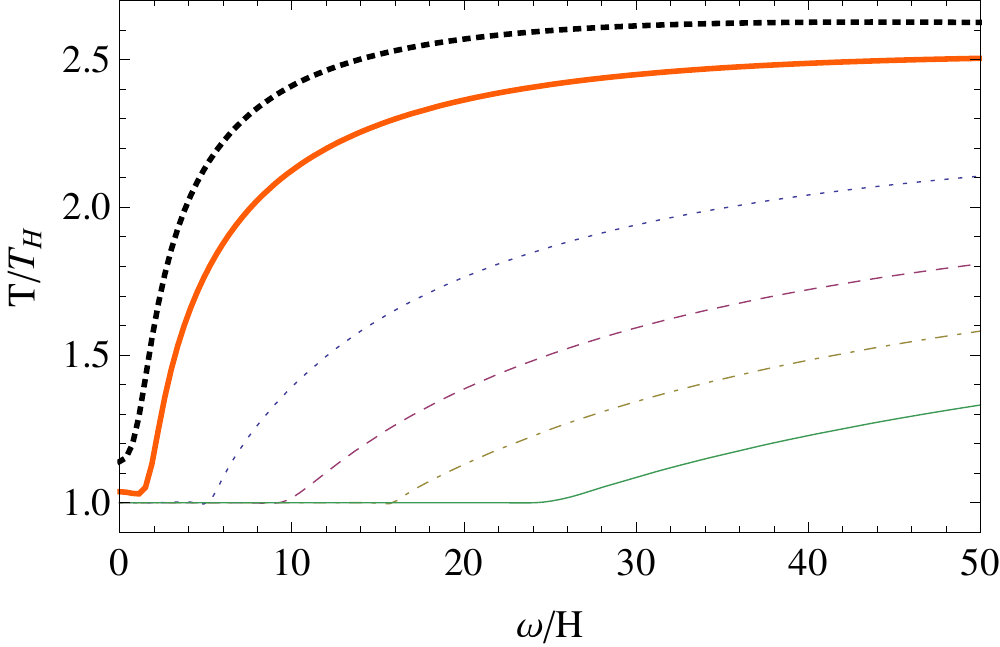}
\includegraphics[width=0.47\linewidth]{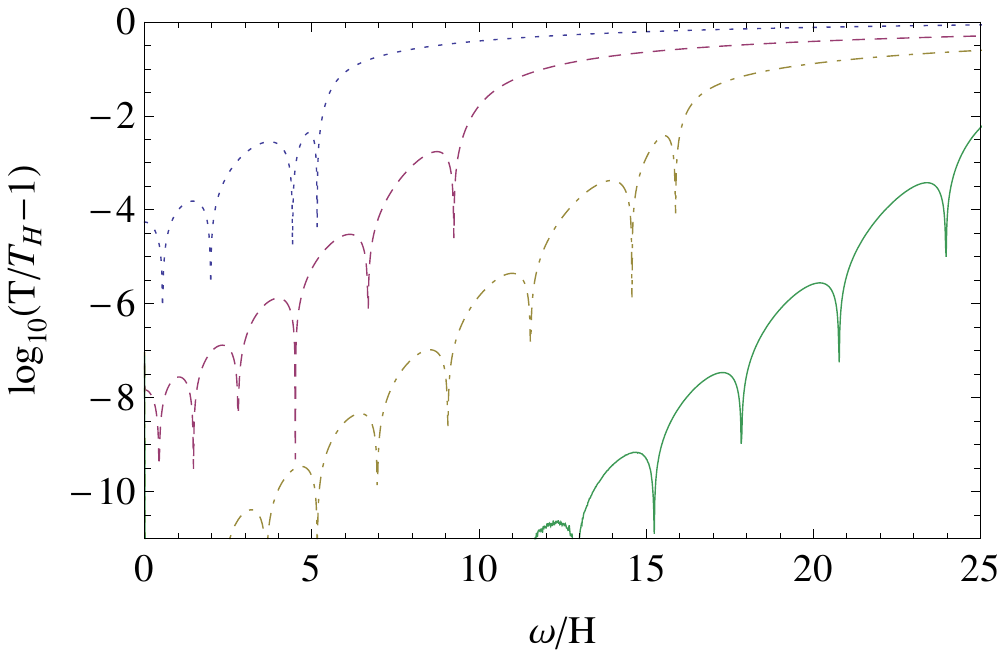} 
\caption{The same ratio as in Fig.~\ref{fig:fig1} (on the left) and Fig.~\ref{fig:figlog} (on the right), for $m=0$ and $\lambda=50$, and for six different positions, namely $HX = 0$(green solid line), $0.3$ (yellow dot dashed), $0.5$ (purple dashed), $0.7$ (blue dotted), $0.9$ (orange thick), and $0.95$ (black thick dots). On the right, the last two curves have not been plotted since they are too far from the other ones. The corresponding values of the acceleration of the detector are $a/H = 0,0.31 ,0.57,1,2 $, and $ 3$. One sees that ${T(\omega,X)}/{T_H}$ becomes larger than $2$ when $X \neq 0$. One also sees that the deviations at fixed $\omega$ increase with $a_X$. }
\label{fig:fig2}
\end{figure}

\begin{SCfigure}[2]
\includegraphics[width=0.47\linewidth]{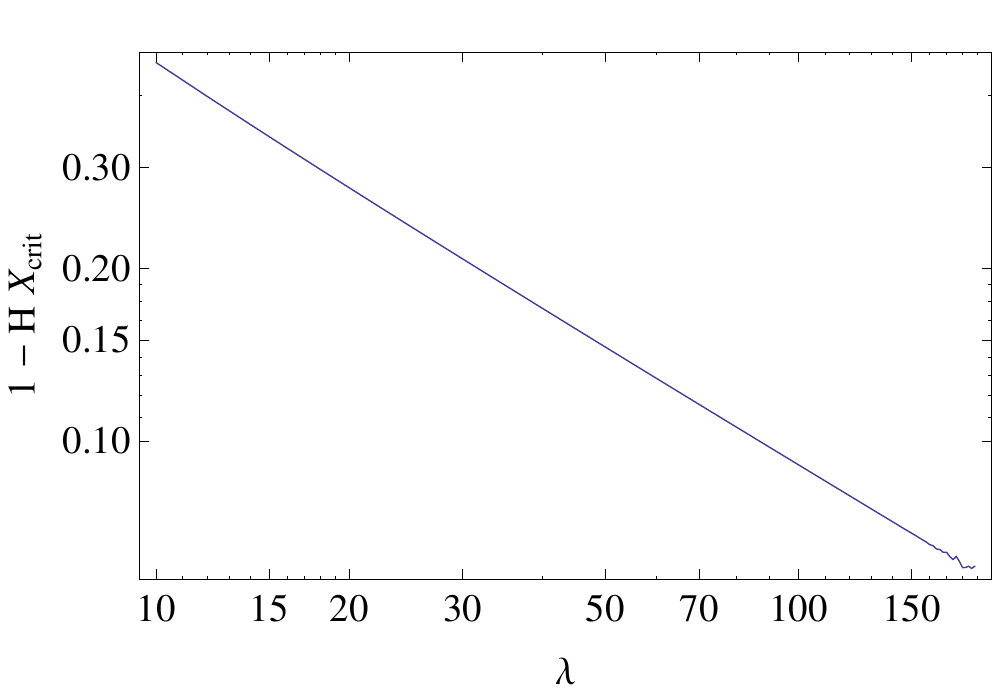}
\caption{Critical value of $HX-1$ defined by $T(X_{\rm crit},\omega=0,\lambda) - T_H = 0.01 T_H$ as a function of $\lambda$. We show that the extension of this region depends on $\Lambda$ with a power equal to $0.675\pm 0.01$. }
\label{fig:xcritoflambda}
\end{SCfigure}

In Fig.~\ref{fig:fig2} we study the $X$ dependence of $T_{\rm gl}(\omega,X) /T_H$. This describes violations of the Tolman global equilibrium law. We see that the transition from the robust regime to the new regime occurs at different critical frequencies when considering detectors following different orbits labeled by $X$. Interestingly, this dependence can be expressed as
\begin{equation}
\omega_{\rm crit}= \frac{\lambda}{2} \left (1 - \frac{a_X}{H} -\frac{a_X^2}{2 H^2} + {\cal O}\left (\frac{a_X}{H}\right )^3 \right ), 
\end{equation} 
where $a_X =H^2 \vert X \vert/\sqrt{1-H^2 X^2} $ is the detector proper acceleration at fixed $X$. In addition, on the left panel and for $\vert HX \vert\geq 0.9$, we notice that the low frequency temperature significantly differs from the standard one. This effect is related to the broadening of the horizon that was observed in~\cite{Finazzi:2010yq,Coutant:2011in}. In those papers, when considering perturbed metric profiles $v =v_{\rm backgrd} + \delta v$, it was found that the asymptotic black hole temperature differs from the standard one when the spatial extension across the horizon of the perturbation $\delta v$ is smaller than $\kappa x \sim (\kappa/\Lambda)^{2/3}$. Here we find that the temperature seen by a particle detector differs from the standard one precisely when it enters this region. In a log-log plot, see Fig.~\ref{fig:xcritoflambda}, we have numerically found that the extension of this region (defined by the locus where the relative temperature difference is 1\%) depends on $\Lambda$ with a power equal to $0.675\pm 0.01$ in agreement with the $2/3$ of the above references. Two lessons are here obtained. First, the near horizon properties can be probed either by perturbing the background metric $v$, or by introducing a local particle detector, with coherent outcomes. Second, since these responses are locally determined, they are common to de Sitter and black holes, see \ref{chap:BHdesitter}.

\begin{figure}[!ht] 
\centering
\includegraphics[width=0.47\linewidth]{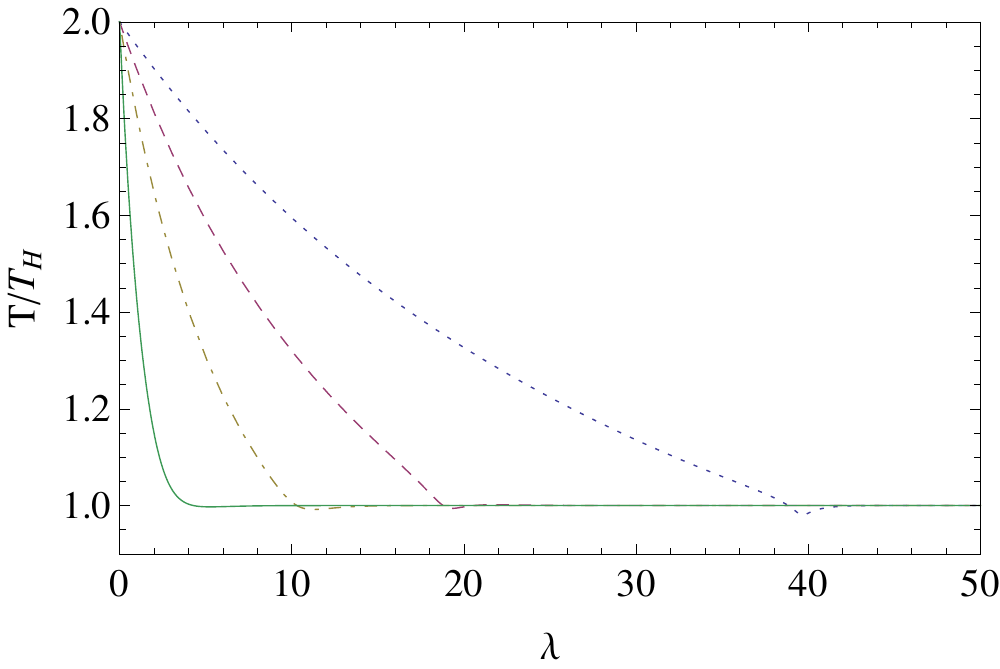}
\includegraphics[width=0.47\linewidth]{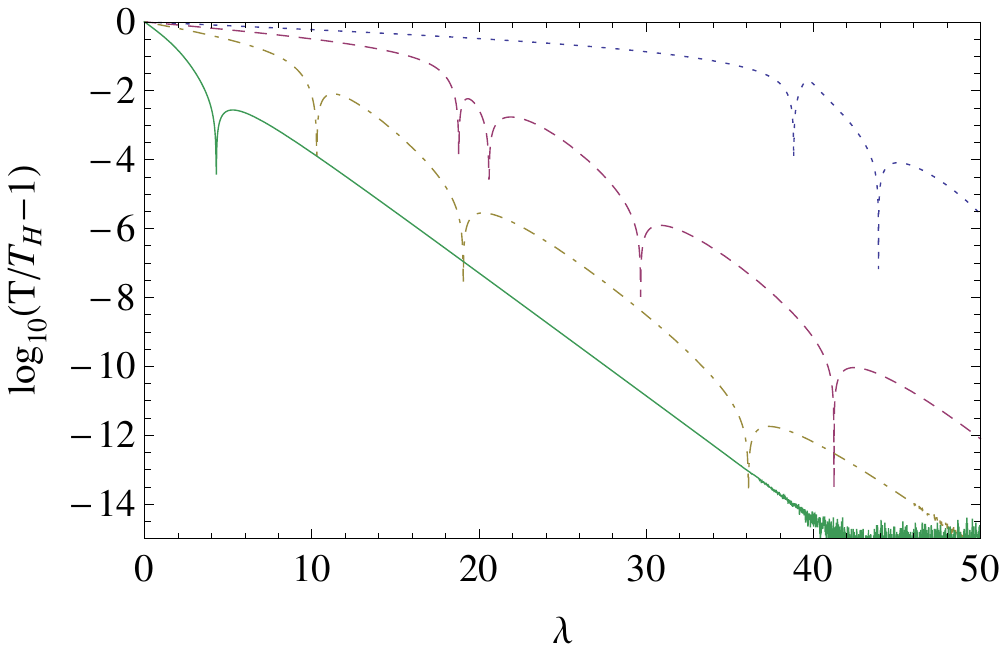}
\caption{\label{fig:fig5} The ratio of the temperature $\frac{T(\omega)}{T_H}$ (on the left) and $\log_{10}\vert T(\omega) /T_H - 1 \vert $ (on the right) as a function of $\lambda$, for $X=0$ and $m=0$, and for four values of $\omega/H$, namely $1$ (green solid line), $5$ (yellow dot dashed), $10$(purple dashed), and $20$(blue dotted). The decay of Eq.~\eqref{eq:pexp} can be observed.}
\end{figure}

Finally it is also interesting to study the behavior of $T_{\rm gl}(\omega)/T_H$ when varying $\lambda$ at fixed $\omega$ and for $X=0$ (see Fig.~\ref{fig:fig5}). When $\lambda$ is large enough, i.e., larger than the critical value $\lambda_{\rm crit} = 2 \omega /H $, the deviations from the standard temperature are extremely small, in agreement to what we saw in Fig.~\ref{fig:figlog}. Instead, for $\lambda \to 0$, $T(\omega, X = 0)/T_H$ always flows to 2, with a slope that depends on the value of $\omega/H$. An examination of these slopes shows that the slope decrease when $\omega$ increases: $dT/d\lambda\vert_{\lambda=0}$ goes from $1.02\pm 0.005$ to $0$. This is the behavior at small $\lambda$. The behavior at large $\lambda$ was given by Eq.~\eqref{eq:pexp}.

\subsection[\texorpdfstring{Asymptotic ${S}$-matrix in the ${\omega}$-representation}{Asymptotic S-matrix in the omega-representation}]{\texorpdfstring{Asymptotic $\boldsymbol{S}$-matrix in the $\boldsymbol{\omega}$-representation}{Asymptotic S-matrix in the omega-representation}}
\label{sec:Smatrix}

In this section, we compute the Bogoliubov transformation between the initial BD modes and the asymptotic $out$ modes in the $\omega$ representation. In this representation, the modes are identified through their spatial asymptotic behavior, and not their temporal one we used in \ref{sec:betacosmo}. Hence the Bogoliubov transformation can be viewed as an $S$-matrix. This is the description which is appropriate to study the mode mixing on an analogue black hole horizon. For more details about mode identification in the $\omega$-representation, we refer to Ref.~\cite{Coutant:2011in}. 

In the present case, at fixed $\omega$ each basis contains $4$ modes. Hence the Bogoliubov coefficients form a $4\times4$ matrix. This matrix is an element of $U(2,2)$ since the two modes $ \phi_{\omega}^U, \phi_{\omega}^V $ have a positive norm, while $ (\phi_{-\omega}^U)^*, (\phi_{-\omega}^V)^*$ have a negative one. In what follows we first study the massless case, and then the massive case $m > H/2$. In both cases we shall see that the $S$-matrix possesses unusual factorization properties that are due to the two symmetries governed by $K_z$ and $K_t$. We shall also see that the elements of this matrix combine the cosmological aspects of \ref{sec:betacosmo} and the stationary thermal-like aspects of \ref{sec:thermidev}

To compute the coefficients of the $S$-matrix, we first need to identify the incoming and outgoing modes. At fixed $\omega$, for quartic dispersion, the general solution of Eq.~\eqref{eq:ommodeeqinP} contains 8 asymptotic branches, 4 for $X\to \infty$, and 4 for $X \to - \infty$. In addition, when forming wave packets in $\omega$, one finds that 4 propagate towards $X = 0$, whereas $4$ propagate away from it. The mode identification is based on this second aspect: The 4 incoming modes, are, by definition, the 4 solutions that only possess one incoming asymptotic branch. These incoming modes are simply given by the Fourier transform of the stationary BD modes $\tilde \phi^{\rm BD}_\omega$, thereby showing that the definitions of $in$ modes based on their temporal behavior and the spatial one are perfectly consistent.

To see this, let us consider as an example $(\phi^{{\rm BD}, U }_{- \omega})^*$. Using Eq.~\eqref{eq:ampfinal}, its asymptotic behavior can be found using~\cite{Wolfram}. Up to an irrelevant overall constant, one finds
\begin{equation}
\begin{split}
(\phi^{{\rm BD}, U}_{- \omega} )^* &\underset{X\to\pm\infty}{\sim} (1\mp 1)\times \ep{\frac{i x^2 }{2}} (\frac{i x^2 }{2})^{-i\frac{\lambda}{4}-\frac34- i \frac{\omega}{2H}} \\
&+ \left\{ Z_{\omega,\lambda,\mu,\pm} \times (\frac{-i x^2 }{2})^{-\frac{1}{4} + i\frac{\omega}{2H}-i\frac{\mu}{2}}+ (\mu \to -\mu) \right\}, 
\end{split}
\label{eq:asmodeinom}
\end{equation}
where $x = HX \sqrt{\lambda}$ and where the coefficient $Z$ is
\begin{equation}
 \begin{split}
 Z_{\omega,\lambda,\mu,\pm} &= \frac{2^{ - i{\mu} -i\frac{\lambda}{4}+i\frac{\omega}{2H}} \Gamma(-i\mu) {\Gamma(\frac{1}{2} - i\frac{\omega}{H}+i{\mu}) } }{ \sqrt{\pi} \Gamma(\frac12 - i\frac{\mu}{2} +i\frac{\lambda}{4}) } \ep{\pm (-i \pi /4+ \pi {\mu}/{2} - \pi{\omega}/{2H})}. 
 \end{split}
\end{equation}
The first term in Eq.~\eqref{eq:asmodeinom} describes the incoming high momentum branch, as can be verified by computing its group velocity $dX/dt = 1/\partial_\omega P_\omega$, where $P_\omega=\partial_X S_\omega$ is the corresponding root of Eq.~\eqref{eq:dispHJeq}. The last two terms describe the $4$ low momentum outgoing branches. One verifies that they propagate away from the static patch, two for $X\to \infty$ and two for $X\to -\infty$. In Fig.~\ref{fig:caracteristics} we schematically represent the space-time pattern associated with a wave packet made with $\phi^{{\rm BD}, U}_\omega$.
 
We now have to identify the $out$ mode basis, i.e., the four unit norm asymptotic outgoing modes. As in \ref{sec:betacosmo}, we treat separately the massless and the massive case. 

\subsubsection{The massless case}

\begin{SCfigure}[2]
\includegraphics[width=0.47\linewidth]{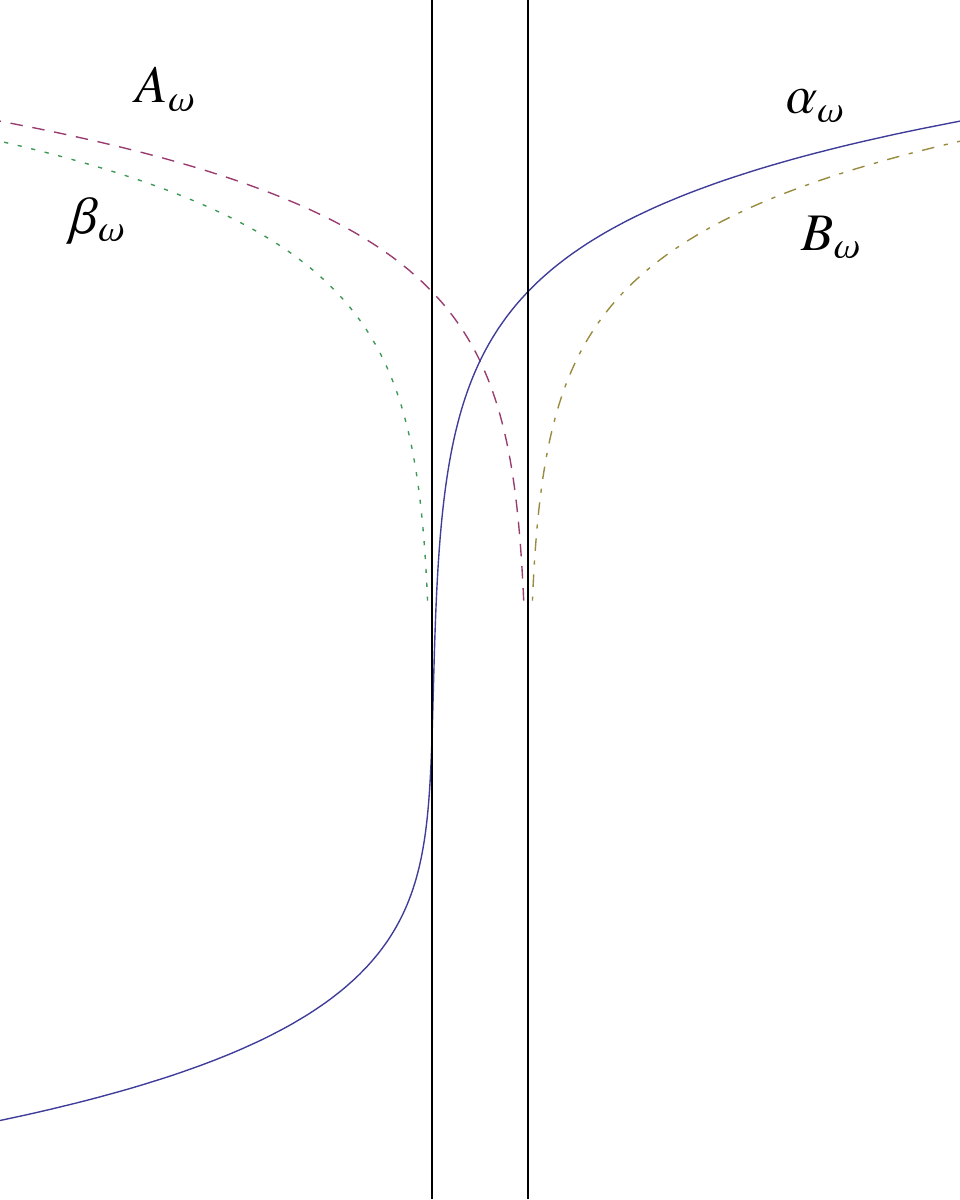}
\caption{
In this figure, unlike the representation of Fig.~\ref{fig:poincaredesitter}, $t=Cst.$ are horizontal lines, and $X=Cst.$ are vertical ones. The two vertical lines represent the Killing horizons at $HX = \pm 1$. An incoming (BD) high momentum positive norm $U$-mode (in thick line) splits into four $out$ modes with low momenta: an outgoing positive norm $U$-mode (blue line), a negative norm $U$-mode (green dots), a positive norm $V$-mode (purple dashes), and a negative norm $V$-mode (yellow dash dots). The respective amplitudes of these four outgoing modes are given in Eq.~\eqref{eq:modedec}. To draw these characteristics, we work with $m=0$, $\omega/H= 5$ and $\Lambda/H= 1000$. The lapse of time spent very close to the horizon is $H \Delta t \sim \ln (\Lambda/\omega)$. It diverges for $\Lambda \to \infty$, in which case one recovers the relativistic behavior, and ultrahigh initial momenta.}
\label{fig:caracteristics}
\end{SCfigure} 

Since asymptotic outgoing modes have low momentum $P$, they obey the two-dimensional relativistic d'Alembert equation. At fixed $\omega$, the equation for the right moving $U$-modes is 
\begin{equation}
 ( i \omega - HX\partial_X ) \phi_\omega^{U} = \partial_X \phi_\omega^{U}.
\end{equation}
For $\omega > 0$, the $out$ $U$-modes of positive and negative unit norm are 
\begin{equation}
\label{eq:phiout}
\begin{split}
 \phi_{\omega,R}^{U, out} &= \theta(1 + H X) \frac{(1 + H X)^{i\omega/H}}{\sqrt{2 \omega/H}}, \\
 (\phi_{-\omega,L}^{U, out})^* &= \theta(-1 - H X) \frac{({-1 - H X })^{i\omega/H}}{\sqrt{2 \omega/H}} .
\end{split}
\end{equation}
The $V$-modes $ \phi_{\omega,L}^{V, out}, ( \phi_{-\omega,R}^{V, out})^*$ are obtained by replacing $X$ by $-X$ in the above. 

We put the 4 modes in a vector in the following order $\Phi_\omega= (\phi_\omega^U, (\phi_{-\omega}^U)^*, \phi_\omega^V, (\phi_{-\omega}^V)^* )$, both for the BD and the $out$ modes, and we define the $S$-matrix by $\Phi^{BD}_\omega = S_\omega \Phi^{out}_\omega$. We find that $S_\omega$ factorizes\footnote{This decomposition is due both to the affine group and to the fact that the infra red is not modified. It will be made explicit in a more general context in \ref{sec:Asc}, see in particular Eq.~\eqref{eq:changeofbase}.} as
\begin{equation}
\begin{split}
S_\omega= \left[ \begin{array}{cccc}
\alpha_k &0 &0 &\beta_k\\
0& \alpha_k ^*&\beta_k^* &0\\
0 &\beta_k &\alpha_k &0\\
\beta_k^* &0 &0 &\alpha_k ^* \\
\end{array}\right] \times 
\left[ \begin{array}{cccc}
\alpha_\omega^H &\beta_\omega^H& 0&0\\
\beta_\omega^H&\alpha_\omega^H &0&0\\
0&0&\alpha_\omega^H &\beta_\omega^H\\
0&0&\beta_\omega^H &\alpha_\omega^H\\
\end{array}\right] .
\label{eq:Smassl}
\end{split}
\end{equation}
Moreover, the Bogoliubov coefficients $\alpha_k, \beta_k$ are those of Eq.~\eqref{eq:betakmassless}, and $\alpha_\omega^H, \beta_\omega^H$ are the standard relativistic coefficients, taken real, and obeying $\beta_\omega^H/\alpha_\omega^H = \ep{- \pi \omega/H} $ and $\vert \alpha_\omega^H\vert ^2 - \vert \beta_\omega^H \vert ^2= 1 $. To get these real coefficients we chose the (arbitrary) phases of the $out$ modes in an appropriate manner. There are only 4 different coefficients in $S_\omega$, and they all have a clear meaning when considering one BD mode. In Fig.~\ref{fig:caracteristics}, we represent the mode 
\begin{equation}
\begin{split}
\phi^{{\rm BD}, U}_\omega &= \alpha_\omega \phi^{out, U}_{\omega,R} + \beta_\omega (\phi^{out, U}_{-\omega,L})^* + A_\omega \phi^{out, V}_{\omega,L} + B_\omega (\phi^{out, V}_{-\omega,R})^* .
\end{split}
\label{eq:modedec} 
\end{equation}
The $\alpha_\omega, \beta_\omega$ coefficients weigh the mode mixing amongst $U$-modes of opposite norm, whereas $A_\omega$ and $B_\omega$ describe respectively the elastic and the anomalous $U-V$ mode mixing. The norm of these four coefficients obey 
\begin{equation}
\begin{array}{rlrl} 
\abs {\alpha_\omega} ^2 &= \abs{\alpha_k}^2 \times (n^H_\omega + 1), &
\abs {\beta_\omega} ^2 &= \abs{\alpha_k}^2 \times n^H_\omega , \\
\abs {A_\omega} ^2 &= \abs{\beta_k}^2 \times n^H_\omega , &
\abs {B_\omega} ^2 &= \abs{\beta_k}^2 \times (n^H_\omega +1) ,
\end{array}
\end{equation}
where $n^H_\omega = 1/(\ep{\omega/T_H} - 1)$ is the Planck spectrum at the standard temperature $T_H$. We see that the deviations from the relativistic spectrum are proportional to $\abs{\alpha_k}^2 - 1 = \abs{\beta_k}^2 \sim \lambda^{-4}$, as those breaking the relativistic $U$-$V$ decoupling. Thus both deviations from the relativistic theory are governed by the cosmological pair creation rates at fixed $\bf k$. We notice that the decay of the deviations from thermality in $1/\lambda^4$ is in agreement with the decay in $1/\omega_{\rm max}^4$ found in a black hole metric when working at fixed $D$, see Fig.~14 of Ref.~\cite{Macher:2009tw}. We also notice that irrespective of $\omega$ and $\Lambda$, the elastic $\abs {A_\omega} $ is the smallest coefficient. We finally emphasize that these extremely simple results are \emph{exact}, and follow from the hypergeometric functions $ _2F _2$ of Eq.~\eqref{eq:ampfinal}.

\subsubsection{The massive case}

As in \ref{sec:betacosmo}, the massive $out$ modes should be handled with care. An orthonormal basis for these $out$ modes is given by the following right modes $(R)$:
\begin{equation}
\begin{split}
\phi^{out}_{R,\omega}= \theta(X) \frac{(HX)^{i\frac{\omega}{H}- i\mu}}{\sqrt{2 \mu H X}},
\quad
\phi^{out *}_{R,-\omega}= \theta(X) \frac{(HX)^{i\frac{\omega}{H}+ i\mu}}{\sqrt{2 \mu (H X)}},
\end{split}
\end{equation} 
together with the $L$-modes obtained by replacing $X$ by $-X$ in the above expressions. We have used this $R$-$L$ separation in the place of the $U$-$V$ one based on the sign of the group velocity, because, for massive modes the asymptotic group velocity with respect to the flow $v = HX$ is no longer well defined. 

We now put the four $out$ modes of frequency $\omega$ in a vector in the following order $\Phi_\omega= (\phi_\omega^R , (\phi_{-\omega}^R)^* , \phi_\omega^L,(\phi_{-\omega}^L)^* )$, while the four $in$ modes are ordered in the same order as in the massless case. Defining again the $S$-matrix by $\Phi^{BD}_\omega = S_\omega \Phi^{out}_\omega$, we obtain 
\begin{equation}
\label{eq:Smassive}
\begin{split}
S_\omega= \left[ \begin{array}{cccc}
\alpha_k &0 &0 &\beta_k\\
0& \alpha_k ^*&\beta_k^* &0\\
0 &\beta_k &\alpha_k &0\\
\beta_k^* &0 &0 &\alpha_k ^* \\
\end{array}\right] \times 
\left[ \begin{array}{cccc}
T_\omega &0& R_\omega&0\\
0&T_{-\omega} &0&R_{-\omega}\\
R_{\omega}&0&T_\omega &0\\
0&R_{-\omega}&0&T_{-\omega} \\
\end{array}\right] .
\end{split}
\end{equation}
On the left matrix, the $\alpha_k, \beta_k$ coefficients are those of Eq.~\eqref{eq:betakmassive}. Hence, as far as this matrix is concerned, we obtain the same structure as in Eq.~\eqref{eq:Smassl}. Instead on the $\omega$-dependent right matrix, the coefficients are 
\begin{equation}
T_{\omega}=\frac{\ep{ - \pi(\mu - \omega/H)/2 }}{\sqrt{2\cosh \pi( \mu - \omega/H) }}, \quad
R_{\omega}=\frac{\ep{\pi(\mu - \omega/H )/2 }}{\sqrt{2\cosh \pi( \mu - \omega/H) }}. 
\end{equation}
They obey the unitarity relation $\abs{T_{\omega}}^2 + \abs{R_{\omega}}^2 = 1$. Hence unlike what was found in Eq.~\eqref{eq:Smassl} the right matrix now describes an elastic scattering between modes of the same norm. As a result, the main difference between the massless and the massive case is that the final occupation number of massive particle no longer diverge as $T_H/\omega$ for $\omega \to 0$. This disappearance of the thermal like divergence was already found in Ref.~\cite{Coutant:2012zh} in black hole metrics.

\section*{Conclusions}
\addcontentsline{toc}{section}{Conclusions}

In this chapter, we obtain two kinds of results, precise mathematical ones characterizing dispersive fields in de Sitter space, and more general ones associated with the observation that thermality is violated when Lorentz invariance is broken at high energy. This work is a completion of Ref.~\cite{Macher:2008yq}, where consequences of dispersion on inflation is studied.

Concerning the first kind, in \ref{sec:setting}, we used the group associated with the two residual symmetries of dispersive fields in de Sitter to provide precise relationships between the two representations of the field, based respectively on the homogeneity and on the stationarity of the settings. The key result is that the homogeneous modes and the stationary ones can be {\it all} expressed in terms of the single BD mode $\chi_{\rm BD}(P)$ and its complex conjugated, where $\chi_{\rm BD}$ obeys Eq.~\eqref{eq:chiWKB}, see Eqs.~\eqref{eq:kp} and~\eqref{eq:omegaP}. For free fields, all observables are thus encoded in that single mode. 

In \ref{sec:thermal} we show that the two-point function computed in the BD vacuum is still stationary and periodic in $\Im (t)$ with period $2\pi/H$, as it is for Lorentz invariant fields. In spite of this, we then show that the BD vacuum is no longer a thermal state when restricted to the static patch. In particular, we show that the temperature function of Eq.~\eqref{eq:tempdef} is, for ultrahigh frequency $\omega /\Lambda \gg 1$, $n$ times the standard one, where $n$ is the highest power of $P^2$ in the dispersion relation of Eq.~\eqref{eq:disprel}. In \ref{sec:thermidev}, by considering the response function of particle detectors with different acceleration, we also show that the Tolman law is violated. Even though the BD vacuum is no longer in thermal equilibrium, we prove that (for free fields at least) it is still the only stationary, regular, and stable state, as it is in relativistic theories~\cite{Marolf:2010nz,Marolf:2011sh,Hollands:2010pr}. In other words, for dispersive fields, there is no (regular) KMS state on de Sitter space. We believe that these properties will remain true when considering interacting fields. Finally we {explain} the origin of the violations of thermality in terms of the loss of the positivity of the stationary Hamiltonian restricted to the static patch. Whereas this operator possesses a spectrum bounded from below for Lorentz invariant theories, this is no longer true for dispersive fields. As a result the ordinary second law of thermodynamics is no longer protected, violations of this law are possible, and the system might develop dynamical instabilities. In this respect the fact that the BD vacuum is shown to be stable in de Sitter becomes a nontrivial result. 

\chapter{Dissipative fields in de Sitter space}
\label{chap:dissipdS}

\section*{Introduction}

 While high-energy dispersion is rather easily introduced (see \ref{chap:dispdS}) and has been studied in many papers both in cosmological settings~\cite{Martin:2000xs,Niemeyer:2000eh,Niemeyer:2001qe,Macher:2008yq} and black hole metrics~\cite{Unruh:1994je,Brout:1995wp}, see e.g., Ref.~\cite{Jacobson:1999zk} for a review, dissipation has received comparatively much less attention. 
 
In this chapter, we study a scalar quantum field $\hat \phi$ that has a standard relativistic behavior at low energy but displays dispersion \textit{and dissipation} at high energy, thereby violating (local) Lorentz invariance. The dissipation is introduced in covariant language so that it can be generalized to any space-time. We then study the effects of dissipation on pair particle production and on the fluxes received by an inertial observer. We also study the fluxes produced by space-time and reaching infinity because it would be experimentally accessible in analogue gravity experiments. This chapter is mainly based on Ref.~\cite{Adamek:2013vw}.

\minitoc
\vfill

\section{Covariant settings for dissipative field}
\label{sec:actionsetting}

\subsection{Action for dissipative field}

When preserving unitarity and general covariance, dissipation is technically more difficult to handle than pure dispersion. To do so in simple terms, following~\cite{Parentani:2007uq}, we introduce dissipation by coupling $\phi$ to some environmental degrees of freedom $\psi$, and the action of the entire system $S_{\rm tot} = S_\phi + S_\psi + S_{\rm int}$ is taken quadratic in $\phi,\psi$, as in models of atomic radiation damping~\cite{Aichelburg1976264} and quantum Brownian motion~\cite{Unruh:1989dd}. Again, for reasons of simplicity, we shall work in $1+1$ dimensions. The reader interested in four-dimensional models may consult~\cite{Adamek:2008mp}, where there is a phenomenological study of inflationary spectra in dissipative models. 

In the present work, we consider dispersion relations that contain both dispersive and dissipative effects. These relations can be parametrized by two real functions $\Gamma,f$ as
\begin{equation}
\label{eq:dispersion}
 \Omega^2 + 2 i \Gamma \Omega= m^2 + P^2+ f = F^2, 
\end{equation}
where $\Gamma(P^2)> 0$ is the damping rate, and $f(P^2) $ describes dispersive effects. To recover a relativistic behavior in the infrared, a typical behavior would be $\Gamma \sim P^2$ and $f \sim P^4$ for $P^2 \to 0$. 
In Eq.~\eqref{eq:dispersion}, $\Omega$ and $P^2$ are, the preferred frequency and momentum defined in \ref{sec:affinefieldth}.

We now consider a unitary model which implements Eq.~\eqref{eq:dispersion}. This model is not unique but can be considered as the simplest one, as shall be made clear below. In covariant terms, the total action $S_{\rm tot} = S_\phi + S_\psi + S_{\rm int}$ is [compare with Eq.~\eqref{eq:dissipHforQM}]
\begin{equation}
\begin{split}
\label{eq:covaction}
S_{\rm tot} = &\frac{1}{2} \int d^2 \mathsf x \sqrt{ - \mathsf g(\mathsf x)} \left[-\mathsf{g}^{\mu\nu} \nabla_\mu \phi \nabla_\nu \phi - m^2 \phi^2 - \phi f \left(-{ \nabla}_s^2 \right) \phi \right]\\
&+ \frac{1}{2} \int d^2 \mathsf x \sqrt{ - \mathsf g(\mathsf x)} \int d\zeta \left[\left( \nabla_u \psi_\zeta\right)^2 - \left(\pi \zeta \right)^2 \psi_\zeta^2\right] \\
& + \int d^2 \mathsf x \sqrt{ - \mathsf g(\mathsf x)} \left[\bigg( \gamma \left( \nabla_s \right) \phi\bigg) \left( \nabla_u \int d\zeta \psi_\zeta \right)\right].
\end{split}
\end{equation}
In the first line, $S_\phi$ is the action for a massive dispersive field. It coincides with the one we took in \ref{chap:dispdS}. In two dimensions, the self-adjoint operator which implements $P^2$ is $-\nabla_s^2 \doteq \nabla_s^\dagger \nabla_s$, where $ \nabla_s = s^\mu \nabla_\mu $ is an anti-self-adjoint operator (when $u$ is a freely falling frame), $ \nabla_s^\dagger = - \nabla_\mu s^\mu$ its adjoint, and $ \nabla_\mu $ the covariant derivative. A four-dimensional version of this model can be found in~\cite{Parentani:2007uq}.

The second line, the action for the $\psi$ field, contains the extra parameter $\zeta$, which can be considered as a momentum in some extra dimension. Its role is to guarantee that the environment degrees of freedom are dense, something necessary to engender dissipative effects when coupling $\psi$ to $\phi$~\cite{Parentani:2007uq,Unruh:1989dd}. The kinetic term of $\psi$ is governed by the anti-self adjoint operator $ \nabla_u \doteq -(u^\mu \nabla_\mu + \nabla_\mu u^\mu) / 2$ which implements $ \Omega = u^\mu p_\mu$. We emphasize that there is no spatial derivative acting on $\psi$. This means that the quanta of $\psi$ are at rest in the preferred frame. This restriction can easily be removed by adding the term $c^2_\psi ( \nabla_s \psi)^2$ which associates to $c_\psi$ the group velocity of the low $\zeta$ quanta. Including this term leads to much more complicated equations because dissipative effects are then described by a nonlocal kernel, as shall be briefly discussed after Eq.~\eqref{eq:Gammaasgamma}. For reasons of simplicity, we shall work with $c_\psi = 0$ which gives a local kernel. Moreover, in homogeneous universes $c_\psi = 0$ also implies that the $\psi$-modes are not parametrically amplified by the cosmological expansion. When working with given functions $\Gamma(P^2)$ and $f(P^2)$, we do not expect that the complications associated with $c_\psi \neq 0$ will qualitatively modify the effective behavior of $\phi$, at least when dissipative scale is well separated from the Hubble scale. 

The interaction between the two fields is given by the action of the third line. The strength and the momentum dependence of the coupling is governed by the function $\gamma(P)$ which has the dimension of a momentum. Its role is to engender the decay rate $\Gamma$ entering Eq.~\eqref{eq:dispersion}. The last two lines possess peculiar properties which have been adopted to obtain simple equations of motion. These are
\begin{subequations}
\begin{align}
\label{eq:eomcovphi}
\left [\nabla_\mu u^\mu u^\nu \nabla_\nu +F^2(- \nabla_s^2) \right ] \phi & = \gamma( \nabla_s^\dagger) \nabla_u \int d\zeta \psi_\zeta, \\
\label{eq:eomcovpsi}
\left [ \nabla_u^2+(\pi \zeta)^2 \right ] \psi_\zeta &= - \nabla_u \gamma( \nabla_s ) \phi.
\end{align}
\end{subequations}

\subsection{Effective equations of motion}

In this section, we use an exact resolution of Eq.~\eqref{eq:eomcovpsi} to get the efffective equations of motion for $\hat\phi$ depending only on the diverse initial conditions on the Cauchy surface ($t \to - \infty$ in de Sitter).

The solution to Eq.~\eqref{eq:eomcovpsi} is 
\begin{equation}
\psi_\zeta(\mathsf x') = \psi_\zeta^0(\mathsf x') - \int d^2 \mathsf x \sqrt{ - \mathsf g(\mathsf x)} G_\zeta(\mathsf x',\mathsf x) \nabla_{u} \gamma(\nabla_s ) \phi(\mathsf x),
\label{eq:psizeta}
\end{equation}
where $\psi_\zeta^0$ is a homogeneous solution, and where the driven solution is governed by $G_\zeta(\mathsf x,\mathsf x')$, the retarded Green function of $\psi_\zeta$. When injecting $\psi_\zeta$ in the rhs of the first equation, one obtains the equation of $\phi$ driven by $\psi_\zeta^0$. The general solution can be written as $\phi = \phi^{\rm dec} + \phi^{\rm dr}$, where the decaying part is a homogeneous solution, and where the driven part is given by
\begin{equation}
\begin{split}
\phi^{\rm dr}(\mathsf x') = \int d^2 \mathsf x \sqrt{ - \mathsf g(\mathsf x)} G_{\rm ret}(\mathsf x',\mathsf x) \gamma(\nabla^\dagger_s) \nabla_u \int d\zeta \psi_\zeta^0(\mathsf x). 
\end{split}
\label{eq:phidr}
\end{equation}
In a general Gaussian $\phi-\psi$ model, the retarded Green function $G_{\rm ret}$ would obey a nonlocal equation, i.e., an integro-differential equation. We have adjusted the properties of $S_\psi$ and $S_{\rm int}$ precisely to avoid this. Two properties are essential. Firstly, at fixed $\zeta$ and along the orbits of $u$, Eq.~\eqref{eq:eomcovpsi} reduces to that of a driven harmonic oscillator. This can be seen by using the coordinates $(t,z)$ defined by $u^\mu \partial_\mu = -\partial_{t \vert z}$ where $z$ is a spatial coordinate which labels the orbits of $u$. Then, $ \nabla_u$ applied to scalars is
\begin{equation}
\label{eq:Duincov}
 \nabla_u =a^{-1/2} \partial_{t\vert z} a^{1/2} ,
\end{equation} 
where $a(t,z) \doteq \ep{ \int^t dt' \Theta(t',z)}$, and where $\Theta \doteq - \nabla_\mu u^\mu$ is the expansion of $u$. Hence the rescaled field 
\begin{equation}
\label{eq:psirenorm}
\Psi_\zeta(t,z) \doteq \sqrt{a(t,z)} \psi_\zeta(t,z)
\end{equation} 
obeys the equation of an oscillator of constant frequency $\omega_\zeta = \pi \abs{\zeta}$, see Eq.~\eqref{eq:eomoscharmQ}. Secondly, when summed over $\zeta$, the retarded Green function of $\psi$ obeys~\cite{Parentani:2007uq}
\begin{equation}
 \nabla_u \int_{-\infty}^{\infty} d\zeta G_\zeta(\mathsf x,\mathsf x') = \delta^2(\mathsf x-\mathsf x') ,
\label{eq:loc}
\end{equation}
where $\delta^2(\mathsf x-\mathsf x')$ is the covariant Dirac delta, i.e., $\int d^2 \mathsf x \sqrt{ - \mathsf g(\mathsf x)} h(\mathsf x) \delta^2(\mathsf x-\mathsf x') = h(\mathsf x') $. Eq.~\eqref{eq:loc} guarantees that the differential operator encoding dissipation is local. Namely, when inserting $\psi_\zeta$ of Eq.~\eqref{eq:psizeta} in Eq.~\eqref{eq:eomcovphi}, one finds
\begin{equation}
\begin{split}
\Box_{\rm diss}& \phi = \gamma( \nabla_s^\dagger) \nabla_u \int d\zeta \psi_\zeta^0 , \\
\end{split}
\end{equation}
with the local differential operator 
\begin{equation}
\label{eq:boxdiss}
\begin{split}
\Box_{\rm diss}& \doteq \left [\nabla_\mu u^\mu u^\nu \nabla_\nu + F^2(- \nabla_s^2) + \gamma( \nabla_s^\dagger) \nabla_{u} \gamma( \nabla_s) \right ] .
\end{split}
\end{equation}

One can now verify that the WKB solutions of $\Box_{\rm diss} \phi = 0$ are governed by a Hamilton-Jacobi action which obeys the dispersion relation of Eq.~\eqref{eq:dispersion} with 
\begin{equation}
\label{eq:Gammaasgamma}
\Gamma = \abs{\gamma}^2 / 2 .
\end{equation}
The reader can also verify that any modification of the actions $S_\psi$ and $S_{\rm int}$ leads to the replacement of $\Box_{\rm diss}$ by a nonlocal operator. When considered in simultaneously homogeneous and static situations, this is not problematic because one can work with Fourier modes in both space and time. However when considered in nonhomogeneous and/or nonstatic backgrounds, it becomes hopeless to solve such an equation by analytical methods. 

In our model, the retarded Green function thus obeys 
\begin{equation}
\begin{split}
\Box_{\rm diss} G_{\rm ret}(\mathsf x,\mathsf x') = \delta^2 (\mathsf x-\mathsf x') ,
\label{eq:locEq}
\end{split}
\end{equation}
and vanishes when $\mathsf x$ is in the past of $\mathsf x'$, where the past is defined with respect to the foliation introduced by the $u$ field. When canonically quantizing $\phi$ and $\psi$, since our action is quadratic in the fields, the commutator $G_\mathrm{c}(\mathsf x,\mathsf x') \doteq [\hat\phi(\mathsf x),\hat\phi(\mathsf x')]$ is independent of $\hat \rho_{\rm tot}$, the state of the entire system. Moreover, it is related to $G_{\rm ret}$ in the usual way
\begin{equation}
\label{eq:GcfromGret}
\begin{split}
-i G_\mathrm{c}(\mathsf x,\mathsf x') = G_{\rm ret}(\mathsf x,\mathsf x') - G_{\rm ret}(\mathsf x',\mathsf x). 
\end{split}
\end{equation}

In this chapter we only consider Gaussian states. This implies~\cite{mandel1995optical,leonhardt1997measuring} that the density matrix $\hat \rho_{\rm tot}$, and all observables, are completely determined by the anti-commutator of $\hat \phi$,
\begin{equation}
\label{eq:Gacdef}
 G_{\rm ac}(\mathsf x,\mathsf x') \doteq \left < \{\hat \phi(\mathsf x),\hat \phi(\mathsf x')\}\right >,
\end{equation}
that of $\hat \psi$, and the mixed one containing $\hat \phi$ and $\hat \psi$. Decomposing the field operator $\hat \phi = \hat \phi^{\rm dec} + \hat \phi^{\rm dr}$, $G_{\rm ac}$ splits into three terms. The first one involves only $\hat \phi^{\rm dec}$, the second contains both $\hat \phi^{\rm dec}$ and $\hat \phi^{\rm dr}$, and the last only $\hat \phi^{\rm dr}$. When assuming that the initial conditions are imposed in the remote past, because of dissipation, only the last one is relevant. Using Eq.~\eqref{eq:phidr}, it is given by
\begin{equation}
\label{eq:acgrennfunction}
G_{\rm ac}^{\rm dr}(\mathsf x,\mathsf x') = \int\int d^2 \mathsf x_1 \sqrt{ - \mathsf g(\mathsf x_1)} d^2 \mathsf x_2 \sqrt{ - \mathsf g(\mathsf x_2)} G_{\rm ret}(\mathsf x,\mathsf x_1) G_{\rm ret}(\mathsf x',\mathsf x_2) N(\mathsf x_1,\mathsf x_2),
\end{equation}
where the noise kernel is
\begin{equation}
\label{eq:disskernel}
\begin{split}
N(\mathsf x,\mathsf x') \doteq & \gamma(\nabla^\dagger_s) \nabla_{u} \gamma(\nabla'^\dagger_s) \nabla'_{u} \int\int d\zeta d\zeta' \left < \{\hat \psi_\zeta^{0}(\mathsf x),\hat \psi_{\zeta'}^{0}(\mathsf x')\}\right >.
\end{split}
\end{equation} 
In \ref{sec:homogen} and \ref{sec:statio}, we compute $G_{\rm ac}^{\rm dr}$ in de Sitter space-time and extract from it pair creation probabilities and Hawking--like effects taking place in de Sitter space for states that are invariant under the affine group, see \ref{sec:affinegroup}. 

\section{Homogeneous picture}
\label{sec:homogen}

\subsection{Dissipation and nonseparability} 
\label{sec:homoentangle}

In this section, we decompose the fields in Fourier modes of fixed $\bk$. This representation is suitable for studying the cosmological pair-creation effects induced by the expansion $a(t) = \ep{Ht} = - 1/H\eta$. 

To express the outcome of dissipation in standard terms, we exploit the fact that Lorentz invariance is recovered in the infrared, for momenta $P = k \ep{-Ht} $ much lower than dispersive and dissipative scales. In this limit, since $\Gamma$ and $f$ of Eq.~\eqref{eq:dispersion} are negligible, the $\bk$ components of $\hat \phi$ decouple from $\hat \psi$, and obey a relativistic wave equation. Hence, the $\bk$ component of the (driven) field operator of Eq.~\eqref{eq:phidr} can be decomposed in the \textit{out} basis as
\begin{equation}
\label{eq:phi_k-decomposition}
\hat \phi_\bk(t) \underset{t\to \infty}{\sim} \hat{\mathrm{a}}_\bk \phi^{\rm rel}_k(t) + \hat{\mathrm{a}}^\dagger_{-\bk} [\phi^{\rm rel}_k(t)]^*,
\end{equation}
where the \textit{out} modes $\phi^{\rm rel}_k$ obey the scalar relativistic wave equation and satisfy the standard positive frequency condition at late time. This means that the (reduced) state of $\hat \phi$ (obtained by tracing over $\hat \psi$) can be asymptotically described in terms of conventional excitations with respect to the asymptotic \textit{out}-vacuum. 

The \textit{out} operators $ \hat{\mathrm{a}}_\bk, \hat{\mathrm{a}}^\dagger_{\bk}$ obey the standard commutation rule $[\hat{\mathrm{a}}_\bk, \hat{\mathrm{a}}^\dagger_{\bk'}] = \delta(\bk - \bk')$. For notational simplicity, we omit the $\delta(\bk-\bk')$ when writing two-point functions because it is common to all of them since we only consider homogeneous states. For instance, $\left < \{\hat \phi_\bk^\dagger, \hat\phi_{\bk'}\}\right > =\delta(\bk-\bk') \times G_{\rm ac}^k$. Using Eq.~\eqref{eq:phi_k-decomposition}, the coefficient of the $\delta$ function is
\begin{equation}
\label{eq:phi-Ga} 
\left.G_{\rm ac}^k(t,t) \right|_{t \rightarrow \infty}= 2 \left[2 n_{k} + 1\right] \vert \phi_k^{\rm rel}(t) \vert^2 + 4 \mathrm{Re} \left [c_{k} \left (\phi_k^{\rm rel}(t)\right )^2 \right ], 
\end{equation}
 where 
\begin{subequations}
\label{eq:nkckdef}
\begin{align}
 n_{k} &\doteq \left <\hat  \hat{\mathrm{a}}^\dagger_\bk \hat{\mathrm{a}}_{\bk}\right > = \left < \hat{\mathrm{a}}^\dagger_{-\bk} \hat{\mathrm{a}}_{-\bk}\right > , \quad c_{k} \doteq  \left < \hat{\mathrm{a}}_{-\bk} \hat{\mathrm{a}}_{\bk}\right >.
 \end{align}
\end{subequations}
The mean number of asymptotic outgoing particles is $n_k > 0$, whereas the complex number $c_k$ characterizes the strength of the correlations between particles of opposite wavenumber. The relative magnitude of this number leads to the notion of nonseparability, see \ref{sec:homosepcriterion}. To characterize the level of coherence, we shall use the parameter $\delta$ of Eq.~\eqref{eq:defdeltacampo}, see Ref.~\cite{Campo:2004sz}. 

\subsection[Invariant states and \texorpdfstring{${P}$}{P} representation]{Invariant states and \texorpdfstring{$\boldsymbol{P}$}{P} representation}

Since the states we consider are invariant under the affine group, $n_k$ and $c_k$ are necessarily independent of $k$. We shall nevertheless keep the label $k$ to remind the reader that we work at fixed $k$ and not at fixed $\omega$ as in the next section. Because of the affine group, 
\begin{eqnarray} 
\chi^{\rm rel}(P) \doteq \sqrt{k} \times \phi^{\rm rel}_k(t), 
\label{eq:phiPout}
\end{eqnarray}
only depends on $P $, where $\phi^{\rm rel}_k(t)$ is the (positive unit norm) \textit{out} mode of Eq.~\eqref{eq:phi_k-decomposition}. The norm of the mode $\chi$ is fixed by the Wronskian, see Eq.~\eqref{eq:unitW}:
\begin{equation}
\label{eq:wronskien}
W(\chi^{\rm rel}) = 2 H \Im[(\chi^{\rm rel})^* \partial_P \chi^{\rm rel}] = 1. 
\end{equation}
Using such $\chi^{\rm rel}$ and Eqs.~\eqref{eq:Prep2ptfn} and~\eqref{eq:phi-Ga}, Eq.~\eqref{eq:acgrennfunction} can be written as
\begin{subequations}
\begin{align}
\label{eq:Gacnandc}
\left.G_{\rm ac}(P,P) \right|_{P \rightarrow 0} & = \frac{2}{H}\left[2 n_k + 1\right] \vert \chi^{\rm rel}(P) \vert^2 + \frac{4}{H} {\rm Re} \left [c_k (\chi^{\rm rel}(P))^2 \right ], \\
\label{eq:GacP}
& = \iint_0^\infty \frac{dP_1}{P_1^2} \frac{dP_2}{P_2^2} G_{\rm ret}(P,P_1)G_{\rm ret}(P,P_2) 
N(P_1,P_2). 
\end{align}
\end{subequations}
In the second line, the noise kernel of Eq.~\eqref{eq:disskernel}, which is also invariant under the affine group for the set of states we are considering, has been written in the $P$-representation using Eq.~\eqref{eq:Prep2ptfn}. To extract $n_k$ and $c_k$ from the above equations, we need to compute $ G_{\rm ret}$ and $N$. 

Using Eq.~\eqref{eq:Prep2ptfn}, Eq.~\eqref{eq:locEq} reads
\begin{equation}
\label{eq:retardedGf}
 \left[H^2 \partial_P^2 - \frac{\gamma(-i P)}{\sqrt{P}} H \partial_P \frac{\gamma(i P)}{ \sqrt{P}} + \frac{F^2}{P^2}\right] G_{\rm ret}(P, P') = \delta(P-P') .
\end{equation}
The unique (retarded) solution can be expressed as
\begin{equation}
\label{eq:Gret-solution}
\begin{split}
 G_{\rm ret} (P,P') = \frac{2}{H} \theta(P' - P) \Im \left( \bar \chi_P \bar \chi^\ast_{P'}\right) \ep{-\mathcal{I}_P^{P'}} ,
 \end{split}
\end{equation}
with the optical depth~\cite{Adamek:2008mp},
\begin{equation}
\label{eq:opticaldepth}
 \mathcal{I}_P^{P'} = \int_{P}^{P'}dP_1 \frac{\Gamma(P_1) }{ H P_1 }.
 \end{equation}
Its role is to limit the integrals over $P_1$ and $P_2$ in Eq.~\eqref{eq:GacP} to low values so that $\mathcal{I}_0^P \lesssim 1$. All information about the state for higher values of $P$ is erased by dissipation. In Eq.~\eqref{eq:Gret-solution} we have used
\begin{equation}
\label{eq:tildephi}
\bar \chi(P) \doteq \ep{ \mathcal{I}_0^{P'}} \chi(P),
\end{equation}
where $\chi$ is a homogeneous damped solution of Eq.~\eqref{eq:retardedGf}, in agreement to the functions introduced before Eq.~\eqref{eq:generaleomchibar}. By construction, $\bar \chi (P)$ obeys the reversible (damping free) equation\footnote{
For high values of $P$, the effective dispersion relation is superluminal if $\partial_P (f - \Gamma^2) > 0$, and subluminal if this quantity is negative. The critical case, $f - \Gamma^2 = 0$, gives rise to a relativistic dispersion. In the case where $F^2 - \Gamma^2$ becomes negative, the mode enters an overdamped regime, see Ref.~\cite{Adamek:2008mp}. To avoid the complications this entails, we will only consider $f - \Gamma^2 \geq 0$.}, see Eq.~\eqref{eq:generaleomchibar}
\begin{equation}
\label{eq:eom-transformed}
 \left[H^2 { P^2}\partial_P^2 + {F^2 - \Gamma^2}\right] \bar \chi(P) = 0,
\end{equation}
and is normalized by Eq.~\eqref{eq:wronskien}. Moreover, we impose that it obeys the \textit{out} positive frequency condition, meaning that in the limit $P\to 0$, it asymptotes to the \textit{out} mode $\chi^{\rm rel}$ of Eq.~\eqref{eq:phiPout}. Hence, comparing Eq.~\eqref{eq:Gacnandc} with Eqs.~\eqref{eq:GacP} and~\eqref{eq:Gret-solution}, we find 
\begin{subequations}
\label{eq:n_k,c_k}
\begin{align}
n_k + \frac12 &= \iint_0^\infty \frac{dP_1}{H P_1^2} \frac{dP_2}{P_2^2} {\rm Re}\left(\bar \chi(P_1) \bar \chi^*(P_2)\right) \ep{-\mathcal{I}_0^{P_1}-\mathcal{I}_0^{P_2}} N(P_1, P_2),\\
 c_k &= \iint_0^\infty\frac{dP_1}{H P_1^2} \frac{dP_2}{P_2^2}\bar \chi^*(P_1) \bar \chi^*(P_2) \ep{-\mathcal{I}_0^{P_1}-\mathcal{I}_0^{P_2}} N(P_1, P_2).
 \end{align}
\end{subequations}
These central equations establish how the environment noise kernel $N$ fixes the late time mean occupation number and the strength of the correlations. 

We now compute $N$. When $u$ is freely falling, the rescaled field $\hat \Psi^0_q$ of Eq.~\eqref{eq:psirenorm} is a dense set of independent harmonic oscillators of constant frequency $\omega_\zeta= \pi |\zeta|$, one at each $z$. The frequency is constant because we set $c_\psi = 0$ in the action for $\psi$, see the discussion after Eq.~\eqref{eq:covaction}. It implies that the positive frequency mode functions are the standard $\ep{-i \omega_\zeta t} / \sqrt{2 \omega_\zeta}$, and that the state of these oscillators remains unaffected by the expansion of the universe. Hence $T_\psi$, the temperature of the environment, is not redshifted. 

We here wish to recall that for relativistic (and dispersive) fields, the vacuum state of zero temperature is the only stationary state which is Hadamard, see \ref{sec:thermal}. Hence, for these fields, the temperature is fixed to zero. This is not the case in our model where any temperature $T_\psi$ is acceptable. In what follows, we shall thus treat $T_\psi$ as a free parameter, and work with homogeneous thermal states. This means that the expectation value of the anticommutator of $\hat \psi^0_q$ is given by
\begin{equation}
\begin{split}
\label{eq:NofDelta}
\left <\{\hat \psi_\zeta^{0}(\mathsf x),\hat \psi_{\zeta'}^{0}(\mathsf x')\}\right > &=\frac{ \delta(z - z')}{\sqrt{a(t) a(t')}} \delta(\zeta-\zeta') \coth\frac{\omega_\zeta}{2 T_\psi} \frac{ \cos\left ( \omega_\zeta \Delta t \right )}{\omega_\zeta}.
\end{split}
\end{equation}
The factor $ \coth({\omega_\zeta}/{2 T_\psi}) =2 n^\psi_{ \zeta} + 1 $ is the standard bosonic thermal distribution. The prefactor $\delta(z - z')/\sqrt{a(t) a(t')}$ comes from the facts that $\hat \psi^0_\zeta$ of Eq.~\eqref{eq:psirenorm} is a dense set of independent oscillators, and that $a(t,z)$ reduces here to the scale factor $a(t)$. To get $N$ of Eq.~\eqref{eq:disskernel} one should differentiate the above and integrate over $\zeta$. The integration gives a distribution which should be understood as Cauchy principal value,
\begin{equation}
\begin{split}
\label{eq:NofDelta2}
\iint d\zeta d\zeta' \nabla_{u} \nabla_{u'}  &\left < \{\hat \psi_\zeta^{0}(\mathsf x), \hat \psi_{\zeta'}^{0}(\mathsf x')\}\right >= - \frac{ \delta(z - z')}{\sqrt{a(t) a(t')}} \times 2 T_\psi \frac{\partial}{\partial{ \Delta t}} \mathtt{P.V.}\coth \left(\pi T_\psi \Delta t\right).
\end{split}
\end{equation}
To be able to re-express Eq.~\eqref{eq:NofDelta2} in the $P$-representation, it is necessary to verify that it is invariant under the affine group. This is easily done using notations of \ref{sec:Prep}. One verifies that the first factor simply equals $\delta(\Delta_2)$, whereas the second term is only a function of $\Delta_1$. Taking into account the derivatives of Eq.~\eqref{eq:disskernel}, in the $P$-representation, the noise kernel at temperature $T_\psi$ reads
\begin{equation}
 \label{eq:noisekernel}
 \begin{split}
N (P, P') = &- \gamma(i P) \gamma(-i P') {2 T_\psi}{\sqrt{P P'}} \frac{\partial}{\partial{\ln\frac{P'}{P}}} \mathtt{P.V.}\coth \left(\frac{\pi T_\psi}{H} \ln\frac{P'}{P}\right).
 \end{split}
\end{equation}
The symbol $\mathtt{P.V.}$ indicates that when evaluated in the integrals of Eq.~\eqref{eq:n_k,c_k}, the nonsingular part should be extracted using a Cauchy principal value prescription on $\ln(P'/P) = H (t- t')$. 

In the high-temperature limit, the double integrals of Eq.~\eqref{eq:n_k,c_k} can be evaluated analytically because $N$ effectively acts as a Dirac delta function. Instead, when working with an environment in its ground state, or at low temperature $T_\psi$, we are not aware of analytical techniques to evaluate these integrals. Hence, to study the impact of dissipation on coherence in (near) vacuum states, we shall numerically integrate Eqs.~\eqref{eq:n_k,c_k}.

\subsection{Numerical Results}

In the forthcoming numerical computations, for simplicity, we work with
\begin{equation}
\label{eq:numdispersion}
f = \frac{P^4}{\Lambda^2}, \quad \Gamma = g^2 \frac{P^2}{2 \Lambda},
\end{equation}
which contain the same ultraviolet momentum scale $\Lambda$. The dimensionless coupling $g^2$ controls the relative importance of dispersive and dissipative effects. In the limit $g^2 \rightarrow 0$, we get the quartic superluminal dispersion studied in \ref{sec:pfour} and in Refs.~\cite{Macher:2008yq,Busch:2012ne}. The critical coupling $g^2_{\rm crit} \doteq 2$, greatly simplifies the calculations, since $f - \Gamma^2 = 0$ guarantees that $\bar \chi(P)$ obeys a relativistic equation, see Eq.~\eqref{eq:eom-transformed}. 

Using a numerically stable procedure to extract the Cauchy principal values like in Ref.~\cite{Adamek:2008mp}, we compute $n_k$ and $c_k$ of Eq.~\eqref{eq:n_k,c_k} in the parameter space $\Lambda$, $g^2$, $m^2$, and $T_\psi$. Since all physical effects only depend on dimensionless ratios, we present the numerical results in terms of $\mu = \sqrt{m^2/H^2-1/4}$, $\lambda = \Lambda/H$, and $\vartheta = T_\psi/H$.

\subsubsection{Massless critical case}
 
\begin{SCfigure}[2][!b]
\includegraphics[width=0.45\linewidth]{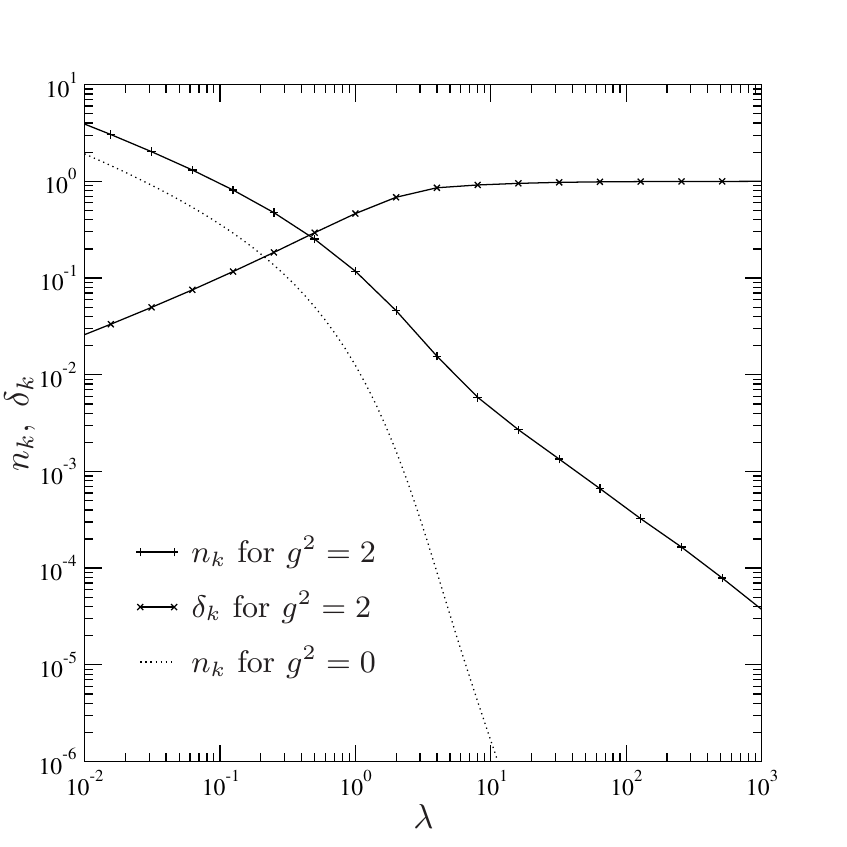}
\caption{\label{fig:massless-critical}
Numerical values for $n_k$ and $\delta_k$ for a massless field with \textit{critical} damping $g= g_\mathrm{crit}$ and quartic superluminal dispersion at the energy scale $\Lambda = H \lambda$. For comparison, we have represented by a dotted line the $n_k$ of the quartic dispersive field (in which case $\delta_k = 0$ identically). Surprisingly, the state is nonseparable, $\delta_k < 1$, for all values of $\lambda$. Moreover, $\delta_k$ decreases when dissipation increases. }
\end{SCfigure}

We begin with the massless case ($m = 0$) and with $g=g_{\rm crit}$. Then Eq.~\eqref{eq:eom-transformed} is particularly simple since the rescaled mode $\bar \chi$ of Eq.~\eqref{eq:tildephi} reduces \textit{for all} $P$ to the \textit{out}-mode $ \bar \chi(P) = {\ep{i P/H}}/{\sqrt{2 H}}$. In this we recover the conformal invariance of the massless field in two dimensions. There usually would be no particle production when it propagates in de Sitter space, however, the conformal invariance being broken by dissipation, pair-creation will take place. 

In Fig.~\ref{fig:massless-critical} we present $n_k$ and $\delta_k$ when the environment is in its ground state ($T_\psi = 0$). For comparison, we also show $n_k$ for quartic dispersion ($g^2 = 0$) which can be computed analytically in the Bunch-Davies vacuum~\cite{Macher:2008yq}. For $\lambda \to \infty$ the number of particles goes to zero as $1/ \lambda$, as is expected since conformal invariance is restored in this limit. Despite dissipation, we find that $\delta_k < 1$ for all values of $\lambda$. This indicates that the state is always nonseparable in the two-mode $k$ basis. In addition, contrary to what might have been expected, the two-mode entanglement is stronger for smaller values of $\lambda$, i.e., stronger dissipative effects. The reason for this has to be found in the fact that $\lambda$ also sets the scale where conformal invariance is broken. 

\begin{figure}[!b]
\begin{minipage}{0.47\linewidth}
\includegraphics[width=1\columnwidth]{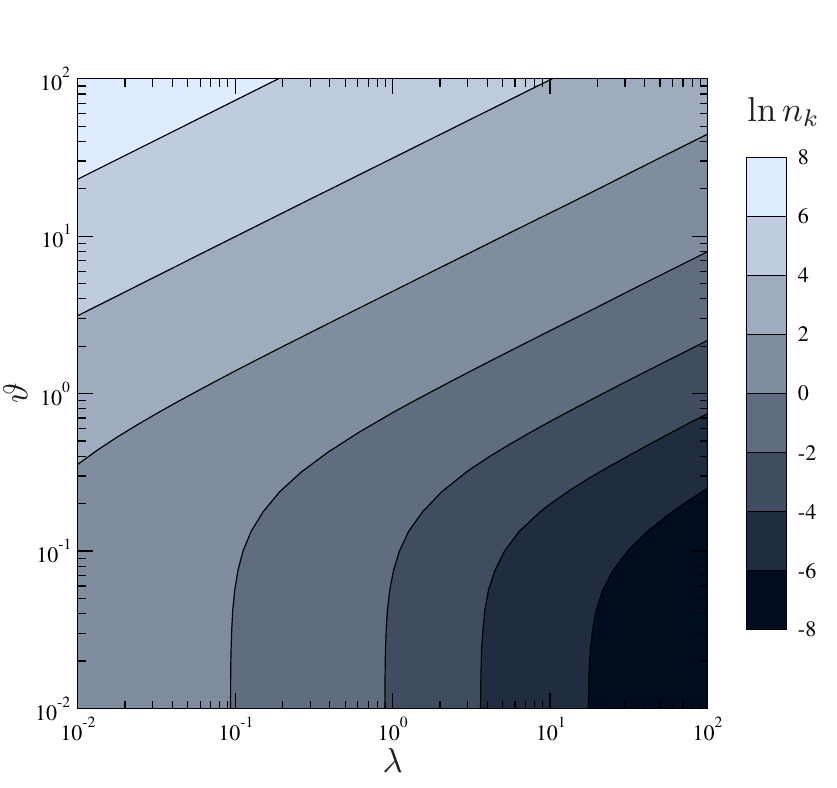}
\end{minipage}
\hspace{0.03\linewidth}
\begin{minipage}{0.47\linewidth}
\includegraphics[width=1\columnwidth]{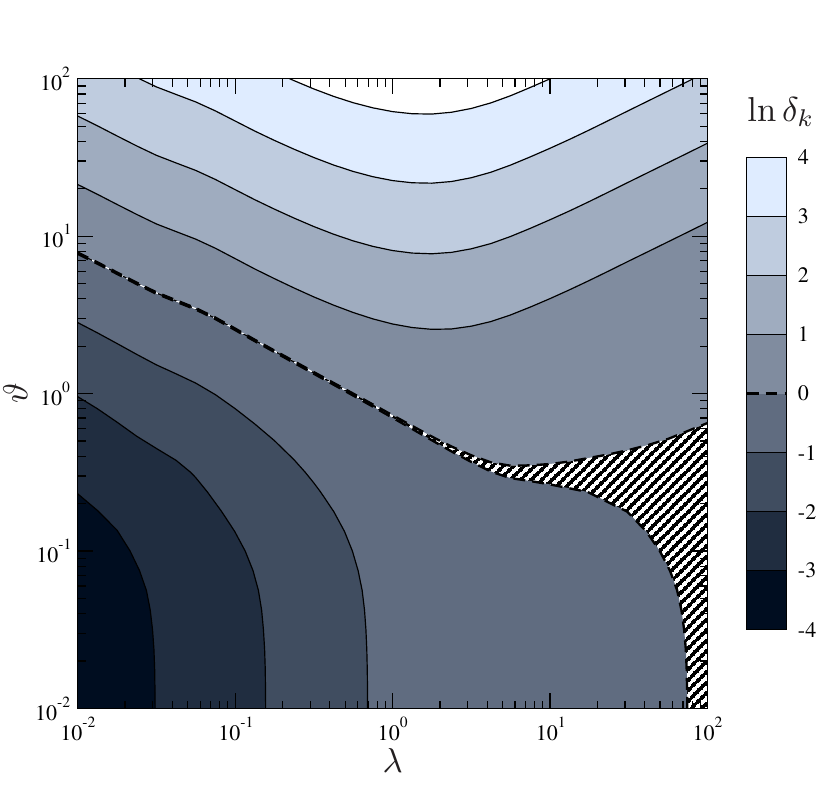}
\end{minipage}
\caption{\label{fig:massless-T} \small
Contour plots of $\ln n_k$ and $\ln \delta_k$ for a massless field with critical coupling $g= g_\mathrm{crit}$ in the parameter space $(\lambda = \Lambda/H,~\vartheta = T_\psi/H)$ . At low temperatures, for $\vartheta = T_\psi/H \lesssim 1/10$, $n_k$ and $\delta_k$ barely depend on $\vartheta$. On the contrary, for high temperatures, $\vartheta \gtrsim 1$, $n_k$ scales as $n_k \propto \vartheta \lambda^{-1/2}$ whereas $\delta_k$ scales as $\delta_k \propto \vartheta \lambda^{-1/2}$ for $\lambda \gtrsim 1$, and $\delta_k \propto \vartheta \lambda^{1/2}$ for $\lambda \lesssim 1$. The hatched region indicates the numerical uncertainty about the threshold value $\delta_k = 1$ found when $n_k \ll 1$. }
\end{figure}

Let us now turn to the effects of the environment temperature $T_\psi$. Figure~\ref{fig:massless-T} shows contour plots of $n_k$ and $\delta_k$ for a massless field with Eq.~\eqref{eq:numdispersion}, again for $g= g_{\rm crit}$. In the limit $\lambda \to \infty$, we observe that $n_k\to 0$ irrespectively of the value of $T_\psi$. This establishes that there is a robustness of the relativistic result in the limit $\lambda \to \infty$ which generalizes that found for dispersive fields, see e.g., Ref.~\cite{Macher:2008yq}. This should not be a surprise since we showed in \ref{sec:thermal} that when $\psi$ is not coupled to $\phi$, i.e., $g^2/\lambda \to 0$, only the BD vacuum is stationary stable state. Moreover, in the high-temperature limit, Eqs.\eqref{eq:n_k,c_k} can be evaluated analytically to give
\begin{equation}
\label{eq:massless-highT}
n_k + \frac{1}{2} \sim \frac{\sqrt{\pi} \vartheta}{\sqrt{\lambda}},\quad
\delta_k \sim \frac{\sqrt{\pi} \vartheta}{\sqrt{\lambda}} \left(1 - \frac{1 + \mathrm{erfi}^2\sqrt{\lambda}}{\ep{2\lambda}}\right),
\end{equation}
where $\mathrm{erfi}$ is the imaginary error function. We compared the corresponding contours with the numerical ones shown in Fig.~\ref{fig:massless-T} and found that they are practically indistinguishable for $\vartheta > 10$.

When considering the effects of $T_\psi$, we observe two regimes. At low temperature ($\vartheta \ll 1$), $n_k$ and $\delta_k$ only depend on $\lambda$ and are basically given by the zero temperature limit shown in Fig.~\ref{fig:massless-critical}. However, at large temperature ($\vartheta \gg 1$), they depend on $\lambda$ and $\vartheta$ according to Eqs.~\eqref{eq:massless-highT}. As expected, the strongest signatures of quantum entanglement, $\delta_k \ll 1$, are found in the region where the breaking of conformal invariance is large (and hence pair-creation is active) and when the environment temperature is small, so that the spontaneous pair-creation events are not negligible with respect to thermally induced events. On the other hand, when the temperature is large, the final state is separable since $\delta_k \gg 1$. In Fig.~\ref{fig:massless-T} (right panel) we see that the threshold case $\delta_k = 1$ is approximatively given by $\vartheta \sim \lambda^{-1/2}$ for $\lambda \lesssim 1$. The hatched region for $\lambda \gtrsim 10$ represents the numerical uncertainty in the region where $n_k$ is much smaller than 1.

\subsubsection{Massive fields}

We remind [see \ref{sec:betacosmo}] that the massless case $m = 0$ is an isolated point in the mass spectrum: a well-defined notion of \textit{out}-quanta requires either $m = 0$ or $\mu^2 > 0$. In the latter case, the asymptotic \textit{out}-modes with positive frequency (see, e.g., Appendix B of Ref.~\cite{Macher:2008yq}) are given by
\begin{equation}
 \bar \chi(P) = \sqrt{\frac{\pi}{2 \sinh \pi {\mu}}} \frac{\sqrt{P}}{H} J_{i {\mu}} (P / H),
\end{equation}
where $J$ denotes the Bessel function of the first kind.

\begin{figure}[!b]
\begin{minipage}{0.47\linewidth}
\centerline{\includegraphics[width=\linewidth]{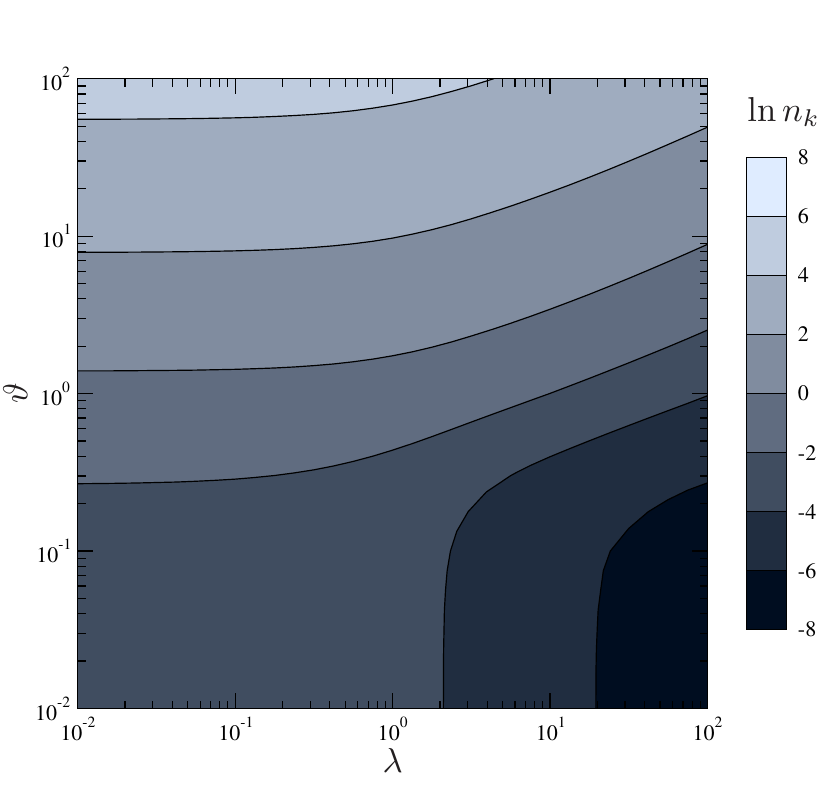}}
\end{minipage}
\hspace{0.03\linewidth}
\begin{minipage}{0.47\linewidth}
\centerline{\includegraphics[width=\linewidth]{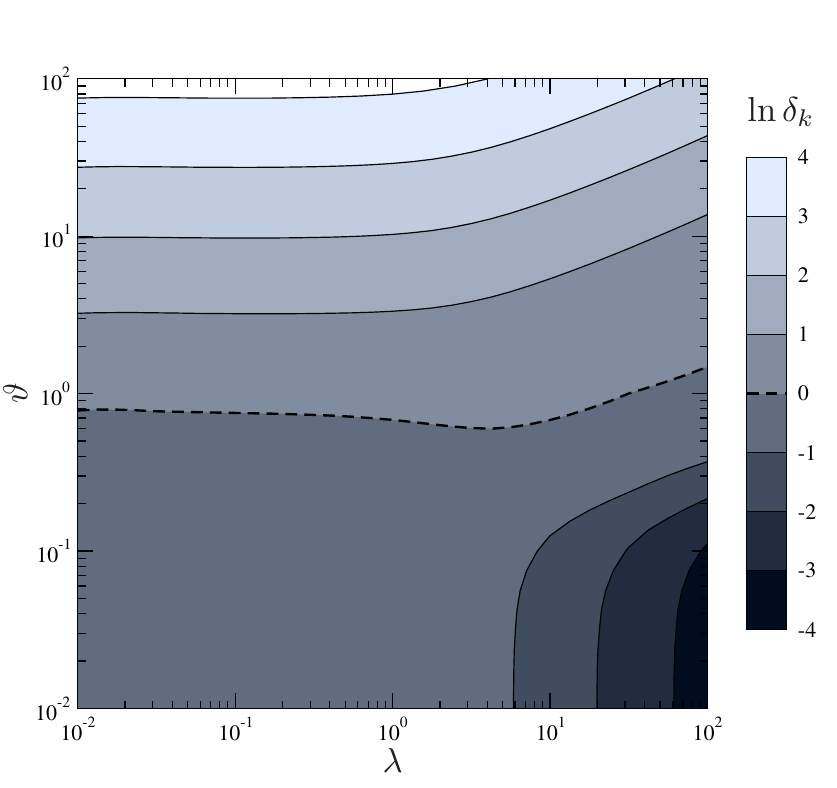}}
\end{minipage}
\caption{\label{fig:massive-T} \small
Contour plots of $\ln n_k$ and $\ln \delta_k$ for a massive field ($\mu = 1$) with critical coupling $g=g_\mathrm{crit} $ in the parameter space $(\lambda = \Lambda/H,~\vartheta = T_\psi/H)$. As in Fig.~\ref{fig:massless-T}, for low temperature $\vartheta \lesssim 0.1$, $n_k$ and $\delta_k$ are independent of the temperature. Instead for $\vartheta \gtrsim 1$ and $\lambda \gg 1$, $n_k$ and $\delta_k$ scale both as $\vartheta \lambda^{-1/2}$. }
\end{figure}

Figure~\ref{fig:massive-T} shows the contour plots of $n_k$ and $\delta_k$ for a massive field with $\mu^2 = 1$ and $g= g_\mathrm{crit}$, in the same parameter space ($\lambda$, $\vartheta$) as in Fig.~\ref{fig:massless-T}. The case of a Lorentz-invariant field in the Bunch-Davies state is recovered in the limit $\lambda \rightarrow \infty$, $\vartheta \rightarrow 0$. Now conformal invariance is already broken by the mass term and therefore $n_k$ remains nonzero in this limit.

At zero temperature, the strongest entanglement (lowest $\delta_k$) is found at large values of $\lambda$, i.e., weak dissipation. This was expected, since dissipation reduces the strength of correlations. However, as in the massless case, the threshold of separability $\delta_k = 1$ is not crossed. 

When increasing the environment temperature $T_\psi$, we see that the strength of correlation is reduced, and separable states are found. The nonseparability criterion $\delta_k < 1$ is therefore only met either when $T_\psi$ is smaller than the Gibbons-Hawking temperature $T_{\rm GH} = H/2\pi$, or when the coupling to the environment is sufficiently weak. Notice also that the behavior at high temperature can again be obtained analytically, the integrals over the Bessel functions becoming hypergeometric functions.

\subsubsection{Role of \texorpdfstring{$g$}{g} in the underdamped regime}

It is also interesting to consider the role of the coupling $g$, see Eq.~\eqref{eq:numdispersion}. As $g^2$ approaches zero, the dissipative scale $2\Lambda/g^2$ is moved deeper into the UV with respect to the dispersive scale which is fixed by $\Lambda$. In the limit $g^2 \rightarrow 0$, the field becomes purely dispersive and $n_k$, $\delta_k$ can be computed analytically~\cite{Macher:2008yq} in the Bunch-Davies vacuum. For $g^2 < 2$ the mode is underdamped. In this case, the solutions to Eq.~\eqref{eq:eom-transformed} which correspond to asymptotic \textit{out}-modes of positive frequency are given by, see Eq.~\eqref{eq:chioupfour} 
\begin{equation}
\bar \chi(P) = {\left(-2 i \tilde \lambda\right)^{\frac{1+ i \mu }{2}} } \sqrt{\frac{1}{2 \mu p}} {{{\rm \bf M}\left(\frac{-i \tilde \lambda}{2} ,\frac{i \mu}{2}, {\frac {i p^{2}}{2 \tilde \lambda}}\right)}},
\end{equation}
where $\tilde{\lambda} \doteq \lambda / \sqrt{4 - g^4}$.

\begin{figure}[ht]
\begin{minipage}{0.47\linewidth}
\centerline{\includegraphics[width=\linewidth]{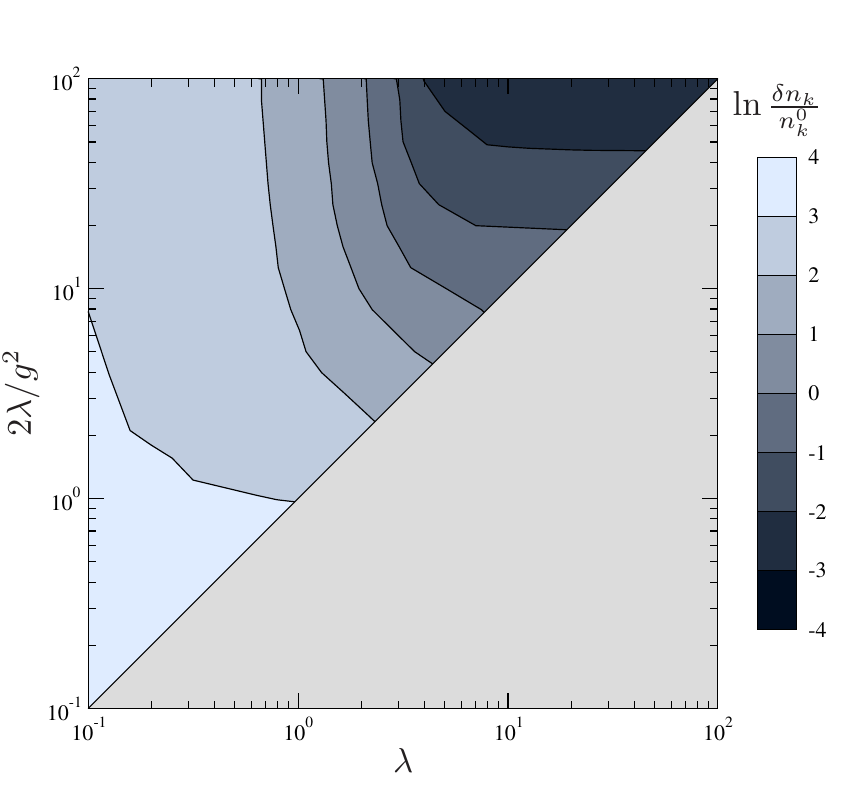}}
\end{minipage}
\hspace{0.03\linewidth}
\begin{minipage}{0.47\linewidth}
\centerline{\includegraphics[width=\linewidth]{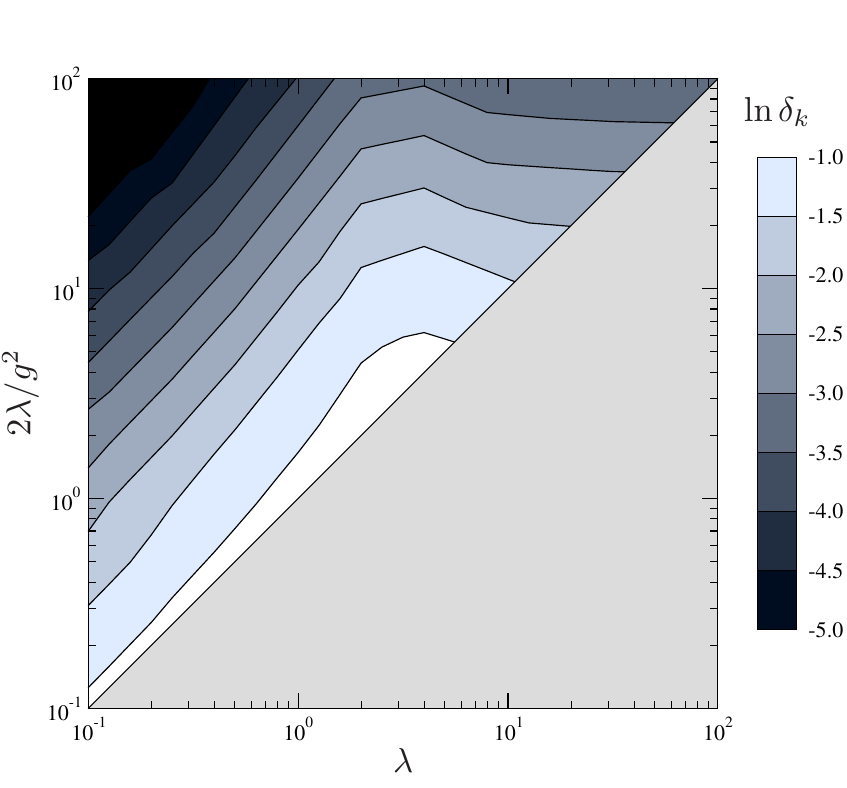}}
\end{minipage}
\caption{\label{fig:massive-vac} \small
Contour plots of $\delta n_k / n_k^0$ and $\delta_k$ for a massive field ($\mu^2 = 1$) in the \textit{underdamped} regime $g^2 \leq g_\mathrm{crit}^2$. The environment is in its ground state ($\vartheta = 0$) and the two axes are the dispersive scale $\lambda$ and the dissipative one $2 \lambda / g^2 \geq \lambda$.}
\end{figure}

Figure~\ref{fig:massive-vac} shows contour plots of $\delta n_k / n_k^0 \doteq (n_k - n_k^0) / n_k^0$ (where $n_k^0$ is the number of particles without dispersion and dissipation) and $\delta_k$ for a massive field in the underdamped regime. Here, we set $T_\psi = 0$, and plot the results in the parameter space spanned by the two (dimensionless) ultraviolet scales: $\lambda$ which characterizes dispersion, and $2 \lambda / g^2$ which is the UV scale of dissipation. The latter is larger than the former in the underdamped regime. The grey areas therefore correspond to the overdamped regime which we did not study. 

In the weak dispersive/dissipative regime $\lambda \gtrsim 10$, it is clear that $\delta n_k$ and $\delta_k$ are both dominated by dissipative effects. For the latter, this is because dispersion alone does not lead to decoherence. For the deviation $\delta n_k$, this follows from the fact that dispersion gives an exponentially small correction to the pair creation process (see Ref.~\cite{Macher:2008yq}), while the corrections due to dissipation are only algebraically small. As a result, the hierarchy of scales does not directly fix the importance of the respective effects.

On the other hand, when dispersion is strong ($\lambda \lesssim 1$) the pair creation process is basically governed by dispersive effects. The correction to the particle number due to dissipation is very small (compared to the dispersive correction). One can also observe that the degree of two-mode entanglement is then basically governed by the separation between the two scales $g^2$, i.e., $\delta_k$ is determined by the strength of dissipation \textit{at the dispersive threshold}, $\left (\Gamma/P\right )\vert_{ P = \Lambda}$.

\section{Stationary picture}
\label{sec:statio}

In the absence of dispersion/dissipation, it is well known that the Bunch-Davies vacuum is a thermal (KMS) state at the Gibbons-Hawking temperature $T_{\rm GH} = H/2\pi$~\cite{birrell1984quantum}. It is also known that this is the temperature seen by any inertial particle detector, and that this is closely related to the Unruh effect found in Minkowski space, and to the Hawking radiation emitted by black holes~\cite{Brout:1995rd}. In the presence of dissipation, while the stationarity of the state of $\phi$ is {\it exactly} preserved when the state of the environment is invariant under the affine group, the thermality of the state is {\it not} exactly preserved. This loss of thermality, which generalizes what was found for dispersive fields, see \ref{sec:thermal}, questions the status of black hole thermodynamics when Lorentz invariance is violated~\cite{Dubovsky:2006vk,Eling:2007qd,Jacobson:2008yc}.

\subsection{Loss of thermality}

To probe the stationary properties of the state, we consider the transition rates of particle detectors at rest with respect to the orbits of $K_t$. This means that the detector is located at fixed $H|X| < 1$ in the coordinates of Eq.~\eqref{eq:dSmetricintwoways}. In this case, the two-point functions only depend on $t-t'$ and can be analyzed at fixed $\omega = i \partial_t\vert_X $, see Eq.~\eqref{eq:fouriert}. (The above restriction on $X$ simply expresses that the trajectory be timelike.)

The transition rates [see Eq.~\eqref{eq:Rs}] are, up to an overall constant, given by Fourier transforms of the Wightman function $G_W$~\cite{Brout:1995rd}. The rates then determine $n_\omega(X)$, the mean number of particles of frequency $\omega > 0$ seen by a detector located at $X$, through
\begin{equation}
\label{eq:nofG}
\begin{split}
\frac{n_\omega (X)}{n_\omega (X)+1} = \frac{G_{W}^\omega(X,X) }{ G_{W}^{-\omega}(X,X) }. 
\end{split}
\end{equation}
To study the deviations with respect to the Gibbons-Hawking temperature $T_H = H/2 \pi$, we generalize the temperature function $T_{\rm gl}(\omega,X)$ introduced in Eq.~\eqref{eq:tempdef} to any state by
\begin{equation}
\frac{n_\omega (X)}{n_\omega (X)+1} = \ep{- \omega/T_{\rm gl}(\omega,X) }.
\end{equation}
It gives the effective temperature seen by the detector, and reduces to the standard notion when it is independent of $\omega$. 

In the following numerical computations, for simplicity, we work at $X=0$ with an inertial detector, with $g= g_{\rm crit}$, $m=0$, and $T_\psi =0$. Since the calculation of the commutator of $\phi$ is much faster and more reliable than that of the anticommutator, instead of using Eq.~\eqref{eq:nofG}, $n_\omega$ shall be numerically computed with
\begin{equation}
\begin{split}
 n_\omega (X)= \frac{G_{W}^\omega(X,X) }{ G_{\rm c}^{\omega}(X,X) }. 
\end{split}
\end{equation}
The denominator is expressed using Eq.~\eqref{eq:GcfromGret}. The numerator is obtained from integrating the retarded green's function $G_{\rm ret}^{\omega}(X, \bP)$ in a mixed $X,P$ representation and the noise of Eq.~\eqref{eq:noisekernel} with $T_\psi \to 0$. In addition, the principal value is replaced by a prescription for the contour of $\ln P/P' = H t $ to be in the upper complex plane. In this we recover the fact that when the anticommutator in the vacuum is $\mathtt{P.V.} (1/t)$, the corresponding vacuum Wightman function is $1/(t-i \epsilon)$.

\begin{figure}[ht]
\begin{minipage}[t]{0.45\linewidth}
\centerline{\includegraphics[width=1\columnwidth]{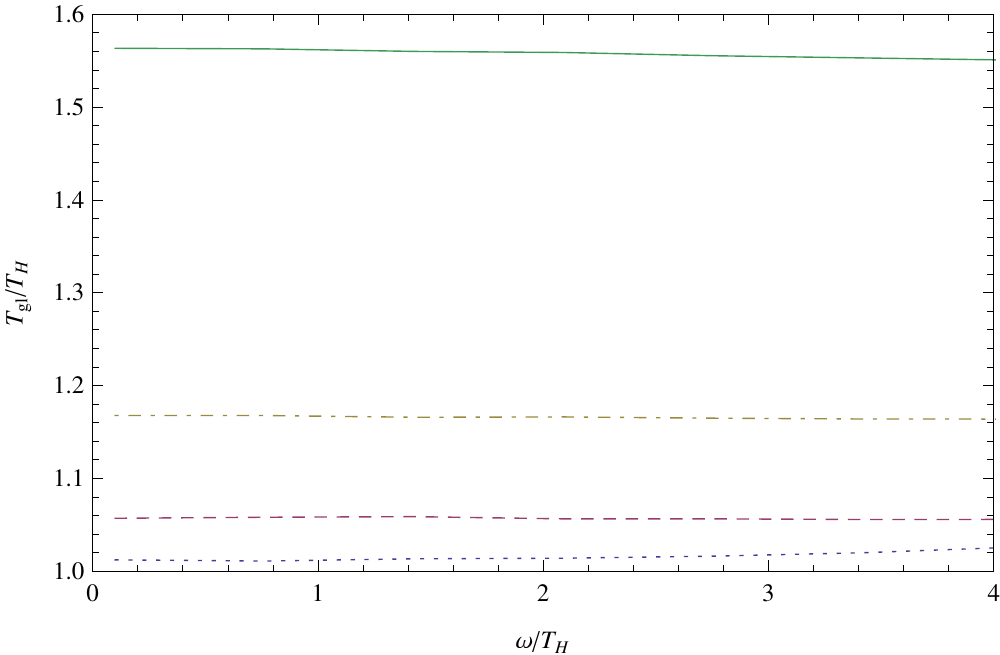}}
\caption{\label{fig:Tofom} \small
Plot of the ratio $T_{\rm gl}(\omega)/T_H$ as a function of $\omega/ T_H$ for various values of $\lambda$. We work with a massless field with $g= g_{\rm crit}$, for a detector localized in the center of the patch ($X=0$), and with $T_\psi= 0$. The values of $\lambda$ are $1$ (continuous), $3$ (dot-dashed), $5$ (dashed), and $10$ (dotted.)}
\end{minipage}
\hspace{0.05\linewidth}
\begin{minipage}[t]{0.45\linewidth}
\centerline{\includegraphics[width=1\columnwidth]{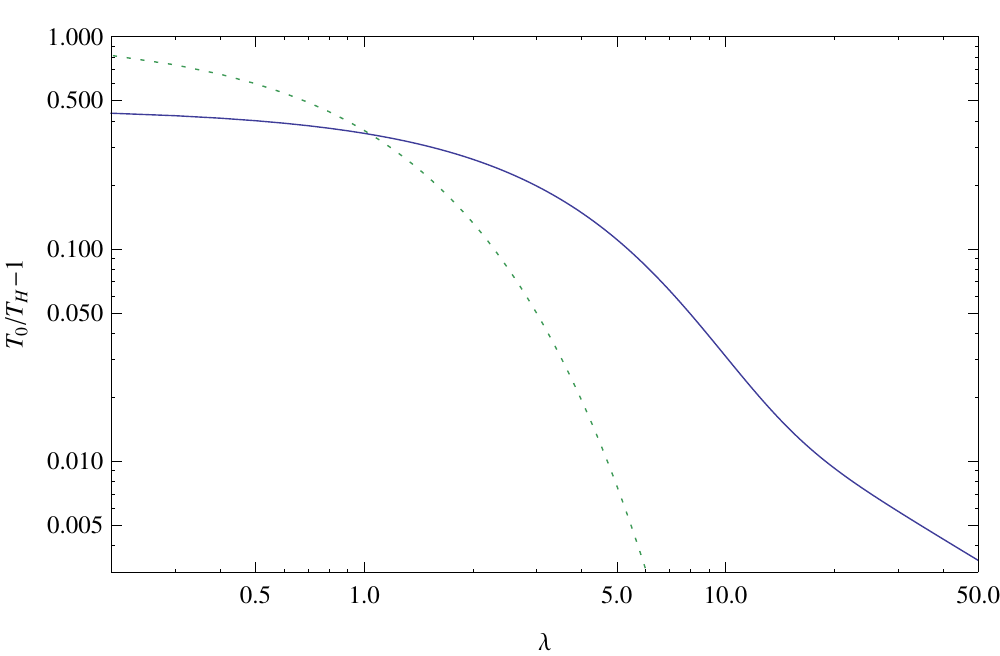}}
\caption{\label{fig:logToflambda} \small
Plot of $T_0/T_H-1$ in logarithmic scales as a function of $\lambda$, where $T_0$ is the low-frequency temperature of massless fields with $g= g_{\rm crit}$ and when $T_\psi = 0$ ($\psi$-vacuum). We have also represented by a dotted curve the same quantity evaluated without dissipation when the state is the Bunch-Davies vacuum.}
\end{minipage}
\end{figure}

In Fig.~\ref{fig:Tofom}, we plot the ratio $T_{\rm gl}(\omega)/T_H$ as a function of $\omega$ for various values of $\lambda$, and for $T_\psi = 0$. We first observe that $T_{\rm gl}(\omega)$ is constant for all frequencies from zero to a few multiples of $T_H$. Hence, the Planckian character of the state is, to a high accuracy, preserved by dissipation, as was found in the presence of dispersion, see \ref{chap:dispdS} and Refs.~\cite{Macher:2009tw,Finazzi:2010yq}. For higher frequencies, i.e., $\omega/T_H > 4$, we were not able to study $T_{\rm gl}(\omega)$ with sufficient accuracy because of the numerical noise associated to $n_\omega < 0.01$. As in the dispersive case, we expect that the temperature function $T_{\rm gl}(\omega)$ is modified for $\omega \gtrsim \Lambda$. 

Second, when $\lambda$ is smaller than 5, i.e., when dissipation is strong, we observe that the temperature is significantly (more than $5\%$) larger than $T_H$. These deviations are further studied in Fig.~\ref{fig:logToflambda}, where we plot the deviations of $T_0$, the low-frequency effective temperature, with respect to $T_H$ as a function of $\lambda$. We observe that the deviation due to dissipation asymptotically follows 
\begin{equation}
\frac{T_0}{T_H}-1 \underset{\lambda \to \infty}\sim (6 \lambda)^{-1} . 
\end{equation}
This law has been verified up to $\lambda = 10^3$. It has to be compared with the deviation due to quartic dispersion studied in \ref{sec:pfour}. This deviation is represented by the dotted curve, and scales as ${T_0^{\rm disp}}/{T_H}-1 \sim \ep{-\pi \lambda/4}$. In other words, the deviation due to (quadratic) dissipation decreases much slower than that due to (quartic) superluminal dispersion. The important lesson for black hole thermodynamical laws is that ultraviolet dispersion and dissipation both destroy the thermality of the state. This lends support to the claim that Lorentz invariance is somehow necessary for these laws to be satisfied.

\subsection{Asymptotic correlations among right movers}
\label{sec:Asc}

As explained in \ref{sec:homoentangle}, at late time, the $\phi$ field decouples from its environment. This allows to use the relativistic \textit{out} basis at fixed $k$ to read out the state of $\phi$. Alternatively, one can also use an \textit{out} basis formed with stationary modes with fixed frequency $\omega$. Indeed, at fixed $\omega$, the momentum $P_\omega \sim |\omega/X| \to 0$ at large $|X|$, and dispersive effects are negligible. Hence $\hat \phi_\omega(X)$, the stationary component of the field operator, decouples from the environment at large $|X|$, and can be analyzed using relativistic modes. As we shall see, this new \textit{out} basis is {\it not} trivially related to the homogeneous one used in \ref{sec:homogen} because it encodes thermal effects at the Gibbons-Hawking temperature. Hence the covariance matrix of the new \textit{out} operators will depend on $n_k$ and $c_k$ of Eq.~\eqref{eq:n_k,c_k}, but also on these thermal effects. At this point we need to explain why we are interested in expressing in a different basis a state which is fully characterized by $n_k$ and $c_k$. The main reason comes from black hole physics. As shall be discussed in the next chapter, when certain conditions are met, the results of this section apply to the Hawking radiation emitted by dissipative fields. 

To compute the covariance matrix in the new basis, we recall some properties of the \textit{relativistic} massless field in de Sitter. First, because of conformal invariance, the field operator splits into two sectors which do not mix, one for the right-moving $U$ modes with $\bk > 0$, and the other for the left-moving $V$ modes with $\bk < 0$. In addition, in de~Sitter, the time-dependence of all homogeneous modes can be expressed through $\chi^{\rm rel}(P)$ of Eq.~\eqref{eq:phiPout}, which here reduces to
\begin{equation}
\begin{split}
\chi^{\rm rel}(P) = \ep{i P/H} /\sqrt{2 H},
 \end{split}
\end{equation}
where $P > 0$. This mode has a unit positive Klein-Gordon norm, as can be verified using the Wronskian condition of Eq.~\eqref{eq:wronskien}.
 
We introduce an intermediate basis constructed with the stationary \enquote{Unruh} modes $\varphi_\omega$~\cite{Unruh:1976db}. In the $P$ representation, they can be written as~\cite{Parentani:2010bn}, see Eq.~\eqref{eq:facto}
\begin{equation}
\begin{split}
\label{eq:unruhmodedef}
 \phi_{\omega}^{\rm rel} = (P/H)^{-i \omega/H -1} \times \chi^{\rm rel}(P) . 
 \end{split}
 \end{equation}
They form an orthonormal and complete mode basis if $\omega \in ]- \infty, \infty[$. The spatial behavior of the $U$ modes is given by
\begin{equation}
\begin{split}
\label{eq:varphiX}
\phi^U_{\omega }(X) = \int_0^\infty \frac{d\bP}{H \sqrt{2\pi}} \ep{i \bP X} \phi_{\omega}^{\rm rel}(P).
\end{split}
\end{equation}
We now introduce the alternative \textit{out} basis formed of stationary modes which are localized on either side of the horizons, henceforth called $R$ and $L$ modes. They behave as Rindler modes in Minkowski space. For $U$-modes, the horizon is located at $HX = - 1$, and these modes are the ones of Eq.~\eqref{eq:phiout} and are defined for $\omega > 0$. They are easily related to the Unruh mode by computing Eq.~\eqref{eq:varphiX}. Indeed, for $\omega > 0$, one gets
\begin{equation}
\begin{split}
\label{eq:UnruhRLchange}
\phi^U_{\omega } = \alpha_\omega^H \phi_{\omega,R}^{U, out} + \beta_\omega^H (\phi_{-\omega,L}^{U, out})^* ,
\end{split}
\end{equation} 
where coefficients $\alpha_\omega^H$ and $\beta_\omega^H$ are the standard Bogoliubov coefficients leading to the Gibbons-Hawking temperature $H/2\pi$. They obey $\abs{\beta_\omega^H/\alpha_\omega^H} = \ep{ - \pi \omega/H}$. Asymptotically in the future and in space, the $U$ part of the field operator can thus be expressed as
\begin{subequations}
\begin{align}
\label{eq:operatorphihomo}
\hat \phi_U (\mathsf x) = \int_0^\infty d \bk & \{ \hat a_{\bk} \ep{i \bk z} \phi^{\rm rel}_k(t) + h.c. \}  \\
\label{eq:operatorphiunruh}
 = \int_{-\infty}^{\infty} d\omega &\{ \hat a_{U}^{ \omega} \ep{- i \omega t} \phi^U_{\omega }(X) + h.c. \}\\
 = \int_{0}^{\infty} d\omega &\{ \hat a_{U, R}^{ \omega} \ep{- i \omega t} \phi_{\omega,R}^{U, out}(X) + (\hat a_{U, L}^{ -\omega})^\dagger \ep{- i \omega t} (\phi_{-\omega,L}^{U, out})^* + h.c. \} 
\end{align}
\end{subequations}
The $V$ part possesses a similar decomposition, and the $V$ modes are obtained from the $U$ ones by replacing $X \to-X$, and $R \leftrightarrow L$. The $\phi_\omega^V$ modes are thus defined on either side of $HX = 1$. 

Using the above equations, the Unruh and the Rindler-like operators of frequency $\vert \omega \vert$ are related by
\begin{equation}
\label{eq:changeofbase}
\begin{split}
\left (
\begin{array}{l}
\hat a_{U, R}^\omega\\
\hat a_{U, L}^{ -\omega \dagger}\\
\hat a_{V, L}^\omega\\
\hat a_{V, R}^{ -\omega \dagger}
\end{array} \right )= \left ( \begin{array}{cccc}
\alpha_\omega^H&\beta_\omega^{H *}&0&0\\
\beta_\omega^{H}& \alpha_\omega^{H *}&0&0\\
0&0&\alpha_\omega^H&\beta_\omega^{H *}\\
0&0&\beta_\omega^{H}& \alpha_\omega^{H *}\\
\end{array}\right ) \times \left (
\begin{array}{l}
\hat a_U^\omega\\
\hat a_U^{- \omega \dagger}\\
\hat a_V^\omega\\
\hat a_V^{- \omega \dagger}\\
\end{array} \right ).
\end{split}
\end{equation}
We considered both $U$ and $V$ modes because our aim is to compute the covariance matrix of the $R$ and $L$ operators in terms of $n_k$ and $c_k$ of Eq.~\eqref{eq:n_k,c_k}, where $c_k$ mixes $U$ and $V$ modes. To do so, we first compute the covariance matrix of the Unruh operators. When working with states that are invariant under the affine group, $n_k$ and $c_k$ of Eq.~\eqref{eq:n_k,c_k} are independent of $k$. This implies that the covariance matrix of the Unruh operators is independent of $\omega$. Indeed, using 
\begin{equation}
\hat a_U^\omega = \int_0^\infty \frac{d \bk}{H} \left (\frac{k}{H}\right )^{ i\omega /H -1/2} \hat a_\bk, 
\end{equation} 
which follows from the Fourier transforms Eqs.~\eqref{eq:operatorphihomo} and~\eqref{eq:operatorphiunruh}, one verifies that the independence of $k$ implies that of $\omega$. As a result, introducing $V_\omega^\dagger = \left ( \hat a_U^{\dagger \omega}, \hat a_U^{-\omega} ,\hat a_V^{\dagger \omega}, \hat a_V^{-\omega} \right )$, the covariance matrix of Unruh operators reads
\begin{equation}
\label{eq:nucudef}
\begin{split}
C&\doteq \left <\left \{ V_\omega \otimes V_{\omega'}^\dagger \right \} \right > \\
&= \delta(\omega - \omega') \times \left [ 2 \left (
\begin{array}{cccc}
n_k & 0 & 0 & c_k \\
0 & n_k & c^*_k &0 \\
0 & c_k & n_k &0 \\
c^*_k & 0 & 0 & n_{k}
\end{array} \right )+{1 }\right ], 
\end{split}
\end{equation}
where $n_k$ and $c_k$ are given in Eq.~\eqref{eq:nkckdef}.

Using the matrix $B_\omega$ of Eq.~\eqref{eq:changeofbase}, and dropping the trivial factor of $\delta(\omega - \omega') $, the covariance matrix of $R$ and $L$ operators is 
\begin{equation}
\begin{split}
C^{RL}_\omega &= B_\omega C B_\omega^\dagger = 2 \left (
\begin{array}{cccc}
n_\omega & c_\omega & m_\omega^* & c_\omega^{UV} \\
c_\omega^* & n_\omega & c_\omega^{UV *} & m_\omega \\
m_\omega & c_\omega^{UV} & n_\omega & c_\omega \\
 c_\omega^{UV *} & m_\omega^* & c_\omega^* & n_\omega 
\end{array} \right )+{1 } , 
\end{split}
\end{equation}
where
\begin{subequations}
\label{eq:nucu}
\begin{align}
2 n_\omega +1 &= \left ( \abs{\alpha_\omega^{H}}^2 + \abs{\beta_\omega^{H} }^2 \right) \left (2 n_k+1\right ), \label{eq:nom} \\
c_\omega&= \alpha_\omega^{H} (\beta_\omega^{H})^* \left (2 n_k+1 \right ) , \label{eq:com} \\
m_\omega &=2 {\rm Re}\left( c_k \alpha_\omega^{H} \beta_\omega^{H} \right), \\
2 c^{UV}_\omega &= (\alpha_\omega^{H} )^2c_k + \left [ (\beta_\omega^{H} )^{2}c_k\right ]^*.
\end{align}
\end{subequations}
The first two coefficients concern separately either the $U$, or the $V$-modes. They fix the spectrum and the strength of the correlations. The last two concern the $U-V$ mode mixing, and are proportional to $c_k$.

\begin{SCfigure}
\includegraphics[width=0.5\linewidth]{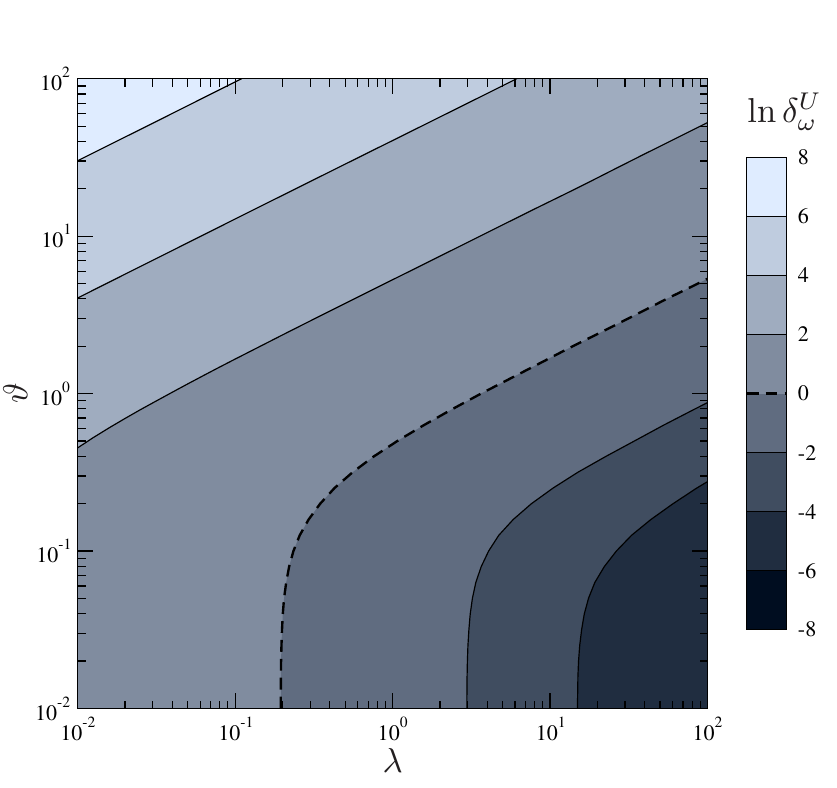}
\caption{\label{fig:massless-n} \small
Figure for $\delta_\omega^U$ with $\omega = H$ for massless field with critical coupling $g=g_{\rm crit}$. In the infalling vacuum, for $T_\psi = 0$, the nonseparability found for the massless relativistic case is preserved as long as $\Lambda/H = \lambda \gtrsim 1/5$. When the environment is characterized by a temperature $T_\psi \neq 0$, the entanglement is preserved as long as $T_\psi \lesssim \sqrt{H \Lambda}/ 2$, as explained in the text.}
\end{SCfigure}

Considering the coherence amongst pairs of $U$-quanta, i.e., ignoring the $V$-modes, as in \ref{sec:homogen}, we use the parameter $\delta$ of Eq.~\eqref{eq:defdeltacampo} 
\begin{equation}
\delta_U^\omega = n_\omega +1 - \abs{c_\omega}^2/n_\omega.
\end{equation}
Using Eq.~\eqref{eq:nucu}, we obtain 
\begin{equation}
\delta_U^\omega = \frac{n_k(n_k+1)}{\left ( |\alpha_\omega^{H}|^2 + |\beta_\omega^{H}|^2 \right) n_k + |\beta_\omega^{H}|^2 }.
\label{eq:dU}
\end{equation}
We see that $\delta_U$ does not depend on $c_k$. This is to be expected since $c_k$ characterizes the correlation between modes of opposite momenta, and since there is no $U-V$ mode mixing for two-dimensional massless fields. More importantly, Eq.~\eqref{eq:dU} is valid irrespectively of the temperature of the environment $T_\psi$. We can thus study how the separability of $U$-quanta is affected by $T_\psi$. The criterion of nonseparability, $\delta_U^\omega < 1$, gives 
\begin{equation}
|\beta_\omega^{H}|^2 = \frac{1}{\ep{\omega/T_{\rm H}} - 1}> \frac{n^2_k(T_\psi)}{2 n_k(T_\psi) +1},
\label{eq:dUT}
\end{equation}
where $n_k(T_\psi)$ is plotted in Fig.~\ref{fig:massless-T}. Using this Figure, in Fig.~\ref{fig:massless-n} we study $\ln \delta_\omega^U$ with $\omega = H$ as a function of $\lambda$ and $\vartheta = T_\psi/H$. At zero temperature $T_\psi = 0$, we see that the pair of $U$-quanta with $\omega = H$ is nonseparable for $\lambda \gtrsim 0.2$, i.e., for a rather strong dissipation since $\Lambda = H/5$. Using Eq.~\eqref{eq:dUT} we see that this is also true for all quanta with $\omega/ H \lesssim 1$. More surprisingly, when $\lambda$ is high enough, this pair is nonseparable {\it even} when $T_\psi > T_H$, i.e., when the environment possesses a temperature higher than the Gibbons-Hawking temperature. Indeed, whenever $T_\psi \lesssim \sqrt{H \Lambda}/2$, the pair is nonseparable, as all pairs with smaller frequency $\omega$.

In other words the quantum entanglement of the low frequency $U$ pairs of quanta is extremely {robust} when working with dissipative fields which are relativistic in the infrared. The robustness essentially follows from the kinematical character of the transformation of Eq.~\eqref{eq:changeofbase} which relates two {\it relativistic} mode bases. It is also due to the fact that $n_k$, the number of $U-V$ pairs created by the cosmological expansion, remains negligible in Eq.~\eqref{eq:nucu} as long as $ 1 \ll \Lambda/H $, and $T_\psi \ll T_H (\Lambda/H)^{1/2} $.

\section*{Conclusions}
\addcontentsline{toc}{section}{Conclusions}

In this chapter we used (a two-dimensional reduction of) the dissipative model introduced in Ref.~\cite{Parentani:2007uq} to compute the spectral properties and the correlations of pairs produced in an expanding de Sitter space. The terms encoding dissipation in Eq.~\eqref{eq:covaction} breaks the (local) Lorentz invariance in the ultraviolet sector. Yet, they are introduced in a covariant manner by using a unit timelike vector field $u$ which specifies the preferred frame. In addition, the unitarity of the theory is preserved by coupling the radiation field $\phi$ to an environmental field $\psi$ composed of a dense set of degrees of freedom taken, for simplicity, at rest with respect to the $u$ field. Again for simplicity, the action is quadratic in $\phi,\psi$, and the spectral density of $\psi$ modes is such that the (exact) retarded Green function of $\phi$ obeys a local differential equation, see Eq.~\eqref{eq:boxdiss} and Eq.~\eqref{eq:locEq}.
 
By exploiting the homogeneous character of the settings, we expressed the final occupation number $n_k$, and the pair-correlation amplitude $c_k$, in terms of the noise kernel $N$ and the retarded Green function, see Eq.~\eqref{eq:n_k,c_k}. Rather than working with integrals over time as usually done, we used the proper momentum $P = k /a(t)$ to parametrize the evolution of field configurations. Hence, Eq.~\eqref{eq:n_k,c_k} can be viewed as flow equations in physical momentum space. 

To precise the analysis of Ref.~\cite{Adamek:2008mp} where consequence of dissipation on inflation is explicitly studied, we numerically computed $n_k$ and $c_k$ in \ref{sec:homogen}. When considering a massless field, $n_k$ and the strength of the correlations are plotted as functions of the scale separation $\Lambda/H$, and the temperature of the environment $T_\psi/H$, in Fig.~\ref{fig:massless-T}. The robustness of the relativistic results is established in the limit of a large ratio $\Lambda/H$. The key result concerns the threshold values of the parameters, see the locus $\delta_k = 1$ on the right panel, for which the final state remains nonseparable, i.e., so entangled that it cannot be described by a stochastic ensemble. This analysis was then extended to massive fields, see Fig.~\ref{fig:massive-T}, and to the consequences of varying the relative importance of dissipative and dispersive effects, see Fig.~\ref{fig:massive-vac}. As expected, the quantum coherence is lost at high coupling, and when the temperature of the environment is high enough. 

In \ref{sec:statio} we exploited the stationarity, and we studied how the thermal distribution characterizing the Gibbons-Hawking effect is affected by dissipation. As in the case of dispersion, see \ref{chap:dispdS}, we found that the thermal character is, to leading order, robust. We also computed the deviations of the effective temperature with respect to the standard one $T_\mathrm{H} = H/2\pi$, see Figs.~\ref{fig:Tofom} and~\ref{fig:logToflambda}. In preparation for the analysis of the Hawking effect, we studied the strength of the asymptotic correlations across the Killing horizon between (right) moving quanta with opposite frequency. Quite remarkably, we found that the pairs remain entangled (the two-mode state remains nonseparable) even for an environment temperature exceeding $T_\mathrm{H} = H/2\pi$, see Fig~\ref{fig:massless-n}. 

In conclusion, even though our results have been derived in $1+1$ dimensions, we believe that very similar results hold in four dimensions, at least for homogeneous cosmological metrics and for spherically symmetric ones, because a change of the dimensionality only affects the low-momentum mode propagation. Hence even if this introduces nontrivial modifications, as grey body factors in black hole metrics, they will not interfere with the high-momentum dissipative effects when the hierarchy of scales $\Lambda/H , \Lambda/\kappa\gg 1$ is found. They can thus be computed separately.

\chapter{Black hole-de Sitter correspondence}
\label{chap:BHdesitter}

\section*{Introduction}

At each point of space-time, because of quantum effects, the vacuum contains pairs of virtual particle-antiparticle. Most of the time, these never become real particles. In some cases however, some energy can be used to convert this virtual pair into a real one. At the neighborhood of a black hole horizon, it can happen that one of these virtual particle fall into the black hole and the other one reaches infinity without being annihilated. The flux of such particles is the Hawking radiation. 

In this chapter, we consider a quantum field theory in a $1+1$ dimensional space-time containing a black hole. We show that many aspects of the radiation escaping it are similar to those found in de Sitter space. We show that the deviations to the Spectrum of the flux is governed by the width of the near horizon region where space time looks like de Sitter. 

\minitoc
\vfill

\section{Black hole space time}
\label{sec:BHspacetime}

\subsection{Generalities}

We here consider the background for a general Lorentz violating QFT in a stationary space time in 1+1 dimensions. We thus have two privileged vector fields in this space time, namely $K_t$ (we suppose it is directed towards the future) the stationary Killing vector field and $u$ the unit timelike vector field generating dispersion and dissipation. As in the previous chapters, we suppose that $u$ has the same symmetries as the space-time, which means that $u$ commutes with $K_t$. Let $s$ be the unit spacelike vector field orthogonal to $u$. This means that the metric reads $g_{\mu \nu} = - u_\mu u_\nu + s_\mu s_\nu$. Because both the metric and $u$ are stationary, so is $s$. $K_t$ and $s$ then form a basis of vectors that commute. There hence exists (locally) a coordinate system ($\tau,X$) such that $K_t = \partial_\tau$ and $s = \partial_X$. By defining the two functions of $X$, $c \doteq K_t^\mu u_\mu$ and $v \doteq - K_t^\mu s_\mu$, one can show that all scalar quantities constructed using $K_t,g,u$ and $s$ only depend on $v,c$ and their derivatives ($v'= s^\mu \partial_\mu v, c'= s^\mu \partial_\mu c, \dots$). The link between the $u$ and the Killing field is
\begin{equation}
\begin{split}
u = \frac{-1}{c} (K_t + v s),
\end{split}
\end{equation} 
and the metric takes a form much similar to the rhs of Eq.~\eqref{eq:dSmetricintwoways}
\begin{equation}
\label{eq:PGBHaccel}
\begin{split}
ds^2 = - c(X)^2 d\tau^2 + (dX - v(X) d\tau)^2.
\end{split}
\end{equation}
This shows that the two functions $v$ and $c$ completely fix the space-time and the preferred frame. Moreover, the accelerations of $u,s$ are given by
\begin{equation}
\begin{split}
\nabla_\mu u_\nu = -\frac{c'}{c} u_\mu s_\nu - \frac{v'}{c} s_\mu s_\nu ,\quad \nabla_\mu s_\nu = -\frac{c'}{c} u_\mu u_\nu - \frac{v'}{c} s_\mu u_\nu.
\end{split}
\end{equation}
And at the level of the algebra, one verifies immediately that
\begin{equation}
\label{eq:commutusBH}
\begin{split}
[u,s] = \frac{c'}{c} u + \frac{v'}{c} s.
\end{split}
\end{equation}
On the other hand, the norm of the Killing field reads $K_t^2 = -c^2 +v^2$. When $\abs{v}$ crosses $c$, the Killing field thus becomes space-like. This is a Killing horizon. It separates the subluminal region $\abs{v}<c $ from the superluminal one $\abs{v}>c$. An important notion in the presence of Killing horizon is the notion of surface gravity. It is defined by the relation $\kappa K_t^\mu = K_t^\nu \nabla_\nu K_t^\mu $ at the horizon. Using our functions $v$ and $c$, it reads
\begin{equation}
\label{eq:defkappa}
\begin{split}
\kappa = \frac{\partial_X \left (v^2 -c^2 \right ) }{2 v}\vert_{\rm horizon}.
\end{split}
\end{equation}
We recover the expression of the surface gravity used in condensed matter models~\cite{Barcelo:2005fc,Macher:2009nz}, and generalize~\cite{Jacobson:2008cx}. In the presence of an horizon situated at $X= X_H$, to simplify notations, we shift the axes and define $x \doteq X-X_H$ so that the horizon is at $x=0$. When $\kappa>0$, this horizon is the boundary of a black hole: no lightlike curve can escape the superluminal region. When $\kappa <0$, the horizon is the boundary of a white hole. Indeed, close to the horizon, the lightlike geodesics are $x = 2 {\rm sgn}(v) t$ for co-propagating modes and $x = x_0 \ep{\kappa t}$ for counter-propagating modes.

\subsection{Freely falling frame}

To get a situation which is closer to that of \ref{sec:dSspace}, we now assume that the preferred frame is freely falling, i.e., $ u^\mu D_\mu u^\nu =0$. This implies that $c$ is a constant and by a redefinition of $K_t$, we chose it to be $c=1$. Then, assuming that the flow is moving to the left, the velocity profile is in the near horizon region (NHR) $v \sim -1 + \kappa x $

The important lesson of the previous section is that under the assumptions of stationarity and freely fallingness, the black hole metric and the preferred frame are completely, and invariantly, determined by $v(x)$. Since the de Sitter background fields of \ref{sec:desitterspace} can be described by the same settings with the extra condition that $v_{dS}$ is linear in $x$, the comparison of dispersive effects associated with a given dispersion relation can be easily done for the solutions of both Eq.~\eqref{eq:ommodeeq} and its Hamilton Jacobi corresponding equation. In particular, we can already predict that the {\it deviations} between de Sitter and the black hole case will be governed by the spatial extension of the black hole NHR where $v$ is approximatively linear in $x$. When 
\begin{equation}
\label{eq:vD}
\begin{split}
v= -1 + D \tanh(\kappa x/D),
\end{split}
\end{equation}
the extension is, roughly speaking, given by $\vert \kappa x \vert = D$.~\footnote{
Even though the parameter $D$ plays no role when computing the Hawking spectrum using relativistic fields, it plays important roles in black hole physics. First, the deviations with respect to the Hawking spectrum due to dispersion are governed by $D$~\cite{Macher:2009tw,Finazzi:2010yq,Coutant:2011in}. Second, the nonlocal correlations across a black hole horizon are also governed by $D$, in that they start to differ from vacuum correlations when $\kappa x \sim D$~\cite{Schutzhold:2010ig,Parentani:2010bn}}
Using this velocity profile, near the horizon, Eq.~\eqref{eq:commutusBH} is given by
\begin{equation}
[u,s] = v'(x) s= \kappa s \left(1- \frac{(\kappa x)^2 }{ D^2} + O(\kappa x)^4\right) ,
\label{eq:usnhr}
\end{equation}
which clearly shows that the deviations with respect to Eq.~\eqref{eq:usc} are governed by $\kappa x/D$. 

\section{Correspondence in the absence of dissipation}
\label{sec:BHdScorrdisp}

We now show that this correspondence is not limited to the background fields, but extends to the {\it dynamics} of dispersive fields. We shall show this first at the classical level (WKB) and then for the quantum field theory. 

\subsection{The characteristics}

\begin{SCfigure}
\includegraphics[width=0.5\linewidth]{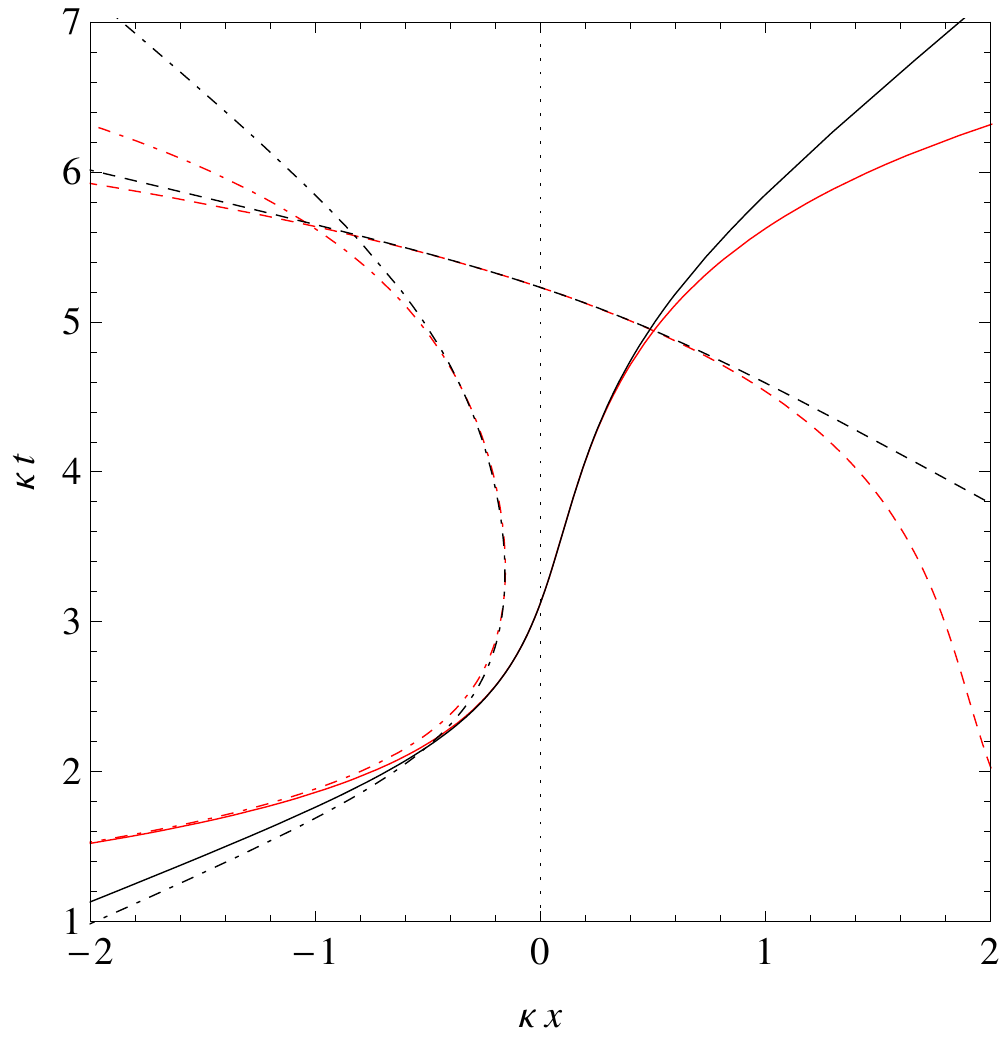}
\caption{The characteristics of a field in the same frame as in Fig.~\ref{fig:caracteristics}. The dispersion relation is superluminal and we draw the characteristics both in the black hole profile of Eq.~\eqref{eq:vD} (black) and in de Sitter (red). Values of the parameters are $D=0.8$, $\Lambda/ \kappa = 30$, $\omega = \kappa$. In dashed, we draw the $V$ mode. In solid lines is represented the $U$ mode of positive frequency. In dot-dash, we show the negative energy mode that is trapped into the black hole. Its turning point corresponds to the place  where the WKB approximation fails. In dotted is represented the horizon $x=0$.}
\label{fig:caracBHdS}
\end{SCfigure}

At the classical level, the correspondence is most clearly seen by considering Hamilton's equations. In particular, irrespective of the choice of $f$ in Eq.~\eqref{eq:disprel}, the time derivative of the momentum $\bP=\partial_x S = s^\mu \partial_\mu S$ obeys
\begin{equation}
\frac{d\bP}{d\tau} = - \frac{1}{\partial_\omega x_\omega(\bP)} = -\bP v'(x_\omega(\bP)) ,
\label{eq:ptau}
\end{equation}
where $x_\omega(\bP)$ is the root of Eq.~\eqref{eq:disprel} at fixed $\omega$, i.e., with $\Omega$ expressed as $\Omega = \omega - v(x) \bP$. We learn here that Eq.~\eqref{eq:ptau} is the dynamical equivalent of Eq.~\eqref{eq:usnhr}. This establishes how the preferred frame algebra imprints the particle's dynamics. Having understood that, as long as $\kappa x \ll D$, Eq.~\eqref{eq:ptau} and Eq.~\eqref{eq:usnhr} guarantee that $\bP$ obeys 
\begin{equation}
\bP(\tau) = \bP_0 \ep{- \kappa \tau}, 
\end{equation}
as in the de Sitter cosmology where $\bP = \bk/a(t)$. It is worth pointing out that this exponential redshift applies for both signs of $\bP$, i.e., for both right and left moving solutions. This correspondence in $P$-space also applies to the classical trajectories in $x$-space. At fixed $\omega$, $x(\tau)$ obeys $\omega - v(x) \bP= \pm F(P)$, where $\bP(\tau)$ is the solution Eq.~\eqref{eq:ptau}, and where $+$ ($-$) describes right moving trajectories. As long as $v \sim -1 + \kappa x$ furnishes a good description of $v$, the dispersive trajectories $x_\omega(\tau)$ in the black hole metric are indistinguishable from those in de Sitter, i.e., $x_\omega(\tau)$ is the same function as $X^{dS}_\omega(t) - 1/H$ for $H = \kappa$ and $t = \tau$, see Fig.~\ref{fig:caracBHdS}.

\subsection{At the level of fields}

The correspondence further extends to dispersive field because the stationary modes $\phi_\omega$ still (exactly) obey Eq.~\eqref{eq:ommodeeq} in the black hole case. Therefore, near the Killing horizon, the black hole Fourier modes $\tilde \phi_\omega(\bP)$ factorize as in Eq.~\eqref{eq:facto}, where $\chi$ will obey Eq.~\eqref{eq:Pmodeeq} with $H = \kappa$. At this point we make two observations. First, Eq.~\eqref{eq:Pmodeeq} resulted in de Sitter from the coexistence of $K_t$ and $K_z$, and their algebra of Eq.~\eqref{eq:alg}. Second, Eq.~\eqref{eq:Pmodeeq} was used in all analytical treatments of the scattering of dispersive modes on a black hole horizon~\cite{Brout:1995wp,Corley:1996ar,Balbinot:2006ua,Unruh:2004zk,Coutant:2011in}. These observations raise several questions: 
\begin{itemize}

\item What is the relevance of this correspondence for the $S$ matrix ? 

\item What is the validity domain of this correspondence in terms of time lapses ?

\item Can we define a field $K_z$ which is approximatively Killing near the horizon ? 

\end{itemize}
The first question is certainly the most important one. As shown in Refs.~\cite{Coutant:2011in,Finazzi:2010yq,Finazzi:2012iu}, in the black hole case, when $\Lambda/\kappa \gg 1$, the {\it leading deviations} from the Planck spectrum at the standard Hawking temperature are governed by inverse powers of the parameter $D$ which enters in Eq.~\eqref{eq:usnhr}. This means that these deviations are in fact defined with respect to the corresponding dispersive spectrum evaluated in de Sitter space. This is perfectly coherent because in de Sitter, the deviations due to dispersion with respect to the relativistic spectrum are very small, see \ref{sec:betacosmo} and \ref{sec:Smatrix}, much smaller than those of the black hole case. In brief, this explains why the parameter $D$ of Eq.~\eqref{eq:usnhr}, which governs the extension of the black hole near horizon region which can be mapped in de Sitter, also governs the spectral deviations of the black hole flux. 

Concerning the second question, as far as space is concerned, the validity range of the linearized expression of $v$ around $K_t^2 = 0$ is trivially fixed by $D$. What is less trivial concerns the lapse of time during which this linearized expression can be used, given the dispersion relation of Eq.~\eqref{eq:disprel}. It is at this level that the separation between the background scale $\kappa \sim \omega$ and the dispersive scale $\Lambda$ enters. When $\Lambda/\omega \gg 1 $, the lapse of time during which the right moving $U$-particles of frequency $\omega$ stay in the NHR scales, for quartic dispersion, as $\kappa \Delta \tau \sim \log(D^{3/2}\Lambda/\omega)$. 

To show this result, one uses the fact that in the limit $\Lambda/\omega \gg 1 $ the solutions of the dispersion relation at $\kappa x = D$ are $\bP \sim \pm \Lambda \sqrt{D}$ for the incoming modes and $\bP \sim \pm \omega /D$ for the outgoing modes. Because in between the wave is in the NHR, $\bP \sim \bP_0 \ep{-\kappa t}$ applies. Making the ratio of the two expressions gives the accumulated redshift from the high initial momentum till the final one. It scales as $\bP_{in}/\bP_{out} \sim \ep{\kappa \Delta \tau }\sim \omega_{\max}^{\rm disp}/\omega$, where
\begin{equation}
\label{eq:omegamaxdisp}
\begin{split}
\omega_{\max}^{\rm disp} = D^{3/2}\Lambda.
\end{split}
\end{equation}
We see that it combines in a nontrivial manner the scale separation and the spatial extension of the NHR. In Ref.~\cite{Finazzi:2012iu} it was explicitly shown that $\kappa \Delta \tau $, the adimensional lapse of time spent in the de Sitter like region, governs the properties of the black hole spectrum.

Having clarified these issues, it is worth returning to geometrical aspects by investigating how a vector field $K_z = \partial_z$ can be introduced in black hole space-times and to what extent it could be considered as an \enquote{approximate Killing field}. It should be first pointed out that, a priori, there exist several ways to introduce a new coordinate $z$. Indeed, in de Sitter, $K^{dS}_z$ obeys several properties that can be used to define the vector field in the black hole case. For instance, the commutator $[u,K^{dS}_z]$ vanishes. Using this property to define $z$, one gets the construction of Ref.~\cite{Parentani:2007uq} that we used in \ref{sec:actionsetting} where the black hole metric reads $ds^2 = - d\tau^2 + a^2 dz^2$, with $a = v(x(\tau,z))/v(z) \sim \ep{-\kappa \tau}$ in the NHR. The disadvantage of this choice is that the lapse of time during which the exponential is found for the \enquote{scale factor} $a$ is much shorter than the lapse $\Delta \tau$ we above discussed. \textit{A posteriori}, it turns out that a better choice is provided by imposing that
Eq.~\eqref{eq:alg} be satisfied:
\begin{equation}
[K_t, K_z] = \kappa K_z .
\label{eq:2KT}
\end{equation}
This implies that $K_z \doteq \ep{ \kappa t} \partial_x $ is the derivative with respect to the new coordinate $z=x \ep{-\kappa t}$. We then have the following commutation relations 
\begin{equation}
[K_z,u] = (v'(x) -\kappa) K_z ,
\end{equation}
and 
\begin{equation}
D_\mu K_{z \nu} +D_\nu K_{z \mu} = \frac{v'(x)-\kappa}{2} (s_\mu u_\nu + u_\mu s_\nu) .
\end{equation}
Since $\kappa = v'(x=0)$, we see that the deviations from the Killingness, i.e., the second equation, and from the homogeneous de Sitterness, the first equation, are both governed by the gradient of $v'$ in the NHR, and not from $\kappa$ itself. It is thus geometrically meaningful, and dynamically relevant, to say that a stationary black hole metric endowed with a freely falling frame possesses, in the NHR, an approximate homogeneous Killing field $K_z$ obeying the affine algebra of Eq.~\eqref{eq:2KT}. 

\section{Analogy in the presence of dissipation}

We now explain when and why the results of \ref{sec:statio} apply to the Hawking radiation emitted by dissipative fields. Our main aim is to establish that the spectrum of Hawking radiation, and the associated long distance correlations across the horizon, are both robust when dissipation occurs at sufficiently high energy with respect to the surface gravity, as was anticipated in Refs.~\cite{Brout:1995wp,Parentani:2007uq}. 

The robustness shall be established by studying the anticommutator of Eq.~\eqref{eq:acgrennfunction}, and showing that its asymptotic behavior is governed by Eqs.~\eqref{eq:nom} and~\eqref{eq:com}.

As in de Sitter case, we shall assume that the environment has the same symmetries as space-time, i.e., is stationary.
This implies that the noise kernel of Eq.~\eqref{eq:disskernel} only depends on $\tau -\tau'$ when evaluated at $x, x'$, along the orbits of the Killing field $K_t$. When these stationary conditions are met, the (driven part of the) anticommutator of $\phi$ is (exactly) given by 
\begin{equation}
\label{eq:GacomXX}
\begin{split}
G^\omega_{\rm ac}(x,x') =\iint dx_1 dx_2 G^\omega_{\rm ret}(x,x_1) [G^\omega_{\rm ret}(x',x_2)]^* N^\omega(x_1,x_2)
\end{split}
\end{equation}
where the two kernels $G^\omega_{\rm ret}$ and $N^\omega$ are now defined in the black hole metric of Eq.~\eqref{eq:PGBHaccel} with $c=1$. 

As a result, to compare the expressions of $G^\omega_{\rm ac}(x,x')$ evaluated in de Sitter and in Eq.~\eqref{eq:PGBHaccel}, it is sufficient to study $G^\omega_{\rm ret}$ and $N^\omega$. To establish the correspondence with controlled approximations, the following four conditions are necessary:
\begin{itemize}
\item the state of $\psi$ should be the same
\item the black hole surface gravity $\kappa = H$
\item the near horizon region should be large enough
\item the dispersive and dissipative scales should both be much larger than $\kappa \sim \omega$. 
\end{itemize}
The first condition is rather obvious and needs no justification. The second and the third conditions concern the metric and the $u$ field. They follow from \ref{sec:BHspacetime}. The fourth condition is a generalization of $\Lambda \gg \omega \sim \kappa $, see \ref{sec:BHdScorrdisp}.

\subsection{The stationary noise kernel}

When considering the model of Eq.~\eqref{eq:covaction} in the metric Eq.~\eqref{eq:PGBHaccel} with $u$ freely falling, the noise kernel $N^\omega$ of Eq.~\eqref{eq:GacomXX} is
\begin{equation}
\label{eq:Nbh}
N^\omega(x_1,x_2) = \gamma(-\partial_1 )\gamma(-\partial_2 ) \left ( -i\omega + \sqrt{v_1} \partial_1 \sqrt{v_1} \right ) \left ( i\omega + \sqrt{v_2} \partial_2 \sqrt{v_2} \right ) \int d\zeta G_{\mathrm{ac}, \psi}^\omega(x_1,x_2, \zeta),
\end{equation} 
where $v_i \doteq v(x_i)$ and $\partial_i \doteq \partial_{x_i}$. The stationary kernel in the integral is the Fourier transform of the anticommutator of $\psi$, see Eq.~\eqref{eq:NofDelta}. To compute it we use the fact that the factor $a(t,z)$ of
Eq.~\eqref{eq:Duincov} is now given by (see Eq.~(55) in Ref.~\cite{Parentani:2007uq} for a three-dimensional radial flow) 
\begin{equation}
\begin{split}
a(\tau ,x) = v(x)/v(z(\tau ,x) ).
\end{split}
\end{equation}
As in Eq.~\eqref{eq:Duincov}, $z$ labels the orbits of $u$. It is here completely fixed by the condition that $z =x$ when $t=0$. Since the orbits are solutions of $dx/dt = v$, $z$ is implicitly given by
\begin{equation}
\begin{split}
\int_{z}^{x} \frac{d x_1}{ v(x_1)} &= t.
 \end{split}
\end{equation}
Using the above equations to re-express the $\delta(z- z')$ of Eq.~\eqref{eq:NofDelta}, one finds 
\begin{equation}
\begin{split}
G_{\mathrm{ac}, \psi}(\Delta \tau, x_1,x_2; \zeta) =& \frac{\delta(\Delta \tau- \int_{x_2}^{x_1} {dx}/{v}) }{\sqrt{v_1 v_2}} \frac{2n_\zeta+1}{\omega_\zeta} \cos\left (\omega_\zeta \Delta t\right ) .
\end{split}
\end{equation}
Its Fourier component with respect to $\Delta \tau$ is trivially 
\begin{equation}
\begin{split}
G_{\mathrm{ac}, \psi}^\omega (x_1,x_2; \zeta) =& \frac{ \ep{i \omega \Delta t_{12} } }{\sqrt{v_1 v_2}} \frac{2n_\zeta+1}{\omega_\zeta} \cos\left (\omega_\zeta \Delta t_{12} \right ), 
\end{split}
\end{equation}
where $\Delta t_{12} = \int_{x_2}^{x_1} {dx}/{v} $ is the lapse of time from $x_2$ to $x_1$ following an orbit $z=cst$ which connects these two points. Since the settings are stationary, these orbits are all the same, as can be seen in Fig.~\ref{fig:caracter}. 

Using Eq.~\eqref{eq:Nbh}, the noise kernel is explicitly given by
\begin{equation}
\begin{split}
N^\omega(x_1,x_2) = &\gamma(-\partial_{1})\gamma(-\partial_{2}) \frac{ \ep{i \omega \Delta t_{12} }}{\sqrt{v_1 v_2}} \int d\zeta {(2n_\zeta+1)} {\omega_\zeta} \cos\left (\omega_\zeta \Delta t_{12} \right ) .
\end{split}
\end{equation}
This kernel is local in that it only depends on $v$ (or equivalently on $\mathsf{g}_{\mu \nu}$ and $u^\mu$) between $x_1$ and $x_2$. Hence, when evaluated in the black hole NHR, it agrees, {\it as an identity}, with the corresponding expression evaluated in de Sitter. 

In conclusion, we notice that this identity follows from our choice of the action of Eq.~\eqref{eq:covaction}. Had we used a more complicated environment, this identity would have been replaced by an approximative correspondence. In that case, the correspondence would have still been accurate if the propagation of $\psi$ had been {adiabatic}. As usual, this condition is satisfied when the degrees of freedom of $\psi$ are \enquote{heavy}, i.e., when their frequency $\omega_\zeta \sim \Lambda \gg \kappa$.

\subsection[The stationary \texorpdfstring{${G_{\rm ret}^\omega}$}{Gret(omega)}]{The stationary \texorpdfstring{$\boldsymbol{G_{\rm ret}^\omega}$}{Gret(omega)}}

The stationary function $G_{\rm ret}^\omega(x,x_1)$ obeys Eq.~\eqref{eq:locEq}, which is a fourth order equation in $\partial_x$ when working with Eq.~\eqref{eq:numdispersion}. Depending on the position of $x$ and $x_1$, its behavior should be analyzed using different techniques. Far away from the horizon, the propagation is well described by WKB techniques since the gradient of $v$ is small. Close to the horizon instead, the WKB approximation fails, as in dispersive theories~\cite{Coutant:2011in}. In this region, the $P$ representation accurately describes the field propagation, and is essentially the same as that taking place in de Sitter. Therefore, the calculation of $G_{\rm ac}^\omega(x,x')$ of Eq.~\eqref{eq:GacomXX} at large distances boils down to connecting the de Sitter--like outcome at high $P$ to the low-momentum WKB modes. As in the case of dispersive fields, the connection entails an inverse Fourier transform from $P$ to $x$ space in the intermediate region \textit{II}, see Fig.~\ref{fig:caracter}, where both descriptions are valid~\cite{Brout:1995wp,Corley:1997pr,Unruh:2004zk,Balbinot:2006ua,Coutant:2011in}. In the present case, these steps are performed at the level of the two-point function rather than being applied to stationary modes. In fact, we shall compute $G_{\rm ac}^\omega$ through 
\begin{equation}
\begin{split}
\label{eq:GacXXP}
G_{\rm ac}^\omega(x,x') &= \int \int^\infty_{-\infty} d\bP_1 d\bP_2 G_{\rm ret}^\omega(x,\bP_1) G_{\rm ret}^{\omega *}(x',\bP_2) N^\omega(\bP_1,\bP_2) ,
\end{split}
\end{equation}
where the two $G_{\rm ret}^\omega$ are expressed in a mixed $x,P$ representation. The early configurations in interaction with the environment are described in $P$ space, while the large distance behavior is expressed in $x$ space.

\subsubsection{Relevant range for \texorpdfstring{$\bP_i$}{Pi}}

Let us give here only the essential points, more details are given in \ref{sec:BHdS}. The validity of the whole procedure relies on a combination of the third and the fourth condition given above, namely ${\rm max}(1, D^{-2}) \ll \Lambda/\kappa $, and is limited to moderate frequencies, i.e., $ 0 < \omega \sim \kappa \ll \Lambda$. 

For simplicity, we consider massless fields. Then $\Lambda/\kappa \gg 1$ guarantees that the infalling $V$ modes essentially decouple from the outgoing $U$ modes because the only source of $U-V$ mixing comes from the ultraviolet sector. Hence, at leading order in $\kappa/\Lambda$, it is legitimate to consider only the $U$ modes. For massive fields with $m \ll \Lambda$, the discussion is more elaborate but the main conclusion is the same: the properties of the Hawking radiation are robust.

For massless fields, at fixed $\omega$, the propagation of the $U$ modes is governed by the effective dispersion relation, see Eq.~\eqref{eq:eom-transformed},
\begin{equation}
\label{eq:HJeqU}
\Omega = \omega - v(x) P = \sqrt{F^2- \Gamma^2} .
\end{equation} 
As long as $P \ll \Lambda$, the $U$ sector of $G_{\rm ret}^\omega$ behaves as for a relativistic field, since $\sqrt{F^2 - \Gamma^2}\sim P (1 + O(P/\Lambda))$. Instead, when $P \gtrsim \Lambda$, the dispersive and dissipative terms weighted by $f$ and $\Gamma$ cannot be neglected in Eq.~\eqref{eq:locEq}. To characterize the transition from these two regimes, we consider the optical depth of Eq.~\eqref{eq:opticaldepth}. When working at fixed $\omega$, one finds 
\begin{equation}
\label{eq:optdepthBH}
\begin{split}
 \mathcal{I}_\omega(x,x_1) &= \int_{P}^{P_1} dP' \frac{\Gamma(P')}{ P' \partial_x v\left [ x_\omega(P') \right ]} = \int^{x}_{x_1} d x' \frac{\Gamma[P_\omega(x')]}{v^\omega_{\rm gr}(x') },
\end{split}
\end{equation}
where $x_\omega(P)$ is the root of~Eq.~\eqref{eq:HJeqU}, as is $P_\omega(x)$ when using $x$ as the variable [see \ref{sec:asympfluxWKB} for the origin of this expression]. The first expression governs $G_{\rm ret}$ in the NHR where $\partial_x v \sim \kappa$ is almost constant, see Eq.~\eqref{eq:Gret-solution}. To leading order in $\Gamma/P \ll 1$, which is satisfied everywhere but very close to the horizon, the second expression governs $G_{\rm ret}$ in $x$ space. Since $v_{\rm gr}^\omega= 1/\partial_\omega P $ is the group velocity in the rest frame, $\mathcal{I}_\omega = \int^t_{t_1} dt' \Gamma(P_\omega)$, where the integral is evaluated along the classical outgoing trajectory. It should be noticed that, when considered in $x$ space, $ \mathcal{I}_\omega$ applies on the right and the left of the horizon. In the R region, $v_{\rm gr} > 0$, while it is negative in L, so that in both cases $\mathcal{I}_\omega > 0$ when $P_1 > P> 0$, i.e., when $P_1$ is in the past of $P$. 

To characterize the retarded Green functions of Eq.~\eqref{eq:GacXXP}, we compute $\mathcal{I}_\omega$ in the mixed representation, in the limit where $P_1$ is large enough so that $x_\omega(P_1)$ is deep inside the NHR, while $x$ is far away from that region. For simplicity, we consider the case of Eq.~\eqref{eq:numdispersion} with $g=g_{\rm crit}$. In this case, only the dissipative effects are significant\footnote{
In the case where $g^2 \ll 1$, dispersive effects are important and may limit the role of dissipation in the NHR. In that case, the decaying part of the field will contribute to Eq.~\eqref{eq:Gacdef}. This situation corresponds to what is found in the surface wave experiments~\cite{Weinfurtner:2010nu,Rousseaux:2007is}.}, and one finds
\begin{equation}
\label{eq:optdepthBHlimit}
\begin{split}
 \mathcal{I}_\omega(x,P_1) \sim \frac{P_1^2}{2\kappa \Lambda} + \frac{\omega^2 |x| }{\Lambda |1+v_{R/L}|^3}, 
\end{split}
\end{equation}
where $v_R$ ($v_L$) is the asymptotic velocity on the right (left) side.
From the second term, we learn that $|\kappa x|$ should be much smaller than $\Lambda/\kappa$ for the Hawking quanta not to be dissipated. Since we work in the regime $\Lambda/\kappa \gg 1$, this condition is easily satisfied. We notice that a similar type of weak damping effect of outgoing modes has been observed in experiments~\cite{Weinfurtner:2010nu}.

From the first term, we learn that $\mathcal{I}_\omega$ gives an upper bound to the domain of $P$ which significantly contributes to Eq.~\eqref{eq:GacXXP}, namely $P^2 \lesssim \Lambda \kappa$, as in de Sitter. A lower bound of this domain is provided by the $\gamma$ factors of Eq.~\eqref{eq:noisekernel}. Using this equation and Eq.~\eqref{eq:prepomega}, the integrand of Eq.~\eqref{eq:GacXXP} scales as 
\begin{equation}
\label{eq:dominPT}
\begin{split}
T(P) \propto & P \Gamma(P) \ep{- 2 \mathcal{I} ( x , P ) } \propto P^3 \ep{-P^2/\Lambda \kappa},
\end{split}
\end{equation}
and its behavior is represented in Fig.~\ref{fig:velocity}. Hence, the relevant domain of $P$, i.e., when $T$ is larger than $10\%$ of its maximum value, scales as 
\begin{equation}
\label{eq:dominP}
\begin{split}
0.36 \sqrt{\kappa \Lambda} = P_{\rm min} \lesssim P \lesssim P_{\rm max} = 2.4 \sqrt{\kappa \Lambda}. 
\end{split}
\end{equation}

\subsubsection{Range of integration in direct space}

Considered in space-time, since $P \sim \ep{-\kappa t}$, this limits the lapse of time during which the coupling to $\psi$ occurs. Interestingly, this lapse is given by $\kappa \Delta t \approx 2$, i.e., two e-folds, irrespective of the value of $\Lambda/\kappa$, and that of $\omega$. It should be also stressed that nothing precise can be said about the domain of $x$, which significantly contributes because the $x$-WKB fails when $P$ is so large. One can simply say that it is roughly characterized by the interval $[- x_{\rm trans}, x_{\rm trans}]$, where $x_{\rm trans} = x_{\omega = \kappa}(P_{\rm min})$ is given by 
\begin{equation}
\label{eq:Xtrans} 
\kappa x_{\rm trans} \sim 3 \sqrt{\kappa / \Lambda}. 
\end{equation}

\begin{SCfigure}[2]
\includegraphics[width=0.45\columnwidth]{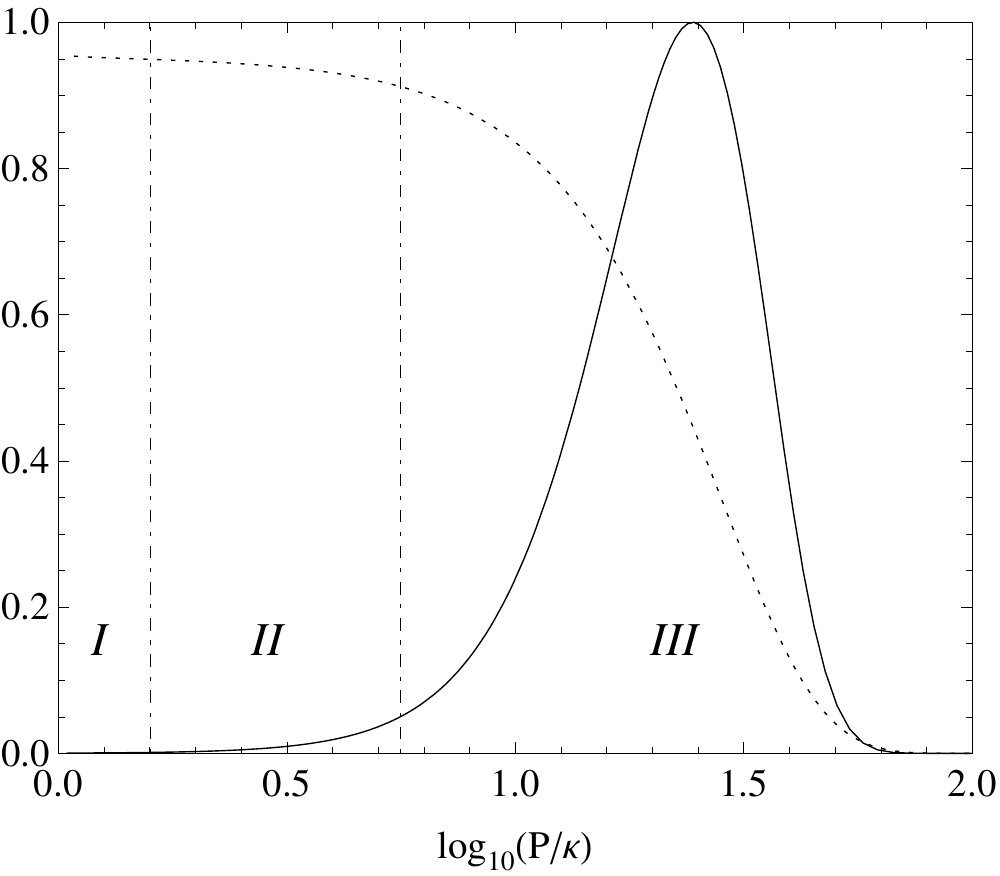}
\caption{\label{fig:velocity}
As a function of $\log_{10} (P/\kappa)$, in a dotted line we plot $\exp\{- \mathcal{I_\omega}(X,P)\}$, the optical depth of Eq.~\eqref{eq:optdepthBH}, evaluated for $\omega/\kappa = 1$, $\Lambda /\kappa = 400$, and $\kappa x = 20$. The solid curve represents $T(P)$ of Eq.~\eqref{eq:dominPT} for the same values, and $D = 0.99$. The left dash-dotted line corresponds to the limit of the NHR: $\kappa x = D/2$, here reexpressed as $P = 2\kappa / D $. For lower $P$, in region \textit{I}, the de Sitter--like $P$ representation fails. The right vertical line indicates the upper limit of the $x$-WKB approximation, see \ref{sec:BHdS}. For the adopted values, the region \textit{II} where the $P$ and the $x$ descriptions are both valid has a finite size. We also see that $T$ vanishes in region~\textit{I}. }
\end{SCfigure}

\begin{SCfigure}
\includegraphics[width=0.5\linewidth]{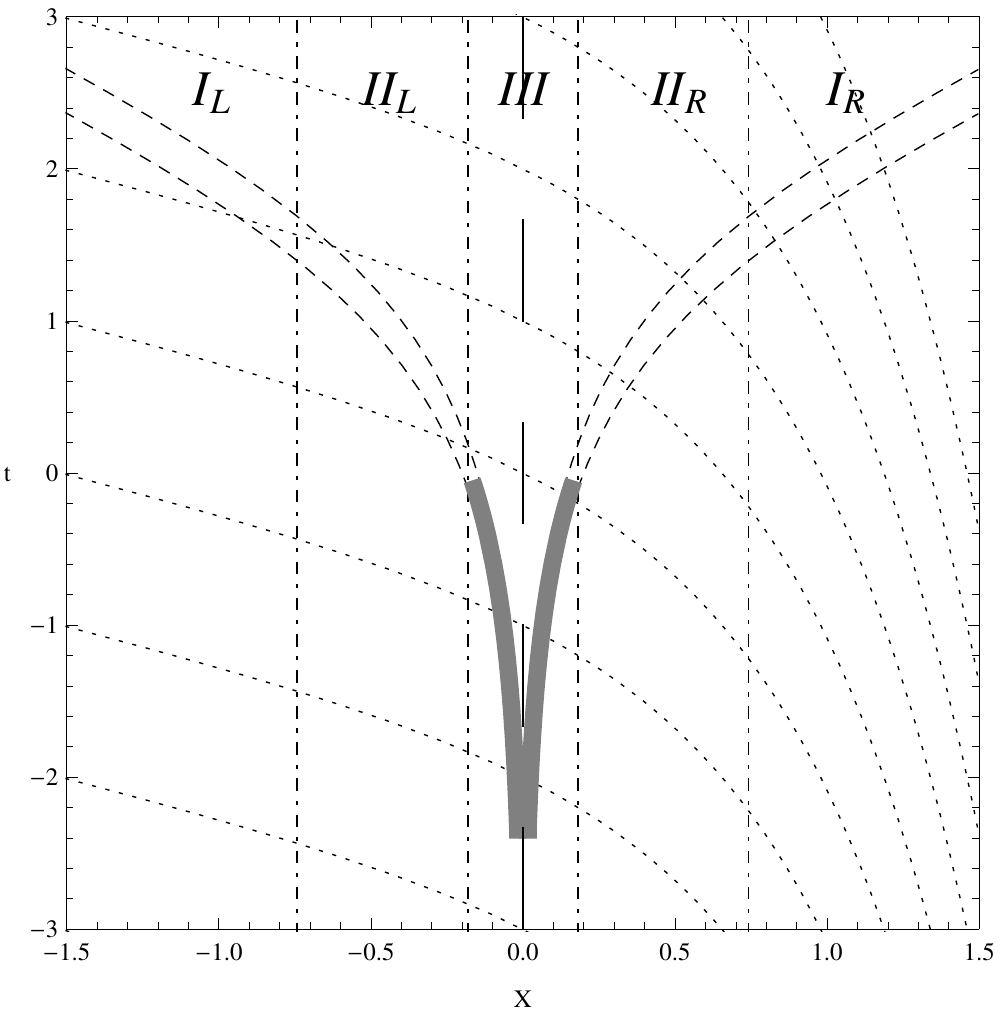}
\caption{\label{fig:caracter} \small
Null outgoing geodesics (dashed lines) on either side of the horizon at $x=0$, and freely falling orbits $z=\mathit{cst}.$ (dotted) in the $\tau,x$ coordinates of Eq.~\eqref{eq:PGBHaccel}. As explained in the text, the nearby geodesics schematically indicate the space-time region where $G_{\rm ac}(\tau,x,\tau_1,x_1)$ is nonvanishing, when $\kappa \tau_1=2.5$, and $\kappa x_1= 1.5$, see Fig.~1 of Ref.~\cite{Parentani:2010bn} for the relativistic case. The two thick solid lines represent the region where the noise kernel contributes to $G_{\rm ac}$, see Eq.~\eqref{eq:GacomXX} and Eq.~\eqref{eq:GacXXP}. In the central region \textit{III}, the propagation is well described in $P$ space, and resembles to that found in de~Sitter.}
\end{SCfigure}

This value defines the central region \textit{III}, see Figs.~\ref{fig:velocity} and~\ref{fig:caracter}. Using the profile of Eq.~\eqref{eq:vD}, $ x_{\rm trans}$ is situated deep inside the NHR when $\kappa / \omega_{\rm max}^{\rm diss} \ll 1$, where the critical frequency $\omega_{\rm max}^{\rm diss}$ is given by
\begin{equation}
\label{eq:ommdiss}
\begin{split}
\omega_{\rm max}^{\rm diss} = \Lambda D^2 .
\end{split}
\end{equation}

Hence, when $\kappa / \omega_{\rm max}^{\rm diss} \ll 1$, the coupling between $\phi$ and $\psi$ is accurately described in the $P$ representation, and takes place in a portion of de Sitter. In addition, the connection between the high- and low-momentum propagation can be safely done in the intermediate region \textit{II}, defined by $\kappa |x_{\rm trans}| \ll \kappa| x | \lesssim D $, see Fig.~\ref{fig:caracter}, where, on the one hand, one is still in a de Sitter--like space since $v$ is still linear in $X$, and, on the other hand, the low-momentum modes can be already well approximated by their WKB expressions. Notice finally that this reasoning only applies for frequencies $\omega \ll \omega_{\rm max}^{\rm diss} $. Indeed, when $\omega = \omega_{\rm max}^{\rm diss}$, dissipation occurs around $\kappa x \sim D$, i.e., no longer in a de Sitter like background. 

These steps are sufficient to establish that the results of \ref{sec:Asc} apply for $\omega \ll \omega_{\rm max}^{\rm diss}$. In particular, Eq.~\eqref{eq:nom} implies that the spectrum of radiation is robust (when the temperature of the environment is low enough, see Fig.~\ref{fig:massless-n}). Namely, to leading order in $\kappa/\Lambda$, the mean occupation number $n_\omega$ of quanta received far away is given by the Planck distribution at the standard relativistic temperature $T_\mathrm{H} = \kappa/2\pi$. As in dispersive settings, the real difficulty is to evaluate the spectral deviations. In this respect, we conjecture that the leading deviations due to dissipation will be suppressed by powers of $\kappa/\omega_{\rm max}^{\rm diss}$. That is, they will be governed by the composite ultraviolet scale of Eq.~\eqref{eq:ommdiss} which depends on the high-energy physics, here with $\Gamma$ quadratic in $P$, and on the extension $D$ of the black hole NHR. This second dependence is highly relevant when $D \ll 1$. 

Together with the robustness of the spectrum, one also has that of the long-distance correlations across the horizon between the Hawking quanta and their partners. These correlations are fixed by the coefficient $c_\omega$ of Eq.~\eqref{eq:com}. To get the space-time properties of the pattern, one should integrate over $\omega$, i.e., perform the inverse Fourier transform of Eq.~\eqref{eq:fouriert}, because it is this integral that introduces the space-time coherence~\cite{Massar:1995im,Brout:1995wp,Parentani:2010bn}. In Fig.~\ref{fig:caracter}, we have schematically represented the anticommutator $G_{\rm ac}(\tau-\tau_1,x,x_1)$ in the $\tau-\tau_1,x$ plane when $x_1$ is taken far away from the horizon.

\begin{itemize}

\item
Far away from the NHR, in regions $I_R$ and $I_L$, for $D \lesssim |\kappa x| $, the characteristics of the field follow null geodesics, see Eq.~\eqref{eq:retism}. Since $v \sim cst.$, they no longer separate from each other. Hence, at large distances, the space-time pattern obtained by fixing one point~\cite{Massar:1995im,Massar:1996tx}, and the equal time correlation pattern~\cite{Balbinot:2007de}, will be the same as those predicted by a relativistic treatment. 

\item
In the two intermediate regions \textit{II}$_R$ and \textit{II}$_L$, for $ x_{\rm trans} \lesssim |\kappa x| \lesssim D $, the characteristics separate from each other following $\delta x \sim \ep{\kappa t}$ since their behavior is already close to the relativistic one. This pattern is obtained by considering two-point functions with one point fixed, or wavepackets~\cite{Brout:1995wp}. It is interesting to notice that it cannot be obtained by considering equal time correlations, since these develop only outside the NHR, for $|\kappa x| \gtrsim D$~\cite{Parentani:2010bn}. Indeed as long as $x$ and $x'$ are in the NHR, the (approximate) de Sitter invariance under $K_z$, see \ref{sec:Prep}, implies that $\int d\omega G^\omega_{ac}$ only depends on $x-x'$.~\footnote{From this observation, we learn that the correspondence between the physics in black hole metrics and in de Sitter is not merely a convenient way to obtain $n_\omega$ and $c_\omega$ in \ref{sec:BHds:ds}. It actually shows up in the NHR when computing observables, such as the mean value or the two-point correlation of $\rho = u^\mu u^\nu T_{\mu \nu}$. Moreover, it ceases when leaving this region. In this sense, the Hawking effect only develops, or separates, from its de Sitter roots for $|\kappa x| \gtrsim D$, and furthermore, this separation is adiabatic.}
 
\item
The central region \textit{III} is the region where the configurations of the $\phi$ field are driven by the noise kernel. In Fig.~\ref{fig:caracter} the two thick solid lines indicate the space-time locus where the interactions involving the configurations selected by $\tau_1, x_1$ are taking place.~\footnote{
To get Fig.~\ref{fig:caracter}, we have filtered out low frequencies. This amounts to considering a wavepacket rather than the two-point function, see Eqs.~(38,39) in Ref.~\cite{Parentani:2010bn}. Had we considered $G_{\rm ac}$, the thick lines would have extended back in time, because the coupling between $\psi$ and $\hat\phi_\omega$ is centered around
$\kappa t_\omega = \ln \omega + cst.$, which fixes the blueshift for $P_\omega \propto \omega$ to reach $\sqrt{\kappa \Lambda}$, see Eq.~\eqref{eq:dominP}.} 
In this central region \textit{III}, the propagation is well described in $P$ space, and corresponds to that found in de Sitter, see Eq.~\eqref{eq:GcXPchiBH} and Eq.~\eqref{eq:ncac}.

\end{itemize}

In brief, when $\kappa/ \omega_{\rm max}^{\rm diss} \ll 1$ and $\omega/ \omega_{\rm max}^{\rm diss} \ll 1$, the nontrivial propagation only occurs deep inside the NHR which is a portion of de~Sitter space [more mathematical argument is given in \ref{sec:BHdS}]. This implies that $n_\omega$ and $c_\omega$ are, to a good approximation, given by their de~Sitter expressions of Eq.~\eqref{eq:nucu}. Given that these (exact) expressions hardly differ from the relativistic ones when $\kappa/\Lambda \ll 1$, we can predict that, when computed in a black hole metric, these two observables are robust whenever the finiteness of the NHR introduces small deviations with respect to the de Sitter case. For $\omega/ \omega_{\rm max}^{\rm diss} \ll 1$, this is guaranteed by $\kappa/ \omega_{\rm max}^{\rm diss} \ll 1$.

\section{Flux and long distance correlations} 
\label{sec:BHdS}

The expressions for the asymptotic flux and the correlation pattern are both encoded in Eq.~\eqref{eq:GacomXX}. To obtain them, we need two things. Firstly, we need to characterize $G_{\rm ret}^\omega$ from the asymptotic region down to the NHR. To this end, we should perform a WKB analysis of the stationary damped modes. Secondly, we need to connect the WKB modes with the high momentum de Sitter-like physics taking place very close to the horizon. 

\subsection{WKB analysis}
\label{sec:asympfluxWKB}

At fixed $\omega$, using Eq.~\eqref{eq:boxdiss}, $\Box_{\rm diss} \phi_{\rm dec} = 0$ implies that the decaying mode $\phi_\omega^{\rm dec}$ obeys 
\begin{equation}
\begin{split}
\bigg [ &\left ( i \omega - \partial_x v \right )\left ( i \omega -v \partial_x \right ) + F^2(-\partial_x^2) - \gamma(- \partial_x) (i \omega - \sqrt{v} \partial_x \sqrt{v}) \gamma( \partial_x) \bigg] \phi_\omega^{\rm dec} =0.
\label{eq:decmodeeq}
 \end{split}
\end{equation}
The mode $\phi_{\rm dec}^\omega$ decays when displacing $x$ along the direction of the group velocity. Hence, on the right of the horizon, the outgoing $U$-mode decays when $x$ increases, while it decreases for decreasing $x< 0$ in the left region, see Fig~\ref{fig:caracter}. Hence, $U$-modes spatially decay on both sides when leaving the horizon.

As in the case of dispersive fields~\cite{Coutant:2011in}, we look for solutions of Eq.~\eqref{eq:decmodeeq} of the form
\begin{equation}
\phi_\omega^{\rm dec}(x) = \ep{ i \int^x dx' Q_\omega(x') },
\end{equation}
where $Q_\omega(x)$ is expanded in powers of the gradient of $v(x)$. To first order, Eq.~\eqref{eq:decmodeeq} gives 
\begin{align}
(\omega - &v(x) Q_\omega + i\Gamma)^2 - (F^2 - \Gamma^2) = -\frac{i}{2} \partial_x \partial_Q \left [ (\omega - v(x) Q_\omega + i\Gamma)^2 - (F^2 - \Gamma^2) \right ],
\end{align}
where the functions $\Gamma > 0$ of Eq.~\eqref{eq:Gammaasgamma} and $F$ are evaluated for $P = Q_\omega$. The leading order solution, the complex momentum $Q^{(0)}_\omega(x) \doteq P^C_\omega(x)$, contains no gradient, and obeys the complex Hamilton-Jacobi equation
\begin{equation}
\label{eq:Q0}
\omega - v(x) P + i\Gamma( P)= \sqrt{F^2( P) - \Gamma^2( P)} \doteq \tilde F(P). 
\end{equation}
As expected, this equation gives Eq.~\eqref{eq:dispersion} since $\Omega = \omega - v P$. To first order in the gradient, we get a total derivative 
\begin{align}
\label{eq:Q1}
Q^{(1)}_\omega = \frac{i}{2} \partial_x \log &\bigg [ \frac{\tilde F(P_\omega^C)}{\partial_\omega P^C_\omega} \bigg ]. 
\end{align}
Combining Eq.~\eqref{eq:Q0} and Eq.~\eqref{eq:Q1}, we obtain the decaying WKB-mode
\begin{equation}
\phi_\omega^{\rm dec}(x) \approx \frac{\ep{ - \mathcal{I}_\omega(x,x_0) } \times \ep{i \int^x_{x_0} dx' P_\omega(x')} }{\sqrt{2 v^C_{\rm gr} \tilde F (P_\omega^C ) }}.
\label{eq:pW}
\end{equation}
To get this expression, we introduced $ v^C_{\rm gr} = 1/\partial_\omega P^C_{\omega} $ which can be conceived as a complex group velocity. We also decomposed $P^C_\omega$ into its real part $P_\omega$, and its imaginary part $P^I_\omega$. The oscillating exponential is the standard expression, while the decaying one is $\int dx P^I_\omega$. The latter is equal to $\mathcal{I}_\omega $ of Eq.~\eqref{eq:optdepthBH} when working to first order in $\Gamma/P$, which is here a legitimate approximation. A preliminary analysis, similar to Eq.~(A12) of Ref.~\cite{Coutant:2011in}, indicates that the corrections to Eq.~\eqref{eq:pW} are bounded by $\mathcal{O} ( \frac{\omega^2}{\Lambda^2 \abs{1+v}^3} + \frac{g^2 \omega}{\Lambda (1+v)^2})$. Hence Eq.~\eqref{eq:pW} gives an accurate description everywhere but in the central region \textit{III} defined by $\kappa x_{\rm trans}$ of Eq.~\eqref{eq:Xtrans}. 

Using Eq.~\eqref{eq:pW}, the $U$-mode contribution to the commutator is, for $\omega > 0$, 
\begin{equation}		
\begin{split}
G_\mathrm{c}^\omega(x,x') &= \theta( \mathcal{I}_\omega(x,x')) \phi_\omega^{\rm dec}(x) 
\left (\phi_\omega^{\rm grw}(x') \right )^* + \theta( \mathcal{I}_\omega(x',x)) \phi_\omega^{\rm grw}(x) \left (\phi_\omega^{\rm dec}(x') \right )^*,
\label{eq:GcU}
\end{split}
\end{equation}
where the growing mode $\phi_\omega^{\rm grw}$ satisfies Eq.~\eqref{eq:decmodeeq} with the opposite sign for the last term which encodes dissipation. The expression for $\omega < 0$ is given by $G_\mathrm{c}^{-\omega} = - (G_\mathrm{c}^\omega)^*$ which follows from the imaginary character of $G_\mathrm{c}$ in $t,x$ space. We used the sign of $\mathcal{I}_\omega $ in Eq.~\eqref{eq:GcU} so that a similar expression is valid on the left of the horizon. Note also that Eq.~\eqref{eq:GcU} cannot be used to estimate $G_\mathrm{c}^\omega$ across the horizon because the WKB approximation fails in region \textit{III}. Note finally that Eq.~\eqref{eq:GcU} is valid only for $\Lambda \abs{x-x'} \gg 1$.

Having characterized in quantitative terms the impact of dissipation, we now work in conditions such that the mode damping is negligible far away from this central region. That is, we work with $x, x'$ obeying 
\begin{equation} 
x_{\rm trans} \ll \vert x \vert \ll \sqrt{\Lambda/\kappa^3},
\label{eq:niced}
\end{equation} 
where the upper limit comes from the neglect of the second term in Eq.~\eqref{eq:optdepthBHlimit}. Under these conditions, the ($out$ part of the) anti-commutator of Eq.~\eqref{eq:GacomXX} is, for $\omega > 0$, given by 
\begin{equation}
\label{eq:Gacnc}
\begin{split}
G_{\rm ac}^\omega (x,x') \approx & (2n_\omega+1) \left [\phi_{\omega,R}^{out, \rm BH} (x) (\phi_{\omega,R}^{out, \rm BH}(x'))^* + (\phi_{-\omega,L}^{out, \rm BH} (x))^* \phi_{-\omega,L}^{out, \rm BH} (x')\right ] \\
&+2 {\rm Re}\left [ c_\omega \phi_{\omega,R}^{out, \rm BH} (x) \phi_{-\omega,L}^{out, \rm BH}(x') \right ],
\end{split}
\end{equation}
where $n_\omega$ and $c_\omega$ are constant because we are far from region \textit{III}, and where the $R$ and $L$ \textit{out} modes live on one side of the horizon and have unit norm. Being undamped, they are either relativistic, or, more generally, dispersive WKB modes. In the former case, they thus behave in the regions of interest, namely $I_{R/L}$ and $\mathit{II}_{R/L}$, as
\begin{equation}
\begin{split}
\phi_{\omega,R}^{out, \rm BH} &\underset{\mathit{II} } \sim \theta(x) \frac{x^{i \omega/\kappa } }{\sqrt{2\omega}} \underset{I} \sim \theta(x) \frac{\ep{i \omega x /(1+v_R)} }{\sqrt{2\omega /(1+v_R)}} , \\
\left (\phi_{-\omega,L}^{out, \rm BH} \right )^* &\underset{\mathit{II}} \sim \theta(-x) \frac{(-x)^{i \omega/\kappa } }{\sqrt{2\omega}} \underset{I}\sim \theta(-x) \frac{ \ep{- i \omega x /|1+v_{L}|} }{\sqrt{\vert 2\omega /(1+v_{L})\vert}}, 
\end{split}
\label{eq:retism}
\end{equation} 
where $v_{R(L)}$ is the asymptotic velocity in the region $R$ ($L$, where $1+v_L < 0$). As in de Sitter, the (positive unit norm) mode $ \phi_{-\omega,L}^{out, \rm BH}$ living in the $L$ region has a negative Killing frequency.

In Eq.~\eqref{eq:Gacnc}, $n_\omega$ and $c_\omega$ are unambiguously defined because the $R/L$ modes are normalized in regions $I_{R/L}$. Thus, they respectively define the spectrum emitted by the black hole, and the $\omega$-contribution of the correlation across the horizon. To compute them, we should find the equivalent of Eq.~\eqref{eq:n_k,c_k}. To this end, we shall use Eq.~\eqref{eq:GacXXP}, and exploit the fact that their values are fixed in the domain of $P$ given in Eq.~\eqref{eq:dominP}. 

\subsection{Connection with de Sitter physics}
\label{sec:BHds:ds}

In Eq.~\eqref{eq:GacXXP}, we need (the $U$-mode contribution of) $G_{\rm ret}^\omega(x,P_1)$ with $\abs{x} \gg x_{\rm trans}$, since we are interested in the far away behavior of $G_{\rm ac}^\omega$, and with $P_1 \gtrsim \sqrt{\kappa \Lambda}$, because the integrand vanishes for lower values of $P$. Since $P_\omega(x) \ll P_1$, the retarded character
of Eq.~\eqref{eq:Gret-solution} is automatically implemented, which means that 
\begin{equation}
\label{eq:GretisGc}
G_{\rm ret}^\omega(x,P_1)= (-i) G_\mathrm{c}^\omega(x,P_1).
\end{equation} 
The commutator $G_\mathrm{c}^\omega(x,P_1)$, on the one hand, obeys Eq.~\eqref{eq:decmodeeq} in $x$, and on the other hand, behaves as in de~Sitter for $P_1 \gtrsim \sqrt{\kappa \Lambda}$, when $\omega_{\rm max}^{\rm diss}$ of Eq.~\eqref{eq:ommdiss} obeys $\kappa/\omega_{\rm max}^{\rm diss}\ll 1$. This second condition means that the high $P_1$ behavior is governed by Eq.~\eqref{eq:Gret-solution} and Eq.~\eqref{eq:prepomega}.
 
For simplicity we consider the massless case of Eq.~\eqref{eq:numdispersion}, when $g^2 = 2$. In this model, in de Sitter, using the Unruh modes of Eq.~\eqref{eq:unruhmodedef}, the $U$-mode contribution is 
\begin{equation}
\begin{split}	
\label{eq:GcXPun}
G_{\rm c, \rm dS}^\omega(x,\bP) = \ep{- \mathcal{I}_0^P } \bigg[& \phi^{U}_\omega (x) (\theta(\bP) \phi_{\omega}^{\rm rel} (P))^* - \left (\phi^{U}_{-\omega} (x)\right )^* (\theta(-\bP)\phi_{-\omega}^{\rm rel} (P)) \bigg ],
\end{split}
\end{equation}
where $\mathcal{I}_0^P$ is given in Eq.~\eqref{eq:opticaldepth}, and where we replaced its lower value $P_\omega(x) \ll \sqrt{\kappa \Lambda}$ by $0$ because $x$ is taken sufficiently large. Using Eq.~\eqref{eq:UnruhRLchange}, we can re-express Eq.~\eqref{eq:GcXPun} in the $R/L$ \textit{out} mode basis. For $\omega>0$ we get
\begin{equation}
\begin{split}
\label{eq:GcXPchi}
G_{\rm c, \rm dS}^\omega(x,\bP) = \ep{- \mathcal{I}_0^P } \bigg[& \phi_{\omega,R}^{out, \rm dS} (x) (\phi_{\omega,R}^{out, \rm dS} (\bP))^* - (\phi_{-\omega,L}^{out, \rm dS} (x))^* \phi_{-\omega,L}^{out, \rm dS} (\bP) \bigg ].
\end{split}
\end{equation}
In this we recover that the commutator possesses the same expression if one uses the \textit{in} (Unruh) or the \textit{out} mode basis. 

Eq.~\eqref{eq:GcXPchi} applies as such to the black hole metric in the regions $\mathit{II}_{R/L}$, $ \kappa x_{\rm trans} \ll \vert \kappa x \vert < D/2$, because $G_{\rm c, \rm BH}$ obeys the same equations, and its normalization is fixed by the equal time commutators. In fact, in these regions the normalized black hole modes $\phi_{\omega,R}^{out, \rm BH}, \phi_{-\omega,L}^{out, \rm BH} $ coincide with the modes $\phi_{\omega,R}^{out, \rm dS}, \phi_{-\omega,L}^{out, \rm dS}$ of Eq.~\eqref{eq:phiout}. Then, the WKB character of $\phi_{\omega,R}^{out, \rm BH}, \phi_{-\omega,L}^{out, \rm BH} $ guarantees that Eq.~\eqref{eq:GcXPchi} applies further away from the horizon, in the regions defined by Eq.~\eqref{eq:niced}. Hence, in these regions, we have 
\begin{equation}
\begin{split}
\label{eq:GcXPchiBH}
G_{\rm c, \rm BH}^\omega(x,\bP)  \approx \ep{- \mathcal{I}_0^P } \bigg[& \phi_{\omega,R}^{out, \rm BH}(x) (\phi_{\omega,R}^{out, \rm dS}(\bP))^* - (\phi_{-\omega,L}^{out, \rm BH} (x))^* \phi_{-\omega,L}^{out, \rm dS} (\bP) \bigg ].
\end{split}
\end{equation}
We kept the de~Sitter modes in $\bP$-space because only $|\bP| \gg {\kappa/\Lambda}$ contribute to Eq.~\eqref{eq:GacXXP}. Using Eq.~\eqref{eq:GretisGc}, inserting the above expression in Eq.~\eqref{eq:GacXXP}, and comparing the resulting expression with Eq.~\eqref{eq:Gacnc}, we get 
\begin{subequations}
\label{eq:ncac}
\begin{align}
(2n_\omega+1) = \int d\bP_1 d\bP_2 & \chi_{R}^{\omega *} (\bP_1) \chi_{R}^{\omega}(\bP_2) \ep{-\mathcal{I}_0^{P_1} -\mathcal{I}_0^{P_2} } N^\omega(P_1,P_2) . \\
2c_\omega = \int d\bP_1 d\bP_2 & \chi_{R}^{\omega *} (\bP_1) \chi_{L}^{-\omega *}(\bP_2) \ep{-\mathcal{I}_0^{P_1} -\mathcal{I}_0^{P_2} } N^\omega(P_1,P_2) . 
\end{align}
\end{subequations}
These expressions are identical to those evaluated in de Sitter. Hence, $n_\omega$ and $c_\omega$ are respectively given by Eqs.~\eqref{eq:nom} and~\eqref{eq:com}. Therefore, to leading order in $\kappa/\Lambda$, and for an environment at zero temperature, $n_\omega$ and $c_\omega$ retain their standard relativistic expressions. 

This means that the state of the outgoing modes when they leave the central region \textit{III}, and propagate freely, is the Unruh vacuum~\cite{Jacobson:1991gr,Jacobson:1993hn,Brout:1995rd}. This can be explicitly checked from Eq.~\eqref{eq:ncac} by re-expressing the \textit{out} modes $\phi_{\omega,R/L}^{out, \rm dS}$ in terms of the Unruh modes of Eq.~\eqref{eq:unruhmodedef}. In this case, one finds that the mean number of Unruh quanta $n_\omega^{\rm Unruh}$ is given by, see Eq.~\eqref{eq:nucudef},
\begin{equation}
\label{eq:Unv} 
\begin{split}
(2n_\omega^{\rm Unruh}+1) = & \iint_0^\infty dP_1 dP_2 (\phi_{\omega }^{\rm rel} (P_1))^* \phi_{\omega}^{\rm rel}(P_2) \ep{-\mathcal{I}_0^{P_1} -\mathcal{I}_0^{P_2} } N^\omega(P_1,P_2) \\
 = & 1 + O(\kappa/\Lambda) . 
\end{split}
\end{equation}
In other words, the role of the double integrals in Eq.~\eqref{eq:ncac} and Eq.~\eqref{eq:Unv}, whose integrand explicitly depends on the actual \enquote{trans-Planckian} physics governed by $\Lambda$, $f(P)$, $\Gamma(P)$, is to implement the Unruh vacuum in dissipative theories. 

\section*{Conclusions}
\addcontentsline{toc}{section}{Conclusions}

In this chapter, we showed that the portion of space close to a black hole of surface gravity $\kappa$ can be mapped into de Sitter space with $H=\kappa$ even in the presence of Lorentz violation. We then used this correspondence to draw conclusions on the flux produced by the black hole, and its link with the asymptotic fluxes produced in de Sitter.

When four conditions are met, we showed that the analysis performed in de~Sitter in \ref{chap:dispdS} and \ref{chap:dissipdS} applies to Hawking radiation. The inequality which ensures the validity of this correspondence is $\kappa/ \omega_{\rm max}^{\rm diss}, \kappa/ \omega_{\rm max}^{\rm disp} \ll 1 $, where $\omega_{\rm max}^{\rm disp},\omega_{\rm max}^{\rm diss}$ are the composite ultraviolet scale of Eqs.~\eqref{eq:omegamaxdisp} and~\eqref{eq:ommdiss}. They depend both on the microscopic scale $\Lambda$, and $D$ which fixes the extension of the black hole near horizon region where the metric and the field $u$ can be mapped into de~Sitter. The validity of the correspondence in turn guarantees that, to leading order, the Hawking predictions are robust -- even if the early propagation completely differs from the relativistic one, see Fig.~\ref{fig:caracter}. This establishes that when leaving the very high momentum $P \sim \Lambda$ (trans-Planckian) region and starting to propagate freely, the outgoing configurations are \enquote{born} in their Unruh vacuum state~\cite{Jacobson:1991gr,Jacobson:1993hn,Brout:1995rd}. The microscopic implementation of this state in dissipative theories is shown in Eq.~\eqref{eq:Unv}. As a result, as in the case of dispersive theories~\cite{Macher:2009tw,Coutant:2011in}, the leading deviations with respect to the relativistic expressions should be suppressed as powers of $\kappa/ \omega_{\rm max}$, i.e., they should be governed by the extension of the black hole NHR which is a portion of de~Sitter space.

\partimage{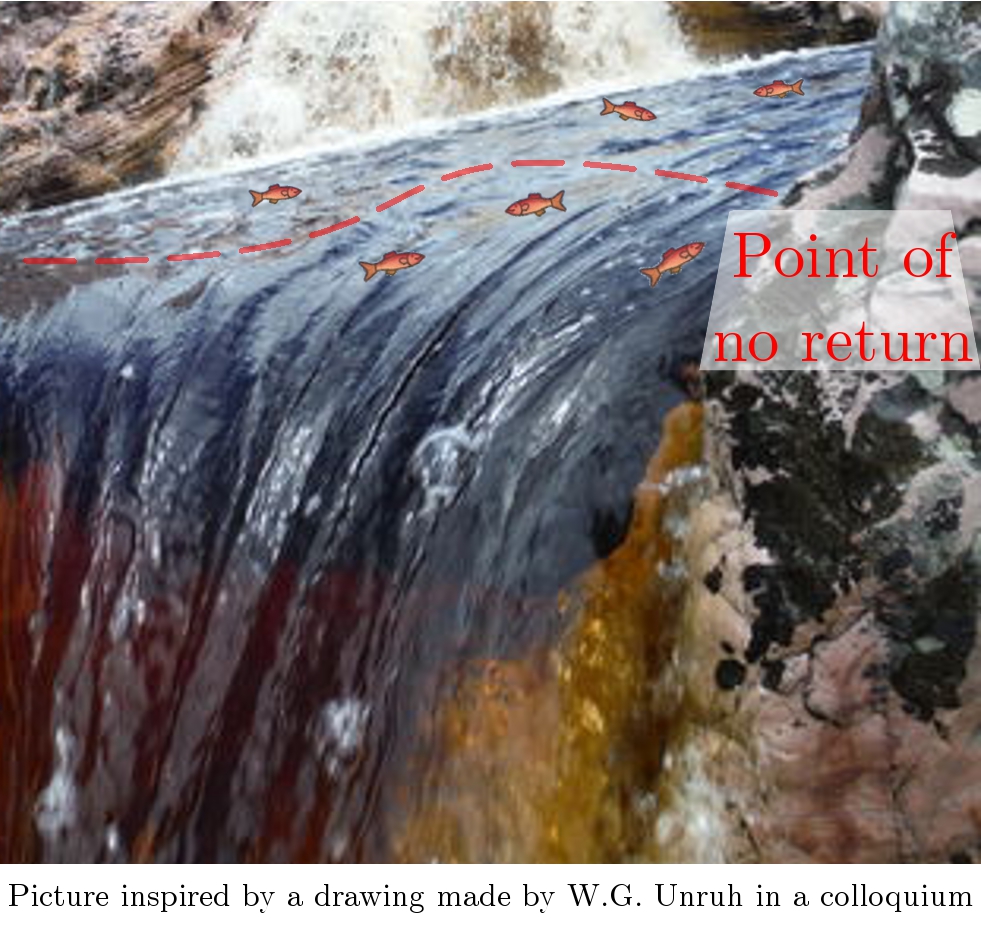}
\partcitation{\textit{After all the ``universe'' is an hypothesis, like the atom, and must be allowed the freedom to have properties and to do things which would be contradictory and impossible for a finite material structure.} \\ Willem de Sitter }
\part{Analogue gravity}
\label{part:analoguegravity}
\chapter*{Introduction}
\phantomsection
\addstarredchapter{Introduction}
\markboth{INTRODUCTION}{INTRODUCTION}

The analogy~\cite{Unruh:1980cg} between sound propagation in nonuniform fluids and light propagation in curved space-times opens the possibility to experimentally test long standing predictions of quantum field theory~\cite{birrell1984quantum}, such as the origin of the large scale structures in our Universe~\cite{Mukhanov:1981xt,mukhanov2005physical} and the Hawking radiation emitted by black holes. In the first case, the cosmic expansion engenders a parametric amplification of homogeneous modes which is very similar to that at the origin of the dynamical Casimir effect (DCE)~\cite{PhysRevLett.109.220401,WilsonDCE,Lahteenmaki12022013}, compare for instance Ref.~\cite{Campo:2003pa} with Refs.~\cite{Fedichev:2003bv,Carusotto:2009re}. Yet, in order to predict with accuracy what should be observed, one must take into account the short distance properties of the medium because the predictions involve short wavelength modes. As a result, the analogy breaks down and a case by case analysis is required. For a modern review on analogue gravity, see Ref.~\cite{Barcelo:2005fc}.

In this part, we shall study some of these fluids. In \ref{chap:BECandAG}, we review what is a Bose condensate and in which sense its vacuum state behaves as quantum vacuum in curved space time. In \ref{chap:dissipBEC}, we consider phonons propagating in such a (homogeneous) atomic condensate. We include the dissipative effects for the phonons due to the interaction of the atoms with their environment. As a consequence, the dispersion relation is of the form
\begin{equation}
\nonumber
\Omega^2 + 2 i \Omega \Gamma(k^2) = c^2 k^2 + f(k^2). 
\end{equation}
In \ref{chap:polariton}, we consider the polariton system in which BEC is replaced by a fluid of light. In one- or two-dimensional microcavity devices, photons acquire an effective mass $m$ because of spatial confinement, while an effective two-body photon-photon interaction can originate from the $\chi^{(3)}$ optical nonlinearity of the cavity medium. As a result, assemblies of many photons in the cavity can display the collective behavior of a Bose-Einstein condensate. However, contrary to the previous case, some loss appears at the level of the condensate and one need to pump the system. As a consequence, the system is out of equilibrium and the phonons are massive excitations. In \ref{chap:thermalBH}, we study the stationary fluxes escaping a black hole in the presence of initially thermal state but neglecting dissipative effects. We also characterize the separability of the final state.

\chapter{Bose condensation and analogue gravity}
\label{chap:BECandAG}

\section*{Introduction}

When one cools down a gas at low enough pressure so that it does not become liquid nor solid, the low energy states get more and more populated while the high energy ones get depleted. As a consequence, one starts to feel quantum effects. If the gas is composed of fermions, the gas becomes degenerate as no more than one particle can be in the same state. One then have $1$ particle in each of the lowest energy states. On the other hand, if it is composed of bosons, the particles form a Bose Einstein condensate.

We here review some basic properties of Bose Einstein condensate. We consider phonons propagating in such a condensate and build effective tools to determine if the state is separable or not (see. \ref{chap:separability}). In \ref{sec:intropolariton}, we consider a particular out of equilibrium condensate.

\minitoc
\vfill

\section{Condensation of an ultracold quantum gas}

\subsection{Notion of condensation}

A gas is composed of particles (atoms, molecules,...). To fix ideas, we consider bosonic atoms. Three typical length scales then enter in competition. The first one is the mean distance between atoms. It is proportional to the power $-1/3$ of the density: $n^{-1/3}$. The second distance is due to the fact that atoms have a size, or, say otherwise, that interaction between atoms occur only if they are at a distance smaller than $R_e$. The third distance is the thermal de Broglie wavelenth due to the temperature $\lambda_T = \sqrt{2 \pi \hbar^2 / m T}$. In the case $\lambda_T \gtrsim n^{-1/3}\gg R_e$, the gas is called ultracold ($\lambda_T \gg R_e$) dilute ($n^{-1/3}\gg R_e$) quantum ($\lambda_T \gtrsim n^{-1/3}$) gas. These are the necessary conditions for the gas to form a Bose condensate~\cite{Shlyapnikovcourse}.

Consider the dilute ultracold case, and neglect interactions. Since the atoms are non relativistic and have bosonic occupation number, the total number of atoms is given as an integral over phase space, as
\begin{equation}
\begin{split}
N = \int \frac{d^3\bx d^3\bk }{(2\pi)^3} \frac{1}{\ep{(\hbar^2 k^2/2m - \mu)/k_B T}-1}.
\end{split}
\end{equation}
The fraction on the right is the thermal density of bosons, $T$ is the temperature of the gas and $\mu < 0$ is the chemical potential. $\hbar^2 k^2/2m$ is the kinematic energy of the atom. In the case, $-\mu \gg k_B T$, one can neglect the $-1$ and one recovers the Boltzmann statistics. However, in such a case, the density is $n \sim \lambda_T^{-3} \ep{\mu/k_B T}$, so that $\lambda_T \ll n^{-1/3}$. In the quantum case, $-\mu \lesssim k_B T$, one cannot neglect the $-1$. In such a case, the density of particles is given by
\begin{equation}
\begin{split}
n = \lambda_T^{-3} \frac{\sqrt{\pi}}{4}\int dz \frac{z^2}{\ep{z^2 - \mu/k_B T}-1} \leq \lambda_T^{-3} \frac{\sqrt{\pi}}{4}\int dz \frac{z^2}{\ep{z^2}-1} \sim 0.51 \lambda_T^{-3}.
\end{split}
\end{equation}
This bound is violated for a quantum gas. This means that a macroscopic number of particle lies in the part of phase space where the density of particles in phase space is infinite, i.e., $1/(\ep{(\hbar^2 k^2/2/m - \mu)/k_B T}-1) =\infty $. This macroscopic number of particles with the same momentum is a Bose-Einstein condensate. We note that in one or two (spatial) dimensions, the integral does not converge since the factor $z^2$ is replaced by $1$ or $z$. The density can hence be as large as one wants by lowering the (absolute value of) chemical potential. No condensation occurs. As a consequence, trying to make condensates on sheets or on a line is impossible\footnote{It becomes possible on sheet when energy is proportional to momentum.}. Only a quasi-condensation can be made~\cite{PhysRevA.67.053615}.

\subsection{Hamiltonian formalism}

At the next order, we consider the two body interaction, but neglect three body terms. The classical Hamiltonian of the system then reads in the presence of a global potential $V(\bx)$:
\begin{equation}
\begin{split}
H = -\frac{1}{2 m } \sum_\alpha p^2_\alpha + V(\bx) + \frac{1}{2} \sum_{\alpha \neq \beta} U(\bx_\alpha - \bx_\beta),
\end{split}
\end{equation}
where the sum runs over all atoms of the gas and where $U$ is the $2$-body interaction potential. 

The quantization procedure is similar to the one of \ref{sec:Quantumparamampli}. We first consider the case of finite volume ${\mathcal{V}}$. Then, in the case with one particle, a (discrete) basis of states is given by the eigenvectors $\left |\Psi_\bp \right >$ of the operator $\hat \bp$: $\hat \bp \left |\Psi_\bp \right > = \bp \left |\Psi_\bp \right >$. With many indistinguishable particles, a basis of states is given by the symmetrized product of such states: $ \sum_\sigma \bigotimes_\alpha \left |\Psi_{\bp_{\sigma(\alpha)}} \right > $, where the sum runs over permutations. Since the same set of momenta $\{\bp_i\}$ appears for each particle, an equivalent way of writing this state is $ \left |N_{\bk_1}, ..., N_{\bk_i},...\right > $, where $N_{\bk_i}$ labels the number of atoms of momentum $\bp_i = \hbar \bk_i$.
For each $N$ (the total number of atom) these states engender a Hilbert space $\mathcal{H}_N$. The sum of all these Hilbert spaces $\mathcal{H} = \sum_N \mathcal{H}_N$ corresponds to the states in the grand canonical ensemble. On this Hilbert space, one defines creation $a_{\bk_i}^\dagger$ and annihilation $a_\bk$ operators as $\hat a_{\bk_i}^\dagger \bigotimes_\alpha \left |\Psi_{\bp_{\alpha}} \right > = \sqrt{N_{\bk_i}+1}\bigotimes_\alpha \left |\Psi_{\bp_{\alpha}} \right > \otimes \left |\Psi_{\hbar \bk_i} \right >$, where the symmetrization is implicit and where $N_{\bk_i}$ is the number of particles in state $\left |\Psi_{\hbar \bk_i} \right >$. In the new language, this reads $\hat a_{\bk_i}^\dagger \left | N_{\bk_1}, ..., N_{\bk_i},... \right > = \sqrt{N_{\bk_i}+1} \left | N_{\bk_1}, ..., N_{\bk_i} +1 ,... \right > $. Using these operators, the Hamiltonian reads
\begin{equation}
\label{eq:hamiltonianBECfirstQ}
\begin{split}
\hat H &= -\frac{\hbar^2}{2 m } \sum_\bk k^2 \hat a_\bk^\dagger \hat a_\bk + \sum_{\bk_1,\bk_2} \int \frac{d\bx}{\mathcal{V}} \ep{i (\bk_1-\bk_2) \bx} V(\bx) \hat a_{\bk_1}^\dagger \hat a_{\bk_2} \\
& + \frac{1}{2} \sum_{\bk_1,\bk_2,\bk_3,\bk_4} \int \frac{d\bx_1 d\bx_2}{{\mathcal V}^2} U(\bx_1-\bx_2) \ep{ -i \bk_1\bx_1 - i\bk_2\bx_2 + i\bk_3\bx_1 + i\bk_4\bx_2} \hat a_{\bk_1}^\dagger \hat a_{\bk_2}^\dagger \hat a_{\bk_3} \hat a_{\bk_4},
\end{split}
\end{equation}
In the limit of large volume, by changing the position variables to their difference and sum, the integral of the second line of Eq.~\eqref{eq:hamiltonianBECfirstQ} reads $\int \frac{d\bx}{\mathcal V} U(\bx) \ep{ -i (\bk_1- \bk_3)\bx } \delta_{\bk_1+\bk_2-\bk_3-\bk_4}$. When the intensity of the coupling $U$ is low, the momentum of the atom hardly changes when it interacts and one can approximate the integral by a constant: $ \frac{g_{\rm at}}{\mathcal V}\delta_{\bk_1+\bk_2-\bk_3-\bk_4}$, where $g_{\rm at} = \int d\bx U(\bx)$. Under these conditions, the Hamiltonian reads
\begin{equation}
\label{eq:HamiltonianBECfirstquantize}
\begin{split}
\hat H = -\frac{\hbar^2}{2 m } \sum_\bk k^2\hat a_\bk^\dagger \hat a_\bk + \sum_{\bk_1,\bk_2} \tilde V(\bk_1-\bk_2) \hat a_{\bk_1}^\dagger \hat a_{\bk_2} + \frac{g_{\rm at} }{2\mathcal{V}} \sum_{\bk_1,\bk_2,\bk_3,\bk_4}\delta_{\bk_1+\bk_2-\bk_3-\bk_4} \hat a_{\bk_1}^\dagger \hat a_{\bk_2}^\dagger \hat a_{\bk_3} \hat a_{\bk_4}
\end{split}
\end{equation}
Even though the gas is non relativistic, because atoms are indistinguishable, it is convenient to introduce a quantum field as $\hat \Phi(\bx) = \sum_\bk \frac{\ep{-i \bk \bx}}{\sqrt{\mathcal{V}}}\hat a_\bk$. The commutation relation on operators $\hat a$ imply the commutation relation on the field: $[\hat \Phi(\bx),\hat \Phi(\bx')]=0$, $[\hat \Phi(\bx),\hat \Phi^\dagger(\bx')] = \delta(\bx- \bx')$. $\hat \Phi$ and $\hat \Phi^\dagger$ are thus canonical variables. Moreover, using such a field, the Hamiltonian reads 
\begin{equation}
\label{eq:HamiltonianBECgen}
\begin{split}
\hat H = \int d\bx\left [ \frac{\hbar^2}{2 m } \partial_\bx \hat \Phi^\dagger(\bx) \partial_\bx \hat \Phi(\bx) + V(\bx) \hat \Phi^\dagger(\bx) \hat \Phi(\bx)+ \frac{g_{\rm at}}{2} \hat \Phi^\dagger(\bx) \hat \Phi^\dagger(\bx) \hat \Phi(\bx) \hat \Phi(\bx) \right ].
\end{split}
\end{equation}
In Heisenberg formalism, the equations of motion for such system are then $i \hbar \partial_t \hat \Phi = [ \hat \Phi,\tilde H]$ [see Eq.~\eqref{eq:eomheisenberggen}]. They give the non linear equation
\begin{equation}
\label{eq:QuantumGPE}
\begin{split}
i \hbar \partial_t \hat \Phi=\frac{-\hbar^2}{2 m } \Delta_\bx \hat \Phi + V \hat \Phi + g_{\rm at} \hat \Phi^\dagger \hat \Phi \hat \Phi,
\end{split}
\end{equation}
where $\Delta_\bx$ is the standard Laplace operator.

When Bose condensation occurs, because a macroscopic (huge) number of atoms are condensed (say at $\bk = 0$\footnote{In the absence of a potential, only $\bk =0 $ can condense and the condensate is homogeneous. In the presence of a potential, this needs not be true.}), one have $\left <\hat a_{\bk=0}\right > \sim \left <\hat a^\dagger_{\bk=0}\right > \gg 1$. To first order, one can hence approximate the operators by their expectation value. This remark inspires the following decomposition (that can always be made for operators):
\begin{equation}
\begin{split}
 \hat \Phi = \Phi_{\rm cond} \left (1+ \hat \phi \right ), \text{ where } \Phi_{\rm cond} = \left < \hat \Phi \right >.
\end{split}
\end{equation}
The interest of such decomposition however is that in the presence of condensate, the expectation values of increasing power of $\hat \phi$ go decreasing. One can thus expand Eq.~\eqref{eq:QuantumGPE} in powers of $\hat \phi$. The zeroth order gives the Gross Pitaevskii equation:
\begin{equation}
\label{eq:GPE}
 \begin{split}
i \hbar \partial_t \Phi_{\rm cond}=\left [\frac{-\hbar^2}{2 m } \Delta_\bx + V + g_{\rm at} \abs{\Phi_{\rm cond}}^2 \right ]\Phi_{\rm cond}.
 \end{split}
 \end{equation} 
Solving the Gross Pitaevskii equation given a potential $V$ is a full problem in itself. However, we shall not consider it in this thesis. Indeed, for all \enquote{nice} [i.e., respecting Eq.~\eqref{eq:conservationequation}] function $\Phi_{\rm cond}$, there exists a potential $V$ (eventually time dependent), such that the function is solution of the corresponding Gross Pitaevskii equation. Rather, we shall interpret the physical quantities entering $\Phi_{\rm cond}$. In the next section, we shall focus on the perturbations $ \hat \phi$ given a background $\Phi_{\rm cond}$.

The first meaningful quantity is the modulus of $\Phi_{\rm cond}$. In term of the original operator, we have for all mesoscopic volume $\mathcal{V}$ where potential is approximately constant $\rho \doteq \abs{\Phi_{\rm cond}}^2 \sim \left < \hat a^\dagger_{\bk=0}\hat a_{\bk=0} \right > / \mathcal{V}$. Matching such volumes, one deduces that $\rho$ is the density of condensed atoms. It is of interest to observe the time dependence of such density. Indeed, using Eq.~\eqref{eq:GPE}, one gets
\begin{equation}
\label{eq:conservationequation}
\begin{split}
\partial_t \rho = \partial_\bx \frac{i \hbar}{2 m} \left [ \Phi_{\rm cond}^* \partial_\bx \Phi_{\rm cond} - \Phi_{\rm cond} \partial_\bx \Phi_{\rm cond}^* \right ].
\end{split}
\end{equation}
The term into the derivative in the right hand side is $-\rho \bv$. $\bv$ is interpreted as the velocity of the condensate, as shall be made clear below.
The general form of $\Phi_{\rm cond}$ thus reads 
\begin{equation}
\label{eq:generalformPhicond}
\begin{split}
\Phi_{\rm cond} = \sqrt{\rho} \ep{ i \int d\bx m \bv/ \hbar} \ep{- i \int \mu(t) dt/\hbar},
\end{split}
\end{equation}
where $\mu$ is a real function of time.

Let now explicit why $\bv$ is the velocity of the condensate. First, under this condition, Eq.~\eqref{eq:conservationequation} reads $\partial_t \rho =- \partial_\bx \rho \bv$, which is the conservation equation for a fluid. Second, when introducing Eq.~\eqref{eq:generalformPhicond} into the Hamiltonian, the energy reads
\begin{equation}
\begin{split}
E= \int d\bx \rho \left [g_{\rm at} \rho +  V  + \frac{1}{2} m v^2 +  \frac{\hbar^2}{8 m} \left (\frac{\partial_\bx \rho}{\rho}\right )^2\right ].
\end{split}
\end{equation}
The prefactor $\int d\bx \rho$ is the sum over all atoms (they have a certain width and are superposed). The term $g_{\rm at} \rho$ is the interaction potential. It is proportional to the density because interaction is contact interaction and because number of atoms is huge so that density of all atoms and density of all atoms but the one we consider are the same. The $V$ term is the external potential. $\frac{1}{2} m v^2$ is the kinetic part of the energy. All these term are classical terms (they correspond to the limit $\hbar \to 0$) and are recovered in fluid mechanics. The last term is a quantum potential. Because it is large only when $\partial_\bx \rho / \rho $ is larger than the size of the atom, it can be conceived as a kind of energy of deformation of the atom.

\section{Phonon field}
\label{sec:phonon}

When expanding Eq.~\eqref{eq:QuantumGPE} around a solution of Eq.~\eqref{eq:GPE}, one gets an equation of motion for $\hat \phi$ which is first order in time. Using Eq.~\eqref{eq:GPE} to remove every dependence in $\partial_t \Phi_{\rm cond}$, one gets to first order in $\hat \phi$:
\begin{equation}
\label{eq:generaleomphonon}
\begin{split}
i \hbar \left ( \partial_t + \bv \partial_\bx \right ) \hat \phi = \frac{-\hbar^2}{2 m \rho } \partial_\bx \rho \partial_\bx\hat \phi + g_{\rm at} \rho (\hat \phi+\hat \phi^\dagger).
\end{split}
\end{equation}
This equation only depends on the effective quantities $\bv, \rho$ and $g_{\rm at}$. Because $V$ can be chosen in such a way that these functions take any value (but respecting Eq.~\eqref{eq:conservationequation}), in the analysis on the perturbation fields $\hat \phi$, we shall mainly consider that the background is given by a profile of $\bv, \rho$ and $g_{\rm at}$, and never specify the potential $V$.

From the canonical commutators of the field $\hat \Phi$, we deduce the canonical ones for the fluctuations $\hat \phi$: 
\begin{equation}
\begin{split}
[\hat \phi(t,\bx),\hat \phi(t,\bx')]=0, \quad [\hat \phi(t,\bx),\hat \phi^\dagger (t,\bx')] = \delta(\bx - \bx') / \rho(t,\bx).
\end{split}
\end{equation}
Moreover, when expanding the Hamiltonian to quadratic order in $\hat \phi$, taking $\Phi_{\rm cond}$ solution of Eq.~\eqref{eq:GPE}, the linear term in $\hat \phi$ vanish and the effective Hamiltonian reads\footnote{Because a non canonical transformation has been made, to get this result, it is necessary to use Lagrangian formalism}
\begin{equation}
\label{eq:HamiltonianBECerturb}
\begin{split}
H = \int d\bx \rho \left [ \frac{\hbar^2}{2 m } \partial_\bx\hat \phi^\dagger \partial_\bx \hat \phi - i\hbar \hat \phi^\dagger \bv\partial_\bx \hat \phi + \frac{g_{\rm at}}{2} \rho (\hat \phi+ \hat \phi^\dagger)^2 \right ].
\end{split}
\end{equation}
The second term is not Hermitian, but associates with the $ i \rho \hat \phi^\dagger \partial_t \hat \phi $ that appears in the Legendre transformation, to give an Hermitian action. 

\subsection{Analogue gravity: characteristics and mode equation}

When solving Eq.~\eqref{eq:generaleomphonon} for $\hat \phi^\dagger$ and inserting it into the hermitian conjugate equation, one gets the self defined second order equation of motion, see Refs.~\cite{Dalfovo:1999zz,Macher:2009nz}.
\begin{equation}
\label{eq:generaleomphononsecondorder}
\begin{split}
\left [-i \left ( \partial_t + \bv \partial_\bx \right ) + \frac{\hbar}{2 m \rho } \partial_\bx \rho \partial_\bx \right ] \frac{m}{ g_{\rm at} \rho }\left [i \left ( \partial_t + \bv \partial_\bx \right ) + \frac{\hbar}{2 m \rho } \partial_\bx \rho \partial_\bx \right ] \hat \phi=\left [\frac{1}{ \rho } \partial_\bx \rho \partial_\bx \right ]\hat \phi.
\end{split}
\end{equation}
This equation is much different from the relativistic version, see e.g., Eq.~\eqref{eq:dispKGeq}. However, at the level of the corresponding Hamilton-Jacobi equation, see Eq.~\eqref{eq:dispHJeq}, it reads
\begin{equation}
\label{eq:BECHJeq}
\begin{split}
 \left ( \partial_t S + \bv \partial_\bx S \right )^2 = \frac{ g_{\rm at} \rho }{m} (\partial_\bx S)^2 + \frac{\hbar^2}{4 m^2 } (\partial_\bx S)^4 ,
\end{split}
\end{equation}
which is identical to its relativistic equivalent. The dispersion relation for such theory is then a superluminal quadratic one, and we recognize that $ c^2 \doteq \frac{ g_{\rm at} \rho }{m}$ is the speed of low energetic excitations, or speed of sound. Moreover, the typical length under which dispersion occurs [the equivalent of $1/ \Lambda$ in Eq.~\eqref{eq:numdispersion}] is called Healing length and is defined by $\xi = \hbar /2 m c$. Using these scales, the dispersion relation is 
\begin{equation}
\begin{split}
\Omega^2 = (\omega- \bv \bk)^2 = c^2 k^2( 1+ \xi^2 k^2) .
\end{split}
\end{equation}
Here, $\Omega$ is the freely falling frequency. We recognize in Eq.~\eqref{eq:BECHJeq} a dispersive theory in a curved background. In particular, the background is the one of Eq.~\eqref{eq:PGBHaccel} (with $\bv$ and $c$ being time dependent since we did not assume stationarity). 

Hence, at the level of characteristics, the theories of a scalar field in curved space-time and the theory of phonon in BEC are indistinguishable. Mode dimensionality and $S$-matrix structure are thus identical. Because of this, similar effects, such as production of pair of phonon in homogeneous settings of production of entangled quanta in analogue black hole (i.e., analogue Hawking radiation) should occur. This is the essence of analogue gravity. On the other hand, because the quantum theory underneath is different (e.g., ordering of operators is not the same), the results may quantitatively differ.

\subsection{Homogeneous case}

In this section, we consider that background is homogeneous, i.e., $\bv, \rho$ and $c$ are constants in space. Going into the comoving frame, we suppose $\bv= 0$. A consequence of the conservation Eq.~\eqref{eq:conservationequation} is that $\hat \rho$ is constant in time. In the general case however, $c$ is a function of time, see Ref.~\cite{PhysRevLett.109.220401} for an experiment realizing such conditions. Under such conditions, we introduce the Fourier transform of the field $ \hat \phi_\bk = \int d\bx/\sqrt{2\pi} \ep{-i \bk \bx} \hat \phi(\bx)$. Using Eq.~\eqref{eq:generaleomphonon}, it is solution of
\begin{equation}
\label{eq:eomhomogephonon}
\begin{split}
i \hbar \partial_t \hat \phi_\bk = \hbar\Omega_k \hat \phi_\bk + m c^2\hat \phi_{-\bk}^\dagger,
\end{split}
\end{equation}
with 
\begin{equation}
\label{eq:OmegakBEC}
\begin{split}
\hbar \Omega_k \doteq \frac{\hbar^2 k^2 }{2 m } + m c^2.
\end{split}
\end{equation}

\subsubsection{Bogoliubov field}

If the system is also stationary ($c$ constant in time), using Eq.~\eqref{eq:eomhomogephonon} and its hermitian conjugate, we diagonalize the system introducing the \enquote{Bogoliubov} field~\cite{Bogolubov}
\begin{equation}
\label{eq:defvarphibogofield}
\begin{split}
\hat \varphi_\bk &\doteq \sqrt{\rho}\left ( u_k \hat \phi_\bk + v_k \hat \phi_{-\bk}^\dagger \right ) ,
\end{split}
\end{equation}
where the parameters $u_k$ and $v_k$ are
\begin{equation}
\label{eq:ukandvk}
\begin{split}
u_k \doteq \frac{\sqrt{\hbar\Omega_k + m c^2}+ \sqrt{\hbar\Omega_k - m c^2}}{2\sqrt{\hbar\omega_k}}, \quad
v_k \doteq \frac{\sqrt{\hbar\Omega_k + m c^2}- \sqrt{\hbar\Omega_k - m c^2}}{2\sqrt{\hbar\omega_k}}, 
\end{split}
\end{equation}
and where 
\begin{equation}
\label{eq:defomegak}
\begin{split}
\hbar^2\omega_k^2 \doteq \hbar^2\Omega_k^2 - m^2 c^4 
\end{split}
\end{equation}
so that $\omega_k^2 = c^2 k^2 ( 1+ \xi^2 k^2) $ is the rhs of the dispersion relation. The Bogoliubov field is then solution of 
\begin{equation}
\label{eq:eomvarphieq}
\begin{split}
i \partial_t \hat \varphi_\bk = \omega_k \hat \varphi_\bk.
\end{split}
\end{equation}
From this equation, we deduce that the sound waves are quantized. Indeed, the solution of Eq.~\eqref{eq:eomvarphieq}
\begin{equation}
\label{eq:defphononoperatorb}
\begin{split}
\hat \varphi_\bk = \hat b_\bk \ep{- i \omega_k t},
\end{split}
\end{equation}
with $[\hat b_\bk, \hat b_{\bk'}^\dagger] = \delta(\bk - \bk') $. These waves are thus quantum particles, called phonons.

The field $\hat \varphi$ is called Bogoliubov field because Eq.~\eqref{eq:defvarphibogofield} is a Bogoliubov transformation in the sense that it mixes positive and negative energy modes and that $\abs{u_k}^2 - \abs{v_k}^2 = 1 $ it has been first introduced in Ref.~\cite{Bogolubov}. The next step is to inverse the Bogoliubov transformation of Eq.~\eqref{eq:defvarphibogofield} to get the original field as
\begin{equation}
\begin{split}
\sqrt{\rho} \hat \phi_\bk = \hat b_\bk u_k \ep{- i \omega_k t} - \hat b_{-\bk}^\dagger v_k \ep{ i \omega_k t} .
\end{split}
\end{equation}
Inserting this decomposition into Eq.~\eqref{eq:HamiltonianBECerturb}, one gets, up to an additional constant
\begin{equation}
\begin{split}
H= \hbar \int d \bk \omega_k b_{\bk}^\dagger b_\bk .
\end{split}
\end{equation}
We are thus dealing with harmonic oscillators and the operator $\hat b_\bk$ diagonalize the Hamiltonian.

In the non stationary case the harmonic oscillator becomes a parametric amplifier and a similar analysis to the one of \ref{chap:paramamptopartprod} occurs. This is due to the fact that (in homogeneous system), a relativistic field is hidden in such system. This emphasizes the analogy with gravitation. We shall make it clear in next section.
 
\subsubsection{Analogy with relativity}

We show here that analogy with relativity is much stronger in homogeneous settings. Indeed, in such a case, one can define, out of the canonical fields, a field that behaves in a relativistic manner. It is given by
\begin{equation}
\label{eq:defchi}
\hat \chi_\bk \doteq - \frac{\hat \phi_\bk + \hat \phi^\dagger_{-\bk}}{\sqrt{2}} \sqrt{\frac{ \rho}{ \xi c k^2}}.
\end{equation}
With such definition, using $\xi c = \hbar/2m$ constant, Eq.~\eqref{eq:eomhomogephonon} guaranties that 
\begin{equation}
\label{eq:partialtchi}
\partial_t \hat \chi_\bk = i \sqrt{\rho \xi c k^2 } \frac{\hat \phi_\bk - \hat \phi^\dagger_{-\bk}}{\sqrt{2}}.
\end{equation}
The commutator of the two fields is thereby canonical: $[\hat \chi_\bk, \partial_t \hat \chi_{\bk'}^\dagger] = i \delta(\bk - \bk')$. Moreover, using once again Eq.~\eqref{eq:eomhomogephonon}, we get for the second derivative of the field
\begin{equation}
\label{eq:eomchinodiss}
\begin{split}
\partial_t^2 \hat \chi_\bk + \omega_k^2 \hat \chi_\bk =0.
\end{split}
\end{equation}
$\hat \chi_\bk$ is hence a field similar to $\hat q$ of \ref{chap:paramamptopartprod}. In particular, in the stationary case, 
\begin{equation}
\label{eq:chiofbexp}
\begin{split}
\hat \chi_\bk = \frac{\hat b_\bk \ep{- i \omega_k t} + \hat b_\bk^\dagger \ep{ i \omega_k t}}{\sqrt{2 \omega_k }}.
\end{split}
\end{equation}

In the non stationary case, when initial and final operators can be identified, there exists a (new) Bogoliubov transformation linking the two, as in Eq.~\eqref{eq:bogosuroperateur}. This transformation corresponds to a time process and is conceptually different from Eq.~\eqref{eq:defvarphibogofield}. When neglecting dispersion ($\xi \to 0$), Eq.~\eqref{eq:eomchinodiss} is similar to the equation of motion of a (rescaled) field in an FLRW space-time, see Eq.~\eqref{eq:HamiltoniantildephiFLRW} and Refs.~\cite{Nation:2011uv,Carusotto:2009re}. The analogy then tells us that the renormalized mass $\tilde m = 0$ and
\begin{equation}
\begin{split}
\omega_k^2(t) = c_{\rm sound}(t)^2 k^2 \leftrightarrow c_{\rm light}^2 k^2 / a(t)^2.
\end{split}
\end{equation}
A varying speed of sound thus corresponds to varying the scale factor $a(t)$ (and not a varying speed of light). A decreasing speed of sound corresponds to an expanding universe.

\section{Observables}

One can define many observables for a condensate. To leading order in the field $\hat \phi$, they all are quadratic operators. In stationary and homogeneous cases, they hence only depend on the two quantities
\begin{equation}
\label{eq:defncphonon}
\begin{split}
n_\bk^b = \left < b_\bk^\dagger b_\bk \right >, \quad c_k^b = \left < b_\bk b_{-\bk} \right >.
\end{split}
\end{equation}
They are respectively the mean occupation number of phonon and their coherence. By making different measurements, one may extract these two quantities and drive conclusions on the separability of the state (see \ref{sec:BECseparability} and \ref{chap:separability}). We shall consider here six observables. Five of them are directly obtained and one is directly related to relativistic observables.

The relativistic observable is used in homogeneous systems, where one can define the $\hat \chi_\bk$ operator of Eq.~\eqref{eq:defchi}. Because the system is Gaussian, the commutator of $\hat \chi_\bk$ is independent of the state. All relevant quantities are then encoded in the anticommutator
\begin{equation}
\label{eq:defGackttBEC}
G_{ac}^k(t,t' ) \doteq \left < \{\hat \chi_\bk(t), \hat \chi_{-\bk}(t')\} \right >.
\end{equation}
Using the results of \ref{chap:paramamptopartprod}, we decompose $\hat \chi_\bk = \hat b_\bk \chi_k + \hat b_\bk^\dagger \chi_k^* $ on a basis of unit norm modes (e.g., initially or finally containing only positive frequencies), thus defining the notion of particle. Using such a decomposition, the anticommutator reads, using Eq.~\eqref{eq:defncphonon},
\begin{equation}
\label{eq:GacchiBEC}
\begin{split}
G_{ac}^k(t,t' ) = \Re \left [ \left (n_\bk^b +\frac{1}{2}\right ) \chi_k^*(t) \chi_k(t') +c_k^b \chi_k(t) \chi_k(t') \right ].
\end{split}
\end{equation}
Note that in such expression, the lhs (and thus the rhs) is independent of the notion of particle. It is once a choice for $\chi_k(t)$ has been made that $n_\bk^b $ and $ c_k^b $ exist.

The five observables usually used in BEC split into two distinct groups. In homogeneous settings, they are related to the anticommutator.

\subsection{Density and phase fluctuation}
\label{sec:rhotheta}

\subsubsection{General definitions}

The first group contains the expectation values of either density fluctuation operator or phase operator. They are defined by
\begin{equation}
\label{eq:defdeltarhotheta}
\begin{split}
\delta \hat \rho &\doteq \hat \Phi^\dagger \hat \Phi - \abs{\Phi_{\rm cond}}^2 \sim \rho\left ( \hat \phi + \hat \phi^\dagger \right ), \\
\hat \theta &\doteq \frac{\Phi_{\rm cond}^* \hat \Phi - \Phi_{\rm cond}\hat \Phi^\dagger }{2 i \rho} \sim \frac{\hat \phi	- \hat \phi^\dagger}{2 i}.
\end{split}
\end{equation}
With these definitions, one shows that to first order
\begin{equation}
\hat \Phi(x,t) = \Phi_{cond} \sqrt{ 1+ \frac{\delta \hat \rho }{ \rho } } \ep{ i \hat \theta},
\end{equation}
hence the name of phase operator.

\subsubsection{Homogeneous settings}

In homogeneous settings, it is possible to relate these three observables to the anticommutator of Eq.~\eqref{eq:defGackttBEC}. Indeed, using Eqs.~\eqref{eq:defchi} and~\eqref{eq:partialtchi}, one gets
\begin{equation}
\begin{split}
\hat \chi_\bk &= - \frac{\delta \hat \rho_\bk}{\sqrt{2 \rho c \xi k^2}} ,\quad \partial_t \hat \chi_\bk = \sqrt{2 \rho\xi c k^2 } \hat \theta_\bk,
\end{split}
\end{equation}
where $\delta \hat \rho_\bk$ and $\theta_\bk$ are the Fourier transform of $\delta \hat \rho$ and $\hat \theta$. The three anticommutators of these fields are thus related to our $G_{ac}$ by 
\begin{equation}
\begin{split}
\left < \{ \delta \hat \rho_\bk(t), \delta \hat \rho_{-\bk}(t')\} \right > &= 2 \rho c \xi k^2 G_{ac}^k(t,t') ,\\
\left < \{\hat \theta_\bk(t), \delta \hat \rho_{-\bk} (t')\}\right > &= - \partial_t G_{ac}^k(t,t') ,\\
\left < \{\hat \theta_\bk(t), \hat \theta_{-\bk}(t')\} \right > &= \frac{ \partial_t \partial_{t'} G_{ac}^k(t,t') }{2 \rho c \xi k^2} . 
\end{split}
\end{equation}
The knowledge of $G_{ac}$ of Eq.~\eqref{eq:defGackttBEC} thus fixes the three of them.

\subsection[The \texorpdfstring{${g_1}$}{g1} and the \texorpdfstring{${g_2}$}{g2}]{The \texorpdfstring{$\boldsymbol{g_1}$}{g1} and the \texorpdfstring{$\boldsymbol{g_2}$}{g2}}

The second group of observables are the so-called $g_1$ and $g_2$ functions~\cite{PhysRevA.67.053615,Dalfovo:1999zz}. They are defined by
\begin{subequations}
\label{eq:defg1g2}
\begin{align}
\label{eq:defg1}
g_1 (\bx,t,\bx',t') &\doteq \left < \hat \Phi^\dagger(\bx,t) \hat \Phi(\bx',t') \right > , \\
\label{eq:defg2}
g_2 (\bx,t,\bx',t') &\doteq \frac{ \left < \hat \Phi^\dagger(\bx,t) \hat \Phi^\dagger(\bx',t') \hat \Phi(\bx',t') \hat \Phi(\bx,t) \right >}{g_1 (\bx,t,\bx,t) g_1 (\bx',t',\bx',t')}.
\end{align}
\end{subequations}
The $g_2$ corresponds to the (normalized and normal-ordered in atomic operators) correlation of the density. It is measured by making a statistical average of the correlations of $\rho(t,\bx)$ over different realizations~\cite{sarchi,Carusotto:2012vz}. The normal ordering comes naturally when the measurement destroys the atom. To picture it in simple terms, if one has only one atom in position $\bx$, the quantity $\left <\hat \rho(\bx) \hat \rho(\bx)\right > \neq 0$ while its normal ordered counterpart is $0$. If the observation destroys the atom, it cannot be observed twice. The $g_1$ is obtained in the interference pattern when making interact the condensate with itself.

In term of the relative fluctuations, to first order, they give
\begin{subequations}
\begin{align}
g_1 (\bx,t,\bx',t') &= \Phi_{\rm cond}(\bx,t)^* \Phi_{\rm cond}(\bx',t')\left [1+ \left < \hat \phi^\dagger(\bx,t) \hat \phi(\bx',t') \right > \right ], \\
g_2 (\bx,t,\bx',t') &= 1+ \left < \hat \phi^\dagger(\bx,t) \left [\hat \phi^\dagger(\bx',t')+ \hat \phi(\bx',t')\right ] + \left [\hat \phi^\dagger(\bx',t') + \hat \phi(\bx',t')\right ] \hat \phi(\bx,t)\right >.
\end{align}
\end{subequations}

One can invert Eq.~\eqref{eq:defdeltarhotheta} to express $\hat \phi$ as a function of $ \delta \hat \rho $ and $\hat \theta$. As a consequence, the $g_1$ and $g_2$ functions read in term of the previous observables
\begin{equation}
\begin{split}
g_1 (\bx,t,\bx',t') &= \Phi_{\rm cond}(\bx,t)^* \Phi_{\rm cond}(\bx',t') \\
&\hspace{1cm}\times \left [1+ \left < \left (\frac{\delta\hat \rho(\bx,t)}{\rho(\bx,t)} - i\hat \theta(\bx,t) \right )\left ( \frac{\delta\hat \rho(\bx',t')}{\rho(\bx',t')} + i\hat \theta(\bx',t') \right ) \right > \right ] , \\
g_2 (\bx,t,\bx',t') &= 1+ \frac{\left < \{\delta\hat \rho(\bx,t) , \delta\hat \rho(\bx',t') \}\right >}{\rho(\bx,t)\rho(\bx',t')} - i \frac{[\hat \theta (\bx,t), \delta\hat \rho (\bx',t')]}{\rho (\bx',t')} 
\end{split}
\end{equation}
They have thus simple expressions in term of $G_{ac}^k$ of Eq.~\eqref{eq:defGackttBEC} in homogeneous settings. Indeed, for $\bk \neq 0$, we have the Fourier transform of the two functions
\begin{equation}
\label{eq:defg1k}
\begin{split}
g_1^k (t,t') &\doteq \ep{ i \int_{t}^{t'} \mu/\hbar} \int d\bx \ep{ -i \bk \cdot\bx}g_1 (\bx,t,\bx'= 0,t') \\
&= \frac{1}{2} \left [ c\xi k^2 + \frac{1}{c\xi k^2} \partial_t \partial_{t'} - i (\partial_t-\partial_{t'})\right ]\left [G_{ac}^k(t,t') + G_c^k(t,t') /2 \right ], 
\end{split}
\end{equation}
where $ G_c^k(t,t')$ is the commutator of $\chi$ (which is imaginary), and
\begin{equation}
\label{eq:g2ofGac}
\begin{split}
g_2^k (t,t') &\doteq \rho \int d\bx \ep{ -i \bk \cdot\bx}g_2 (\bx,t,\bx'= 0,t') \\
& = 2 \left [ c\xi k^2 G_{ac} ^k(t,t') - i \partial_t G_c^k(t,t') \right ].
\end{split}
\end{equation}
Using Eq.~\eqref{eq:GacchiBEC}, we see that a measurement of $g_1^k(t,t'=t)$, or $g_2^k(t,t'=t)$, for various $t$ is sufficient to extract the different $n_\bk^b$ and $c_k^b$. 

\subsection{Measure of separability}
\label{sec:BECseparability}

Before we build a tool to determine whether a state is separable or not, we need to determine what we mean when we say that the state of the condensate is separable. We introduced in \ref{chap:separability} very general tools and we noticed that a state is not separable in itself, but that it depends on the notion of particle that we consider. At first view, in a condensate, particles are the atoms. However, we saw in \ref{sec:phonon} that in homogeneous and stationary systems, an other notion of particle exists, i.e., the phonons. We thus have two (or more if many phonon interpretations exist) notions of separability.

In the analogy with gravitation, the equivalent of propagating \enquote{lightlike particles} are the phonon, see Eq.~\eqref{eq:chiofbexp} or Eq.~35 in~\cite{Macher:2009nz}. From the analogue gravity point of view, when trying to determine the separability of Hawking flux or of pair particle production, the right notion is thus the separability of the phonon state. 

To have at our disposal a well defined phonon, we shall suppose that the system is (at least to a good approximation) homogeneous and stationary. With such assumptions, the operators $\hat b_\bk$, the mean occupation number $n_\bk^b$ and the coherence $c_k^b$ are uniquely defined. Using the results of \ref{chap:separability}, the separability condition for the phonons of momentum $k$ reads, see Eq.~\eqref{eq:Camposeparable}
\begin{equation}
\label{eq:sepcondphononsnc}
\abs{c_k^b}^2 \leq n_\bk^b n_{-\bk}^b.
\end{equation}
When considering the phonon of given frequency $\omega$, the dimensionality of the space of solutions may be higher, see \ref{sec:scalarpdt}. As a consequence, the separability criterion is not uniquely defined and many bipartite systems can be considered. This case is considered in more details in \ref{chap:thermalBH}.

We now consider how Eq.~\eqref{eq:sepcondphononsnc} translates into the observables. Because it is much simple in two of them, we only consider these two. First, using Eq.~\eqref{eq:GacchiBEC}, the equal time anticommutator reads
\begin{equation}
\label{eq:Gacofnandc}
G_{ac}^k(t,t'=t) = \frac{n_\bk^b + n_{-\bk}^b + 1 + 2 \abs{c_k^b} \cos\left ( 2 \omega_k t + \arg c_k^b \right ) }{2 \omega_k} .
\end{equation}
In isotropic media ($n_\bk^b = n_{-\bk}^b$), the measure of such quantity is hence sufficient to determine whether the state is separable or not. Second, using Eq.~\eqref{eq:g2ofGac}, the equal time $g_2$ function reads
\begin{equation}
g_2^k (t,t'=t) = c\xi k^2 \frac{n_\bk^b + n_{-\bk}^b + 1 + 2 \abs{c_k^b} \cos\left ( 2 \omega_k t + \arg c_k^b \right ) }{ \omega_k} - 1 .
\end{equation}
In isotropic media, the measure of such quantity is thus also enough to determine the nonseparability. Moreover, at large values of $k$, because $\omega_k \sim c\xi k^2$, we have $g_2^k (t,t'=t)\sim 2 n_k^b + 2 \abs{c_k^b}^2 \cos\left ( 2 \omega_k t + \arg c_k^b \right )$.

To summarize, we have
\begin{itemize}
\item
For states being both homogeneous and isotropic, whenever there exists a time such that $\omega_k G_{ac}^k(t,t'=t) < 1/ 2$, the phonon state of momentum $k$ is nonseparable.
\item
For states being both homogeneous and isotropic, whenever there exists a time such that $ g_2^k (t,t'=t) < c\xi k^2/\omega_k -1 $, the phonon state of momentum $k$ is nonseparable. Moreover, the rhs is $0$ in the large $k$ limit.
\end{itemize}

One could also work at fixed $t$ and vary $t'$. In this case, one would get a periodic behavior of frequency $\omega_{\rm f} /2\pi$ for $G_{ac}$. However, at fixed $t$, $\omega_{\rm f} G_{ac} (t, t')$ is now centered on $0$, and the maxima vary from $n+1/2 -\abs c$ to $n+1/2 +\abs c$ according to the value of the fixed time $t$. Hence, the nonseparability of the state reveals itself in the fact that there exists some values of $t$ such that $\omega_{\rm f} G_{ac} (t, t')$ remains smaller than $1/2$ for all $t'$. 

\section{A particular system, polaritons}
\label{sec:intropolariton}

\subsection{General description of the system}

To get Bose condensation, we saw that three ingredients are necessary and sufficient. First, the particles should be bosons. Second, they should have a mass. And third, they should interact. Photon are bosons, but they are neither massive nor self interacting. However, when putting them between two mirrors, the transverse momentum (which is quantized) plays the role of an effective mass for the lower dimensional effective theory. Indeed, the dispersion relation reads
\begin{equation}
\begin{split}
E^2 = \hbar^2 c^2 \bk_\perp^2 + \hbar^2  c^2 \bk_\parallel^2.
\end{split}
\end{equation}
In this expression, the first term plays the role of a mass term and the second of momentum. On the other hand, if between the two mirrors, one replaces the vacuum by some non linear media, the photon interact with electrons that are self interacting. This introduces some effective interaction between photons. At the quantum level, the particle that propagates is a linear superposition of an electron-hole pair (called exciton) and a photon. This superposition is named polariton or fluid of light\cite{Carusotto:2012vz}.

\begin{SCfigure}[2][b]
\includegraphics[width=0.47\linewidth]{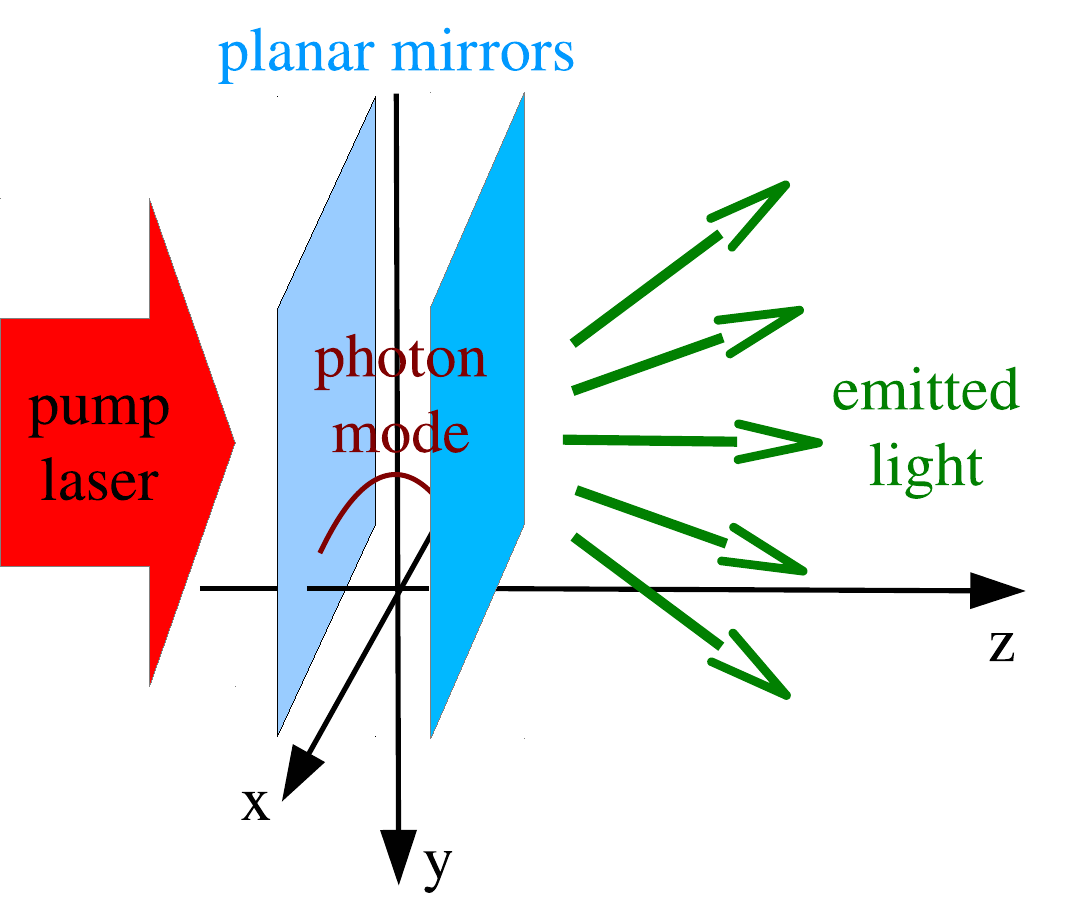}
\caption{Sketch of the planar microcavity system under consideration.}
\label{fig:sketch}
\end{SCfigure} 

In physical systems, because mirrors are never perfectly reflecting, there are inherent losses that one compensates by pumping the system. A sketch of the planar microcavity system we are considering is shown in Fig.~\ref{fig:sketch}. A comprehensive review of its rich physics can be found in Ref.~\cite{Carusotto:2012vz}. Here we shortly summarize the main features that are important for our discussion. In the simplest configuration, light is confined in a cavity material of refractive index $n_0$ sandwitched between two high-quality plane-parallel metallic or dielectric mirrors spaced by a distance $\ell_z$. Photon propagation along the $z$-axis is then quantized as $q_z=\pi M/ \ell_z$, $M$ being a positive integer. For each longitudinal mode $M$, the frequency dispersion of the mode as a function of the in-plane wavevector $\bk$ has the form
\begin{equation}
E_{\rm cav}(k) = \frac{\hbar c}{n_0} \sqrt{q_z^2 + k^2}\simeq E_0^{\rm bare} + \frac{\hbar^2 k^2}{2 m},
\end{equation}
where the effective mass $m$ of the photon and the rest energy $E_0^{\rm bare}$ are related by the relativistic-like expression
\begin{equation}
m= \frac{\hbar q_z}{c/ n_0}= \frac{E_0^{\rm bare}}{c^2/n_0^2}.
\end{equation}
Neglecting for simplicity polarization degrees of freedom, we can define the creation and destruction operators $\hat a^\dagger_{\bk}$ and $\hat a_{\bk}$ for each mode of wavevector $\bk$ and their real-space counterparts $\hat \Phi(\bx)$ as around Eq.~\eqref{eq:HamiltonianBECfirstquantize}. 

\subsection{Hamiltonian of the system}

In term of the quantum field operator $\hat \Phi(\bx)$, the isolated cavity Hamiltonian can be written in the form, see Eq.~\eqref{eq:HamiltonianBECgen}
\begin{equation}
H_0 = \int d\bx \left [ E_0^{\rm bare} \hat \Phi^\dagger \hat \Phi+ \frac{\hbar^2}{2m} (\nabla_\bx\hat \Phi^\dagger) (\nabla_\bx\hat \Phi) + \frac{g_{\rm at}}{2} \hat \Phi^\dagger \hat \Phi^\dagger \hat \Phi \hat \Phi \right].
\end{equation}
The first two terms describe the photon rest energy and its effective (kinetic) mass, respectively. The last term accounts for a Kerr optical nonlinearity of the cavity medium which is essential to have sizable photon-photon interactions. The $g_{\rm at}$ coefficient quantifying the interaction strength is proportional to the material $\chi^{(3)}$: explicit expressions can be found in Ref.~\cite{Carusotto:2012vz}.

In addition to its conservative internal dynamics ruled by $H_0$, the cavity is coupled to external baths including, e.g., the radiative coupling to the propagating photon modes outside the cavity via the (small) transmittivity of the mirrors. A typical way of modeling the dissipative effects due to this environment is based on a Hamiltonian formalism where the environment is described by a phenomenological quantum field $ \hat{ \Psi}_\zeta$. The bath and interaction Hamiltonians have the form
\begin{subequations}
\label{eq:hamiltonian_bath}
\begin{align}
H_{\rm bath} &= \int d\bx \int d\zeta \omega_\zeta \hat{ \Psi}_\zeta^\dagger(\bx) \hat{ \Psi}_\zeta(\bx) \\
H_{\rm int} &= \int d\bx \int d\zeta g_\zeta \left ( \hat \Phi^\dagger(\bx) \hat {\Psi}_\zeta(\bx) + \hat \Phi(\bx) \hat {\Psi}_\zeta ^\dagger(\bx) \right ).
\end{align}
\end{subequations}
This Hamiltonian form is the non relativistic version of the way we introduced dissipation, see. Eq.~\eqref{eq:covaction}. Here, the bath operators $ \hat{ \Psi}_\zeta$ obey the non relativistic ETC $[\hat{ \Psi}_\zeta(\bx,t),\hat{ \Psi}_{\zeta'}^\dagger (\bx',t)] = \delta(\bx-\bx') \delta(\zeta - \zeta')$. In the case of radiative loss processes from a planar microcavity, the $\hat{ \Psi}_\zeta(\bx,t)$ operator corresponds to the destruction operator of extra-cavity photons and the $\zeta$ quantum number indicates the value of the normal component of the extra-cavity wavevector. While more realistic descriptions of the microcavity device can be used to obtain first principle predictions for the $g_\zeta$ coupling constant~\cite{PhysRevA.74.033811}, in the present part we consider it as a (real-valued) model parameter to be adjusted so as to reproduce the experimentally observed photon decay rate. In the quantum optical literature, approaches to dissipation based on Hamiltonians of the form of Eq.~\eqref{eq:hamiltonian_bath} go under the name of input-output formalism~\cite{PhysRevA.74.033811,d1995quantum,gardiner2004quantum}. With respect to equivalent descriptions based on the truncated Wigner~\cite{Gerace:2012an,Koghee2013}, this method has the main advantage that unitarity is manifestly preserved as $\hat \Phi$ and $ \hat{ \Psi}_\zeta$ are treated on an equal footing.

As a last step, we have to include the Hamiltonian term describing the coherent pumping of the cavity by an incident laser field of (normalized) amplitude $F(\bx,t)$,
\begin{equation}
H_F = \int d\bx \left[\hat \Phi(\bx) F^*(\bx,t) + \hat \Phi^\dagger(\bx) F(\bx,t)\right].
 \end{equation}
For a monochromatic pump of frequency $\omega_P$, wavevector $\bk_P$, and a very wide waist, we can perform a plane wave approximation and write
\begin{equation}
 F(\bx,t)=F_0 e^{-i\omega_P t} e^{i\bk\cdot\bx},
\end{equation}
where $\bk$ is the projection of $\bk_P$ along the cavity plane. 

To summarize, the total Hamiltonian has the form
\begin{equation}
\label{eq:hamiltonian}
\begin{split}
H = \int d\bx &\left [ E_0^{\rm bare} \hat \Phi^\dagger \hat \Phi+ \frac{\hbar^2}{2m} (\nabla_\bx\hat \Phi^\dagger) (\nabla_\bx\hat \Phi) + \frac{g_{\rm at}}{2} \hat \Phi^\dagger \hat \Phi^\dagger \hat \Phi \hat \Phi \right. \\
& + \left. \hat \Phi F^* + \hat \Phi^\dagger F + \int d\zeta \omega_\zeta \hat{ \Psi}_\zeta^\dagger \hat{ \Psi}_\zeta + g_\zeta \left ( \hat \Phi^\dagger \hat {\Psi}_\zeta + \hat \Phi \hat {\Psi}_\zeta ^\dagger \right )\right ].
\end{split}
\end{equation}

\section*{Conclusions}
\addcontentsline{toc}{section}{Conclusions}

In this chapter we review how a gas of atoms condensate when one lowers its temperature. We see that the condensate can be described in term of a quantum field and that its Hamiltonian is the Hamiltonian of a free field  in an external potential to which one adds a quartic interaction term. We use the fact that a macroscopic part of the gas condensates to make a perturbative expansion of the field. The first order fluctuations are quantized sound waves, or phonons. We see that these phonons behave to linear order as a free field in a curved space-time, the geometry of which is fixed by the condensate. We also compute observables that are specific to condensates and express the separability condition in term of these. We see that for the $g_2$, the separability condition is simply expressed as a lower bound on the equal time $g_2$. In the final section, we compute the Hamiltonian for a slightly different system that shall be studied in more details in \ref{chap:polariton}. For a system that is pumped and dissipated, one needs to add to the Hamiltonian both the pump term and the term describing the environment responsible for dissipation. 

\chapter{Dissipative phonons}
\label{chap:dissipBEC}

\section*{Introduction}

When considering phonons propagating in a BEC, a linear analysis is generally performed. However, the total Hamiltonian remains fourth order in the phonon operator. These cubic and quartic terms induce two body interactions. The dominant ones are three body and are possible because of dispersion. As announced in \ref{sec:kinematicsofdissip}, they induce dissipation.

In this chapter, we shall consider phonons living on an homogeneous BEC. Using a formalism similar to the one of \ref{chap:dissipdS}, we introduce dissipation for the phonons. We use a large class of dissipative rates $\Gamma$ so that any (local in time) dissipative rate can be recovered. We then consider a homogeneous condensate subject to a sudden change. As a first step, we compute the observables in the absence of dissipation, so as to see the effects of initial temperature on the nonseparability. As a final step, we include dissipation. The separability of the final state is studied. This chapter is mainly based on~\cite{Busch:2013sma}.

\minitoc
\vfill

\section{Actions for dissipative phonons}

\subsection{Effective Hamiltonian for relative density fluctuations}

To get dissipation at the level of phonon, we shall allow atoms to interact with an environment. Because the operation that allows to describe fluctuations from the atom field is not a canonical transformation, we shall use the action formalism. The action for the second quantized field describing a free dilute ultracold atomic gas is, see Eq.~\eqref{eq:HamiltonianBECgen} or~\cite{Dalfovo:1999zz}
\begin{equation}
\label{eq:SPhi1}
\begin{split}
S_\Phi = \int dt d\bx \bigg[ i \hat \Phi^\dagger \partial_t \hat \Phi &- \frac{1}{2m} \partial_\bx \hat \Phi^\dagger \partial_\bx \hat \Phi - V \hat \Phi^\dagger \hat \Phi - \frac{g_{at}}{2} \hat \Phi^\dagger \hat \Phi^\dagger \hat \Phi \hat \Phi\bigg]. 
\end{split}
\end{equation}
We shall take the relativistic environment of Eq.~\eqref{eq:covaction} (with preferred time being the lab time). We want the interaction term to conserve the number of atoms (for the case where atoms get dissipated, see \ref{chap:polariton}). We shall thus chose
\begin{subequations}
\begin{align}
\label{eq:SPsi}
S_\Psi &= \frac12 \int dt d\bx \int_{-\infty}^\infty d\zeta \left \{ \abs{\partial_t \bar \Psi_\zeta}^2 - \abs{\pi \zeta \bar \Psi_\zeta}^2 \right \},\\
S_{\it int} &= - \int dt d\bx \left \{ [(\hat \Phi^{\dagger})^{\alpha} \hat \Phi^\alpha] \tilde g ( \partial_\bx) \partial_t (\int d\zeta \bar \Psi_\zeta) \right \}.
\label{eq:Sint2}
\end{align}
\end{subequations}
Notice that $S_\Psi$ contains no spatial gradients, and that it does not depend on $\rho$. Hence, the kinematics of the environment degrees of freedom is independent of both $k$ and $\rho$. This is a simplifying approximation. In fact, as in \ref{chap:dissipdS}, our philosophy is to choose the simplest model that possesses some key properties. These are as follows: unitarity of the whole system, standard action for the phonons, and ability to reproduce most of the decay rates. From such a system, equations of motion read
\begin{subequations}
\begin{align}
\label{eq:eomincondfieldPhi}
i\partial_t \hat \Phi &= \left ( -\frac{1}{2m} \partial_\bx^2 + V + g_{at}\hat \Phi^\dagger \hat \Phi \right )\hat \Phi + \alpha (\hat \Phi^{\dagger})^{\alpha-1} \hat \Phi^{\alpha} \tilde g(\partial_\bx) \partial_t (\int d\zeta \bar \Psi_\zeta),\\
(\partial_t^2 &+ \omega_\zeta^2) \bar \Psi_\zeta = \partial_t \tilde g ( -\partial_\bx) [(\hat \Phi^{\dagger})^{\alpha} \hat \Phi^\alpha].
\end{align}
\end{subequations}
The condensation of $\hat \Phi$ thus engenders the condensation of the environment. We thus write it as
\begin{equation}
\begin{split}
\bar \Psi_\zeta = \Psi_\zeta^{ \rm cond} + \hat \Psi_\zeta.
\end{split}
\end{equation}
At first order, Eq.~\eqref{eq:eomincondfieldPhi}, implies a modified Gross-Pitaevskii equation, see Eq.~\eqref{eq:GPE}:
\begin{equation}
\begin{split}
i\partial_t \Phi_{\rm cond} = \mathcal{H}_{{\rm cond}} \Phi_{\rm cond} ,
\end{split}
\end{equation}
with 
\begin{equation}
\begin{split}
\mathcal{H}_{{\rm cond}} = \frac{-\partial_\bx^2}{2m} &+ V + g_{at}\abs{\Phi_{\rm cond}}^2 + \alpha \abs{\Phi_{\rm cond}}^{\alpha-1} \tilde g(\partial_\bx) \partial_t (\int d\zeta \bar \Psi_\zeta^{ \rm cond} ) .
\end{split}
\end{equation}
Since this last term does not contain any derivative acting on $\Phi_{\rm cond}$, the conservation Eq.~\eqref{eq:conservationequation} remains unchanged. It gives rise, in the quadratic part of Eq.~\eqref{eq:SPhi1}, to a term that is to be added to the standard action, see Eq.~\eqref{eq:HamiltonianBECerturb}
\begin{equation}
\label{eq:addedterm}
\begin{split}
\delta S_{\phi } = \int dt d\bx \alpha \phi^\dagger \phi \rho^{\alpha} \tilde g(\partial_\bx) \partial_t (\int d\zeta \Psi_\zeta^{ \rm cond} ).
\end{split}
\end{equation}
On the other hand, Eq.~\eqref{eq:Sint2} gives two contributions:
\begin{equation}
\begin{split}
S_{int}^{(1)} &= - \int dt d\bx \bigg \{ [ \frac{\alpha(\alpha -1)}{2}(\hat \phi^{\dagger})^{2} + \alpha^2 \hat \phi^{\dagger} \hat \phi + \frac{\alpha(\alpha -1)}{2} \hat \phi^2] \rho^\alpha \tilde g ( \partial_\bx) \partial_t (\int d\zeta \Psi_\zeta^{ \rm cond} ) \bigg \},\\
S_{int}^{(2)} &= - \int dt d\bx \left \{ \alpha \rho^\alpha[\hat \phi^{\dagger}+ \hat \phi] \tilde g ( \partial_\bx) \partial_t (\int d\zeta \hat \Psi_\zeta) \right \}.
\end{split}
\end{equation}
The first one is quadratic in $\phi$ and zeroth order in $ \hat \Psi_\zeta$. It combines with Eq.~\eqref{eq:addedterm} to give 
\begin{equation}
 \begin{split}
\delta S_{\phi }+ S_{int}^{(1)} = \int dt d\bx &\frac{\alpha(\alpha-1)}{2} (\phi^\dagger + \phi)^2 \rho^{\alpha} \tilde g(\partial_\bx) \partial_t (\int d\zeta \Psi_\zeta^{ \rm cond} ) .
 \end{split}
 \end{equation} 
The role of this term is to modify the speed of sound which is now given by 
\begin{equation}
\begin{split}
m c^2 \doteq g_{at} \rho + \frac{\alpha(\alpha-1)}{2} \rho^{\alpha-1} \tilde g(\partial_\bx) \partial_t (\int d\zeta \Psi_\zeta^{ \rm cond} ). 
\end{split}
\end{equation}
With this new definition, the free Hamiltonian is still of the form of Eq.~\eqref{eq:HamiltonianBECerturb}
\begin{equation}
\begin{split}
H_\phi = \int d\bx \rho \big \{ \frac{ 1}{2 m} \partial_\bx\hat \phi^\dagger \partial_\bx \hat \phi - i\hbar \hat \phi^\dagger \bv\partial_\bx \hat \phi  + m c^2 (\hat \phi+\hat \phi^\dagger)^2 \big \},
\end{split}
\end{equation}
while the free Hamiltonian for the environment is simply $\int d\bx d\zeta (\pi\zeta)^2 \hat \Psi_\zeta^\dagger \hat \Psi_\zeta$.
The second contribution $S_{int}^{(2)}$ gives rise to the interaction Hamiltonian 
\begin{equation}
\begin{split}
H_{\it int} &= \int d\bx \frac{g}{\sqrt{\xi}} \left \{ (\rho \xi)^\alpha (\xi \partial_\bx)^n (\hat \phi + \hat \phi^\dagger) \partial_t (\int d\zeta \hat \Psi_\zeta) \right \}.
\label{eq:Sint}
\end{split}
\end{equation}
where we posed $ \alpha \tilde{g}(-\partial_\bx) = g \xi^{\alpha+1/2} (\xi \partial_\bx)^n$. A linear superposition of such interaction Hamiltonian with different $n$ and $\alpha$ may be made so as to get, for the dissipative rate, any function of the momentum.

\subsection{Field equations and effective dispersion relation}

Because the condensate is homogeneous, it is appropriate to work with the Fourier components at fixed wave vector $\bk = - i\partial_\bx$, where $\bk$ is real. Then the total Hamiltonian splits into sectors that do not interact: $H_T = \int d\bk H_\bk$, with $H_\bk = H_{-\bk}^\dagger$. In the rest frame of the condensate, at fixed $\bk$, the field equations are, see Eq.~\eqref{eq:eomhomogephonon}
\begin{subequations}
\label{eq:eomphipsi}
\begin{align}
\label{eq:eomphi}
i\partial_t \hat \phi_\bk &= \Omega_k \hat \phi_\bk + m c^2 \hat \phi_{-\bk}^\dagger + \frac{\gamma_\bk }{\sqrt{\rho}} \partial_t \int d\zeta \hat \Psi_{\zeta,\bk}, \\
\label{eq:eompsi}
[\partial_t^2 + (\pi \zeta )^2 ] \hat \Psi_{\zeta,\bk} &= \partial_t\left \{ \gamma_\bk^* \sqrt{\rho} (\hat \phi_\bk + \hat \phi^\dagger_{-\bk})\right \} \doteq \hat {j}_{\Psi,\bk}.
\end{align}
\end{subequations}
where $\Omega_k$ was given in Eq.~\eqref{eq:OmegakBEC} and where
\begin{equation}
\begin{split}
 \gamma_\bk &= g (\rho \xi)^{\alpha - 1/2} (i \xi \bk)^n , 
\label{eq:gamma}
\end{split}
\end{equation}
is the effective dimensionless coupling for the wave number $k$. The solution of Eq.~\eqref{eq:eompsi} is [see Eq.~\eqref{eq:psizeta}]
\begin{equation}
\label{eq:psidecom}
\begin{split}
\hat \Psi_{\zeta,\bk} = \hat \Psi_{\zeta,\bk}^0 + \int dt' R^0_\zeta(t,t') \hat{ j}_{\Psi,\bk}(t'),
\end{split}
\end{equation}
where $\hat \Psi_{\zeta,\bk}^0$ is a homogeneous solution we shall describe below, and where $R^0_\zeta$ is the retarded Green function. It is independent of $\bk $ because the action $S_\Psi$ contains no spatial gradient. When summing over $\zeta$, it obeys, see Eq.~\eqref{eq:loc} or Refs.~\cite{Unruh:1989dd,Parentani:2007uq} 
\begin{equation}
\label{eq:localgretofpsi}
\begin{split}
\partial_t \left (\int d\zeta R^0_\zeta(t,t') \right ) = \delta(t-t').
\end{split}
\end{equation}
This guarantees that the kernel encoding dissipation is local.
 Indeed, inserting Eq.~\eqref{eq:psidecom} into Eq.~\eqref{eq:eomphi}, and using Eq.~\eqref{eq:localgretofpsi}, Eq.~\eqref{eq:eomphi} gives 
\begin{equation}
\label{eq:eqofmot}
\begin{split}
 i\partial_t \hat \phi_\bk &= \Omega_k \hat \phi_\bk + m c^2 \hat \phi_{-\bk}^\dagger + \gamma_\bk \partial_t \hat \Psi_\bk^0 /\sqrt{ \rho} + \gamma_\bk \partial_t\left \{ \gamma_\bk^* (\hat \phi_\bk + \hat \phi_{-\bk}^\dagger)\right \} ,
\end{split}
\end{equation}
where $\hat \Psi_\bk^0 = \int d\zeta \hat \Psi_{\zeta,\bk}^0$. As announced, the last term in the r.h.s. is local in time. Basically all other choices of $S_\Psi$ and $S_{\it int}$ would give a nonlocal kernel.\footnote{
In this respect, we notice that one can generalize our model so as to deal with an \textit{ab initio} computed nonlocal dissipative kernel $D(t-t')$. To do so, one should compute its Fourier transform $D(\omega)$ (which is $1/(\omega+i\epsilon)$ in our model) and replace $\int d\zeta \hat \Psi_\zeta$ by $\int d\zeta \sqrt{\omega_\zeta D(\omega_\zeta)} \hat \Psi_\zeta$ in Eq.~\eqref{eq:Sint}.} 
In these cases, Eq.~\eqref{eq:eqofmot} would be an integralo-differential equation. These more complicated models do not seem appropriate to efficiently calculate the consequences of dissipation on phonon correlation functions.

To get the effective dispersion relation, we consider Eq.~\eqref{eq:eqofmot} when all background quantities are constant and when $\Psi^0_\bk = 0$. We get
\begin{equation}
\label{eq:disprelBEC}
\begin{split}
& (\Omega + i \Gamma_k )^2 = \tilde \omega_k^2 ,
\end{split}
\end{equation}
where 
\begin{equation}
\label{eq:omGamma}
\begin{split}
 \tilde \omega_k^2 = \omega_k^2 - \Gamma_k^2, \quad \Gamma_k = \abs{\gamma_\bk}^2 k^2 c \xi . 
\end{split}
\end{equation}
and where $\omega_k$ was defined in Eq.~\eqref{eq:defomegak}. Using Eq.~\eqref{eq:gamma}, we verify that the second equation delivers
\begin{equation}
\begin{split}
\Gamma_k = g^2 {(c / \xi)} (\rho \xi)^{2\alpha - 1} {(k \xi)^{2+2 n}} . 
\label{eq:Gammapheno}
\end{split}
\end{equation} 
Hence, by choosing $g$, $\alpha$ and $n$ our model shall be able to reproduce many ab initio computed decay rates. As expected, we also verify that for $\abs{\gamma_\bk}^2 \to 0$, one recovers the standard Bogoliubov dispersion for all $k$. 

In Eq.~\eqref{eq:Gammapheno}, the coupling constant $ g$ is dimensionless, $\rho$ is the condensed atom density, and $\xi$ a short distance length which corresponds to the healing length in atomic Bose gases. The powers $n$ and $\alpha$ can be chosen to reproduce effects computed from first principles. For instance, in Bose gases, two types of dissipative effects are found: The first one, called Beliaev decay~\cite{pitaevskii2003bose}, scales with $\alpha=0$ and $n=3/2$, while the second one, the Landau decay~\cite{pethick2002bose}, depends on the temperature, has also $\alpha=0$, and is proportional to $ck \sqrt{1 + k^2 \xi^2}$.

In what follows, the quantities $g,c,\xi$ depend on time, while preserving the homogeneity. Hence the nontrivial dynamics will occur within two mode sectors $\{\bk, - \bk\}$. 

\subsection{Time-dependent settings}

In all this chapter, because we work with a relativistic environment, and because we work in homogeneous time-dependent systems we shall use the relativistic field defined in Eq.~\eqref{eq:defchi}. Using such field, $S_\bk$, the action of the $\bk$ sector, reads 
\begin{equation}
\begin{split}
S_\bk= \frac{1}{2} \int dt &\Big\{\vert \partial_t \hat \chi_\bk \vert^2 - \omega_k^2 \vert \hat \chi_\bk \vert^2 + \int d\zeta {\vert \partial_t \hat \Psi_{\zeta,\bk}\vert^2 } - (\pi \zeta)^2 \vert {\hat \Psi_{\zeta,\bk} \vert^2 }\\
&+ 2 \hat \chi^\dagger_{\bk} \sqrt{2 \Gamma_k} \partial_t \int d\zeta\hat \Psi_{\zeta,\bk} \Big\}~ ,
\end{split}
\end{equation}
where $c$, $\xi$, and $\Gamma$ are arbitrary (positive) time-dependent functions, and where a phase $( -i sgn(\bk))^n$ has been absorbed in $ \hat \Psi_{\zeta,\bk}$. In an atomic condensate, $c$ and $\xi$ are related by $c \xi = 1/2m = cst$.

From the above action, or from Eq.~\eqref{eq:eqofmot}, we get the equation for $\chi_\bk$:
\begin{equation}
\label{eq:eomchi}
\left [ (\partial_t + \Gamma_k)^2 + \tilde \omega_k^2\right ] \hat \chi_\bk = \sqrt{2 \Gamma_k }\partial_t \hat \Psi_\bk^0 .
\end{equation}
The general solution can be written as
\begin{equation}
\begin{split}
\hat \chi_\bk(t) &= \hat \chi_\bk^{dec}(t;t_0) + \hat \chi_\bk^{dr}(t;t_0), 
\label{eq:decomp}
\end{split}
\end{equation}
where the driven part $\hat \chi_\bk^{dr}(t;t_0)$ and its temporal derivative vanish at $t=t_0$. The decaying part $\hat \chi_\bk^{dec}(t;t_0)$ is thus the solution of the homogeneous equation which obeys the ETC at that time. Hence it possesses the following decomposition:
\begin{equation}
\label{eq:chidec}
\begin{split}
\hat \chi_\bk^{dec}(t;t_0) &= \ep{- \int_{t_0}^t\Gamma_k dt' } \left ( \hat b_\bk \bar \chi_k(t) + \hat b^\dagger_{-\bk} \bar \chi_{k}^*(t) \right ), 
\end{split}
\end{equation}
where the destruction and creation operators $\hat b_\bk, \hat b^\dagger_{-\bk}$ obey the standard canonical commutators $[\hat b_{\bk},\hat b^\dagger_\bk]=1$ and correspond to the phonic operators of Eq.~\eqref{eq:chiofbexp} in the decoupling limit $\Gamma \to 0$, and where $\bar \chi_k$ is a solution of [compare with Eq.~\eqref{eq:generaleomchibar}]
 \begin{equation}
 \label{eq:homoeqvarphi}
\begin{split}
(\partial_t ^2 + \omega_k^2 ) \bar \chi_k = 0, 
\end{split}
\end{equation}
of unit Wronskian $i(\bar \chi_k^* \partial_t \bar \chi_k -\bar \chi_k \partial_t \bar \chi_k^* ) =1$. The usefulness of this decomposition is twofold. On the one hand, $t_0$ can be conceived as the initial time when the state is fixed. The operators $\hat a^\dagger_{\bk},\hat a_\bk $ can then be used to specify the particle content of this state. On the other hand, Eq.~\eqref{eq:decomp} and Eq.~\eqref{eq:chidec} furnish an \enquote{instantaneous} particle representation around any time $t_0$. Indeed, in the limit $\Gamma/\omega \ll 1$ and $\Gamma (t-t_0)\ll 1$, the contribution of $\hat \chi_\bk^{dr}(t;t_0)$ can be neglected, and $\hat \chi_\bk (t)\sim \hat \chi_\bk^{dec}(t;t_0)$ behaves as a standard canonical field since the prefactor of Eq.~\eqref{eq:chidec} is approximately equal to $1$. We shall return to this in \ref{sec:nbofpartdissip}.

The driven part of Eq.~\eqref{eq:decomp} is given by [see Eq.~\eqref{eq:phidr}]
\begin{equation}
\begin{split}
\hat \chi_\bk^{dr}(t,t_0) = \int_{t_0}^\infty dt' G_{\rm ret}^\chi (t,t';k) \sqrt{2 \Gamma_k(t') } \partial_{t'} \hat \Psi_\bk^0(t'),
\end{split}
\end{equation}
where $G_{\rm ret}^\chi$ is the retarded Green function of Eq.~\eqref{eq:eomchi}. Using the unit Wronskian solution of Eq.~\eqref{eq:homoeqvarphi}, it can be expressed as, see Eq.~\eqref{eq:Gret-solution}
\begin{equation}
\label{eq:Gret}
\begin{split}
G_{\rm ret}^\chi(t,t';k) = \theta (t - t') \ep{-\int_{t'}^{t} \Gamma_k dt} \times 2 \Im(\bar \chi_k(t)\bar \chi_k^*(t')) .
\end{split}
\end{equation}

Since $\hat \chi$ is a canonical and linear field, the standard relation between the commutator and the retarded Green function holds, see Eq.~\eqref{eq:GcfromGret}. In consequence, when the state of the system is Gaussian and homogeneous, the reduced state of $\chi$ is completely fixed by its anticommutator. Because of Eq.~\eqref{eq:decomp}, it contains three terms
\begin{equation}
\label{eq:Gact0}
\begin{split}
G_{ac}(t,t';k ) &\doteq \left < \{\hat \chi_\bk(t), \hat \chi_{-\bk}(t')\} \right > = G_{ac}^{dec} +G_{ac}^{dr} +G_{ac}^{mix} .
\end{split}
\end{equation}
The first one decays and is governed by $\hat \chi^{dec}$
\begin{equation}
\begin{split}
G_{ac}^{dec}(t,t';k) = \left < \{\hat \chi_\bk^{dec}(t), \hat \chi_{-\bk}^{dec}(t')\}\right >.
\end{split}
\end{equation}
The second one is driven and governed by $\hat \Psi^0$
\begin{equation}
\label{eq:Gacasintegrals}
\begin{split}
G_{ac}^{dr}(t,t';k) = &\int_{t_0}^\infty d\tau d\tau' \sqrt{4\Gamma_k(\tau)\Gamma_k(\tau')}G_{\rm ret}^\chi(t,\tau) G_{\rm ret}^\chi(t',\tau' ) \partial_\tau \partial_{\tau'} \left < \{\hat \Psi_\bk^0(\tau), \hat \Psi_{-\bk}^0(\tau')\}\right >.
\end{split}
\end{equation}
The third one describes the correlations between $\chi$ and $\Psi$. It is nonzero either when the initial state is not factorized as $\hat \rho = \hat \rho_\chi \otimes \hat \rho_\Psi$, or when the two fields have interacted. It is given by twice the symmetrization of 
\begin{equation}
\label{eq:Gacmix}
\begin{split}
\widetilde G_{ac}^{mix}(t,t';k) \doteq \int_{t_0}^\infty d\tau & \sqrt{2\Gamma_k(\tau) }G_{\rm ret}^\chi(t',\tau) \partial_\tau \left < \{ \hat \chi_\bk^{dec}(t),\hat \Psi_{-\bk}^0(\tau)\}\right >.
\end{split}
\end{equation}
When the state is prepared at an early time $ \Gamma (t -t_0)\gg 1$, only the driven term significantly contributes to Eq.~\eqref{eq:Gact0}, which means that the system would have thermalized with the bath. 

At fixed $\bk$ and $\zeta$, $\hat \Psi_{\zeta,\bk}^0$, the homogeneous solution of Eq.~\eqref{eq:eomphipsi} is a complex harmonic oscillator of pulsation $ \omega_\zeta = \pi \abs{\zeta}$. It thus reads
\begin{equation}
\label{eq:psi0fock}
\hat \Psi_{\zeta, \bk}^0 (t) = \frac{\ep{- i \omega_\zeta t} \hat c_{\zeta, \bk} +\ep{ i \omega_\zeta t} \hat c_{-\zeta, -\bk}^\dagger }{\sqrt{2 \omega_\zeta}}, 
\end{equation}
where $\hat c_{\zeta, \bk}$ and $\hat c_{\zeta,\bk}^\dagger $ are standard destruction and creation operators. In the following sections we shall work with thermal baths at temperature $T$. In such states, the noise kernel entering Eq.~\eqref{eq:Gacasintegrals} is, 
\begin{equation}
\label{eq:Trpsipsi}
\left < \{\hat \Psi_\bk ^0(\tau),\hat \Psi_{-\bk}^0(\tau')\}\right > = \int_{0}^\infty \frac{d\omega_\zeta }{ \pi } \frac{\coth\left (\frac{\omega_\zeta}{2 k_B T} \right )}{2 \omega_\zeta} \cos[\omega_\zeta (\tau - \tau')] .
\end{equation}
We notice that it does not depend on $k$. 


We study the time evolution of the phonon state when making two assumptions. First, we consider states which are prepared a long time before the experiment, so that Eq.~\eqref{eq:Gact0} is given by Eq.~\eqref{eq:Gacasintegrals}, with $t_0 = -\infty$. Second, we suppose a sudden change of the condensed atoms occurs at time $t=0$. In \ref{app:modulDCE}, we study the case where the change is modulated in time at a given frequency. The speed of sound $c$ and the effective coupling $\gamma_\bk$ entering Eq.~\eqref{eq:eomchi} will change on a similar time scale. Since approximating the change of the sound speed by a step function only modifies the response for very high $k$, as can be seen in Ref.~\cite{Carusotto:2009re}, for simplicity, we shall work with an instantaneous change for $c$. For $\gamma_\bk$ instead, we shall use a continuous profile because an instantaneous change would lead to divergences, as we shall see below. Hence we shall work with
\begin{equation}
\label{eq:grad}
\begin{split}
 c(t) &=c_{\rm f} + (c_{\rm in} -c_{\rm f}) \theta(-t) ,\\
 \gamma_\bk (t) &= \gamma_{\rm f} + (\gamma_{\rm in} - \gamma_{\rm f}) h(\kappa t), 
\end{split}
\end{equation}
where $h(\kappa t)$ is a smoothing out function which goes from $1$ to $0$ around $t= 0$ in a time lapse of the order of $1/\kappa$. 
At this point, it should be noticed that any physical system, such as an atomic Bose condensate described by Eq.~\eqref{eq:GPE}, would only respond after a finite amount of time of the order of $\xi/c$.
Hence the function $h$ of Eq.~\eqref{eq:grad} is physically meaningful, and $\kappa$ should be of the order of $c/\xi$.

Considering the asymptotic values of these profiles, we shall use the following notations
\begin{equation}
\label{eq:defGammaomegainandf}
\begin{split}
\Gamma_{\rm in / f} &\doteq \gamma_{\rm in / f}^2 (c \xi) k^2 ,\\
\tilde \omega_{\rm in / f}^2 &\doteq c_{\rm in / f}^2 k^2 + (c \xi)^2 k^4 - \Gamma_{\rm in / f}^2, 
\end{split}
\end{equation}
see Eq.~\eqref{eq:omGamma}. 

Before considering dissipation, we study the dispersive case, first, to analyze the reduction of the correlations due to stimulated processes and, second, to know the outcome in the dissipation free case, so as to be able to isolate the consequences of dissipation.

\section{The dispersive case}
\label{sec:nodiss}

\subsection{Analytic study}

\begin{SCfigure}[2]
\includegraphics[width=0.47\linewidth]{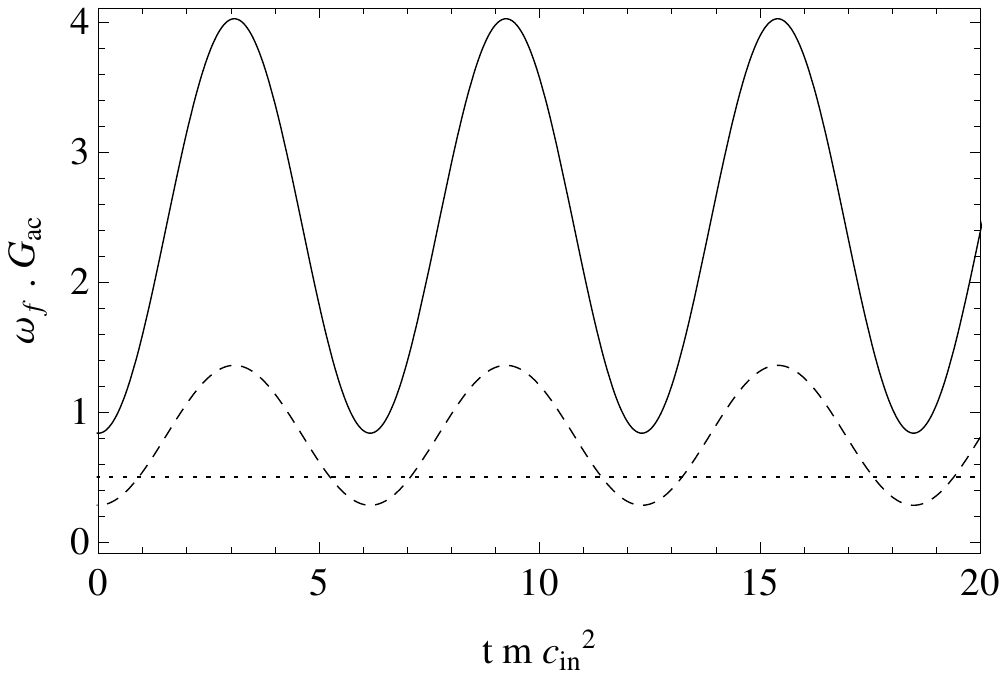}
\caption{We represent the product $\omega_{\rm f} \times G_{ac}(t,t'=t)$, where the anticommutator $G_{ac}$ is given in Eq.~\eqref{eq:gacofnc}, as a function of the adimensionalized time $t m c_{\rm in}^2$, for $ k = m c_{\rm in} $, and for two values of the temperature, namely, $ T_{\xi_{\rm in}} /2 $ (dashed line) and $ 2 T_{\xi_{\rm in}} $ (solid line); see Eq.~\eqref{eq:defTxiin}. The value of the jump is $c_{\rm f}/ c_{\rm in} = 0.1$. As explained in the text, the dotted line gives the threshold value $1/2$ which distinguishes nonseparable states. When increasing the temperature, the contribution of the stimulated amplification with respect to the spontaneous one is larger. As a result, the coherence is reduced; i.e., the minima of $\omega_{\rm f} \times G_{ac} $ are increased.} 
\label{fig:Gacoft}
\end{SCfigure}

\begin{figure}[htb] 
\begin{minipage}{0.47\linewidth}
\includegraphics[width=1 \linewidth]{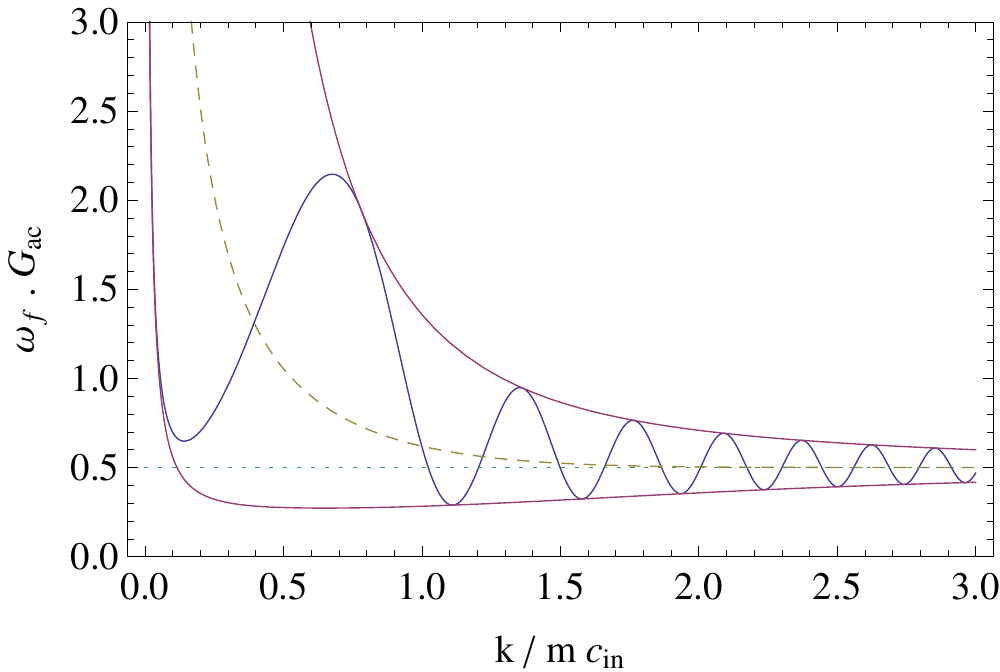}
\end{minipage}
\hspace{0.03\linewidth}
\begin{minipage}{0.47\linewidth}
\includegraphics[width=1 \linewidth]{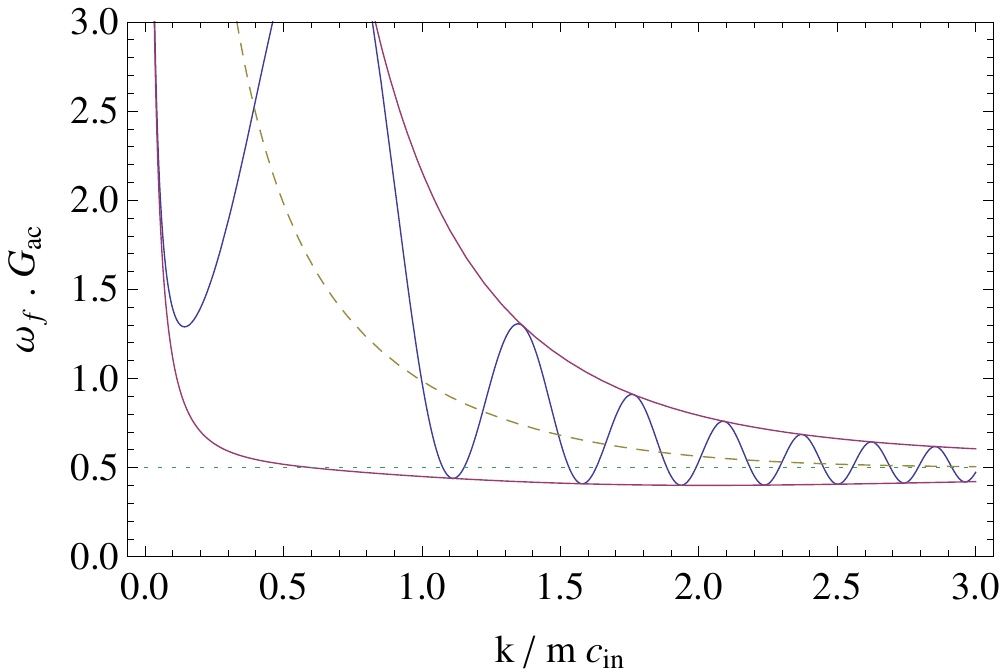}
\end{minipage}
\caption{The anticommutator of Eq.~\eqref{eq:gacofnc} multiplied by $\omega_{\rm f}$ as a function of $k /m c_{\rm in}$, when evaluated at equal time $t=t' = 5 / m c_{\rm in}^2$, for $c_{\rm f} = c_{\rm in}/10$, on the left panel for $T = T_{\xi_{\rm in}}/2$ and on the right for $T=T_{\xi_{\rm in}}$ (solid blue oscillating curves). The dashed yellow line in the middle gives the value of $\omega_{\rm f} G_{ac} = n^{in} + 1/2$ before the sudden change. The envelopes of the minima and maxima are indicated by solid purple lines. One clearly sees that the domain of $k$, where the lower line is below the threshold value $1/2$, is reduced when increasing the temperature. Notice that all curves asymptote to $1/2$ because in the limit $k\to \infty$, one has $n_k=c_k=0$. }
\label{fig:Gacofk}
\end{figure} 

\begin{figure}
\begin{minipage}{0.43\linewidth}
\includegraphics[width=1\linewidth]{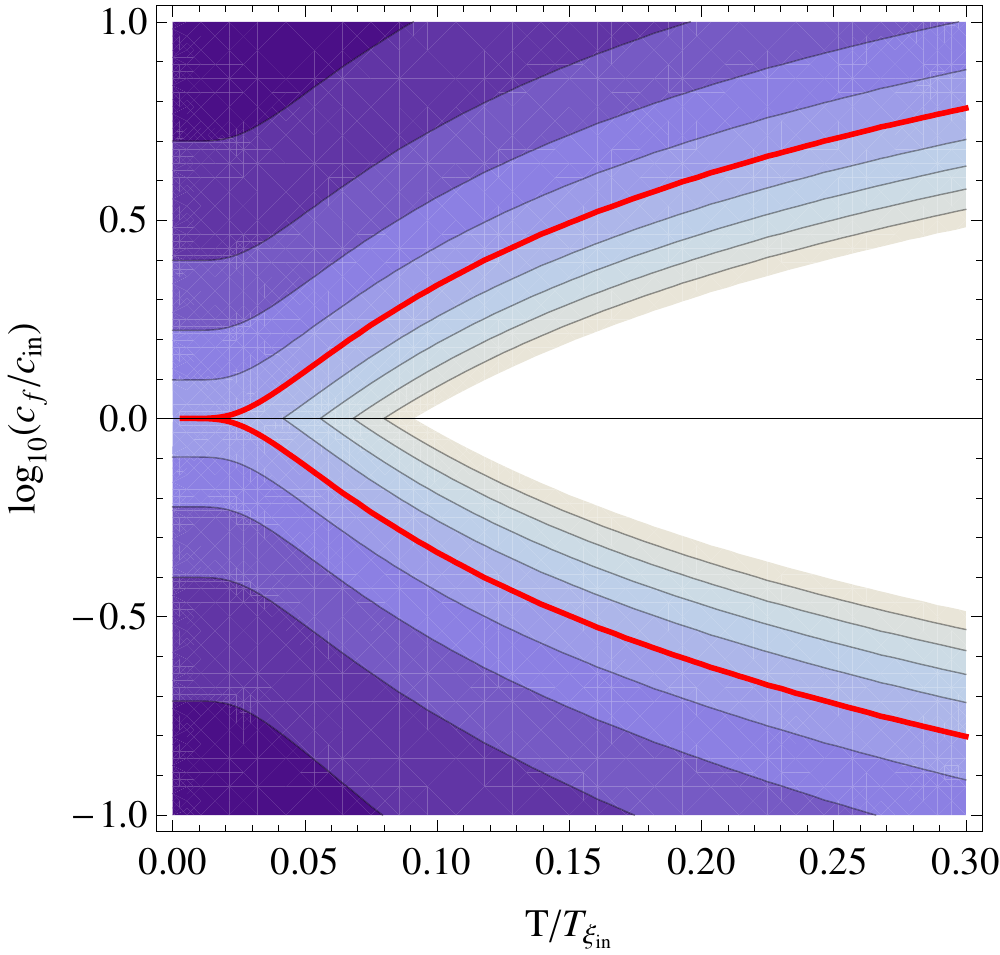}
\end{minipage}
\hspace{0.01\linewidth}
\begin{minipage}{0.09\linewidth}
\includegraphics[width = 1 \linewidth]{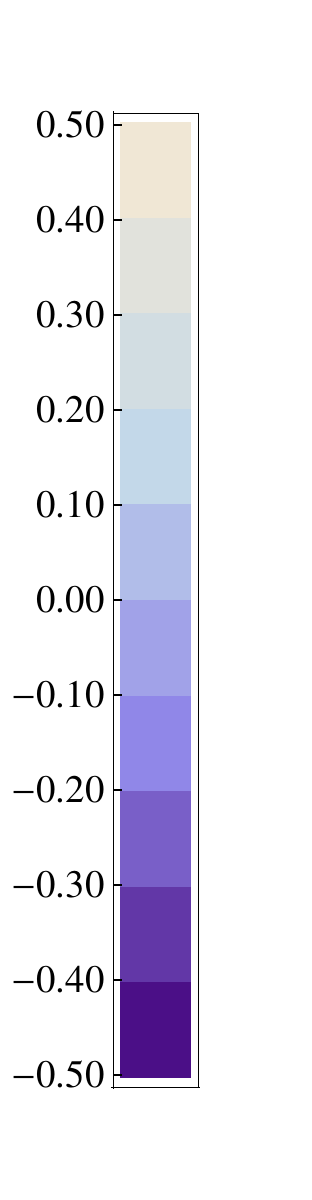}
\vspace{0.5cm} 
\end{minipage}
\begin{minipage}{0.43\linewidth}
\includegraphics[width=1\linewidth]{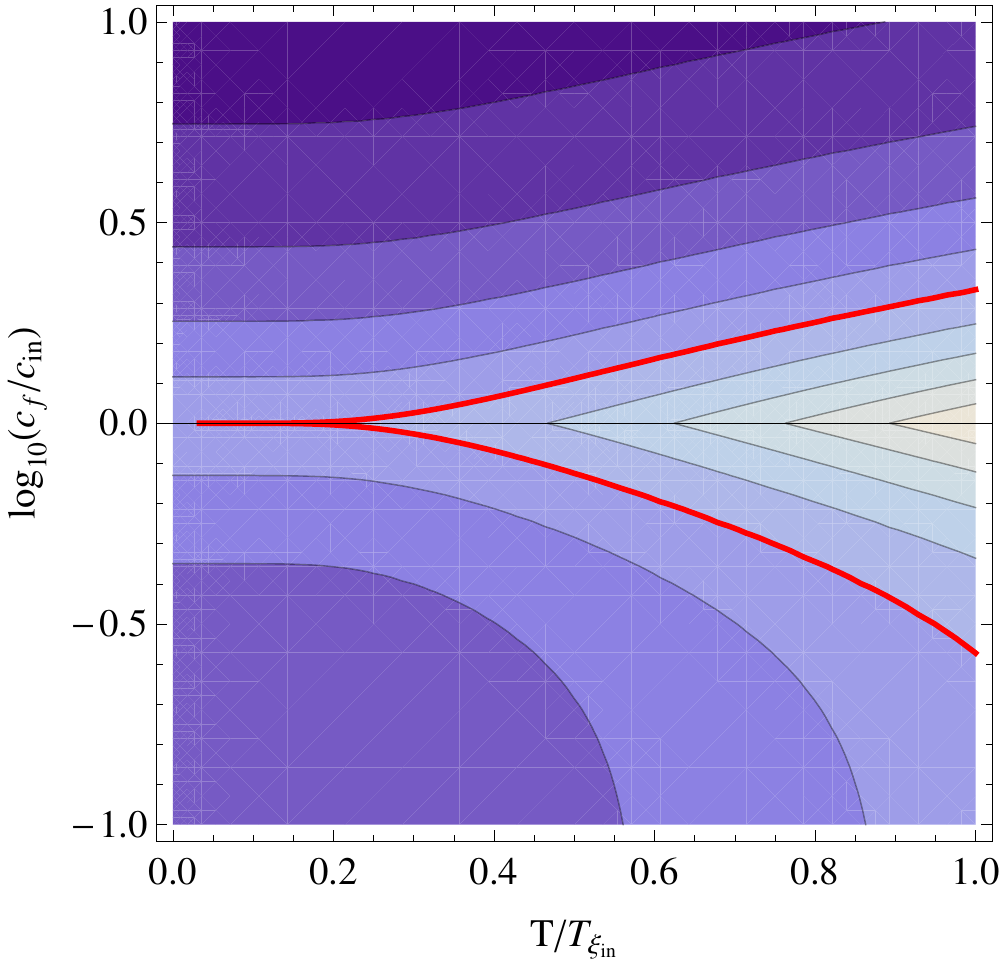} 
\end{minipage}
\caption{Contour plot of $\Delta^{out}$ induced by a sudden variation of the sound speed, as a function of temperature $T/T_{\xi_{\rm in}}$ and the logarithm of the ratio $c_{\rm f}/c_{\rm in}$. In the left panel, $k = m c_{\rm in}/10$ is in the hydrodynamical regime, and in the right panel, $k = m c_{\rm in}$. The threshold value $\Delta^{out} = 0$ is indicated by a thick line. One clearly sees the competition between the height of the jump governed by $c_{\rm f}/c_{\rm in}$ which increases the coherence, i.e., reduces the value of $\Delta$, and the initial occupation number which increases $\Delta$. One also sees that the state of a higher momentum mode stays nonseparable for higher temperatures. }
\label{deltadispgradino1}
\end{figure} 

In the absence of dissipation, phonon excitations can be analyzed before and after the jump using a standard particle interpretation. Hence, the consequences of the jump are all encoded in the Bogoliubov coefficients $\alpha,\beta$ entering
\begin{equation}
\begin{split}
\chi_{in} &= \alpha \chi_{out} + \beta \chi_{out}^* ,
\label{eq:inout}
\end{split}
\end{equation}
which relate the $in$ mode to the $out$ mode. These modes have a positive unit Wronskian and are equal to $\ep{-i \omega_{\rm in / f} t} \big/ \sqrt{2 \omega_{\rm in / f} }$ for $t<0$ or $t>0$ respectively. Using Eq.~\eqref{eq:homoeqvarphi}, one verifies that the modes are $\mathcal{C}^1$ across the jump. From the junction conditions, one finds the Bogoliubov coefficients~\cite{Carusotto:2009re} 
\begin{equation}
\label{eq:bogodisp}
\alpha = \frac{\omega_{\rm f}+\omega_{\rm in}}{2\sqrt{\omega_{\rm f} \omega_{\rm in}}}, \quad \beta = \frac{\omega_{\rm f}-\omega_{\rm in}}{2\sqrt{\omega_{\rm f} \omega_{\rm in}}}.
\end{equation}

To prepare the comparison with the dissipative case, the initial phonon state is taken to be a thermal bath at temperature $T$. This means that the initial mean occupation number is $n^{in} = 1/({\ep{\omega_{\rm in} / T} -1 })$ and that $c^{in}$, the initial correlation term between $\bk$ and $-\bk$, vanishes. After the jump, the mean occupation number and the correlation term are 
\begin{subequations}
\label{eq:outnc} 
\begin{align}
n^{out}&= n^{out}_{\rm spont.}+ n^{out}_{\rm stim.} = \abs{\beta}^2 + \frac{\abs{\alpha}^2 + \abs{\beta}^2 }{\ep{\omega_{\rm in} /T} -1 } ,\\
c^{out}&= c^{out}_{\rm spont.}+ c^{out}_{\rm stim.} = \alpha \beta + \frac{2 \alpha \beta }{\ep{\omega_{\rm in} /T} -1 }.
\end{align}
\end{subequations}
These two quantities completely fix the late time behavior of the anticommutator of the relativistic field $ G_{ac}(t,t') = (n^{in}+1/2){\rm Re }\{\chi_{in}(t) \chi_{in}^*(t')\}$. In fact, for $t, t' > 0$, one has 
\begin{equation}
\label{eq:gacofnc}
\begin{split}
 G_{ac}(t,t') =& (n^{out}+1/2) \frac{\cos \left ( \omega_{\rm f} (t-t')\right ) }{\omega_{\rm f}} + \Re \left (c^{out} \frac{ \ep{-i \omega_{\rm f} (t+t')}}{\omega_{\rm f}}\right ).
\end{split}
\end{equation}
Using this expression, it is clear that the contribution of the stimulated amplification [weighted by $n^{in} = 1/({\ep{\omega_{\rm in} /T} -1 })$] and that of spontaneous processes are not easy to distinguish. As said in \ref{sec:homosepcriterion}, to be able to do so, it is useful to introduce the parameter $\Delta$ of Eq.~\eqref{eq:defDeltalinear}. 

In the present case, the value of $\Delta$ associated with Eqs.~\eqref{eq:outnc} is
\begin{equation}
\begin{split}
\Delta^{out}&= \Delta^{out}_{\rm spont.}+ \Delta^{out}_{\rm stim.} = -(\abs{\alpha}- \abs{\beta}) \abs{\beta} + \frac{(\abs{\alpha} - \abs{\beta})^2 }{\ep{\omega_{\rm in} /T} -1 }.
\label{eq:deltaout}
\end{split}
\end{equation}
At fixed $\abs{\beta/\alpha}$, the threshold value of nonseparability $\Delta=0$ defines a critical temperature $T_C$. It is given by (see also Ref.~\cite{Bruschi:2013tza})
\begin{equation}
\label{eq:Tcdef}
\begin{split}
\abs{\beta/\alpha} = \ep{- \omega_{\rm in}/ T_C} . 
\end{split}
\end{equation}
In Eq.~\eqref{eq:deltaout}, one clearly sees the competition between the squeezing of the state due to the sudden jump governed by $\beta$ which reduces the value of $\Delta$, and the initial occupation number which increases its value. It remains to extract this information from Eq.~\eqref{eq:gacofnc}. 

\subsection{Numeric study}

To this end, we plot in Fig.~\ref{fig:Gacoft} the product $\omega_{\rm f} G_{ac}(t,t'=t)$ of Eq.~\eqref{eq:gacofnc} as a function of time, for two different values of $T$, respectively, half and twice the temperature
\begin{equation}
\label{eq:defTxiin}
\begin{split}
 T_{\xi_{\rm in}} \doteq m c_{\rm in}^2 = \frac{1}{4 m \xi_{\rm in}^2} , 
\end{split}
\end{equation} 
fixed by the initial value of the healing length. We obtain two perfect sinusoidal curves since the mode $\chi_k$ freely propagates after the sudden change. As stated after Eq.~\eqref{eq:Gacofnandc}, the minima of the curves correspond to $\Delta +1/2$. Therefore, if in an experiment, the minimal value of $\omega_{\rm f} G_{ac}$ is measured with enough precision to be less than $1/2$, one can assert that the state is nonseparable (in the absence of dissipation).

To complete the analysis, we study how these results depend on the wave number $k$. We work with the standard Bogoliubov dispersion relation [see Eq.~\eqref{eq:disprelBEC}] with $\Gamma = 0$, $ c \xi = 1/2 m$ constant, and with $c_{\rm f}/c_{\rm in} = 1/10$. In Fig.~\ref{fig:Gacofk} we plot the anticommutator $\omega_{\rm f} \times G_{ac}(t,t'=t)$ as a function of $k$ for two temperatures, namely, $T_{\xi_{\rm in}}/2$ (left panel) and $T_{\xi_{\rm in}}$ (right panel). We first see that the modes with lower $k$ are more amplified than those with higher $k$. As expected from Eq.~\eqref{eq:Tcdef}, when looking at the lower envelope, we also see that the coherence level is higher (the minima of $G_{ac}$ lower) when working with a smaller temperature, and/or with higher $k$, i.e., with rarer events governed by a smaller initial occupation number $n_{in}$. To quantify this effect, and possibly also to guide future experiments, we characterize the domain where the resulting state is nonseparable, i.e., where $\Delta^{out} < 0$. To this end, in Fig.~\ref{deltadispgradino1}, we plot $\Delta^{out}$ as a function of $T/T_{\xi_{\rm in}}$ and the ratio $c_{\rm f}/c_{\rm in}$. We consider two values of $k$, namely one in the hydrodynamical regime, and one of the order of the inverse healing length. This clearly confirms that at higher momenta, states are more likely to be nonseparable. Moreover, one sees that for a wave number smaller than the healing length, at a temperature $\sim T_{\xi_{\rm in}}$ of Eq.~\eqref{eq:defTxiin}, in order to obtain a nonseparable state, $c_{\rm f} / c_{\rm in}$ should be either larger than $3$ or smaller than $1/3$.

To illustrate these aspects with a concrete example, we consider the experiment of Ref.~\cite{PhysRevLett.109.220401}. The relevant values are $T = 6.05 T_{\xi_{\rm in}}$ and $k \sim 2.15 m c_{\rm in}$, so that the initial number of particles is of the order of $3$. On the other hand, one has $c_{\rm f} / c_{\rm in} \sim 2^{1/4}$. (To get these numbers, we used $T=200nK$, $\omega/2\pi = 2kHz$, $m= 7. 10^{-27} kg$ and $c_{\rm in}=8mm/s$.) The corresponding value of the coherence level is $ \Delta \sim 1.4 $. Hence the state is separable. In order to reach $\Delta =0$, one should either increase the ratio $c_{\rm f} / c_{\rm in} \sim 6 $, or work with a lower temperature, of the order of $0.6 T_{\xi_{\rm in}}$. 

\section{The dissipative case}

In the presence of dissipation, the mode interpretation involving the Bogoliubov coefficients of Eq.~\eqref{eq:inout} is no longer valid. In fact, the state of $\chi$ is now characterized by $G_{ac}^{dr}$ of Eq.~\eqref{eq:Gacasintegrals}, which is governed by the retarded Green function and the noise kernel. The separability of the state should thus be deduced from its properties.

When the environment state is a thermal state, using Eq.~\eqref{eq:Trpsipsi}, Eq.~\eqref{eq:Gacasintegrals} can be expressed as 
\begin{equation}
\label{eq:Gacdr}
\begin{split}
G_{ac}^{dr}(t,t')& = \int \frac{d\omega_\zeta}{2\pi} \omega_\zeta \coth\left (\frac{\omega_\zeta}{2T}\right ) \widetilde G_{\rm r}(t,\omega_\zeta) \widetilde G_{\rm r}(t',-\omega_\zeta) , 
\end{split}
\end{equation}
where we introduced the Fourier transform 
\begin{equation}
\label{eq:buildingblock}
\begin{split}
\widetilde G_{\rm r}(t,\omega_\zeta)&\doteq \int_{- \infty }^\infty d\tau \ep{i \omega_\zeta \tau} \sqrt{\Gamma(\tau)} G_{\rm ret}^\chi(t,\tau),\\
\end{split}
\end{equation}
of the retarded Green function of Eq.~\eqref{eq:Gret}. In the following, we compute Eq.~\eqref{eq:Gacdr}, which is easier to handle than Eq.~\eqref{eq:Gacasintegrals}, in two different cases. In the first one, there is no dissipation after the sudden change, i.e., $\gamma_{\rm f}=0$ in Eq.~\eqref{eq:grad}. In the second one, $\Gamma$ is constant.

\subsection{Turning off dissipation after the jump}

When $\Gamma_{\rm f}=0$ for $\kappa t \gg 1$, we have the possibility of using the standard particle interpretation to read the asymptotic state. In fact, inserting Eq.~\eqref{eq:Gret} into Eq.~\eqref{eq:buildingblock}, using Eq.~\eqref{eq:grad} and $\kappa t \gg 1$, one gets 
\begin{equation}
\label{eq:TFGretGammaf0}
\begin{split}
\widetilde G_{\rm r}(t,\omega_\zeta) = \frac{ \sqrt{\Gamma_{\rm in}}}{2  \omega_{\rm f}} \left [ \ep{i \omega_{\rm f} t} R( \omega_\zeta) - \ep{-i \omega_{\rm f} t}R^*( -\omega_\zeta) \right ],
\end{split} 
\end{equation}
where
\begin{equation}
\label{eq:defR}
\begin{split}
R(\omega_\zeta) \doteq \sqrt{2 \omega_{\rm f}} \int_{- \infty }^\infty d\tau h(\kappa \tau ) \ep{i \omega_{\zeta} \tau } \ep{ - \int_\tau^\infty \Gamma } \bar \chi^{out}(\tau).
\end{split}
\end{equation}
The function $\bar \chi^{out}( \tau)$ is the standard out mode: it is the positive unit Wronskian mode of Eq.~\eqref{eq:homoeqvarphi} which is positive frequency at asymptotically late time. The time dependence of Eq.~\eqref{eq:TFGretGammaf0} guarantees that Eq.~\eqref{eq:Gacdr} has exactly the form of Eq.~\eqref{eq:gacofnc}. The final occupation number $n^{out}$ and correlations $c^{out}$ are found to be
\begin{equation}
\label{eq:nandcdiverge}
\begin{split}
 n^{out} + \frac{1}{2} &= \frac{\Gamma_{\rm in}}{ \omega_{\rm f}} \int_{0}^\infty \frac{ d\omega_\zeta}{\pi} \omega_\zeta \coth (\frac{\omega_\zeta}{2T}) \left (\abs{R(\omega_\zeta)}^2+\abs{R(-\omega_\zeta)}^2 \right ) , \\
 c^{out *} &= 2\frac{ \Gamma_{\rm in} }{ \omega_{\rm f}} \int_{0}^\infty \frac{ d\omega_\zeta}{\pi} \omega_\zeta \coth (\frac{\omega_\zeta}{2 T}) R(\omega_\zeta) R(-\omega_\zeta) .
\end{split}
\end{equation}
We notice that these expressions are similar to those of Eq.~\eqref{eq:outnc}, and that $R^*(\pm \omega_\zeta)$ plays the role of a density (in $\omega_\zeta$) of $\alpha$ and $\beta$, respectively. In fact, when taking the limit $\Gamma_{\rm in} \to 0$ in Eq.~\eqref{eq:nandcdiverge}, one recovers the dispersive expressions of Eq.~\eqref{eq:outnc}. 

We can now explain why we introduced the function $h$ in Eq.~\eqref{eq:grad}. For $\kappa\to \infty$, $h(\kappa t)$ becomes the step function $\theta(-t)$. In this limit, Eq.~\eqref{eq:defR} gives 
\begin{equation}
R(\omega_\zeta) = \frac{ \omega_{\rm f} +( \omega_\zeta-i \Gamma_{\rm in}) }{\tilde \omega_{\rm in}^2 - ( \omega_\zeta- i \Gamma_{\rm in} )^2} + \mathcal{O } \left (\frac1\kappa\right ),
\end{equation}
which indicates that $R$ behaves as $1/\omega_\zeta$ for $\omega_\zeta \to \infty$. Hence both $n^{out}$ and $c^{out}$ of Eq.~\eqref{eq:nandcdiverge} would logarithmically diverge. The divergences arise from the fact that the environment field $\hat \Psi$ contains arbitrary high frequencies $\omega_\zeta$. To regulate the divergences, several avenues can be envisaged. One could either introduce a $\zeta$-dependent coupling in Eq.~\eqref{eq:Sint} or cut off the high frequency $\omega_\zeta$ spectrum in Eq.~\eqref{eq:SPsi}. However, these would spoil the locality of Eq.~\eqref{eq:localgretofpsi}. For this (mathematical) reason, we prefer to use $h(\kappa t)$ of Eq.~\eqref{eq:grad}. Moreover, taking an instantaneous change in $\Gamma(t)$ would remove the $\mathcal{C}^1$ character of $G_{ac}(t,t')$ found in the dispersive case; see the discussion after Eq.~\eqref{eq:inout}. 

\subsubsection{Approximating Eq.~\texorpdfstring{\eqref{eq:defR}}{44}}

We now give an approximate value of Eq.~\eqref{eq:defR} both for a general profile $h$ and when applied to the particular case
\begin{equation}
\label{eq:chosenf}
\begin{split}
h (z) = \left \{
\begin{array}{ll}
1 & \mbox{ if } z<0,\\
1-z & \mbox{ if } 0<z<1,\\
0 & \mbox{ if } 1<z.
\end{array} \right .
\end{split}
\end{equation}
This shall be used to obtain the following figures. To do so, we first consider that $h$ is constant for negative times.\footnote{This simplifies the study, but is not necessary. When it is not the case, we should apply the same treatment to positive and negative times.} 
Hence, the part of the integral that runs on negative times is easy to handle and gives
\begin{equation}
\label{eq:negtimeR}
\begin{split}
&\int_{-\infty}^0 d\tau h(\kappa \tau ) \ep{i \omega_{\zeta} \tau } \ep{ - \frac{\Gamma_{\rm in}}{\kappa} \int_{\kappa \tau}^\infty h^2(z) dz } \sqrt{2 \omega_{\rm f}} \bar \chi^{out}( \tau) \sim \ep{ - \frac{\Gamma_{\rm in}}{\kappa} \int_{0}^\infty h^2 } \frac{ \omega_{\rm f} +( \omega_\zeta-i \Gamma_{\rm in}) }{ \tilde \omega_{\rm in}^2 - ( \omega_\zeta- i \Gamma_{\rm in} )^2}.
\end{split}
\end{equation}
Here, we neglected the effect of the change of $\tilde \omega$ due to the change in $\Gamma$ for positive times on $\varphi^{out}$. The deviation is generically of order $\Gamma_{\rm in}^2 / \omega_{\rm f} \kappa $.\footnote{
Note that for low values of $\kappa$, a WKB approximation gives the same first term, with an upper bound on the approximation given by $\Gamma^2/\omega^2 \left (1+ \kappa / \omega\right )$.}
To get this bound, we can write, for positive times, $\bar \chi^{out} = \ep{- i \omega_{\rm f}\tau}/ \sqrt{2\omega_{\rm f}} + \bar \chi_1$, and perturbatively in $\bar \chi_1$, get to $\abs{\bar \chi_1} \sqrt{2\omega_{\rm f}} < \Gamma_{\rm in}^2 / \omega_{\rm f} \kappa \int_0^\infty h^2 $. This bound can be relaxed order by order by solving Eq.~\eqref{eq:homoeqvarphi} with a source. This is not the goal of this section. For positive times, we now make the same approximation and get
\begin{equation}
\begin{split}
\int_{0}^\infty d\tau h(\kappa \tau ) \ep{i \omega_{\zeta} \tau } \ep{ - \frac{\Gamma_{\rm in}}{\kappa} \int_{\kappa \tau}^\infty h^2(z) dz } \sqrt{2 \omega_{\rm f}} \bar \chi^{out}( \tau) \sim \frac{1}{\kappa}\int_{0}^\infty d\tau h(\tau) \ep{ - \frac{\Gamma_{\rm in}}{\kappa} \int_{\tau}^\infty h^2 } \ep{i (\omega_{\zeta}- \omega_{\rm f}) \tau }.
\end{split}
\end{equation}
We can now compute this last integral perturbatively in $\Gamma / \kappa$. To get coherent results and to get rid of the $1/ \omega_{\zeta}$ term, the same expansion is necessary in Eq.~\eqref{eq:negtimeR}.

Since for the cases we consider (i.e., sudden change, $\Gamma \ll \omega \ll \kappa$), $ \Gamma_{\rm in} / \kappa > \Gamma_{\rm in}^2 / \omega_{\rm f} \kappa > \Gamma_{\rm in}^2 / \kappa^2 $, the expansion in $\Gamma_{\rm in} / \kappa$ should be done to first order maximum. To this order, $R$ becomes
\begin{equation}
\label{eq:Rforanyf}
\begin{split}
R&= \bigg[\left (1 { - \frac{\Gamma_{\rm in}}{\kappa} \int_{0}^\infty h^2 } \right ) \frac{ \omega_{\rm f} +( \omega_\zeta-i \Gamma_{\rm in}) }{\tilde \omega_{\rm in}^2 - ( \omega_\zeta- i \Gamma_{\rm in} )^2}+\int_{0}^\infty d\tau h(\kappa \tau) \left(1 - \frac{\Gamma_{\rm in}}{\kappa} \int_{\tau}^\infty h^2 \right ) \ep{i (\omega_{\zeta}- \omega_{\rm f}) \tau } \bigg]\\
&\hspace{1cm} \times \left [1 + \mathcal{O} \left ( \frac{\Gamma_{\rm in}^2 }{ \omega_{\rm f} \kappa }\right )\right ].
\end{split}
\end{equation}
When working with $h$ of Eq.~\eqref{eq:chosenf}, one obtains 
\begin{equation}
\label{eq:Rforchosenf}
\begin{split}
R&\sim \frac{ \omega_{\rm f} +( \omega_\zeta-i \Gamma_{\rm in}) }{\tilde \omega_{\rm in}^2 - ( \omega_\zeta- i \Gamma_{\rm in} )^2} + \frac{ h_1\left (\frac{i (\omega_\zeta -\omega_{\rm f})}{\kappa }\right ) }{\omega_\zeta -\omega_{\rm f}} - \frac{\Gamma_{\rm in}}{3 \kappa} \left [ \frac{ \omega_{\rm f} +( \omega_\zeta-i \Gamma_{\rm in}) }{\tilde \omega_{\rm in}^2 - ( \omega_\zeta- i \Gamma_{\rm in} )^2}+\frac{ h_4\left (\frac{i (\omega_\zeta -\omega_{\rm f})}{\kappa }\right )}{\omega_\zeta -\omega_{\rm f}} \right ] .
\end{split}
\end{equation}
where $ \ep{x} = \sum_{k=0}^{n} x^k/k! - h_n (x) x^n/n! $ defines the remainder term of order $n$, $h_n(x)$, of the Taylor expansion of the exponential function. We notice that $h_n(x)\sim x$ for $x\to 0$ so that $R$ is regular at $\omega_\zeta = \omega_{\rm f}$. Moreover, at large $x$, $h_n (x) \sim -\ep{x} n! /x^n +  1+ \mathcal{O}(1/x)$ so that $R \sim 1/ \omega_\zeta^2$ at large $\omega_\zeta$.

To complete the study, we checked the validity of Eq.~\eqref{eq:Rforchosenf} by numerically evaluating $R$. When considering only the first two terms of Eq.~\eqref{eq:Rforchosenf}, we observed that the relative error is smaller than $\Gamma / \kappa $, as predicted. When including the last term, the relative error remains smaller than $\Gamma^2 / \kappa \omega_{\rm f}$.

\begin{SCfigure}[2]
\begin{minipage}{0.40\linewidth} 
\includegraphics[width = 1\linewidth]{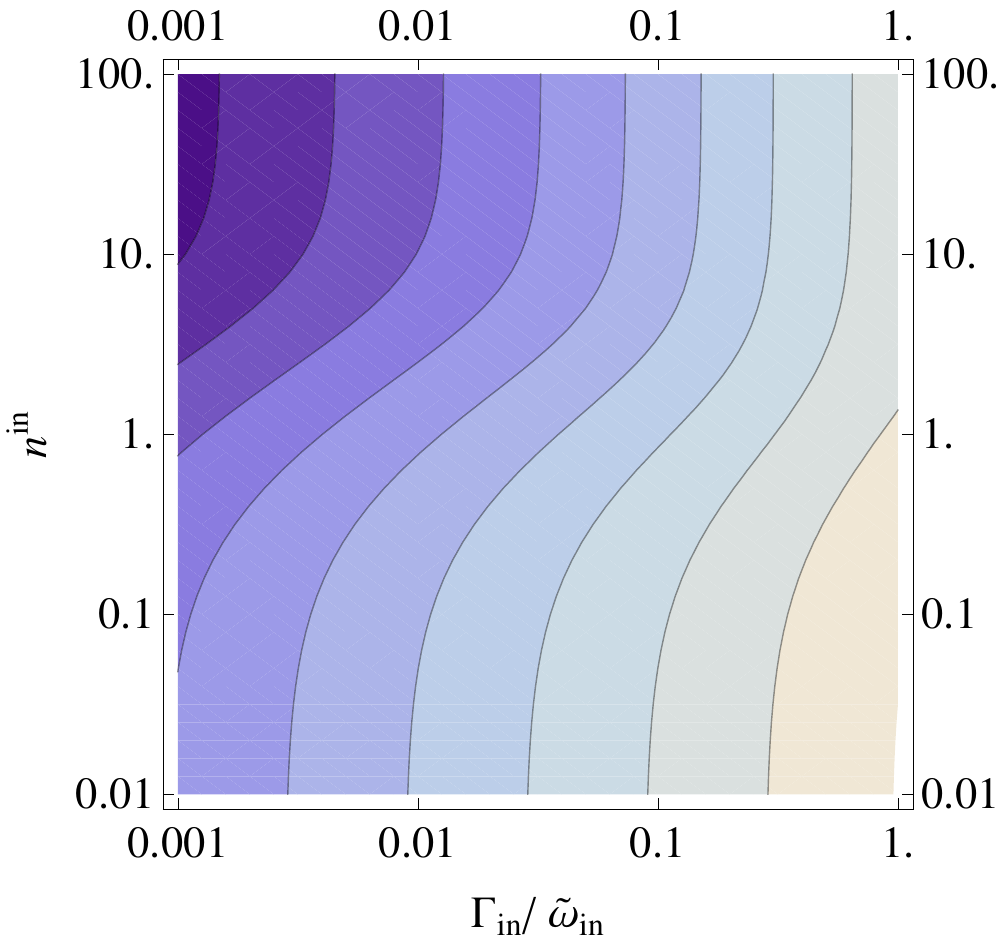} 
\end{minipage}
\begin{minipage}{0.1\linewidth}
\includegraphics[width = 1\linewidth]{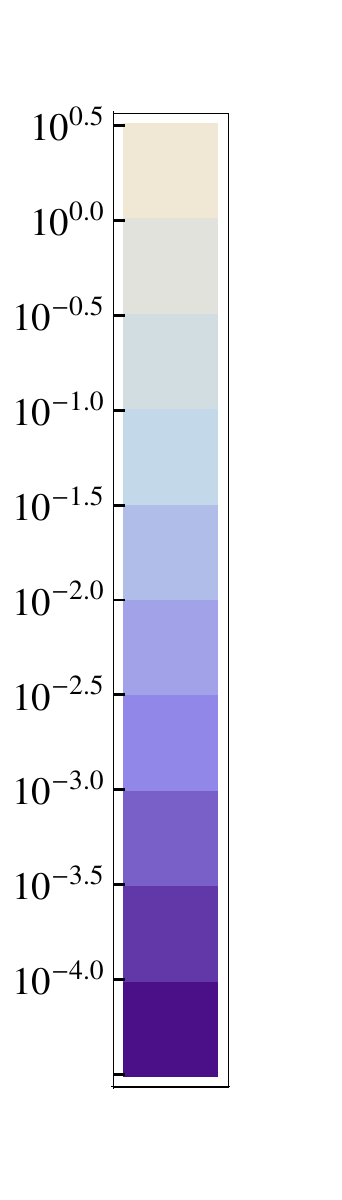}
\end{minipage}
\caption{
The ratio $\Delta n_r$ of Eq.~\eqref{eq:rapsp} for a jump $\omega_{\rm f} / \tilde \omega_{\rm in} = 0.1$ and $\kappa = 10 \tilde \omega_{\rm in}$ is represented in the plane $n^{in}, \Gamma_{\rm in}/ \tilde \omega_{\rm in}$. For low initial occupation number, we see that $\Delta n_r $ is proportional to dissipative rate $\Gamma_{\rm in}/\tilde \omega_{\rm in} $, whereas it evolves from linear to quadratic for high numbers. As it is explained in the text, these observations are confirmed by analytic expressions.
}
\label{fig:contourdeltanovern}
\end{SCfigure}

\subsubsection{Spectral deviations due to dissipation}

\begin{figure}[htb]
\includegraphics[width=0.47\linewidth]{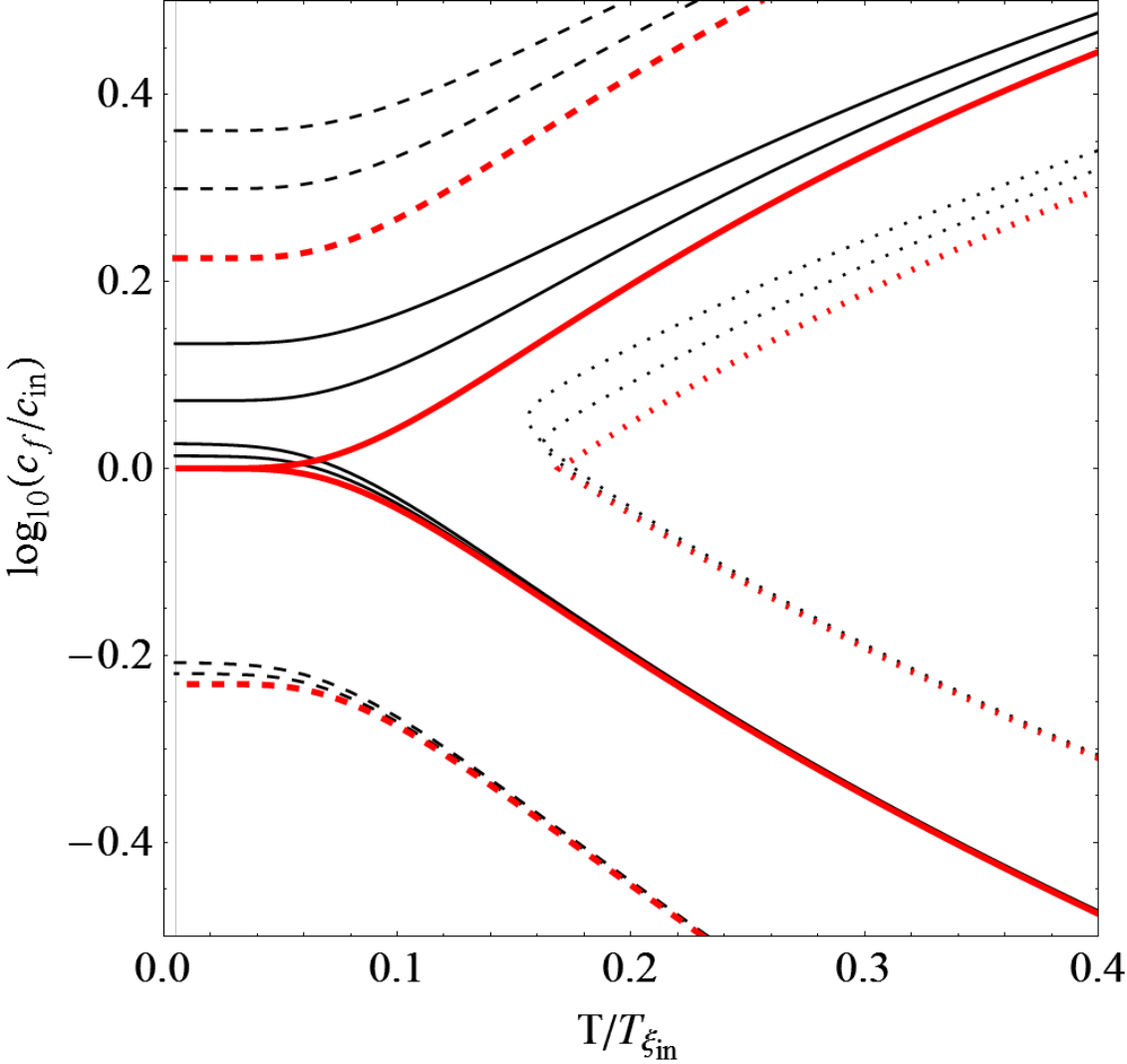}
\hspace{0.03\linewidth}
\includegraphics[width=0.47\linewidth]{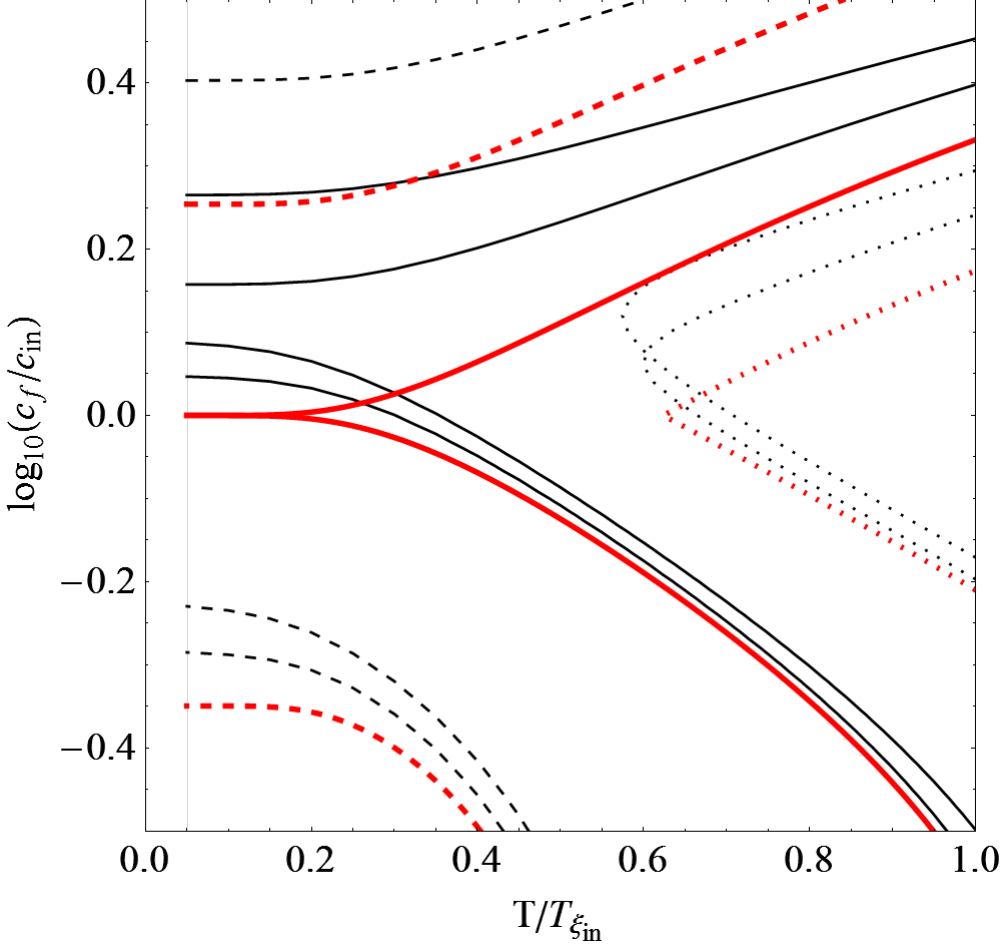} 
\caption{We represent the lines of $\Delta = -0.2$ (dashed), $0$(solid) and $0.2$ (dotted) in the plane $\{ T/T_{\xi_{\rm in}}, \log_{10} ( c_{\rm f} / c_{\rm in}) \}$, on the left panel for $k = 0.3 mc_{\rm in}$ and on the right one for $k = mc_{\rm in}$. $\gamma_\bk^2$ takes three values: from $0$, i.e., the dispersive case shown by the thick red line, to ${0.5}$, and $\kappa = 10 m c_{\rm in}^2 $. For temperatures that are not too low, we observe that the value of $\Delta$ is robust when $c_{\rm f} < c_{\rm in}$ (which corresponds to an expanding universe), while it increases when $c_{\rm f} > c_{\rm in}$.} 
\label{fig:deltadissip}
\end{figure}

We study how $n^{out}$ of Eq.~\eqref{eq:nandcdiverge} depends on the dissipation rate $\Gamma_{\rm in}$. To this end, we study its difference with the dispersive occupation number $n^{out}_{disp}$ of Eq.~\eqref{eq:outnc} evaluated with the same values for the temperature $T$, and the initial and the final frequency [see Eq.~\eqref{eq:defGammaomegainandf}]. In Fig.~\ref{fig:contourdeltanovern} we represent the relative change 
\begin{equation}
\begin{split}
\Delta n_r \doteq \frac{n^{out}(\Gamma_{\rm in}) - n^{out}_{disp}}{n^{out}_{disp}+1/2} , 
\label{eq:rapsp}
\end{split}
\end{equation}
as a function of the initial occupation number $n^{ in}$ and the ratio $ \Gamma_{\rm in} / \tilde \omega_{\rm in}$. We work with a jump $\omega_{\rm f} / \tilde \omega_{\rm in} = 0.1$, and with $\kappa = 10 \tilde \omega_{\rm in}$. [We use this parametrization because, at fixed $k$, Eq.~\eqref{eq:eomchi} only depends on $\omega(t)$ and $\Gamma(t)$. Hence the healing length and $k$ need not be specified.] For these values, we find two regimes. First, for a low occupation number, we observe that the deviation $\Delta n_r$ linearly depends on $ \Gamma_{\rm in}$. An analytical treatment based on Eq.~\eqref{eq:Rforanyf} reveals that, for small $ \Gamma_{\rm in}/ \omega_{\rm in}$ and large $\kappa/ \omega_{\rm in}$, $\Delta n_r$ behaves as 
\begin{equation}
\begin{split}
\Delta n_r = \left (\frac{ \Gamma_{\rm in} }{\omega_{\rm in}}\right) \frac{1} {n^{in}+1/2 } \times g(\kappa/\omega_{\rm in}) , 
\label{eq:rapsp2}
\end{split}
\end{equation}
where $g(\kappa/\omega_{\rm in}) $ is a rather complicated function\footnote{
It is given by 
\begin{equation}
\begin{split}
g(\kappa/\omega_{\rm in}) = \frac{1}{\pi} \left ( \frac{ 2 \log \left (\frac{\kappa_{\rm eff} }{ \omega_{\rm in}}\right ) }{1 + (\omega_{\rm f} /\omega_{\rm in})^2} -1\right ),
\label{eq:rapsp2pres}
\end{split}
\end{equation}
where $\kappa_{\rm eff}$ is the effective slope of the profile $h(\kappa t)$. It is given by
\begin{equation}
\begin{split}
\kappa_{\rm eff} &\doteq \kappa \exp \left [{- \gamma - \int dt dt'\left ( \partial_t h \right ) \left ( \partial_{t'} h \right ) \log(\kappa \abs{t-t'}) } \right ] ,
\label{eq:keff}
\end{split} 
\end{equation}
where $\gamma$ is the Euler constant. The interesting part of these equations is that they apply for any $h(\kappa t)$ when $\kappa/\omega_{\rm in} \gg 1$. Moreover, $\kappa_{\rm eff}$ also governs the logarithmic growth of $\abs c$.}.
We did numerically checked that Eqs.~\eqref{eq:rapsp2} and~\eqref{eq:rapsp2pres} apply for $n_{in} \lesssim 5$ and $\kappa \gtrsim 10 \omega_{\rm in}$. Hence, under these conditions, $\Delta n_r$ depends on $\kappa_{\rm eff}$ of Eq.~\eqref{eq:keff}, but not on the exact shape of $h$ of Eq.~\eqref{eq:grad}. 

Second, for high occupation numbers and high $\kappa \Gamma / \omega_{\rm in}^2$, we observe in Fig.~\ref{fig:contourdeltanovern} a quadratic dependence in $\Gamma $. Lowering $\kappa \Gamma / \omega_{\rm in}^2$, we observe a transition from this quadratic behavior to a linear one. These numerical observations are in agreement with the analytical result 
\begin{equation}
\label{eq:deltan2}
\begin{split}
 \Delta n_r \sim \left (\frac{\Gamma_{\rm in}}{\omega_{\rm in}}\right )^2 \frac{\omega_{\rm in}^2-\omega_{\rm f}^2}{\omega_{\rm in}^2+\omega_{\rm f}^2} + \mathcal{O}\left ( \frac{\Gamma_{\rm in}}{\kappa}\right ), 
\end{split}
\end{equation}
which applies in the limit $\Gamma_{\rm in} \ll \omega \ll \kappa \ll T$. 

In brief, Eqs.~\eqref{eq:rapsp2} and~\eqref{eq:deltan2} establish how $n^{out}$ of Eq.~\eqref{eq:nandcdiverge} converges towards the dispersive occupation number of Eq.~\eqref{eq:outnc} when $\Gamma_{\rm in} / \omega_{\rm in} \ll 1$ and $\kappa \gg \omega_{\rm in}$. A similar analysis can be done for the coefficient $c^{out}$ of Eq.~\eqref{eq:nandcdiverge}, and it gives similar results.

\subsubsection{Final coherence level}
\begin{SCfigure}[2]
\includegraphics[width=0.47\linewidth]{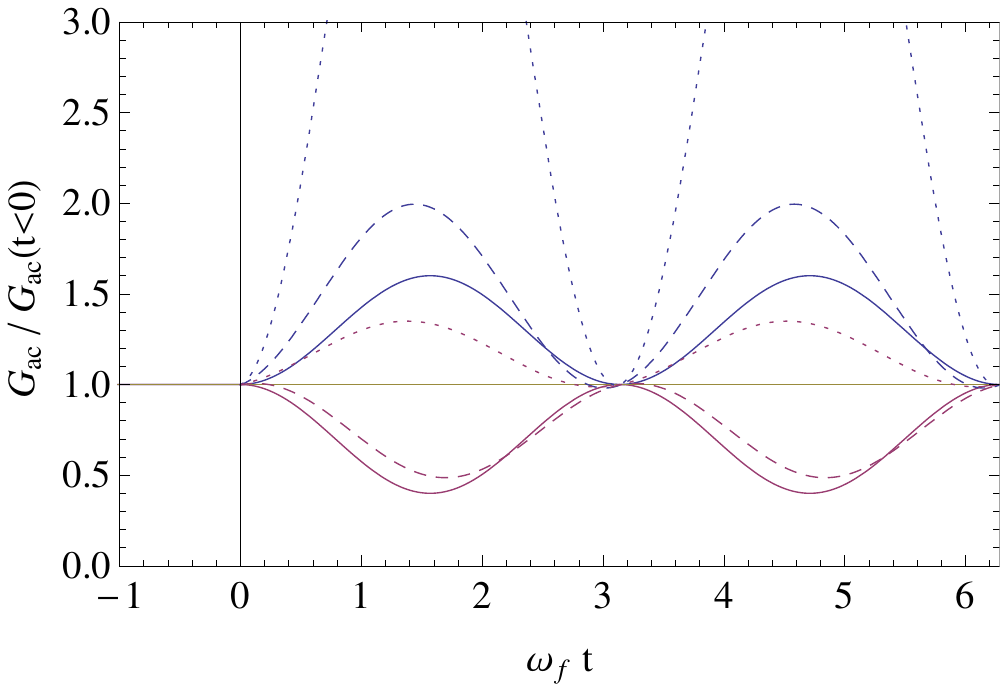}
\caption{We represent the anticommutator $G_{ac}$ normalized to its value before the jump, as a function of $\omega _{\rm f }t $, and for $T =T_{\xi_{\rm in}}$ and $k = 2 m c_{\rm in}$. The three upper curves (in blue) represent an expanding universe ($c_{\rm f} = 0.5c_{\rm in}$), while the three lower (in purple) ones represent a contracting universe ($c_{\rm f} = 2c_{\rm in}$). The solid lines show the two dispersive cases, the dashed lines represent the dissipative cases with $\gamma^2=0.15$ and $\kappa = 100 m c_{\rm in }^2$, and the dotted lines show the results when $\kappa$ is increased to $10^4 m c_{\rm in }^2$ in order to see the logarithmic growth of $n$ and $c$ of Eq.~\eqref{eq:outnc}. In all cases, $G_{ac}$ is $\mathcal{C}^{1}$ across the jump. Since the coherence is based on the minima of $G_{ac}$, the value of $\Delta$ is robust when $c_{\rm f}/c_{\rm in} < 1$, whereas it necessarily increases when $c_{\rm f}/c_{\rm in} > 1$. }
\label{fig:morerobust}
\end{SCfigure}

In Fig.~\ref{fig:deltadissip}, we represent the coherence level $\Delta$ of Eq.~\eqref{eq:defDeltalinear} as a function of the temperature and $c_{\rm f}/c_{\rm in}$, for two different values of $k$, namely, $k/ m c_{\rm in}= 0.3$ and $1$, and three values of $\gamma^2$, namely, $0, 0.25$ and $0.5$. The value of $\kappa$ is $\kappa = 10 m c_{\rm in}^2$. As one might have expected, we observe a continuous deviation of $\Delta$ for increasing values of the coupling $\gamma^2_\bk$. More surprisingly, we observe that increases of $c(t)$, $c_{\rm f} > c_{\rm in}$, and decreases, $c_{\rm f} < c_{\rm in}$, behave very differently. In the first case, there is a large increase of $\Delta$, which implies a loss of coherence. On the contrary, in the second case, the value of $\Delta$ is robust, and some marginal gain of coherence can even be found. To validate these observations, we studied the behavior of $\Delta$ for different profiles of the smoothing function $h(\kappa t)$. Whenever $c_{\rm f} / c_{\rm in}$ is not too close to 1, we obtained similar results, thereby showing that the choice of the profile of $h$ does not significantly matter. Instead, when $c_{\rm f} / c_{\rm in} \sim 1$, the behavior of $\Delta$ is less universal.

In order to understand the different behaviors of $c_{\rm f} > c_{\rm in}$ and $c_{\rm f} < c_{\rm in}$, we represent in Fig.~\ref{fig:morerobust} the anticommutator normalized to its value at the jump $G_{ac}(t=t') / G_{t=t'=0}$ as a function of $\omega_{\rm f} t$, and for different values of the steepness parameter $\kappa$. We first notice that this function is $\mathcal{C}^1$ across the jump, as were the modes in dispersive theories. This implies that one extremum of the anticommutator coincides with the value before the jump. In the absence of dissipation, one easily verifies that it is a minimum for $c_{\rm f} < c_{\rm in}$ and a maximum otherwise. For weak dissipation, by continuity in $\Gamma$, this must still be the case. Hence, when $c_{\rm f} < c_{\rm in}$, the minima of the anticommutator are fixed by the initial state. Instead, for $c_{\rm f} > c_{\rm in}$, they are fixed by the intensity of the jump and the injection of energy from the environment. As clearly seen in the figure, this injection increases with $\kappa$ and explains why coherence is more robust when $c_{\rm f} < c_{\rm in}$ (i.e., in expanding universes).

\subsection{Constant dissipation rate}
\label{sec:cstdissiprate}

In this section, we study our model when the dissipative rate $\Gamma$ is constant, as it is found for instance in polariton systems~\cite{Gerace:2012an,Koghee2013} and in Josephson metamaterial~\cite{Lahteenmaki12022013}. In this case, there is no unambiguous notion of ($out$) quanta, even though the anticommutator of Eq.~\eqref{eq:Gacdr} is well defined for all $t, t'$. Nevertheless, provided $\Gamma/\omega$ is low enough, we shall see that an approximate reading of the final state can be reached in term of the instantaneous particle representation based on Eq.~\eqref{eq:chidec}.

Because $\gamma$ is constant in Eq.~\eqref{eq:grad}, there is a simplification with respect to the previous subsection: no regularization is now needed since Eq.~\eqref{eq:Gacdr} is finite. Moreover, the retarded Green function of Eq.~\eqref{eq:Gret} is exactly known. It is given by
\begin{equation}
G_{\rm ret}^\chi = \ep{-\Gamma (t-t')} \times \left \{
\begin{array}{ll}
\theta(t-t') \frac{\sin \tilde \omega (t-t')}{\tilde \omega}& \mbox{ for } t',t <0 \mbox{ or } t'>0, \\
 \frac{\sin (\tilde \omega_{\rm f} t) \cos(\tilde \omega_{\rm in} t')}{\tilde \omega_{\rm f}} - & \frac{\cos (\tilde \omega_{\rm f} t) \sin(\tilde \omega_{\rm in} t')}{\tilde  \omega_{\rm in} } \\
 &\hspace{-1cm}\mbox{ for } t' <0\mbox{ and } t >0. \\
\end{array}\right .
\end{equation}
Hence, after the jump of $c$, for positive times, the Fourier transform of Eq.~\eqref{eq:buildingblock} gives 
\begin{equation}
\begin{split}
\widetilde G_{\rm r}(t,\omega_\zeta)=& \frac{\sqrt{ \Gamma} \ep{i\omega_\zeta t}}{\tilde \omega_{\rm f}^2 - ( \omega_\zeta- i \Gamma )^2} + \sqrt{\Gamma} \frac{ \ep{-\Gamma t+i\tilde \omega_{\rm f} t}}{2 \tilde \omega_{\rm f}} \bigg[\frac{ \tilde \omega_{\rm f} +( \omega_\zeta-i \Gamma) }{\tilde \omega_{\rm in}^2 - ( \omega_\zeta- i \Gamma)^2} - \frac{1 }{\tilde \omega_{\rm f} - (\omega_\zeta - i\Gamma )} \bigg] \\
&+(\tilde \omega_{\rm f} \to - \tilde \omega_{\rm f}) .
\end{split} 
\end{equation}
This means that we (exactly) know the integrand of Eq.~\eqref{eq:Gacdr}. The integral can be performed by analytic methods (by evaluating the residues of poles), and then recognizing the infinite sum as a finite sum of hypergeometric functions. We do not present the analytic result here since it is long and not instructive. Instead, the main results are presented below. 

\subsubsection{Two-point correlation function}

\begin{figure}[htb]
\begin{minipage}{0.47\linewidth}
\includegraphics[width=1\linewidth]{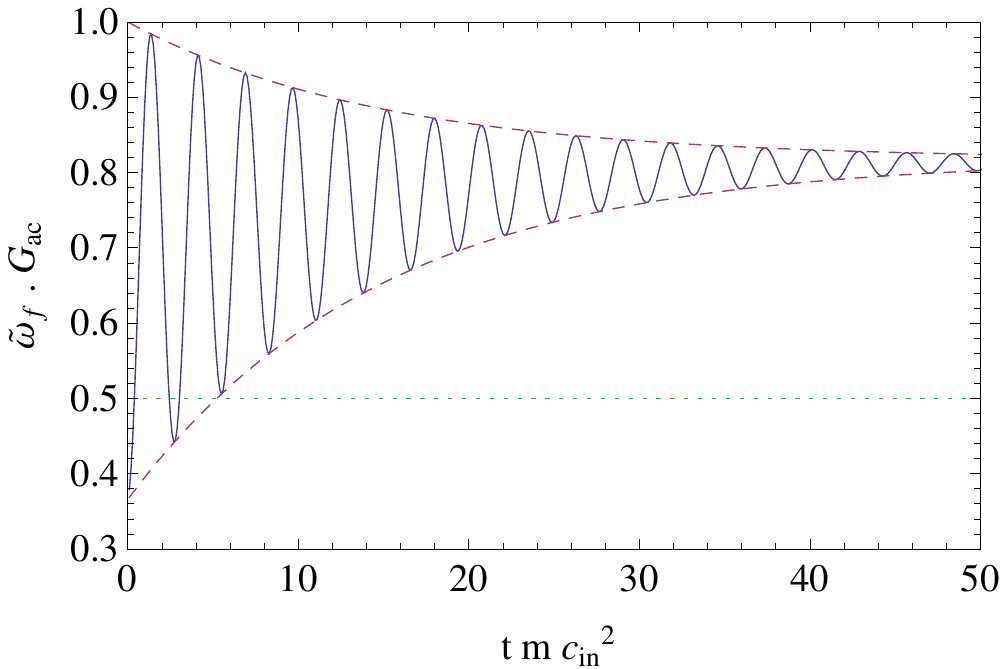}
\end{minipage}
\hspace{0.03\linewidth}
\begin{minipage}{0.47\linewidth}
\includegraphics[width=1\linewidth]{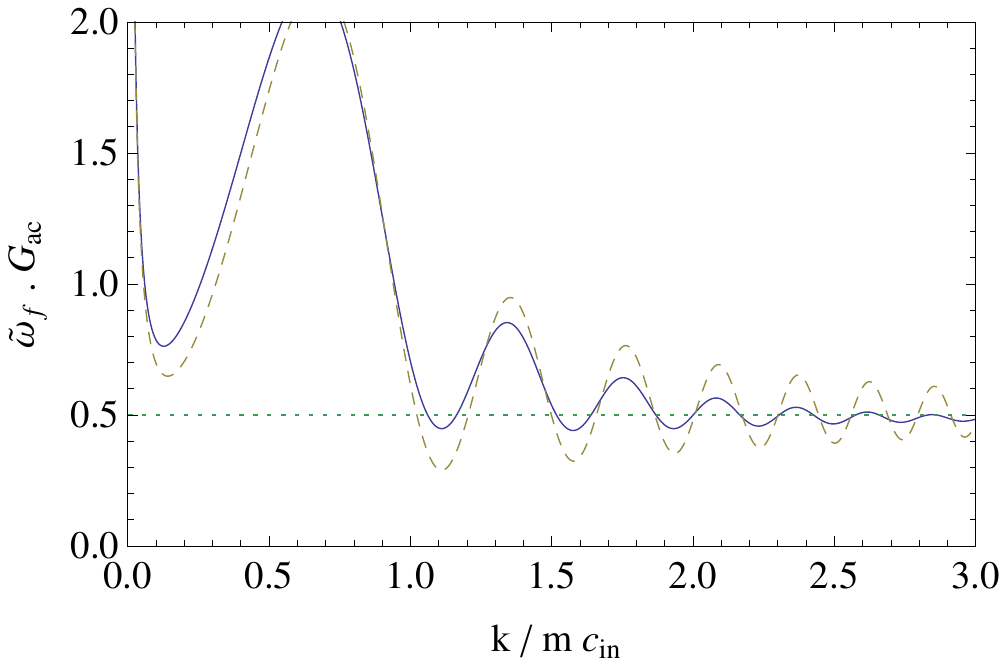}
\end{minipage}
\caption{We represent the product $\tilde \omega_{\rm f} \times G_{ac}(t,t'=t)$ where $G_{ac}$ is given in Eq.~\eqref{eq:gacofnc}, on the left panel, as a function of the adimensionalized time $t m c_{\rm in}^2$ for $k=1.5 m c_{\rm in}$, $\gamma_\bk^2 =0.03$ and $T = 0.8 T_{\xi_{\rm in}}$, and on the right panel, as a function of $k/ m c_{\rm in}$ for $ t = 5/ m c_{\rm in}^2$, $\gamma_\bk^2 =0.05$ and $T = 0.5 T_{\xi_{\rm in}}$. The dashed yellow line on the right panel is the case with no dissipation $\gamma_\bk^2 =0$. In all cases, the height of the jump is $c _{\rm f} / c_{\rm in} =0.1$.}
\label{fig:Gofk1}
\end{figure}

To discover the effects of dissipation, in Fig.~\ref{fig:Gofk1} we plot $\tilde \omega_{\rm f} \times G_{ac} (k,t=t') $ both as a function of time, as in Fig.~\ref{fig:Gacoft}, and as a function of the wave number $k$, as in Fig.~\ref{fig:Gacofk}. When considered as a function of $t$, we observe that the oscillations take place in a narrowing envelope. As expected, the latter follows an exponential decay in $e^{-2\Gamma t}$ towards the equilibrium value $\tilde \omega_{\rm f} G_{ac}^{eq} = \tilde \omega_{\rm f} G_{ac}(t=t'\to \infty)$. This simple behavior implies that the nonseparability of the state is quickly lost at high temperatures. Indeed, a rough estimate of the lapse of time for the decoherence to happen is of the order $ (2 \Gamma n^{eq})^{-1}$, where $n^{eq}$ is the mean occupation number at equilibrium. Hence, when $n^{eq} \gg 1$, the time for the loss of coherence is smaller than the dissipative time $1/\Gamma$ by a factor $1/2 n^{eq}$. When considered as a function of $k$, on the right panel, we observe damped oscillations. For large $k$, they are more damped than those of the dispersive case (represented by a dashed line) since the decay rate $\Gamma \propto k^2$. 

\begin{SCfigure}[2]
\includegraphics[width=0.47\linewidth]{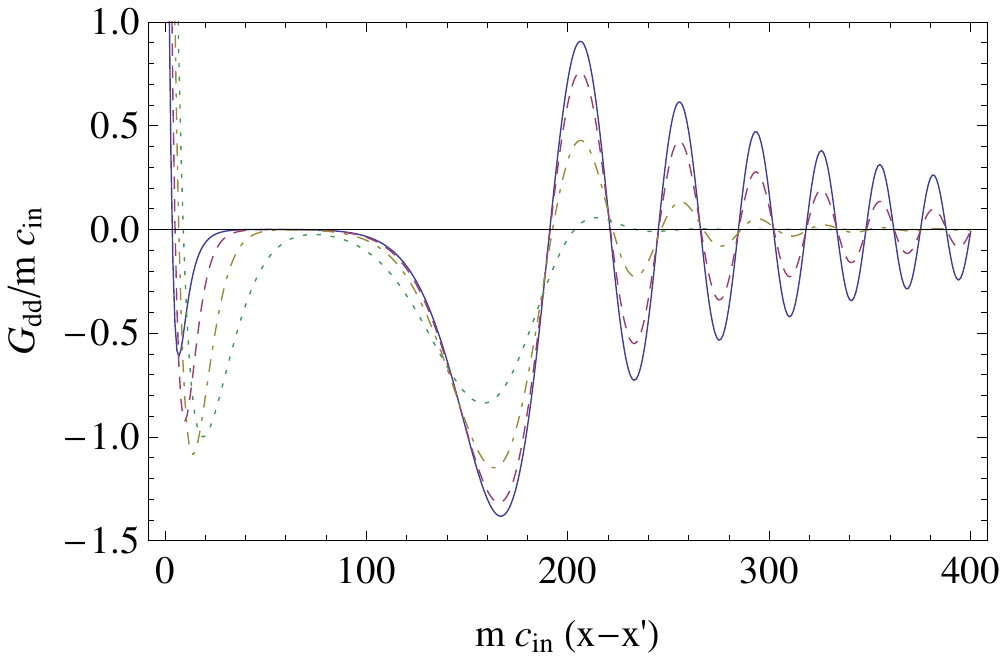} 
\caption{We draw the equal time density-density correlation of Eq.~\eqref{eq:defd} as a function of $ m c_{\rm in} \abs{\bx-\bx'}$, for an elongated ($1D$) condensate and  $c_{\rm f}/c_{\rm in} =0.1$, and $T = T_{\xi_{\rm in}} $. We take $t = 7.5 /mc_{\rm f}^2$ and four values of $\gamma_\bk^2$, namely, $10^{-2}$ (solid blue line), $10^{-1.5}$ (dashed purple line), $10^{-1}$ (dot-dashed yellow line) and $10^{-0.5}$ (green dotted line). The parameter $n$ of Eq.~\eqref{eq:gamma} is $n=0$. We observe a peak at $x=x'$ that is broadened by dissipation, and a series of peaks propagating away from the center with a group velocity higher than $c_{\rm f}$. The faster they propagate, the more damped they are since we have $\Gamma \sim k^2$. } 
\label{fig:deltaGofx}
\end{SCfigure}

To further study the effects of increasing $\Gamma$, in Fig.~\ref{fig:deltaGofx}, we represent the equal time density-density two-point function (see Ref.~\cite{Fedichev:2003bv} and \ref{sec:rhotheta})
\begin{equation}
\label{eq:defd}
\begin{split}
G_{dd} &\doteq \frac{\left < \delta \hat \rho(t,\bx) \delta \hat \rho(t,\bx') \right > }{\rho} = \int \frac{d\bk}{\pi } \ep{ i \bk (\bx-\bx')} c \xi k^2 G_{ac}^k(t,t'=t) ,
\end{split}
\end{equation} 
for the case of one dimensional (elongated) condensate and four values of $\gamma^2$, namely, $\log_{10}(\gamma^2) = - 2, -1.5, -1$ and $-0.5$. We take a rather large $t = 7.5 /mc_{\rm f}^2$ to see the propagation of the phonon waves. As expected we observe a peak centered on $x=x'$ plus a series of peaks for $\abs{x-x'} > 2 c_{\rm f} t$. The first one is present even in vacuum and is amplified by the mean occupation numbers $n_k > 0$. We see that it is broadened by dissipation. The series of peaks for $\abs{x-x'} > 2 c_{\rm f} t$ is due to the fact that the phonons are produced in pairs. As in inflationary cosmology~\cite{Campo:2003pa}, their amplitudes are fixed by the $c_k$ coefficients. These peaks propagate at different speeds because of dispersion. The fastest are more damped because dispersion is anomalous, and because dissipation goes in $k^2$. We also notice that the first propagating peak is negative when working with $c_{\rm f} < c_{\rm in}$. (For $c_{\rm f} > c_{\rm in}$ instead, it would be positive.) We conclude by noticing that this plot gives no indication of whether the state is separable or not, mainly because $G_{dd}$ mixes different two-mode sectors labeled by $k$, some of which being nonseparable, but not all.

\subsubsection{Approximate particle interpretation and separability}
\label{sec:nbofpartdissip}

To interpret the properties of $G_{ac}$, we now use the instantaneous particle representation based on $\hat \chi^{dec}$ of Eq.~\eqref{eq:chidec}. Even though the anti-commutator $G_{ac}$ is well-defined, for dissipative systems, there is no unique (canonical) way of defining the concept of particle. Hence the mean occupation number $n$ and the correlation term $c$ are somehow ambiguous. The issue is twofold as it requires one to treat separately the equilibrium and the out of equilibrium parts of $G_{ac}$. 

First, a close examination of the out of equilibrium part $\delta G_{ac} \doteq G_{ac} - G_{ac}^{eq} $ reveals that it contains nonoscillating terms which {\it cannot} be expressed as the anticommutator of $\hat \chi^{dec}$ in Eq.~\eqref{eq:chidec}. In fact the equal time anti-commutator of $\hat \chi^{dec}$ decays as $\ep{- 2 \Gamma t }$, whereas the non oscillating terms decay as $\ep{- 2 \pi t T}$. Hence, when $\Gamma < \pi T$, these extra terms can be neglected for $t, t' \gg 1/(\pi T- \Gamma)$. When these conditions are fulfilled, one can define $\delta n (t_0)$ and $\delta c (t_0)$, the out of equilibrium value of the occupation number and the coherence at $t_0$, by
\begin{equation}
\label{eq:deltaGac}
\begin{split}
\delta G_{ac}(t,t') & \sim \delta n (t_0)\chi^{dec} (t;t_0) [\chi^{dec} (t';t_0)]^* + \delta c(t_0) \chi^{dec} (t;t_0) \chi^{dec} (t';t_0) +cc,
\end{split}
\end{equation}
where $\chi^{dec} (t;t_0) = \ep{- \Gamma (t-t_0)} \ep{ - i\tilde \omega_{\rm f} (t-t_0)} / \sqrt{2 \tilde \omega_{\rm f}}$ is the decaying solution of Eq.~\eqref{eq:chidec} which contains only positive frequency and which is unit Wronskian at $t=t_0$. Since $\delta G_{ac}(t,t')$ is independent of $t_0$, one immediately deduces that
\begin{equation}
\delta n (t_0) = \delta n(0) \ep{- 2 \Gamma t_0} ,\quad \vert \delta c(t_0) \vert = 
\vert \delta c(0) \vert \ep{- 2 \Gamma t_0} .
\end{equation}
This matches the behavior of the envelope observed in Fig.~\ref{fig:Gofk1}. In the limit of small dissipation, one finds that the initial values obey
\begin{equation}
\begin{split}
\delta n(0) &= \delta n^{disp} + \mathcal{O} \left (\frac{\Gamma}{ \omega_{\rm in}}+\frac{\Gamma}{ T} \right ),\\
\delta c(0) &= c^{disp}+ \mathcal{O} \left (\frac{\Gamma}{ \omega_{\rm in}}+\frac{\Gamma}{ T} \right ),
\end{split}
\end{equation}
where $\delta n^{disp} = n^{out} - n^{eq}$ and $c^{disp}=c^{out}$ are the corresponding quantities evaluated with the dispersive case $\gamma = 0$. The values of $n^{out}$ and $c^{out}$ are given in \ref{sec:nodiss}, and $n^{eq}$ is the mean occupation number in a thermal bath.

Second, a similar analysis of the equilibrium part of $G_{ac}$ gives
\begin{equation}
\begin{split}
G_{ac}^{eq}(t,t') & = \ep{-\Gamma\abs{t-t'}} G_{ac}^{th,disp}(t,t') + {\mathcal O} (\frac{\Gamma}{\omega_{\rm f}}),
\end{split}
\end{equation}
where $G_{ac}^{th,disp}$ is the corresponding dispersive anticommutator in a thermal state. It is worth noticing that in the presence of dissipation, the rescaled anticommutator can be smaller than $1/2$, as can be seen in Fig.~\ref{fig:Gofk1} for high $k$. This is because the $\phi$ field is still interacting with the environment. Yet, in the limit $\Gamma \abs{t-t'} \ll 1$ and $\Gamma/ \omega_{\rm f} \ll 1$, $G_{ac}^{eq}$ is indistinguishable from $G_{ac}^{th,disp}$. We can then work with $2 n^{eq}+1 = { \coth[ \omega_{\rm f} /2T ]} $. Doing so, we get an imprecision of the order of $\Gamma / \omega_{\rm f}$. 

\begin{SCfigure}[2]
\includegraphics[width=0.47\linewidth]{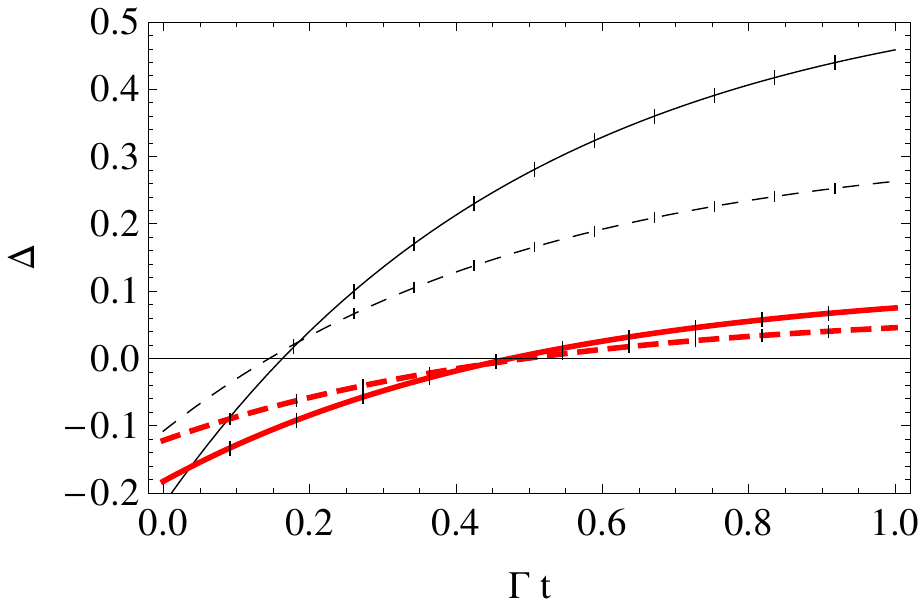} 
\caption{ We represent the coherence level $\Delta$ as a function of $\Gamma t$, for $T= T_{\xi_{\rm in}}/2$ and $\gamma_\bk ^2 =0.01$. We consider two values of $c_{\rm f} / c_{\rm in} = 0.1$ (solid line) or $0.5$ (dashed line), and two values of $k / m c_{\rm in} = 1$ (black line) or $1.5$ (thick red line). The imprecision in the value of $\Delta$ is indicated by vertical bars. In the present weakly dissipative cases, the spread of $\Delta$ is of the order $0.02$. Therefore, the moment where the nonseparability of the state is lost is known with some precision. }
\label{fig:deltadissipoft}
\end{SCfigure}

Having identified $n = \delta n + n^{eq}$ and $ c$, we can compute the coherence level $\Delta$ of Eq.~\eqref{eq:defDeltalinear}, which of course inherits the imprecision coming from $n^{eq}$. In Fig.~\ref{fig:deltadissipoft}, we represent $\Delta_\bk$ and its imprecision as a function of time for four different cases. As already discussed, we notice that the nonseparability of the state is lost in a time much smaller than $1/\Gamma$. We also notice that when $\Gamma/\omega$ is low enough, the imprecision in $\Delta$ does not significantly affect our ability to predict when nonseparability will be lost. In brief, for low values of $\Gamma/\omega$ and $\Gamma/T$, the anticommutator $G_{ac}$ can be reliably interpreted at any time $t_0$ using the instantaneous particle representation based on $\hat \chi^{dec}(t,t_0)$ of Eq.~\eqref{eq:chidec}.

\section*{Conclusions}
\addcontentsline{toc}{section}{Conclusions}

In this chapter, we computed the spectral properties $n_k$ and the coherence coefficient $c_k$ of quasiparticles produced when a sudden change is applied to a one-dimensional homogeneous system. We took into account both the effects of an initial temperature and the fact that the quasiparticles are coupled to a reservoir of modes, something which induces dissipative effects. For definiteness, the quasiparticles are taken to be Bogoliubov phonons propagating in an atomic condensate. Yet our results should apply, mutatis mutandis, to all weakly dissipative systems (as long as the background is not dissipated).

For simplicity, we work with a quadratic action. Importantly, this allows us to compute the anticommutator for nonequal times [see Eqs.~\eqref{eq:Gact0} to~\eqref{eq:Gacmix}], something which is not generally done when using the truncated Wigner method~\cite{0953-4075-35-17-301}, but which could be very useful for future experiments. Because our system is coupled to a bath, $n_k$ and $c_k$ are {\it a posteriori} extracted from the anticommutator of the phonon field [see Eq.~\eqref{eq:nandcdiverge} and~\eqref{eq:deltaGac}]. Then, to distinguish classical correlations from quantum entanglement, we used the fact that negative values of the parameter $\Delta_k$ of Eq.~\eqref{eq:defDeltalinear} correspond to nonseparable states (for the Gaussian states we consider).

When neglecting dissipative effects, we studied the competition between the squeezing of the quasiparticles state, which is induced by the sudden change, and the initial temperature, which increases the contribution of stimulated effects [see Eq.~\eqref{eq:deltaout}]. In Fig.~\ref{fig:Gacoft}, one clearly sees that the value of the minima of the equal time anticommutator allows one to know if the state is separable or not. The outcome of the competition is summarized in Fig.~\ref{deltadispgradino1}, where the coherence parameter $\Delta$ is plotted as a function of the sudden change of the sound speed and the temperature. We applied our analysis to the recent experiment of Ref.~\cite{PhysRevLett.109.220401} and concluded that one should either increase the change of $c$ or work with a lower temperature in order to obtain a nonseparable state.

We then included dissipative effects. When there is no (significant) dissipation after the sudden change, we showed in Fig.~\ref{fig:contourdeltanovern} how the final number of particles is progressively affected by increasing dissipation. When the initial occupation number is low, the deviations are linear in the decay rate $\Gamma$, whereas they are quadratic for occupation numbers $n^{\rm in } \gtrsim 5$. Interestingly, we observed in Fig.~\ref{fig:deltadissip} that dissipation, on the one hand, hardly affects the coherence parameter $\Delta$ when the sudden change is due to a decrease of the sound speed (something which corresponds to an expanding universe when using the analogy with gravity) and, on the other hand severely reduces the coherence when the sound speed increases. This discrepancy is further studied in Fig.~\ref{fig:morerobust} which illustrates the key role played by the ${\cal C}^1$ character of the anticommutator across the sudden change. 

We also studied the case when the dissipative rate is constant. In this case, the main effect of dissipation on the anticommutator is the expected damping towards the equilibrium value, see Fig.~\ref{fig:Gofk1}. As a result, for high occupation numbers, the nonseparability of the state is lost in a time much shorter than the inverse decay rate, see Fig.~\ref{fig:deltadissipoft}. In spite of the fact that the quasiparticles are still coupled to the environment, we showed that a reliable study of this loss can be performed for weakly dissipative systems, i.e., with $\Gamma/\omega \ll 1$. On the contrary, for strongly dissipative systems, i.e., rapidly decaying quasiparticles, we believe the notion of nonseparability cannot be meaningfully implemented.

\chapter{Dissipative condensates}
\label{chap:polariton}

\section*{Introduction}

In an atomic Bose condensate, the number of atoms is (to a very good approximation) conserved. As a consequence, to make a measurement on the state, one modifies (and sometimes destroy) the condensate. On the other hand, there exist some systems, such as polariton systems, see \ref{sec:intropolariton} where the particles forming the condensate are dissipated. When measuring the outgoing flux, one then measures the properties of the system. This is the main advantage of such a system: it is not necessary to destroy the condensate to observe it. 

In this chapter, we first review the equations of motion for phonon in such a system. We then study in details the stationary state so that we identify its main properties. In particular, since the system is out of equilibrium, the fluctuation dissipation theorem is violated. In \ref{sec:DCE}, we compute the reaction of the phonon state to some sudden change. To avoid technical difficulties, the sudden change is fine tuned so that the condensate is stationary. This chapter is mainly based on~\cite{Busch:2013gna}.

\minitoc
\vfill

\section{The physical system and the equations of motion}

We use here the polariton system of \ref{sec:intropolariton}, with the Hamiltonian given in Eq.~\eqref{eq:hamiltonian}. Moreover, we shall restrict to the case of a monochromatic pump normally incident on the cavity, which gives a vanishing in-plane $\bk=0$ and therefore a spatially homogeneous and isotropic pump amplitude $F(\bx,t)=F_0(t) e^{-i\omega_p t}$: this pump configuration injects into the cavity a photon fluid that is spatially homogeneous and at rest. Even though the coherent pump acts on the single $\bk=0$ mode, because of the interaction term, the field dynamics involve the whole continuum of in-plane $\bk$ modes.

From Eq.~\eqref{eq:hamiltonian}, the equations of motion are 
\begin{subequations}
\begin{align}
\label{eq:eompsipol}
i\partial_t \hat \Phi & = \left ( E_0^{\rm bare} - \frac{\partial_\bx^2}{2m} + g \hat \Phi^\dagger \hat \Phi\right ) \hat \Phi + \int d\zeta g_\zeta \hat {\Psi}_\zeta +F , \\
\label{eq:eomX}
i \partial_t \hat {\Psi}_\zeta& = \omega_\zeta \hat {\Psi}_\zeta+ g_\zeta \hat \Phi . 
\end{align}
\end{subequations}
The solution of Eq.~\eqref{eq:eomX} can be written as
\begin{equation}
\label{eq:solX}
 \begin{split}
\hat {\Psi}_\zeta(\bx,t) &= \hat {\Psi}_\zeta^0(\bx,t) -i \int dt' \theta(t-t') \ep{ - i\omega_\zeta (t-t')} g_\zeta \hat \Phi (\bx,t').
 \end{split}
 \end{equation} 
The first term is the homogeneous solution (contrary to Eq.~\eqref{eq:psi0fock}, the environment is not relativistic)
\begin{equation}
 \hat {\Psi}_\zeta^0 (\bx,t) =\hat c(\bx, \zeta) \ep{- i \omega_\zeta t}. 
\end{equation}
Here $\hat c(\bx,\zeta)$ is the destruction operator of a environment quantum of energy $\omega_\zeta$ localized at $\bx$. It obeys the canonical commutator 
\begin{equation}
[ \hat c(\bx',\zeta'), \hat c^\dagger(\bx,\zeta) ] =\delta(\bx-\bx') \delta(\zeta- \zeta') . 
\end{equation}
 
Introducing the right hand side of Eq.~\eqref{eq:solX} in Eq.~\eqref{eq:eompsipol} gives the effective equation of motion for the photon field,
\begin{equation}
\label{eq:eomdissip}
\begin{split}
i\partial_t \hat \Phi &= \left ( E_0^{\rm bare} - \frac{\partial_\bx^2}{2m} + g \hat \Phi^\dagger \hat \Phi\right) \hat \Phi - i \int dt' D(t-t') \hat \Phi (t') + \int d\zeta g_\zeta \hat {\Psi}_\zeta^0 + F =0 .
\end{split}
\end{equation} 
The non local dissipative kernel is [compare to Eq.~\eqref{eq:localgretofpsi}]
\begin{equation}
\label{eq:disskernelBEC}
\begin{split}
 D(t-t') &\doteq \theta(t-t') \int d\zeta g_\zeta^2\ep{ - i\omega_\zeta (t-t')} ,
\end{split}
\end{equation} 
and its Fourier transform is 
\begin{equation}
\begin{split}
 \tilde D(\omega) &= \int d\zeta g_\zeta^2 \frac{i}{\omega - \omega_\zeta + i \epsilon} . 
\label{eq:tDom}
\end{split}
\end{equation} 
Because of the high frequency of the pump as compared to the time-scale of the hydrodynamic evolution of the fluid, we shall see below that $D(t-t')$ can be well approximated by a local kernel within a sort of {\em Markov} approximation and correspondingly $\tilde D(\omega)$ can be approximated by a constant value independent of $\omega$.

Under a weak-interaction assumption, we perform the usual dilute gas approximation~\cite{pitaevskii2003bose} and we split the field operator as the sum $\hat \Phi = \Phi_{\rm cond} ( 1 + \hat \phi)$ of a (large) coherent component $\Phi_{\rm cond}(\bx,t)$ corresponding to the condensate and a (small) quantum fluctuation field $ \hat \phi(\bx,t)$. Including the new terms stemming from pumping and from losses, the mean field $\Phi_{\rm cond}(\bx,t)$ can be shown to obey a generalized Gross-Pitaevskii-Langevin equation of the form
\begin{equation}
\label{eq:GPEgeneral}
\begin{split}
i\partial_t \Phi_{\rm cond}(\bx,t) = &\left ( E_0^{\rm bare} - \frac{\partial_x^2}{2m} + g \abs{\Phi_{\rm cond}}^2\right ) \Phi_{\rm cond}(\bx,t)\\
& - i \int dt' D(t-t') \Phi_{\rm cond} (\bx,t') + F(\bx,t).
\end{split}
\end{equation}
When assuming that the pump is almost monochromatic $F(\bx,t)=F_0(\bx,t) e^{- i \omega_p t}$ and $\Phi_{\rm cond}(\bx,t) =\bar \Phi_0(\bx,t) e^{- i \omega_p t}$ with $\bar \Phi_0(\bx,t)$ and $F_0(\bx,t)$ slowly varying functions of time, it is appropriate to extract the temporally local part of the dissipative kernel and to rewrite Eq.~\eqref{eq:GPEgeneral} as 
\begin{equation}
\begin{split}
\label{eq:GPEomegae}
i\partial_t \bar \Phi_0(\bx,t) = F_0(\bx,t) + \left ( E_0^{\rm bare} + \Delta E - \omega_p - \frac{\partial_x^2}{2m} + g \abs{\bar \Phi_0}^2 - i\Gamma \right) \bar \Phi_0(\bx,t) \\
 - i \int dt' D(t-t') e^{i \omega_p (t-t')} \left [\bar \Phi_0(\bx,t') - \bar \Phi_0(\bx,t) \right ] ,
\end{split}
\end{equation}
where the real and imaginary parts of $\tilde D( \omega_p)=\Gamma+i\Delta E$ defined in Eq.~\eqref{eq:tDom}, respectively give the decay rate $\Gamma$ and a (small) shift $\Delta E $ of the photon frequency, see ref.~\cite{Caldeira:1981rx}. Explicitly, one has 
\begin{equation}
\begin{split}
\Delta E &= - \int d\zeta g_\zeta^2 \mathtt{P.V.}\frac{1}{ (\omega_\zeta- \omega_{\rm p}) } , \quad \Gamma = \int d\zeta g_\zeta^2 \pi \delta(\omega_\zeta- \omega_{\rm p} ) ,
\end{split}
\end{equation} 
where $\mathtt{P.V.}$ is the Cauchy principal value. In the following, all formulas will be written in terms of the effective cavity photon frequency, $E_0=E_0^{\rm bare}+\Delta E$. The ratio $ \omega_p /\Gamma =Q_p$ gives the quality factor of the cavity: in typical microcavity systems one has $Q_p \lesssim 10^5$. For a given $\omega_p$, various choices of $g_\zeta$ giving the same value for $\Gamma$ should be considered at this level as physically equivalent. 

\section{The stationary state} 

\begin{SCfigure}[2]
\includegraphics[width=0.47 \linewidth]{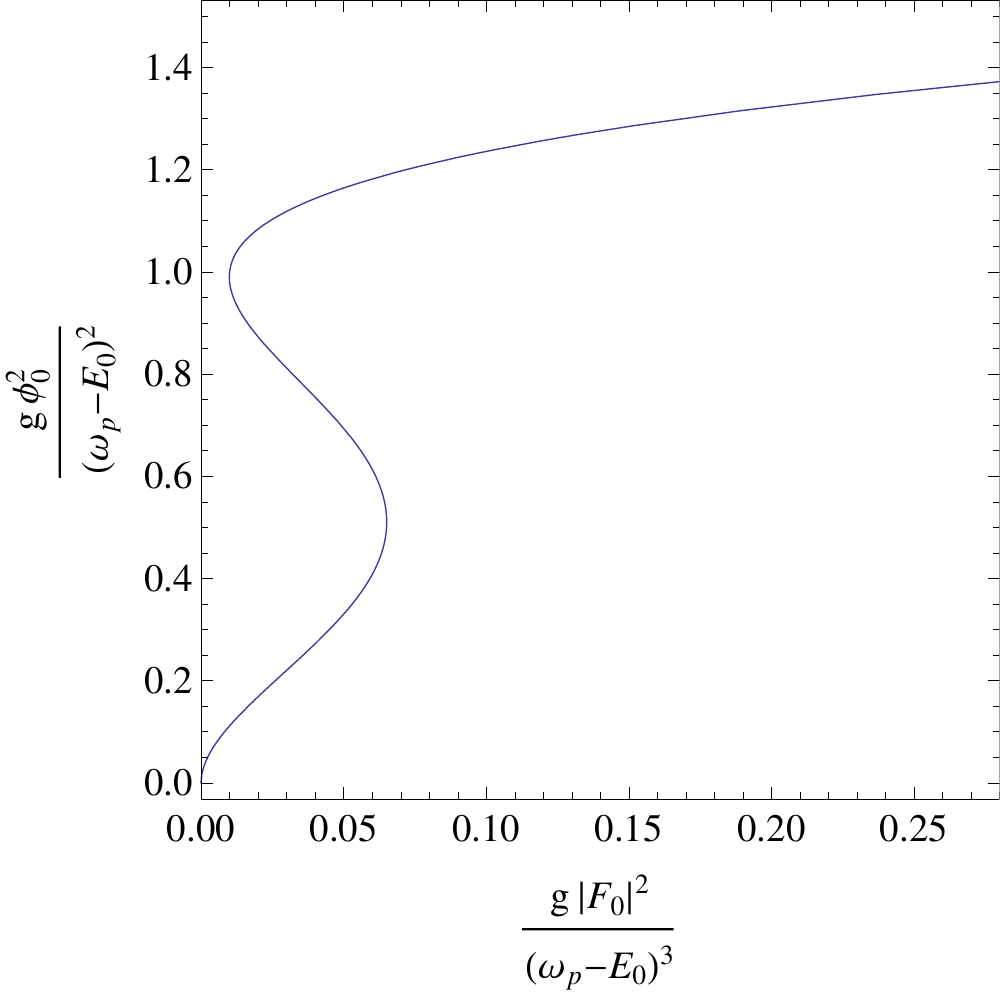}
\caption{We represent the density of the condensate, solution of Eq.~\eqref{eq:GPEhomo} as a function of the pump intensity. The relevant parameter in such a figure is the distance from the bistability loop to the vertical axes fixed by $\Gamma / (\omega_p - E_0)$ (here $0.1$). We only consider condensates such that $g {\mathtt \Phi}_0^2 \gtrsim \omega_p - E_0$, i.e., in the upper branch of the bistability loop, in the stable region.}
\label{fig:bistabilityloop}
\end{SCfigure}

We begin our discussion of quantum fluctuations in a stationary state under a spatially homogeneous and monochromatic pump at frequency $\omega_p$, $F(\bx,t)=F_0 e^{- i \omega_p t}$ with a constant pump amplitude $F_0$. In this case, we can safely assume the coherent component $\Phi_{\rm cond}$ of the photon field to be itself spatially homogeneous and monochromatically oscillating at $\omega_p$, $\Phi_{\rm cond}(\bx,t) = {\mathtt \Phi}_0 \ep{ - i \omega_p t}$. Using Eq.~\eqref{eq:GPEomegae}, it is immediate to see that ${\mathtt \Phi}_0$ obeys the state equation
\begin{equation}
\label{eq:GPEhomo}
\begin{split}
\left [ \omega_p - E_0 - g \abs{{\mathtt \Phi}_0}^2 + i\Gamma \right ] {\mathtt \Phi}_0 = F_0.
\end{split}
\end{equation}
In the following, we shall assume that the phase of the pump $F_0$ is chosen in a way to give a real and positive ${\mathtt \Phi}_0>0$. As we are interested in a stable configuration where the phonon mass is the smallest, we will follow previous work on analog models based on superfluids of light in microcavities~\cite{Gerace:2012an} and concentrate our attention on the case of a pump frequency blue-detuned with respect to the bare photon frequency $\omega_p>E_0$, where the dependence of the internal intensity $\mathtt{\Phi}_0^2$ on the pump intensity $|F_0|^2$ shows a bistability loop, see Fig.~\ref{fig:bistabilityloop} or Refs.~\cite{PhysRevLett.93.166401,Carusotto:2012vz}. More specifically, we shall concentrate on the upper branch of the bistability loop, where interactions have shifted the effective photon frequency $E_0+g\mathtt{\Phi}_0^2$ to the blue side of the pump laser, $E_0+g\mathtt{\Phi}_0^2 \geq \omega_p$. Exact resonance $\omega_p=E_0+g\mathtt{\Phi}_0^2$ is found at the end-point of the upper branch of the bistability loop: as we shall see shortly, only this point corresponds to a vanishing phonon mass. The more complex physics of quantum fluctuations under a monochromatic pump in the vicinity of the so-called \enquote{magic angle} was discussed in~\cite{sarchi} for pump intensities spanning across the optical parametric oscillation threshold~\cite{Ciuti:SST2003}.

\subsection{The equation of motion} 
\label{sec:eompolaritons}

Eqs.~\eqref{eq:eomdissip} and~\eqref{eq:GPEgeneral} determine the equation for linear perturbations $\hat \phi$. Taking into account the spatial homogeneity of the mean-field solution $\Phi_{\rm cond}$, we use the relative perturbation $\hat \phi_\bk $ at given wave number $\bk$. Using a Markovian\footnote{
The exact equation is given in Eq.~\eqref{eq:perturb}. Since the characteristic frequency $\omega_k$ of phonon modes is much lower than $\omega_p$, the non-local part of the dissipative term of Eq.~\eqref{eq:disskernelBEC} can be neglected as it gives corrections proportional to $\omega_k/\omega_p$. In fact, using Eq.~\eqref{eq:tDom}, $[\tilde D(\omega_p + \omega) - \tilde D(\omega_p)] \sim \tilde D(\omega_p) \omega/ \omega_p$ for typical dissipation baths, which is much smaller in magnitude than $\tilde D(\omega_p)$ when $\omega\ll \omega_p$.} 
approximation to neglect the non-local part of the dissipative term, one obtains the following quantum Langevin equation of motion [compare with Eq.~\eqref{eq:eomphi}]
\begin{equation}
\label{eq:2x2eom}
\begin{split}
i(\partial_t + \Gamma )\hat \phi_\bk &= \Omega_k \hat \phi_\bk+ m c^2 \hat \phi_{-\bk}^\dagger +
\frac{\hat S_\bk}{{\mathtt \Phi}_0}.
\end{split}
\end{equation}
Its conservative part shows interesting differences from the case of atomic condensates: While the interaction energy has the same form as in Eq.~\eqref{eq:eomhomogephonon}
\begin{equation}
 mc^2\doteq g \mathtt{\Phi}_0^2,
\label{eq:inten}
\end{equation}
the detuning coefficient multiplying $\hat \phi_\bk$ in Eq.~\eqref{eq:2x2eom} keeps track of the pump frequency $\omega_p$. It is given by
\begin{equation}
\Omega_k \doteq \frac{k^2}{2m} - \omega_p + E_0 + 2 m c^2 ,
\label{eq:defOmega}
\end{equation}
instead of Eq.~\eqref{eq:OmegakBEC}. It allows for a larger variety of Bogoliubov dispersions~\cite{PhysRevLett.93.166401,Carusotto:2012vz}. The eigenmodes of the deterministic part of the linear problem described by Eq.~\eqref{eq:2x2eom} are in fact characterized by the dispersion Eq.~\eqref{eq:defomegak} and collective phonon destruction operator of the form of Eq.~\eqref{eq:defvarphibogofield} where $u_k$ and $v_k$ are given by Eq.~\eqref{eq:ukandvk}. Using Eq.~\eqref{eq:defOmega}, we get to the explicit expression
\begin{equation}
\begin{split}
 \tilde \omega_k^2 = M (M+2m) c^4 + k^2 c^2 (1+M/m) + \frac{k^4}{4m^2} ,
\label{eq:disprelpol}
\end{split}
\end{equation}
in terms of the mass parameter $M$ defined by
\begin{equation}
M c^2 = E_0 + m c^2 - \omega_p\geq 0.
\end{equation}
We put a $\tilde{ }$ on $ \tilde \omega_k^2 $ to emphasize that the dispersion relation is the one of Eq.~\eqref{eq:disprelBEC}, or equivalently $ \Omega^2 + 2 i \Gamma \Omega = \omega_k^2 $, with $ \omega_k $ and $ \tilde \omega_k $ linked by Eq.~\eqref{eq:omGamma}. The presence of a finite phonon rest energy is a crucial difference as compared to the equilibrium case where phonons are always massless. The phonon mass is however dramatically suppressed $M\ll m$ when the pump frequency approaches resonance with the (interaction-shifted) cavity mode, $\omega_p \simeq E_0 + g {\mathtt \Phi}_0^2$, that is, when the operating point approaches the leftmost end-point of the upper branch of the bistability loop. In this limit, $M\to 0$ and the dispersion exactly recovers the usual Bogoliubov dispersion of equilibrium Bose condensates~\cite{pitaevskii2003bose}, with massless phonons and a low-frequency speed of sound equal to $c$, see under Eq.~\eqref{eq:defomegak}. 

As usual for quantum Langevin equations, the equation of motion~\eqref{eq:2x2eom} also involves a decay term proportional to $\Gamma$ and an effective quantum source term
\begin{equation}
\label{eq:Sdef}
\hat S_\bk(t) \doteq \int d\zeta g_\zeta \hat c(\bk,\zeta) \ep{- i (\omega_\zeta -\omega_p)t}
\end{equation}
summarizing quantum fluctuations in the initial state of the environment, assumed to be decorrelated from the system. In the Markovian limit $\omega\ll \omega_p$, the Langevin quantum noise operator $\hat S_\bk(t)$ satisfies the bosonic commutation relations of a destruction operator 
\begin{subequations}
\begin{align}
 [\hat S_\bk(t) , \hat S^\dagger_{\bk'}(t')] & = \int d\zeta g_\zeta^2 \ep{-i (\omega_\zeta-\omega_p) (t-t')} \sim 2 \Gamma \delta(t-t') \delta(\bk - \bk'), \\
 [\hat S_\bk(t) , \hat S_{\bk'}(t')] &= 0.
\end{align}
\end{subequations}
We further assume that the environment is initially in an equilibrium thermal state with low temperature $T_e \ll \omega_p$. As the characteristic phonon frequencies $\tilde \omega_k$ are also much smaller than $\omega_p$, we can safely approximate the expectation values by the following expressions
\begin{subequations}
\begin{align}
\left < \hat S_\bk(t)\hat S_{\bk'}(t') \right > & = 0 , \\
\left < \hat S_\bk (t)\hat S_{\bk'}^\dagger(t')\right > &\simeq 2\Gamma \delta(t-t') \delta(\bk-\bk'), \label{eq:whitenoise} \\
\left < \hat S_\bk^\dagger (t)\hat S_\bk(t')\right > &= \frac{2 \Gamma}{\ep{ {\omega_p}/{T_{\rm e}}}-1} \delta(t-t') \delta(\bk-\bk')\simeq 0 .
\end{align}
\end{subequations}
This means that the environment is a vacuum white noise bath with a flat frequency distribution. 

\subsection{Quantum fluctuations in the steady state}
\label{sec:Qflofsteadystate}

In the present stationary case, the Bogoliubov transformation of Eq.~\eqref{eq:defvarphibogofield} is time independent. In terms of the phonon operator $\hat \varphi_{\bk}$, Eq.~\eqref{eq:2x2eom} then becomes,
\begin{equation}
\begin{split}
i(\partial_t + \Gamma ) \hat \varphi_{\bk} = \tilde \omega_k \hat \varphi_\bk + \left (u_k \hat S_\bk - v_k \hat S_{-\bk}^\dagger\right ).
\label{eq:phoeom}
\end{split}
\end{equation}
Because of $\omega_p > 0$, the creation operator $\hat S_{-\bk}^\dagger$ contains positive frequency. Indeed, using Eq.~\eqref{eq:Sdef}, one gets
\begin{equation}
\begin{split}
\int dt \ep{i \omega t} \hat S_{-\bk}^\dagger &= 2 \pi \int d\zeta g_\zeta \hat c_0^\dagger(\bk,\zeta) \delta (\omega + \omega_\zeta -\omega_p) , 
\label{eq:Spositf}
\end{split}
\end{equation}
which only vanishes for $ \omega > \omega_p$, that is, far outside the frequency range involved in the phonon dynamics. As a result, the quantum fluctuations of the environment heat up the phonon state even when $\hat \Psi^0_\zeta$ is in its vacuum state, i.e., when the environment state is annihilated by $\hat S_\bk$.

As in Eq.~\eqref{eq:decomp}, the solution of Eq.~\eqref{eq:phoeom} has the following structure
\begin{equation} 
\begin{split}
\label{eq:Phidecom}
\hat \varphi_\bk(t) &= \hat \varphi_\bk^{\rm dec}(t;t_0) + \hat \varphi_\bk^{\rm dr}(t;t_0) .
\end{split} 
\end{equation}
The decaying part is 
\begin{equation}
\label{eq:phidec}
\begin{split}
 \hat \varphi_\bk^{\rm dec}(t;t_0) = \hat b_\bk \ep{- \Gamma (t-t_0)} \ep{- i \tilde \omega_k (t-t_0)} ,
\end{split}
\end{equation}
where the $\hat b_\bk$ operator destroys a phonon at time $t_0$ and obeys the canonical commutator $[\hat b_\bk, \hat b_{\bk'}^\dagger] = \delta(\bk - \bk')$. The driven part is 
\begin{equation}
\label{eq:phidrBEC}
 \begin{split}
 \hat \varphi_\bk^{\rm dr}(t;t_0) = -i \int_{t_0}^t dt' &\ep{- \Gamma (t-t')} \ep{- i \tilde \omega_k (t-t')} \left (u_k \hat S_\bk (t') - v_k \hat S_{-\bk}^\dagger(t')\right ).
 \end{split}
 \end{equation}
One verifies that $\hat \varphi_\bk$ of Eq.~\eqref{eq:Phidecom} obeys the usual equal time commutators 
\begin{subequations}
\begin{align}
[\hat \varphi_\bk(t), \hat \varphi_{\bk'}^\dagger (t) ] &= \delta(\bk - \bk'), \\
\left[\hat \varphi_\bk(t), \hat \varphi_{\bk'}(t)\right] &= 0,
\end{align}
\end{subequations}
as an identity, irrespectively of the choice of $t_0$. More precisely, the two-time commutators are given by:
\begin{subequations}
\begin{align}
\label{eq:comm1}
[ \hat \varphi_\bk(t) , \hat \varphi_{\bk'}^\dagger (t') ] &= \ep{- \Gamma\abs{t-t'}} \ep{- i\tilde \omega_k (t-t')} \delta(\bk - \bk'), \\
[ \hat \varphi_\bk(t) , \hat \varphi_{-\bk'} (t') ] &= \mathcal{O}\left ( \frac{\tilde \omega_k}{\omega_p}\right ) \delta(\bk - \bk'). 
\end{align}
\end{subequations}
In the Markov limit under consideration here $\tilde \omega_k\ll \omega_p$, the latter commutator is negligible. Once the stationary state has been reached (i.e., in the $t_0\to -\infty$ limit), the decaying part $\hat \varphi_\bk^{\rm dec}$ is also negligible and $\hat\varphi_\bk$ is given by $\hat\varphi_\bk^{\rm dr}$ of Eq.~\eqref{eq:phidrBEC}.

\subsubsection{ Two-point functions in the steady state}

The statistical properties of the phonon field are summarized by two point correlation functions 
\begin{equation}
\begin{split}
\label{eq:Gphidef}
G^{\varphi^\dagger \varphi}(t,t'; \bk) & \doteq \left < \hat \varphi_{\bk} ^\dagger (t)\hat \varphi_{\bk} (t') \right > ,\\
G^{\varphi \varphi}(t,t'; \bk) &\doteq \left < \hat \varphi _{-\bk} (t)\hat \varphi_{\bk} (t')\right > , 
\end{split}
\end{equation} 
which are directly related to the physically observable second-order coherence function, see Eq.~\eqref{eq:defg2}, describing the correlations of density fluctuations of the in-cavity photon field. The Fourier transform of $g_2 (\bx,t,\bx',t')$ [see Eq.~\eqref{eq:g2ofGac}] is related to the so-called structure factor of the fluid and provides direct information on the $\bk$ component of the density fluctuations~\cite{pitaevskii2003bose}. To quadratic order in $\hat \phi$ it is equal to 
\begin{equation}
g_{2}^k (t,t') =  2 \left (u_k-v_k \right )^2 \Re[G(t,t', \bk)] -2 v_k(u_k-v_k) \ep{- \Gamma\abs{t-t'}} \cos[\tilde \omega_k (t-t')] ,
\label{eq:g2k}
\end{equation}
where 
\begin{equation}
\begin{split}
G(t,t', \bk) \doteq G^{\varphi^\dagger\varphi}(t,t', \bk) + G^{\varphi \varphi}(t,t', \bk)
\label{eq:RGdef}
\end{split}
\end{equation}
is linked to the anticommutator of $\hat \chi$ we used in \ref{chap:dissipBEC} by
\begin{equation}
\begin{split}
\Re[G(t,t', \bk)] = \omega_{\rm f} G_{ac} - \frac{\cos[\tilde \omega_k (t-t')]}{2}.
\end{split}
\end{equation}
We thus see that a measurement of $g_2$ provides complete information on $\Re[G]$: the second term of Eq.~\eqref{eq:g2k} is in fact state-independent, as it is equal to the real part of the commutator in Eq.~\eqref{eq:comm1} multiplied by some known factor. 

Using Eqs.~\eqref{eq:phidrBEC} and~\eqref{eq:whitenoise}, one obtains the two-point functions of Eq.~\eqref{eq:Gphidef} in the stationary state:
\begin{equation}
\begin{split} 
\label{eq:Gphieq}
G_{\rm st}^{\varphi^\dagger\varphi}(t,t'; \bk) &= n^b_{k,\rm st} \ep{- \Gamma \abs{t-t'}} \ep{ i \tilde \omega_k(t-t')},\\
G_{\rm st}^{\varphi\varphi}(t,t'; \bk) &= \bar c^b_{k,\rm st} \ep{- (\Gamma +i \tilde \omega_k) \abs{t-t'}} ,\\
\end{split}
\end{equation}
where 
\begin{equation}
\label{eq:nbarceq}
\begin{split}
n^b_{k,\rm st} =v_k^2,\quad \bar c ^b_{k,\rm st} =\frac{ u_k v_k \Gamma}{\Gamma + i \tilde \omega_k}.
\end{split}
\end{equation}
Roughly speaking, these quantities give the mean occupation and the correlation function in the phonon point of view. As shown by a more careful analysis, these identifications are subjected to some inherent imprecision, see the discussion below in \ref{sec:weakdissip} and \ref{sec:cstdissiprate}. 

An alternative description of this state in terms of the photon variables (instead of the phonon ones) can be obtained using the Bogoliubov transformation Eq.~\eqref{eq:defvarphibogofield}. The mean occupation number and the correlations of photon operators are equal to
\begin{equation}
\label{eq:nka}
\begin{split}
n_k^{a,{\rm st}} = 2 u_k^2 v_k^2 \frac{\tilde \omega_k^2}{\Gamma^2+\tilde \omega_k^2} , \quad c_k^{a,{\rm st}} = i\tilde \omega_k u_k v_k \left[ \frac{v_k^2 }{ \Gamma -i \tilde \omega_k}- \frac{u_k^2}{ \Gamma +i \tilde \omega_k} \right].
\end{split}
\end{equation}
These quantities are accessible from the intensity pattern of the far-field emission from the cavity and its coherence properties: the presence of a non-vanishing emission $n_k^{a,{\rm st}}$ at a wavevector distinct from the coherent pump at $\bk=0$ stems from parametric processes analogous to the ones taking place in parametric down-conversion experiments. The non-vanishing correlation $c_k^{a,{\rm st}}\neq 0$ is a signature of the two-mode squeezed nature of this emission~\cite{d1995quantum,gardiner2004quantum,PhysRevB.63.041303}.

From the photon momentum distribution Eq.~\eqref{eq:nka}, it is immediate to calculate the first-order coherence function defined in Eq.~\eqref{eq:defg1}. For simplicity, we restrict our attention to $g_1$ evaluated at equal times $t=t'$,
\begin{equation}
\begin{split}
g_1 (\bx,\bx',t'=t) ={\mathtt \Phi}_0^2+\int \frac{d\bk}{(2\pi)^d} e^{i\bk\cdot(\bx'-\bx)} n_\bk^{a,{\rm st}}.
\end{split}
\end{equation}
In this equation, $d$ is the dimensionality of the fluid along the cavity: while standard planar cavities as the one sketched in Fig.~\ref{fig:sketch} have $d=2$, effective one-dimensional $d=1$ fluids can be created with an additional in-plane confinement~\cite{Carusotto:2012vz}. The modified Bogoliubov coefficients $u_k, v_k$ which appear in Eq.~\eqref{eq:nka} are given in Eq.~\eqref{eq:ukandvk} and the frequency $\tilde \omega_k$ in Eq.~\eqref{eq:disprelpol}. Using these expressions, a straightforward calculation gives
\begin{equation}
n_k^{a,{\rm st}} = \frac{m^2 c^4}{2(\Gamma^2+\tilde \omega_k^2)}
\end{equation}
which is regular in the $k\to 0$ limit both because of the (small) phonon mass $M(M+2m)c^4$ in Eq.~\eqref{eq:disprelpol} and because of losses. 

Using isotropy, it is immediate to see from what precedes that for any dimensionality $d$, the $|\bx-\bx'|\to\infty$ long distance limit of $g_1$ shows a condensate plus an exponentially decaying term,
\begin{equation}
g_1 (\bx,\bx',t'=t) \simeq {\mathtt \Phi}_0^2 + A e^{-|\bx-\bx'|/\ell_c} ,
\end{equation}
with a coherence length
\begin{equation}
\ell_c=\left[\frac{c^2(1+M/m)}{\Gamma^2+M(M+2m)c^4}\right]^{1/2}.
\end{equation}
This shows that in the present case, thanks to the presence of the coherent pump, the long-distance coherence of the photon \enquote{condensate} is robust against fluctuations independently of the dimensionality. For a recent discussion of the long distance coherence of $g_1$ under an incoherent pump, we refer to~\cite{Chiocchetta:arXiv1302.6158,Altman:arXiv1311.0876}. 

\subsubsection{Dissipationless limit}

\begin{figure}[htb]
\begin{minipage}[t]{0.47\linewidth}
\includegraphics[width=1\linewidth]{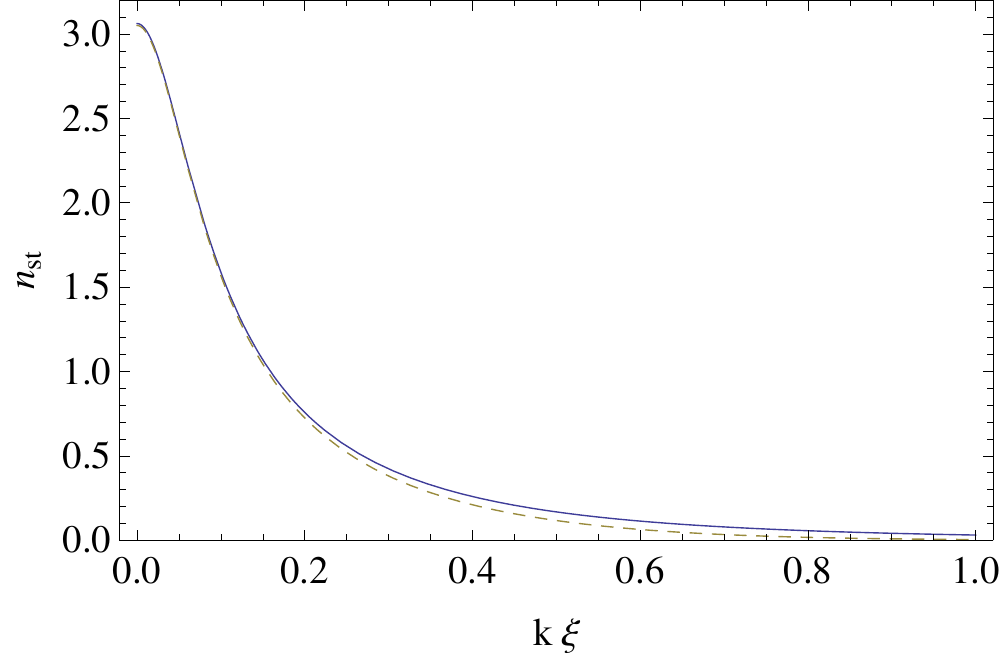}
\caption{Mean occupation number of phonons $n_{k,\rm st}^b = v_k^2$ for a low phonon mass parameter $M/m = 0.01$ in the driven-dissipative steady state (solid), and in a thermal state at temperature $T= 1/2 m c^2 $ (dashed). Both curves are independent of the dissipative rate $\Gamma$. Whatever the value of the mass parameter $M$, one can show that the {\em absolute} deviation between the two curves is always smaller than $0.052$ which is reached for $\Omega_k \sim 1.5 m c^2$. On the other hand, the {\em relative} difference between the two occupation numbers becomes large at high momenta, see Fig.~\ref{fig:nandbarc}.} 
\label{fig:ninofk}
\end{minipage}
\hspace{0.03\linewidth}
\begin{minipage}[t]{0.47\linewidth}
\includegraphics[width=1\linewidth]{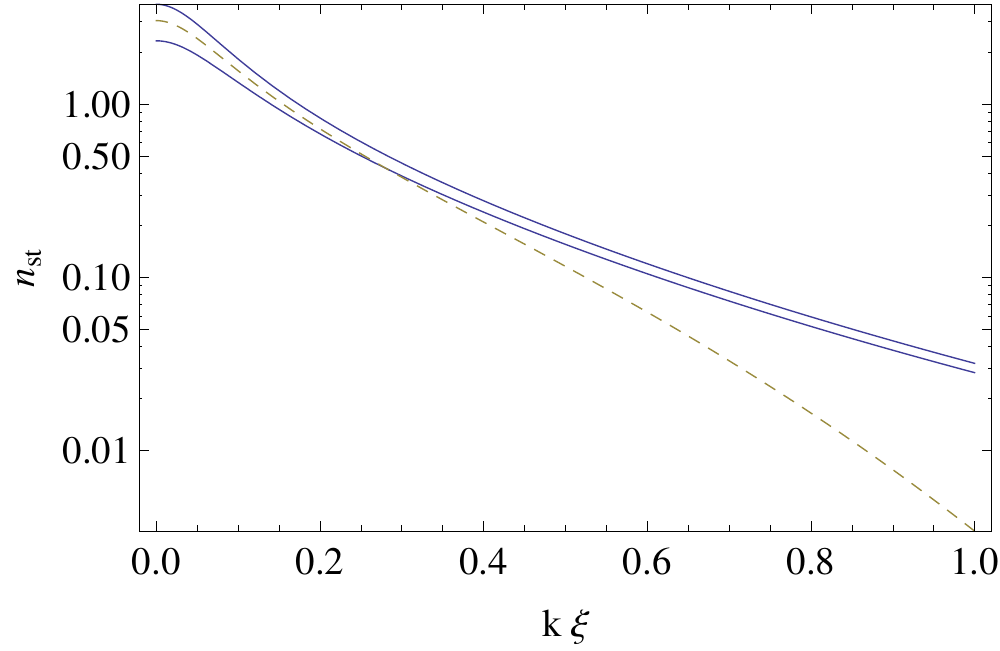}
\caption{ Intrinsic imprecision in the measurement of the mean occupation number of phonons. The solid curves show $ n_k^b= n_{k,\rm st}^b \pm \abs{\bar c_{k,\rm st}^b}$ involved in Eq.~\eqref{eq:RG4}. The dashed line represents the thermal distribution at $T= 1/2 m c^2 $: while $n_k^b$ decays as $1/k^2$ for large momenta in the non-equilibrium stationary state, the thermal distribution decays according to a much faster Boltzmann law as $\ep{- (k /mc)^2 }$. Same system parameters as in Fig.~\ref{fig:ninofk}: phonon mass $M/m = 0.01 $ and dissipation rate $\Gamma/m c^2 = 0.03 $.}
\label{fig:nandbarc}
\end{minipage}
\end{figure}

To understand the physical implications of these results, we first consider the $\Gamma \to 0$ limit where dissipation tends to zero. Note that because of the presence of the pump the system does not recover a standard thermodynamical equilibrium state in this limit, but maintain a non-equilibrium character. More details on this crucial fact are given in \ref{sec:FD}. In this case, the effective phonon state is incoherent, since $G_{\rm st}^{\varphi\varphi} = \bar c ^b_{k,\rm st} = 0$, as in standard thermal equilibrium. The state is thus fully characterized by the finite value of Eq.~\eqref{eq:nbarceq} of the mean phonon occupation number $n^b_{k}$. Interestingly, even in the $\Gamma \to 0$ limit, the stationary state of the system differs from a standard thermodynamical equilibrium state, as it is manifest in the phonon occupation distribution not following the Planck distribution. Nonetheless, as can be seen in Fig.~\ref{fig:ninofk}, the state it is very close to a thermal state at temperature $k_B T_{\rm st} = m c^2/2 $ fixed by the interaction energy, see Eq.~\eqref{eq:inten}. This is our first result: Because of the unusual presence of positive frequency in $\hat S_{-\bk}^\dagger$, see Eq.~\eqref{eq:Spositf}, the phonon field is effectively heated up even when the environment is in its vacuum state. In Fig.~\ref{fig:ninofk} and in subsequent figures, the wavevector $k$ is adimensionalized by making use of the healing length defined by $\xi = 1/ 2m c $ (since $\hbar = 1$). 

From a physical point of view, it is important to note the conceptual difference of this result with respect to the quantum depletion of the Bose condensate as predicted for the ground state of equilibrium Bogoliubov theory~\cite{pitaevskii2003bose}: the finite occupation number $n^b_{k}$ refers here to phonon quasiparticle excitations, while standard quantum depletion refers to the underlying particles (in our case, photons). Along the same lines, one should not confuse the finite phonon occupation in the present driven-dissipative stationary state, with the finite photon occupation in the ground state of a microcavity device in the ultra-strong light-matter coupling, as discussed in~\cite{bastard,PhysRevA.74.033811}.

To better understand the physical meaning of the two different photon and phonon descriptions of the same state, we shall apply the separability criterion in two distinct ways, either to photon or to phonon operators: the results are not expected to coincide as photon and phonon operators are related by Eq.~\eqref{eq:defvarphibogofield} which is a $U(1,1)$ transformation mixing creation and destruction operators. 

From the phonon point of view, the stationary state of the system is manifestly separable in the $\Gamma\to 0$ limit as phonons are fully incoherent, $\bar c ^b_{k,\rm st} = 0$. On the contrary, the same state is nonseparable from the photon point of view since 
\begin{equation}
\abs{c_k^a}^2 = n_k^a( n_k^a + 1/2) ,
\end{equation}
violates the separability bound Eq.~\eqref{eq:Camposeparable}. Note that this result is not peculiar to the driven-dissipative case, but is also found in the ground state of equilibrium Bogoliubov theory. It has a straightforward physical interpretation if one reminds that the finite-$\bk$ photons originate from a parametric scattering process where two pump photons at $\bk_p=0$ scatter into the $\pm\bk$ states.

Even though the state is nonseparable only from the photon point of view, we can explicitly verify that the entropy of the state agrees in the two points of view, as it is expected from the invariance of entropy under $U(1,1)$ transformations. This is straightforwardly done knowing that the entropy is equal to 
\begin{equation}
 S= 2[(\bar n +1) \log (\bar n +1) - \bar n \log \bar n ],
\end{equation}
in terms of $\bar n$ defined by $(\bar n +1/2)^2 \doteq (n+1/2)^2 - \abs{c^2}$~\cite{Campo:2005sy}.

\subsubsection{Weak dissipation} 
\label{sec:weakdissip}

Having understood the state properties in the limit $\Gamma \to 0$, we now turn to the case of small but finite dissipative rates, $\Gamma \ll\omega_k$. The main change is that the correlation function Eq.~\eqref{eq:Gphieq} is now $G_{\rm st}^{\varphi \varphi} \neq 0$. While its $t-t'$ time dependence correctly expresses stationarity of the state, it dramatically differs from the usual $\tilde \omega_k(t+t')$ one describing correlations of real phonon pairs at wavevectors $\pm \mathbf{k}$~\cite{Carusotto:2009re}. This means that $G_{\rm st}^{\varphi \varphi} \neq 0$ {\it cannot} be straightforwardly interpreted as describing {\em real} pairs of phonons with opposite momenta. Still, because of the quantum fluctuations associated to the dissipation processes, there are non-trivial correlations $G_{\rm st}^{\varphi \varphi} \neq 0$ between phonon modes of opposite wavevectors $\pm\bk$. This is a second main result of this chapter.

The presence of a non-zero correlation $\bar c ^b_{k,\rm st}\neq 0$ in the stationary state has important consequences when one attempts to measure the occupation number $n^b_{k,\rm st}$ via a measurement of $g_2$ and thus of $\Re[G]$. To be specific, let us consider an experiment where $\Re[G(t,t', \bk)]$ is measured for various values of the interval $\tau=t'-t$. Provided $\tau$ is short enough, $\Gamma \tau \ll 1$, one gets
\begin{equation}
\begin{split}
\Re[G_{\rm st}(t,t+\tau, \bk)] \sim n^b_{k,\rm st} \cos(\tilde \omega_k \tau) + \Re (\bar c^b_{k,\rm st}\ep{- i \tilde \omega_k \tau }).
\end{split}
\end{equation}
For very small dissipation rates $\Gamma \to 0$, correlations are negligible: as a result, the l.h.s. divided by $\cos( \tilde \omega_k \tau)$ directly provides information on the mean number of particles $n^b_{k, \rm st}$. When proceeding in the same way in the presence of a significant dissipation, the same procedure gives 
\begin{equation}
\begin{split}
\tilde{n}^b_{k, \rm st} = n^b_{k,\rm st} + \Re (\bar c ^b_{k,\rm st}) + \Im (\bar c^b_{k,\rm st}) \tan( \tilde \omega_k \tau), 
\label{eq:RG4}
\end{split}
\end{equation}
which shows periodic deviations in $\tau$ around an average value $ n^b_{k,\rm st} + \Re (\bar c ^b_{k,\rm st})$; note that this average still differs from $n^b_{k,\rm st}$ by a systematic error proportional to $\bar c ^b_{k,\rm st}$, see Fig.~\ref{fig:nandbarc}.

\subsection{Fluctuation dissipation relation} 
\label{sec:FD}

We saw in \ref{sec:Qflofsteadystate} that the stationary state of phonons when the photon fluid is in its steady state in contact with the environment is not thermal. This might appear at a first glance as quite surprising since under very general conditions, systems weakly interacting with a large stationary reservoir reach a thermal equilibrium state as it is guaranteed by fluctuation-dissipation (FD) relations~\cite{landau1996statistical,Anglin:1992uq}. In this section, we shall see the violation of the FD relation stems from the fact that our system is externally driven by the coherent laser pump with a finite frequency $\omega_p$.

To show that this violation is not due to some approximation, we use the exact Heisenberg equation of motion without performing the Markov approximation we used in \ref{sec:eompolaritons}. From Eqs.~\eqref{eq:eomdissip} and~\eqref{eq:GPEgeneral}, in the place of Eq.~\eqref{eq:2x2eom}, the exact equation for linear perturbations is
\begin{equation}
\label{eq:perturb}
\begin{split}
i(\partial_t &+ \Gamma )\hat \phi_\bk(t) = \Omega_k \hat \phi_\bk(t)+ m c^2 \hat \phi_{-\bk}^\dagger(t) +
\frac{\hat S_\bk(t)}{\abs{{\mathtt \Phi}_0}} \\
& - i \int dt' D(t-t') \ep{i \omega_p(t-t')} \left ( \hat \phi_\bk(t') - \hat \phi_\bk(t) \right ) .
\end{split}
\end{equation}
Using Eq.~\eqref{eq:Sdef} to express $\hat S_\bk$ in terms of the destruction operators $\hat c(\bk,\zeta)$ of the environment, and working in Fourier transform to exploit the stationarity of the situation, the equation takes the form\footnote{For Eq.~\eqref{eq:perturb}, $\hat O_1 (\omega) = \omega + i \Gamma + i\tilde D(\omega_p+\omega)- i\tilde D(\omega_p) - \Omega_k$ and $\hat O_2 (\omega)= -mc^2$}
\begin{equation}
\begin{split}
O_1 (\omega) \hat \phi_\bk^\omega &+ O_2(\omega) \left (\hat \phi_{-\bk}^{-\omega}\right )^\dagger = \int d\zeta g_\zeta \delta({\omega + \omega_p - \omega_\zeta}) \hat c(\bk,\zeta) .
\end{split}
\end{equation}
Using the complex conjugated equation for $-\omega,-\bk $ to eliminate $(\phi_{-\bk}^{-\omega})^\dagger$, we get 
\begin{equation}
\label{eq:solphiomega}
\begin{split}
\bigg( O_1 (\omega) O_1^*(-\omega) -O_2(\omega) O_2^*(-\omega) \bigg) \hat \phi_\bk^\omega = &\int d\zeta g_\zeta 
\bigg( \delta({\omega +\omega_p - \omega_\zeta}) O_1^*(-\omega) \hat c(\bk,\zeta) \\
&\quad \hspace{1cm} + \delta({\omega - \omega_p + \omega_\zeta}) O_2(\omega) \hat c(-\bk,\zeta)^\dagger \bigg ) .
\end{split}
\end{equation}
Making the Bogoliubov transformation of Eq.~\eqref{eq:defvarphibogofield} to get the equation for the phonon field $\hat \varphi_\bk^\omega$ simply amounts to replace in the above equation $O_i$ by $u_k O_i - v_k O_{3-i}$, for $i \in \{1,2\}$. Hence, the same type of expression applies to $\hat \varphi_\bk^\omega$, or, more generally, to any linear superposition (even $\omega$ dependent) of $\hat \phi_\bk^\omega $ and $(\hat \phi_{-\bk}^{-\omega})^\dagger $.

We now remind the reader that the FD relation trivially applies at the level of the operators of the environment. Namely, when working in a thermal state, one has
\begin{equation}
\begin{split}
\frac{\left < \left \{ \hat c(\bk,\zeta), \hat c(\bk,\zeta)^\dagger \right \} \right >}{\left [ \hat c(\bk,\zeta), \hat c(\bk,\zeta)^\dagger \right ]} = \frac{ \coth \frac{\beta \omega_\zeta }{2}}{2}, 
\label{eq:FDenv}
\end{split}
\end{equation}
as can be immediately verified by computing the commutator and the expectation value of the anti-commutator of $\hat c(\bk,\zeta)$ and $\hat c(\bk',\zeta')^\dagger $.

When the pump frequency $\omega_p = 0$, the situation is simple: Because of the Dirac $\delta$ function in Eq.~\eqref{eq:solphiomega}, and because the energy of the environment modes $\omega_\zeta $ is positive for all $\zeta$, for $\omega>0$, $\hat \phi_\bk^\omega$ is only driven by the destruction operator $\hat c(\bk,\zeta)$ with $\omega_\zeta = \omega$. Then, using Eq.~\eqref{eq:FDenv}, a direct evaluation gives 
\begin{equation}
\begin{split}
\frac{ \left < \left \{ \hat \phi_\bk^\omega,(\hat \phi_\bk^{\omega})^\dagger \right \} \right >}{\left [ \hat \phi_\bk^\omega,(\hat \phi_\bk^{\omega})^\dagger \right ]} = \frac{ \coth \frac{\beta \omega}{2}}{2} ,
\end{split}
\end{equation}
irrespectively of the values of $O_1(\omega)$ {\it and} $O_2(\omega)$. This is the standard FD relation. 

When $\omega_p \neq 0$, to get a concise expression, as in Ref.~\cite{Caldeira:1981rx}, it is useful to introduce the effective density of states $J(\omega)$ through $d\zeta g_\zeta^2 = d\omega_\zeta J(\omega_\zeta)$. A direct evaluation then gives 
\begin{equation}
\label{eq:FDviolated}
\begin{split}
&\frac{ \left < \{ \hat \phi_\bk^\omega, \hat \phi_\bk^{\dagger,\omega} \} \right >}{[ \hat \phi_\bk^\omega, \hat \phi_\bk^{\dagger,\omega} ]} \\
&= \frac{ \abs{ O_1(-\omega) }^2 J(\omega_p + \omega) \coth\beta(\omega_p+\omega)/2 - \abs{O_2 (\omega)}^2 J(\omega_p - \omega) \coth\beta(\omega_p - \omega)/2 }{ 2\left (\abs{O_1(-\omega)}^2 J(\omega_p + \omega) - \abs{O_2 (\omega)}^2 J(\omega_p - \omega) \right )} .\\
\end{split}
\end{equation}
We see that a FD relation is recovered only if $O_2(\omega) J(\omega_p - \omega) = 0$. In such a case, the argument in the $\coth$ in the right hand side of Eq.~\eqref{eq:FDviolated} is displaced as if there were a chemical potential $\mu = - \omega_p$.

For a general environment, all frequencies $\omega_\zeta$ are positive and cover the whole $\omega>0$ region, so $J(\omega)$ vanishes only for $\omega<0$. As a result, the $O_2(\omega) J(\omega_p - \omega) = 0$ condition requires either working at very high frequencies $\omega > \omega_p$ outside the region of interest for quantum hydrodynamics, or having $O_2(\omega) = 0$, that is a vanishing interaction between photons.

\section{Phonon pair production by a sudden modulation} 
\label{sec:DCE}

In the previous section, we studied the quantum fluctuations in a stationary state under a monochromatic continuous wave pump. In this section we shall extend the discussion to the case when a sudden change is imposed on the system and pairs of phonons are expected to be generated at the time of the fast modulation via processes that are closely analogous to the cosmological pair creation effect in the early universe (see \ref{chap:dissipdS} and Refs.~\cite{Fedichev:2003bv,Campo:2003pa}) and to the dynamical Casimir effect (see \ref{chap:dissipBEC} and Refs.~\cite{Carusotto:2009re,PhysRevLett.109.220401}). 

\subsection{The modified state} 

To facilitate analytical calculations, we will restrict our attention here to a very idealized model inspired by Ref.~\cite{Carusotto:2008ep}, where the spatially homogeneous and stationary condensate wave function of amplitude $\mathtt{\Phi}_{0} $ remains an exact solution of Eq.~\eqref{eq:GPEomegae} at all times. As compared to atomic gases, this requirement is a bit more subtle in the present non-equilibrium case as the photon density is related to the pump intensity by the more complicate state equation~\eqref{eq:GPEhomo}. A possible strategy to fulfill this condition might consist of assuming that $\Gamma$, $m$, the pump amplitude $F_0$ and its frequency $\omega_p$ remain constant while $g$ and $E_0$ suddenly change at $t = 0$ keeping $E_0(t) + g(t) \mathtt{\Phi}_{0}^2$ constant.\footnote{
In this case, the change is specified by one parameter. Two-parameter changes can be considered by changing both $\Gamma$ and $F_0$ while keeping their ratio constant. This can still be generalized by changing both $\omega_p$ and $E_0$ while keeping $\omega_p - E_0$ constant. In all cases, the gluing of the background across the jump is easily done.} 
While we agree that such modulations are quite unrealistic in state-of-the art experiments, still the predicted phonon pair production process appears to be conceptually identical to the one taking place in the more realistic but more complex configurations where the condensate wavefunction is itself varying as in, e.g., Ref.~\cite{Koghee2013}.

As a result of the modulation, the phonon frequency $\tilde \omega_k(t)$ of Eq.~\eqref{eq:disprelpol} experiences a sudden change (the subscript ${\rm in/f}$ refers to its value at times $t \gtrless 0$)
\begin{equation}
\begin{split}
\tilde \omega_k(t) &= \omega_{k, {\rm in}} + \theta(t) (\omega_{k, {\rm f}} - \omega_{k, {\rm in}}) ,
\end{split}
\end{equation}
that directly reflects onto the Bogoliubov operators: while the photon operator $\hat \phi$ in Eq.~\eqref{eq:2x2eom} is continuous at $t=0$, the phononic ones $\hat \varphi$ defined in Eq.~\eqref{eq:defvarphibogofield} experience the following sudden jump, see Eq.~\eqref{eq:bogodisp} and Ref.~\cite{Carusotto:2009re}
\begin{equation}
\label{eq:bogojump}
\begin{split}
\hat \varphi_{\bk}(t=0^+) &= \alpha_{k} \hat \varphi_{\bk}(t=0^-) + \beta_{k}\hat \varphi^\dagger_{-\bk}(t=0^-) ,\\
\alpha_{k}& = u_{k,{\rm f}} u_{k,{\rm in}}-v_{k,{\rm f}} v_{k,{\rm in}} = \frac{\omega_{\rm f} + \omega_{\rm in}}{2\sqrt{\omega_{\rm f} \omega_{\rm in}}}, \\
\beta_{k} &= v_{k,{\rm f}} u_{k,{\rm in}} -u_{k,{\rm f}} v_{k,{\rm in}}= \frac{\omega_{\rm f} - \omega_{\rm in}}{2\sqrt{\omega_{\rm f} \omega_{\rm in}}},
\end{split}
\end{equation} 
where the second equality of the last two lines follow from the constancy of $\Omega_k-mc^2$. 

Hence, for positive times $t$ and with, we have $ \hat \varphi_\bk^{\rm dec}(t) $ given by Eq.~\eqref{eq:phidec} with $t_0 =0^+$, $\hat b_\bk= \hat \varphi_{\bk}(t=0^+) $, and $ \hat \varphi_\bk^{\rm dr}(t) $ given by Eq.~\eqref{eq:phidrBEC}. Using the fact that the source term $\hat S$ has a white noise profile, at all times $t>0$ one has
\begin{equation}
\begin{split}
\left <\hat \varphi_\bk^{\rm dec} \hat \varphi_{-\bk}^{\rm dr} \right > = \left < \hat \varphi_\bk^{\rm dec} (\hat \varphi_{\bk}^{\rm dr})^\dagger \right >=0.
\end{split}
\end{equation}
For $t,t'>0$, after the jump, the two-point correlation functions defined in Eq.~\eqref{eq:Gphidef} then have the form
\begin{equation}
\label{eq:diisnb}
\begin{split}
G^{\varphi^\dagger, \varphi}_{\rm DCE}(t,t', \bk)& = \left ( n_{k,\rm f}^b \ep{- \Gamma \abs{t-t'}} + \delta n_{k}^b \ep{- \Gamma (t+t')} \right ) \ep{ i \omega_{k,{\rm f}}(t-t')}\\
G^{\varphi, \varphi}_{\rm DCE}(t,t', \bk)& = \bar c_{k,\rm f}^b \ep{- (\Gamma +i \omega_{k,+}) \abs{t-t'}} + c_k^b \ep{- (\Gamma + i \omega_{k,{\rm f}})(t+t')} .
\end{split}
\end{equation}

Four independent and constant quantities are identified in Eq.~\eqref{eq:diisnb} through the time dependence of their associated exponential factor, namely 
\begin{subequations}
\label{eq:4q}
\begin{align}
\label{eq:nfk}
n_{k,\rm f}^b &=v_{k,{\rm f}}^2 \\
\label{eq:barcfk}
\bar c_{k,\rm f}^b &= \frac{u_{k,{\rm f}} v_{k,{\rm f}} \Gamma}{\Gamma + i \omega_{k,{\rm f}}} \\
\delta n_{k}^b&= \left < \hat \varphi_{\bk,{\rm f}}^\dagger \hat \varphi_{\bk,{\rm f}} \right > - n_{k,\rm f}^b , \\
c_k^b &=\left < \hat \varphi_{-\bk,{\rm f}} \hat \varphi_{\bk,{\rm f}} \right >- \bar c_{k,\rm f}^b .
\end{align}
\end{subequations}
The first two quantities, $n_{k,\rm f}^b$ and $\bar c_{k,\rm f}^b$, give the final values once the stationary state is again reached for the new parameters after the jump: They have the same physical interpretation as $n_{k,\rm st}^b$ and $\bar c_{k,\rm st}^b$ defined in Eq.~\eqref{eq:nbarceq} and discussed at length in the previous section. Instead, $\delta n_{k}^b$ and $c_k^b$ govern the time dependence of the correlation functions in response to the jump in the parameters. They involve two traces taken at a time $t=0^+$ which are, see Eq.~\eqref{eq:bogojump} 
\begin{equation}
\begin{split}
\left < \hat \varphi_{\bk,{\rm f}}^\dagger \hat \varphi_{\bk,{\rm f}} \right> &= \left (\alpha_{k} ^2 +\beta_{k} ^2\right ) n_{k,\rm in}^b+\beta_{k} ^2 + 2\alpha_{k} \beta _{k} \Re\left ( \bar c_{k, \rm in}^b \right ) , \\
\left < \hat \varphi_{-\bk,{\rm f}} \hat \varphi_{\bk,{\rm f}} \right >&=\alpha_{k} ^2 \bar c_{k, \rm in}^b+\beta_{k} ^2 (\bar c_{k,\rm in}^b)^* +\alpha_{k} \beta_{k} \left(2 n_{k,\rm in}^b+1\right), \\
\end{split}
\end{equation}
where $n_{k,\rm in}^b$ and $\bar c_{k,\rm in}^b$ are given by Eqs.~\eqref{eq:nfk} and~\eqref{eq:barcfk} with $\rm f$ subscript replaced by $\rm in$. These are the initial stationary values as predicted by Eq.~\eqref{eq:nbarceq}. More specifically, $\delta n_k^b$ is involved in the only decaying term in Eq.~\eqref{eq:diisnb} which oscillates $\ep{i\omega_{k,{\rm f}} (t-t')}$: physically, its equal-time value $\delta n_k^b(t) = \delta n_k^b \ep{- 2 \Gamma t}$ describes the number of extra photons with respect to $n_{k,\rm f}^b$ that are generated by the jump and still present at time $t$. $c_k^b$ is instead involved in the only term which is rotating as $\ep{-i \omega_{k,{\rm f}} (t+t')}$: its equal-time value $c_k^b(t) = c_k^b \ep{- 2 (\Gamma+i\omega_{k,{\rm f}}) t}$ gives the instantaneous correlation between these extra phonons: as it is illustrated in Fig.~\ref{fig:nkmck}, these non-trivial correlations can produce nonseparability at the level of phonons. Studies of this physics for lossless systems were reported in~\cite{Campo:2005sy,Campo:2005sv,Campo:2008ju,Bruschi:2013tza,finazzi2013}.

For completeness, it is useful to give explicit expression of the corresponding quantities in the photon (rather than phonon) point of view. Using again the Bogoliubov transformation Eq.~\eqref{eq:defvarphibogofield}, one gets
\begin{subequations}
\begin{align}
n_k^a(t) = n_{k,{\rm f}}^{a,\rm st} + &e^{-2\Gamma t} \left\{ (u_{k,{\rm f}}^2 + v_{k,{\rm f}}^2) \delta n_k^b - 2 u_{k,{\rm f}}v_{k,{\rm f}} \Re [c_b e^{-2i\omega_{k,{\rm f}} t}] \right\} \\
c_k^a(t)= c_{k,{\rm f}}^{a,\rm st} + &e^{-2\Gamma t} \left\{ u_{k,{\rm f}}^2 c_b e^{-2i\omega_{k,{\rm f}} t} + v_{k,{\rm f}}^2 c_b^* e^{2i\omega_{k,{\rm f}} t}-2 u_{k,{\rm f}}v_{k,+} \delta n_k^b \right\}.
\end{align}
\end{subequations}
As expected, the Bogoliubov transformation is responsible for temporal oscillations in these photonic quantities in response to the jump: as compared to the atomic case see Ref.~\cite{Carusotto:2009re}, oscillations are now damped at the loss rate $\Gamma$ and tend to their static values $n_{k,{\rm f}}^{a,\rm st}$ and $c_{k,{\rm f}}^{a,\rm st}$ for the final parameters after the jump.

\subsection{The observables}

This last section is devoted to a discussion of possible strategies aiming to extract the four quantities (i.e., six real quantities) of Eq.~\eqref{eq:4q} from accurate measurements of the coherence functions of the cavity photon field. Two of them $n_{k, \rm f}^b, c_{k, \rm f}^b,$ characterize the final stationary state, while $ \delta n_k^b, {c_k^b}$ characterize the decaying properties of the state. Knowledge of the following four real quantities, $|{c_k^b}|, \delta n_k^b, n_{k, \rm f}^b$ and $|{\bar c_{k,\rm f}^b}|$ allows to assess the nonseparability of the phonon state.

\subsubsection{The anticommutator \texorpdfstring{$G_{ac}$}{Gac} }

As a first example, we consider the equal time combination analogous to Eq.~\eqref{eq:RGdef},
\begin{equation}
 \Re[G_{\rm DCE}(t,t, \bk)] = n^b_{k,\rm f} + \delta n_k^b e^{-2\Gamma t} + \Re[c_k^b e^{2 (\Gamma+i\omega_{k,{\rm f}}) t}+ \bar c_{k,\rm f}^b]. 
\label{eq:RG2}
 \end{equation} 
For underdamped phonon modes such that $\omega_{k,{\rm f}}>\Gamma$, this quantity oscillates between maxima and minima given by $n^b_{k,\rm f} + \delta n^b_{k}(t) + \Re[\bar c_{k,\rm f}^b ] \pm \abs{c_k^b(t)}$: the function $c_k^b(t) = c_k^b e^{- 2\Gamma t}$ can thus be extracted from the amplitude of oscillations. The mid-point of the oscillations provides instead information on $n_k^b(t) = n^b_{k,\rm f}+ \Re(\bar c_{k,\rm f}^b)+ \delta n_{k}^b e^{- 2\Gamma t} $. This quantity can be taken as a operative definition of the mean occupation number. If one wishes to extract the $\delta n_{k}^b(t) = \delta n_{k}^b e^{- 2\Gamma t} $ contribution from the correlation correction, one has just to measure $\Re[G_{\rm DCE}(t,t)]$ for different times $t$: since it is the only term that possess this time dependence, $\delta n_{k}^b(t)$ is therefore well defined. Hence, the only quantity affected by $\bar c_{k,\rm f}^b$ is $n_{k,\rm f}^b$.

\begin{SCfigure}[2][htb]
\includegraphics[width=0.47\linewidth]{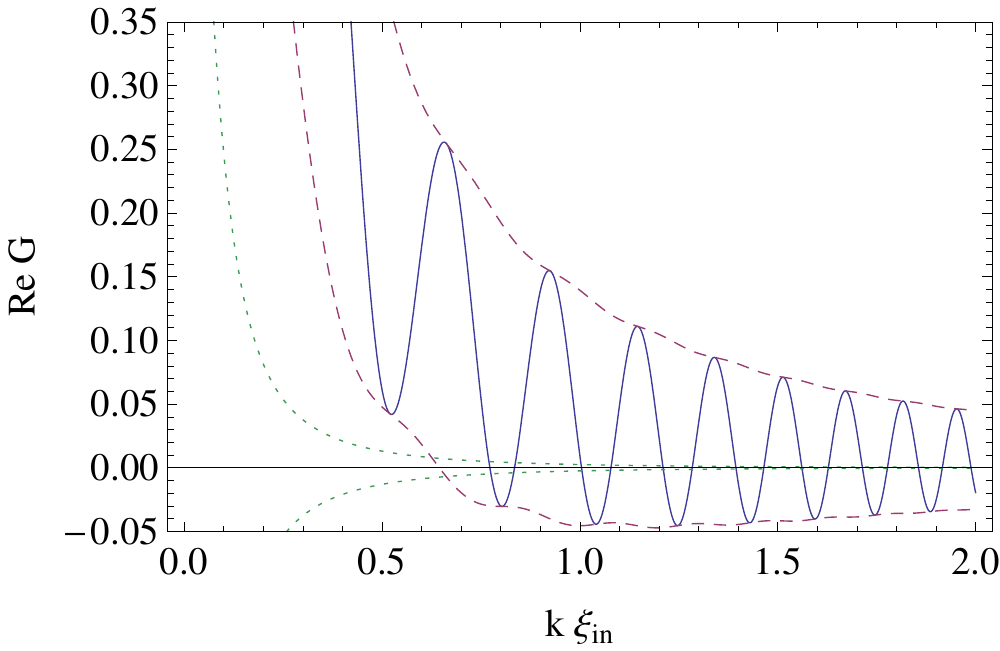}
\caption{Nonseparability of the phonon state after a sudden jump. The oscillating solid blue line shows the equal time function $\Re[G_{\rm DCE}(t,t,\mathbf{k})]$ defined in Eq.~\eqref{eq:RG2} for a time $t =3/m c_{\rm in}^2$ as a function of the (normalized) phonon momentum $\xi_{\rm in} k$. Nonseparable phonon states are found wherever the lower envelope (dashed line), indicating $G_{DCE}^{\varphi^\dagger\varphi}(t,t,\bk)-|G_{DCE}^{\varphi\varphi}(t,t,\bk)|$, goes below $0$. In the present case, the intrinsic imprecision $\pm |{\bar c_{k,\rm f}^b}|$ (dotted lines) does not significantly affect the identification of nonseparable states. System and jump parameters: $M_{\rm in}/{ m } = 0.01$, $c_{\rm f}^2/c_{\rm in}^2 = 2$, and $\Gamma = 0.03 m c_{\rm in}^2$. }
\label{fig:nkmck}
\end{SCfigure}

In terms of $\Re[G_{\rm DCE}(t,t, \bk)]$, the nonseparability condition of Eq.~\eqref{eq:Camposeparable} applied to phonon states, $\abs{c_k^b(t)}\geq n_k^b(t)$, is simply reexpressed as $\Re[G_{\rm DCE}(t,t, \bk)] \leq 0$ up to error terms of order $\mathcal{O}[\bar c_{k, {\rm f}}^{b}]$. The simplicity of this condition arises from the fact that $G_{\rm DCE}(t,t, \bk)$ is the expectation value of normal ordered products of phonon operators $\hat b_\bk, \hat b_{-\bk}^\dagger$ of Eq.~\eqref{eq:phidec}. The condition and its intrinsic imprecision are visually represented in Fig.~\ref{fig:nkmck} by the two dotted lines.

\subsubsection{The \texorpdfstring{$g_2$}{g2}}

\begin{figure}[htb]
\begin{minipage}{0.47\linewidth}
\includegraphics[width=1\linewidth]{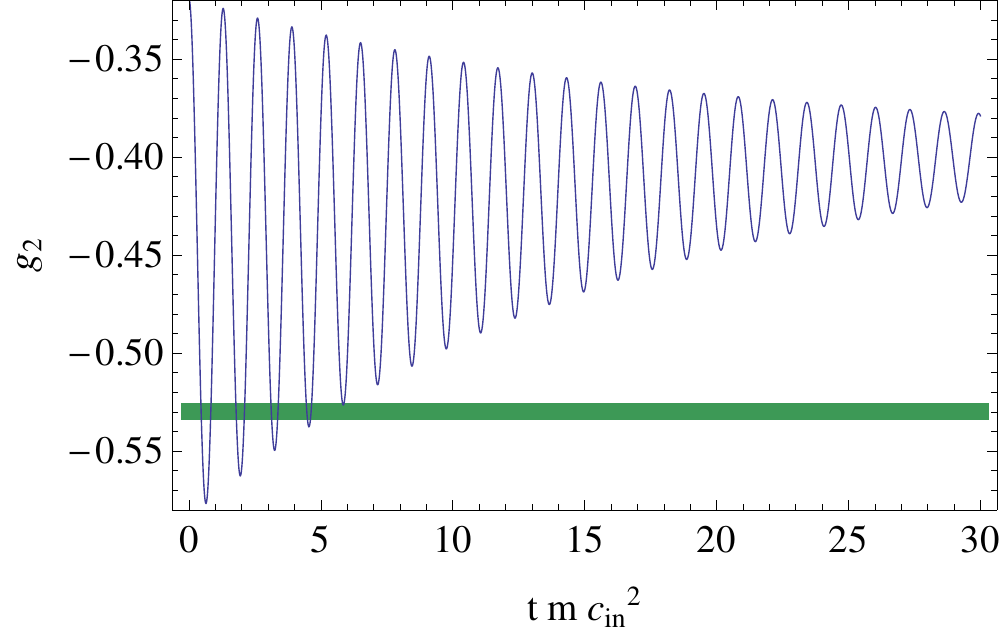}
\caption{Time-evolution in Fourier-space of the second-order coherence of the photon field. The solid blue line shows the equal time $g_{2, \bk}(t,t')$ at $t=t'$ for a given $k \xi_{\rm in} = 0.75 $. In this case, the initial and final values of the phonon occupation are respectively $n^b_{{\rm st},k}\simeq 0.06$ and $n^b_{{\rm f},k}\simeq 0.14$. As in Fig.~\ref{fig:Gofk1}, an exponential convergence towards the final value is apparent. The horizontal green line represents the phononseparability threshold: Nonseparability is found as long as the lower envelope of the oscillating solid line stays below the horizontal line, whose thickness shows the intrinsic imprecision $\pm\abs{\bar c_{k,\rm f}^b}$ of the mean occupation number. Same system and jump parameters as in Fig.~\ref{fig:nkmck}.}
\label{fig:g2oft}
\end{minipage}
\hspace{0.03\linewidth}
\begin{minipage}{0.47\linewidth}
\includegraphics[width=1\linewidth]{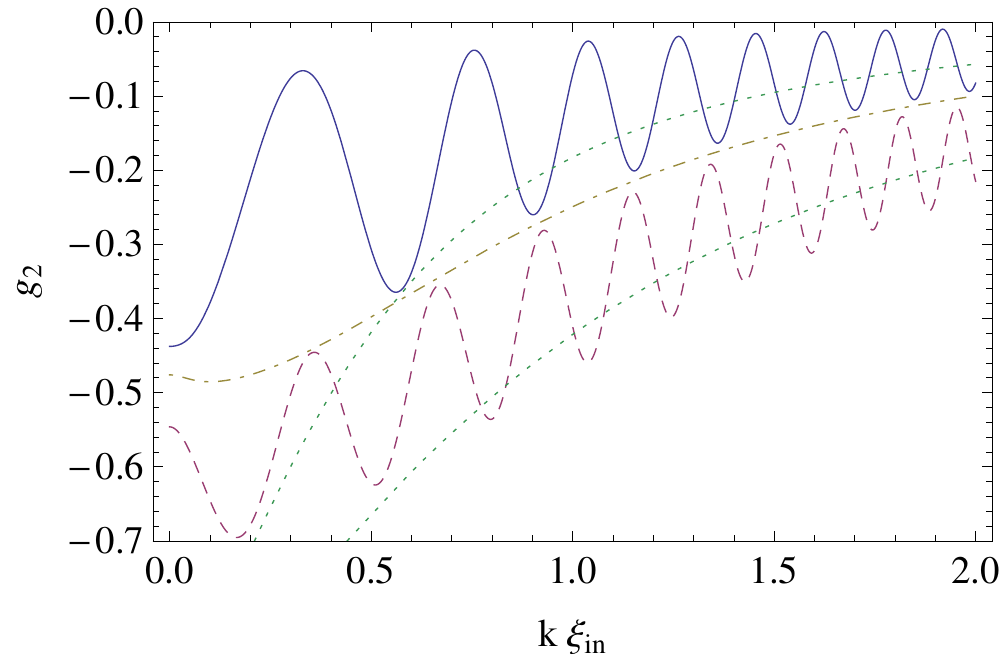}
\caption{Fourier-space second-order coherence of the photon field. The two oscillating curves represent the equal time $g_{2, \bk}(t,t)$ as a function of normalized momentum $k \xi_{\rm in}$ at a time $t = 3/ m c_{\rm in}^2$ after a jump characterized by $c_{\rm f}^2/ c_{\rm in}^2 = 2$ for the upper (blue solid) curve, and $c_{\rm f}^2/ c_{\rm in}^2 = 1/2$ for the lower (purple dashed) one. The two green dotted curves give the separability thresholds of these two cases: nonseparability is found whenever the lower envelope (not represented here) of an oscillating solid line goes below the corresponding dotted curve. The middle yellow dot-dashed curve is the (common) value of $g_{2, \bk}$ before the sudden change. Same system parameters as in Fig.~\ref{fig:nkmck}.}
\label{fig:g2ofk}
\end{minipage}
\end{figure}

In practice, optical measurements typically involve coherence function of a field. In our case, the second-order coherence $g_2$ is most important as it is the simplest to analyze. Inserting the expectation values of Eq.~\eqref{eq:diisnb} into Eq.~\eqref{eq:g2k}, we immediately identify the stationary and the decaying contributions
\begin{equation}
\begin{split}
g_{2 ,\bk} (t,t') &= \ep{- \Gamma\abs{t-t'}} g_{2 ,\bk}^{\rm st} (t,t') + \ep{- \Gamma(t+t')} g_{2 ,\bk}^{\rm dec} (t,t').
\end{split}
\label{eq:g2k_split}
\end{equation}
The time dependence of $g_{2 ,\bk}^{\rm st}$ and $g_{2 ,\bk}^{\rm dec}$ is of the form 
\begin{subequations}
\begin{align}
g_{2 ,\bk}^{\rm st} (t,t')&=A_1 \cos\left [ \omega_{k,{\rm f}}\abs{t-t'} + \theta_1 \right ],\\
g_{2 ,\bk}^{\rm dec} (t,t') &= A_2 \cos\left [ \omega_{k,{\rm f}}(t-t') \right ]+A_3 \cos\left [ \omega_{k,{\rm f}}(t+t') + \theta_3 \right ] ,
\end{align}
\end{subequations}
where the three constants are
\begin{equation}
\begin{split}
A_1 \ep{- i \theta_1} &= 2(u_k-v_k)^2 \left [n_{k,\rm f}^b + \bar c_{k,\rm f}^b \right ] - 2 v_k (u_k-v_k) , \\
A_2&= 2(u_k-v_k)^2 \delta n_{k}^b , \\
A_3 \ep{- i \theta_3} &=2(u_k-v_k)^2 c_k^b . 
\end{split}
\end{equation}
From measurements of $g_{2 ,\bk} (t,t')$ at different times $t,t'$, we can thus extract 5 real quantities (out of the $6$ physical ones), namely $\Re[c_k^b], \Im[c_k^b], \delta n_{k}^b, \Im \bar c_{k,\rm f}^b$ and $n_{k,\rm f}^b +\Re[ \bar c_{k,\rm f}^b]$. To disentangle $n_{k,\rm f}^b $ from $\Re[\bar c_{k,\rm f}^b]$, another observable, such as the $\bk$ component of the $g_{1}$ function, is needed. 

In Fig.~\ref{fig:g2oft} we represent the equal time $g_{2 ,\bk}$ as a function of $t$, for a given wave number $k \xi_{\rm in} =0.75$ and the same parameters of the previous figures: for these values, the initial value oscillates with amplitude $A_3 = 0.26$ around the mean value $A_1 \cos(\theta_1) + A_2 = -0.45 $. Its final value is $A_1 \cos(\theta_1) = -0.4$. The threshold value of nonseparability is reached when the minimum of the $g_2$ crosses $ -0.53 \pm 0.005$. Neglecting for simplicity the intrinsic imprecision due to $\pm |\bar{c}^b_{k,\rm f}|$, as in \ref{chap:dissipBEC}, losses make nonseparability to disappear within a time of the order
\begin{equation}
\begin{split}
t_{\rm loss} \doteq \log\left ( (\abs{c_k^b}-\delta n_k)/n_{k,\rm f} \right )/2\Gamma \lesssim 1/4 n_{k,\rm f} \Gamma ,
\label{eq:tloss}
\end{split}
\end{equation}
where the last inequality follows from Eq.~\eqref{eq:Campotoujoursvrai} and applies when $2n_{k,\rm f}\gg 1$. In the present case, $t_{\rm loss} \Gamma \simeq 0.16 $. 

In Fig.~\ref{fig:g2ofk}, we represent the $k$ dependence of the equal time $g_{2, \bk}$ function, at a time $t= 3 / m c_{\rm in}^2$ in two different cases: when ${c_{\rm f}^2}/{c_{\rm in}^2} = 2$ as in the former figure, but also when ${c_{\rm f}^2}/{c_{\rm in}^2} = 1/2$, i.e., when the final sound speed is divided by $\sqrt{2}$ rather than multiplied by $\sqrt{2}$. In both cases, we use the same system parameters as in the former Figure. We observe two oscillating functions, the minima of the upper one close to the maxima of the lower one. Their common value is $g^{\rm st}_{2, \bk}$ evaluated before the jump, see Fig.~\ref{fig:morerobust} for more details. It is represented by a dotted line. From the two envelopes of each curve, we can measure the width on the oscillations $A_3 \ep{- 2 \Gamma t}$, which gives the $k$ dependence of the strength of the correlations, and the average value $A_1 \cos(\theta_1) + A_2$. The two dashed curves in Fig.~\ref{fig:g2ofk} are the corresponding thresholds of nonseparability. In both cases, there is a large domain of $k$ where the state is nonseparable.

\begin{SCfigure}[2]
\includegraphics[width=0.47\linewidth]{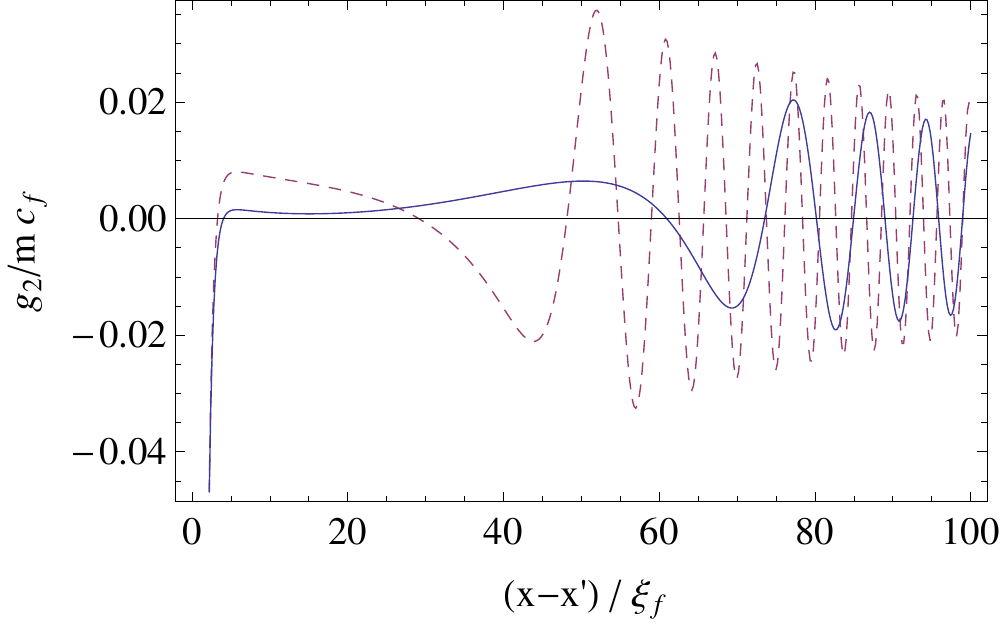}
\caption{We plot the equal time function $g_2(t,t, \bx-\bx')$ as a function of the (normalized) spatial distance $\bx-\bx'$ at $t= 12/ mc_{\rm in}^2$ (purple dashed), and $18/mc_{\rm in}^2$ (blue solid). In addition to the negative peak at $x=x'$ due to repulsive interactions, the phonon pairs generated at the jump are visible in the series of moving fringes with spatially decreasing spatial period. Given the small value of $\Gamma/m c^2_{\rm in} = 0.03$, dissipative effects have a minor effect on the profiles shown here. Separability features are hard to ascertain from this figure. Same system and jump parameters as in Fig.~\ref{fig:nkmck}.}
\label{fig:g2ofx}
\end{SCfigure}

To complete the study of the $g_2$, we represent in Fig.~\ref{fig:g2ofx} its spatial dependence on $\bx-\bx'$ after integration over $\bk$. For the sake of simplicity, we restrict the study of $g_2$ to a one-dimensional geometry where photons are strongly confined also along the $y$ direction. None of the qualitative features is however expected to be modified when going to higher dimensions. In addition to the negative peak at $\bx=\bx'$ due to the repulsive interparticle interactions, we see a propagating correlation pattern which is governed by the group velocity $v_{\rm gr} = \partial_k \tilde \omega_k$ where $\tilde \omega_k$ is given in Eq.~\eqref{eq:disprelpol}. As in the case of equilibrium condensates, see Fig.~\ref{fig:deltaGofx} and Ref.~\cite{Carusotto:2009re}, the fast oscillations at large separations are due to the superluminal form of the dispersion relation Eq.~\eqref{eq:disprelpol} in the high momentum region. Low momenta modes $k^2/2m < M c^2 $ propagate with a smaller velocity because of the small but finite phonon mass and are responsible for the long-wavelength oscillations that are visible at small $\bx-\bx'$. It is worth observing that dissipation introduces an extra dissipative length scale $L_d = c/ \Gamma$ in addition to the usual healing length $\xi =1/mc $: for the parameters of the figures, we have $L_d/\xi_{\rm f} \sim 40$, which means that dissipation affects the profiles of $g_2$ only at large distances.

\subsubsection{The \texorpdfstring{$g_1$}{g1}}

For the sake of completeness, we conclude the section with a study of the first-order coherence function $g_1$ as defined in Eq.~\eqref{eq:defg1k}. It describes the photon momentum distribution (for $t=t'$) and the photon coherence in momentum space (for generic $t\neq t'$). Using Eq.~\eqref{eq:defvarphibogofield}, this quantity is given for $\bk \neq 0 $ by
\begin{equation}
\begin{split}
g_{1 ,\bk} (t,t') =  u_k^2 G^{\varphi^\dagger \varphi}(t,t'; \bk) &+ v_k^2 G^{\varphi^\dagger \varphi}(t',t; \bk) - 2 u_k v_k \Re \left [G^{\varphi \varphi}(t,t'; \bk)\right ] \\
&+v_k^2 \ep{- \Gamma\abs{t-t'}} \ep{-i \tilde \omega_k (t-t')}  .
\end{split}
\end{equation}
When considering the state after a sudden change, the $g_1$ splits analogously $g_2$ in Eq.~\eqref{eq:g2k_split} into its stationary and its decaying parts, 
\begin{equation}
\begin{split}
g_{1 ,\bk} (t,t')  =& \ep{- \Gamma\abs{t-t'}} g_{1 ,\bk}^{\rm st} (t,t') + \ep{- \Gamma(t+t')} g_{1 ,\bk}^{\rm dec} (t,t').
\end{split}
\end{equation}
Using Eq.~\eqref{eq:diisnb}, the two components define $4$ independent quantities
\begin{subequations}
\begin{align}
g_{1 ,\bk}^{\rm st} &(t,t') = \Re[ B_1 \ep{ -i \omega_{k,+} \abs{t-t'}}] + i B_2 \sin \left [ \omega_{k,+} (t-t')\right ], \\
\begin{split}
g_{1 ,\bk}^{\rm dec} &(t,t') = B_3 (u_{k,+}^2+v_{k,+}^2) \cos\left [ \omega_{k,+}(t-t') \right ] + \Re [B_4 \ep{ -i \omega_{k,+} (t+t')}] -i B_3 \sin\left [ \omega_{k,+}(t-t') \right ]. 
 \end{split}
\end{align}
\end{subequations}
given by
\begin{equation}
\begin{split}
B_1 &= u_{k,+}^2 n_{k,\rm f}^b +v_{k,+}^2 (n_{k,\rm f}^b + 1) - 2 u_{k,+} v_{k,+} \bar c_{k,\rm f}^b , \\
B_2 &= n_{k,\rm f}^b- v_{k,+}^2 , \\
B_3 &= \delta n_{k}^b ,\\
B_4 &= -2 u_{k,+} v_{k,+} c_k^b .
\end{split}
\end{equation}
These encode the $6$ independent real quantities which characterize the correlation functions of Eq.~\eqref{eq:diisnb}. Hence, unlike the $g_{2,\bk}$, the $g_{1 ,\bk}$ fully characterizes the bipartite state $\bk,-\bk$. 

\begin{SCfigure}[2]
\includegraphics[width=0.47\linewidth]{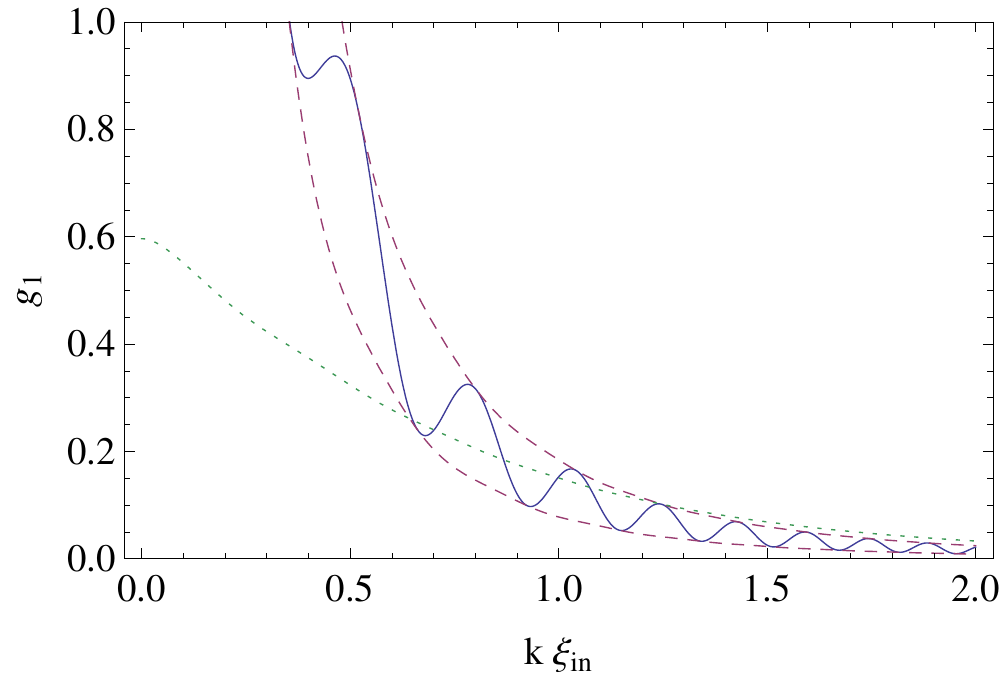}
\caption{Momentum distribution of the cavity photons. The equal-time $g_{1, \bk}(t,t)$ is plotted in momentum space at $t =3/ m c_{\rm in}^2$ (solid blue line). Dashed purple lines indicate its lower and upper envelopes. The phonon state is nonseparable whenever the lower envelope goes below the dotted green line indicating the separability condition Eq.~\eqref{eq:nnsep}. Same system and jump parameters as in Fig.~\ref{fig:nkmck} } 
\label{fig:g1ofk}
\end{SCfigure}

In Fig.~\ref{fig:g1ofk}, we represent the equal time $g_{1 ,\bk} (t,t)$, for $t m c_{\rm in}^2= 3 $ and for the same parameters as in the previous figures. Contrary to what was found for $g_{2, \bk}$, the separability threshold of Eq.~\eqref{eq:Camposeparable} ($n_k^b = \abs{c_k^b}$) does not simply enter in $g_{1, \bk}$. In fact, to extract it, we need both the upper and lower envelopes of $g_{1 ,\bk} (t,t)$, called respectively $U_k(t)$ and $L_k(t)$. Violation of the inequality 
\begin{equation}
\begin{split}
\label{eq:nnsep}
L_k(t) > \frac{(u_{k,+}-v_{k,+})^2 U_k(t) + 2 v_{k,+}^2}{(u_{k,+}+v_{k,+})^2} ,
\end{split}
\end{equation}
implies that the phonon state is nonseparable, i.e., $n_k^b(t) < \abs{c_k^b(t)}$. In the figure, the ratio of Eq.~\eqref{eq:nnsep} is represented by a dotted green line. We again see the large domain of $k$ where the phonon state is nonseparable, namely $k \xi_{\rm in} > 0.6$. 

\subsection{Time evolution of separability criterion}

We now show that the nonseparability criterion of Eq.~\eqref{eq:Camposeparable} based on the phonon $\hat \varphi_\bk$ operators is equivalent to the violation of a Cauchy-Schwartz (CS) inequality for phonon operators. We consider the modified equal-time second-order correlation which is obtained from the standard photonic one $g_{2,\bk}(t,t')$ by
\begin{equation}
\begin{split}
g_{2,\bk}^b(t,t') &\doteq g_{2,\bk}(t,t')+2 v_k(u_k-v_k)\Re [\hat\varphi_\bk(t) , \hat \varphi_\bk^\dagger(t')]\\
&= 2 \left (u_k-v_k \right )^2 \Re[G(t,t', \bk)].
\end{split} 
\end{equation} 
The subtraction of the contribution of the commutator in this expression is equivalent to taking the normal ordering with respect to the phonon operators $\hat b_\bk$ of Eq.~\eqref{eq:phidec}, hence the $b$ superscript in the above notation. In terms of this quantity, the CS inequality reads 
\begin{equation}
\label{eq:CS}
\begin{split}
\mathcal{D}_k(t,t')=\frac{g_{2 ,\bk}^b (t,t) g_{2 ,\bk}^b (t',t')-|{g_{2 ,\bk}^b (t,t')}|^2}{4 \left (u_k-v_k \right )^4} \geq 0 :
\end{split}
\end{equation}
No violation of Eq.~\eqref{eq:CS} can occur in classical statistical physics. In the absence of dissipation, the phonon mean occupation number $n_k^b$ and correlation term $c_k^b$ are both well defined, and constant before and after a sudden jump. Using these two quantities, one obtains
\begin{equation}
\begin{split}
\mathcal{D}_\bk(t,t') = \left [ (n_k^b)^2 - \abs{c_k^b}^2\right ] \sin^2[\tilde \omega_k (t-t')]. 
\end{split}
\end{equation}
Hence, the CS inequality is violated if and only if the state is nonseparable (when $\sin[\tilde \omega_k (t-t')]\neq 0$).

\begin{SCfigure}
\includegraphics[width = 0.47 \linewidth]{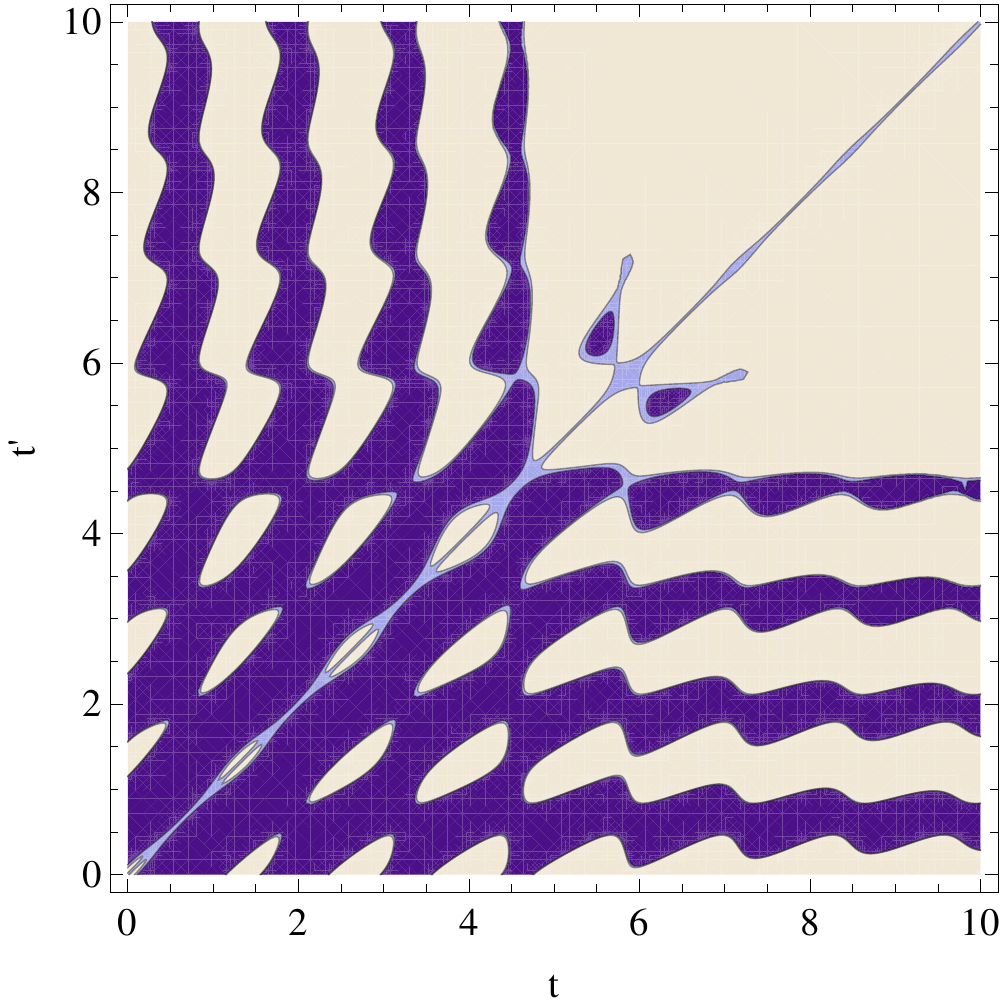}
\caption{Plot of the Cauchy-Schwarz criterion Eq.~\eqref{eq:CS} as a function of $t,t'$ for $k= 1.5 m c_{\rm in}$. Dark blue regions indicate values below $-10^{-4}$ that significantly violate the inequality. White regions indicate values larger than $10^{-4}$. Light blue regions indicate values close to $0$. Same system and jump parameters as in Fig.~\ref{fig:nkmck}.}
\label{fig:2DCS}
\end{SCfigure}

In the presence of dissipation, as discussed in \ref{sec:weakdissip} and in \ref{sec:cstdissiprate}, the coupling of the phonon field $\hat\varphi$ to an environment introduces intrinsic ambiguities in the definition of nonseparability. Nevertheless decomposition of $\hat\varphi$ at any time $t$ over {\it instantaneous} destruction and creation operators $\hat b_\bk, \hat b^\dagger_{-\bk}$ allows to show that for $\Gamma(t - t') \ll 1$, the above relation between the sign of $\mathcal{D}_\bk(t,t')$ and the nonseparability criterion based on these operators remains valid to leading order in $\Gamma/\omega_k$. Accepting this inherent uncertainty of order $\Gamma/\omega_k \ll 1$, one can then follow how nonseparability is progressively lost as time goes on. This physics is illustrated in fig~\ref{fig:2DCS}: the quantity in Eq.~\eqref{eq:CS} displays three different behaviors depending on $t,t'$ compared to the characteristic time $t_{\rm loss}$ of Eq.~\eqref{eq:tloss}. 
\begin{enumerate}
\item
For $(t,t')\gg t_{\rm loss}$, no violation is observed, because the state is separable, as expected.

\item
For $t \ll t_{\rm loss} \ll t' $, Eq.~\eqref{eq:CS} can only be violated for $t$ such that $g_{2,\bk}^b(t,t)<0$, because $g_{2,\bk}^b(t',t')>0$. Along a constant-$t'$ cut, the reader will recognize the behavior already seen in Fig.~\ref{fig:g2oft}.

\item
For $t,t' \ll t_{\rm loss}$ and $ \Gamma/ \omega_k^2 \ll |t-t'| \lesssim 1/ \omega_k $, Eq.~\eqref{eq:CS} is violated. This is the most robust regime for nonseparability.
\end{enumerate}
Note that besides the transition from point $2$ to point $3$, there is always a narrow band $ |t-t'| \ll \Gamma/ \omega_k^2$ where the inequality is never violated with a positiveness of order $\Gamma / \omega_k$. As a result, a two-time measurement of $g_{2,\bk}^b(t\neq t')$ is required to identify nonseparable states. 

We conclude this section with a short discussion of the standard momentum-space CS inequality for photon $\hat a_\bk$ operators, see~\cite{deNova:2012hm} or~\cite{PhysRevLett.108.260401} for its atomic counterpart. The momentum-space second-order photon coherence\footnote{
Note that this quantity should not be confused with $g_{2,\bk}(t,t')$ defined in Eq.~\eqref{eq:g2k} as the Fourier transform of the real-space $g_2(\bx,t,\bx',t')$.} is defined as
\begin{equation}
\begin{split}
 \mathcal{G}_2(\bk,\bk') = \left < a^\dagger_\bk a^\dagger_{\bk'} a_\bk a_{\bk'}\right >.
\end{split}
\end{equation}
The explicit form of the CS inequality is:
\begin{equation}
\label{eq:CSfora}
\begin{split}
[\mathcal{G}_2(\bk, \bk')]^2 \leq \mathcal{G}_2(\bk,\bk) \mathcal{G}_2(\bk',\bk') , 
\end{split}
\end{equation}
Physically, this quantity describes the correlations between the fluctuations of the photon occupation numbers in the modes $\bk$ and $\bk'$. Thanks to the Gaussian nature of the state, we can apply Wick theorem and expand this expression in terms of quadratic operators. For homogeneous states, we get 
\begin{equation}
 \begin{split}
 \mathcal{G}_2(\bk,\bk') &= \delta_{\bk+\bk'}\abs{c_k^a}^2 + \delta_{\bk-\bk'} (n_k^a)^2+ n^a_k n^a_{k'}.
 \end{split}
 \end{equation} 
For $\bk' = - \bk$, the CS condition Eq.~\eqref{eq:CSfora} is then equivalent to the separability condition Eq.~\eqref{eq:Camposeparable} applied to photon operators.

\section*{Conclusions}
\addcontentsline{toc}{section}{Conclusions}

In this chapter, we studied the quantum fluctuations in coherently pumped and spatially homogeneous photon fluids in planar microcavities. Our attention was focused on the simplest case of a quasi-resonant coherent pump at normal incidence on the microcavity, where the photon fluid is at rest and the effective mass of phonon excitations on top of the photon fluid is very small.

When the pump is monochromatic and stationary, the system reaches a stationary state: most remarkably, even if the environment is in its vacuum state, the stationary state of the photon gas is not a vacuum state, but contains a finite occupation of (almost) incoherent phonons. Even though the phonon distribution qualitatively resembles a Planck law at an effective temperature of the order of the interaction energy in the fluid, the non-equilibrium nature of the system leads to quantitatively significant deviations and to violations of the standard fluctuation-dissipation relations.

When the system parameters are suddenly modulated in time, entangled pairs of extra phonons are created in the fluid via processes that are the analog of cosmological pair production or the dynamical Casimir effect. Due to the dissipation, these phonons eventually decay while the system relaxes to a new stationary state. Accurate information on the properties of these extra phonons can be obtained from measurements of the time-dependence of the first- and second-order coherence functions of the cavity photons, which can be used to assess the quantum nonseparability of the phonon state after the jump.

\chapter{Separability of analogue black hole radiation}
\label{chap:thermalBH}

\section*{Introduction}

Apart from dynamical Casimir effect, the second main prediction of analogue gravity is the analogue Hawking radiation. However, as in the previous chapters, there is a competition between the spontaneous quanta emission from vacuum and initially present thermal fluctuations. The challenge is then to identify separable states.

In this chapter, we use a phenomenological description of the analogue Hawking radiation to characterize the separability of the outgoing fluxes. This allows us to consider a very wide class of situations. We first identify the relevant parameters governing the separability of the state. Then, using a superluminal dispersion relation, we identify in which regimes the final state is separable. As a final step, we quickly study the subluminal case and the separability of the flux emitted by a white hole. This chapter is mainly based on~\cite{Busch:2014bza}.

\minitoc
\vfill

\section{The system}
\label{sec:system}

We consider here a generic stationary system presenting a black hole profile and dispersion but no dissipation. It can either be the astrophysical black hole of Eq.~\eqref{eq:PGBHaccel} with the field of Eq.~\eqref{eq:Actiondesitterdisp}, or an analogue black hole in BEC, with perturbations governed by Eq.~\eqref{eq:generaleomphononsecondorder}. To fix notations and ideas, we shall suppose that the dispersion relation does not depend on space and reads
\begin{equation}
\label{eq:disprelsuper}
\begin{split}
F^2(k) &= k^2 ( 1+ \frac{k^2}{\Lambda^2}) .
\end{split}
\end{equation}
The subluminal case will be briefly studied in \ref{sec:sublum}. The only non trivial dynamics is then encoded in the velocity of the flow, $v(x)$, where $v$ is either the velocity of the condensate, or the parameter in the metric.

Because the flow is stationary, the solution of the mode equation splits into $\omega$ sectors: $\hat \phi_\omega(x) = \int dt/\sqrt{2\pi} e^{ i \omega t} \hat \phi(t,x)$ which can be studied separately~\cite{Brout:1995wp}. When the flow is asymptotically uniform on both sides, the incoming modes $\phi^{\rm in, a}_\omega$, with a single branch with group velocity pointing toward the horizon, are well defined and asymptotically superpositions of plane waves. (The index $a$ refers to the dimensionality of the set of solutions at fixed $|\omega|$.) The same apply to the outgoing modes $\phi_\omega^{\rm out, a}$, with the group velocity pointing now away from the horizon. More details about this identification and the dimensionality of the set of modes can be found in Refs.~\cite{Macher:2009nz,Macher:2009tw}.

In brief, because of superluminal dispersion, there is a threshold value $\omega_{\rm max}$ above which there is no pair creation. For $0< \omega< \omega_{\rm max}$, there are three independent modes. They are called $\phi_{\omega}^{\rm in, u}, \,\phi_{\omega}^{\rm in, v}, (\phi_{-\omega}^{\rm in, u})^*$ for the $in$ modes (and similarly for the $out$ ones). The first two have positive norm and describe respectively counter- and copropagating quasiparticles, see Fig.~\ref{fig:caracBHdS}. The third one, $(\phi_{-\omega}^{\rm in, u})^*$, has a negative norm, and describes the incoming negative frequency partner trapped in the supersonic region. These three modes are scattered in the near horizon region. As a result, the outgoing modes are nontrivially related to the incident ones. The $S$-matrix relating the (normalized) $in$ modes to the $out$ ones is thus an element of $U(1,2)$. Following Ref.~\cite{Macher:2009nz}, we name its coefficients 
\begin{equation}
\begin{split}
\left ( \begin{array}{l}
\hat a_\omega^{u } \\
(\hat a_{-\omega}^{u} )^\dagger\\
\hat a_\omega^{v} 
\end{array} \right ) = \left ( \begin{array}{lll}
\alpha_\omega & \beta_\omega^*& A_\omega \\
\beta_{-\omega}& \alpha_{-\omega}^* & B_\omega \\
\tilde A_\omega & \tilde B_\omega^* & \alpha^v_\omega 
\end{array} \right ) \left ( \begin{array}{l}
\hat a_\omega^{u,\rm in}\\
  (\hat a_{-\omega}^{u,\rm in})^\dagger \\
 \hat a_\omega^{v,\rm in}
\end{array}\right ).
\label{eq:Bogcoef}
\end{split}
\end{equation}
To avoid ponderous notation, only the superscript \enquote{in} will be written. The superscript \enquote{out} is thus implied. The three independent pairs of destruction and creation operators associated with the three $in$ (or three $out$) modes obey the canonical commutation relations. 
 
When $\hat \rho$, the state of the quantum field, is stationary and Gaussian, it factorizes into three-mode sectors of fixed $|\omega|$. In this chapter we shall only consider such states. Then because the $S$-matrix of Eq.~\eqref{eq:Bogcoef} only mixes modes with the same $|\omega|$, the factorization equally applies to the description of $\hat \rho$ in terms of $in$, or $out$ quasiparticle content. We shall study the latter since we aim to identify the cases where the state after the scattering is nonseparable. 

For $\omega >0$ and at late time, each three-mode state is fully characterized by six numbers 
\begin{equation}
\begin{split}
\label{eq:noutcoutdef}
n^{u}_{\pm\omega} &= \left < (\hat a_{\pm\omega}^{u})^\dagger \hat a_{\pm\omega}^{u}\right >,\quad  n^v_\omega = \left < (\hat a_\omega^{v})^\dagger \hat a_\omega^{v} \right >,\\
c^{u u/v}_\omega &= \left < \hat a_{-\omega}^{u} \hat a_\omega^{u/v}\right >,\quad\ 
d^{uv}_\omega = \left < (\hat a_\omega^{u})^\dagger \hat a_\omega^{v}\right >.
\end{split}
\end{equation}
The interpretation of the three (real and positive) final occupation numbers $n^a_\omega$ is straightforward and standard. The two $c_\omega$ are complex, and their norms quantify the strength of the statistical correlations between outgoing quasiparticles of opposite energy, namely between the $uu$ pairs of counterpropagating $out$ modes $(\phi_{\omega}^{\rm out, u}, \phi_{-\omega}^{\rm out, u})$, and the $uv$ pairs $(\phi_{\omega}^{\rm out, v}, \phi_{-\omega}^{\rm out, u})$. They generalize the $c_k$ term defined in Eq.~\eqref{eq:defncphonon}. In the present case, we thus have two differences, see Eq.~\eqref{eq:defDeltasquares}
\begin{subequations}
\label{eq:NSepuv}
\begin{align}
\Delta_\omega^{uu}  \doteq  n_{-\omega}^{u} n_\omega^{u} - \abs{c_\omega^{uu}}^2 , \\
\Delta_\omega^{uv}  \doteq  n_{-\omega}^{u} n_\omega^{v} - \abs{c_\omega^{uv}}^2 .
\end{align}
\end{subequations}
We showed in \ref{chap:separability} that if one of them is negative, the state is nonseparable. The last coefficient of Eq.~\eqref{eq:noutcoutdef}, $d^{uv}_{\omega}$, characterizes the strength of the correlations between $u$ and $v$ modes which have been elastically scattered. Thus, it results from the analogue \enquote{greybody} factors. These correlations are never strong enough to violate classical inequalities; see \ref{sec:homosepcriterion}. Hence, they shall no longer be mentioned. 

To be able to determine if the final state is nonseparable, it is necessary to know the initial state and the coefficients of the $S$-matrix. Our aim is not so much to perform the calculation in a particular realization; rather, we aim to characterize the domains in parameter space where the state is nonseparable. To this end, we need to identify the independent parameters which span this space, and to adopt a phenomenological description of their behaviors.

As a first step, we assume that the initial state is incoherent. It is thus characterized by the three initial occupations numbers $n^{\rm in, u}_{\omega}, n^{\rm in, v}_{\omega}, n^{\rm in, u}_{-\omega}$ since the three  correlation terms initially vanish. Physically, this is a very legitimate assumption, as it means that the three modes are not correlated prior to being scattered. In this case, using Eq.~\eqref{eq:Bogcoef}, Eq.~\eqref{eq:noutcoutdef} gives
\begin{equation}
\label{eq:noutcoutSmatrix}
\begin{split}
n^{u}_{\omega}  = & \abs{\alpha_\omega}^2 n_{\omega}^{u, \rm in}+\abs{\beta_{\omega}}^2 (n_{-\omega}^{u, \rm in}+1)+\abs{ A_\omega}^2 n_{\omega}^{v, \rm in}, \\
n^{v}_{\omega} = & \abs{ \alpha^v_\omega}^2 n_{\omega}^{v, \rm in}+\abs{\tilde B_\omega}^2 (n_{-\omega}^{u, \rm in}+1)+\abs{\tilde A_\omega}^2 n_{\omega}^{u, \rm in} , \\
n^{u}_{-\omega} = & \abs{\alpha_{-\omega}}^2 n_{-\omega}^{u, \rm in} + \abs{\beta_{-\omega}}^2 (n_{\omega}^{u, \rm in}+1)+\abs{ B_\omega}^2 (n_{\omega}^{v, \rm in}+1) , \\
c^{uu}_\omega = & \alpha_\omega \beta_{-\omega}^* (n_{\omega}^{u, \rm in}+\frac{1}{2})+\alpha_{-\omega} \beta_{\omega}^* (n_{-\omega}^{u, \rm in}+\frac{1}{2})+A_\omega  B_\omega^* (n_{\omega}^{v, \rm in}+\frac{1}{2}) , \\
c^{uv}_\omega = & \tilde A_\omega \beta_{-\omega}^* (n_{\omega}^{u, \rm in}+\frac{1}{2})+\alpha_{-\omega} \tilde B_\omega^* (n_{-\omega}^{u, \rm in}+\frac{1}{2})+ \alpha^v_\omega  B_\omega^* (n_{\omega}^{v, \rm in}+\frac{1}{2}) , \\
d^{uv}_\omega = & \tilde A_\omega \alpha_{\omega}^* n_{\omega}^{u, \rm in}+\beta_{\omega} \tilde B_\omega^* (n_{-\omega}^{u, \rm in}+1)+ \alpha^v_\omega  A_\omega^* n_{\omega}^{v, \rm in} . 
\end{split}
\end{equation}
When working in the $in$ vacuum, $n^{u}_{\omega} = \abs{\beta_{\omega}}^2,\, n^{v}_{\omega} = \abs{\tilde B_{\omega}}^2 $ and $n^{u}_{-\omega} = \abs{\beta_{-\omega}}^2+ \abs{B_{\omega}}^2$ respectively give the mean number of the $u$ quanta {\it spontaneously} emitted to the right (the Hawking quanta), that of the $v$ quanta emitted to the left, and that of their negative energy partners. When the initial state is not the vacuum, the terms weighted by $n^{\rm in, a}_{\omega}$ give the {\it induced} contributions. One then sees that the norms $\abs{ A_\omega}^2, \abs{\tilde A_\omega}^2$ respectively quantify the greybody factors, i.e., the reflection of $v$ quanta into $u$ ones, and vice versa. 

Notice that unlike what is found for $2 \times 2$ $S$-matrices, one has  $\abs{\beta_{\omega}}^2 \neq \abs{\beta_{-\omega}}^2$, $\abs{B_\omega}^2 \neq \abs{\tilde B_\omega}^2$, and  $\abs{A_\omega}^2 \neq \abs{\tilde A_\omega}^2$. Yet, as shown by numerical simulations~\cite{Macher:2009nz,Macher:2009tw}, the relative difference between $\abs{\beta_{\omega}}^2$ and $\abs{\beta_{-\omega}}^2$ is generally small. Instead the differences $A_\omega - \tilde A_\omega$ and $B_\omega - \tilde B_\omega$ diverge in general when $\omega \to 0$. We shall return to this important point below. Notice finally that the coefficients  $\abs{\beta_{-\omega}}^2$ and $\abs{B_{\omega}}^2$ considered separately give the mean number of $u$ and $v$ quanta emitted by the corresponding white hole flow; see \ref{sec:whitehole} for more details. 

\subsection{Parametrization of the scattering}

We now show that, for stationary incoherent states, only four independent parameters of $S$ enter Eq.~\eqref{eq:NSepuv}. We shall work with the four squared norms $\abs{\beta_{\omega}}^2$, $\abs{\beta_{-\omega}}^2$, $\abs{ A_\omega}^2$ and $\abs{ B_\omega}^2$.

The $U(1,2)$ character of the $S$ matrix imposes the relations from the normality of lines and columns,
\begin{equation}
\label{eq:5modulusfixed}
\begin{split}
\abs{ \alpha^v_\omega}^2 &=  1+\abs{B_{\omega}}^2  - \abs{A_{\omega}}^2 , \\
\abs{\alpha_\omega}^2 &= 1+\abs{ \beta_\omega}^2 - \abs{A_\omega}^2 , \\
\abs{\alpha_{-\omega}}^2 &= 1+\abs{ \beta_{-\omega}}^2  + \abs{B_\omega}^2,  \\
\abs{\tilde A_\omega}^2 &=\abs{\beta_{-\omega}}^2  -\abs{ \beta_\omega}^2  + \abs{A_\omega}^2 , \\
\abs{\tilde B_\omega}^2 &=  \abs{ \beta_{-\omega}}^2   - \abs{\beta_{\omega}}^2 + \abs{B_\omega}^2,
\end{split}
\end{equation}
and from the orthogonality of the lines
\begin{subequations}
\label{eq:2linesortho}
\begin{align}
\tilde A_\omega \beta_{-\omega}^*-\alpha_{-\omega} \tilde B_\omega^* + \alpha^v_\omega  B_\omega^* =0,\\
\alpha_\omega \beta_{-\omega}^* -\alpha_{-\omega} \beta_{\omega}^* + A_\omega  B_\omega^* =0.
\end{align}
\end{subequations}
The number of real independent quantities is then reduced from eighteen (nine complex numbers) to nine, which can be taken to be five phases and the above four norms. This choice is convenient because the five phases drop out from Eq.~\eqref{eq:NSepuv}. 

In addition, Eqs.~\eqref{eq:2linesortho} and the positivity of the r.h.s. of Eqs.~\eqref{eq:5modulusfixed}, imply some inequality among the four norms.\footnote{
The origin of this fact is the following: Eqs.~\eqref{eq:2linesortho} define two triangles in complex plane. Hence, one length cannot be larger than the sum of the two others. In addition to the nine real parameters of the $S$-matrix, one finds that there is an extra multiplicity two. It produces the symmetrical triangles with respect to the real axis. This extra multiplicity has no influence in the sequel since the initial state is incoherent.} 
These constraints are equivalent to 
\begin{subequations}
\begin{align}
\abs{A_{\omega}}^2 \leq 1+ \abs{B_{\omega}}^2 , \\
\label{eq:boundsonbeta}
\beta_{\omega}^{\rm min} \leq \abs{\beta_{\omega}} \leq \beta_{\omega}^{\rm max} , 
\end{align}
\end{subequations}
where
\begin{equation}
\begin{split}
\beta_{\omega}^{\rm min/max}&\doteq \abs{\frac{\abs{A_\omega B_\omega \alpha_{-\omega}} \pm\abs{\alpha_\omega^v\beta_{-\omega}}}{1+\abs{B_\omega}^2 }}.
\end{split}
\end{equation}
In this expression, $\alpha_{-\omega}$ and $\alpha_\omega^v$ are implicit expressions of $A_\omega, B_\omega$ and $\beta_{-\omega}$. To implement the right condition of Eq.~\eqref{eq:boundsonbeta} and reduce the number of independent norms to three, we impose
\begin{equation}
\label{eq:betaomimposed}
\begin{split}
\abs{\frac{\beta_{\omega}}{\beta_{-\omega}}} =\frac{\abs{\alpha_\omega^v} + \abs{A_\omega B_\omega }}{1+\abs{B_\omega}^2 } = \frac{1-\abs{A_\omega}^2 }{\abs{\alpha_\omega^v} - \abs{A_\omega B_\omega }}.
\end{split}
\end{equation}
The left condition of Eq.~\eqref{eq:boundsonbeta} is then equivalent to $ \abs{\beta_{\omega}\beta_{-\omega}}  \geq \abs{A_\omega B_\omega }^2 /4\abs{\alpha_\omega^v} $. To implement this inequality, we introduce $\abs{\beta^0_{\omega}}^2$ by
\begin{equation}
\label{eq:betaimposed2}
\begin{split}
 \abs{\beta_{\omega}\beta_{-\omega}} = \abs{\beta^0_{\omega}}^2 + \abs{A_\omega B_\omega }^2 /4\abs{\alpha_\omega^v}.
\end{split}
\end{equation}
When $A_\omega$ and $ B_\omega$ vanish, one has $\abs{\beta_{\omega}}^2 = \abs{\beta_{-\omega}}^2 =  \abs{\beta^0_{\omega}}^2$.

In conclusion, our parametrization of the relevant coefficients of the $S$ matrix is based on $\abs{ A_\omega}^2 ,\abs{ B_\omega}^2$ and $\abs{\beta^0_{\omega}}^2$. Eqs.~\eqref{eq:betaomimposed} and~\eqref{eq:betaimposed2} then fix $\abs{\beta_{\pm \omega}}^2$.

\subsection{Parametrization of dispersive spectra}

So far we worked at fixed $\omega$. To characterize the spectrum, we need to parametrize the $\omega$-dependence
of $\abs{ A_\omega}^2 ,\abs{ B_\omega}^2$ and $\abs{\beta^0_{\omega}}^2$. To this end, we consider the flow profile of Eq.~\eqref{eq:vD}:
\begin{equation}
\label{eq:vpluscBH}
\begin{split}
v(x)  = - 1 + D \tanh \left ( \frac{\kappa x}{D} \right ).
\end{split}
\end{equation}
The parameter $D$ fixes the asymptotic values of $v+1$ on either side. In the present case they are equal and opposite. For more general asymmetric profiles we refer to Refs.~\cite{Finazzi:2012iu,Robertson:2011xp}. The frequency $\kappa$ fixes the surface gravity, and determines the temperature of the black hole radiation $T_H \doteq \kappa/(2\pi)$ in the {\it Hawking regime}, i.e., when dispersive effects are negligible because $\Lambda /\kappa \gg 1$. We work in units where $c = \hbar = k_B= 1$. When leaving this regime, numerical and analytical studies~\cite{Coutant:2011in} have established that $D$ also matters. In particular, when the coupling to the counterpropagating mode is small, i.e., if $\abs{ A_\omega}^2,\abs{ B_\omega}^2 \ll 1$, the spectrum of $u$ quanta spontaneously emitted $n^u_\omega = \abs{\beta^0_{\omega}}^2$ remains remarkably Planckian, even though the effective temperature, hereafter called $T_{\rm hor}$, is significantly modified. It is well approximated by~\cite{Robertson:2012ku,Finazzi:2012iu}
\begin{equation}
T_{\rm hor} \doteq T_H\tanh \left ( { T_\infty}/{T_H} \right ), \quad 
T_\infty \doteq \frac{\Lambda D^{3/2}}{(2+D) \sqrt{2-D}}.
\label{eq:Thor}
\end{equation}
The effective temperature $T_{\rm hor}$ thus interpolates between the {\it Hawking regime} for low $T_H/T_\infty$, and the {\it dispersive regime} where it asymptotes to $T_\infty$ for $T_H/T_\infty \gg 1$. 

Numerical studies have also shown that  $\abs{ A_\omega}^2$ and $\abs{ B_\omega}^2$ are not fully determined by $\kappa, \Lambda, D$. They depend on the exact properties of the wave equation, and on the background profiles. Yet, they are generally smaller than $\abs{\beta^0_{\omega}}^2$, and remain finite for $\omega \to 0$. To implement these numerical observations, we shall work with 
\begin{equation}
\label{eq:ABbetaomegadep}
\abs{ A_\omega}^2 = \frac{2A^2}{\ep{\omega/T_{\rm hor}}+1},\quad
\abs{ B_\omega}^2 = \frac{2B^2}{\ep{\omega/T_{\rm hor}}+1},\quad
\abs{\beta^0_{\omega}}^2=\frac{1}{\ep{\omega/T_{\rm hor}}-1},
\end{equation}
where the constants $A^2$ and $B^2$ fix the overall norm of the two couplings between the counterpropagating mode with the Hawking mode and its partner.

It should be noticed that Eqs.~\eqref{eq:ABbetaomegadep} and~\eqref{eq:5modulusfixed} correctly imply that only $A_\omega, B_\omega$ and $\alpha_\omega^v$ are regular in the limit $\omega \to 0$, whereas the squared norms of the six other coefficients diverge as $1/\omega$. It can be shown that this interesting property follows from the normalization in $1/\sqrt{\omega}$ of the low momentum modes.

\subsection{Initial state}

\begin{figure}[htb]
\begin{minipage}{0.47\linewidth}
\includegraphics[width=1\linewidth]{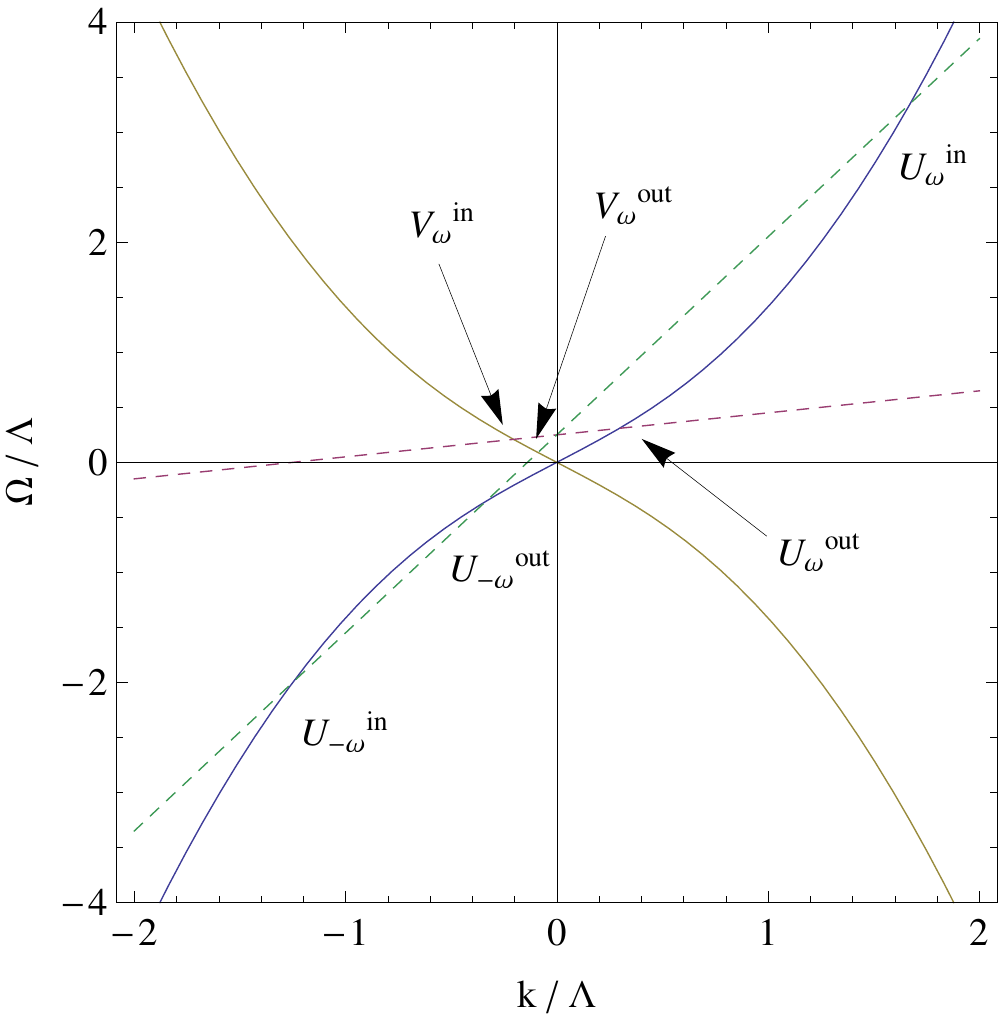}
\end{minipage}
\hspace{0.03\linewidth}
\begin{minipage}{0.47\linewidth}
\includegraphics[width=1\linewidth]{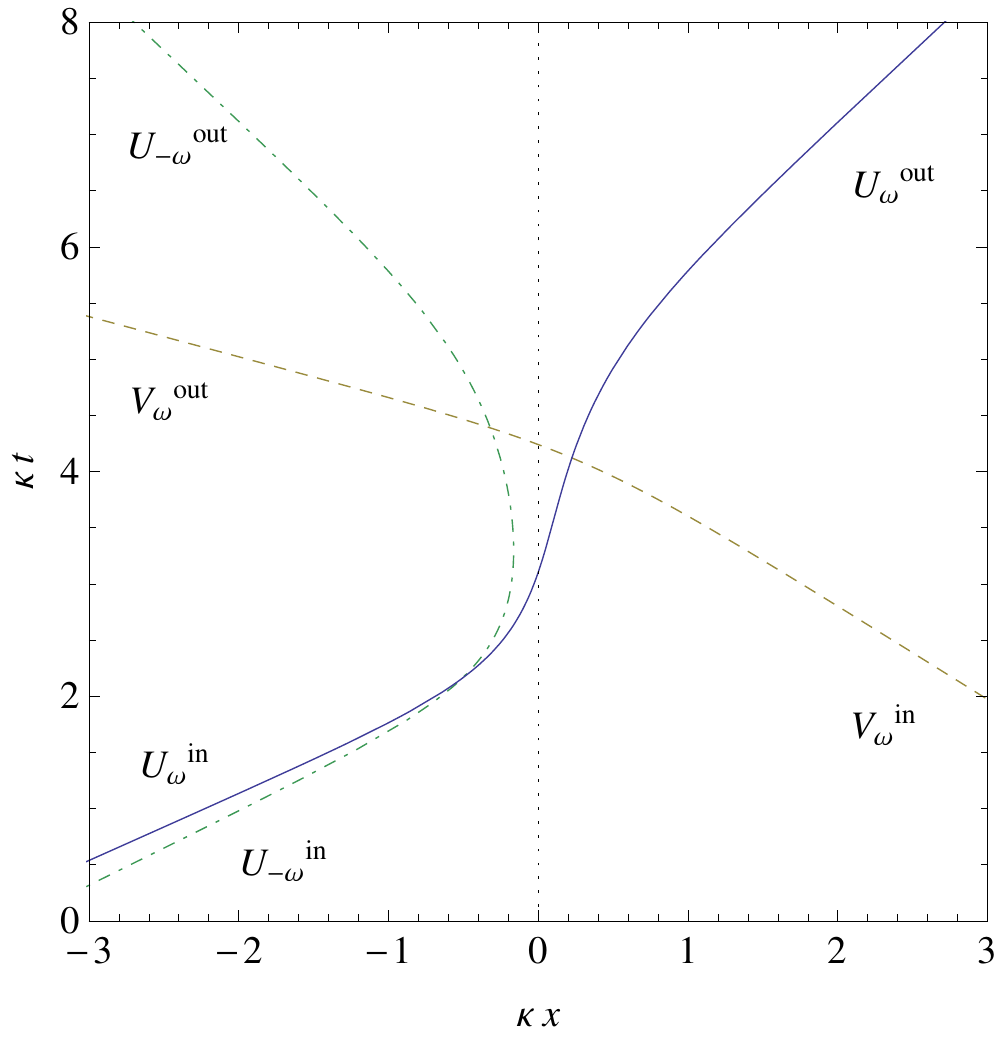}
\end{minipage}
\caption{Dispersion relation in the frame of the fluid $\Omega (k) $ (left). In blue is the $U$ branch with counter-propagating modes, in yellow the $V$ branch with co-propagating modes. In dash we represent the lines $\Omega = \omega - v k $, with $v=-1+D$ for the subsonic region (purple) and $v = -1- D$ for the supersonic region (green). Modes of frequency $\omega $ are situated at the intersection of the solid and dash curves. Negative norm modes correspond to modes with negative frequency in the frame of the fluid $\Omega $ and are indicated by the subscript $-\omega$. The $in $ modes are differentiated from the $out$ modes because they have a group velocity directed towards the black hole in the frame of the lab. Parameters are $\omega = \Lambda/4 $ and $D = 0.8$. On the right, the space-time pattern of the same modes. In blue the positive norm $U$ mode, in green dot dashed is the  negative norm trapped $U$ mode, in yellow dashed is the positive norm $V$ mode. Parameters are $\omega = \Lambda/40 $, $ \omega= \kappa $ and $D = 0.8$.}
\label{fig:disprel}
\end{figure}

To compute Eq.~\eqref{eq:noutcoutSmatrix} we also need the initial mean occupation numbers $n_\omega^{\rm in, a}$, where the superscript $a$ labels the three modes. As in Ref.~\cite{Macher:2009nz}, we assume that far from the horizon the initial state is a thermal bath at some global temperature $T_{\rm in}$ {\it in the frame of the fluid}. This means that the three $n_\omega^{\rm in, a}$ are given by
\begin{equation}
\label{eq:ninofOmega}
\begin{split}
n_\omega^{\rm in, a} = \frac{1}{\exp{(\Omega^{\rm in, a}_\omega/T_{\rm in})} -1},
\end{split}
\end{equation}
where $\Omega^{\rm in, a}_\omega$ is the asymptotic value of the comoving frequency of the corresponding asymptotic $in$ mode. Its value is given by 
\begin{equation}
\begin{split}
\Omega^{\rm in, a}_\omega= \omega - v_{\rm as}^a k^{\rm in, a}_\omega ,
\end{split}
\end{equation}
where $k^{\rm in, a}_\omega $ is the corresponding wave vector, and where $v_{\rm as}^a$ is the asymptotic value of $v$ evaluated on the left or right side. In the present case, one has $v_{\rm as}^a = -1 \pm D $. The $-$ sign is associated to the $u$ modes, and the $+$ sign to the $v$ mode, see Fig.~\ref{fig:disprel}. 

In the low frequency limit, $\omega/\Lambda \ll 1 $, the expressions of $\Omega^{\rm in, a}_\omega$ can be analytically computed~\cite{Macher:2009nz}. Using them, one obtains
\begin{equation}
\label{eq:ninofmuT}
\begin{split}
n_{ \pm \omega}^{u,\rm in} &\sim \frac{1}{\exp\left [ (\mu \mp \omega) /T_{\rm in}^u  \right ] -1}, \\
n_{\omega}^{v, \rm in} &\sim \frac{1}{\exp\left ( \omega/T_{\rm in}^v  \right ) -1}.
\end{split}
\end{equation}
The chemical potential of $u$-modes, $\mu$, and the redshifted temperatures are 
\begin{equation}
\begin{split}
\frac{\mu}{ \Lambda } &= \frac{(1+ D) (D (2 + D))^{3/2}}{1+4 D +2 D^2},\\
{T_{\rm in}^u } &= {  T_{\rm in}}  \frac{ D(2 + D)}{1+4  D +2  D^2}, \\ 
T_{\rm in}^v &= T_{\rm in} ( 2  - D).
\label{eq:Tuvin}
\end{split}
\end{equation}
The leading quantity governing $n_{ \pm \omega}^{u,\rm in}$ is $\mu /T_{\rm in}^u$. It scales as $\Lambda D^{1/2}/T_{\rm in}$. When $T_{\rm in} \ll \Lambda D^{1/2}$, the redshift is so important that the $u$-modes are effectively in their ground state, as in relativistic settings. 

\subsection{Summary} 
\label{sec:sixparameters}

Our parametrization of the final state, see Eq.~\eqref{eq:noutcoutSmatrix}, is based on six dimensionless quantities, namely 
\begin{equation}
\label{eq:sixparameters}
\begin{split}
\omega/ T_{\rm hor},\  D,\ \Lambda/ \kappa,\ A,\ B, \mbox{ and } T_{\rm in } / T_{\rm hor}. 
\end{split}
\end{equation}
The first ratio is the frequency in the units of the effective temperature, which itself depends on the surface gravity $\kappa$, the dispersive wave-number $\Lambda$, and the height of the velocity profile $D$, see Eq.~\eqref{eq:Thor}. The parameters $A$ and $B$ respectively quantify the greybody factors and the pair creation of $uv$ pairs. The last ratio gives the initial temperature in the units of the effective temperature. 

Notice that these parameters are not independent, as $T_{\rm hor}$ depends on $D$. We have adopted this set, precisely because the residual dependence on $D$ at fixed $T_{\rm hor}$ is very weak. Hence, $D$ can be effectively fixed. As we shall see below, the other five parameters are all relevant. We believe they effectively provide a complete description of the system, at least when the flow profile is smooth enough, i.e., $v(x)$ is monotonic and characterized by a single length related to the surface gravity in the vicinity of the horizon. When these conditions are not met, resonant effects~\cite{Zapata:2011ze,deNova:2012hm} related to the black hole laser effect~\cite{Coutant:2009cu,Finazzi:2010nc,Finazzi:2010yq} could play an important role and must be separately described.

\section{Domains of nonseparability} 
\label{sec:domainsofnonsep}

Using Eqs.~\eqref{eq:5modulusfixed} and~\eqref{eq:2linesortho}, Eq.~\eqref{eq:NSepuv} can be expressed in terms of the three initial occupation numbers. In agreement with Eq.~(9) in Ref~\cite{deNova:2012hm}, we obtain
\begin{equation}
\label{eq:Sepsimplified}
\begin{split}
\Delta_\omega^{uu} =& \abs{\alpha_\omega^v}^2 n_{\omega}^{u, \rm in} n_{-\omega}^{u, \rm in} + \abs{\tilde B_\omega}^2 n_{\omega}^{u, \rm in} n_{\omega}^{v, \rm in} + \abs{\tilde A_\omega}^2 n_{-\omega}^{u, \rm in} n_{\omega}^{v, \rm in} + \abs{B_\omega}^2 n_{\omega}^{u, \rm in} +  \abs{\beta_{-\omega}}^2 n_{\omega}^{v, \rm in}\\
& - \abs{\beta_{\omega}}^2 (1+ n_{\omega}^{v, \rm in} + n_{\omega}^{u, \rm in} + n_{-\omega}^{u, \rm in}) , \\
\Delta_\omega^{uv} =& \abs{A_\omega}^2 n_{\omega}^{u, \rm in} n_{-\omega}^{u, \rm in} + \abs{\beta_{\omega}}^2 n_{\omega}^{u, \rm in} n_{\omega}^{v, \rm in} + \abs{\alpha_\omega}^2 n_{-\omega}^{u, \rm in} n_{\omega}^{v, \rm in} + \abs{B_\omega}^2 n_{\omega}^{u, \rm in} +  \abs{\beta_{-\omega}}^2 n_{\omega}^{v, \rm in} \\
&- \abs{\tilde B_\omega}^2 (1+ n_{\omega}^{v, \rm in} + n_{\omega}^{u, \rm in} + n_{-\omega}^{u, \rm in}) .
\end{split}
\end{equation}
We notice that the above expressions are much more complicated than the corresponding ones in homogeneous and isotropic situations. One here looses the neat separation of the contributions of the spontaneous and the induced channels of Eq.~\eqref{eq:deltaout}.

We also notice that the maximum value of $\Delta_\omega^{uu}$ and $\Delta_\omega^{vv}$ is bounded by $ n_{-\omega}^{u}$. Indeed Heisenberg uncertainties guarantee; see Eq.~\eqref{eq:Campotoujoursvrai}
\begin{equation}
\begin{split}
\abs{c^{u u/v}_\omega}^2 \leq n^u_{-\omega} (n^{u/v}_\omega  +1),
\end{split}
\end{equation}
for both the $uu$ and the $uv$ channels. It is thus useful to introduce the relative quantities of Eq.~\eqref{eq:defdeltacampo}; see Ref.~\cite{Campo:2008ju}
\begin{equation}
\label{eq:Nsepuvrelat}
\begin{split}
\delta_\omega^{uu} \doteq  \frac{\Delta_\omega^{uu}}{ n_{-\omega}^{u} }+1,  \quad
\delta_\omega^{uv} \doteq  \frac{\Delta_\omega^{uv}}{  n_{-\omega}^{u} }+1 .
\end{split}
\end{equation}
which are both positive, irrespective of the state $\hat \rho$.

In what follows, we study the domains of negativity of $\Delta_\omega^{uu}$ and $\Delta_\omega^{vv}$ by making use of the parametrization of \ref{sec:system}. In a first time, we consider the dispersion relation of Eq.~\eqref{eq:disprelsuper}. The sub-luminal case is briefly studied in \ref{sec:sublum}. To identify the domains of nonseparability, we shall mainly use figures.

\subsection[Nonseparability of \texorpdfstring{$uu$}{uu} pairs]{Nonseparability of \texorpdfstring{\boldmath{$uu$}}{uu} pairs}

\subsubsection{The dependence in \texorpdfstring{$\omega$}{omega}}

\begin{figure}[htb]
\begin{minipage}[t]{0.47\linewidth}
\includegraphics[width= 1 \linewidth]{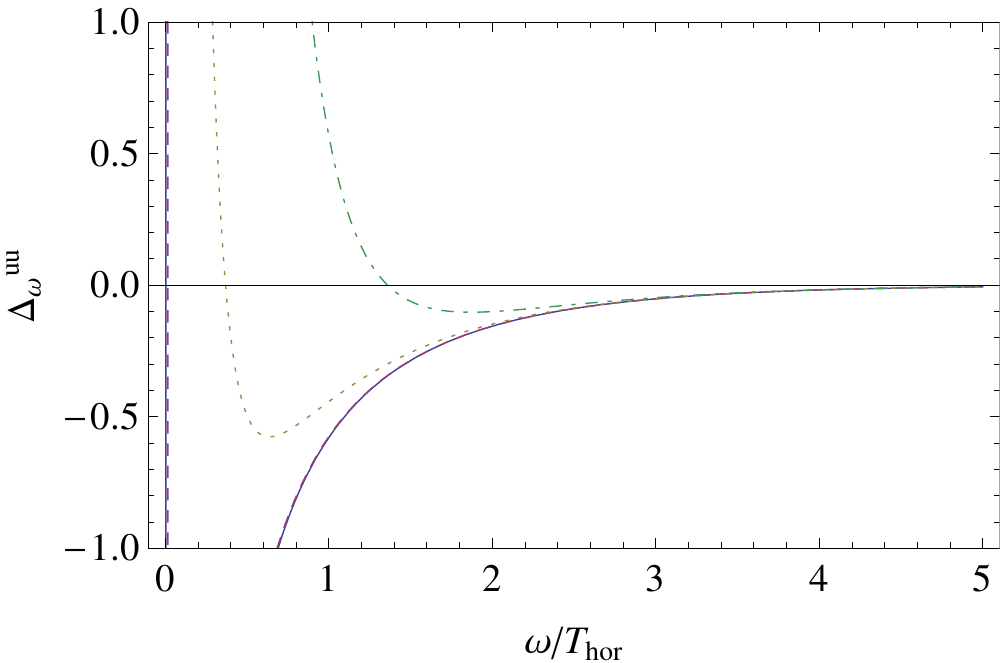}
\end{minipage}
\hspace{0.03\linewidth}
\begin{minipage}[t]{0.47\linewidth}
\includegraphics[width= 1 \linewidth]{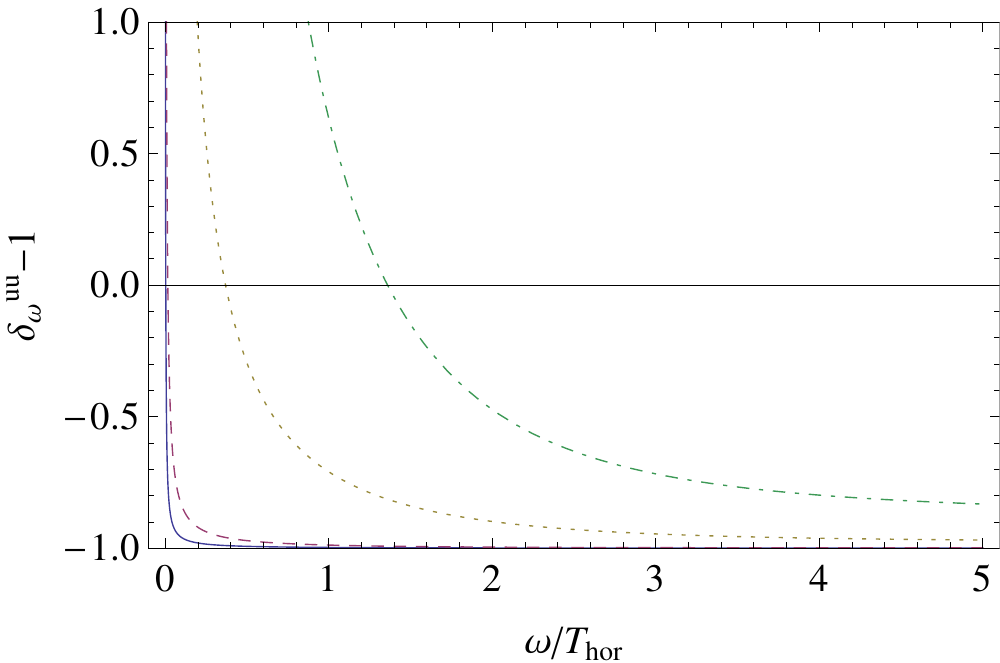}
\end{minipage}
\caption{The quantities $\Delta_\omega^{uu}$ of Eq.~\eqref{eq:Sepsimplified} (left panel) and $\delta_\omega^{uu}$ of Eq.~\eqref{eq:Nsepuvrelat} (right panel) are represented as functions of $\omega/T_{\rm hor}$ for a low initial temperature $T_{\rm in} = \Lambda/20$, for $\Lambda = 10 \kappa$, $D=1/2$, and for three values of $A = 4 B$, namely $B= 0.01$ (solid blue), $0.02$ (dashed purple), $0.1$ (dotted yellow), and $0.25$ (dot-dashed green). One clearly sees that low frequency modes are separable, and that increasing $A = 4 B$ monotonously reduces the domain of nonseparability.}
\label{fig:deltauofomegalowtemp}
\end{figure}

In Fig.~\ref{fig:deltauofomegalowtemp}, we study the $\omega/T_{\rm hor}$ dependence of $\Delta_\omega^{uu}$ and $\delta_\omega^{uu}$ in the low initial temperature regime, for $T_{\rm in} = \Lambda/20$. We consider three  different values of the coefficients $A,B$ of Eq.~\eqref{eq:ABbetaomegadep}, with a fixed ratio $A/B = 4$. At large frequency, $\omega/T_{\rm hor} \gtrsim 2$, the state is nonseparable independently of the values of $A$ and $B$. On the other hand, at low frequency, the state is always separable, even though decreasing $A$ and $B$ clearly increases the domain of nonseparability. The minimum value of $\Delta_\omega^{uu}$ is reached for $\omega/T_{\rm hor}\sim 1$. Instead, the minimum of $\delta_\omega^{uu}$ is reached for $\omega \to \infty$. In brief, for low initial temperatures, the low frequency sector contains many pairs but they are separable, the high frequency sector contains very few pairs which are highly nonseparable, and the intermediate regime contains few of them which are barely nonseparable. 

\begin{figure}[htb]
\begin{minipage}[t]{0.47\linewidth}
\includegraphics[width= 1 \linewidth]{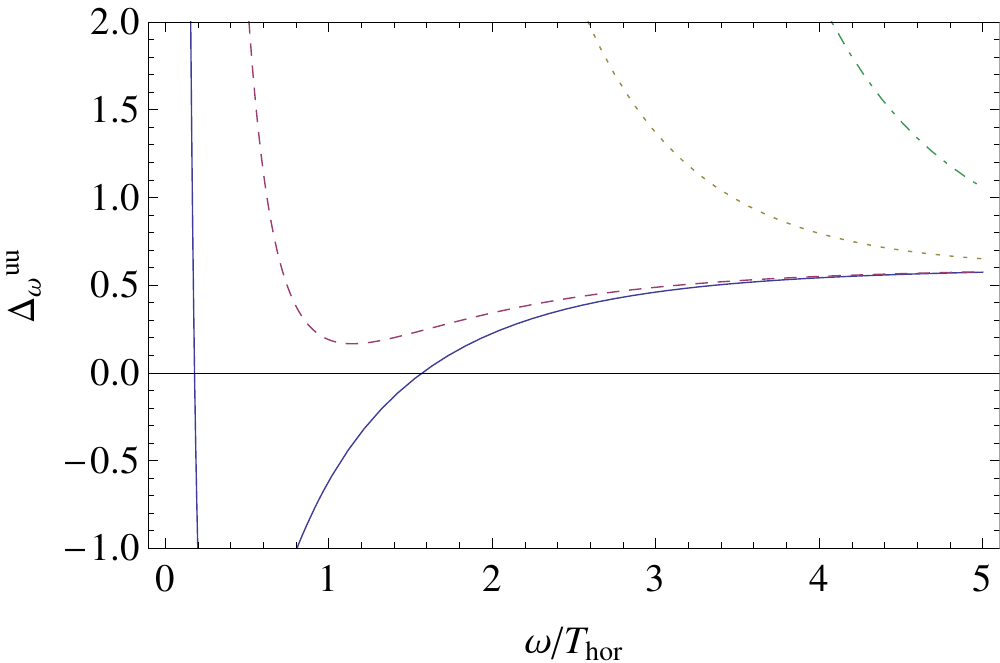}
\end{minipage}
\hspace{0.03\linewidth}
\begin{minipage}[t]{0.47\linewidth}
\includegraphics[width= 1 \linewidth]{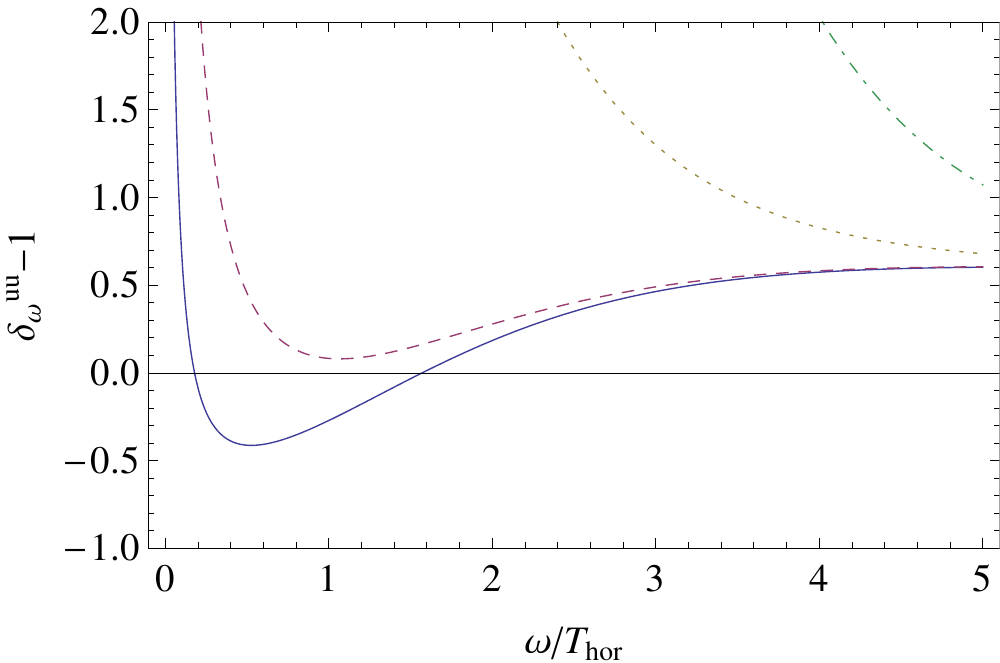}
\end{minipage}
\caption{We represent the same functions as in Fig.~\ref{fig:deltauofomegalowtemp}, for the same parameters, but for a higher initial temperature $T_{\rm in} = 2 \Lambda$. As expected, when compared to Fig.~\ref{fig:deltauofomegalowtemp}, one observes a reduction of the nonseparability domains. As found at low temperature, increasing $A = 4 B$ still reduces the domain of nonseparability. However, the high frequency sectors are now separable because $T_{\rm in}^u > 2 T_{\rm hor}$, as discussed below Eq.~\eqref{eq:Deltaulowomega2}. }
\label{fig:deltauofomegalargetemp}
\end{figure}

In Fig.~\ref{fig:deltauofomegalargetemp}, we study the same functions for a much larger initial temperature: $T_{\rm in} = 2 \Lambda$. In this case, the induced effects are much more important than above. As a result, the final state becomes separable even at large frequency. In fact, according to the value of $A$ and $B$, three different regimes show up. When $A,B$ are low enough, there still exists a finite range in $\omega$ where the state is nonseparable. When increasing $A,B$, this domain disappears, but $\Delta_\omega^{uu}$ still possesses a local minimum. When further increasing $A$ and $B$, $\Delta_\omega^{uu}$ becomes a monotonically decreasing function of $\omega$. Hence, as expected, increasing the initial temperature severely restricts the nonseparability of the state, or even completely suppresses it. 

It is of value to study analytically the asymptotic behaviors. The infrared behavior is dominated by the values of $A$ and $B$, as can be seen from 
\begin{equation}
\label{eq:Deltaulowomega}
\begin{split}
\Delta_\omega^{uu} \underset{\omega \to 0}{\sim } \frac{T_{\rm hor} T_{\rm in}^v}{\omega^2} \left ( n_{\omega}^{u, \rm in} + n_{-\omega}^{u, \rm in}+1 \right ) \left (  \gamma_+  B - A   \gamma_- \right )^2 ,
\end{split}
\end{equation}
where $\gamma_\pm$ is the limit of $\sqrt{\omega \abs{\beta_{\pm \omega}}^2 / T_{\rm hor} }$ for $\omega \to 0$. Eq.~\eqref{eq:Deltaulowomega} diverges as $\omega \to 0$ and is positive defined. This implies the separability of the low frequency regime.~\footnote{There is a noticeable exception: when $A=B$, the leading divergence in $1/\omega^2$ is absent. As a result, the domain of nonseparability further extends at low frequency. This is the case studied in Refs.~\cite{Unruh:1994je,Brout:1995wp,Macher:2009tw} where high frequency dispersion is added on the two-dimensional massless scalar equation. } 
In the large frequency regime $ \mu > \omega  \gg  T_{\rm in},T_{\rm hor} $, we obtain 
\begin{equation}
\begin{split}
\Delta_\omega^{uu} \sim \frac{e^{-2 \mu /T_{\rm in}^u}-  e^{-{\omega }/{T_{\rm hor}}}}{1-e^{-\abs{\mu-\omega  }/T_{\rm in}^u}}.
\label{eq:Deltaulowomega2}
\end{split}
\end{equation}
This is negative when $\omega T_{\rm in}^u \lesssim 2 \mu  T_{\rm hor}$. Hence, since $\omega<\mu$, the UV sector is nonseparable if $2  T_{\rm hor} \gtrsim T_{\rm in}^u $. With $D=1/2$ and $\Lambda = 10 \kappa$, this limit corresponds to $ T_{\rm in} \sim \Lambda/11$, independently of the values of $A$ and $B$. In addition, we observed that changing $D$ at fixed $T_{\rm hor}$ and $T^u_{in}$ of Eq.~\eqref{eq:Tuvin} has basically no effect.

In brief, the state in the infrared sector is generically separable because of the divergent contribution governed by the coefficients $A$ and $B$. Instead, the separability in the ultraviolet sector critically depends on $T_{\rm in}^u/T_{\rm hor}$. The intermediate regime is nonseparable if $A,B$ are low enough. 

Having characterized how $\Delta_\omega^{uu}$ depends on $\omega/T_{\rm hor}$, we now study how it depends on $A,B,T_{\rm hor}$ and $T_{\rm in}$. We shall establish that only two types of behaviors are found, depending on the ratio $T_H/T_{\infty}$, see Eq.~\eqref{eq:Thor}. In this we extend what was found in the spectral analysis of Ref.~\cite{Finazzi:2012iu}. When $T_H/T_{\infty} \lesssim 1/3$, one lives in the Hawking regime, with small dispersive effects. Instead, when  $T_H/T_{\infty} \gtrsim 3$, one finds the dispersive regime where the surface gravity plays no significant role. 

\subsubsection{Hawking regime} 

\begin{figure}[htb]
\begin{minipage}[t]{0.47\linewidth}
\includegraphics[width= 1 \linewidth]{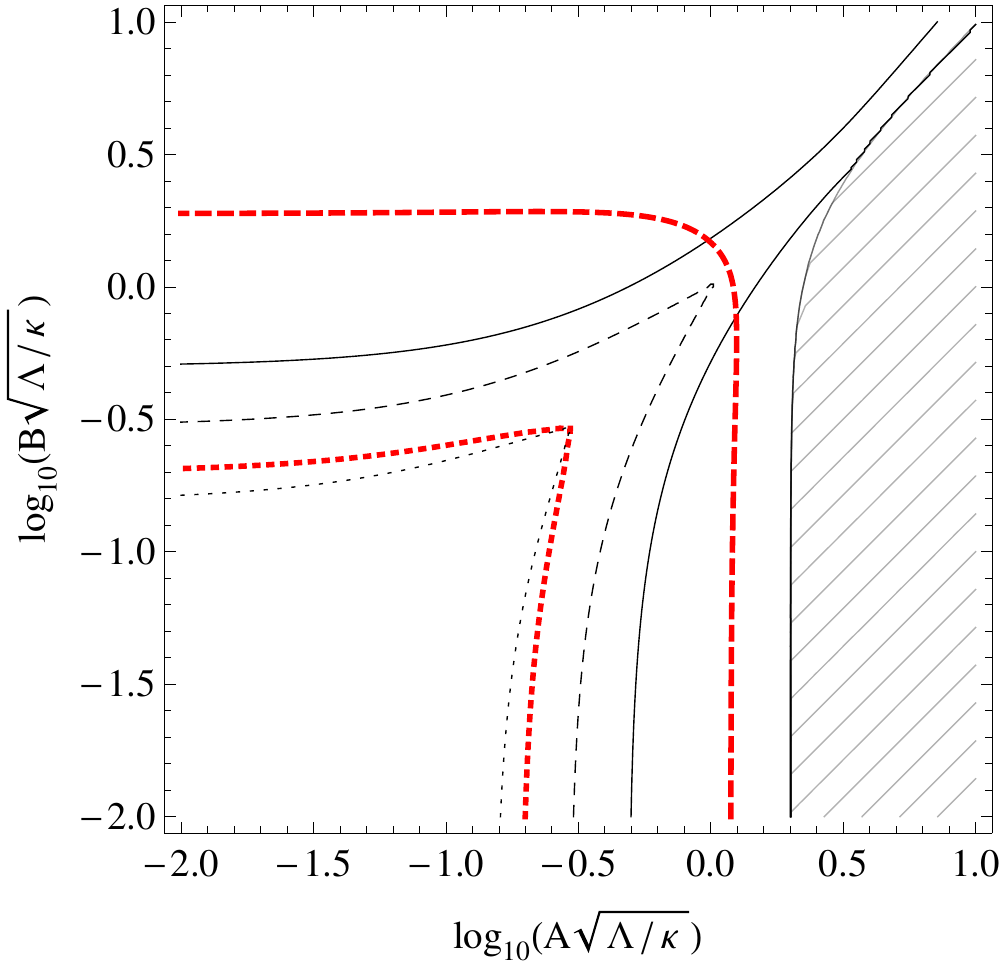}
\caption{The minimum value of $\Delta_\omega^{uu}$ over $\omega$ in the plane of $\log_{10} A \sqrt{\Lambda/\kappa}$, $\log_{10} B\sqrt{\Lambda/\kappa}$, for $D = 1/2$, $\Lambda/\kappa = 4$, and for three initial temperatures $ 5 T_{\rm in}^u /2 \mu = 1/3$ (Solid), $1$ (Dashed) or $3$ (Dotted). The thick red line is $\min \Delta_\omega^{uu} = 0$, and indicates the limit of nonseparability. The black line gives $\min \Delta_\omega^{uu} = -0.5$. It indicates the domain where the nonseparability is significant. We have $T_H  \sim T_{\infty}/3$, so we work at the edge of the {\it Hawking regime}. The dashed region represents the forbidden region where $\abs{\alpha^v}^2 = 1+\abs{B}^2-\abs{A}^2 <0$. }
\label{fig:minsepuhawk}
\end{minipage}
\hspace{0.03\linewidth}
\begin{minipage}[t]{0.47\linewidth}
\includegraphics[width=1\linewidth]{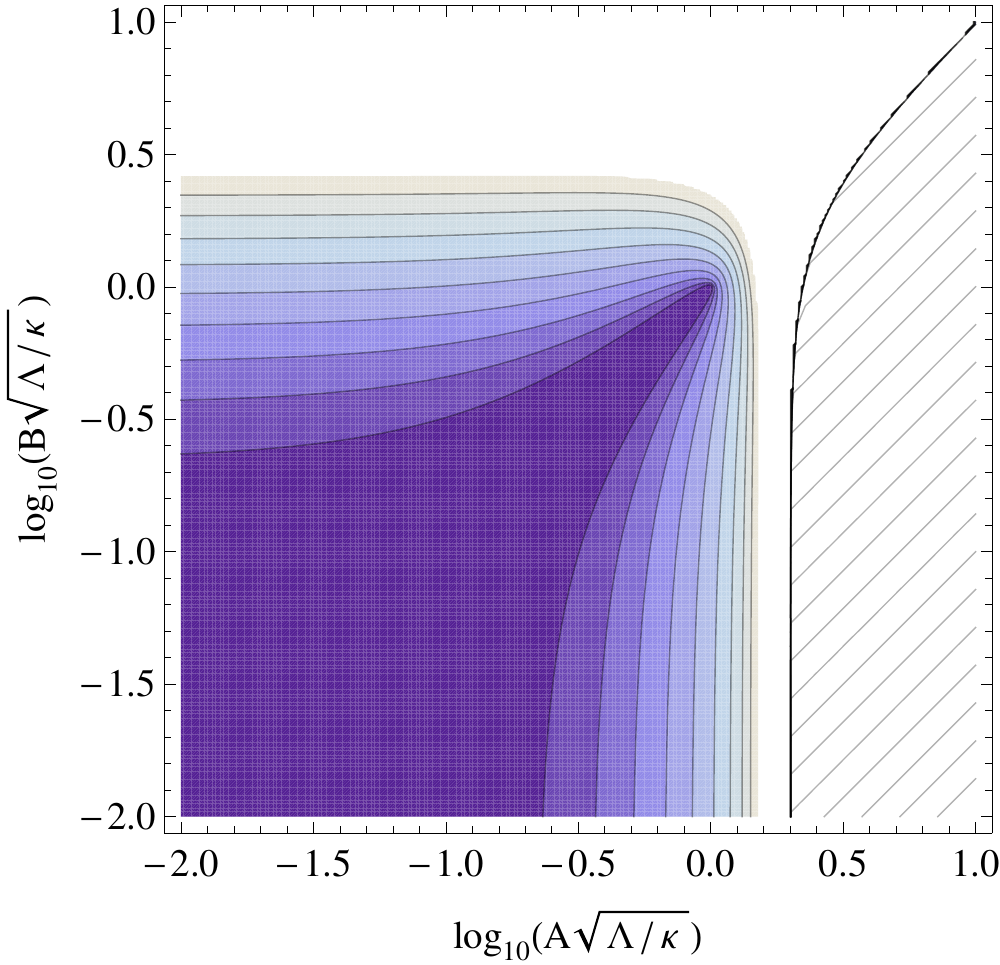}
\caption{The value of $\omega/T_{\rm hor}$ that minimizes $\Delta_\omega^{uu}$ in the same plane as in Fig.~\ref{fig:minsepuhawk}, for the same parameters, and for the intermediate temperature $5 T_{\rm in}^u = 2 \mu $. The lines of constant $\omega/T_{\rm hor}$ go by steps of $1/2$ from $1/2$ (dark blue) to $4.5$ (clear blue). The most relevant sector is $\omega/T_{\rm hor}\lesssim 1$, see Eq.~\eqref{eq:ABbetaomegadep}. } 
\label{fig:omegaminu}
\end{minipage}
\end{figure}

We first work at the edge of the Hawking regime, with $T_H/T_{\infty} = 1/3$. Reducing this ratio, which means reducing $\kappa/\Lambda$, does not affect the properties of Fig.~\ref{fig:minsepuhawk}. Hence what follows applies to the entire Hawking regime.

To eliminate $\omega$, we consider the minimum value of $\Delta_\omega^{uu}$ for $\omega < 5 T_{\rm hor}$. (It is pointless to consider higher values since the pair production rates are exponentially suppressed in that regime.) We shall consider two values of the minimum, namely $\min_\omega \Delta_\omega^{uu}= 0$ and $ = -0.5$. The first one gives the limit of nonseparability, whereas the second curve indicates the domain where the nonseparability is significant, and therefore more likely to be observed in an experiment. In Fig.~\ref{fig:minsepuhawk}, both curves $\min_\omega \Delta_\omega^{uu}= 0$ and $ = -0.5$ are represented in the plane of $\log_{10} (A\sqrt{\Lambda/\kappa})$ and $\log_{10}( B\sqrt{\Lambda/\kappa})$, and for three different initial temperatures, namely $ 5 T_{\rm in}^u /2 \mu  = 1/3$, $1$ or $3$. After several tries, we have adopted these axis and this parametrization of the initial temperature, because changing $D$ and $\Lambda$ at fixed 
\begin{equation}
\begin{split}
\frac{ T_{\rm in}^u }{2\mu}, \, {A}{\sqrt{\Lambda/\kappa} }, \,  {B}{\sqrt{\Lambda/\kappa} },
\label{eq:3hawk}
\end{split}
\end{equation}
has no significant influence on the curves. This means that in the Hawking regime, the minimal value of $\Delta_{uu}$ only depends on these three composite scales. This is the first important result of this chapter. 

The other lesson from Fig.~\ref{fig:minsepuhawk} is that $A$ and $B$ should be both smaller than $\sim \sqrt{T_{\rm in}^u /{6\mu}} \times\sqrt{\kappa/\Lambda}$ for the state to be significantly entangled, i.e., $\Delta_\omega^{uu} < - .5$. When this condition is met, the state can be found entangled even when the initial temperature $T_{\rm in}$ is significantly larger than the horizon temperature $T_{\rm hor}$. To give an example, when $T_{\rm in} = \Lambda = 10 T_{\rm hor}$, the state is nonseparable if $ A, B \lesssim 1/10 $. In addition, one also sees that $A\sim B$ enhances the nonseparability of the state. This is because the $1/\omega^2$ divergence of Eq.~\eqref{eq:Deltaulowomega} is reduced when $A\sim B$. 

To complete the information and also guide future experiments, in Fig.~\ref{fig:omegaminu} we represent the value of $\omega/T_{\rm hor}$ that minimizes $\Delta_\omega^{uu}$ for the same parameters as those of Fig.~\ref{fig:minsepuhawk}, and for the middle temperature $T_{\rm in}^u / \mu = 2/5$. Notice that a rough characterization of the curves can be obtained by considering the asymptotic behaviors of Eqs.~\eqref{eq:Deltaulowomega} and~\eqref{eq:Deltaulowomega2}, and by minimizing their sum. The symmetry with respect to interchanging $A$ and $B$ is then explained. In addition, when increasing the initial temperature $T_{\rm in}$, we learn that one should increase the value of $\omega/T_{\rm hor}$ in order to minimize $\Delta_\omega^{uu}$. Roughly speaking, one gets $(\omega/T_{\rm hor})^3\sim T_{\rm in}/T_{\rm hor}$.

\subsubsection{The dispersive regime}

We now proceed in the same way for the dispersive regime. We work at the edge of this domain with $T_H/T_{\infty} = 3$. We have verified that what follows applies in the whole dispersive regime $T_H/T_{\infty} > 3$. 

As for the Hawking regime, we extract the $\omega$ dependence by taking the minimum of $\Delta_\omega^{uu}$ over $\omega$, for $\omega < \omega_{\max}$, where $\omega_{\max}$ is the maximum value for which the negative norm mode exists~\cite{Macher:2009tw}. Since we are in the dispersive regime, the approximate expression of Eq.~\eqref{eq:ninofmuT} is no longer valid, even though $\mu$ and $T_{\rm in}^u$ of Eq.~\eqref{eq:Tuvin} are still well defined. We thus use the exact expression of Eq.~\eqref{eq:ninofOmega} in this section. 

\begin{SCfigure}[2][b!]
\includegraphics[width= 0.47\linewidth]{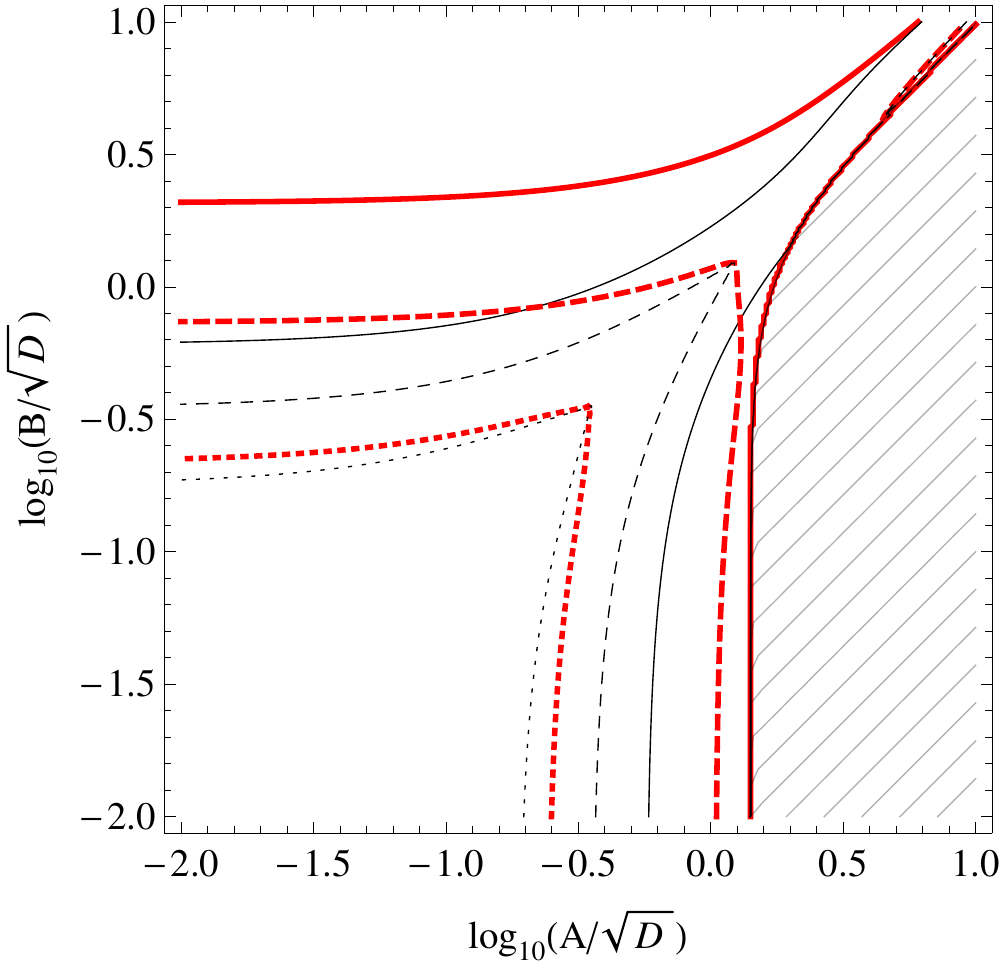} 
\caption{As in Fig.~\ref{fig:minsepuhawk}, we represent the minimum value of $\Delta_\omega^{uu}$ for three initial temperatures $ 5 T_{\rm in}^u /2\mu = 1/3$ (Solid), $1$ (Dashed) or $3$ (Dotted), and for $D = 1/2$. Here, we work in the {\it dispersive regime} since $\kappa /\Lambda = 10/4 $, and $T_H / T_{\infty} \sim 3$. As explained in the text, the two coordinates now are $\log_{10} A/\sqrt{D}$ and $\log_{10} B/\sqrt{D}$. The black line represents $\min \Delta_\omega^{uu} = -0.5$, and the thick red line $\min \Delta_\omega^{uu} = 0$. Notice the similarity of the present figure with Fig.~\ref{fig:minsepuhawk}. It indicates that the crossover, around $\Lambda/\kappa D \sim 1$, from the Hawking regime to the dispersive one is smooth.}
\label{fig:minsepudisp}
\end{SCfigure}

As in Fig.~\ref{fig:minsepuhawk}, in Fig.~\ref{fig:minsepudisp} we draw the constant values $\min_\omega \Delta_\omega^{uu}$ equal to $0$ and $-0.5$, to respectively get the nonseparability, and the significantly nonseparable, domains. In the present case, the axes have been chosen to be $\log_{10} (A/\sqrt{D}), \, \log_{10} (B/\sqrt{D})$, because, when adopting them, we observed that changing $D$ and $\Lambda$ at fixed $T_{\rm in}^u /\mu$ and $\{A,B\}/\sqrt{D}$ has no significant effect. As a result, in the dispersive regime, the minimal value of $\Delta_{uu}$ only depends on the following three composite scales:
\begin{equation}
\begin{split}
\frac{T_{\rm in}^u}{2 \mu }, \, \frac{A}{\sqrt{D }}, \,  \frac{B}{\sqrt{D} }. 
\label{eq:3disp}
\end{split}
\end{equation}
This is the second important result of this chapter. Notice that these three ratios differ from those of Eq.~\eqref{eq:3hawk}. 
We also notice that Fig.~\ref{fig:minsepudisp} is very similar to Fig.~\ref{fig:minsepuhawk}. This means that the crossover, around $\Lambda/\kappa D \sim 1$, from the Hawking regime to the dispersive one is rather smooth. 

As a result, when taken together, Figs.~\ref{fig:minsepuhawk} and~\ref{fig:minsepudisp} offer a full characterization of the nonseparability domains when the initial temperature belongs to the domain $0.1\lesssim T_{\rm in}/\mu \lesssim 1  $. The main conclusion of this section is that separability of the state depends mainly on three quantities. The first one is the initial temperature in the unit of the chemical potential $\mu$. This was expected since this ratio governs the initial distribution of $u$-quasiparticles. The other two are the $A$ and $B$ parameters encoding the coupling with the third mode. To obtain domains with well-defined scaling properties, these dimensionless parameters should be rescaled by $\sqrt{\Lambda/\kappa}$ in the Hawking regime and by $\sqrt{D}$ in the dispersive one\footnote{
We point here that the quartic superluminal dispersion relation has only been used in Eqs.~\eqref{eq:Thor} and~\eqref{eq:ninofmuT}. From this we can deduce that the above results should hardly be modified had we used another type of superluminal dispersion relation. This is due to the fact that we here parametrize the elements of the $S$-matrix in a manner which is basically independent of the dispersion relation once $T_{\rm hor}$ is given, see Eq.~\eqref{eq:ABbetaomegadep}.}.

\subsection[Nonseparability of \texorpdfstring{$uv$}{uv} pairs]{Nonseparability of \texorpdfstring{\boldmath{$uv$}}{uv} pairs}

\subsubsection{The dependence in \texorpdfstring{$\omega$}{omega}}

\begin{figure}
\begin{minipage}[t]{0.47\linewidth}
\includegraphics[width= 1 \linewidth]{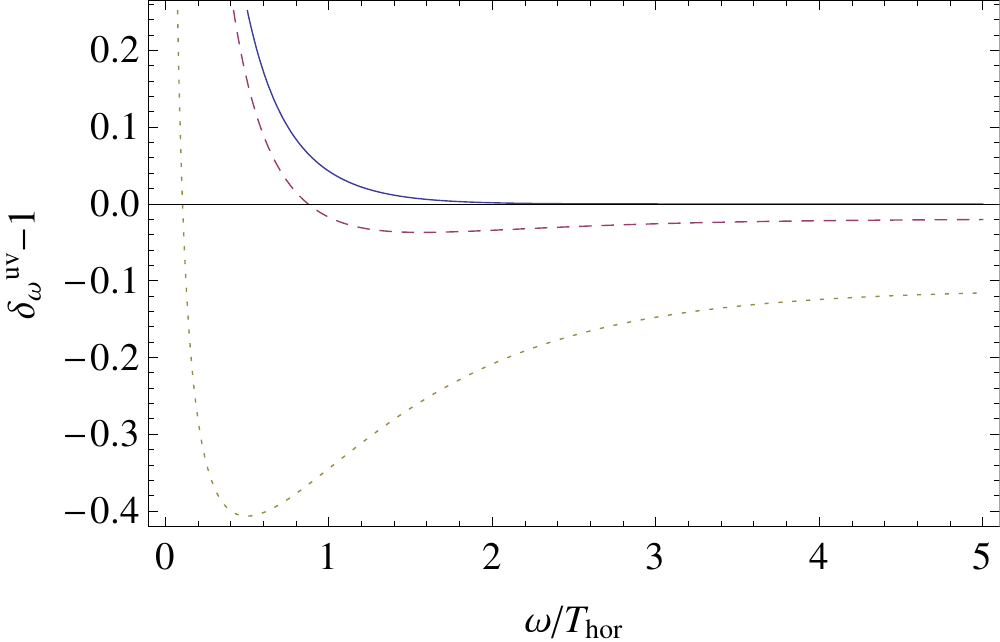}
\caption{The relative quantity $\delta_\omega^{uv}$ of Eq.~\eqref{eq:Nsepuvrelat} as a function of $\omega/T_{\rm hor}$ for a low temperature $T_{\rm in} = \Lambda /300 $, for three values of $B= 0.01$ (solid blue), $0.1$ (dashed purple), and $0.25$ (dotted yellow), for $A = 4 B$, $\Lambda = 10 \kappa$, and $D=1/2$. For $uv$ pairs, increasing $B$ now increases the nonseparability since $B^2$ governs their creation rate.}
\label{fig:vlowtempofomega}
\end{minipage}
\hspace{0.03\linewidth}
\begin{minipage}[t]{0.47\linewidth}
\includegraphics[width= 1 \linewidth]{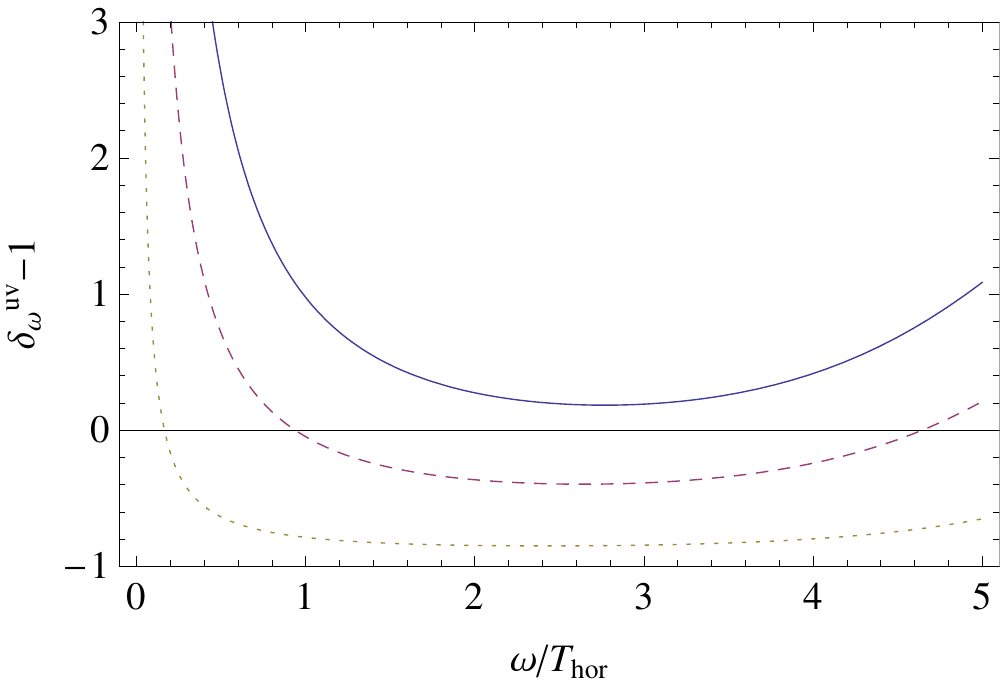}
\caption{The same relative quantity $\delta_\omega^{uv}$ as a function of $\omega/T_{\rm hor}$ for a high temperature $T_{\rm in} = \Lambda /3 $, for three values of $B= 0.1$ (solid blue), $1$ (dashed purple), and $2$ (dotted yellow), for  $A = B/4$, $\Lambda = 10 \kappa$, and $D=1/2$. With respect to Fig~\ref{fig:vlowtempofomega}, the values of $B$ have been increased by a factor of $\sim 10$. }
\label{fig:vlargetempofomega}
\end{minipage}
\end{figure}
 
Because the analysis is rather similar to that of the previous section, we only give the main results. In Fig.~\ref{fig:vlowtempofomega}, we first study $\delta_\omega^{uv}$ as a function of $\omega/T_{\rm hor}$ for three different values of $A,B$. As was found Fig.~\ref{fig:deltauofomegalowtemp}, we observe that the low frequency sector is always separable. 

When increasing the initial temperature, as expected, the value of $\delta_\omega^{uv}$ increases and the state becomes separable for all $\omega$. We then need to increase the pair creation rate $B^2$ to get nonseparable states, see Fig.~\ref{fig:vlargetempofomega}. We observe that the low frequency sector remains as it was at low temperature. On the contrary, for high frequency, the state is now separable. Yet, when $B$ is large enough, there exists an intermediate regime where the state remains nonseparable. In this regime, the nonseparability depends on a competition between the coupling $B$ and the initial temperature.

These observations can be verified analytically. First, the low frequency behavior is
\begin{equation}
\label{eq:Deltauvlowom}
\begin{split}
\Delta_\omega^{uv} \underset{\omega \to 0}{\sim}  \frac{ T_{\rm hor} T_v }{\omega^2 } \gamma_+^2 \left ( n_{\omega}^{u, \rm in} + n_{-\omega}^{u, \rm in}+1 \right ).
\end{split}
\end{equation}
We obtain a behavior similar to that of $uu$ pairs given in Eq.~\eqref{eq:Deltaulowomega}. However, in the present case, $\Delta_\omega^{uv} $ remains positive even when $A=B$. Second, at large frequency, the behavior of $\Delta_\omega^{uv}$ is very different to that of $\Delta_\omega^{uu}$. Indeed, when $\mu> \omega \gg T_{\rm in},T_{\rm hor}$, we have
\begin{equation}
\label{eq:Deltauvlargeom}
\begin{split}
\Delta_\omega^{uv} \sim&  \frac{ e^{-\omega /T_{\rm in}^v }+\ep{ \abs{\omega -\mu }/T_{\rm in}^u-\omega /T_{\rm hor}} (e^{-\omega /T_{\rm in}^v } - 2 B^2) }{e^{ \abs{\omega -\mu }/T_{\rm in}^u}-1}.\\
\end{split}
\end{equation}
The source of nonseparability is the term proportional to $B^2$. This makes perfect sense since $B^2$ fixes the production of $uv$ pairs. With more precision, the large frequency ($\omega = \mu$) behavior is nonseparable only if
\begin{equation}
\label{eq:vUVnonsep}
\begin{split}
2 B^2 \gtrsim \ep{\mu/T_{\rm hor} - \mu /T_{\rm in}^v},
\end{split}
\end{equation}
which requires a very low initial temperature in order to be satisfied. When Eq.~\eqref{eq:vUVnonsep} is not fulfilled, the state is separable at large $\omega$. However, nonseparability is possible when
\begin{equation}
\begin{split}
2 B^2 \gtrsim  \ep{ - \mu T_{\rm hor} / T_{\rm in}^v (T_{\rm hor}+ T_{\rm in}^u )}
\end{split}
\end{equation}
for frequencies obeying 
\begin{equation}
\begin{split}
\omega \lesssim \frac{ \mu + T_{\rm in}^u \log(2B^2) }{1 + T_{\rm in}^u /T_{\rm hor}- T_{\rm in}^u/T_{\rm in}^v} .
\end{split}
\end{equation}
This is the order of magnitude that we observe in Fig.~\ref{fig:vlargetempofomega}.

\subsubsection{The parametric dependence}

\begin{figure}[htb]
\begin{minipage}[t]{0.47\linewidth}
\includegraphics[width= 1 \linewidth]{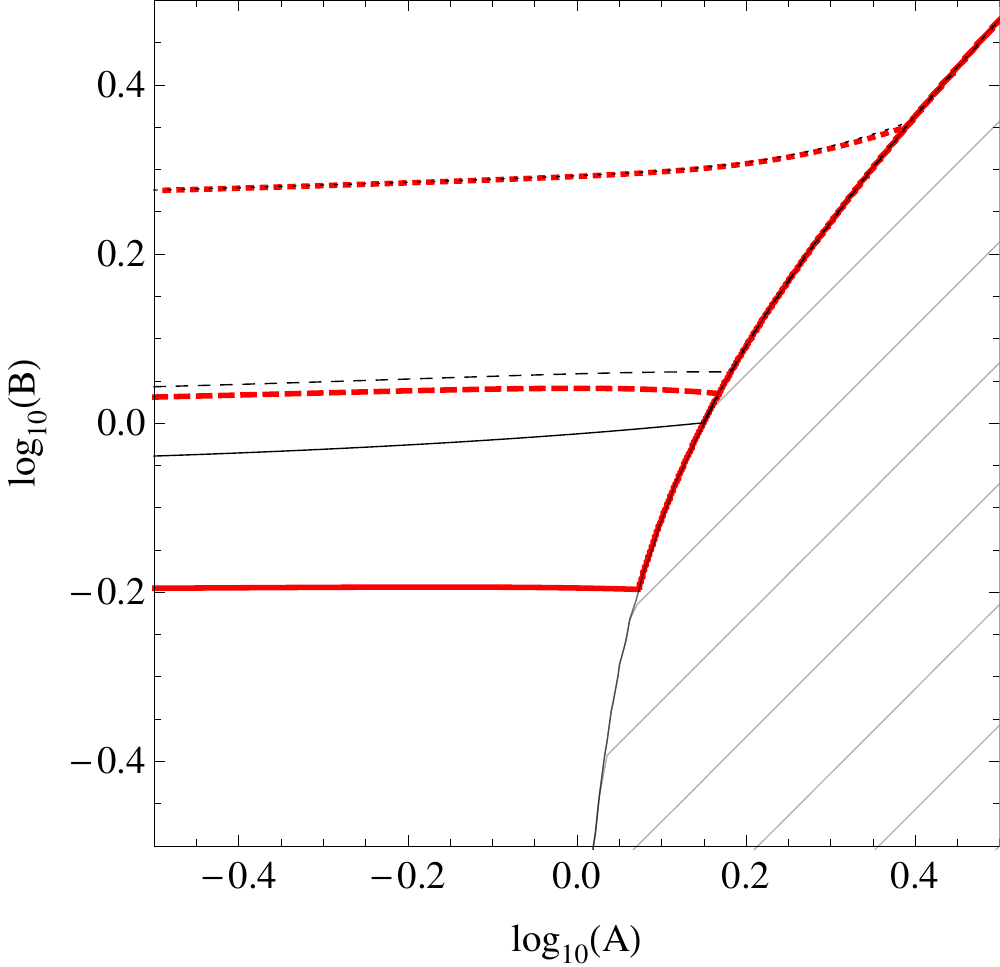}
\caption{The minimum over $\omega$ of $\Delta_\omega^{uv}$ for $\Lambda = 4 \kappa$, $D=1/2$ and for a temperature $ T = 1/3 \Lambda \sqrt{D}$ (Dotted), $\Lambda\sqrt{D}$ (Dashed) or $3 \Lambda \sqrt{D}$ (solid). The dashed region represents the region with $\abs{\alpha^v}^2 = 1+\abs{B}^2-\abs{A}^2 <0$. The line $\min \Delta = 0$ is indicated in thick red. The line $\min \Delta = -0.5$ is indicated in black.}
\label{fig:minsepv}
\end{minipage}
\hspace{0.03\linewidth}
%
\begin{minipage}[t]{0.47\linewidth}
\includegraphics[width= 1 \linewidth]{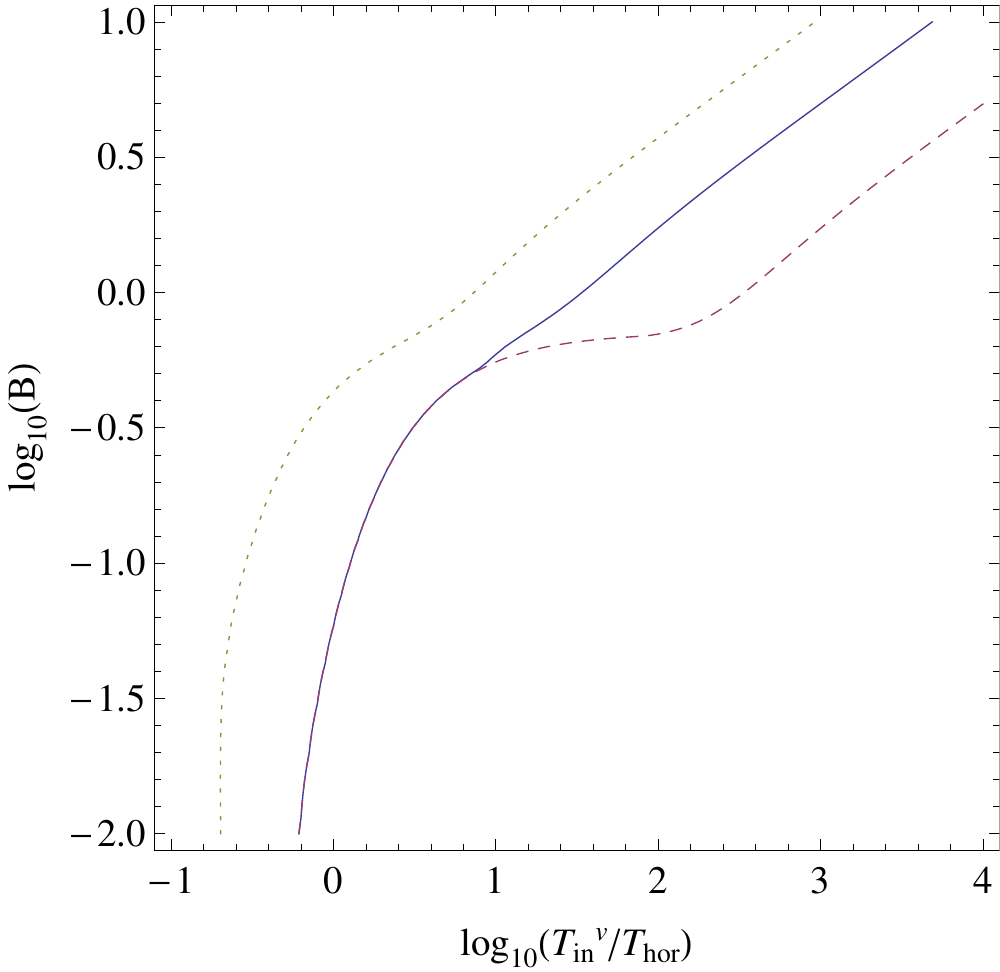}
\caption{The threshold of $uv$ nonseparability $\Delta_\omega^{uv} = 0$ in the $T_{in}^v/T_{\rm hor},B$ plane for $3$ values of $\Lambda\sqrt{D}/\kappa$, i.e., $40$ (dashed purple), $4$ (solid blue), $0.4$ (dotted yellow). In the Hawking regime, there exists a critical temperature, which controls the separability of the $UV$ sector. This critical temperature disappears in the dispersive regime. At a larger temperature, nonseparability is found if $B^2> T_{\rm in}^v / \Lambda\sqrt{D}$. When going further in the dispersive regime, $\Delta_\omega^{uv} = 0$ no longer evolves, and remains along the dotted curve. } 
\label{fig:enrab}
\end{minipage}
\end{figure}

As for $uu$ pairs, we now determine what is the domain of the parameter space where the state is nonseparable. We represent in Fig.~\ref{fig:minsepv} the minimum over $\omega$ of $\Delta_\omega^{uv}$ for different temperatures in the $A,B$ plane. The first observation is that $A$ plays no role. This can be seen from the asymptotic behaviors of Eqs.~\eqref{eq:Deltauvlowom} and~\eqref{eq:Deltauvlargeom}. A closer analysis reveals that the relevant parameters governing the nonseparability of $uv$ pairs are 
\begin{equation}
\label{eq:3uv}
\begin{split}
\frac{T_{\rm in}^v}{T_{\rm hor}}, B, \frac{\Lambda \sqrt{D}}{\kappa}.
\end{split}
\end{equation}
This is the third important result of this chapter.

Given that observation, we represent in Fig.~\ref{fig:enrab} the nonseparability threshold $\Delta_\omega^{uv} = 0$ in the $B$, $T_{\rm in}/T_{\rm hor}$ plane for different values of $\Lambda/ \kappa$. We observe first, that in the Hawking regime, there is a critical temperature $T_{\rm in}^{\rm crit} \sim  T_{\rm hor} $ below which the state is always nonseparable. This limit is due to the $UV$ behavior of the spectrum. Indeed, we see from Eq.~\eqref{eq:vUVnonsep} that when $\mu$ is large (i.e., deep in the Hawking regime) and $T_{\rm in}^v < T_{\rm hor}$, the states with $\omega \sim \mu$ are nonseparable for all values of $B$. The second observation is that this critical temperature decreases as we leave the Hawking regime. This is because the nonseparable regime $\omega \gg T_{\rm hor}$ no longer exists when $T_{\rm hor} \sim \omega_{\max}$. At higher temperature, the nonseparability criterion becomes $ B^2 \gtrsim T_{\rm in}/\Lambda$.

To summarize, the state is nonseparable when $T_{\rm in}^v \lesssim T_{\rm hor} $ or $T_{\rm in}^v \lesssim B^2 \Lambda $.

\section{Different cases}

\subsection{Subluminal dispersion relation}
\label{sec:sublum}

\begin{SCfigure}[2][htb]
\includegraphics[width= 0.47 \linewidth]{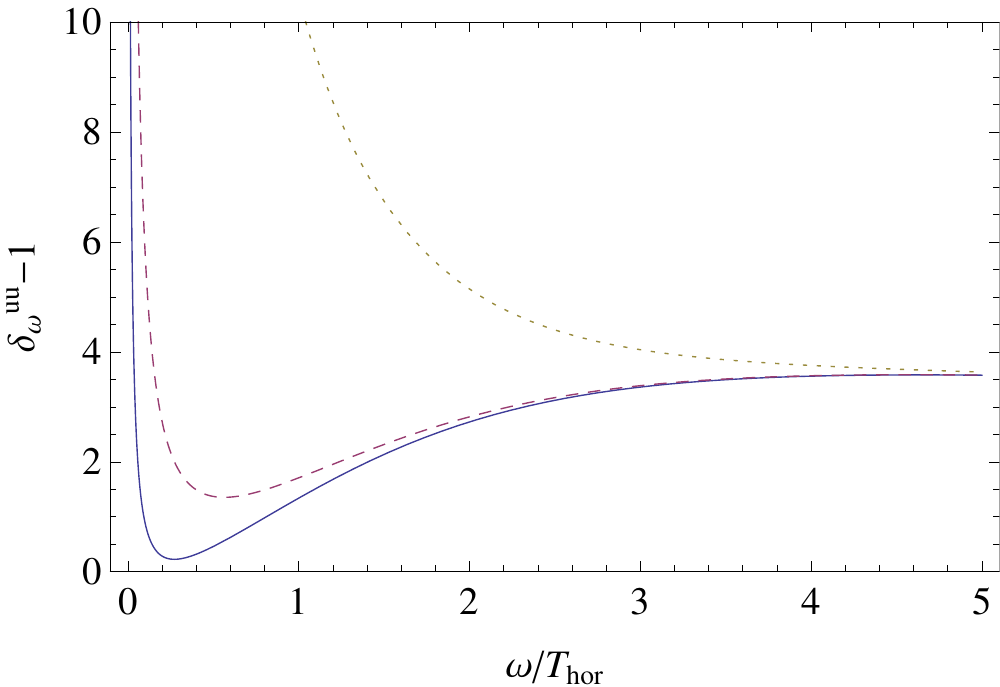}
\caption{The relative quantity $\delta_\omega^{uu}$ of Eq.~\eqref{eq:Nsepuvrelat} for the subluminal dispersion relation of Eq.~\eqref{eq:drsub} as a function of $\omega/T_{\rm hor}$, for a high initial temperature $T_{\rm in} = 2  \Lambda$, for three values of $B= 0.01$ (solid blue), $0.02$ (dashed purple), and $0.09$ (dotted yellow), $A = 4 B$, $\Lambda = 10 \kappa$, and $D=1/2$.
} 
\label{fig:deltauofomegalargetempsub}
\end{SCfigure}

We briefly consider the sub-luminal dispersion relation, 
\begin{equation}
\begin{split}
\label{eq:drsub}
F^2(k) &= c^2 ( k^2 - \frac{k^4}{\Lambda^2}),
\end{split}
\end{equation}
in order to present the main differences with the nonseparability of the super-luminal case considered in \ref{sec:domainsofnonsep}. In Fig.~\ref{fig:deltauofomegalargetempsub}, as in the right panel of Fig.~\ref{fig:deltauofomegalargetemp}, we represent the relative quantity $\delta_\omega^{uu}$ for a high initial temperature. For such temperature, we see that the state is slightly less entangled than in the superluminal case. The origin of this is due to the fact that $in$ modes come from the sub-sonic side of the horizon. As a result, for a given initial temperature $T_{\rm in}$, the effective $u$-temperature $T_{\rm in}^u$ of Eq.~\eqref{eq:ninofmuT} is larger than that found when the $in$ modes come from the supersonic side. In other words the initial distribution of $u$-quanta is less red-shifted for sub than super-luminal dispersion. This implies that the contribution of stimulated emission is higher, and this reduces the domains of nonseparability.

The low temperature behavior of $\delta_\omega^{uu}$ is much less sensitive to the sign of the dispersion relation because in that case, the nonseparability threshold is mainly governed by the coupling of the $v$ modes. Because there is no novel aspect in this case, we do not represent it. In addition, similar effects are also observed concerning the nonseparability of $uv$ pairs. Hence, these need not to be studied separately.

\subsection{Analogue white holes}
\label{sec:whitehole}

We consider the white hole flow obtained by replacing $v(x)$ by $-v(x)$, where the flow profile $v(x)$ describes a black hole, for an example see Eq.~\eqref{eq:vpluscBH}. In this case, as explained in Ref.~\cite{Macher:2009tw}, $S_{WH}$, the $S$-matrix in the white hole flow is simply given by the inverse of the corresponding black hole one given in Eq.~\eqref{eq:Bogcoef}. Because of unitarity, $S_{WH}$ is of the form $S_{WH} = T S^\dagger T$, where $T= {\rm diag}(1,-1,1)$. In terms of the black hole coefficients, $S_{WH}$ reads
\begin{equation}
S_{WH} = \left (\begin{array}{llll}
\alpha_\omega^* &- \beta_{-\omega}^* &\tilde A_\omega^* \\
 - \beta_\omega & \alpha_{-\omega}  &- \tilde B_\omega \\
A_\omega^*      &- B_\omega^*        & (\alpha^v_\omega)^* 
\end{array} \right ).
\end{equation}

When considering an incoherent initial state characterized by three initial occupation numbers,
Eq.~\eqref{eq:Sepsimplified} becomes
\begin{equation}
\begin{split}
\label{eq:D-WH}
\Delta_\omega^{uu} &= \abs{\alpha_\omega^v}^2 n_{ \omega, {\rm WH}}^{u,\rm in}  n_{ -\omega, {\rm WH}}^{u,\rm in}  + \abs{ B_\omega}^2 n_{ \omega, {\rm WH}}^{u,\rm in}  n_{\omega, {\rm WH}}^{v, \rm in} + \abs{ A_\omega}^2 n_{ -\omega, {\rm WH}}^{u,\rm in}  n_{\omega, {\rm WH}}^{v, \rm in} \\
& + \abs{\tilde B_\omega}^2 n_{ \omega, {\rm WH}}^{u,\rm in}  +  \abs{\beta_{\omega}}^2 n_{\omega, {\rm WH}}^{v, \rm in} - \abs{\beta_{-\omega}}^2 (1+ n_{\omega, {\rm WH}}^{v, \rm in} + n_{ \omega, {\rm WH}}^{u,\rm in}  + n_{ -\omega, {\rm WH}}^{u,\rm in} )\\
\Delta_\omega^{uv} &= \abs{\tilde A_\omega}^2 n_{ \omega, {\rm WH}}^{u,\rm in}  n_{ -\omega, {\rm WH}}^{u,\rm in}  + \abs{\beta_{-\omega}}^2 n_{ \omega, {\rm WH}}^{u,\rm in}  n_{\omega, {\rm WH}}^{v, \rm in} + \abs{\alpha_\omega}^2 n_{ -\omega, {\rm WH}}^{u,\rm in}  n_{\omega, {\rm WH}}^{v, \rm in} \\
& + \abs{\tilde B_\omega}^2 n_{ \omega, {\rm WH}}^{u,\rm in}  +  \abs{\beta_{\omega}}^2 n_{\omega, {\rm WH}}^{v, \rm in} - \abs{ B_\omega}^2 (1+ n_{\omega, {\rm WH}}^{v, \rm in} + n_{ \omega, {\rm WH}}^{u,\rm in}  + n_{ -\omega, {\rm WH}}^{u,\rm in} ).
\end{split}
\end{equation} 
On the other hand, working again with the thermal initial state of Eq.~\eqref{eq:ninofOmega}, 
in the place of Eq.~\eqref{eq:ninofmuT}, the initial distributions are
\begin{equation}
\begin{split}
n_{\omega, {\rm WH}}^{v, \rm in} \sim \frac{1}{\exp\left (  \omega / T_{\rm in, WH}^v \right ) -1}, \\ 
n_{ \pm \omega, {\rm WH}}^{u,\rm in} \sim \frac{1}{\exp\left (  \omega /T_{\rm in, WH}^{u} \right ) -1} ,
\end{split}
\end{equation}
where 
\begin{equation}
\begin{split}
T_{\rm in, WH}^{u} =  T_{\rm in} D, \quad 
T_{\rm in, WH}^v = T_{\rm in} (2  +  D) . 
\end{split}
\end{equation}
These two effective temperatures are independent of the dispersion relation because the three incoming modes are now low momentum ones. As a result, for white hole flows, the dispersive scale $\Lambda$ only enters in the final distributions only through the effective temperature $T_{\rm hor}$ of Eq.~\eqref{eq:Thor}, which is the same for the black and the white hole flows $\pm v(x)$.

When considering the entanglement of the quasiparticles emitted by a white hole, one expects that it will be weaker than that of the corresponding black hole. The reason is clear: in white hole flows, stimulated effects dominate in over the spontaneous channel because low frequency excitations are blueshifted (at fixed $\omega$, the final value of the wave number $k_\omega$ is larger than the incoming one). As a result, the quantities of Eq.~\eqref{eq:D-WH} diverge in the low frequency limit as
\begin{equation}
\begin{split}
\omega^2 \times \Delta_\omega^{uu} &\underset{\omega\to 0} \sim   \alpha_v^2 (T_{\rm in, WH}^{u})^2 + (A^2 + B^2) T_{\rm in, WH}^{u} T_{\rm in, WH}^v \\
&- T_{\rm hor} \left [ ( \gamma_- - \gamma_+ ) T_{\rm in, WH}^v  + (\gamma_- + \gamma_+ )T_{\rm in, WH}^{u}  \right ]\\
\omega^3 \times \Delta_\omega^{uv} &\underset{\omega\to 0}\sim T_{\rm in, WH}^{u} T_{\rm hor}  \left [ ( \gamma_- - \gamma_+ ) T_{\rm in, WH}^{u} + (\gamma_-+\gamma_+)  T_{\rm in, WH}^v \right ] , \\
\end{split}
\end{equation}
where $\gamma_\pm$ are the two quantities defined after Eq.~\eqref{eq:Deltaulowomega}. 

\begin{SCfigure}[2][htb]
\includegraphics[width= 0.47 \linewidth]{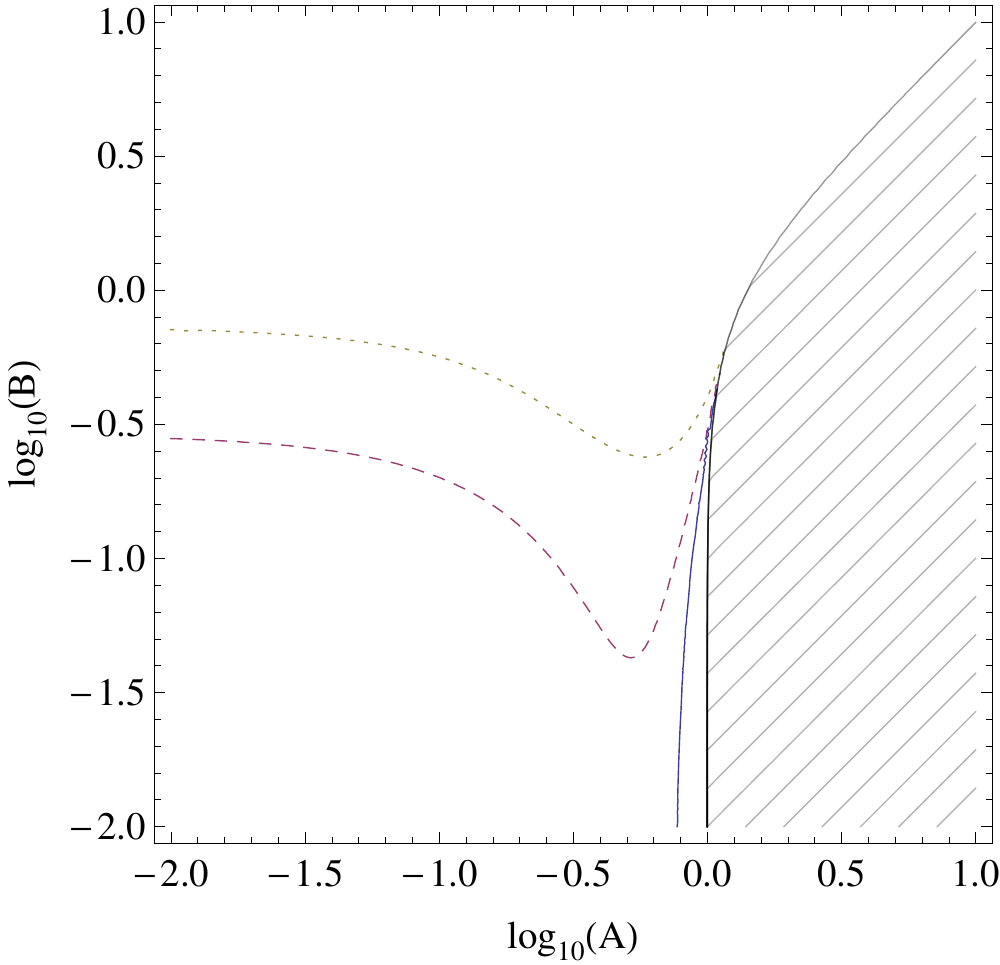}
\caption{The limit of nonseparability for the limit $\omega \to 0$ as a function of $A,B$ in a white hole flow. Parameters are $D= 1/2 $ and $T_{\rm in, WH}^{u} = 2 T_{\rm hor}  \times 0.5$(dotted yellow), $0.8$ (dashed purple) and $1$ (solid blue).}
\label{fig:lowomegauunsep}
\end{SCfigure}

The main consequence of these equations is that the nonseparability can be effectively studied by considering the low frequency limit. In Fig.~\ref{fig:lowomegauunsep}, we represent the limit of nonseparability $\Delta_{\omega}^{uu} = 0$ at $\omega = 0$ in the $(A ,B)$ plane. For $T_{\rm in, WH}^{u} < 2 T_{\rm hor}$, we observe that the state is nonseparable in a very large domain of the plane. Instead, for $T_{\rm in, WH}^{u} \geq 2 T_{\rm hor}$, only a small domain remains nonseparable. The transition between the two regimes is rapid since changing the value of the temperature by $20\%$ is sufficient to obtain nonseparability for $B \lesssim 0.3$. We verified that these conclusions are not significantly modified by relaxing the condition $\omega \to 0$, and taking the minimum of $\delta_{\omega}^{uu}$ as was done in \ref{sec:domainsofnonsep}. For $T_{\rm in, WH}^{u} < 2 T_{\rm hor}$, we observed an increase of the nonseparability domain. Instead for $T_{\rm in, WH}^{u} \geq 2 T_{\rm hor}$, we did not observe any increase. 

\begin{figure}[bht]
\begin{minipage}[t]{0.47\linewidth}
\includegraphics[width= 1 \linewidth]{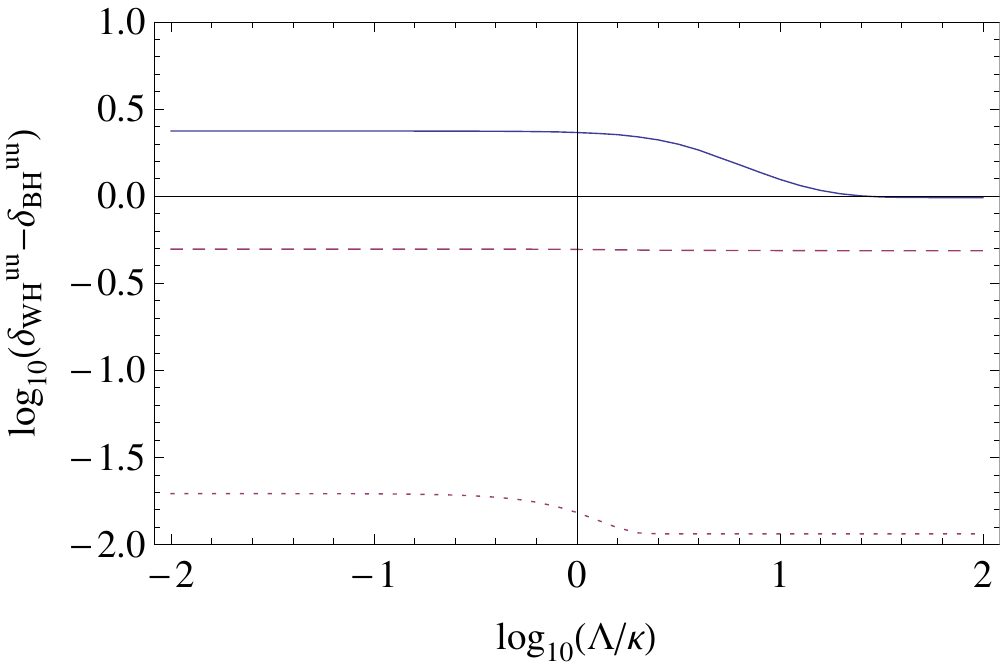}
\end{minipage}
\hspace{0.03\linewidth}
\begin{minipage}[t]{0.47\linewidth}
\includegraphics[width= 1 \linewidth]{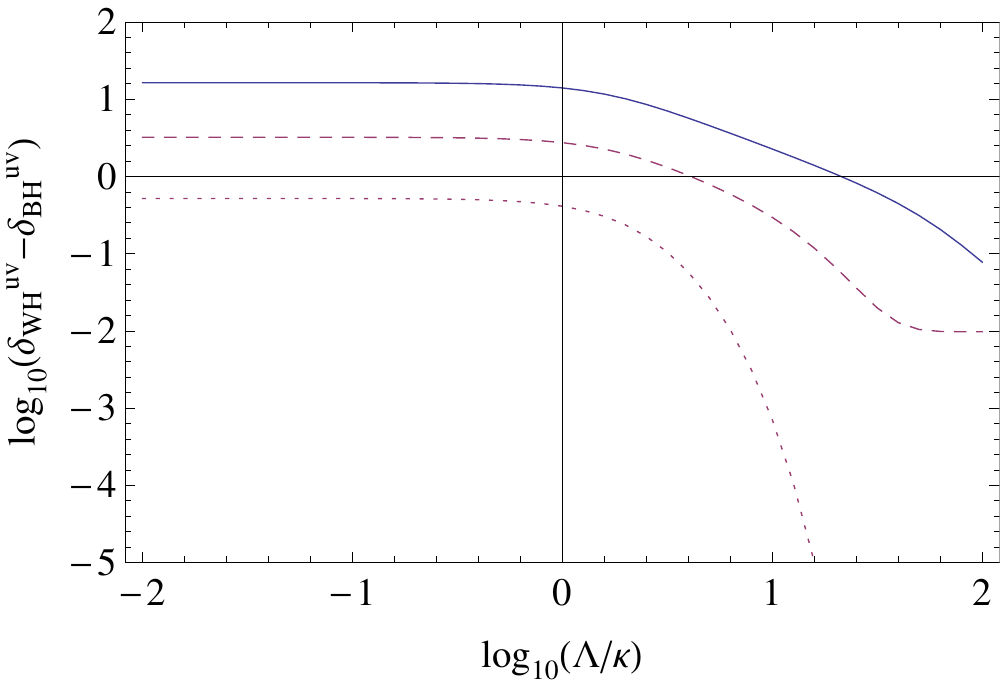}
\end{minipage}
\caption{The difference of the minima over $\omega$ of $\delta_{\omega}^{uu}$ (left panel) and $\delta_{\omega}^{uv}$ (right panel) computed in a white hole and in a black hole, for three different temperatures $T_{\rm in } = 2 T_{\rm hor} \times 0.2$ (dotted yellow), $1$ (dashed purple), $5$ (solid blue). The parameters are $D=1/2, A=B =0.1$. One sees that increasing the initial temperature further increases the difference between the values of $\delta_{\omega}^{uu}$ and $\delta_{\omega}^{uv}$, which means that the nonseparable character of the state is more rapidly lost for white holes than black holes.}
\label{fig:deltauuBHminusWH}
\end{figure}

To conclude this section, it is interesting to compare the values of $\delta_{\omega}^{uu}$ computed in a white hole and in a black hole for the same initial state. In the left panel of Fig.~\ref{fig:deltauuBHminusWH} (resp. the right panel) we represent the difference of the minima over $\omega$ of $\delta_\omega^{uu}$ (resp. $\delta_\omega^{uv}$) of Eq.~\eqref{eq:Nsepuvrelat} between the black hole and the white hole case. We used the sames values of the parameters $D=1/2, A=B =0.1$, $T_{\rm in}/T_{\rm hor}$ and plot the dependence of the function in $\Lambda/ \kappa$. We observe that $\delta$ is generically higher in the white hole flow than in the corresponding black hole one.

\section*{Conclusions}
\addcontentsline{toc}{section}{Conclusions}

We analyzed the strength of the correlations characterizing the two types of pairs that are emitted by a stationary black hole flow. To distinguish classical correlations associated with stimulated effects from genuine quantum entanglement due to spontaneous effects, we worked at fixed $\omega$, and used the criterion of nonseparability of the state which implies that one of the differences of Eq.~\eqref{eq:NSepuv} should be negative, see \ref{chap:separability}. 

In \ref{sec:system}, we studied the generic properties of the $S$-matrix on an analogue black hole horizon in order to adopt a parametrization of the spectra that takes into account the (super-luminal) dispersion relation. We then combined these parameters with those characterizing the three initial distributions of quasiparticles which are scattered on the horizon. The set of six parameters we used is described in \ref{sec:sixparameters}. 

In \ref{sec:domainsofnonsep}, we first studied the dependence in $\omega$ of $\Delta_\omega^{uu}$, the difference of Eq.~\eqref{eq:NSepuv} which characterizes the pairs of Hawking quanta. As expected, in the infrared sector, the state is separable, because stimulated effects dominate over spontaneous ones. We also observed that the domain of nonseparability critically depends on the strength of the couplings between the spectator third mode and the two modes under study. We then studied how the minimum value of $\Delta_\omega^{uu}$ over $\omega$ depends on the five parameters we adopted, see Eq.~\eqref{eq:sixparameters}. We showed that the domain of nonseparability only depends on three combinations of these parameters. In addition, two of the three combinations possess two different forms. The parameter that distinguishes the two regimes is the ratio of the temperatures $T_H/T_{\infty}$, see Eq.~\eqref{eq:Thor}. When it is smaller than $1/3$, dispersive effects are small, and one effectively works in a regime close to the relativistic Hawking one. Instead, when it is larger than $3$, one works in a regime where the surface gravity plays no significant role. We also showed that the crossover from one regime to the other is rather smooth. Combining the scalings in the two regimes, we obtained a rather complete characterization of the domains of nonseparability, which we hope will be useful to guide future experiments to identify the appropriate range of parameters where the spontaneous Hawking effect dominates over stimulated ones. 

The main lesson is that in both regimes, the nonseparability threshold critically depends on the initial temperature of the system (something which was expected), but also on the intensity of the coupling with the third spectator mode. When the latter is small enough, the final state can be quantum mechanically entangled even when the initial temperature is higher than the black hole temperature. For completeness, we also studied the quantity $\Delta_\omega^{uv}$ which governs the nonseparability of the other type of pairs emitted by an analogue black hole. The relevant parameters are completely different.

In \ref{sec:sublum}, we briefly studied the modifications when replacing the superluminal dispersion relation by a subluminal one. No significant change is observed besides the fact that stimulated effects are slightly more important for subluminal dispersion, as the redshift of the initial distribution is less pronounced. In \ref{sec:whitehole}, we compared the entanglement obtained in a white hole flow with that of a black hole one. We showed that white holes are less appropriate to look for quantum entanglement because stimulated effects are much more important.

\chapter*{Conclusions and perspectives}
\phantomsection
\addstarredchapter{Conclusions and perspectives}
\markboth{\MakeUppercase{Conclusions and perspectives}}{\MakeUppercase{Conclusions and perspectives}}

In this thesis we studied effects induced by a nontrivial and nonquantum metric (either gravitational or effective in analogue systems) on quantum matter. From the gravity point of view, we introduced non trivial dispersion and some dissipation by the addition of a background timelike vector field to the metric. In \ref{part:basics}, we reviewed the basic elements of quantum field theory in curved space-time and the separability criterion. These tools allowed us to build many theories. In \ref{part:desitter}, we studied in details a rather simple curved space-time, i.e., the de Sitter space-time. We saw that even though dispersion and dissipation modify the ultraviolet behavior of the theory, most of the predictions of the standard theory are robust to the introduction of dispersion and dissipation. From the cosmological point of view, this completes the inflationary works of Refs.~\cite{Macher:2008yq,Adamek:2008mp}. Moreover, using the correspondence between de Sitter and black holes, we showed that the Hawking flux remains thermal to a good approximation even in the presence of dispersion and dissipation. We also characterized an upper bound to the leading deviations. In \ref{part:analoguegravity}, we considered many condensed matter systems that are analogue to quantum field theory in curved space-time and characterized in each case the domain in parameter space where the predictions of quantum field theory in curved space-time should be manifest. In particular, we considered the analogue of pair particle production, i.e., dynamical Casimir effect in Bose-Einstein condensates in the presence of a generic dissipative rate, and we studied the effect of the initial temperature on analogue Hawking radiation.

This work is part of a global program, the aim of which is to measure the quantum properties of vacuum and to validate Hawking's predictions in analogue experiments. From a more theoretical point of view, it joins quantum gravity theories which contain ultraviolet dispersion, such as Horava theory~\cite{Horava:2009uw} and Einstein-Aether theory~\cite{Eling:2004dk}.

Even though many different subjects have been studied, this work only considers a subclass of aspects of quantum field theory in curved space-time. As an example, we could go further including interactions. We also only considered bosonic matter fields and more specifically scalar fields.  Moreover, the back-reaction of matter fields on the background fields, i.e., on space time and preferred time may be a sequel to this work. On the analogue gravity side, an extension of this work could consist in the explicit study of Hawking effect in presence of dissipation, or more generally in a non trivial background. In particular, when considering white hole flows, experiments have shown the intriguing presence of a non trivial standing waves.

\newpage\thispagestyle{plain}
%
%
%

\chapter*{Acknowledgments}
\phantomsection
\addstarredchapter{Acknowledgments}
\markboth{\uppercase{Acknowledgments}}{\uppercase{Acknowledgments}}

I am grateful to my PhD supervisor, the physical intuition of whom impressed me from the first month of my thesis. He knew how to transmit some of his knowledge and made me discover what research is. 
I also thank Iacopo Carusotto and Julian Adamek with whom I shared a publication. I thank them and every researcher with whom I had long discussions for the time they gave me. 
I also thank the jury of this PhD for the interest they show in my work. A particular attention is reserved for the referees who helped me improve this manuscript.
To the PhD students of LPT, I wish to address my gratitude for the working atmosphere we could have in the laboratory. 
I also thank the Cosmology group, for their long coffee break. Thank to you, I learned how to answer someone's knocking without without lifting my head from my calculation.
I am very grateful for the \textit{Corps de l'armement} for they allowed me to make a PhD in a subject that far from their main objectives.
But this thesis would not have been the same without the Platypus braxx band and all its members that allowed me to spend musical evenings and week ends. I am also particularly grateful to my family, my parents and Alexandra for the support and the trust they show every day. 
To conclude, I wish to thank every single person who dedicate their life to physics and who push forward the boundaries of human knowledge. It is such a shame they don't have the recognition they deserve.

 \appendix
\clearpage
\phantomsection
\addstarredchapter{\appendixtocname}
\appendixpage 

\chapter{Modulated DCE}
\label{app:modulDCE}

\section*{Introduction}

There have been two recent experiments able to test the nonseparability condition: first using an atomic BEC~\cite{PhysRevLett.109.220401} and second, in a Josephson metamaterial~\cite{Lahteenmaki12022013}. Interestingly, in the first case, the condition was not met, while it was met in the second. In both cases, the experimental teams used a periodic modulation of some parameter over an extended period of time. 

In this appendix we study theoretically the reaction of a system to a long-lasting modulation with the aim of understanding under which conditions the final state will be nonseparable. The main change with respect to \ref{chap:dissipBEC} and \ref{chap:polariton} is in the expression of the Bogoliubov coefficients. In a first time, we present the basic equations governing the time evolution of the modes of a homogeneous system in response to an arbitrary modulation in time. We then characterize the nonseparability of the state after a sinusoidal modulation. As a final step, we introduce dissipation through its main effects. This appendix is mainly based on Ref.~\cite{Busch:2014vda}

\minitoc

\section{Time-dependence in homogeneous media}
\label{sec:tempchangeeom}

In this section we consider the effects of time dependence on a quantum system. While the nature of the time dependence is left unspecified, we shall restrict our attention to homogeneous media, allowing the entire analysis to be done at fixed wave vector $\bk$. As a result, the dimensionality of the system drops out, and need not be specified. To fix the notation and the concepts, we shall work in an atomic Bose condensate~\cite{Dalfovo:1999zz,PhysRevLett.109.220401}. However, the following analysis is easily adapted to other media, such as polariton systems~\cite{Carusotto:2012vz} and Josephson metamaterials~\cite{WilsonDCE,Lahteenmaki12022013}. It is also applicable to pair creation in cosmological models, such as primordial inflation~\cite{Starobinsky:1979ty,Mukhanov:1981xt}; in particular, the time variation we shall study in \ref{sec:numanalisys} is very similar to that occurring during the preheating phase at the end of inflation~\cite{Kofman:1997yn}. 

\subsection{Equations of motion}

In a condensed dilute gas, linear density perturbations obey the Bogoliubov-de~Gennes equation~\cite{Dalfovo:1999zz}. At fixed $\bk$, in units where $\hbar = 1$, one obtains neglecting dissipation [see Eqs.~\eqref{eq:eomphi} and~\eqref{eq:2x2eom}]
\begin{equation}
\label{eq:eomphinondiss}
i\partial_t \hat \phi_\bk = \Omega_k \hat \phi_\bk + m c^2 \hat \phi_{-\bk}^\dagger .
\end{equation}
As in Refs.~\cite{PhysRevA.67.033602,Finazzi:2010nc}, we shall describe Eq.~\eqref{eq:eomphinondiss}, as well as its corresponding Hermitian conjugate equation with $\bk \to -\bk$, as a matrix equation:
\begin{equation}
\begin{split}
i\partial_t \left [ \begin{array}{ll}
 \hat \phi_\bk\\
\hat \phi_{-\bk}^\dagger
\end{array}\right ] &= \left [ \begin{array}{ll}
 \Omega_k & mc^2 \\
-mc^2 & -\Omega_k
\end{array}\right ] \times
\left [ \begin{array}{ll}
 \hat \phi_\bk\\
\hat \phi_{-\bk}^\dagger
\end{array}\right ].
\end{split}
\end{equation}
To clearly identify the effects that are due to a temporal change of $\Omega_k$ or $c$, we perform the standard Bogoliubov transformation, see Eq.~\eqref{eq:defvarphibogofield}
\begin{equation}
\label{eq:defbogofield}
\begin{split}
\left [ \begin{array}{ll}
 \hat \varphi_\bk\\
\hat \varphi_{-\bk}^\dagger
\end{array}\right ] &= \left [ \begin{array}{ll}
u_k & v_k\\
v_k & u_k
\end{array}\right ]\times \left [ \begin{array}{ll}
 \hat \phi_\bk\\
\hat \phi_{-\bk}^\dagger
\end{array}\right ] ,
\end{split}
\end{equation}
where $u_k$ and $v_k$ are given by Eq.~\eqref{eq:ukandvk}
and the frequency $\omega_k$ is given by Eq.~\eqref{eq:defomegak}.
When $\Omega_k$ and/or $c$ vary in time, so too do $u_k$, $v_k$, and $\omega_k$. Using the fact that $u_k^2-v_k^2=1$, straightforward algebraic manipulation leads to the following equation of motion\footnote{
Equation~\eqref{eq:eomvarphi} is very similar to the equation governing the photon field in a cavity of modulated Josephson metamaterial in Ref.~\cite{Lahteenmaki12022013}, although due to the stationarity of that inhomogeneous system (in a rotating frame), the correlations are between opposite frequencies rather than opposite wave vectors.}
for the Bogoliubov operators\footnote{Note that, as in \ref{chap:dissipBEC}, we could also have considered $\hat{\chi}_{\bk} \propto \hat{\phi}_{\bk}+\hat{\phi}_{-\bk}^{\dagger}$ which obeys $\left[\partial_{t}^{2}+\omega_{k}^2(t)\right]\hat{\chi}_{\bk}=0$. For a sinusoidal modulation of $\omega_{k}^{2}(t)$, this is (up to a coordinate transformation) the Mathieu equation~\cite{Abramowitz}, which also plays a role in preheating cosmological scenarios~\cite{Kofman:1997yn}. } $\varphi_\bk$ and $\hat \varphi_{-\bk}^\dagger$:
\begin{equation}
\label{eq:eomvarphi}
\begin{split}
i\partial_t \left [ \begin{array}{ll}
 \hat \varphi_\bk\\
 \hat \varphi_{-\bk}^\dagger
\end{array}\right ] &= \left [ \begin{array}{ll}
 \omega_k & i \frac{\dot u_k}{v_k} \\
i \frac{\dot u_k}{v_k} 
& - \omega_k
\end{array}\right ] \times
\left [ \begin{array}{ll}
 \hat \varphi_\bk\\
 \hat \varphi_{-\bk}^\dagger
\end{array}\right ] ,
\end{split}
\end{equation}
where $\dot u_k = \partial_t u_k$.

In stationary systems, one recovers the standard diagonal matrix governed by $\omega_k$. In that case, the fields are trivially related to the (canonical) phonon creation and annihilation operators, see Eq.~\eqref{eq:defphononoperatorb}. In term of vectors, this equation reads 
\begin{equation}
\label{eq:statiovarphi}
\begin{split}
\left [ \begin{array}{ll}
\hat \varphi_\bk\\
 \hat \varphi_{-\bk}^\dagger
\end{array}\right ]
&= \left [ \begin{array}{ll}
\ep{- i \omega_k t}\\
0
\end{array}\right ] \hat b_\bk + \left [ \begin{array}{ll}
0\\
\ep{ i \omega_k t}
\end{array}\right ](\hat b_{-\bk})^\dagger .
\end{split}
\end{equation}
When the system is stationary for asymptotic early times, the initial operators $\hat b_\bk^{\rm in}$ and $(b_{-\bk}^{\rm in})^\dagger$ are well defined and related at early times to the field operators by the above equation. The same is true when the system becomes stationary for asymptotic late times, where the late behavior of the fields $\varphi_\bk$ and $\hat \varphi_{-\bk}^\dagger$ defines the final operators $\hat b_\bk^{\rm out}$ and $(b_{-\bk}^{\rm out})^\dagger$. Then, because the field equation is linear, the two sets of asymptotic operators are related by an overall Bogoliubov transformation, see Eq.~\eqref{eq:bogosuroperateur}
\begin{equation}
\label{eq:bogoliubov}
\begin{split}
\hat b_\bk^{\rm out} = \alpha_k^{\rm as} \hat b_\bk^{\rm in} + ( \beta_k^{\rm as})^* (\hat b_{-\bk}^{\rm in})^\dagger ,
\end{split}
\end{equation}
where the requirement that both the initial and final operators satisfy the bosonic commutation relations imposes the condition $\abs{\alpha_k^{\rm as}}^{2} - \abs{\beta_k^{\rm as}}^{2} = 1$.

The asymptotic $in$ operators define a two-component mode $W_k^{\rm in}(t)$ via the commutator
\begin{equation}
 W_k^{\rm in}(t) \doteq \left [ 
 \begin{array}{ll}
\left [\hat \varphi_\bk(t), (b^{\rm in}_\bk)^\dagger\right ]  \\  
 \null [\hat \varphi^{\dagger}_{-\bk}(t) , (b^{\rm in}_\bk)^\dagger  ]
\end{array} \right ] .
\label{phimode}
\end{equation}
To simplify the notation, we shall no longer write the subscript $\bk$ since all equations shall be defined for a fixed value of $k = \abs{\bk}$. Equation~\eqref{phimode} implies that the mode doublet $W^{\rm in}(t)$ is the solution of Eq.~\eqref{eq:eomvarphi} with initial conditions $W^{\rm in}\underset{t \to -\infty}\sim \left [ \begin{array}{ll}
\ep{- i \omega t}\\
0
\end{array}\right ]$. 
For large times it behaves as 
\begin{equation}
\begin{split}
W^{\rm in} \underset{t \to +\infty}\sim \left [ \begin{array}{ll}
\alpha^{\rm as} \ep{- i \omega t} \\
\beta^{\rm as} \ep{ i \omega t} 
\end{array}\right ].
\end{split}
\end{equation}
Following the standard method~\cite{Massar:1997en} to evaluate the Bogoliubov coefficients $\alpha^{\rm as}$ and $\beta^{\rm as}$, we introduce the functions $\alpha(t)$ and $\beta(t)$ through the expression
\begin{equation}
\begin{split}
W^{\rm in}(t) = \left [ \begin{array}{ll}
\alpha(t) \, \ep{- i \int^t \omega dt'} \\
\beta(t) \, \ep{ i \int^t \omega dt'} 
\end{array}\right ] .
\end{split}
\end{equation}
By definition, their initial values are $1$ and $0$, and their late-time values coincide (up to a phase) with $\alpha^{\rm as}$ and $\beta^{\rm as}$. They obey the first-order coupled equations 
\begin{equation}
\label{eq:eomalphabeta}
\partial_t \alpha = \frac{\dot u}{v} \ep{ 2i \int \omega dt} \beta , \quad
\partial_t \beta = \frac{\dot u}{v} \ep{- 2i \int \omega dt} \alpha .
\end{equation}
One then verifies that the zeroth order solution, i.e., constant values for $\alpha$ and $\beta$, corresponds to the WKB approximation. One also verifies that corrections are associated with nonadiabatic transitions, and are here interpreted as creation of phonon pairs with opposite wave vectors.

It is interesting to notice that
\begin{equation}
\begin{split}
\frac{\dot u }{v}= \frac{\dot \omega }{ 2\omega} - \frac{\partial_t(\Omega - mc^2)}{2(\Omega - mc^2)}.
\end{split}
\end{equation}
In the following, we shall assume that $\Omega - mc^2$ is constant.\footnote{In atomic BEC, this corresponds to a constant atomic mass. This is not necessarily true in systems like polaritons or atoms on a lattice, as pointed out to us by C.~Westbrook~\cite{RevModPhys.78.179}.} In this case, the Bogoliubov coefficients are governed solely by the time evolution of $\omega$.

\subsection{Frequency modulation}

\begin{SCfigure}[1][htb]
\includegraphics[width=0.5\linewidth]{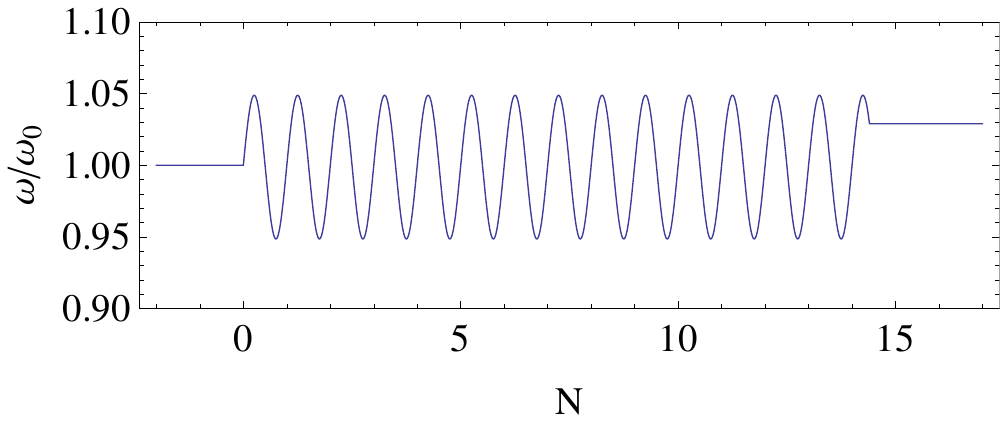}
\caption{Here is plotted an example of the frequency modulation of Eq.~\eqref{modul}, square-rooted so as to give $\omega/\omega_0$. The horizontal axis represents $N =\omega_p t/2\pi$. The amplitude $A=0.1$, and the total number of oscillations $N=14.4$.}
\label{fig:omegaoft}
\end{SCfigure}

In the following, we consider an extended coherent modulation of the system that induces a corresponding modulation of $\omega^2$. More precisely, $\omega^2$ is assumed to be constant for negative times, then follows a sinusoidal oscillation of duration $\Delta t$, and settles on a constant final value for all later times:
\begin{equation}
\omega^2(t)/\omega_0^2 = \left \{
\begin{array}{ll}
1 &\text{ if } t < 0 ,\\
1 + A \sin \omega_p t &\text{ if } 0 < t < \Delta t , \\
1 + A \sin \omega_p \Delta t &\text{ if } \Delta t < t .
\label{modul}
\end{array}\right .
\end{equation}
This function is illustrated in Fig.~\ref{fig:omegaoft}. It defines three dimensionless parameters that are all relevant in the following: the relative peak-to-peak\footnote{Assuming $A \ll 1$, the square root of the second line of Eq.~\eqref{modul} gives $\omega_k(t)/\omega_0 \approx 1 + \left(A/2\right) \sin \omega_p t$, so that the relative amplitude of the frequency modulation (as opposed to that of the {\it squared} frequency) is $A/2$.} amplitude $A$ of the frequency modulation, the number of oscillations $N$, and a resonance parameter we call $R$. Explicitly, these are defined by
\begin{equation}
\label{eq:3param}
N \doteq \omega_p \Delta t / 2\pi ,\quad
AR/4 \doteq \left( 2\omega_0 - \omega_p \right)/\omega_p. 
\end{equation}
Notice that $R$ combines in a particular way the detuning $\left( 2\omega_0 - \omega_p \right)/\omega_p$ and the amplitude $A$: it describes the relative frequency gap from resonance, scaled by $A$ so that it depends on this distance as a fraction of the ``width'' of the frequency modulation. Notice also that $N$ is not necessarily an integer.

For convenience, in the rest of the paper, we shall use $ N_t = \omega_p t / 2\pi $ as a dimensionless time parameter. Since $\alpha$ and $\beta$ are continuous in time, we can think of $\alpha(N)$ and $\beta(N)$ either as their final values after a modulation of length $N$, or as their instantaneous values at $N_t=N$ during a modulation of indeterminate length. These equivalent points of view allow us to use the same notation for $N$ and $N_t$, and also for $\abs{\beta^{\rm as}(N)}$ and $\abs{\beta(t = \Delta t_N)}$.

\section{First effects of modulation}
\label{sec:firsteffect}

Here we apply the concepts described in \ref{sec:tempchangeeom} to the temporal modulation of Eq.~\eqref{modul}, solving Eqs.~\eqref{eq:eomalphabeta} to find the behavior of the Bogoliubov coefficient $\abs{\beta^{\rm as}}$ and of the nonseparability parameter $\Delta_{\rm out}$ of Eq.~\eqref{eq:defDeltalinear}.

\subsection{Analysis of \texorpdfstring{$\boldsymbol{\abs{\beta}}$}{modulus of beta}} 
\label{sec:numanalisys}

\subsubsection{Numerical analysis}

\begin{figure}[htb]
\begin{minipage}[t]{0.47\linewidth}
\includegraphics[width=1\linewidth]{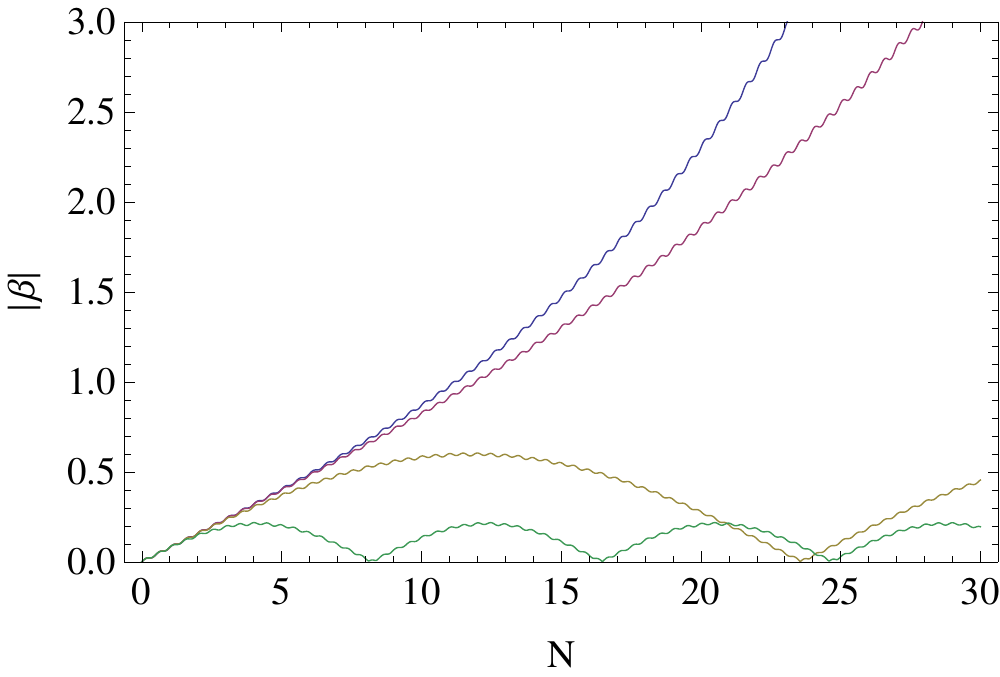}
\caption{Here is plotted $\abs{\beta}$ as a function of $N $ for various values of $R$, from top to bottom: $0$ (blue), $0.5$ (purple), $2$ (yellow), and $5$ (green). For all plots, the modulation amplitude $A=0.1$. After an initial linear growth for all curves, those with $R<1$ grow exponentially with $N$, while those with $R>1$ rise and fall periodically, the amplitude and period being approximately proportional to $1/R$. We also note the small rapid oscillations occurring on top of the long-time behavior.}
\label{fig:betaofN}
\end{minipage}
\hspace{0.03\linewidth}
\begin{minipage}[t]{0.47\linewidth}
\includegraphics[width=1\linewidth]{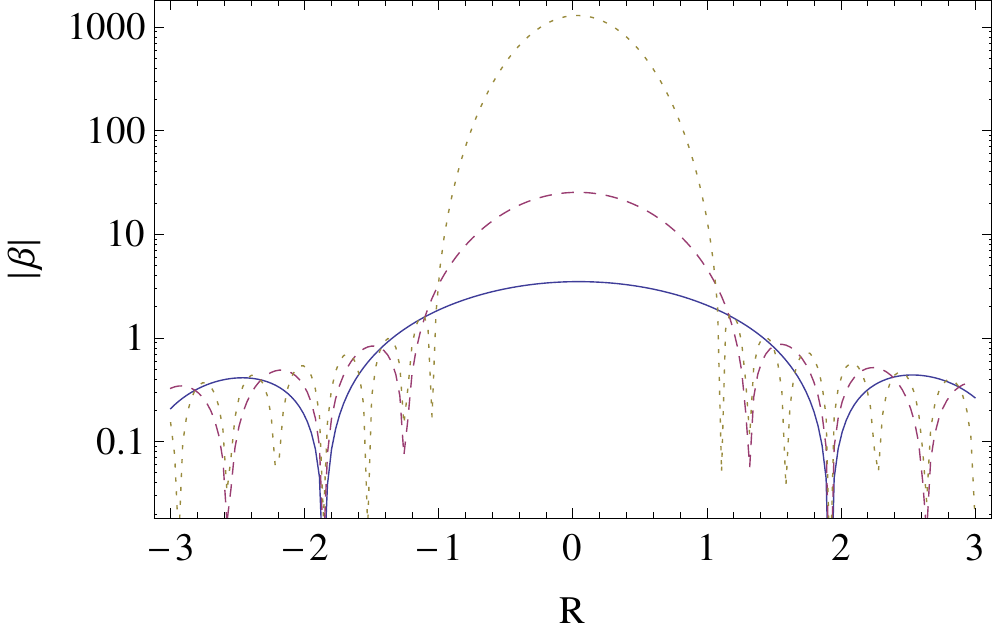}
\caption{Here is shown a logarithmic plot of $\abs{\beta}$ as a function of $R$ for various values of $N$: $25$ (blue solid), $50$ (purple dash), and $100$ (yellow dotted). For all plots, the modulation amplitude $A=0.1$. We clearly see the emergence of a central peak with increasing $N$, extending from $R=-1$ to $1$. For $\abs{R}>1$, the curves oscillate in a complicated way, as for small values of $\abs{\beta}$ the small rapid oscillations become more important. Because of these, $\abs{\beta}$ need not exactly vanish at the completion of a cycle. The number of long-time oscillations increases in proportion to $N$, and their maxima trace out an envelope corresponding to the $1/\sqrt{R^2-1}$ behavior of their maxima.}
\label{fig:betaofres}
\end{minipage}
\end{figure}

\begin{SCfigure}[1][htb]
\includegraphics[width=0.5\linewidth]{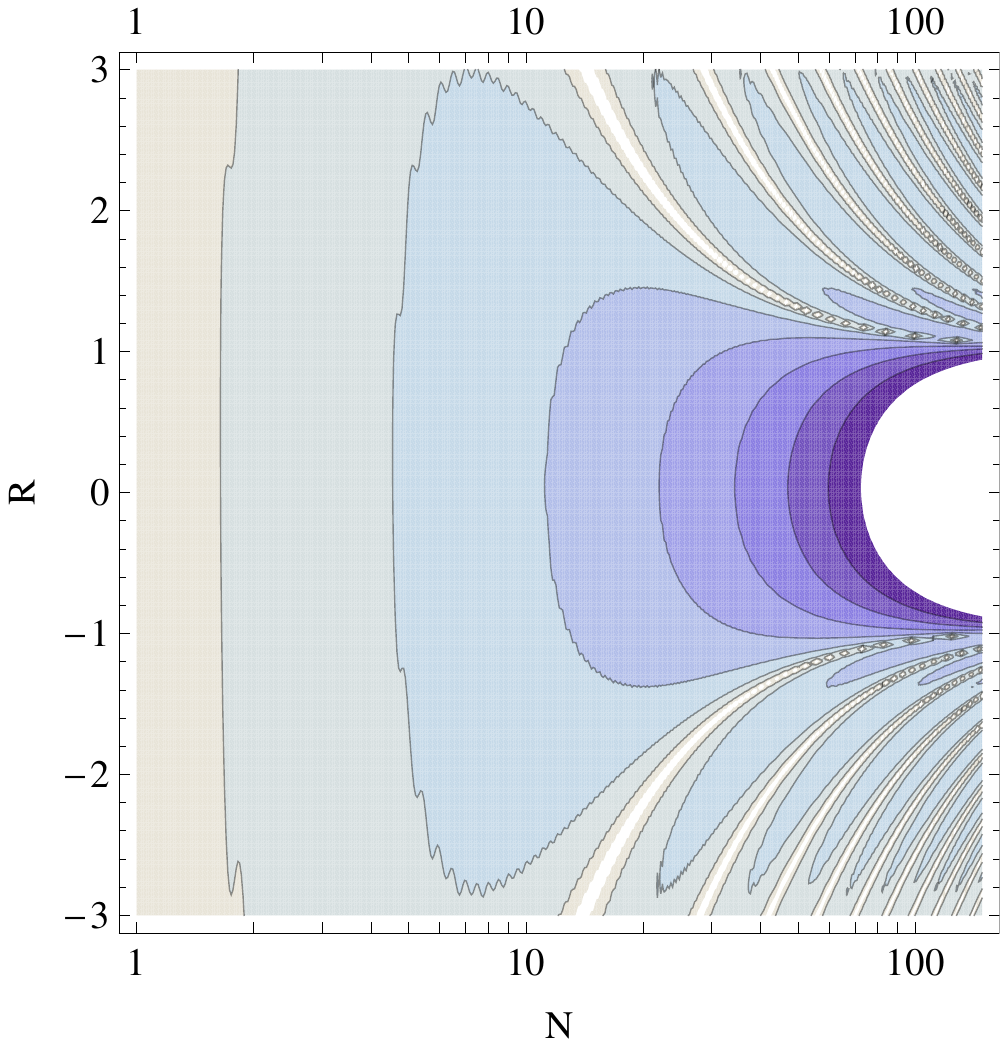}
\caption{Here is shown a contour plot of $\ln \abs{\beta}$ as a function of $N$ and $R$. As in Figs.~\ref{fig:betaofN} and~\ref{fig:betaofres}, the modulation amplitude $A=0.1$. The contour values are $5$ (the maximum shown, at the boundary between dark blue and white), $4$, and decrease by steps of $1$. We clearly see, with increasing $N$, the emergence of an exponentially growing resonant regime for $\abs{R}<1$ and an oscillating nonresonant regime for $\abs{R}>1$. Note also that the contours themselves are not exactly smooth, having jagged edges due to the short-time oscillations of $\abs{\beta}$.}
\label{fig:betaofresN}
\end{SCfigure}

To reveal the various effects of the parametric amplification induced by Eq.~\eqref{modul}, we first numerically study the behavior of the norm of $\abs{\beta}$, whose square gives the final occupation number when working in the $in$ vacuum. In this section, we provide only a qualitative description, but the following observations are all explained in the analytical study presented in \ref{app:analytic}.

In Fig.~\ref{fig:betaofN}, we represent $\abs{\beta}$ as a function of adimensionalized time $N$, for various values of the parameter $R$. Each of the curves is a superposition of a large long-time variation, and a small short-time variation. The former is by far the most significant contribution, and shall be discussed below. The origin of the latter is provided in Eq.~\eqref{eq:betaoffres2term}. We notice that, after an initial linear growth whose rate is independent of $R$, the dependence of $\abs{\beta}$ on $N$ falls into one of two types, depending on the value of $R$:
\begin{enumerate}[i.]
\item For $\abs{R}>1$, $\abs{\beta}$ eventually drops below the linear curve, heading back towards zero and falling into a periodic oscillation. This is the off-resonant regime.
\item For $\abs{R}<1$, $\abs{\beta}$ eventually climbs above the linear curve, tending towards exponential growth. This is the resonant regime, and is due to stimulated amplification of formerly spontaneously created quanta.
\end{enumerate}

We thus get a sense from Fig.~\ref{fig:betaofN} of the importance of the parameter $R$. To investigate this further, in Fig.~\ref{fig:betaofres} we represent $\abs{\beta}$ in logarithmic scale as a function of $R$ for various values of $N$. We observe a central peak around $R=0$ of increasing height and decreasing width, though the width saturates with increasing $N$ such that it extends from $R=-1$ to $1$. This peak corresponds to the resonant regime, and as the evolution of $\abs{\beta}$ becomes exponential for large $N$ we find that the height of the peak in $\log \abs{\beta}$ varies linearly with $N$. For $\abs{R}>1$, we observe oscillations with a fixed maximum value for each value of $R$, such that the maxima trace out an envelope that is a smooth function of $R$. This is the nonresonant regime, where $\abs{\beta}$ never rises above a fixed maximum value.

For completeness, in Fig.~\ref{fig:betaofresN} we combine the above figures in a contour plot, where the contours are lines of constant $\abs{\beta}$ in the $(N,R)$-plane. One can clearly see the emergence of the resonant peak for $\abs{R}<1$ with increasing $N$, as well as the oscillatory behavior for $\abs{R}>1$.

\subsubsection{Analytical properties}
\label{app:analytic}

As can be seen from Figs~\ref{fig:betaofN} and~\ref{fig:betaofres}, the dependence of $\abs{\beta}$ on the parameters $N$, $A$ and $R$ is rather complicated. Yet, the essential features can be obtained analytically, as we now show.

To simplify the following equations, we use the adimensional time $\tau$ and the detuning parameter $r$ given by
\begin{equation}
\label{eq:tau-r}
\begin{split}
\tau \doteq \omega_p t, 
\quad r \doteq \frac{ 2\omega_0 - \omega_p}{\omega_p} = AR/4 .
\end{split}
\end{equation}
We also assume that $A \ll 1$, so the relative modulation of $\omega$ is small. Then Eqs.~\eqref{eq:eomalphabeta} simplify and become
\begin{subequations}
\label{eq:eomalphabetaapprox1}
\begin{align}
\label{eq:eomapproxalpha}
\partial_{\tau } \alpha &\approx \frac{A}{8} \left[ e^{i r \tau } + e^{i(2- r)\tau } \right] \beta ,\\
\label{eq:eomapproxbeta}
\partial_{\tau } \beta &\approx \frac{A}{8} \left[ e^{-i r\tau } + e^{-i(2-r)\tau } \right] \alpha .
\end{align}
\end{subequations}
To solve these equations, two cases will be separately considered: in the first, the modulation is nonresonant so $\beta \ll 1$ for all times; in the second, the modulation is close to resonant so $A R /4 = r \ll 1$.

\subsubsubsection{Non-resonant case}

When $\beta$ is very small, unitarity $\abs{\alpha}^2 =1+ \abs{\beta}^2 $ implies that $\abs{\alpha}$ remains close to $1$. Equation~\eqref{eq:eomapproxalpha} then guarantees that the phase of $\alpha$ is slowly varying in time, so $\partial_t(\beta / \alpha) \sim \partial_t(\beta) / \alpha$. Since we seek only the magnitude $\abs{\beta}$, we shall not consider this phase. We thus have
\begin{equation}
\label{eq:dtbetaoffres}
\begin{split}
\partial_{\tau } \beta &\approx \frac{A}{8} \left[ e^{-i r\tau } + e^{-i(2-r)\tau } \right] .
\end{split}
\end{equation}
This is trivially solved by
\begin{equation}
\label{eq:betaoffres2term}
\begin{split}
\beta(t) &\approx \frac{-A}{8} \left [ \frac{e^{-i \tau r}-1}{ r} + \frac{e^{-i \tau (2+r)}-1}{2 +r} \right ].
\end{split}
\end{equation}
This equation correctly describes two effects that are visible in Figs.~\ref{fig:betaofN} and~\ref{fig:betaofres}: The first term describes long-time variations of large magnitude, while the second describes short-time variations of small magnitude.

\subsubsubsection{Close to resonance}

We now suppose that we are close to resonance so $r \ll 1$. In such a case, a rotating wave approximation can be performed so that we neglect terms oscillating with frequency $2 \omega_0 + \omega_p$. Under such circumstances, the Bogoliubov coefficients are solutions of 
\begin{equation}
\label{eq:eomalphabetaapprox}
\partial_{\tau } \alpha \approx \frac{A}{8} e^{i r \tau } \beta ,\quad
\partial_{\tau } \beta \approx \frac{A}{8} e^{-i r \tau } \alpha .
\end{equation}
Imposing the initial conditions $\alpha=1$ and $\beta=0$ at $t=0$, the exact solutions of these equations are
\begin{subequations}
\label{eq:alphabetasinh}
\begin{align}
\alpha(t) &\sim \ep{i \frac{r}{2} \tau } \left [ \cosh \left ( \frac{A}8 \sqrt{1-R^2}\tau \right ) - i R\frac{\sinh \left ( \frac{A}8 \sqrt{1-R^2}\tau \right )}{\sqrt{1-R^2}} \right ] , \\
\beta(t) &\sim \ep{-i \frac{r}{2} \tau } \frac{\sinh \left ( \frac{A}8 \sqrt{1-R^2}\tau \right )}{\sqrt{1-R^2}}.
 \label{eq:betasinh}
\end{align}
\end{subequations} 
Several comments should be made. First, for low values of $\tau =\omega_p t$, we have $\abs{\beta} \propto A \tau $. This explains the fact that all curves of Fig.~\ref{fig:betaofN} are initially linear with a growth rate that is independent of $R$. 

Second, Eqs.~\eqref{eq:alphabetasinh} reveal the crucial role played by $R$, which did not appear in Eqs.~\eqref{eq:eomapproxalpha}. The value of $R$ delineates the two behaviors that we observed, and characterizes the transition from one to the other occurring at $\abs{R}=1$. When $\abs{R}>1$, the square root is imaginary and $\beta$ oscillates in time, with a maximum given by $\abs{\beta}_{\rm max} \sim 1/\sqrt{R^2-1}$. This is the off-resonant behavior. In addition, the fact that $\abs{\beta}_{\rm max}$ depends only on $R$ explains the envelope traced out with increasing $N$ in Fig.~\ref{fig:betaofres}. In contrast, when $\abs{R} < 1$, $\beta$ grows exponentially, as can be clearly seen in the low-$R$ curves of Fig.~\ref{fig:betaofN}. This is the resonant regime, and the fact that it occurs over a finite range of $R$ explains the finite width of the growing part of the spectrum seen in Figs.~\ref{fig:betaofres} and~\ref{fig:betaofresN}. Indeed, at large times, we find $\ln \abs{\beta} \sim (\pi NA/4) \sqrt{1-R^2} - \ln(2 \sqrt{1-R^2})$. The critical case is $\abs{R}=1$. In this case, under the assumptions we used, $\beta$ grows linearly in time.

Third, in the limit $R \gg 1$, $r = A R/4 \ll 1$, Eq.~\eqref{eq:betasinh} gives
\begin{equation}
\begin{split}
\abs{\beta} \sim \frac{\sin \left [ r \tau /2 \right ]}{ R }, 
\end{split}
\end{equation}
which corresponds to the first term of Eq.~\eqref{eq:betaoffres2term}. There is thus a perfect compatibility of the two descriptions in this intermediate range $1 \ll R \ll 4/A$ where they overlap. 

Finally, we can substitute the expressions of Eq.~\eqref{eq:alphabetasinh} into the right-hand side of Eqs.~\eqref{eq:eomalphabetaapprox1}, yielding improved solutions that are relevant close to resonance and include the rapidly oscillating terms. In fact, iterating this operation gives a perturbative expansion for the solutions of Eqs.~\eqref{eq:eomalphabetaapprox1}, of which Eqs.~\eqref{eq:alphabetasinh} are the lowest-order terms.

\subsection{Dependence on temperature and final entanglement}

\begin{figure*}
\begin{minipage}{0.47\linewidth}
\includegraphics[width=1\linewidth]{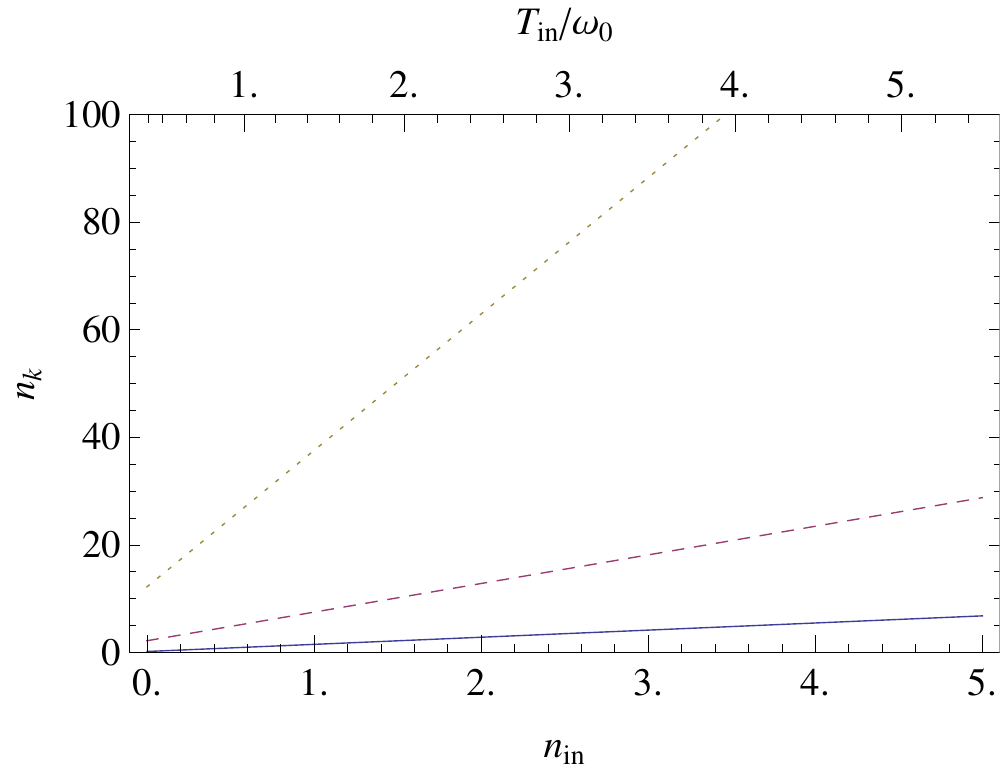}
\end{minipage}
\hspace{0.03\linewidth}
\begin{minipage}{0.47\linewidth}
\includegraphics[width=1\linewidth]{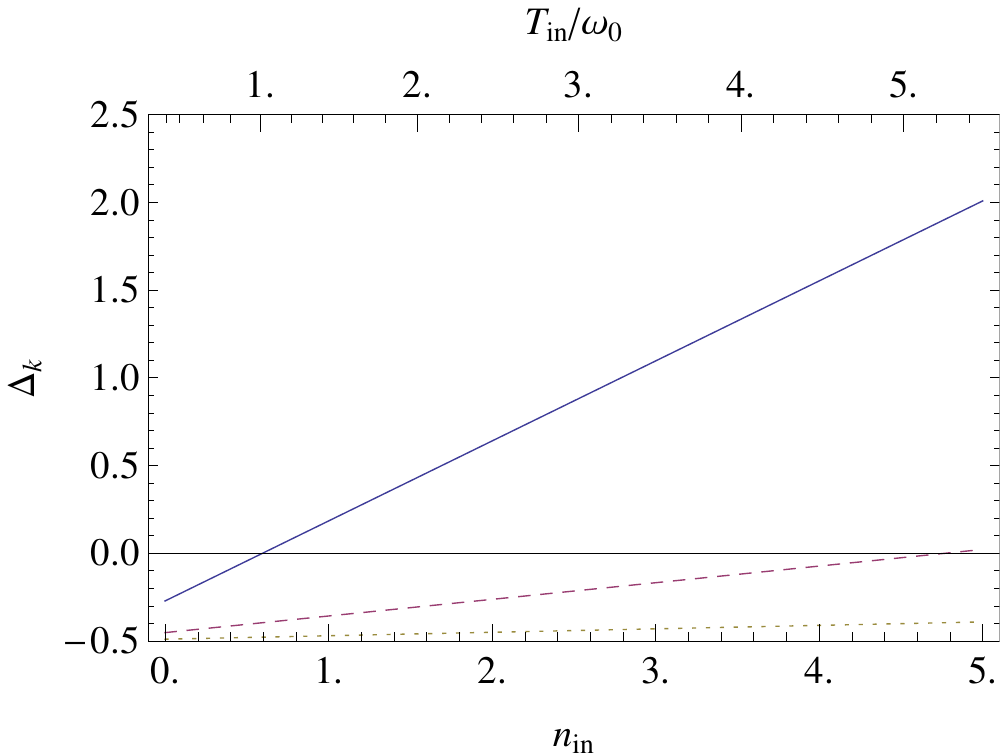}
\end{minipage}
\caption{Here are plotted, for exact resonance $R=0$, the final occupation number $n_{\rm out}$ (left figure) and $\Delta$ (right figure) as functions of the initial occupation number $n_{\rm in}$ (and the initial temperature in units of the mean frequency $\omega_0$) for various values of $N$: $5$ (blue solid), $15$ (dashed purple), and $25$ (dotted yellow). It is clear that both $n_{\rm out}$ and $\Delta$ increase with temperature in an approximately linear fashion. However, whereas increasing $N$ raises both the intercept and slope of the $n_{\rm out}$-curves, it has the opposite effect on the $\Delta$-curves, yielding a nonseparable state over a progressively wider range of initial temperatures.}
\label{fig:NDeltaofT}
\end{figure*}

\begin{SCfigure}[1][htb]
\includegraphics[width=0.5\linewidth]{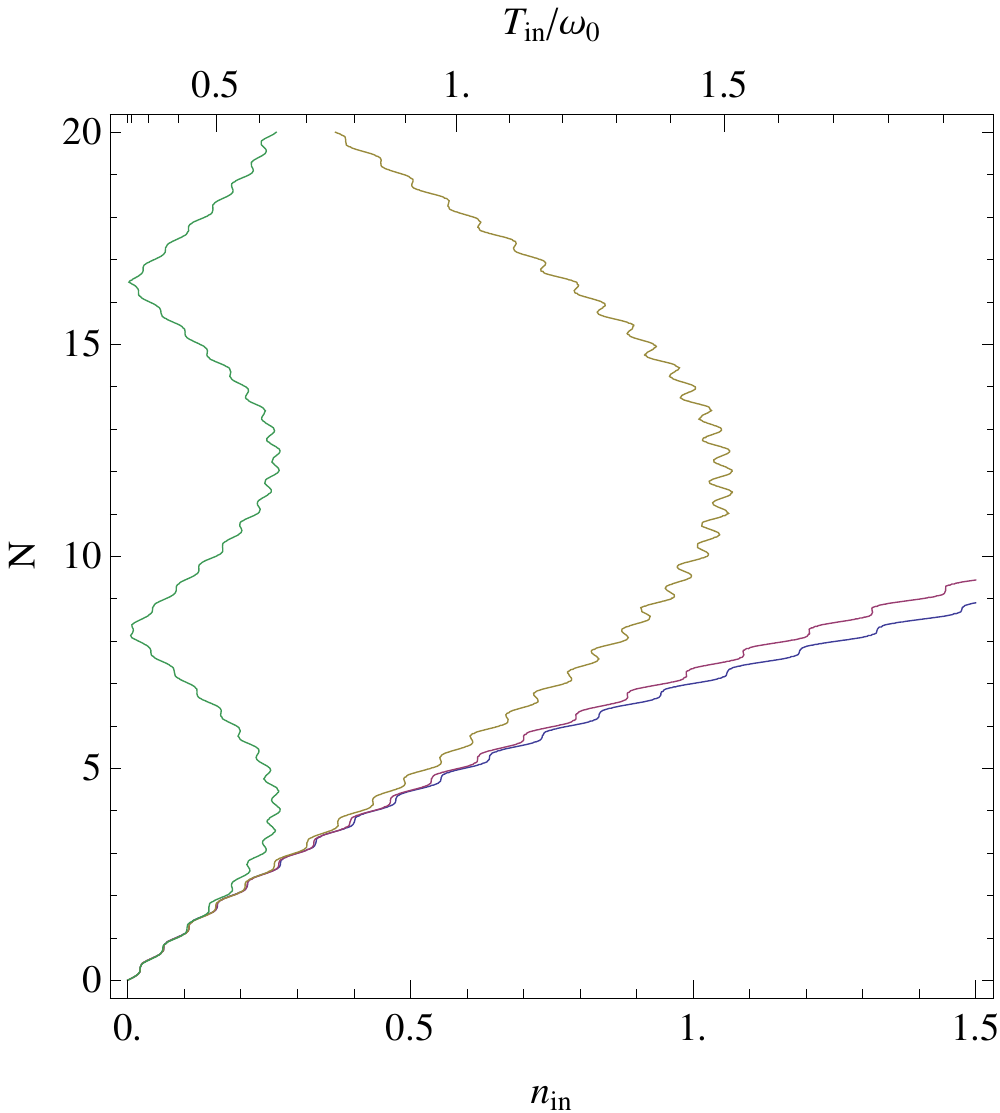}
\caption{Plotted here are loci of the separability threshold $\Delta=0$ in the $(T_{\rm in},N)$-plane for various values of $R$, from right to left: $0$ (blue), $0.99$ (purple), $2$ (yellow), and $5$ (green). As for previous plots, we work with $A=0.1$. Note once again the splitting of the large-$N$ behavior into resonant ($\abs{R}<1$) and nonresonant ($\abs{R}>1$) regimes. For $R<1$, the curves are seen to extend indefinitely towards higher values of $n_{\rm in}$, meaning that it is possible, for any initial temperature, to reach a nonseparable state if $N$ is made large enough. On the other hand, for $R>1$, the curves reach some maximum value of $n_{\rm in}$ and then turn around, so the state can only be made nonseparable for temperatures below some maximum value, and even then the system will oscillate between separable and nonseparable states. Again we notice the lack of smoothness of the curves due to the short-time oscillations of $\abs{\beta}$.}
\label{fig:DeltaofNT}
\end{SCfigure}

We now include the effects of a nonzero initial temperature -- or, equivalently, a non-zero initial occupation number -- on both the final occupation number $n_{\rm out}$ and the separability parameter $\Delta_{\rm out}$.  
To do so, the initial occupation number is supposed to follow the thermal distribution:
\begin{equation}
\label{eq:thermal}
n_{th}(\omega/T) = \frac{1}{e^{\omega/T}-1} .
\end{equation}
 $n_{\rm out}$ and $\Delta_{\rm out}$ are then given by Eqs.~\eqref{eq:outnc} and~\eqref{eq:deltaout}.
In Fig.~\ref{fig:NDeltaofT}, $n_{\rm out}$ and $\Delta_{\rm out}$ are plotted as functions of $n_{\rm in}$ and $T/\omega_0$ for a system exactly at resonance ($R=0$) and for various values of $N$. We observe that $n_{\rm out}$ increases both with initial temperature and with $N$, while $\Delta$ -- which is sensitive to the division of $n_{\rm out}$ into spontaneous and stimulated contributions -- increases with initial temperature but \textit{decreases} with $N$. This is in accordance with Eq.~\eqref{eq:deltaout} since $\abs{\beta}$ increases with the duration $N$.

In Fig.~\ref{fig:DeltaofNT}, we represent the nonseparability threshold in the $(N,T)$-plane -- that is, the locus where $\Delta=0$ -- for various values of $R$. Notice that $\Delta$ is positive to the right of the curves since it always increases with $n_{\rm in}$. In the case of resonance ($\abs{R}<1$), we observe that whatever the initial temperature, $\Delta$ becomes negative and the state becomes nonseparable for $N$ larger than some value. By contrast, in the nonresonant case ($\abs{R}>1$), there exists a temperature above which the state is separable for all values of $N$. This critical temperature depends on $R$ and is generically lower than $\omega_0$. The analytic treatment of \ref{app:analytic} gives $T_{\max} \sim \omega_0/ \ln(R)$.

To conclude this section, we consider the experiment of Ref.~\cite{PhysRevLett.109.220401}, the results of which triggered the present analysis. From the data, we estimate that the peak-to-peak amplitude $A \sim 0.1$ and that the duration $N \sim 50$. (In fact, this is only an upper bound on $A$. It is actually the frequency of the trapping potential that is modulated with this amplitude, and estimating the corresponding amplitude for the mode frequencies is rather nontrivial.) We have not been able to determine with precision the appropriate value for $R$. In principle, if one works exactly at the resonance $R = 0$, the phonon state would be nonseparable for $n_{\rm in} \lesssim 1200 $. Instead, when working with $R = 1$, nonseparability would occur only for $n_{\rm in} \lesssim 40$. These findings seem to overestimate the observed intensity of the correlations. A possible explanation for this is the neglect of weak dissipative effects. To study this possibility, in \ref{sec:dissip} we include weak dissipation while the system is being modulated. We shall see that weak dissipation is sufficient to ruin the nonseparability reached by the system in the absence of dissipation, thereby possibly explaining what was reported in Ref.~\cite{PhysRevLett.109.220401}.

\section{Weak dissipation} 
\label{sec:dissip}

In this section, we introduce a weak dissipative rate $\Gamma$ -- where ``weak'' means $\Gamma/\omega_0 \ll 1$ -- and study its effects on quasiparticle creation and entanglement.

\subsection{General model of weak dissipation}
\label{app:dissip}

In the presence of dissipation, the notion of Bogoliubov coefficients is no longer well defined, as the system is coupled to some environment. As a result, the state of the system can no longer be characterized by $\alpha(t)$ and $\beta(t)$ as in the non-dissipative case. Instead, the mean occupation number $n$ and the correlation $c$ can still be defined for all times when dissipation is weak enough, see \ref{chap:dissipBEC} and \ref{chap:polariton}. In this regime, the separability parameter $\Delta$ remains related to these by Eq.~\eqref{eq:defDeltalinear}. 

We adopt a simple effective approach to dissipation, inspired by the results of \ref{chap:dissipBEC} and \ref{chap:polariton} in which it was incorporated using Hamiltonian models that respect unitarity. In these chapters, only single sudden changes were considered, and it was found that the Bogoliubov coefficients -- which can be \textit{locally} defined in the vicinity of the change when dissipation is weak enough -- respond to the sudden change as if dissipation were not present. In fact, the main effect of weak dissipation observed was the expected exponential damping of the system towards an equilibrium state. 

These observations are here implemented by considering a series of infinitesimal double steps of duration $dt$. In each double step, the system evolves according to two processes:
\begin{itemize}

\item 
a non-dissipative modulation linking $[n(t)$, $c(t)]$ to an intermediate $[\tilde n(t)$, $\tilde c(t)]$ by an infinitesimal Bogoliubov transformation $\delta S$, which can be derived from the local behavior of $\alpha(t)$ and $\beta(t)$ in the absence of dissipation; 

\item 
an exponential damping due to dissipation which carries $[\tilde n(t)$, $\tilde c(t)]$ to $[ n(t+dt)$, $ c(t+dt)]$.

\end{itemize}

The non-dissipative modulation over $dt$ gives rise to a small change in the quantum amplitude operators as in Eq.~\eqref{eq:bogoliubov}:
\begin{equation}
\label{eq:infoperators}
\hat{b}_{\bk}(t+dt) = \delta\alpha \, \hat{b}_{\bk}(t) + \left(\delta\beta\right)^{*} \hat{b}^{\dagger}_{-\bk}(t) ,
\end{equation}
where $\delta\alpha-1$ and $\delta\beta$ are proportional to $dt$. As described above, this is the same infinitesimal transformation that acts locally in the absence of dissipation, so if $S(t)$ is the finite Bogoliubov transformation relating the instantaneous amplitude operators $\hat{b}_{\bk}(t)$ and $\hat{b}^{\dagger}_{-\bk}(t)$ to $\hat{b}^{in}_{\bk}$ and $\left(\hat{b}^{in}_{-\bk}\right)^{\dagger}$ in the non-dissipative case, then $\delta S = S(t+dt) S^{-1}(t)$. In terms of $\alpha(t)$ and $\beta(t)$, this is equivalent to 
\begin{equation}
\label{eq:infbog}
\begin{split}
\delta \alpha &= \alpha(t+dt) \alpha^*(t) - \beta(t+dt)^* \beta(t) \sim 1+ (\dot\alpha \alpha^* - \dot\beta^* \beta)dt  , \\
\delta \beta &= \beta(t+dt) \alpha(t)^* - \alpha(t+dt)^* \beta(t) \sim ( \dot\beta \alpha^* -\dot\alpha^* \beta)dt .
\end{split}
\end{equation}
Moreover, unitarity requires that $\abs{\delta\alpha}^{2}-\abs{\delta\beta}^{2}=1$, so the difference $\abs{\delta\alpha}^{2} - 1$ is second-order in $dt$. Thus we are led to the following equations for the resulting changes in $n$ and $c$, see Eq.~\eqref{eq:outnc}:
\begin{equation}
\label{eq:infmod}
\begin{split}
\tilde n(t) &= \abs{\delta \beta}^2 + \left ( \abs{\delta \beta}^2+ \abs{\delta \alpha}^2 \right ) n(t) + 2 \Re\left (\delta \alpha \delta \beta c(t) \right ) \sim n(t) + 2 \Re\left [\delta \beta c(t) \right ] , \\
\tilde c(t) &= \delta \alpha \delta \beta^{*} + 2 \delta \alpha \delta \beta^{*} n(t) + \delta \alpha^2 c(t) + \left(\delta \beta^{*}\right)^2 c^{*}(t)  \sim \delta \beta^{*}[ 1 + 2 n(t) ] + \delta \alpha^2 c(t) .
\end{split}
\end{equation}
In the second part of the double step, we account for the process of weak dissipation, which is described by
\begin{equation}
\label{eq:infdiss}
\begin{split}
n(t+dt) &= n_{eq}(t) + [\tilde n(t)-n_{eq}(t)] \ep{- 2 \Gamma dt}\sim \tilde n(t) - 2 \Gamma [\tilde n(t)-n_{eq}(t) ] dt , \\
c(t+dt) &= \tilde c(t) \ep{- 2 \Gamma dt}\sim \tilde c(t) - 2 \Gamma \tilde c(t) dt ,
\end{split}
\end{equation}
where $n_{eq}(t)=n_{eq}(\omega_k(t))$ is the mean occupation number when the system is in equilibrium, typically the thermal distribution of Eq.~\eqref{eq:thermal}.
When coupled to such incoherent states, the equilibrium value of the coherence parameter $c_{eq}$ vanishes, as is assumed in Eqs.~\eqref{eq:infdiss}.

To first order in $dt$, Eqs.~\eqref{eq:infbog} to~\eqref{eq:infdiss} combine to give
\begin{equation}
\label{eq:eomncdissip}
\begin{split}
(\partial_t+ 2 \Gamma)n &= 2 \Gamma n_{eq} + 2 \Re \left [ ( \alpha^* \partial_t \beta -\beta \partial_t \alpha^* ) c \right ] , \\
(\partial_t+ 2 \Gamma)c &= ( \alpha \partial_t \beta^* -\beta^* \partial_t \alpha ) (1+2n) + 2 (\alpha^* \partial_t \alpha - \beta \partial_t \beta^*) c .
\end{split}
\end{equation}
These are equivalent to Eqs.~\eqref{eq:eomdissipncexact} when the non-dissipative equations for $\alpha(t)$ and $\beta(t)$, Eqs.~\eqref{eq:eomalphabeta}, are substituted in Eqs.~\eqref{eq:eomncdissip}, and the unitarity condition $\abs{\alpha}^2 -\abs{\beta}^2 =1$ is used. One can check that in the limit $\Gamma \to 0$, Eqs.~\eqref{eq:outnc} satisfy the above equations. One can also check that, when there is no modulation or after the modulation has ended -- i.e., when $\alpha$, $\beta$ and $n_{eq}$ are constant in time -- Eqs.~\eqref{eq:eomncdissip} imply that $n$ and $c$ decay exponentially toward their equilibrium values $n_{eq}$ and $0$. 

Furthermore, Eqs.~\eqref{eq:eomncdissip} yield the following simple equation for the evolution of the effective number $\bar n = n(n+1) - \abs{c}^2$ that fixes the value of the entropy~\cite{Campo:2005sy}:
\begin{equation}
\begin{split}
\partial_t \bar n = 2 \Gamma \left [ 2 \abs{c}^2 + (1+2n)(n_{eq} - n) \right ] .
\end{split}
\end{equation}
The evolution of $\bar n(t)$ determined by this equation governs the entropy exchanges between the system and its environment.

\subsection{Analytic study}

We now use the model built in \ref{app:dissip}, and in particular Eq.~\eqref{eq:eomncdissip} to the modulated evolution of Eq.~\eqref{modul}. Using Eq.~\eqref{eq:eomalphabeta}, Eq.~\eqref{eq:eomncdissip} gives
\begin{equation}
\label{eq:eomdissipncexact}
\begin{split}
(\partial_t+ 2 \Gamma)n &= 2 \Gamma n_{eq} + \frac{\dot u}{v} \Re \left [ \ep{-2 i \int \omega dt} c \right ] , \\
(\partial_t+ 2 \Gamma)c &= \frac{\dot u}{v} \ep{2 i \int \omega dt} (1+2n) .
\end{split}
\end{equation}
Here, $n_{eq}(t) = n_{eq}(\omega_k(t))$, where $n_{eq}(\omega)$ is the mean occupation number at frequency $\omega$ when the system reaches equilibrium in the limit $t \rightarrow \infty$. It is determined by the state of the environment, and is typically given by $n_{eq}(\omega) = n_{th}(\omega/T)$, for given temperature $T$ and where $n_{th}(\omega/T)$ is the thermal distribution of Eq.~\eqref{eq:thermal}.

We work in a regime where the relative modulation of the mode frequency is small, and where it is not too far from resonance; in \ref{app:analytic}, we saw that this regime corresponds to $A\ll 1 $, $AR/4 \ll 1 $.  Thus we can approximate $ \frac{\dot u}{v} \ep{-2 i \int \omega dt} \sim \frac{A}{4} \omega_p \ep{- i AR \omega_p t/4} $, and
if we also average over the rapid oscillations so that these can be neglected, Eqs.~\eqref{eq:eomdissipncexact} become 
\begin{equation}
\label{eq:eomncdissipapprox}
\begin{split}
(\partial_t+ 2 \Gamma)n &= 2 \Gamma n_{eq} + \frac{A}{4} \omega_p \Re \left[ \ep{-i AR \omega_p t/4} c \right] ,\\
(\partial_t+ 2 \Gamma)c &=\frac{\omega_p A}{8} \ep{i AR\omega_p t /4 } (1+2n) . 
\end{split}
\end{equation}
Neglecting the time dependence of $n_{eq}$ and working at exact resonance $R=0$, it can be derived from Eqs.~\eqref{eq:eomncdissipapprox} that, for $\Gamma/\omega_0>A/4$, $n$ grows and saturates at
\begin{equation}
\label{eq:nsaturate}
\begin{split}
n_{max} = \frac{32 \Gamma^2 n_{eq} + A^2 \omega_0^2}{32 \Gamma^2 - 2 A^2 \omega_0^2}.
\end{split}
\end{equation}
On the other hand, for $\Gamma/\omega_0 < A/4$, $n$ grows exponentially, albeit at a slower rate due to dissipation. In both these cases, $\Delta$ decays exponentially towards the limiting value
\begin{equation}
\label{eq:Deltasaturate}
\begin{split}
\Delta_{min} = \frac{8 \Gamma n_{eq} - A \omega_0}{8 \Gamma + 2 A \omega_0}.
\end{split}
\end{equation}
Thus, the state eventually becomes nonseparable if $2n_{eq} \Gamma/\omega_0 < A/4$. Note that this condition for nonseparability is independent of the condition for exponential growth of $n$.  More precisely, if $n_{eq}<1/2$, there exists a regime where $n$ saturates and the final state is nonseparable, whereas if $n_{eq}>1/2$, there exists a regime in which the final state is separable and $n$ grows exponentially. 

\subsection{Numerical analysis}

\begin{figure}[b!]
\includegraphics[width=0.47\linewidth]{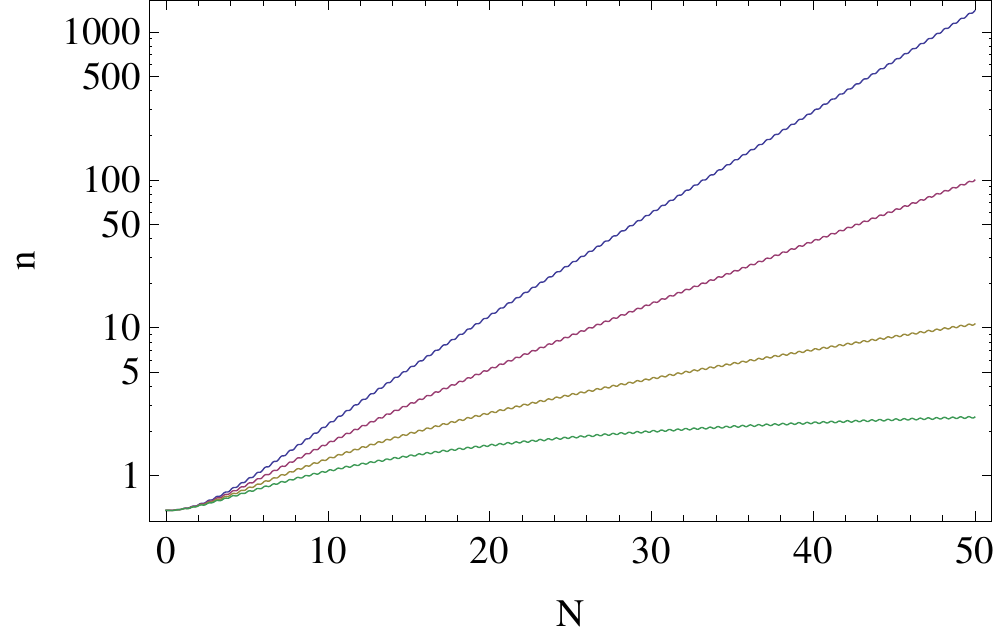}
\hspace{0.03\linewidth}
\includegraphics[width=0.47\linewidth]{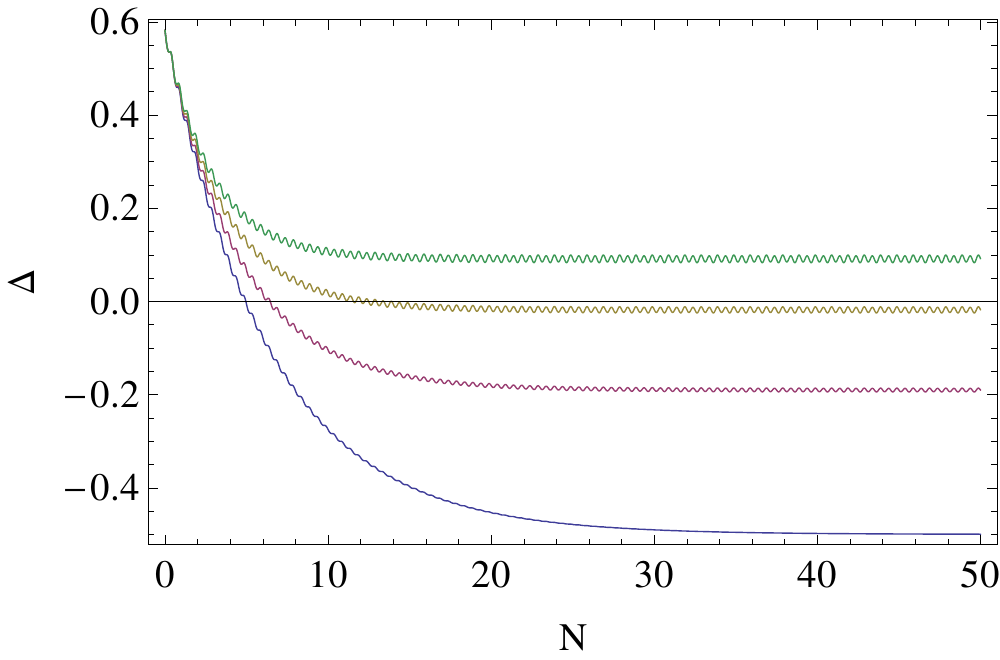}
\caption{Here are plotted, at exact resonance $R=0$ and for an initial thermal bath with $T_{\rm in}=\omega_0$, the mean occupation number (left figure) and the separability parameter $\Delta$ (right figure) immediately after the end of the frequency modulation. The amplitude is as before taken to be $A=0.1$. The various curves correspond to different dissipative rates $\Gamma/\omega_0$ (from larger to smaller $n$ and from lower to higher $\Delta$): $0$ (blue), $0.01$ (purple), $0.02$ (yellow) and $0.03$ (green). The rate of increase of $n$ is seen to be reduced by larger dissipation rates; but, in accordance with the prediction of \ref{sec:dissip}, it approaches exponential growth for $\Gamma/\omega_0 < A/4 = 0.025$ and saturates for $\Gamma/\omega_0>0.025$. Similarly, $\Delta$ approaches a limiting value which increases with the dissipation rate, and as predicted in \ref{sec:dissip} the final state is nonseparable only when $\Gamma/\omega_0 < A/8n_{eq} \approx 0.021$.}
\label{fig:ndissiphighT}
\end{figure}

\begin{figure}[htb]
\includegraphics[width=0.47\linewidth]{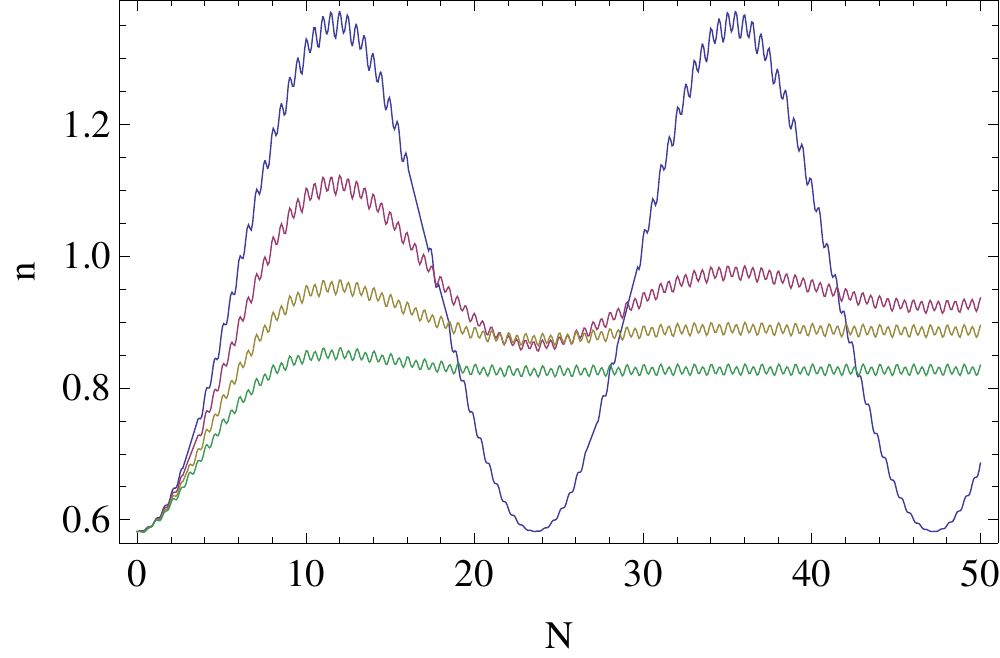}
\hspace{0.03\linewidth}
\includegraphics[width=0.47\linewidth]{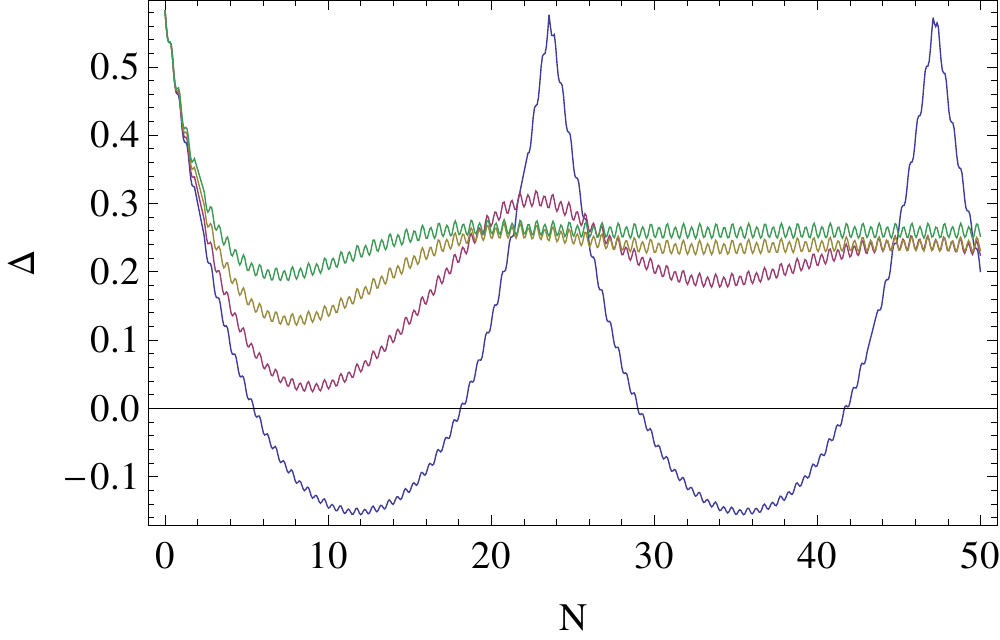}
\caption{Here are shown, at $R=2$ and for an initial thermal bath with $T=\omega_0$, the mean occupation number (left figure) and separability parameter $\Delta$ (right figure) immediately after the end of the frequency modulation. As before, the amplitude $A=0.1$. The various curves correspond to different values of the dissipative rate $\Gamma/\omega_0$, from larger to smaller first oscillation: $0$ (blue), $0.01$ (purple), $0.02$ (yellow) and $0.03$ (green). We note a smoothing out of the oscillations with increasing dissipative rate, and eventually their disappearance, as in overdamped systems. $n$ and $\Delta$ are seen to approach limiting values, which (respectively) decrease and slightly increase with increasing $\Gamma/\omega_0$.}
\label{fig:ndissipoffres}
\end{figure}

\begin{SCfigure}[1][htb]
\includegraphics[width=0.47\linewidth]{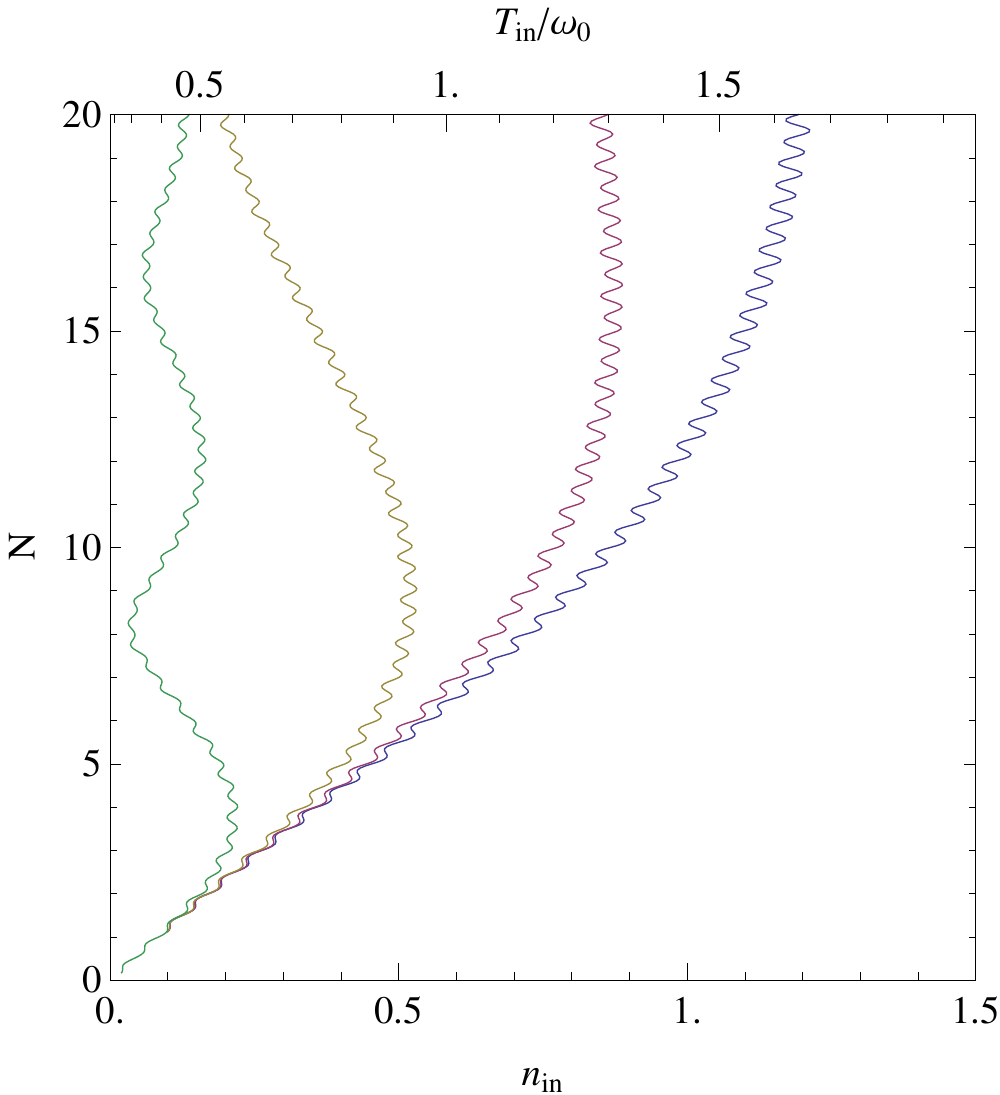}
\caption{Plotted here are loci of the separability threshold $\Delta=0$ in the $(T_{\rm in},N)$-plane for the same parameters as in Fig.~\ref{fig:DeltaofNT}. The value of the dissipative rate is $\Gamma/ \omega_0 =0.01$. The effect of dissipation is here manifest as it introduces a critical temperature above which state always is separable. If one is under this critical temperature, the minimal time needed to reach nonseparability is increased.}
\label{fig:DeltaofNTdiss}
\end{SCfigure}

For the results presented here, we take for initial and equilibrium values of $n$ the thermal distribution of Eq.~\eqref{eq:thermal} and for equation of motion Eq.~\eqref{eq:eomdissipncexact}. In Fig.~\ref{fig:ndissiphighT} we represent, for various dissipative rates $\Gamma/\omega_0$, the time evolution of $n$ and $\Delta$ at exact resonance $R=0$ and with a temperature for the environment $T=\omega_0$. We observe that the mean occupation number decreases with $\Gamma/\omega_0$, the deviation becoming larger with increasing $N$; and that, for large enough dissipative rates, $n$ is seen to saturate at large $N$. Correspondingly, the coherence is reduced (i.e., $\Delta$ increases), and for large enough $\Gamma/\omega_0$ the limiting value of $\Delta$ is positive so that the state never becomes nonseparable. As an example, with the numbers of Ref.~\cite{PhysRevLett.109.220401}, 
namely for $T_{\rm in}/ \omega_0 = 1$ and $N = 50 $, a weak dissipation of $ \Gamma / \omega_0 =2\% $ is almost sufficient to ruin the nonseparability which is found in the absence of dissipation. To be more explicit, in the absence of dissipation, $\Delta =- 0.4995 $, while for $\Gamma / \omega_0 =2\%$, $\Delta = -0.018 $. Nonseparability is lost for $\Gamma / \omega_0 \sim 2.1\%$. 

Slightly off resonance, with $0 < \abs{R} < 1$, the differences with respect to the resonant case are similar to those of the non dissipative case.  See Fig.~\ref{fig:betaofres} for the behavior of the mean occupation number $n_{out}=\abs{\beta}^2$ and Fig.~\ref{fig:DeltaofNT} for the behavior of the separability parameter $\Delta$ for $0<\abs{R}<1$ in the absence of dissipation.  As can be seen there, $n_{out}$ falls with $R$ and $\Delta$ increases with $R$.

In the nonresonant case, with $R>1$, dissipation is observed to dampen the oscillations in $n$ and $\Delta$, both of which approach limiting values that (respectively) decrease and increase with increasing $\Gamma/\omega_0$; see Fig.~\ref{fig:ndissipoffres}. It is even possible to reach an overdamped regime where no oscillations occur. As in the absence of dissipation, on top of this long range behavior some small and rapid oscillations of frequency near $2\omega_p$ occur. These do not decay when the system reaches a near-equilibrium state, as can be verified by considering the near-stationary solution of Eqs.~\eqref{eq:eomncdissipapprox} when the rapid oscillations are taken into account. 

As in the non dissipative case, in Fig.~\ref{fig:DeltaofNTdiss}, we represent the nonseparability threshold in the $(N,T)$-plane for various values of $R$, and for a dissipative rate $\Gamma/ \omega_0 = 1\%$. As in Fig.~\ref{fig:DeltaofNT}, $\Delta$ is positive to the right of the curves since it always increases with $T_{\rm in}$. We observe that regime where nonseparability is reached is much reduced by dissipation. In particular, even in the resonant regime, there exists a maximum temperature above which the final state is always separable.

We conclude this section by applying our results to the experiment described in Ref.~\cite{Lahteenmaki12022013}. We find that the relevant parameters are $n_{eq}=0.0056$, $A \approx 0.048$ and $\Gamma/\omega_0 > 0.009$. (We can only give a lower bound for $\Gamma/\omega_0$ because it is acknowledged that there is additional source of dissipation -- probably two-photon dissipation -- that is not accounted for.) Assuming the experiment is performed very close to resonance, we take $R=0$, so Eqs.~\eqref{eq:nsaturate} and~\eqref{eq:Deltasaturate} are applicable. Since $A/8n_{eq} \approx 1 > \Gamma/\omega_0$, we conclude that $-1/4 \leq \Delta < 0$ and the state is nonseparable. This is in agreement with the results of Ref.~\cite{Lahteenmaki12022013} which reports $\Delta = \left(2^{-0.32}-1\right)/2 \approx -0.1$. The behavior of $n$, however, is more difficult to ascertain since $A/4$ is slightly above the lower bound of $\Gamma/\omega_0$.  We expect that the additional dissipative effects will take $\Gamma/\omega_0$ above $A/4$, so that $n$ should saturate. 

\section*{Conclusions}
\addcontentsline{toc}{section}{Conclusions}

We have considered the spectrum of quasiparticles and their degree of entanglement due to a sinusoidal modulation of the (squared) frequency in a homogeneous quantum system. For definiteness, the system under consideration was taken to be an atomic Bose-Einstein condensate. The modulation was found to be describable by three parameters: its length, its amplitude, and the detuning of its frequency from resonance (at twice the mean mode frequency). The final amount of spontaneous creation, described by the magnitude of the Bogoliubov coefficient $\abs{\beta}$, is found to have a complicated dependence on these three parameters, while the behavior of the separability parameter $\Delta_{\rm out}$ was given in Eq.~\eqref{eq:deltaout}.

A key observation, in accordance with similar results seen in Ref.~\cite{Kofman:1997yn}, is the existence of a finite width of ``resonant'' frequencies. Averaging out the small rapid oscillations that are superimposed on a large long-time behavior, the spontaneous contribution to resonant quasiparticle modes grows exponentially with the duration of the modulation, and for any initial temperature, the final state can be made nonseparable if the modulation lasts long enough. For off-resonant modes, however, the spontaneous contribution rises and falls periodically, never reaching above some maximum value. At the level of entanglement, this has the effect that, for off-resonant modes, there is a temperature above which the final state is always separable, no matter the length of the modulation.

Finally, we evaluated the consequences of weakly dissipative effects. We demonstrated that the nonseparability of the final state can be significantly reduced and even destroyed when these are taken into account. It is thus clear that weak dissipation could play an important role in the experimental attempts to establish nonseparability of the final state. These considerations have been illustrated by considering two recent experiments.

\clearpage
\phantomsection
\addcontentsline{toc}{chapter}{Bibliography}


\bibliographystyle{../../latex/biblio/alpharevtitle}
\bibliography{../../latex/biblio/bibliopubli}

\end{document}

%
%